\documentclass[11pt]{article} 

\usepackage{textcomp, amssymb}  
\usepackage{anysize}            
\usepackage{amsfonts}
\usepackage{amsmath}
\usepackage{dsfont}
\usepackage{MnSymbol}
\usepackage{mathtools}
\usepackage{graphicx}
\usepackage{epsfig}
\usepackage{epstopdf}
\usepackage{lmerge}
\usepackage{tikz}
\usepackage{amsthm}
\usepackage{ifpdf}

\usetikzlibrary{calc,decorations.pathmorphing,shapes,graphs}

\newcounter{sarrow}
\newcommand\xrsquigarrow[1]{%
\stepcounter{sarrow}%
\mathrel{\begin{tikzpicture}[baseline= {( $ (current bounding box.south) + (0,-0.5ex) $ )}]
\node[inner sep=.5ex] (\thesarrow) {$\scriptstyle #1$};
\path[draw,<-,decorate,
  decoration={zigzag,amplitude=0.7pt,segment length=1.2mm,pre=lineto,pre length=4pt}]
    (\thesarrow.south east) -- (\thesarrow.south west);
    \end{tikzpicture}}%
}

\newcounter{sarrow1}

\newcommand\xnrsquigarrow[1]{%
\stepcounter{sarrow1}%
\mathrel{\begin{tikzpicture}[baseline= {( $ (current bounding box.south) + (0,-0.5ex) $ )}]
\node[inner sep=.5ex] (\thesarrow) {$\scriptstyle #1$};
\path[draw,<-,decorate,
  decoration={zigzag,amplitude=0.7pt,segment length=1.2mm,pre=lineto,pre length=4pt}]
    (\thesarrow1.south east) -- (\thesarrow1.south west);
    $\slashedarrowfill@\relbar\relbar/$
    \end{tikzpicture}}%
}

\makeatletter
\def\slashedarrowfill@#1#2#3#4#5{%
  $\m@th\thickmuskip0mu\medmuskip\thickmuskip\thinmuskip\thickmuskip
   \relax#5#1\mkern-7mu%
   \cleaders\hbox{$#5\mkern-2mu#2\mkern-2mu$}\hfill
   \mathclap{#3}\mathclap{#2}%
   \cleaders\hbox{$#5\mkern-2mu#2\mkern-2mu$}\hfill
   \mkern-7mu#4$%
}
\def\rightslashedarrowfillb@{%
  \slashedarrowfill@\relbar\relbar/\rightarrow}
\newcommand\xnrightarrow[2][]{%
  \ext@arrow 0055{\rightslashedarrowfillb@}{#1}{#2}}

\def\rightslashedarrowfille@{%
  \slashedarrowfill@\relbar\relbar/\twoheadrightarrow}
\newcommand\xntworightarrow[2][]{%
  \ext@arrow 0055{\rightslashedarrowfille@}{#1}{#2}}

\def\rightslashedarrowfillg@{%
  \slashedarrowfill@\relbar\relbar{\raisebox{.12em}{}}\twoheadrightarrow}
\newcommand\xtworightarrow[2][]{%
  \ext@arrow 0055{\rightslashedarrowfillg@}{#1}{#2}}

\def\rightslashedarrowfillx@{%
  \slashedarrowfill@\Relbar\Relbar/\rightrightarrows}
\newcommand\xnTworightarrow[2][]{%
  \ext@arrow 0055{\rightslashedarrowfillx@}{#1}{#2}}

\def\rightslashedarrowfilly@{%
  \slashedarrowfill@\Relbar\Relbar{\raisebox{.12em}{}}\rightrightarrows}
\newcommand\xTworightarrow[2][]{%
  \ext@arrow 0055{\rightslashedarrowfilly@}{#1}{#2}}

\pgfdeclareshape{slash underlined}
{
  \inheritsavedanchors[from=rectangle] 
  \inheritanchorborder[from=rectangle]
  \inheritanchor[from=rectangle]{north}
  \inheritanchor[from=rectangle]{north west}
  \inheritanchor[from=rectangle]{north east}
  \inheritanchor[from=rectangle]{center}
  \inheritanchor[from=rectangle]{west}
  \inheritanchor[from=rectangle]{east}
  \inheritanchor[from=rectangle]{mid}
  \inheritanchor[from=rectangle]{mid west}
  \inheritanchor[from=rectangle]{mid east}
  \inheritanchor[from=rectangle]{base}
  \inheritanchor[from=rectangle]{base west}
  \inheritanchor[from=rectangle]{base east}
  \inheritanchor[from=rectangle]{south}
  \inheritanchor[from=rectangle]{south west}
  \inheritanchor[from=rectangle]{south east}
  \inheritanchorborder[from=rectangle]
  \foregroundpath{
    \southwest \pgf@xa=\pgf@x \pgf@ya=\pgf@y
    \northeast \pgf@xb=\pgf@x \pgf@yb=\pgf@y
    \pgf@xc=\pgf@xa
    \advance\pgf@xc by .5\pgf@xb
    \pgf@yc=\pgf@ya
    \advance\pgf@xc by -1.3pt
    \advance\pgf@yc by -1.8pt
    \pgfpathmoveto{\pgfqpoint{\pgf@xc}{\pgf@yc}}
    \advance\pgf@xc by  2.6pt
    \advance\pgf@yc by  3.6pt
    \pgfpathlineto{\pgfqpoint{\pgf@xc}{\pgf@yc}}
    \pgfpathmoveto{\pgfqpoint{\pgf@xa}{\pgf@ya}}
    \pgfpathlineto{\pgfqpoint{\pgf@xb}{\pgf@ya}}
 }
}
\tikzset{nomorepostaction/.code=\let\tikz@postactions\pgfutil@empty}

\newtheorem{theorem}{Theorem}[section]
\newtheorem{definition}[theorem]{Definition}

\begin{document}

\begin{titlepage}
\thispagestyle{empty}

\hrule
\begin{center}
{\bf\LARGE Reversible Quantum Process Algebra}

\vspace{0.7cm}
--- Yong Wang ---

\vspace{2cm}
\begin{figure}[!htbp]
 \centering
 \includegraphics[width=1.0\textwidth]{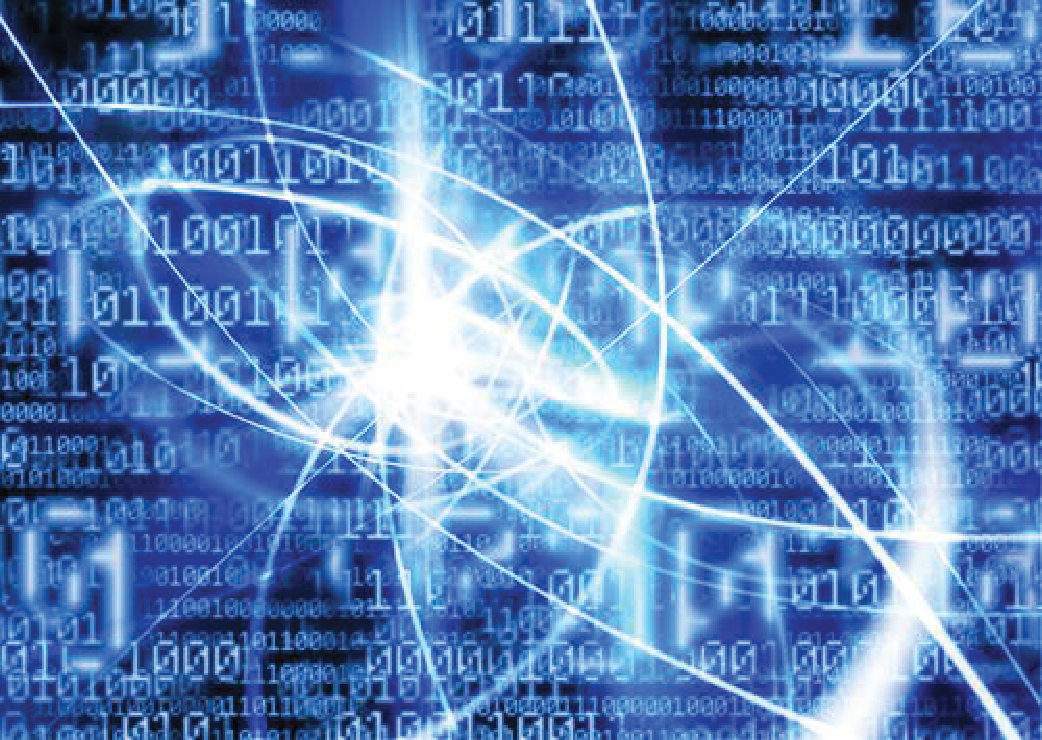}
\end{figure}

\end{center}
\end{titlepage}

\newpage 

\setcounter{page}{1}\pagenumbering{roman}

\tableofcontents

\newpage

\setcounter{page}{1}\pagenumbering{arabic}

        \section{Introduction}

Truly concurrent process algebras are generalizations to the traditional process algebras for true concurrency, CTC \cite{CTC} to CCS \cite{CC} \cite{CCS}, APTC \cite{ATC} to ACP \cite{ACP},
$\pi_{tc}$ \cite{PITC} to $\pi$ calculus \cite{PI1} \cite{PI2}, APPTC \cite{APPTC} to probabilistic process algebra \cite{PPA} \cite{PPA2} \cite{PPA3}. And we also did some work on 
reversible process algebra \cite{APRTC} and probabilistic truly concurrent process algebra \cite{APPTC}.

In quantum process algebras, there are several well-known work \cite{PSQP} \cite{QPA} \cite{QPA2} \cite{CQP} \cite{CQP2} \cite{qCCS} \cite{BQP} \cite{PSQP} \cite{SBQP}, and we ever
did some work \cite{QPA11} \cite{QPA12} \cite{QPA13} to unify quantum and classical computing under the framework of ACP \cite{ACP} and probabilistic process algebra \cite{PPA}.

Now, it is the time to utilize reversible truly concurrent process algebras APRTC \cite{APRTC} and APPTC \cite{APPTC} to model quantum computing and unify quantum and classical computing 
in this book.
This book is organized as follows. In chapter \ref{bg}, we introduce the preliminaries. In chapter \ref{qaprtc} and \ref{aqaprtc}, we introduce the utilization of APRTC to unify quantum
and classical computing and its usage in verification of quantum communication protocols. In chapter \ref{aprptc2}, \ref{qaprptc2} and \ref{aqaprptc}, we introduce APRPTC, and the utilization of APRPTC to unifying
quantum and classical computing and its usage in verification of quantum communication protocols.

\newpage\section{Backgrounds}\label{bg}

To make this book self-satisfied, we introduce some preliminaries in this chapter, including some introductions on operational semantics, proof techniques, reversible truly concurrent process algebra
\cite{ATC} which is based on reversible truly concurrent operational semantics, and also probabilistic truly concurrent process algebra and probabilistic truly concurrent
operational semantics, and also operational semantics for quantum computing.

\subsection{Operational Semantics}\label{OS}

The semantics of $ACP$ is based on bisimulation/rooted branching bisimulation equivalences, and the modularity of $ACP$ relies on the concept of conservative extension, for the
conveniences, we introduce some concepts and conclusions on them.

\begin{definition}[Bisimulation]
A bisimulation relation $R$ is a binary relation on processes such that: (1) if $p R q$ and $p\xrightarrow{a}p'$ then $q\xrightarrow{a}q'$ with $p' R q'$; (2) if $p R q$ and
$q\xrightarrow{a}q'$ then $p\xrightarrow{a}p'$ with $p' R q'$; (3) if $p R q$ and $pP$, then $qP$; (4) if $p R q$ and $qP$, then $pP$. Two processes $p$ and $q$ are bisimilar,
denoted by $p\sim_{HM} q$, if there is a bisimulation relation $R$ such that $p R q$.
\end{definition}

\begin{definition}[Congruence]
Let $\Sigma$ be a signature. An equivalence relation $R$ on $\mathcal{T}(\Sigma)$ is a congruence if for each $f\in\Sigma$, if $s_i R t_i$ for $i\in\{1,\cdots,ar(f)\}$, then
$f(s_1,\cdots,s_{ar(f)}) R f(t_1,\cdots,t_{ar(f)})$.
\end{definition}

\begin{definition}[Branching bisimulation]
A branching bisimulation relation $R$ is a binary relation on the collection of processes such that: (1) if $p R q$ and $p\xrightarrow{a}p'$ then either $a\equiv \tau$ and $p' R q$ or there is a sequence of (zero or more) $\tau$-transitions $q\xrightarrow{\tau}\cdots\xrightarrow{\tau}q_0$ such that $p R q_0$ and $q_0\xrightarrow{a}q'$ with $p' R q'$; (2) if $p R q$ and $q\xrightarrow{a}q'$ then either $a\equiv \tau$ and $p R q'$ or there is a sequence of (zero or more) $\tau$-transitions $p\xrightarrow{\tau}\cdots\xrightarrow{\tau}p_0$ such that $p_0 R q$ and $p_0\xrightarrow{a}p'$ with $p' R q'$; (3) if $p R q$ and $pP$, then there is a sequence of (zero or more) $\tau$-transitions $q\xrightarrow{\tau}\cdots\xrightarrow{\tau}q_0$ such that $p R q_0$ and $q_0P$; (4) if $p R q$ and $qP$, then there is a sequence of (zero or more) $\tau$-transitions $p\xrightarrow{\tau}\cdots\xrightarrow{\tau}p_0$ such that $p_0 R q$ and $p_0P$. Two processes $p$ and $q$ are branching bisimilar, denoted by $p\approx_{bHM} q$, if there is a branching bisimulation relation $R$ such that $p R q$.
\end{definition}

\begin{definition}[Rooted branching bisimulation]
A rooted branching bisimulation relation $R$ is a binary relation on processes such that: (1) if $p R q$ and $p\xrightarrow{a}p'$ then $q\xrightarrow{a}q'$ with $p'\approx_{bHM} q'$;
(2) if $p R q$ and $q\xrightarrow{a}q'$ then $p\xrightarrow{a}p'$ with $p'\approx_{bHM} q'$; (3) if $p R q$ and $pP$, then $qP$; (4) if $p R q$ and $qP$, then $pP$. Two processes $p$ and $q$ are rooted branching bisimilar, denoted by $p\approx_{rbHM} q$, if there is a rooted branching bisimulation relation $R$ such that $p R q$.
\end{definition}

\begin{definition}[Conservative extension]
Let $T_0$ and $T_1$ be TSSs (transition system specifications) over signatures $\Sigma_0$ and $\Sigma_1$, respectively. The TSS $T_0\oplus T_1$ is a conservative extension of $T_0$ if
the LTSs (labeled transition systems) generated by $T_0$ and $T_0\oplus T_1$ contain exactly the same transitions $t\xrightarrow{a}t'$ and $tP$ with $t\in \mathcal{T}(\Sigma_0)$.
\end{definition}

\begin{definition}[Source-dependency]
The source-dependent variables in a transition rule of $\rho$ are defined inductively as follows: (1) all variables in the source of $\rho$ are source-dependent; (2) if
$t\xrightarrow{a}t'$ is a premise of $\rho$ and all variables in $t$ are source-dependent, then all variables in $t'$ are source-dependent. A transition rule is source-dependent if
all its variables are. A TSS is source-dependent if all its rules are.
\end{definition}

\begin{definition}[Freshness]
Let $T_0$ and $T_1$ be TSSs over signatures $\Sigma_0$ and $\Sigma_1$, respectively. A term in $\mathbb{T}(T_0\oplus T_1)$ is said to be fresh if it contains a function symbol from
$\Sigma_1\setminus\Sigma_0$. Similarly, a transition label or predicate symbol in $T_1$ is fresh if it does not occur in $T_0$.
\end{definition}

\begin{theorem}[Conservative extension]\label{TCE}
Let $T_0$ and $T_1$ be TSSs over signatures $\Sigma_0$ and $\Sigma_1$, respectively, where $T_0$ and $T_0\oplus T_1$ are positive after reduction. Under the following conditions,
$T_0\oplus T_1$ is a conservative extension of $T_0$. (1) $T_0$ is source-dependent. (2) For each $\rho\in T_1$, either the source of $\rho$ is fresh, or $\rho$ has a premise of the
form $t\xrightarrow{a}t'$ or $tP$, where $t\in \mathbb{T}(\Sigma_0)$, all variables in $t$ occur in the source of $\rho$ and $t'$, $a$ or $P$ is fresh.
\end{theorem}

\subsection{Proof Techniques}\label{PT}

In this subsection, we introduce the concepts and conclusions about elimination, which is very important in the proof of completeness theorem.

\begin{definition}[Elimination property]
Let a process algebra with a defined set of basic terms as a subset of the set of closed terms over the process algebra. Then the process algebra has the elimination to basic terms
property if for every closed term $s$ of the algebra, there exists a basic term $t$ of the algebra such that the algebra$\vdash s=t$.
\end{definition}

\begin{definition}[Strongly normalizing]
A term $s_0$ is called strongly normalizing if does not an infinite series of reductions beginning in $s_0$.
\end{definition}

\begin{definition}
We write $s>_{lpo} t$ if $s\rightarrow^+ t$ where $\rightarrow^+$ is the transitive closure of the reduction relation defined by the transition rules of an algebra.
\end{definition}

\begin{theorem}[Strong normalization]\label{SN}
Let a term rewriting system (TRS) with finitely many rewriting rules and let $>$ be a well-founded ordering on the signature of the corresponding algebra. If $s>_{lpo} t$ for each
rewriting rule $s\rightarrow t$ in the TRS, then the term rewriting system is strongly normalizing.
\end{theorem}

\subsection{Reversible Truly Concurrent Process Algebra -- APRTC}

\begin{definition}[Prime event structure with silent event]
Let $\Lambda$ be a fixed set of labels, ranged over $a,b,c,\cdots$ and $\tau$. A ($\Lambda$-labelled) prime event structure with silent event $\tau$ is a tuple
$\mathcal{E}=\langle \mathbb{E}, \leq, \sharp, \lambda\rangle$, where $\mathbb{E}$ is a denumerable set of events, including the silent event $\tau$. Let
$\hat{\mathbb{E}}=\mathbb{E}\backslash\{\tau\}$, exactly excluding $\tau$, it is obvious that $\hat{\tau^*}=\epsilon$, where $\epsilon$ is the empty event. Let
$\lambda:\mathbb{E}\rightarrow\Lambda$ be a labelling function and let $\lambda(\tau)=\tau$. And $\leq$, $\sharp$ are binary relations on $\mathbb{E}$, called causality and conflict
respectively, such that:

\begin{enumerate}
  \item $\leq$ is a partial order and $\lceil e \rceil = \{e'\in \mathbb{E}|e'\leq e\}$ is finite for all $e\in \mathbb{E}$. It is easy to see that
  $e\leq\tau^*\leq e'=e\leq\tau\leq\cdots\leq\tau\leq e'$, then $e\leq e'$.
  \item $\sharp$ is irreflexive, symmetric and hereditary with respect to $\leq$, that is, for all $e,e',e''\in \mathbb{E}$, if $e\sharp e'\leq e''$, then $e\sharp e''$.
\end{enumerate}

Then, the concepts of consistency and concurrency can be drawn from the above definition:

\begin{enumerate}
  \item $e,e'\in \mathbb{E}$ are consistent, denoted as $e\frown e'$, if $\neg(e\sharp e')$. A subset $X\subseteq \mathbb{E}$ is called consistent, if $e\frown e'$ for all
  $e,e'\in X$.
  \item $e,e'\in \mathbb{E}$ are concurrent, denoted as $e\parallel e'$, if $\neg(e\leq e')$, $\neg(e'\leq e)$, and $\neg(e\sharp e')$.
\end{enumerate}
\end{definition}

\begin{definition}[Configuration]
Let $\mathcal{E}$ be a PES. A (finite) configuration in $\mathcal{E}$ is a (finite) consistent subset of events $C\subseteq \mathcal{E}$, closed with respect to causality
(i.e. $\lceil C\rceil=C$). The set of finite configurations of $\mathcal{E}$ is denoted by $\mathcal{C}(\mathcal{E})$. We let $\hat{C}=C\backslash\{\tau\}$.
\end{definition}

A consistent subset of $X\subseteq \mathbb{E}$ of events can be seen as a pomset. Given $X, Y\subseteq \mathbb{E}$, $\hat{X}\sim \hat{Y}$ if $\hat{X}$ and $\hat{Y}$ are isomorphic as
pomsets. In the following of the paper, we say $C_1\sim C_2$, we mean $\hat{C_1}\sim\hat{C_2}$.

\begin{definition}[Forward-reverse (FR) pomset transitions and forward-reverse (FR) step]
Let $\mathcal{E}$ be a PES and let $C\in\mathcal{C}(\mathcal{E})$, $\emptyset\neq X\subseteq \mathbb{E}$, $\mathcal{K}\subseteq \mathbb{N}$, and $X[\mathcal{K}]$ denotes that for each
$e\in X$, there is $e[m]\in X[\mathcal{K}]$ where $(m\in\mathcal{K})$, which is called the past of $e$, and we extend $\mathbb{E}$ to $\mathbb{E}\cup\{\tau\}\cup\mathbb{E}[\mathcal{K}]$.
If $C\cap X[\mathcal{K}]=\emptyset$ and $C'=C\cup X[\mathcal{K}], X\in\mathcal{C}(\mathcal{E})$, then $C\xrightarrow{X} C'$ is called a forward pomset transition from $C$ to $C'$, and
$C'\xtworightarrow{X[\mathcal{K}]} C$ is called a reverse pomset transition from $C'$ to $C$. When the events in $X$ are pairwise concurrent, we say that $C\xrightarrow{X}C'$ is a
forward step and $C'\xtworightarrow{X[\mathcal{K}]} C$ is a reverse step.
\end{definition}

\begin{definition}[Weak forward-reverse (FR) pomset transitions and weak forward-reverse (FR) step]
Let $\mathcal{E}$ be a PES and let $C\in\mathcal{C}(\mathcal{E})$, and $\emptyset\neq X\subseteq \hat{\mathbb{E}}$, $\mathcal{K}\subseteq \mathbb{N}$, and $X[\mathcal{K}]$ denotes that
for each $e\in X$, there is $e[m]\in X[\mathcal{K}]$ where $(m\in\mathcal{K})$, which is called the past of $e$. If $C\cap X[\mathcal{K}]=\emptyset$ and
$\hat{C'}=\hat{C}\cup X[\mathcal{K}], X\in\mathcal{C}(\mathcal{E})$, then $C\xRightarrow{X} C'$ is called a weak forward pomset transition from $C$ to $C'$, where we define
$\xRightarrow{e}\triangleq\xrightarrow{\tau^*}\xrightarrow{e}\xrightarrow{\tau^*}$ and $\xRightarrow{X}\triangleq\xrightarrow{\tau^*}\xrightarrow{e}\xrightarrow{\tau^*}$, for every
$e\in X$. And $C'\xTworightarrow{X[\mathcal{K}]} C$ is called a weak reverse pomset transition from $C'$ to $C$, where we define
$\xTworightarrow{e[m]}\triangleq\xtworightarrow{\tau^*}\xtworightarrow{e[m]}\xtworightarrow{\tau^*}$,
$\xTworightarrow{X[\mathcal{K}]}\triangleq\xtworightarrow{\tau^*}\xtworightarrow{e[m]} \xtworightarrow{\tau^*}$, for every $e\in X$ and $m\in\mathcal{K}$. When the events in $X$ are
pairwise concurrent, we say that $C\xRightarrow{X}C'$ is a weak forward step and $C'\xTworightarrow{X[\mathcal{K}]} C$ is a weak reverse step.
\end{definition}

We will also suppose that all the PESs in this book are image finite, that is, for any PES $\mathcal{E}$ and $C\in \mathcal{C}(\mathcal{E})$, and $a\in \Lambda$,
$\{e\in \mathbb{E}|C\xrightarrow{e} C'\wedge \lambda(e)=a\}$ and $\{e\in\hat{\mathbb{E}}|C\xRightarrow{e} C'\wedge \lambda(e)=a\}$, and $a\in \Lambda$,
$\{e\in \mathbb{E}|C'\xtworightarrow{e[m]} C\wedge \lambda(e)=a\}$ and $\{e\in\hat{\mathbb{E}}|C'\xTworightarrow{e[m]} C\wedge \lambda(e)=a\}$ are finite.

\begin{definition}[Forward-reverse (FR) pomset, step bisimulation]
Let $\mathcal{E}_1$, $\mathcal{E}_2$ be PESs. An FR pomset bisimulation is a relation $R\subseteq\mathcal{C}(\mathcal{E}_1)\times\mathcal{C}(\mathcal{E}_2)$, such that (1)
if $(C_1,C_2)\in R$, and $C_1\xrightarrow{X_1}C_1'$ then $C_2\xrightarrow{X_2}C_2'$, with $X_1\subseteq \mathbb{E}_1$, $X_2\subseteq \mathbb{E}_2$, $X_1\sim X_2$ and $(C_1',C_2')\in R$,
and vice-versa; (2) if $(C_1',C_2')\in R$, and $C_1'\xtworightarrow{X_1[\mathcal{K}_1]}C_1$ then $C_2'\xtworightarrow{X_2[\mathcal{K}_2]}C_2$, with $X_1\subseteq \mathbb{E}_1$,
$X_2\subseteq \mathbb{E}_2$, $\mathcal{K}_1,\mathcal{K}_2\subseteq\mathbb{N}$, $X_1\sim X_2$ and $(C_1,C_2)\in R$, and vice-versa. We say that $\mathcal{E}_1$, $\mathcal{E}_2$ are FR
pomset bisimilar, written $\mathcal{E}_1\sim_p^{fr}\mathcal{E}_2$, if there exists an FR pomset bisimulation $R$, such that $(\emptyset,\emptyset)\in R$. By replacing FR pomset
transitions with FR steps, we can get the definition of FR step bisimulation. When PESs $\mathcal{E}_1$ and $\mathcal{E}_2$ are FR step bisimilar, we write
$\mathcal{E}_1\sim_s^{fr}\mathcal{E}_2$.
\end{definition}

\begin{definition}[Weak forward-reverse (FR) pomset, step bisimulation]
Let $\mathcal{E}_1$, $\mathcal{E}_2$ be PESs. A weak FR pomset bisimulation is a relation $R\subseteq\mathcal{C}(\mathcal{E}_1)\times\mathcal{C}(\mathcal{E}_2)$, such that (1) if
$(C_1,C_2)\in R$, and $C_1\xRightarrow{X_1}C_1'$ then $C_2\xRightarrow{X_2}C_2'$, with $X_1\subseteq \hat{\mathbb{E}_1}$, $X_2\subseteq \hat{\mathbb{E}_2}$, $X_1\sim X_2$ and
$(C_1',C_2')\in R$, and vice-versa; (2) if $(C_1',C_2')\in R$, and $C_1'\xTworightarrow{X_1[\mathcal{K}_1]}C_1$ then $C_2'\xTworightarrow{X_2[\mathcal{K}_2]}C_2$, with
$X_1\subseteq \hat{\mathbb{E}_1}$, $X_2\subseteq \hat{\mathbb{E}_2}$, $\mathcal{K}_1,\mathcal{K}_2\subseteq\mathbb{N}$, $X_1\sim X_2$ and $(C_1,C_2)\in R$, and vice-versa. We say that
$\mathcal{E}_1$, $\mathcal{E}_2$ are weak FR pomset bisimilar, written $\mathcal{E}_1\approx_p^{fr}\mathcal{E}_2$, if there exists a weak FR pomset bisimulation $R$, such that
$(\emptyset,\emptyset)\in R$. By replacing weak FR pomset transitions with weak FR steps, we can get the definition of weak FR step bisimulation. When PESs $\mathcal{E}_1$ and
$\mathcal{E}_2$ are weak FR step bisimilar, we write $\mathcal{E}_1\approx_s^{fr}\mathcal{E}_2$.
\end{definition}

\begin{definition}[Forward-reverse (FR) (hereditary) history-preserving bisimulation]
An FR history-preserving (hp-) bisimulation is a posetal relation $R\subseteq\mathcal{C}(\mathcal{E}_1)\overline{\times}\mathcal{C}(\mathcal{E}_2)$ such that (1) if $(C_1,f,C_2)\in R$,
and $C_1\xrightarrow{e_1} C_1'$, then $C_2\xrightarrow{e_2} C_2'$, with $(C_1',f[e_1\mapsto e_2],C_2')\in R$, and vice-versa, (2) if $(C_1',f',C_2')\in R$, and
$C_1'\xtworightarrow{e_1[m]} C_1$, then $C_2'\xtworightarrow{e_2[n]} C_2$, with $(C_1,f'[e_1[m]\mapsto e_2[n]],C_2)\in R$, and vice-versa. $\mathcal{E}_1,\mathcal{E}_2$ are FR
history-preserving (hp-) bisimilar and are written $\mathcal{E}_1\sim_{hp}^{fr}\mathcal{E}_2$ if there exists an FR hp-bisimulation $R$ such that $(\emptyset,\emptyset,\emptyset)\in R$.

An FR hereditary history-preserving (hhp-)bisimulation is a downward closed FR hp-bisimulation. $\mathcal{E}_1,\mathcal{E}_2$ are FR hereditary history-preserving (hhp-)bisimilar and
are written $\mathcal{E}_1\sim_{hhp}^{fr}\mathcal{E}_2$.
\end{definition}

\begin{definition}[Weak forward-reverse (FR) (hereditary) history-preserving bisimulation]
A weak FR history-preserving (hp-) bisimulation is a weakly posetal relation $R\subseteq\mathcal{C}(\mathcal{E}_1)\overline{\times}\mathcal{C}(\mathcal{E}_2)$ such that (1) if
$(C_1,f,C_2)\in R$, and $C_1\xRightarrow{e_1} C_1'$, then $C_2\xRightarrow{e_2} C_2'$, with $(C_1',f[e_1\mapsto e_2],C_2')\in R$, and vice-versa, (2) if $(C_1',f',C_2')\in R$, and
$C_1'\xTworightarrow{e_1[m]} C_1$, then $C_2'\xTworightarrow{e_2[n]} C_2$, with $(C_1,f'[e_1[m]\mapsto e_2[n]],C_2)\in R$, and vice-versa. $\mathcal{E}_1,\mathcal{E}_2$ are weak FR
history-preserving (hp-) bisimilar and are written $\mathcal{E}_1\approx_{hp}^{fr}\mathcal{E}_2$ if there exists a weak FR hp-bisimulation $R$ such that
$(\emptyset,\emptyset,\emptyset)\in R$.

A weak FR hereditary history-preserving (hhp-) bisimulation is a downward closed weak FR hp-bisimulation. $\mathcal{E}_1,\mathcal{E}_2$ are weak FR hereditary history-preserving (hhp-)
bisimilar and are written $\mathcal{E}_1\approx_{hhp}^{fr}\mathcal{E}_2$.
\end{definition}

\begin{definition}[Forward-reverse branching pomset, step bisimulation]
Assume a special termination predicate $\downarrow$, and let $\surd$ represent a state with $\surd\downarrow$. Let $\mathcal{E}_1$, $\mathcal{E}_2$ be PESs. A FR branching pomset
bisimulation is a relation $R\subseteq\mathcal{C}(\mathcal{E}_1)\times\mathcal{C}(\mathcal{E}_2)$, such that:
 \begin{enumerate}
   \item if $(C_1,C_2)\in R$, and $C_1\xrightarrow{X}C_1'$ then
   \begin{itemize}
     \item either $X\equiv \tau^*$, and $(C_1',C_2)\in R$;
     \item or there is a sequence of (zero or more) $\tau$-transitions $C_2\xrightarrow{\tau^*} C_2^0$, such that $(C_1,C_2^0)\in R$ and $C_2^0\xRightarrow{X}C_2'$ with
     $(C_1',C_2')\in R$;
   \end{itemize}
   \item if $(C_1,C_2)\in R$, and $C_2\xrightarrow{X}C_2'$ then
   \begin{itemize}
     \item either $X\equiv \tau^*$, and $(C_1,C_2')\in R$;
     \item or there is a sequence of (zero or more) $\tau$-transitions $C_1\xrightarrow{\tau^*} C_1^0$, such that $(C_1^0,C_2)\in R$ and $C_1^0\xRightarrow{X}C_1'$ with
     $(C_1',C_2')\in R$;
   \end{itemize}
   \item if $(C_1,C_2)\in R$ and $C_1\downarrow$, then there is a sequence of (zero or more) $\tau$-transitions $C_2\xrightarrow{\tau^*}C_2^0$ such that $(C_1,C_2^0)\in R$ and
   $C_2^0\downarrow$;
   \item if $(C_1,C_2)\in R$ and $C_2\downarrow$, then there is a sequence of (zero or more) $\tau$-transitions $C_1\xrightarrow{\tau^*}C_1^0$ such that $(C_1^0,C_2)\in R$ and
   $C_1^0\downarrow$;
   \item if $(C_1',C_2')\in R$, and $C_1'\xtworightarrow{X[\mathcal{K}]}C_1$ then
   \begin{itemize}
     \item either $X[\mathcal{K}]\equiv \tau^*$, and $(C_1,C_2')\in R$;
     \item or there is a sequence of (zero or more) $\tau$-transitions $C_2'\xtworightarrow{\tau^*} C_2'^0$, such that $(C_1',C_2'^0)\in R$ and
     $C_2'^0\xTworightarrow{X[\mathcal{K}]}C_2$ with $(C_1,C_2)\in R$;
   \end{itemize}
   \item if $(C_1',C_2')\in R$, and $C_2'\xtworightarrow{X}C_2$ then
   \begin{itemize}
     \item either $X[\mathcal{K}]\equiv \tau^*$, and $(C_1',C_2)\in R$;
     \item or there is a sequence of (zero or more) $\tau$-transitions $C_1'\xtworightarrow{\tau^*} C_1'^0$, such that $(C_1'^0,C_2')\in R$ and
     $C_1'^0\xTworightarrow{X[\mathcal{K}]}C_1$ with $(C_1,C_2)\in R$;
   \end{itemize}
   \item if $(C_1',C_2')\in R$ and $C_1'\downarrow$, then there is a sequence of (zero or more) $\tau$-transitions $C_2'\xtworightarrow{\tau^*}C_2'^0$ such that $(C_1',C_2'^0)\in R$
   and $C_2'^0\downarrow$;
   \item if $(C_1',C_2')\in R$ and $C_2'\downarrow$, then there is a sequence of (zero or more) $\tau$-transitions $C_1'\xtworightarrow{\tau^*}C_1'^0$ such that $(C_1'^0,C_2')\in R$
   and $C_1'^0\downarrow$.
 \end{enumerate}

We say that $\mathcal{E}_1$, $\mathcal{E}_2$ are FR branching pomset bisimilar, written $\mathcal{E}_1\approx_{bp}^{fr}\mathcal{E}_2$, if there exists a FR branching pomset
bisimulation $R$, such that $(\emptyset,\emptyset)\in R$.

By replacing FR pomset transitions with FR steps, we can get the definition of FR branching step bisimulation. When PESs $\mathcal{E}_1$ and $\mathcal{E}_2$ are FR branching step
bisimilar, we write $\mathcal{E}_1\approx_{bs}^{fr}\mathcal{E}_2$.
\end{definition}

\begin{definition}[Forward-reverse (FR) rooted branching pomset, step bisimulation]
Assume a special termination predicate $\downarrow$, and let $\surd$ represent a state with $\surd\downarrow$. Let $\mathcal{E}_1$, $\mathcal{E}_2$ be PESs. A rooted branching FR pomset
bisimulation is a relation $R\subseteq\mathcal{C}(\mathcal{E}_1)\times\mathcal{C}(\mathcal{E}_2)$, such that:
 \begin{enumerate}
   \item if $(C_1,C_2)\in R$, and $C_1\xrightarrow{X}C_1'$ then $C_2\xrightarrow{X}C_2'$ with $C_1'\approx_{bp}C_2'$;
   \item if $(C_1,C_2)\in R$, and $C_2\xrightarrow{X}C_2'$ then $C_1\xrightarrow{X}C_1'$ with $C_1'\approx_{bp}C_2'$;
   \item if $(C_1',C_2')\in R$, and $C_1'\xtworightarrow{X[\mathcal{K}]}C_1$ then $C_2'\xtworightarrow{X[\mathcal{K}]}C_2$ with $C_1\approx_{bp}^{fr}C_2$;
   \item if $(C_1',C_2')\in R$, and $C_2'\xtworightarrow{X[\mathcal{K}]}C_2$ then $C_1'\xtworightarrow{X[\mathcal{K}]}C_1$ with $C_1\approx_{bp}^{fr}C_2$;
   \item if $(C_1,C_2)\in R$ and $C_1\downarrow$, then $C_2\downarrow$;
   \item if $(C_1,C_2)\in R$ and $C_2\downarrow$, then $C_1\downarrow$.
 \end{enumerate}

We say that $\mathcal{E}_1$, $\mathcal{E}_2$ are FR rooted branching pomset bisimilar, written $\mathcal{E}_1\approx_{rbp}^{fr}\mathcal{E}_2$, if there exists a FR rooted branching
pomset bisimulation $R$, such that $(\emptyset,\emptyset)\in R$.

By replacing FR pomset transitions with FR steps, we can get the definition of FR rooted branching step bisimulation. When PESs $\mathcal{E}_1$ and $\mathcal{E}_2$ are FR rooted branching
step bisimilar, we write $\mathcal{E}_1\approx_{rbs}^{fr}\mathcal{E}_2$.
\end{definition}

\begin{definition}[Forward-reverse (FR) branching (hereditary) history-preserving bisimulation]
Assume a special termination predicate $\downarrow$, and let $\surd$ represent a state with $\surd\downarrow$. A FR branching history-preserving (hp-) bisimulation is a weakly posetal
relation $R\subseteq\mathcal{C}(\mathcal{E}_1)\overline{\times}\mathcal{C}(\mathcal{E}_2)$ such that:

 \begin{enumerate}
   \item if $(C_1,f,C_2)\in R$, and $C_1\xrightarrow{e_1}C_1'$ then
   \begin{itemize}
     \item either $e_1\equiv \tau$, and $(C_1',f[e_1\mapsto \tau^{e_1}],C_2)\in R$;
     \item or there is a sequence of (zero or more) $\tau$-transitions $C_2\xrightarrow{\tau^*} C_2^0$, such that $(C_1,f,C_2^0)\in R$ and $C_2^0\xrightarrow{e_2}C_2'$ with
     $(C_1',f[e_1\mapsto e_2],C_2')\in R$;
   \end{itemize}
   \item if $(C_1,f,C_2)\in R$, and $C_2\xrightarrow{e_2}C_2'$ then
   \begin{itemize}
     \item either $e_2\equiv \tau$, and $(C_1,f[e_2\mapsto \tau^{e_2}],C_2')\in R$;
     \item or there is a sequence of (zero or more) $\tau$-transitions $C_1\xrightarrow{\tau^*} C_1^0$, such that $(C_1^0,f,C_2)\in R$ and $C_1^0\xrightarrow{e_1}C_1'$ with
     $(C_1',f[e_2\mapsto e_1],C_2')\in R$;
   \end{itemize}
   \item if $(C_1,f,C_2)\in R$ and $C_1\downarrow$, then there is a sequence of (zero or more) $\tau$-transitions $C_2\xrightarrow{\tau^*}C_2^0$ such that $(C_1,f,C_2^0)\in R$ and
   $C_2^0\downarrow$;
   \item if $(C_1,f,C_2)\in R$ and $C_2\downarrow$, then there is a sequence of (zero or more) $\tau$-transitions $C_1\xrightarrow{\tau^*}C_1^0$ such that $(C_1^0,f,C_2)\in R$ and
   $C_1^0\downarrow$;
   \item if $(C_1',f',C_2')\in R$, and $C_1'\xtworightarrow{e_1[m]}C_1$ then
   \begin{itemize}
     \item either $e_1[m]\equiv \tau$, and $(C_1,f'[e_1[m]\mapsto \tau^{e_1[m]}],C_2')\in R$;
     \item or there is a sequence of (zero or more) $\tau$-transitions $C_2'\xtworightarrow{\tau^*} C_2'^0$, such that $(C_1',f',C_2'^0)\in R$ and $C_2'^0\xtworightarrow{e_2[n]}C_2$
     with $(C_1,f'[e_1[m]\mapsto e_2[n]],C_2)\in R$;
   \end{itemize}
   \item if $(C_1',f',C_2')\in R$, and $C_2'\xtworightarrow{e_2[n]}C_2$ then
   \begin{itemize}
     \item either $e_2[n]\equiv \tau$, and $(C_1',f'[e_2[n]\mapsto \tau^{e_2[n]}],C_2)\in R$;
     \item or there is a sequence of (zero or more) $\tau$-transitions $C_1'\xtworightarrow{\tau^*} C_1'^0$, such that $(C_1'^0,f',C_2')\in R$ and $C_1'^0\xtworightarrow{e_1[m]}C_1$
     with $(C_1,f[e_2[n]\mapsto e_1[m]],C_2)\in R$;
   \end{itemize}
   \item if $(C_1',f',C_2')\in R$ and $C_1'\downarrow$, then there is a sequence of (zero or more) $\tau$-transitions $C_2'\xtworightarrow{\tau^*}C_2'^0$ such that
   $(C_1',f',C_2'^0)\in R$ and $C_2'^0\downarrow$;
   \item if $(C_1',f',C_2')\in R$ and $C_2'\downarrow$, then there is a sequence of (zero or more) $\tau$-transitions $C_1'\xtworightarrow{\tau^*}C_1'^0$ such that
   $(C_1'^0,f',C_2')\in R$ and $C_1'^0\downarrow$.
 \end{enumerate}

$\mathcal{E}_1,\mathcal{E}_2$ are FR branching history-preserving (hp-)bisimilar and are written $\mathcal{E}_1\approx_{bhp}^{fr}\mathcal{E}_2$ if there exists a FR branching
hp-bisimulation $R$ such that $(\emptyset,\emptyset,\emptyset)\in R$.

A FR branching hereditary history-preserving (hhp-)bisimulation is a downward closed FR branching hp-bisimulation. $\mathcal{E}_1,\mathcal{E}_2$ are FR branching hereditary
history-preserving (hhp-)bisimilar and are written $\mathcal{E}_1\approx_{bhhp}^{fr}\mathcal{E}_2$.
\end{definition}

\begin{definition}[Forward-reverse (FR) rooted branching (hereditary) history-preserving bisimulation]
Assume a special termination predicate $\downarrow$, and let $\surd$ represent a state with $\surd\downarrow$. A FR rooted branching history-preserving (hp-) bisimulation is a weakly
posetal relation $R\subseteq\mathcal{C}(\mathcal{E}_1)\overline{\times}\mathcal{C}(\mathcal{E}_2)$ such that:

 \begin{enumerate}
   \item if $(C_1,f,C_2)\in R$, and $C_1\xrightarrow{e_1}C_1'$, then $C_2\xrightarrow{e_2}C_2'$ with $C_1'\approx_{bhp}C_2'$;
   \item if $(C_1,f,C_2)\in R$, and $C_2\xrightarrow{e_2}C_2'$, then $C_1\xrightarrow{e_1}C_1'$ with $C_1'\approx_{bhp}C_2'$;
   \item if $(C_1',f',C_2')\in R$, and $C_1'\xtworightarrow{e_1[m]}C_1$, then $C_2'\xtworightarrow{e_2[n]}C_2$ with $C_1\approx_{bhp}^{fr}C_2$;
   \item if $(C_1',f',C_2')\in R$, and $C_2'\xtworightarrow{e_2[n]}C_2$, then $C_1'\xtworightarrow{e_1[m]}C_1$ with $C_1\approx_{bhp}^{fr}C_2$;
   \item if $(C_1,f,C_2)\in R$ and $C_1\downarrow$, then $C_2\downarrow$;
   \item if $(C_1,f,C_2)\in R$ and $C_2\downarrow$, then $C_1\downarrow$.
 \end{enumerate}

$\mathcal{E}_1,\mathcal{E}_2$ are FR rooted branching history-preserving (hp-)bisimilar and are written $\mathcal{E}_1\approx_{rbhp}^{fr}\mathcal{E}_2$ if there exists a FR rooted
branching hp-bisimulation $R$ such that $(\emptyset,\emptyset,\emptyset)\in R$.

A FR rooted branching hereditary history-preserving (hhp-)bisimulation is a downward closed FR rooted branching hp-bisimulation. $\mathcal{E}_1,\mathcal{E}_2$ are FR rooted branching
hereditary history-preserving (hhp-)bisimilar and are written $\mathcal{E}_1\approx_{rbhhp}^{fr}\mathcal{E}_2$.
\end{definition}

\subsubsection{Basic Algebra for Reversible True Concurrency}

In the following, let $e_1, e_2, e_1', e_2'\in \mathbb{E}$, and let variables $x,y,z$ range over the set of terms for true concurrency, $p,q,s$ range over the set of closed terms. The
predicate $Std(x)$ denotes that $x$ contains only standard events (no histories of events) and $NStd(x)$ means that $x$ only contains histories of events. The set of axioms of BARTC
consists of the laws given in Table \ref{AxiomsForBARTC5}.

\begin{center}
    \begin{table}
        \begin{tabular}{@{}ll@{}}
            \hline No. &Axiom\\
            $A1$ & $x+ y = y+ x$\\
            $A2$ & $(x+ y)+ z = x+ (y+ z)$\\
            $A3$ & $x+ x = x$\\
            $A41$ & $(x+ y)\cdot z = x\cdot z + y\cdot z\quad Std(x),Std(y), Std(z)$\\
            $A42$ & $x\cdot (y+z) = x\cdot y + x\cdot z\quad NStd(x),NStd(y), NStd(z)$\\
            $A5$ & $(x\cdot y)\cdot z = x\cdot(y\cdot z)$\\
        \end{tabular}
        \caption{Axioms of BARTC}
        \label{AxiomsForBARTC5}
    \end{table}
\end{center}

\begin{definition}[Basic terms of BARTC]\label{BTBARTC5}
The set of basic terms of BARTC, $\mathcal{B}(BARTC)$, is inductively defined as follows:
\begin{enumerate}
  \item $\mathbb{E}\subset\mathcal{B}(BARTC)$;
  \item if $e\in \mathbb{E}, t\in\mathcal{B}(BARTC)$ then $e\cdot t\in\mathcal{B}(BARTC)$;
  \item if $e[m]\in \mathbb{E}, t\in\mathcal{B}(BARTC)$ then $t\cdot e[m]\in\mathcal{B}(BARTC)$;
  \item if $t,s\in\mathcal{B}(BARTC)$ then $t+ s\in\mathcal{B}(BARTC)$.
\end{enumerate}
\end{definition}

\begin{theorem}[Elimination theorem of BARTC]\label{ETBARTC5}
Let $p$ be a closed BARTC term. Then there is a basic BARTC term $q$ such that $BARTC\vdash p=q$.
\end{theorem}

In this subsection, we will define a term-deduction system which gives the operational semantics of BARTC. We give the forward operational transition rules of operators $\cdot$ and $+$
as Table \ref{SETRForBARTC5} shows, and the reverse rules of operators $\cdot$ and $+$ as Table \ref{RSETRForBARTC5} shows. And the predicate $\xrightarrow{e}e[m]$ represents
successful forward termination after forward execution of the event $e$, the predicate $\xtworightarrow{e[m]}e$ represents successful reverse termination after reverse execution of
the event history $e[m]$.

\begin{center}
    \begin{table}
        $$\frac{}{e\xrightarrow{e}e[m]}$$
        $$\frac{x\xrightarrow{e}e[m]}{x+ y\xrightarrow{e}e[m]} \quad\frac{x\xrightarrow{e}x'}{x+ y\xrightarrow{e}x'} \quad\frac{y\xrightarrow{e}e[m]}{x+ y\xrightarrow{e}e[m]} \quad\frac{y\xrightarrow{e}y'}{x+ y\xrightarrow{e}y'}$$
        $$\frac{x\xrightarrow{e}e[m]}{x\cdot y\xrightarrow{e} e[m]\cdot y} \quad\frac{x\xrightarrow{e}x'}{x\cdot y\xrightarrow{e}x'\cdot y}$$
        \caption{Forward single event transition rules of BARTC}
        \label{SETRForBARTC5}
    \end{table}
\end{center}

\begin{center}
    \begin{table}
        $$\frac{}{e[m]\xtworightarrow{e[m]}e}$$
        $$\frac{x\xtworightarrow{e[m]}e}{x+ y\xtworightarrow{e[m]}e} \quad\frac{x\xtworightarrow{e[m]}x'}{x+ y\xtworightarrow{e[m]}x'} \quad\frac{y\xtworightarrow{e[m]}e}{x+ y\xtworightarrow{e[m]}e} \quad\frac{y\xtworightarrow{e[m]}y'}{x+ y\xtworightarrow{e[m]}y'}$$
        $$\frac{y\xtworightarrow{e[m]}e}{x\cdot y\xtworightarrow{e[m]} x\cdot e} \quad\frac{y\xtworightarrow{e[m]}y'}{x\cdot y\xtworightarrow{e[m]}x\cdot y'}$$
        \caption{Reverse single event transition rules of BARTC}
        \label{RSETRForBARTC5}
    \end{table}
\end{center}

The forward pomset transition rules are shown in Table \ref{PTRForBARTC5}, and reverse pomset transition rules are shown in Table \ref{RPTRForBARTC5}, different to single event
transition rules, the pomset transition rules are labeled by pomsets, which are defined by causality $\cdot$ and conflict $+$.

\begin{center}
    \begin{table}
        $$\frac{}{X\xrightarrow{X}X[\mathcal{K}]}$$
        $$\frac{x\xrightarrow{X}X[\mathcal{K}]}{x+ y\xrightarrow{X}X[\mathcal{K}]} (X\subseteq x)\quad\frac{x\xrightarrow{X}x'}{x+ y\xrightarrow{X}x'} (X\subseteq x) \quad\frac{y\xrightarrow{Y}Y[\mathcal{K}]}{x+ y\xrightarrow{Y}Y[\mathcal{K}]} (Y\subseteq y)\quad\frac{y\xrightarrow{Y}y'}{x+ y\xrightarrow{Y}y'}(Y\subseteq y)$$
        $$\frac{x\xrightarrow{X}X[\mathcal{K}]}{x\cdot y\xrightarrow{X} X[\mathcal{K}]\cdot y} (X\subseteq x)\quad\frac{x\xrightarrow{X}x'}{x\cdot y\xrightarrow{X}x'\cdot y} (X\subseteq x)$$
        \caption{Forward pomset transition rules of BARTC}
        \label{PTRForBARTC5}
    \end{table}
\end{center}

\begin{center}
    \begin{table}
        $$\frac{}{X[\mathcal{K}]\xtworightarrow{X[\mathcal{K}]}X}$$
        $$\frac{x\xtworightarrow{X[\mathcal{K}]}X}{x+ y\xtworightarrow{X[\mathcal{K}]}X} (X\subseteq x)\quad\frac{x\xtworightarrow{X[\mathcal{K}]}x'}{x+ y\xtworightarrow{X[\mathcal{K}]}x'} (X\subseteq x) \quad\frac{y\xtworightarrow{Y[\mathcal{K}]}Y}{x+ y\xtworightarrow{Y[\mathcal{K}]}Y} (Y\subseteq y)\quad\frac{y\xtworightarrow{Y[\mathcal{K}]}y'}{x+ y\xtworightarrow{Y[\mathcal{K}]}y'}(Y\subseteq y)$$
        $$\frac{y\xtworightarrow{Y[\mathcal{K}]}Y}{x\cdot y\xtworightarrow{Y[\mathcal{K}]} x\cdot Y} (Y\subseteq y)\quad\frac{y\xtworightarrow{Y[\mathcal{K}]}y'}{x\cdot y\xtworightarrow{Y[\mathcal{K}]}x\cdot y'} (Y\subseteq y)$$
        \caption{Reverse pomset transition rules of BARTC}
        \label{RPTRForBARTC5}
    \end{table}
\end{center}

\begin{theorem}[Congruence of BARTC with respect to FR truly concurrent bisimulation equivalences]
FR truly concurrent bisimulation equivalence $\sim_p^{fr}$, $\sim_s^{fr}$, $\sim_{hp}^{fr}$, $\sim_{hhp}^{fr}$ are all congruences with respect to BARTC.
\end{theorem}

\begin{theorem}[Soundness of BARTC modulo FR truly concurrent bisimulation equivalence]
Let $x$ and $y$ be BARTC terms.

\begin{enumerate}
  \item If $BARTC\vdash x=y$, then $x\sim_p^{fr} y$;
  \item If $BARTC\vdash x=y$, then $x\sim_s^{fr} y$;
  \item If $BARTC\vdash x=y$, then $x\sim_{hp}^{fr} y$;
  \item If $BARTC\vdash x=y$, then $x\sim_{hhp}^{fr} y$.
\end{enumerate}
\end{theorem}

\begin{theorem}[Completeness of BARTC modulo FR truly concurrent bisimulation equivalence]
Let $p$ and $q$ be closed BARTC terms, then,

\begin{enumerate}
  \item if $p\sim_p^{fr} q$ then $p=q$;
  \item if $p\sim_s^{fr} q$ then $p=q$;
  \item if $p\sim_{hp}^{fr} q$ then $p=q$;
  \item if $p\sim_{hhp}^{fr} q$ then $p=q$.
\end{enumerate}
\end{theorem}

\subsubsection{Algebra for Parallelism in Reversible True Concurrency}

The forward transition rules for parallelism $\parallel$ are shown in Table \ref{TRForParallel5}, and the reverse transition rules for $\parallel$ are shown in Table
\ref{RTRForParallel5}.

\begin{center}
    \begin{table}
        $$\frac{x\xrightarrow{e_1}e_1[m]\quad y\xrightarrow{e_2}e_2[m]}{x\parallel y\xrightarrow{\{e_1,e_2\}}e_1[m]\parallel e_2[m]}
        \quad\frac{x\xrightarrow{e_1}x'\quad y\xrightarrow{e_2}e_2[m]}{x\parallel y\xrightarrow{\{e_1,e_2\}}x'\parallel e_2[m]}$$
        $$\frac{x\xrightarrow{e_1}e_1[m]\quad y\xrightarrow{e_2}y'}{x\parallel y\xrightarrow{\{e_1,e_2\}}e_1[m]\parallel y'}
        \quad\frac{x\xrightarrow{e_1}x'\quad y\xrightarrow{e_2}y'}{x\parallel y\xrightarrow{\{e_1,e_2\}}x'\between y'}$$
        $$\frac{x\xrightarrow{e_1}e_1[m]\quad y\xrightarrow{e_2}e_2[m] \quad(e_1\leq e_2)}{x\leftmerge y\xrightarrow{\{e_1,e_2\}}e_1[m]\leftmerge e_2[m]}
        \quad\frac{x\xrightarrow{e_1}x'\quad y\xrightarrow{e_2}e_2[m] \quad(e_1\leq e_2)}{x\leftmerge y\xrightarrow{\{e_1,e_2\}}x'\leftmerge e_2[m]}$$
        $$\frac{x\xrightarrow{e_1}e_1[m]\quad y\xrightarrow{e_2}y' \quad(e_1\leq e_2)}{x\leftmerge y\xrightarrow{\{e_1,e_2\}}e_1[m]\leftmerge y'}
        \quad\frac{x\xrightarrow{e_1}x'\quad y\xrightarrow{e_2}y' \quad(e_1\leq e_2)}{x\leftmerge y\xrightarrow{\{e_1,e_2\}}x'\between y'}$$
        \caption{Forward transition rules of parallel operator $\parallel$}
        \label{TRForParallel5}
    \end{table}
\end{center}

\begin{center}
    \begin{table}
        $$\frac{x\xtworightarrow{e_1[m]}e_1\quad y\xtworightarrow{e_2[m]}e_2}{x\parallel y\xtworightarrow{\{e_1[m],e_2[m]\}}e_1\parallel e_2}
        \quad\frac{x\xtworightarrow{e_1[m]}x'\quad y\xtworightarrow{e_2[m]}e_2}{x\parallel y\xtworightarrow{\{e_1[m],e_2[m]\}}x'\parallel e_2}$$
        $$\frac{x\xtworightarrow{e_1[m]}e_1\quad y\xtworightarrow{e_2[m]}y'}{x\parallel y\xtworightarrow{\{e_1[m],e_2[m]\}}e_1\parallel y'}
        \quad\frac{x\xtworightarrow{e_1[m]}x'\quad y\xtworightarrow{e_2[m]}y'}{x\parallel y\xtworightarrow{\{e_1[m],e_2[m]\}}x'\between y'}$$
        $$\frac{x\xtworightarrow{e_1[m]}e_1\quad y\xtworightarrow{e_2[m]}e_2 \quad(e_1\leq e_2)}{x\leftmerge y\xtworightarrow{\{e_1[m],e_2[m]\}}e_1\leftmerge e_2}
        \quad\frac{x\xtworightarrow{e_1[m]}x'\quad y\xtworightarrow{e_2[m]}e_2 \quad(e_1\leq e_2)}{x\leftmerge y\xtworightarrow{\{e_1[m],e_2[m]\}}x'\leftmerge e_2}$$
        $$\frac{x\xtworightarrow{e_1[m]}e_1\quad y\xtworightarrow{e_2[m]}y' \quad(e_1\leq e_2)}{x\leftmerge y\xtworightarrow{\{e_1[m],e_2[m]\}}e_1\leftmerge y'}
        \quad\frac{x\xtworightarrow{e_1[m]}x'\quad y\xtworightarrow{e_2[m]}y' \quad(e_1\leq e_2)}{x\leftmerge y\xtworightarrow{\{e_1[m],e_2[m]\}}x'\between y'}$$
        \caption{Reverse transition rules of parallel operator $\parallel$}
        \label{RTRForParallel5}
    \end{table}
\end{center}

The forward and reverse transition rules of communication $\mid$ are shown in Table \ref{TRForCommunication5} and Table \ref{RTRForCommunication5}.

\begin{center}
    \begin{table}
        $$\frac{x\xrightarrow{e_1}e_1[m]\quad y\xrightarrow{e_2}e_2[m]}{x\mid y\xrightarrow{\gamma(e_1,e_2)}\gamma(e_1,e_2)[m]}
        \quad\frac{x\xrightarrow{e_1}x'\quad y\xrightarrow{e_2}e_2[m]}{x\mid y\xrightarrow{\gamma(e_1,e_2)}\gamma(e_1,e_2)[m]\cdot x'}$$
        $$\frac{x\xrightarrow{e_1}e_1[m]\quad y\xrightarrow{e_2}y'}{x\mid y\xrightarrow{\gamma(e_1,e_2)}\gamma(e_1,e_2)[m]\cdot y'}
        \quad\frac{x\xrightarrow{e_1}x'\quad y\xrightarrow{e_2}y'}{x\mid y\xrightarrow{\gamma(e_1,e_2)}\gamma(e_1,e_2)[m]\cdot x'\between y'}$$
        \caption{Forward transition rules of communication operator $\mid$}
        \label{TRForCommunication5}
    \end{table}
\end{center}

\begin{center}
    \begin{table}
        $$\frac{x\xtworightarrow{e_1[m]}e_1\quad y\xtworightarrow{e_2[m]}e_2}{x\mid y\xtworightarrow{\gamma(e_1,e_2)[m]}\gamma(e_1,e_2)} \quad\frac{x\xtworightarrow{e_1[m]}x'\quad y\xtworightarrow{e_2[m]}e_2}{x\mid y\xtworightarrow{\gamma(e_1,e_2)[m]}\gamma(e_1,e_2)\cdot x'}$$
        $$\frac{x\xtworightarrow{e_1[m]}e_1\quad y\xtworightarrow{e_2[m]}y'}{x\mid y\xtworightarrow{\gamma(e_1,e_2)[m]}\gamma(e_1,e_2)\cdot y'} \quad\frac{x\xtworightarrow{e_1[m]}x'\quad y\xtworightarrow{e_2[m]}y'}{x\mid y\xtworightarrow{\gamma(e_1,e_2)[m]}\gamma(e_1,e_2)\cdot x'\between y'}$$
        \caption{Reverse transition rules of communication operator $\mid$}
        \label{RTRForCommunication5}
    \end{table}
\end{center}

The conflict elimination is also captured by two auxiliary operators, the unary conflict elimination operator $\Theta$ and the binary unless operator $\triangleleft$. The forward and reverse transition rules for $\Theta$ and $\triangleleft$ are expressed by ten transition rules in Table \ref{TRForConflict5} and Table \ref{RTRForConflict5}.

\begin{center}
    \begin{table}
        $$\frac{x\xrightarrow{e_1}e_1[m]\quad (\sharp(e_1,e_2))}{\Theta(x)\xrightarrow{e_1}e_1[m]} \quad\frac{x\xrightarrow{e_2}e_2[n]\quad (\sharp(e_1,e_2))}{\Theta(x)\xrightarrow{e_2}e_2[n]}$$
        $$\frac{x\xrightarrow{e_1}x'\quad (\sharp(e_1,e_2))}{\Theta(x)\xrightarrow{e_1}\Theta(x')} \quad\frac{x\xrightarrow{e_2}x'\quad (\sharp(e_1,e_2))}{\Theta(x)\xrightarrow{e_2}\Theta(x')}$$
        $$\frac{x\xrightarrow{e_1}e_1[m] \quad y\nrightarrow^{e_2}\quad (\sharp(e_1,e_2))}{x\triangleleft y\xrightarrow{\tau}\surd}
        \quad\frac{x\xrightarrow{e_1}x' \quad y\nrightarrow^{e_2}\quad (\sharp(e_1,e_2))}{x\triangleleft y\xrightarrow{\tau}x'}$$
        $$\frac{x\xrightarrow{e_1}e_1[m] \quad y\nrightarrow^{e_3}\quad (\sharp(e_1,e_2),e_2\leq e_3)}{x\triangleleft y\xrightarrow{e_1}e_1[m]}
        \quad\frac{x\xrightarrow{e_1}x' \quad y\nrightarrow^{e_3}\quad (\sharp(e_1,e_2),e_2\leq e_3)}{x\triangleleft y\xrightarrow{e_1}x'}$$
        $$\frac{x\xrightarrow{e_3}e_3[l] \quad y\nrightarrow^{e_2}\quad (\sharp(e_1,e_2),e_1\leq e_3)}{x\triangleleft y\xrightarrow{\tau}\surd}
        \quad\frac{x\xrightarrow{e_3}x' \quad y\nrightarrow^{e_2}\quad (\sharp(e_1,e_2),e_1\leq e_3)}{x\triangleleft y\xrightarrow{\tau}x'}$$
        \caption{Forward transition rules of conflict elimination}
        \label{TRForConflict5}
    \end{table}
\end{center}

\begin{center}
    \begin{table}
        $$\frac{x\xtworightarrow{e_1[m]}e_1\quad (\sharp(e_1,e_2))}{\Theta(x)\xtworightarrow{e_1[m]}e_1} \quad\frac{x\xtworightarrow{e_2[n]}e_2\quad (\sharp(e_1,e_2))}{\Theta(x)\xtworightarrow{e_2[n]}e_2}$$
        $$\frac{x\xtworightarrow{e_1[m]}x'\quad (\sharp(e_1,e_2))}{\Theta(x)\xtworightarrow{e_1[m]}\Theta(x')} \quad\frac{x\xtworightarrow{e_2[n]}x'\quad (\sharp(e_1,e_2))}{\Theta(x)\xtworightarrow{e_2[n]}\Theta(x')}$$
        $$\frac{x\xtworightarrow{e_1[m]}e_1 \quad y\xntworightarrow{e_2[n]}\quad (\sharp(e_1,e_2))}{x\triangleleft y\xtworightarrow{\tau}\surd}
        \quad\frac{x\xtworightarrow{e_1[m]}x' \quad y\xntworightarrow{e_2[n]}\quad (\sharp(e_1,e_2))}{x\triangleleft y\xtworightarrow{\tau}x'}$$
        $$\frac{x\xtworightarrow{e_1[m]}e_1 \quad y\xntworightarrow{e_3[l]}\quad (\sharp(e_1,e_2),e_2\geq e_3)}{x\triangleleft y\xtworightarrow{e_1[m]}e_1}
        \quad\frac{x\xtworightarrow{e_1[m]}x' \quad y\xntworightarrow{e_3[l]}\quad (\sharp(e_1,e_2),e_2\geq e_3)}{x\triangleleft y\xtworightarrow{e_1[m]}x'}$$
        $$\frac{x\xtworightarrow{e_3[l]}e_3 \quad y\xntworightarrow{e_2[n]}\quad (\sharp(e_1,e_2),e_1\geq e_3)}{x\triangleleft y\xtworightarrow{\tau}\surd}
        \quad\frac{x\xtworightarrow{e_3[l]}x' \quad y\xntworightarrow{e_2[n]}\quad (\sharp(e_1,e_2),e_1\geq e_3)}{x\triangleleft y\xtworightarrow{\tau}x'}$$
        \caption{Reverse transition rules of conflict elimination}
        \label{RTRForConflict5}
    \end{table}
\end{center}

\begin{theorem}[Congruence theorem of APRTC]
FR truly concurrent bisimulation equivalences $\sim_{p}^{fr}$, $\sim_s^{fr}$, $\sim_{hp}^{fr}$ and $\sim_{hhp}^{fr}$ are all congruences with respect to APRTC.
\end{theorem}

\begin{definition}[Basic terms of APRTC]\label{BTAPTC5}
The set of basic terms of APRTC, $\mathcal{B}(APRTC)$, is inductively defined as follows:
\begin{enumerate}
  \item $\mathbb{E}\subset\mathcal{B}(APRTC)$;
  \item if $e\in \mathbb{E}, t\in\mathcal{B}(APRTC)$ then $e\cdot t\in\mathcal{B}(APRTC)$;
  \item if $e[m]\in \mathbb{E}, t\in\mathcal{B}(APRTC)$ then $t\cdot e[m]\in\mathcal{B}(APRTC)$;
  \item if $t,s\in\mathcal{B}(APRTC)$ then $t+ s\in\mathcal{B}(APRTC)$;
  \item if $t,s\in\mathcal{B}(APRTC)$ then $t\leftmerge s\in\mathcal{B}(APRTC)$.
\end{enumerate}
\end{definition}

We design the axioms of parallelism in Table \ref{AxiomsForParallelism5}, including algebraic laws for parallel operator $\parallel$, communication operator $\mid$, conflict elimination operator $\Theta$ and unless operator $\triangleleft$, and also the whole parallel operator $\between$. Since the communication between two communicating events in different parallel branches may cause deadlock (a state of inactivity), which is caused by mismatch of two communicating events or the imperfectness of the communication channel. We introduce a new constant $\delta$ to denote the deadlock, and let the atomic event $e\in \mathbb{E}\cup\{\delta\}$.

\begin{center}
    \begin{table}
        \begin{tabular}{@{}ll@{}}
            \hline No. &Axiom\\
            $A6$ & $x+ \delta = x$\\
            $A7$ & $\delta\cdot x =\delta(\textrm{Std(x)})$\\
            $RA7$ & $x\cdot \delta =\delta(\textrm{NStd(x)})$\\
            $P1$ & $x\between y = x\parallel y + x\mid y$\\
            $P2$ & $x\parallel y = y \parallel x$\\
            $P3$ & $(x\parallel y)\parallel z = x\parallel (y\parallel z)$\\
            $P4$ & $x\parallel y=x\leftmerge y+y\leftmerge x$\\
            $P5$ & $(e_1\leq e_2)\quad e_1\leftmerge (e_2\cdot y) = (e_1\leftmerge e_2)\cdot y$\\
            $RP5$ & $(e_1[m]\leq e_2[m])\quad e_1[m]\leftmerge (y\cdot e_2[m]) = y\cdot(e_1[m]\leftmerge e_2[m])$\\
            $P6$ & $(e_1\leq e_2)\quad (e_1\cdot x)\leftmerge e_2 = (e_1\leftmerge e_2)\cdot x$\\
            $RP6$ & $(e_1[m]\leq e_2[m])\quad (x\cdot e_1[m])\leftmerge e_2[m] = x\cdot(e_1[m]\leftmerge e_2[m])$\\
            $P7$ & $(e_1\leq e_2)\quad (e_1\cdot x)\leftmerge (e_2\cdot y) = (e_1\leftmerge e_2)\cdot (x\between y)$\\
            $RP7$ & $(e_1[m]\leq e_2[m])\quad(x\cdot e_1[m])\leftmerge (y\cdot e_2[m]) = (x\between y)\cdot(e_1[m]\leftmerge e_2[m])$\\
            $P8$ & $(x+ y)\leftmerge z = (x\leftmerge z)+ (y\leftmerge z)(\textrm{Std(x)},\textrm{Std(y)},\textrm{Std(z)})$\\
            $RP8$ & $x\leftmerge (y+ z) = (x\leftmerge y)+ (x\leftmerge z)(\textrm{NStd(x)},\textrm{NStd(y)},\textrm{NStd(z)})$\\
            $P9$ & $\delta\leftmerge x = \delta(\textrm{Std(x)})$\\
            $RP9$ & $x\leftmerge \delta = \delta(\textrm{NStd(x)})$\\
            $C1$ & $e_1\mid e_2 = \gamma(e_1,e_2)$\\
            $RC1$ & $e_1[m]\mid e_2[m] = \gamma(e_1,e_2)[m]$\\
            $C2$ & $e_1\mid (e_2\cdot y) = \gamma(e_1,e_2)\cdot y$\\
            $RC2$ & $e_1[m]\mid (y \cdot e_2[m]) =y\cdot \gamma(e_1,e_2)[m]$\\
            $C3$ & $(e_1\cdot x)\mid e_2 = \gamma(e_1,e_2)\cdot x$\\
            $RC3$ & $(x \cdot e_1[m])\mid e_2[m] =x\cdot \gamma(e_1,e_2)[m]$\\
            $C4$ & $(e_1\cdot x)\mid (e_2\cdot y) = \gamma(e_1,e_2)\cdot (x\between y)$\\
            $RC4$ & $(x \cdot e_1[m])\mid (y \cdot e_2[m]) =(x\between y)\cdot \gamma(e_1,e_2)[m]$\\
            $C5$ & $(x+ y)\mid z = (x\mid z) + (y\mid z)$\\
            $C6$ & $x\mid (y+ z) = (x\mid y)+ (x\mid z)$\\
            $C7$ & $\delta\mid x = \delta$\\
            $C8$ & $x\mid\delta = \delta$\\
            $CE1$ & $\Theta(e) = e$\\
            $RCE1$ & $\Theta(e[m]) = e[m]$\\
            $CE2$ & $\Theta(\delta) = \delta$\\
            $CE3$ & $\Theta(x+ y) = \Theta(x)\triangleleft y + \Theta(y)\triangleleft x$\\
            $CE4$ & $\Theta(x\cdot y)=\Theta(x)\cdot\Theta(y)$\\
            $CE5$ & $\Theta(x\parallel y) = ((\Theta(x)\triangleleft y)\parallel y)+ ((\Theta(y)\triangleleft x)\parallel x)$\\
            $CE6$ & $\Theta(x\mid y) = ((\Theta(x)\triangleleft y)\mid y)+ ((\Theta(y)\triangleleft x)\mid x)$\\
        \end{tabular}
        \caption{Axioms of parallelism}
        \label{AxiomsForParallelism5}
    \end{table}
\end{center}

\begin{center}
    \begin{table}
        \begin{tabular}{@{}ll@{}}
            \hline No. &Axiom\\
            $U1$ & $(\sharp(e_1,e_2))\quad e_1\triangleleft e_2 = \tau$\\
            $RU1$ & $(\sharp(e_1[m],e_2[n]))\quad e_1[m]\triangleleft e_2[n] = \tau$\\
            $U2$ & $(\sharp(e_1,e_2),e_2\leq e_3)\quad e_1\triangleleft e_3 = e_1$\\
            $RU2$ & $(\sharp(e_1[m],e_2[n]),e_2[n]\geq e_3[l])\quad e_1[m]\triangleleft e_3[l] = e_1[m]$\\
            $U3$ & $(\sharp(e_1,e_2),e_2\leq e_3)\quad e3\triangleleft e_1 = \tau$\\
            $RU3$ & $(\sharp(e_1[m],e_2[n]),e_2[n]\geq e_3[l])\quad e3[l]\triangleleft e_1[m] = \tau$\\
            $U4$ & $e\triangleleft \delta = e$\\
            $U5$ & $\delta \triangleleft e = \delta$\\
            $U6$ & $(x+ y)\triangleleft z = (x\triangleleft z)+ (y\triangleleft z)$\\
            $U7$ & $(x\cdot y)\triangleleft z = (x\triangleleft z)\cdot (y\triangleleft z)$\\
            $U8$ & $(x\parallel y)\triangleleft z = (x\triangleleft z)\parallel (y\triangleleft z)$\\
            $U9$ & $(x\mid y)\triangleleft z = (x\triangleleft z)\mid (y\triangleleft z)$\\
            $U10$ & $x\triangleleft (y+ z) = (x\triangleleft y)\triangleleft z$\\
            $U11$ & $x\triangleleft (y\cdot z)=(x\triangleleft y)\triangleleft z$\\
            $U12$ & $x\triangleleft (y\parallel z) = (x\triangleleft y)\triangleleft z$\\
            $U13$ & $x\triangleleft (y\mid z) = (x\triangleleft y)\triangleleft z$\\
        \end{tabular}
        \caption{Axioms of parallelism (continuing)}
        \label{AxiomsForParallelism52}
    \end{table}
\end{center}

Based on the definition of basic terms for APRTC (see Definition \ref{BTAPTC5}) and axioms of parallelism (see Table \ref{AxiomsForParallelism5}), we can prove the elimination theorem
of parallelism.

\begin{theorem}[Elimination theorem of FR parallelism]\label{ETParallelism5}
Let $p$ be a closed APRTC term. Then there is a basic APRTC term $q$ such that $APRTC\vdash p=q$.
\end{theorem}

\begin{theorem}[Generalization of the algebra for parallelism with respect to BARTC]
The algebra for parallelism is a generalization of BARTC.
\end{theorem}

\begin{theorem}[Soundness of parallelism modulo FR truly concurrent bisimulation equivalence]
Let $x$ and $y$ be APRTC terms.

\begin{enumerate}
  \item If $APRTC\vdash x=y$, then $x\sim_{p}^{fr} y$;
  \item If $APRTC\vdash x=y$, then $x\sim_{s}^{fr} y$;
  \item If $APRTC\vdash x=y$, then $x\sim_{hp}^{fr} y$;
  \item If $APRTC\vdash x=y$, then $x\sim_{hhp}^{fr} y$.
\end{enumerate}
\end{theorem}

\begin{theorem}[Completeness of parallelism modulo FR truly concurrent bisimulation equivalence]
Let $p$ and $q$ be closed APRTC terms, then,

\begin{enumerate}
  \item if $p\sim_{p}^{fr} q$ then $p=q$;
  \item if $p\sim_{s}^{fr} q$ then $p=q$;
  \item if $p\sim_{hp}^{fr} q$ then $p=q$;
  \item if $p\sim_{hhp}^{fr} q$ then $p=q$.
\end{enumerate}
\end{theorem}

The mismatch of two communicating events in different parallel branches can cause deadlock, so the deadlocks in the concurrent processes should be eliminated. Like $APTC$ \cite{ATC},
we also introduce the unary encapsulation operator $\partial_H$ for set $H$ of atomic events, which renames all atomic events in $H$ into $\delta$. The whole algebra including
parallelism for true concurrency in the above subsections, deadlock $\delta$ and encapsulation operator $\partial_H$, is called Reversible Algebra for Parallelism in True Concurrency,
abbreviated APRTC.

The forward transition rules of encapsulation operator $\partial_H$ are shown in Table \ref{TRForEncapsulation5}, and the reverse transition rules of encapsulation operator
$\partial_H$ are shown in Table \ref{RTRForEncapsulation5}.

\begin{center}
    \begin{table}
        $$\frac{x\xrightarrow{e}e[m]}{\partial_H(x)\xrightarrow{e}\partial_H(e[m])}\quad (e\notin H)
        \quad\frac{x\xrightarrow{e}x'}{\partial_H(x)\xrightarrow{e}\partial_H(x')}\quad(e\notin H)$$
        \caption{Forward transition rules of encapsulation operator $\partial_H$}
        \label{TRForEncapsulation5}
    \end{table}
\end{center}

\begin{center}
    \begin{table}
        $$\frac{x\xtworightarrow{e[m]}e}{\partial_H(x)\xtworightarrow{e[m]}e}\quad (e\notin H)
        \quad\frac{x\xtworightarrow{e}x'}{\partial_H(x)\xtworightarrow{e}\partial_H(x')}\quad(e\notin H)$$
        \caption{Reverse transition rules of encapsulation operator $\partial_H$}
        \label{RTRForEncapsulation5}
    \end{table}
\end{center}

Based on the transition rules for encapsulation operator $\partial_H$ in Table \ref{TRForEncapsulation5} and Table \ref{RTRForEncapsulation5}, we design the axioms as Table
\ref{AxiomsForEncapsulation5} shows.

\begin{center}
    \begin{table}
        \begin{tabular}{@{}ll@{}}
            \hline No. &Axiom\\
            $D1$ & $e\notin H\quad\partial_H(e) = e$\\
            $RD1$ & $e\notin H\quad\partial_H(e[m]) = e[m]$\\
            $D2$ & $e\in H\quad \partial_H(e) = \delta$\\
            $RD2$ & $e\in H\quad \partial_H(e[m]) = \delta$\\
            $D3$ & $\partial_H(\delta) = \delta$\\
            $D4$ & $\partial_H(x+ y) = \partial_H(x)+\partial_H(y)$\\
            $D5$ & $\partial_H(x\cdot y) = \partial_H(x)\cdot\partial_H(y)$\\
            $D6$ & $\partial_H(x\leftmerge y) = \partial_H(x)\leftmerge\partial_H(y)$\\
        \end{tabular}
        \caption{Axioms of encapsulation operator}
        \label{AxiomsForEncapsulation5}
    \end{table}
\end{center}

\begin{theorem}[Conservativity of APRTC with respect to the algebra for parallelism]
APRTC is a conservative extension of the algebra for parallelism.
\end{theorem}

\begin{theorem}[Congruence theorem of encapsulation operator $\partial_H$]
Truly concurrent bisimulation equivalences $\sim_p^{fr}$, $\sim_s^{fr}$, $\sim_{hp}^{fr}$ and $\sim_{hhp}^{fr}$ are all congruences with respect to encapsulation operator $\partial_H$.
\end{theorem}

\begin{theorem}[Elimination theorem of APRTC]\label{ETEncapsulation5}
Let $p$ be a closed APRTC term including the encapsulation operator $\partial_H$. Then there is a basic APRTC term $q$ such that $APRTC\vdash p=q$.
\end{theorem}

\begin{theorem}[Soundness of APRTC modulo FR truly concurrent bisimulation equivalence]
Let $x$ and $y$ be APRTC terms including encapsulation operator $\partial_H$.

\begin{enumerate}
  \item If $APRTC\vdash x=y$, then $x\sim_{p}^{fr} y$;
  \item If $APRTC\vdash x=y$, then $x\sim_{s}^{fr} y$;
  \item If $APRTC\vdash x=y$, then $x\sim_{hp}^{fr} y$;
  \item If $APRTC\vdash x=y$, then $x\sim_{hhp}^{fr} y$.
\end{enumerate}
\end{theorem}

\begin{theorem}[Completeness of APRTC modulo FR truly concurrent bisimulation equivalence]
Let $p$ and $q$ be closed APRTC terms including encapsulation operator $\partial_H$, then,

\begin{enumerate}
  \item if $p\sim_{p}^{fr} q$ then $p=q$;
  \item if $p\sim_{s}^{fr} q$ then $p=q$;
  \item if $p\sim_{hp}^{fr} q$ then $p=q$;
  \item if $p\sim_{hhp}^{fr} q$ then $p=q$.
\end{enumerate}
\end{theorem}

\subsubsection{Recursion}

\begin{definition}[Recursive specification]
A recursive specification is a finite set of recursive equations

$$X_1=t_1(X_1,\cdots,X_n)$$
$$\cdots$$
$$X_n=t_n(X_1,\cdots,X_n)$$

where the left-hand sides of $X_i$ are called recursion variables, and the right-hand sides $t_i(X_1,\cdots,X_n)$ are process terms in $APTC$ with possible occurrences of the recursion
variables $X_1,\cdots,X_n$.
\end{definition}

\begin{definition}[Solution]
Processes $p_1,\cdots,p_n$ are a solution for a recursive specification $\{X_i=t_i(X_1,\cdots,X_n)|i\in\{1,\cdots,n\}\}$ (with respect to truly concurrent bisimulation equivalences
$\sim_s^{fr}$($\sim_p^{fr}$, $\sim_{hp}^{fr}$, $\sim_{hhp}^{fr}$)) if $p_i\sim_s^{fr} (\sim_p^{fr}, \sim_{hp}^{fr},\sim{hhp}^{fr})t_i(p_1,\cdots,p_n)$ for $i\in\{1,\cdots,n\}$.
\end{definition}

\begin{definition}[Guarded recursive specification]
A recursive specification

$$X_1=t_1(X_1,\cdots,X_n)$$
$$...$$
$$X_n=t_n(X_1,\cdots,X_n)$$

is guarded if the right-hand sides of its recursive equations can be adapted to the form by applications of the axioms in $APTC$ and replacing recursion variables by the right-hand
sides of their recursive equations,

$$(a_{11}\leftmerge\cdots\leftmerge a_{1i_1})\cdot s_1(X_1,\cdots,X_n)+\cdots+(a_{k1}\leftmerge\cdots\leftmerge a_{ki_k})\cdot s_k(X_1,\cdots,X_n)+(b_{11}\leftmerge\cdots\leftmerge b_{1j_1})+\cdots+(b_{1j_1}\leftmerge\cdots\leftmerge b_{lj_l})$$

where $a_{11},\cdots,a_{1i_1},a_{k1},\cdots,a_{ki_k},b_{11},\cdots,b_{1j_1},b_{1j_1},\cdots,b_{lj_l}\in \mathbb{E}$, and the sum above is allowed to be empty, in which case it
represents the deadlock $\delta$.
\end{definition}

\begin{definition}[Linear recursive specification]
A recursive specification is linear if its recursive equations are of the form

$$(a_{11}\leftmerge\cdots\leftmerge a_{1i_1})X_1+\cdots+(a_{k1}\leftmerge\cdots\leftmerge a_{ki_k})X_k+(b_{11}\leftmerge\cdots\leftmerge b_{1j_1})+\cdots+(b_{1j_1}\leftmerge\cdots\leftmerge b_{lj_l})$$

where $a_{11},\cdots,a_{1i_1},a_{k1},\cdots,a_{ki_k},b_{11},\cdots,b_{1j_1},b_{1j_1},\cdots,b_{lj_l}\in \mathbb{E}$, and the sum above is allowed to be empty, in which case it
represents the deadlock $\delta$.
\end{definition}

\begin{center}
    \begin{table}
        $$\frac{t_i(\langle X_1|E\rangle,\cdots,\langle X_n|E\rangle)\xrightarrow{\{e_1,\cdots,e_k\}}e_1[m]\leftmerge\cdots\leftmerge e_k[m]}{\langle X_i|E\rangle\xrightarrow{\{e_1,\cdots,e_k\}}e_1[m]\leftmerge\cdots\leftmerge e_k[m]}$$
        $$\frac{t_i(\langle X_1|E\rangle,\cdots,\langle X_n|E\rangle)\xrightarrow{\{e_1,\cdots,e_k\}} y}{\langle X_i|E\rangle\xrightarrow{\{e_1,\cdots,e_k\}} y}$$
        $$\frac{t_i(\langle X_1|E\rangle,\cdots,\langle X_n|E\rangle)\xtworightarrow{\{e_1[m],\cdots,e_k[m]\}}e_1\leftmerge\cdots\leftmerge e_k}{\langle X_i|E\rangle\xtworightarrow{\{e_1[m],\cdots,e_k[m]\}}e_1\leftmerge\cdots\leftmerge e_k}$$
        $$\frac{t_i(\langle X_1|E\rangle,\cdots,\langle X_n|E\rangle)\xtworightarrow{\{e_1,\cdots,e_k\}} y}{\langle X_i|E\rangle\xtworightarrow{\{e_1,\cdots,e_k\}} y}$$
        \caption{Transition rules of guarded recursion}
        \label{TRForGR}
    \end{table}
\end{center}

The $RDP$ (Recursive Definition Principle) and the $RSP$ (Recursive Specification Principle) are shown in Table \ref{RDPRSP21}.

\begin{center}
\begin{table}
  \begin{tabular}{@{}ll@{}}
\hline No. &Axiom\\
  $RDP$ & $\langle X_i|E\rangle = t_i(\langle X_1|E,\cdots,X_n|E\rangle)\quad (i\in\{1,\cdots,n\})$\\
  $RSP$ & if $y_i=t_i(y_1,\cdots,y_n)$ for $i\in\{1,\cdots,n\}$, then $y_i=\langle X_i|E\rangle \quad(i\in\{1,\cdots,n\})$\\
\end{tabular}
\caption{Recursive definition and specification principle}
\label{RDPRSP21}
\end{table}
\end{center}

\begin{theorem}[Conservitivity of APRTC with guarded recursion]
APRTC with guarded recursion is a conservative extension of APRTC.
\end{theorem}

\begin{theorem}[Congruence theorem of APRTC with guarded recursion]
Truly concurrent bisimulation equivalences $\sim_p^{fr}$, $\sim_s^{fr}$ and $\sim_{hp}^{fr}$ are all congruences with respect to APRTC with guarded recursion.
\end{theorem}

\begin{theorem}[Elimination theorem of APRTC with linear recursion]\label{ETRecursion5}
Each process term in APRTC with linear recursion is equal to a process term $\langle X_1|E\rangle$ with $E$ a linear recursive specification.
\end{theorem}

\begin{theorem}[Soundness of APRTC with guarded recursion]\label{SAPRTCR5}
Let $x$ and $y$ be APRTC with guarded recursion terms. If $APRTC\textrm{ with guarded recursion}\vdash x=y$, then
\begin{enumerate}
  \item $x\sim_{s}^{fr} y$;
  \item $x\sim_p^{fr} y$;
  \item $x\sim_{hp}^{fr} y$;
  \item $x\sim_{hhp}^{fr} y$.
\end{enumerate}
\end{theorem}

\begin{theorem}[Completeness of APRTC with linear recursion]\label{CAPRTCR5}
Let $p$ and $q$ be closed APRTC with linear recursion terms, then,
\begin{enumerate}
  \item if $p\sim_{s}^{fr} q$ then $p=q$;
  \item if $p\sim_p^{fr} q$ then $p=q$;
  \item if $p\sim_{hp}^{fr} q$ then $p=q$;
  \item if $p\sim_{hhp}^{fr} q$ then $p=q$.
\end{enumerate}
\end{theorem}

\subsubsection{Abstraction}

\begin{definition}[Guarded linear recursive specification]
A recursive specification is linear if its recursive equations are of the form

$$(a_{11}\leftmerge\cdots\leftmerge a_{1i_1})X_1+\cdots+(a_{k1}\leftmerge\cdots\leftmerge a_{ki_k})X_k+(b_{11}\leftmerge\cdots\leftmerge b_{1j_1})+\cdots+(b_{1j_1}\leftmerge\cdots\leftmerge b_{lj_l})$$

where $a_{11},\cdots,a_{1i_1},a_{k1},\cdots,a_{ki_k},b_{11},\cdots,b_{1j_1},b_{1j_1},\cdots,b_{lj_l}\in \mathbb{E}\cup\{\tau\}$, and the sum above is allowed to be empty, in which case
it represents the deadlock $\delta$.

A linear recursive specification $E$ is guarded if there does not exist an infinite sequence of $\tau$-transitions
$\langle X|E\rangle\xrightarrow{\tau}\langle X'|E\rangle\xrightarrow{\tau}\langle X''|E\rangle\xrightarrow{\tau}\cdots$.
\end{definition}

To abstract away from the internal implementations of a program, and verify that the program exhibits the desired external behaviors, the silent step $\tau$ and abstraction operator $\tau_I$ are introduced, where $I\subseteq \mathbb{E}$ denotes the internal events. The transition rule of $\tau$ is shown in Table \ref{TRForTau5}. In the following, let the atomic event $e$ range over $\mathbb{E}\cup\{\delta\}\cup\{\tau\}$, and let the communication function $\gamma:\mathbb{E}\cup\{\tau\}\times \mathbb{E}\cup\{\tau\}\rightarrow \mathbb{E}\cup\{\delta\}$, with each communication involved $\tau$ resulting in $\delta$.

\begin{center}
    \begin{table}
        $$\frac{}{\tau\xrightarrow{\tau}\surd}$$
        $$\frac{}{\tau\xtworightarrow{\tau}\surd}$$
        \caption{Transition rule of the silent step}
        \label{TRForTau5}
    \end{table}
\end{center}

\begin{theorem}[Conservitivity of $RAPTC$ with guarded linear recursion silent step]
$RAPTC$ with guarded linear recursion silent step is a conservative extension of $RAPTC$.
\end{theorem}

\begin{theorem}[Congruence theorem of $RAPTC$ with silent step]
Rooted branching FR truly concurrent bisimulation equivalences $\approx_{rbp}^{fr}$, $\approx_{rbs}^{fr}$ and $\approx_{rbhp}^{fr}$ are all congruences with respect to $RAPTC$ with silent step.
\end{theorem}

We design the axioms for the silent step $\tau$ in Table \ref{AxiomsForTau5}.

\begin{center}
\begin{table}
  \begin{tabular}{@{}ll@{}}
\hline No. &Axiom\\
  $B1$ & $e\cdot\tau=e$\\
  $RB1$ & $\tau\cdot e[m]=e[m]$\\
  $B2$ & $e\cdot(\tau\cdot(x+y)+x)=e\cdot(x+y)$\\
  $RB2$ & $((x+y)\cdot\tau+x)\cdot e[m]=(x+y)\cdot e[m]$\\
  $B3$ & $x\leftmerge\tau=x$\\
\end{tabular}
\caption{Axioms of silent step}
\label{AxiomsForTau5}
\end{table}
\end{center}

\begin{theorem}[Elimination theorem of APRTC with silent step and guarded linear recursion]\label{ETTau5}
Each process term in APRTC with silent step and guarded linear recursion is equal to a process term $\langle X_1|E\rangle$ with $E$ a guarded linear recursive specification.
\end{theorem}

\begin{theorem}[Soundness of APRTC with silent step and guarded linear recursion]\label{SAPRTCTAU5}
Let $x$ and $y$ be APRTC with silent step and guarded linear recursion terms. If APRTC with silent step and guarded linear recursion $\vdash x=y$, then
\begin{enumerate}
  \item $x\approx_{rbs}^{fr} y$;
  \item $x\approx_{rbp}^{fr} y$;
  \item $x\approx_{rbhp}^{fr} y$;
  \item $x\approx_{rbhhp}^{fr} y$.
\end{enumerate}
\end{theorem}

\begin{theorem}[Completeness of APRTC with silent step and guarded linear recursion]\label{CAPRTCTAU5}
Let $p$ and $q$ be closed APRTC with silent step and guarded linear recursion terms, then,
\begin{enumerate}
  \item if $p\approx_{rbs}^{fr} q$ then $p=q$;
  \item if $p\approx_{rbp}^{fr} q$ then $p=q$;
  \item if $p\approx_{rbhp}^{fr} q$ then $p=q$;
  \item if $p\approx_{rbhhp}^{fr} q$ then $p=q$.
\end{enumerate}
\end{theorem}

The unary abstraction operator $\tau_I$ ($I\subseteq \mathbb{E}$) renames all atomic events in $I$ into $\tau$. APRTC with silent step and abstraction operator is called $APRTC_{\tau}$. The transition rules of operator $\tau_I$ are shown in Table \ref{TRForAbstraction5}.

\begin{center}
    \begin{table}
        $$\frac{x\xrightarrow{e}\surd}{\tau_I(x)\xrightarrow{e}\surd}\quad e\notin I
        \quad\frac{x\xrightarrow{e}x'}{\tau_I(x)\xrightarrow{e}\tau_I(x')}\quad e\notin I$$

        $$\frac{x\xrightarrow{e}\surd}{\tau_I(x)\xrightarrow{\tau}\surd}\quad e\in I
        \quad\frac{x\xrightarrow{e}x'}{\tau_I(x)\xrightarrow{\tau}\tau_I(x')}\quad e\in I$$

        $$\frac{x\xtworightarrow{e[m]}e}{\tau_I(x)\xtworightarrow{e[m]}e}\quad e[m]\notin I
        \quad\frac{x\xtworightarrow{e[m]}x'}{\tau_I(x)\xtworightarrow{e[m]}\tau_I(x')}\quad e[m]\notin I$$

        $$\frac{x\xtworightarrow{e[m]}\surd}{\tau_I(x)\xtworightarrow{\tau}\surd}\quad e[m]\in I
        \quad\frac{x\xtworightarrow{e[m]}x'}{\tau_I(x)\xtworightarrow{\tau}\tau_I(x')}\quad e[m]\in I$$
        \caption{Transition rule of the abstraction operator}
        \label{TRForAbstraction5}
    \end{table}
\end{center}

\begin{theorem}[Conservitivity of $APRTC_{\tau}$ with guarded linear recursion]
$APRTC_{\tau}$ with guarded linear recursion is a conservative extension of APRTC with silent step and guarded linear recursion.
\end{theorem}

\begin{theorem}[Congruence theorem of $APRTC_{\tau}$ with guarded linear recursion]
FR Rooted branching truly concurrent bisimulation equivalences $\approx_{rbp}^{fr}$, $\approx_{rbs}^{fr}$, $\approx_{rbhp}^{fr}$ and $\approx_{rbhhp}^{fr}$ are all congruences with respect to $APRTC_{\tau}$.
\end{theorem}

We design the axioms for the abstraction operator $\tau_I$ in Table \ref{AxiomsForAbstraction5}.

\begin{center}
\begin{table}
  \begin{tabular}{@{}ll@{}}
\hline No. &Axiom\\
  $TI1$ & $e\notin I\quad \tau_I(e)=e$\\
  $RTI1$ & $e[m]\notin I\quad \tau_I(e[m])=e[m]$\\
  $TI2$ & $e\in I\quad \tau_I(e)=\tau$\\
  $RTI2$ & $e[m]\in I\quad \tau_I(e[m])=\tau$\\
  $TI3$ & $\tau_I(\delta)=\delta$\\
  $TI4$ & $\tau_I(x+y)=\tau_I(x)+\tau_I(y)$\\
  $TI5$ & $\tau_I(x\cdot y)=\tau_I(x)\cdot\tau_I(y)$\\
  $TI6$ & $\tau_I(x\parallel y)=\tau_I(x)\parallel\tau_I(y)$\\
\end{tabular}
\caption{Axioms of abstraction operator}
\label{AxiomsForAbstraction5}
\end{table}
\end{center}

\begin{theorem}[Soundness of $APRTC_{\tau}$ with guarded linear recursion]\label{SAPRTCABS5}
Let $x$ and $y$ be $APRTC_{\tau}$ with guarded linear recursion terms. If $APRTC_{\tau}$ with guarded linear recursion $\vdash x=y$, then
\begin{enumerate}
  \item $x\approx_{rbs}^{fr} y$;
  \item $x\approx_{rbp}^{fr} y$;
  \item $x\approx_{rbhp}^{fr} y$;
  \item $x\approx_{rbhp}^{fr} y$.
\end{enumerate}
\end{theorem}

\begin{definition}[Cluster]\label{CLUSTER5}
Let $E$ be a guarded linear recursive specification, and $I\subseteq \mathbb{E}$. Two recursion variable $X$ and $Y$ in $E$ are in the same cluster for $I$ iff there exist sequences of transitions $\langle X|E\rangle\xrightarrow{\{b_{11},\cdots, b_{1i}\}}\cdots\xrightarrow{\{b_{m1},\cdots, b_{mi}\}}\langle Y|E\rangle$ and $\langle Y|E\rangle\xrightarrow{\{c_{11},\cdots, c_{1j}\}}\cdots\xrightarrow{\{c_{n1},\cdots, c_{nj}\}}\langle X|E\rangle$, or $\langle X|E\rangle\xtworightarrow{\{b_{11}[m],\cdots, b_{1i}[m]\}}\cdots\xtworightarrow{\{b_{m1}[m],\cdots, b_{mi}[m]\}}\langle Y|E\rangle$ and $\langle Y|E\rangle\xtworightarrow{\{c_{11}[n],\cdots, c_{1j}[n]\}}\cdots\xtworightarrow{\{c_{n1}[n],\cdots, c_{nj}[n]\}}\langle X|E\rangle$, where $b_{11},\cdots,b_{mi},c_{11},\cdots,c_{nj}, b_{11}[m],\cdots,b_{mi}[m],c_{11}[n],\cdots,c_{nj}[n]\in I\cup\{\tau\}$.

$a_1\parallel\cdots\parallel a_k$, or $(a_1\parallel\cdots\parallel a_k) X$, or $a_1[m]\parallel\cdots\parallel a_k[m]$, or $X (a_1[m]\parallel\cdots\parallel a_k[m])$ is an exit for the cluster $C$ iff: (1) $a_1\parallel\cdots\parallel a_k$, or $(a_1\parallel\cdots\parallel a_k) X$, or $a_1[m]\parallel\cdots\parallel a_k[m]$, or $X (a_1[m]\parallel\cdots\parallel a_k[m])$ is a summand at the right-hand side of the recursive equation for a recursion variable in $C$, and (2) in the case of $(a_1\parallel\cdots\parallel a_k) X$, and $X (a_1[m]\parallel\cdots\parallel a_k[m])$ either $a_l, a_l[m]\notin I\cup\{\tau\}(l\in\{1,2,\cdots,k\})$ or $X\notin C$.
\end{definition}

\begin{center}
\begin{table}
  \begin{tabular}{@{}ll@{}}
\hline No. &Axiom\\
  CFAR & If $X$ is in a cluster for $I$ with exits \\
           & $\{(a_{11}\leftmerge\cdots\leftmerge a_{1i})Y_1,\cdots,(a_{m1}\leftmerge\cdots\leftmerge a_{mi})Y_m, b_{11}\leftmerge\cdots\leftmerge b_{1j},\cdots,b_{n1}\leftmerge\cdots\leftmerge b_{nj}\}$, \\
           & then $\tau\cdot\tau_I(\langle X|E\rangle)=$\\
           & $\tau\cdot\tau_I((a_{11}\leftmerge\cdots\leftmerge a_{1i})\langle Y_1|E\rangle+\cdots+(a_{m1}\leftmerge\cdots\leftmerge a_{mi})\langle Y_m|E\rangle+b_{11}\leftmerge\cdots\leftmerge b_{1j}+\cdots+b_{n1}\leftmerge\cdots\leftmerge b_{nj})$\\
           & Or exists,\\
           & $\{Y_1(a_{11}[m]\leftmerge\cdots\leftmerge a_{1i}[m1]),\cdots,Y_m(a_{m1}[mm]\leftmerge\cdots\leftmerge a_{mi}[mm]),$\\
           & $b_{11}[n1]\leftmerge\cdots\leftmerge b_{1j}[n1],\cdots,b_{n1}[nn]\leftmerge\cdots\leftmerge b_{nj}[nn]\}$, \\
           & then $\tau_I(\langle X|E\rangle)\cdot\tau=$\\
           & $\tau_I(\langle Y_1|E\rangle(a_{11}[m1]\leftmerge\cdots\leftmerge a_{1i}[m1])+\cdots+\langle Y_m|E\rangle(a_{m1}[mm]\leftmerge\cdots\leftmerge a_{mi}[mm])$\\
           & $+b_{11}[n1]\leftmerge\cdots\leftmerge b_{1j}[n1]+\cdots+b_{n1}[nn]\leftmerge\cdots\leftmerge b_{nj}[nn])\cdot\tau$\\
\end{tabular}
\caption{Cluster fair abstraction rule}
\label{CFAR5}
\end{table}
\end{center}

\begin{theorem}[Soundness of CFAR]\label{SCFAR5}
CFAR is sound modulo rooted branching FR truly concurrent bisimulation equivalences $\approx_{rbs}^{fr}$, $\approx_{rbp}^{fr}$, $\approx_{rbhp}^{fr}$ and $\approx_{rbhhp}^{fr}$.
\end{theorem}

\begin{theorem}[Completeness of $APRTC_{\tau}$ with guarded linear recursion and CFAR]\label{CCFAR5}
Let $p$ and $q$ be closed $APRTC_{\tau}$ with guarded linear recursion and CFAR terms, then,
\begin{enumerate}
  \item if $p\approx_{rbs}^{fr} q$ then $p=q$;
  \item if $p\approx_{rbp}^{fr} q$ then $p=q$;
  \item if $p\approx_{rbhp}^{fr} q$ then $p=q$;
  \item if $p\approx_{rbhhp}^{fr} q$ then $p=q$.
\end{enumerate}
\end{theorem}

\subsection{Probabilistic Truly Concurrent Process Algebra -- APPTC}

The theory $APPTC$ (Algebra of Probabilistic Processes for True Concurrency) has four modules: $BAPTC$ (Basic Algebra for Probabilistic True Concurrency), $APPTC$ (Algebra for Parallelism
in Probabilistic True Concurrency), recursion and abstraction.

\begin{definition}[Probabilistic prime event structure with silent event]\label{PPES}
Let $\Lambda$ be a fixed set of labels, ranged over $a,b,c,\cdots$ and $\tau$. A ($\Lambda$-labelled) prime event structure with silent event $\tau$ is a quintuple
$\mathcal{E}=\langle \mathbb{E}, \leq, \sharp, \sharp_{\pi}, \lambda\rangle$, where $\mathbb{E}$ is a denumerable set of events, including the silent event $\tau$. Let
$\hat{\mathbb{E}}=\mathbb{E}\backslash\{\tau\}$, exactly excluding $\tau$, it is obvious that $\hat{\tau^*}=\epsilon$, where $\epsilon$ is the empty event.
Let $\lambda:\mathbb{E}\rightarrow\Lambda$ be a labelling function and let $\lambda(\tau)=\tau$. And $\leq$, $\sharp$, $\sharp_{\pi}$ are binary relations on $\mathbb{E}$,
called causality, conflict and probabilistic conflict respectively, such that:

\begin{enumerate}
  \item $\leq$ is a partial order and $\lceil e \rceil = \{e'\in \mathbb{E}|e'\leq e\}$ is finite for all $e\in \mathbb{E}$. It is easy to see that
  $e\leq\tau^*\leq e'=e\leq\tau\leq\cdots\leq\tau\leq e'$, then $e\leq e'$.
  \item $\sharp$ is irreflexive, symmetric and hereditary with respect to $\leq$, that is, for all $e,e',e''\in \mathbb{E}$, if $e\sharp e'\leq e''$, then $e\sharp e''$;
  \item $\sharp_{\pi}$ is irreflexive, symmetric and hereditary with respect to $\leq$, that is, for all $e,e',e''\in \mathbb{E}$, if $e\sharp_{\pi} e'\leq e''$, then $e\sharp_{\pi} e''$.
\end{enumerate}

Then, the concepts of consistency and concurrency can be drawn from the above definition:

\begin{enumerate}
  \item $e,e'\in \mathbb{E}$ are consistent, denoted as $e\frown e'$, if $\neg(e\sharp e')$ and $\neg(e\sharp_{\pi} e')$. A subset $X\subseteq \mathbb{E}$ is called consistent, if $e\frown e'$ for all
  $e,e'\in X$.
  \item $e,e'\in \mathbb{E}$ are concurrent, denoted as $e\parallel e'$, if $\neg(e\leq e')$, $\neg(e'\leq e)$, and $\neg(e\sharp e')$ and $\neg(e\sharp_{\pi} e')$.
\end{enumerate}
\end{definition}

\begin{definition}[Configuration]
Let $\mathcal{E}$ be a PES. A (finite) configuration in $\mathcal{E}$ is a (finite) consistent subset of events $C\subseteq \mathcal{E}$, closed with respect to causality
(i.e. $\lceil C\rceil=C$). The set of finite configurations of $\mathcal{E}$ is denoted by $\mathcal{C}(\mathcal{E})$. We let $\hat{C}=C\backslash\{\tau\}$.
\end{definition}

A consistent subset of $X\subseteq \mathbb{E}$ of events can be seen as a pomset. Given $X, Y\subseteq \mathbb{E}$, $\hat{X}\sim \hat{Y}$ if $\hat{X}$ and $\hat{Y}$ are
isomorphic as pomsets. In the following of the paper, we say $C_1\sim C_2$, we mean $\hat{C_1}\sim\hat{C_2}$.

\begin{definition}[Pomset transitions and step]
Let $\mathcal{E}$ be a PES and let $C\in\mathcal{C}(\mathcal{E})$, and $\emptyset\neq X\subseteq \mathbb{E}$, if $C\cap X=\emptyset$ and $C'=C\cup X\in\mathcal{C}(\mathcal{E})$,
then $C\xrightarrow{X} C'$ is called a pomset transition from $C$ to $C'$. When the events in $X$ are pairwise concurrent, we say that $C\xrightarrow{X}C'$ is a step.
\end{definition}

\begin{definition}[Probabilistic transitions]
Let $\mathcal{E}$ be a PES and let $C\in\mathcal{C}(\mathcal{E})$, the transition $C\xrsquigarrow{\pi} C^{\pi}$ is called a probabilistic transition from $C$ to $C^{\pi}$.
\end{definition}

\begin{definition}[Weak pomset transitions and weak step]
Let $\mathcal{E}$ be a PES and let $C\in\mathcal{C}(\mathcal{E})$, and $\emptyset\neq X\subseteq \hat{\mathbb{E}}$, if $C\cap X=\emptyset$ and
$\hat{C'}=\hat{C}\cup X\in\mathcal{C}(\mathcal{E})$, then $C\xRightarrow{X} C'$ is called a weak pomset transition from $C$ to $C'$, where we define
$\xRightarrow{e}\triangleq\xrightarrow{\tau^*}\xrightarrow{e}\xrightarrow{\tau^*}$. And $\xRightarrow{X}\triangleq\xrightarrow{\tau^*}\xrightarrow{e}\xrightarrow{\tau^*}$,
for every $e\in X$. When the events in $X$ are pairwise concurrent, we say that $C\xRightarrow{X}C'$ is a weak step.
\end{definition}

We will also suppose that all the PESs in this book are image finite, that is, for any PES $\mathcal{E}$ and $C\in \mathcal{C}(\mathcal{E})$ and $a\in \Lambda$,
$\{\langle C,s\rangle\xrsquigarrow{\pi} \langle C^{\pi},s\rangle\}$,
$\{e\in \mathbb{E}|\langle C,s\rangle\xrightarrow{e} \langle C',s'\rangle\wedge \lambda(e)=a\}$ and
$\{e\in\hat{\mathbb{E}}|\langle C,s\rangle\xRightarrow{e} \langle C',s'\rangle\wedge \lambda(e)=a\}$ is finite.

A probability distribution function (PDF) $\mu$ is a map $\mu:\mathcal{C}\times\mathcal{C}\rightarrow[0,1]$ and $\mu^*$ is the cumulative probability distribution function (cPDF).

\begin{definition}[Probabilistic pomset, step bisimulation]\label{PPSB}
Let $\mathcal{E}_1$, $\mathcal{E}_2$ be PESs. A probabilistic pomset bisimulation is a relation $R\subseteq\mathcal{C}(\mathcal{E}_1)\times\mathcal{C}(\mathcal{E}_2)$, such that (1) if
$(C_1,C_2)\in R$, and $C_1\xrightarrow{X_1}C_1'$ then $C_2\xrightarrow{X_2}C_2'$, with $X_1\subseteq \mathbb{E}_1$, $X_2\subseteq \mathbb{E}_2$, $X_1\sim X_2$ and $(C_1',C_2')\in R$,
and vice-versa; (2) if $(C_1,C_2)\in R$, and $C_1\xrsquigarrow{\pi}C_1^{\pi}$ then $C_2\xrsquigarrow{\pi}C_2^{\pi}$ and $(C_1^{\pi},C_2^{\pi})\in R$, and vice-versa; (3) if $(C_1,C_2)\in R$,
then $\mu(C_1,C)=\mu(C_2,C)$ for each $C\in\mathcal{C}(\mathcal{E})/R$; (4) $[\surd]_R=\{\surd\}$. We say that $\mathcal{E}_1$, $\mathcal{E}_2$ are probabilistic pomset bisimilar, written $\mathcal{E}_1\sim_{pp}\mathcal{E}_2$,
if there exists a probabilistic pomset bisimulation $R$, such that
$(\emptyset,\emptyset)\in R$. By replacing probabilistic pomset transitions with steps, we can get the definition of probabilistic step bisimulation. When PESs $\mathcal{E}_1$ and $\mathcal{E}_2$ are probabilistic step
bisimilar, we write $\mathcal{E}_1\sim_{ps}\mathcal{E}_2$.
\end{definition}

\begin{definition}[Posetal product]
Given two PESs $\mathcal{E}_1$, $\mathcal{E}_2$, the posetal product of their configurations, denoted $\mathcal{C}(\mathcal{E}_1)\overline{\times}\mathcal{C}(\mathcal{E}_2)$,
is defined as

$$\{(C_1,f,C_2)|C_1\in\mathcal{C}(\mathcal{E}_1),C_2\in\mathcal{C}(\mathcal{E}_2),f:C_1\rightarrow C_2 \textrm{ isomorphism}\}.$$

A subset $R\subseteq\mathcal{C}(\mathcal{E}_1)\overline{\times}\mathcal{C}(\mathcal{E}_2)$ is called a posetal relation. We say that $R$ is downward closed when for any
$(C_1,f,C_2),(C_1',f',C_2')\in \mathcal{C}(\mathcal{E}_1)\overline{\times}\mathcal{C}(\mathcal{E}_2)$, if $(C_1,f,C_2)\subseteq (C_1',f',C_2')$ pointwise and $(C_1',f',C_2')\in R$,
then $(C_1,f,C_2)\in R$.

For $f:X_1\rightarrow X_2$, we define $f[x_1\mapsto x_2]:X_1\cup\{x_1\}\rightarrow X_2\cup\{x_2\}$, $z\in X_1\cup\{x_1\}$,(1)$f[x_1\mapsto x_2](z)=
x_2$,if $z=x_1$;(2)$f[x_1\mapsto x_2](z)=f(z)$, otherwise. Where $X_1\subseteq \mathbb{E}_1$, $X_2\subseteq \mathbb{E}_2$, $x_1\in \mathbb{E}_1$, $x_2\in \mathbb{E}_2$.
\end{definition}

\begin{definition}[Probabilistic (hereditary) history-preserving bisimulation]\label{PHHPB}
A probabilistic history-preserving (hp-) bisimulation is a posetal relation $R\subseteq\mathcal{C}(\mathcal{E}_1)\overline{\times}\mathcal{C}(\mathcal{E}_2)$ such that (1) if $(C_1,f,C_2)\in R$,
and $C_1\xrightarrow{e_1} C_1'$, then $C_2\xrightarrow{e_2} C_2'$, with $(C_1',f[e_1\mapsto e_2],C_2')\in R$, and vice-versa; (2) if $(C_1,f,C_2)\in R$, and
$C_1\xrsquigarrow{\pi}C_1^{\pi}$ then $C_2\xrsquigarrow{\pi}C_2^{\pi}$ and $(C_1^{\pi},f,C_2^{\pi})\in R$, and vice-versa; (3) if $(C_1,f,C_2)\in R$,
then $\mu(C_1,C)=\mu(C_2,C)$ for each $C\in\mathcal{C}(\mathcal{E})/R$; (4) $[\surd]_R=\{\surd\}$. $\mathcal{E}_1,\mathcal{E}_2$ are probabilistic history-preserving
(hp-)bisimilar and are written $\mathcal{E}_1\sim_{php}\mathcal{E}_2$ if there exists a probabilistic hp-bisimulation $R$ such that $(\emptyset,\emptyset,\emptyset)\in R$.

A probabilistic hereditary history-preserving (hhp-)bisimulation is a downward closed probabilistic hp-bisimulation. $\mathcal{E}_1,\mathcal{E}_2$ are probabilistic hereditary history-preserving (hhp-)bisimilar and are
written $\mathcal{E}_1\sim_{phhp}\mathcal{E}_2$.
\end{definition}

\begin{definition}[Weakly posetal product]
Given two PESs $\mathcal{E}_1$, $\mathcal{E}_2$, the weakly posetal product of their configurations, denoted $\mathcal{C}(\mathcal{E}_1)\overline{\times}\mathcal{C}(\mathcal{E}_2)$,
is defined as

$$\{(C_1,f,C_2)|C_1\in\mathcal{C}(\mathcal{E}_1),C_2\in\mathcal{C}(\mathcal{E}_2),f:\hat{C_1}\rightarrow \hat{C_2} \textrm{ isomorphism}\}.$$

A subset $R\subseteq\mathcal{C}(\mathcal{E}_1)\overline{\times}\mathcal{C}(\mathcal{E}_2)$ is called a weakly posetal relation. We say that $R$ is downward closed when for any
$(C_1,f,C_2),(C_1',f,C_2')\in \mathcal{C}(\mathcal{E}_1)\overline{\times}\mathcal{C}(\mathcal{E}_2)$, if $(C_1,f,C_2)\subseteq (C_1',f',C_2')$ pointwise and $(C_1',f',C_2')\in R$,
then $(C_1,f,C_2)\in R$.

For $f:X_1\rightarrow X_2$, we define $f[x_1\mapsto x_2]:X_1\cup\{x_1\}\rightarrow X_2\cup\{x_2\}$, $z\in X_1\cup\{x_1\}$,(1)$f[x_1\mapsto x_2](z)=
x_2$,if $z=x_1$;(2)$f[x_1\mapsto x_2](z)=f(z)$, otherwise. Where $X_1\subseteq \hat{\mathbb{E}_1}$, $X_2\subseteq \hat{\mathbb{E}_2}$, $x_1\in \hat{\mathbb{E}}_1$,
$x_2\in \hat{\mathbb{E}}_2$. Also, we define $f(\tau^*)=f(\tau^*)$.
\end{definition}

\begin{definition}[Weak pomset transitions and weak step]
Let $\mathcal{E}$ be a PES and let $C\in\mathcal{C}(\mathcal{E})$, and $\emptyset\neq X\subseteq \hat{\mathbb{E}}$, if $C\cap X=\emptyset$ and
$\hat{C'}=\hat{C}\cup X\in\mathcal{C}(\mathcal{E})$, then $C\xRightarrow{X} C'$ is called a weak pomset transition from $C$ to $C'$, where we define
$\xRightarrow{e}\triangleq\xrightarrow{\tau^*}\xrightarrow{e}\xrightarrow{\tau^*}$. And $\xRightarrow{X}\triangleq\xrightarrow{\tau^*}\xrightarrow{e}\xrightarrow{\tau^*}$,
for every $e\in X$. When the events in $X$ are pairwise concurrent, we say that $C\xRightarrow{X}C'$ is a weak step.
\end{definition}

\begin{definition}[Probabilistic branching pomset, step bisimulation]\label{PBPSB}
Assume a special termination predicate $\downarrow$, and let $\surd$ represent a state with $\surd\downarrow$. Let $\mathcal{E}_1$, $\mathcal{E}_2$ be PESs. A probabilistic branching pomset
bisimulation is a relation $R\subseteq\mathcal{C}(\mathcal{E}_1)\times\mathcal{C}(\mathcal{E}_2)$, such that:

 \begin{enumerate}
   \item if $(C_1,C_2)\in R$, and $C_1\xrightarrow{X}C_1'$ then
   \begin{itemize}
     \item either $X\equiv \tau^*$, and $(C_1',C_2)\in R$;
     \item or there is a sequence of (zero or more) probabilistic transitions and $\tau$-transitions $C_2\rightsquigarrow^*\xrightarrow{\tau^*} C_2^0$, such that $(C_1,C_2^0)\in R$ and $C_2^0\xRightarrow{X}C_2'$ with
     $(C_1',C_2')\in R$;
   \end{itemize}
   \item if $(C_1,C_2)\in R$, and $C_2\xrightarrow{X}C_2'$ then
   \begin{itemize}
     \item either $X\equiv \tau^*$, and $(C_1,C_2')\in R$;
     \item or there is a sequence of (zero or more) probabilistic transitions and $\tau$-transitions $C_1\rightsquigarrow^*\xrightarrow{\tau^*} C_1^0$, such that $(C_1^0,C_2)\in R$ and $C_1^0\xRightarrow{X}C_1'$ with
     $(C_1',C_2')\in R$;
   \end{itemize}
   \item if $(C_1,C_2)\in R$ and $C_1\downarrow$, then there is a sequence of (zero or more) probabilistic transitions and $\tau$-transitions $C_2\rightsquigarrow^*\xrightarrow{\tau^*} C_2^0$ such that $(C_1,C_2^0)\in R$
   and $C_2^0\downarrow$;
   \item if $(C_1,C_2)\in R$ and $C_2\downarrow$, then there is a sequence of (zero or more) probabilistic transitions and $\tau$-transitions $C_1\rightsquigarrow^*\xrightarrow{\tau^*} C_1^0$ such that $(C_1^0,C_2)\in R$
   and $C_1^0\downarrow$.
   \item if $(C_1,C_2)\in R$,then $\mu(C_1,C)=\mu(C_2,C)$ for each $C\in\mathcal{C}(\mathcal{E})/R$;
   \item $[\surd]_R=\{\surd\}$.
 \end{enumerate}

We say that $\mathcal{E}_1$, $\mathcal{E}_2$ are probabilistic branching pomset bisimilar, written $\mathcal{E}_1\approx_{pbp}\mathcal{E}_2$, if there exists a probabilistic branching pomset bisimulation $R$,
such that $(\emptyset,\emptyset)\in R$.

By replacing probabilistic branching pomset transitions with steps, we can get the definition of probabilistic branching step bisimulation. When PESs $\mathcal{E}_1$ and $\mathcal{E}_2$ are probabilistic branching step bisimilar,
we write $\mathcal{E}_1\approx_{pbs}\mathcal{E}_2$.
\end{definition}

\begin{definition}[Probabilistic rooted branching pomset, step bisimulation]\label{PRBPSB}
Assume a special termination predicate $\downarrow$, and let $\surd$ represent a state with $\surd\downarrow$. Let $\mathcal{E}_1$, $\mathcal{E}_2$ be PESs. A branching pomset
bisimulation is a relation $R\subseteq\mathcal{C}(\mathcal{E}_1)\times\mathcal{C}(\mathcal{E}_2)$, such that:

 \begin{enumerate}
   \item if $(C_1,C_2)\in R$, and $C_1\rightsquigarrow C_1^{\pi}\xrightarrow{X}C_1'$ then $C_2\rightsquigarrow C_2^{\pi}\xrightarrow{X}C_2'$ with $C_1'\approx_{pbp}C_2'$;
   \item if $(C_1,C_2)\in R$, and $C_2\rightsquigarrow C_2^{\pi}\xrightarrow{X}C_2'$ then $C_1\rightsquigarrow C_1^{\pi}\xrightarrow{X}C_1'$ with $C_1'\approx_{pbp}C_2'$;
   \item if $(C_1,C_2)\in R$ and $C_1\downarrow$, then $C_2\downarrow$;
   \item if $(C_1,C_2)\in R$ and $C_2\downarrow$, then $C_1\downarrow$.
 \end{enumerate}

We say that $\mathcal{E}_1$, $\mathcal{E}_2$ are probabilistic rooted branching pomset bisimilar, written $\mathcal{E}_1\approx_{prbp}\mathcal{E}_2$, if there exists a probabilistic rooted branching pomset
bisimulation $R$, such that $(\emptyset,\emptyset)\in R$.

By replacing probabilistic pomset transitions with steps, we can get the definition of probabilistic rooted branching step bisimulation. When PESs $\mathcal{E}_1$ and $\mathcal{E}_2$ are probabilistic rooted branching step
bisimilar, we write $\mathcal{E}_1\approx_{prbs}\mathcal{E}_2$.
\end{definition}

\begin{definition}[Probabilistic branching (hereditary) history-preserving bisimulation]\label{PBHHPB}
Assume a special termination predicate $\downarrow$, and let $\surd$ represent a state with $\surd\downarrow$. A probabilistic branching history-preserving (hp-) bisimulation is a weakly
posetal relation $R\subseteq\mathcal{C}(\mathcal{E}_1)\overline{\times}\mathcal{C}(\mathcal{E}_2)$ such that:

 \begin{enumerate}
   \item if $(C_1,f,C_2)\in R$, and $C_1\xrightarrow{e_1}C_1'$ then
   \begin{itemize}
     \item either $e_1\equiv \tau$, and $(C_1',f[e_1\mapsto \tau],C_2)\in R$;
     \item or there is a sequence of (zero or more) probabilistic transitions and $\tau$-transitions $C_2\rightsquigarrow^*\xrightarrow{\tau^*} C_2^0$, such that $(C_1,f,C_2^0)\in R$ and $C_2^0\xrightarrow{e_2}C_2'$ with
     $(C_1',f[e_1\mapsto e_2],C_2')\in R$;
   \end{itemize}
   \item if $(C_1,f,C_2)\in R$, and $C_2\xrightarrow{e_2}C_2'$ then
   \begin{itemize}
     \item either $X\equiv \tau$, and $(C_1,f[e_2\mapsto \tau],C_2')\in R$;
     \item or there is a sequence of (zero or more) probabilistic transitions and $\tau$-transitions $C_1\rightsquigarrow^*\xrightarrow{\tau^*} C_1^0$, such that $(C_1^0,f,C_2)\in R$ and $C_1^0\xrightarrow{e_1}C_1'$ with
     $(C_1',f[e_2\mapsto e_1],C_2')\in R$;
   \end{itemize}
   \item if $(C_1,f,C_2)\in R$ and $C_1\downarrow$, then there is a sequence of (zero or more) probabilistic transitions and $\tau$-transitions $C_2\rightsquigarrow^*\xrightarrow{\tau^*} C_2^0$ such that $(C_1,f,C_2^0)\in R$
   and $C_2^0\downarrow$;
   \item if $(C_1,f,C_2)\in R$ and $C_2\downarrow$, then there is a sequence of (zero or more) probabilistic transitions and $\tau$-transitions $C_1\rightsquigarrow^*\xrightarrow{\tau^*} C_1^0$ such that $(C_1^0,f,C_2)\in R$
   and $C_1^0\downarrow$;
   \item if $(C_1,C_2)\in R$,then $\mu(C_1,C)=\mu(C_2,C)$ for each $C\in\mathcal{C}(\mathcal{E})/R$;
   \item $[\surd]_R=\{\surd\}$.
 \end{enumerate}

$\mathcal{E}_1,\mathcal{E}_2$ are probabilistic branching history-preserving (hp-)bisimilar and are written $\mathcal{E}_1\approx_{pbhp}\mathcal{E}_2$ if there exists a probabilistic branching hp-bisimulation
$R$ such that $(\emptyset,\emptyset,\emptyset)\in R$.

A probabilistic branching hereditary history-preserving (hhp-)bisimulation is a downward closed probabilistic branching hhp-bisimulation. $\mathcal{E}_1,\mathcal{E}_2$ are probabilistic branching hereditary history-preserving
(hhp-)bisimilar and are written $\mathcal{E}_1\approx_{pbhhp}\mathcal{E}_2$.
\end{definition}

\begin{definition}[Probabilistic rooted branching (hereditary) history-preserving bisimulation]\label{PRBHHPB}
Assume a special termination predicate $\downarrow$, and let $\surd$ represent a state with $\surd\downarrow$. A probabilistic rooted branching history-preserving (hp-) bisimulation is a
posetal relation $R\subseteq\mathcal{C}(\mathcal{E}_1)\overline{\times}\mathcal{C}(\mathcal{E}_2)$ such that:

 \begin{enumerate}
   \item if $(C_1,f,C_2)\in R$, and $C_1\rightsquigarrow C_1^{\pi}\xrightarrow{e_1}C_1'$, then $C_2\rightsquigarrow C_2^{\pi}\xrightarrow{e_2}C_2'$ with $C_1'\approx_{pbhp}C_2'$;
   \item if $(C_1,f,C_2)\in R$, and $C_2\rightsquigarrow C_2^{\pi}\xrightarrow{e_2}C_1'$, then $C_1\rightsquigarrow C_1^{\pi}\xrightarrow{e_1}C_2'$ with $C_1'\approx_{pbhp}C_2'$;
   \item if $(C_1,f,C_2)\in R$ and $C_1\downarrow$, then $C_2\downarrow$;
   \item if $(C_1,f,C_2)\in R$ and $C_2\downarrow$, then $C_1\downarrow$.
 \end{enumerate}

$\mathcal{E}_1,\mathcal{E}_2$ are probabilistic rooted branching history-preserving (hp-)bisimilar and are written $\mathcal{E}_1\approx_{prbhp}\mathcal{E}_2$ if there exists a probabilistic rooted branching
hp-bisimulation $R$ such that $(\emptyset,\emptyset,\emptyset)\in R$.

A probabilistic rooted branching hereditary history-preserving (hhp-)bisimulation is a downward closed probabilistic rooted branching hhp-bisimulation. $\mathcal{E}_1,\mathcal{E}_2$ are probabilistic rooted branching
hereditary history-preserving (hhp-)bisimilar and are written $\mathcal{E}_1\approx_{prbhhp}\mathcal{E}_2$.
\end{definition}

\subsubsection{Basic Algebra for Probabilistic True Concurrency}

In this section, we will discuss the algebraic laws for prime event structure $\mathcal{E}$, exactly for causality $\leq$, conflict $\sharp$ and probabilistic conflict $\sharp_{\pi}$.
We will follow the conventions of process algebra, using $\cdot$ instead of $\leq$, $+$ instead of $\sharp$ and $\boxplus_{\pi}$ instead of $\sharp_{\pi}$. The resulted algebra is called
Basic Algebra for Probabilistic True Concurrency, abbreviated $BAPTC$.

In the following, the variables $x,x',y,y',z,z'$ range over the collection of process terms, $s,s',t,t',u,u'$ are closed terms, $\tau$ is the special constant silent step, $\delta$
is the special constant deadlock, $A$ is the collection of atomic actions, atomic actions $a,b\in A$, $A_{\delta}=A\cup\{\delta\}$, $A_{\tau}=A\cup\{\tau\}$.
$\rightsquigarrow$ denotes probabilistic transition, and action transition labelled by an atomic action $a\in A$, $\xrightarrow{a}$ and $\xrightarrow{a}\surd$.
$x\xrightarrow{a}p$ means that by performing action $a$ process $x$ evolves into $p$; while $x\xrightarrow{a}\surd$ means that $x$ performs an $a$ action and then terminates.
$p\rightsquigarrow x$ denotes that process $p$ chooses to behave like process $x$ with a non-zero probability $\pi >0$.

\begin{center}
    \begin{table}
        \begin{tabular}{@{}ll@{}}
            \hline No. &Axiom\\
            $A1$ & $x+ y = y+ x$\\
            $A2$ & $(x+ y)+ z = x+ (y+ z)$\\
            $A3$ & $x+ x = x$\\
            $A4$ & $(x+ y)\cdot z = x\cdot z + y\cdot z$\\
            $A5$ & $(x\cdot y)\cdot z = x\cdot(y\cdot z)$\\
            $PA1$ & $x\boxplus_{\pi} y=y\boxplus_{1-\pi} x$\\
            $PA2$ & $x\boxplus_{\pi}(y\boxplus_{\rho} z)=(x\boxplus_{\frac{\pi}{\pi+\rho-\pi\rho}}y)\boxplus_{\pi+\rho-\pi\rho} z$\\
            $PA3$ & $x\boxplus_{\pi}x=x$\\
            $PA4$ & $(x\boxplus_{\pi}y)\cdot z=x\cdot z\boxplus_{\pi}y\cdot z$\\
            $PA5$ & $(x\boxplus_{\pi}y)+z=(x+z)\boxplus_{\pi}(y+z)$\\
        \end{tabular}
        \caption{Axioms of $BAPTC$}
        \label{AxiomsForBAPTC}
    \end{table}
\end{center}

\begin{definition}[Basic terms of $BAPTC$]
The set of basic terms of $BAPTC$, $\mathcal{B}(BAPTC)$, is inductively defined as follows:

\begin{enumerate}
  \item $\mathbb{E}\subset\mathcal{B}(BAPTC)$;
  \item if $e\in \mathbb{E}, t\in\mathcal{B}(BAPTC)$ then $e\cdot t\in\mathcal{B}(BAPTC)$;
  \item if $t,s\in\mathcal{B}(BAPTC)$ then $t+ s\in\mathcal{B}(BAPTC)$;
  \item if $t,s\in\mathcal{B}(BAPTC)$ then $t\boxplus_{\pi} s\in\mathcal{B}(BAPTC)$.
\end{enumerate}
\end{definition}

\begin{theorem}[Elimination theorem of $BAPTC$]
Let $p$ be a closed $BAPTC$ term. Then there is a basic $BAPTC$ term $q$ such that $BAPTC\vdash p=q$.
\end{theorem}

In this subsection, we will define a term-deduction system which gives the operational semantics of $BAPTC$. Like the way in \cite{PPA}, we also introduce the counterpart $\breve{e}$
of the event $e$, and also the set $\breve{\mathbb{E}}=\{\breve{e}|e\in\mathbb{E}\}$.

We give the definition of PDFs of $BAPTC$ in Table \ref{PDFBAPTC}.

\begin{center}
    \begin{table}
        $$\mu(e,\breve{e})=1$$
        $$\mu(x\cdot y, x'\cdot y)=\mu(x,x')$$
        $$\mu(x+y,x'+y')=\mu(x,x')\cdot \mu(y,y')$$
        $$\mu(x\boxplus_{\pi}y,z)=\pi\mu(x,z)+(1-\pi)\mu(y,z)$$
        $$\mu(x,y)=0,\textrm{otherwise}$$
        \caption{PDF definitions of $BAPTC$}
        \label{PDFBAPTC}
    \end{table}
\end{center}

We give the operational transition rules for operators $\cdot$, $+$ and $\boxplus_{\pi}$ as Table \ref{SETRForBAPTC} shows. And the predicate $\xrightarrow{e}\surd$ represents
successful termination after execution of the event $e$.

\begin{center}
    \begin{table}
        $$\frac{}{e\rightsquigarrow\breve{e}}$$
        $$\frac{x\rightsquigarrow x'}{x\cdot y\rightsquigarrow x'\cdot y}$$
        $$\frac{x\rightsquigarrow x'\quad y\rightsquigarrow y'}{x+y\rightsquigarrow x'+y'}$$
        $$\frac{x\rightsquigarrow x'}{x\boxplus_{\pi}y\rightsquigarrow x'}\quad \frac{y\rightsquigarrow y'}{x\boxplus_{\pi}y\rightsquigarrow y'}$$
        $$\frac{}{\breve{e}\xrightarrow{e}\surd}$$
        $$\frac{x\xrightarrow{e}\surd}{x+ y\xrightarrow{e}\surd} \quad\frac{x\xrightarrow{e}x'}{x+ y\xrightarrow{e}x'} \quad\frac{y\xrightarrow{e}\surd}{x+ y\xrightarrow{e}\surd} \quad\frac{y\xrightarrow{e}y'}{x+ y\xrightarrow{e}y'}$$
        $$\frac{x\xrightarrow{e}\surd}{x\cdot y\xrightarrow{e} y} \quad\frac{x\xrightarrow{e}x'}{x\cdot y\xrightarrow{e}x'\cdot y}$$
        \caption{Single event transition rules of $BAPTC$}
        \label{SETRForBAPTC}
    \end{table}
\end{center}

\begin{theorem}[Congruence of $BAPTC$ with respect to probabilistic truly concurrent bisimulation equivalences]
Probabilistic truly concurrent bisimulation equivalences $\sim_{pp}$, $\sim{ps}$, $\sim_{php}$, and $\sim_{phhp}$ are all congruences with respect to $BAPTC$.
\end{theorem}

\begin{theorem}[Soundness of $BAPTC$ modulo probabilistic truly concurrent bisimulation equivalences]
Let $x$ and $y$ be $BAPTC$ terms.

\begin{enumerate}
  \item If $BAPTC\vdash x=y$, then $x\sim_{pp} y$;
  \item If $BAPTC\vdash x=y$, then $x\sim_{ps} y$;
  \item If $BAPTC\vdash x=y$, then $x\sim_{php} y$;
  \item If $BAPTC\vdash x=y$, then $x\sim_{phhp} y$.
\end{enumerate}
\end{theorem}

\begin{theorem}[Completeness of $BAPTC$ modulo probabilistic truly concurrent bisimulation equivalences]
Let $p$ and $q$ be closed $BAPTC$ terms.

\begin{enumerate}
  \item If $p\sim_{pp} q$ then $p=q$;
  \item If $p\sim_{ps} q$ then $p=q$;
  \item If $p\sim_{php} q$ then $p=q$;
  \item If $p\sim_{phhpp} q$ then $p=q$.
\end{enumerate}
\end{theorem}

\subsubsection{Algebra for Parallelism in Probabilistic True Concurrency}

We design the axioms of parallelism in Table \ref{AxiomsForPParallelism}, including algebraic laws for parallel operator $\parallel$, communication operator $\mid$, conflict elimination
operator $\Theta$ and unless operator $\triangleleft$, and also the whole parallel operator $\between$. Since the communication between two communicating events in different parallel
branches may cause deadlock (a state of inactivity), which is caused by mismatch of two communicating events or the imperfectness of the communication channel. We introduce a new
constant $\delta$ to denote the deadlock, and let the atomic event $e\in \mathbb{E}\cup\{\delta\}$.

\begin{center}
    \begin{table}
        \begin{tabular}{@{}ll@{}}
            \hline No. &Axiom\\
            $A3$ & $e+e=e$\\
            $A6$ & $x+ \delta = x$\\
            $A7$ & $\delta\cdot x =\delta$\\
            $P1$ & $(x+x=x,y+y=y)\quad x\between y = x\parallel y + x\mid y$\\
            $P2$ & $x\parallel y = y \parallel x$\\
            $P3$ & $(x\parallel y)\parallel z = x\parallel (y\parallel z)$\\
            $P4$ & $(x+x=x,y+y=y)\quad x\parallel y = x\leftmerge y + y\leftmerge x$\\
            $P5$ & $(e_1\leq e_2)\quad e_1\leftmerge (e_2\cdot y) = (e_1\leftmerge e_2)\cdot y$\\
            $P6$ & $(e_1\leq e_2)\quad (e_1\cdot x)\leftmerge e_2 = (e_1\leftmerge e_2)\cdot x$\\
            $P7$ & $(e_1\leq e_2)\quad (e_1\cdot x)\leftmerge (e_2\cdot y) = (e_1\leftmerge e_2)\cdot (x\between y)$\\
            $P8$ & $(x+ y)\leftmerge z = (x\leftmerge z)+ (y\leftmerge z)$\\
            $P9$ & $\delta\leftmerge x = \delta$\\
            $C10$ & $e_1\mid e_2 = \gamma(e_1,e_2)$\\
            $C11$ & $e_1\mid (e_2\cdot y) = \gamma(e_1,e_2)\cdot y$\\
            $C12$ & $(e_1\cdot x)\mid e_2 = \gamma(e_1,e_2)\cdot x$\\
            $C13$ & $(e_1\cdot x)\mid (e_2\cdot y) = \gamma(e_1,e_2)\cdot (x\between y)$\\
            $C14$ & $(x+ y)\mid z = (x\mid z) + (y\mid z)$\\
            $C15$ & $x\mid (y+ z) = (x\mid y)+ (x\mid z)$\\
            $C16$ & $\delta\mid x = \delta$\\
            $C17$ & $x\mid\delta = \delta$\\
            $PM1$ & $x\parallel (y\boxplus_{\pi} z)=(x\parallel y)\boxplus_{\pi}(x\parallel z)$\\
            $PM2$ & $(x\boxplus_{\pi} y)\parallel z=(x\parallel z)\boxplus_{\pi}(y\parallel z)$\\
            $PM3$ & $x\mid (y\boxplus_{\pi} z)=(x\mid y)\boxplus_{\pi}(x\mid z)$\\
            $PM4$ & $(x\boxplus_{\pi} y)\mid z=(x\mid z)\boxplus_{\pi}(y\mid z)$\\
            $CE18$ & $\Theta(e) = e$\\
            $CE19$ & $\Theta(\delta) = \delta$\\
            $CE20$ & $\Theta(x+ y) = \Theta(x)\triangleleft y + \Theta(y)\triangleleft x$\\
            $PCE1$ & $\Theta(x\boxplus_{\pi} y) = \Theta(x)\triangleleft y \boxplus_{\pi} \Theta(y)\triangleleft x$\\
            $CE21$ & $\Theta(x\cdot y)=\Theta(x)\cdot\Theta(y)$\\
            $CE22$ & $\Theta(x\leftmerge y) = ((\Theta(x)\triangleleft y)\leftmerge y)+ ((\Theta(y)\triangleleft x)\leftmerge x)$\\
            $CE23$ & $\Theta(x\mid y) = ((\Theta(x)\triangleleft y)\mid y)+ ((\Theta(y)\triangleleft x)\mid x)$\\
            $U24$ & $(\sharp(e_1,e_2))\quad e_1\triangleleft e_2 = \tau$\\
            $U25$ & $(\sharp(e_1,e_2),e_2\leq e_3)\quad e_1\triangleleft e_3 = e_1$\\
            $U26$ & $(\sharp(e_1,e_2),e_2\leq e_3)\quad e3\triangleleft e_1 = \tau$\\
            $PU1$ & $(\sharp_{\pi}(e_1,e_2))\quad e_1\triangleleft e_2 = \tau$\\
            $PU2$ & $(\sharp_{\pi}(e_1,e_2),e_2\leq e_3)\quad e_1\triangleleft e_3 = e_1$\\
            $PU3$ & $(\sharp_{\pi}(e_1,e_2),e_2\leq e_3)\quad e_3\triangleleft e_1 = \tau$\\
            $U27$ & $e\triangleleft \delta = e$\\
            $U28$ & $\delta \triangleleft e = \delta$\\
            $U29$ & $(x+ y)\triangleleft z = (x\triangleleft z)+ (y\triangleleft z)$\\
            $PU4$ & $(x\boxplus_{\pi} y)\triangleleft z = (x\triangleleft z)\boxplus_{\pi} (y\triangleleft z)$\\
            $U30$ & $(x\cdot y)\triangleleft z = (x\triangleleft z)\cdot (y\triangleleft z)$\\
            $U31$ & $(x\leftmerge y)\triangleleft z = (x\triangleleft z)\leftmerge (y\triangleleft z)$\\
            $U32$ & $(x\mid y)\triangleleft z = (x\triangleleft z)\mid (y\triangleleft z)$\\
            $U33$ & $x\triangleleft (y+ z) = (x\triangleleft y)\triangleleft z$\\
            $PU5$ & $x\triangleleft (y\boxplus_{\pi} z) = (x\triangleleft y)\triangleleft z$\\
            $U34$ & $x\triangleleft (y\cdot z)=(x\triangleleft y)\triangleleft z$\\
            $U35$ & $x\triangleleft (y\leftmerge z) = (x\triangleleft y)\triangleleft z$\\
            $U36$ & $x\triangleleft (y\mid z) = (x\triangleleft y)\triangleleft z$\\
        \end{tabular}
        \caption{Axioms of parallelism}
        \label{AxiomsForPParallelism}
    \end{table}
\end{center}

\begin{definition}[Basic terms of $APPTC$]\label{BTAPPTC}
The set of basic terms of $APPTC$, $\mathcal{B}(APPTC)$, is inductively defined as follows:
\begin{enumerate}
  \item $\mathbb{E}\subset\mathcal{B}(APPTC)$;
  \item if $e\in \mathbb{E}, t\in\mathcal{B}(APPTC)$ then $e\cdot t\in\mathcal{B}(APPTC)$;
  \item if $t,s\in\mathcal{B}(APPTC)$ then $t+ s\in\mathcal{B}(APPTC)$;
  \item if $t,s\in\mathcal{B}(APPTC)$ then $t\boxplus_{\pi} s\in\mathcal{B}(APPTC)$;
  \item if $t,s\in\mathcal{B}(APPTC)$ then $t\parallel s\in\mathcal{B}(APPTC)$.
\end{enumerate}
\end{definition}

\begin{theorem}[Elimination theorem of parallelism]\label{ETPParallelism}
Let $p$ be a closed $APPTC$ term. Then there is a basic $APPTC$ term $q$ such that $APPTC\vdash p=q$.
\end{theorem}

We give the definition of PDFs of $APPTC$ in Table \ref{PDFAPPTC}.

\begin{center}
    \begin{table}
        $$\mu(\delta,\breve{\delta})=1$$
        $$\mu(x\between y,x'\parallel y'+x'\mid y')=\mu(x,x')\cdot\mu(y,y')$$
        $$\mu(x\parallel y,x'\leftmerge y+y'\leftmerge x)=\mu(x,x')\cdot \mu(y,y')$$
        $$\mu(x\leftmerge y, x'\leftmerge y)=\mu(x,x')$$
        $$\mu(x\mid y,x'\mid y')=\mu(x,x')\cdot \mu(y,y')$$
        $$\mu(\Theta(x),\Theta(x'))=\mu(x,x')$$
        $$\mu(x\triangleleft y, x'\triangleleft y)=\mu(x,x')$$
        $$\mu(x,y)=0,\textrm{otherwise}$$
        \caption{PDF definitions of $APPTC$}
        \label{PDFAPPTC}
    \end{table}
\end{center}

We give the transition rules of APTC in Table \ref{TRForAPPTC1}, \ref{TRForAPPTC}, it is suitable for all truly concurrent behavioral equivalence, including probabilistic pomset bisimulation,
probabilistic step bisimulation, probabilistic hp-bisimulation and probabilistic hhp-bisimulation.

\begin{center}
    \begin{table}
        $$\frac{x\rightsquigarrow x'\quad y\rightsquigarrow y'}{x\between y\rightsquigarrow x'\parallel y'+x'\mid y'}$$
        $$\frac{x\rightsquigarrow x'\quad y\rightsquigarrow y'}{x\parallel y\rightsquigarrow x'\leftmerge y+y'\leftmerge x}$$
        $$\frac{x\rightsquigarrow x'}{x\leftmerge y\rightsquigarrow x'\leftmerge y}$$
        $$\frac{x\rightsquigarrow x'\quad y\rightsquigarrow y'}{x\mid y\rightsquigarrow x'\mid y'}$$
        $$\frac{x\rightsquigarrow x'}{\Theta(x)\rightsquigarrow \Theta(x')}$$
        $$\frac{x\rightsquigarrow x'}{x\triangleleft y\rightsquigarrow x'\triangleleft y}$$
        \caption{Probabilistic transition rules of APPTC}
        \label{TRForAPPTC1}
    \end{table}
\end{center}

\begin{center}
    \begin{table}
        $$\frac{x\xrightarrow{e_1}\surd\quad y\xrightarrow{e_2}\surd}{x\parallel y\xrightarrow{\{e_1,e_2\}}\surd} \quad\frac{x\xrightarrow{e_1}x'\quad y\xrightarrow{e_2}\surd}{x\parallel y\xrightarrow{\{e_1,e_2\}}x'}$$
        $$\frac{x\xrightarrow{e_1}\surd\quad y\xrightarrow{e_2}y'}{x\parallel y\xrightarrow{\{e_1,e_2\}}y'} \quad\frac{x\xrightarrow{e_1}x'\quad y\xrightarrow{e_2}y'}{x\parallel y\xrightarrow{\{e_1,e_2\}}x'\between y'}$$
        $$\frac{x\xrightarrow{e_1}\surd\quad y\xrightarrow{e_2}\surd \quad(e_1\leq e_2)}{x\leftmerge y\xrightarrow{\{e_1,e_2\}}\surd} \quad\frac{x\xrightarrow{e_1}x'\quad y\xrightarrow{e_2}\surd \quad(e_1\leq e_2)}{x\leftmerge y\xrightarrow{\{e_1,e_2\}}x'}$$
        $$\frac{x\xrightarrow{e_1}\surd\quad y\xrightarrow{e_2}y' \quad(e_1\leq e_2)}{x\leftmerge y\xrightarrow{\{e_1,e_2\}}y'} \quad\frac{x\xrightarrow{e_1}x'\quad y\xrightarrow{e_2}y' \quad(e_1\leq e_2)}{x\leftmerge y\xrightarrow{\{e_1,e_2\}}x'\between y'}$$
        $$\frac{x\xrightarrow{e_1}\surd\quad y\xrightarrow{e_2}\surd}{x\mid y\xrightarrow{\gamma(e_1,e_2)}\surd} \quad\frac{x\xrightarrow{e_1}x'\quad y\xrightarrow{e_2}\surd}{x\mid y\xrightarrow{\gamma(e_1,e_2)}x'}$$
        $$\frac{x\xrightarrow{e_1}\surd\quad y\xrightarrow{e_2}y'}{x\mid y\xrightarrow{\gamma(e_1,e_2)}y'} \quad\frac{x\xrightarrow{e_1}x'\quad y\xrightarrow{e_2}y'}{x\mid y\xrightarrow{\gamma(e_1,e_2)}x'\between y'}$$
        $$\frac{x\xrightarrow{e_1}\surd\quad (\sharp(e_1,e_2))}{\Theta(x)\xrightarrow{e_1}\surd} \quad\frac{x\xrightarrow{e_2}\surd\quad (\sharp(e_1,e_2))}{\Theta(x)\xrightarrow{e_2}\surd}$$
        $$\frac{x\xrightarrow{e_1}x'\quad (\sharp(e_1,e_2))}{\Theta(x)\xrightarrow{e_1}\Theta(x')} \quad\frac{x\xrightarrow{e_2}x'\quad (\sharp(e_1,e_2))}{\Theta(x)\xrightarrow{e_2}\Theta(x')}$$
        $$\frac{x\xrightarrow{e_1}\surd\quad (\sharp_{\pi}(e_1,e_2))}{\Theta(x)\xrightarrow{e_1}\surd} \quad\frac{x\xrightarrow{e_2}\surd\quad (\sharp_{\pi}(e_1,e_2))}{\Theta(x)\xrightarrow{e_2}\surd}$$
        $$\frac{x\xrightarrow{e_1}x'\quad (\sharp_{\pi}(e_1,e_2))}{\Theta(x)\xrightarrow{e_1}\Theta(x')} \quad\frac{x\xrightarrow{e_2}x'\quad (\sharp_{\pi}(e_1,e_2))}{\Theta(x)\xrightarrow{e_2}\Theta(x')}$$
        $$\frac{x\xrightarrow{e_1}\surd \quad y\nrightarrow^{e_2}\quad (\sharp(e_1,e_2))}{x\triangleleft y\xrightarrow{\tau}\surd}
        \quad\frac{x\xrightarrow{e_1}x' \quad y\nrightarrow^{e_2}\quad (\sharp(e_1,e_2))}{x\triangleleft y\xrightarrow{\tau}x'}$$
        $$\frac{x\xrightarrow{e_1}\surd \quad y\nrightarrow^{e_3}\quad (\sharp(e_1,e_2),e_2\leq e_3)}{x\triangleleft y\xrightarrow{e_1}\surd}
        \quad\frac{x\xrightarrow{e_1}x' \quad y\nrightarrow^{e_3}\quad (\sharp(e_1,e_2),e_2\leq e_3)}{x\triangleleft y\xrightarrow{e_1}x'}$$
        $$\frac{x\xrightarrow{e_3}\surd \quad y\nrightarrow^{e_2}\quad (\sharp(e_1,e_2),e_1\leq e_3)}{x\triangleleft y\xrightarrow{\tau}\surd}
        \quad\frac{x\xrightarrow{e_3}x' \quad y\nrightarrow^{e_2}\quad (\sharp(e_1,e_2),e_1\leq e_3)}{x\triangleleft y\xrightarrow{\tau}x'}$$
        $$\frac{x\xrightarrow{e_1}\surd \quad y\nrightarrow^{e_2}\quad (\sharp_{\pi}(e_1,e_2))}{x\triangleleft y\xrightarrow{\tau}\surd}
        \quad\frac{x\xrightarrow{e_1}x' \quad y\nrightarrow^{e_2}\quad (\sharp_{\pi}(e_1,e_2))}{x\triangleleft y\xrightarrow{\tau}x'}$$
        $$\frac{x\xrightarrow{e_1}\surd \quad y\nrightarrow^{e_3}\quad (\sharp_{\pi}(e_1,e_2),e_2\leq e_3)}{x\triangleleft y\xrightarrow{e_1}\surd}
        \quad\frac{x\xrightarrow{e_1}x' \quad y\nrightarrow^{e_3}\quad (\sharp_{\pi}(e_1,e_2),e_2\leq e_3)}{x\triangleleft y\xrightarrow{e_1}x'}$$
        $$\frac{x\xrightarrow{e_3}\surd \quad y\nrightarrow^{e_2}\quad (\sharp_{\pi}(e_1,e_2),e_1\leq e_3)}{x\triangleleft y\xrightarrow{\tau}\surd}
        \quad\frac{x\xrightarrow{e_3}x' \quad y\nrightarrow^{e_2}\quad (\sharp_{\pi}(e_1,e_2),e_1\leq e_3)}{x\triangleleft y\xrightarrow{\tau}x'}$$
        \caption{Action transition rules of APPTC}
        \label{TRForAPPTC}
    \end{table}
\end{center}

\begin{theorem}[Generalization of the algebra for parallelism with respect to $BAPTC$]
The algebra for parallelism is a generalization of $BAPTC$.
\end{theorem}

\begin{theorem}[Congruence of $APPTC$ with respect to probabilistic truly concurrent bisimulation equivalences]
Probabilistic truly concurrent bisimulation equivalences $\sim_{pp}$, $\sim_{ps}$, $\sim_{php}$ and $\sim_{phhp}$ are all congruences with respect to $APPTC$.
\end{theorem}

\begin{theorem}[Soundness of parallelism modulo probabilistic truly concurrent bisimulation equivalences]
Let $x$ and $y$ be $APPTC$ terms.

\begin{enumerate}
  \item If $APPTC\vdash x=y$, then $x\sim_{pp} y$;
  \item If $APPTC\vdash x=y$, then $x\sim_{ps} y$;
  \item If $APPTC\vdash x=y$, then $x\sim_{php} y$;
  \item If $APPTC\vdash x=y$, then $x\sim_{phhp} y$.
\end{enumerate}
\end{theorem}

\begin{theorem}[Completeness of parallelism modulo probabilistic truly concurrent bisimulation equivalences]
Let $p$ and $q$ be closed $APPTC$ terms.

\begin{enumerate}
  \item If $p\sim_{pp} q$ then $p=q$;
  \item If $p\sim_{ps} q$ then $p=q$;
  \item If $p\sim_{php} q$ then $p=q$;
  \item If $p\sim_{phhp} q$ then $p=q$.
\end{enumerate}
\end{theorem}

We give the definition of PDFs of encapsulation in Table \ref{PDFEncap22}.

\begin{center}
    \begin{table}
        $$\mu(\partial_H(x),\partial_H(x'))=\mu(x,x')$$
        $$\mu(x,y)=0,\textrm{otherwise}$$
        \caption{PDF definitions of $\partial_H$}
        \label{PDFEncap22}
    \end{table}
\end{center}

The transition rules of encapsulation operator $\partial_H$ are shown in Table \ref{TRForPEncapsulation22}.

\begin{center}
    \begin{table}
        $$\frac{x\rightsquigarrow x'}{\partial_H(x)\rightsquigarrow\partial_H(x')}$$
        $$\frac{x\xrightarrow{e}\surd}{\partial_H(x)\xrightarrow{e}\surd}\quad (e\notin H)\quad\quad\frac{x\xrightarrow{e}x'}{\partial_H(x)\xrightarrow{e}\partial_H(x')}\quad(e\notin H)$$
        \caption{Transition rules of encapsulation operator $\partial_H$}
        \label{TRForPEncapsulation22}
    \end{table}
\end{center}

Based on the transition rules for encapsulation operator $\partial_H$ in Table \ref{TRForPEncapsulation22}, we design the axioms as Table \ref{AxiomsForPEncapsulation22} shows.

\begin{center}
    \begin{table}
        \begin{tabular}{@{}ll@{}}
            \hline No. &Axiom\\
            $D1$ & $e\notin H\quad\partial_H(e) = e$\\
            $D2$ & $e\in H\quad \partial_H(e) = \delta$\\
            $D3$ & $\partial_H(\delta) = \delta$\\
            $D4$ & $\partial_H(x+ y) = \partial_H(x)+\partial_H(y)$\\
            $D5$ & $\partial_H(x\cdot y) = \partial_H(x)\cdot\partial_H(y)$\\
            $D6$ & $\partial_H(x\leftmerge y) = \partial_H(x)\leftmerge\partial_H(y)$\\
            $PD1$ & $\partial_H(x\boxplus_{\pi}y)=\partial_H(x)\boxplus_{\pi}\partial_H(y)$\\
        \end{tabular}
        \caption{Axioms of encapsulation operator}
        \label{AxiomsForPEncapsulation22}
    \end{table}
\end{center}

\begin{theorem}[Conservativity of $APPTC$ with respect to the algebra for parallelism]
$APPTC$ is a conservative extension of the algebra for parallelism.
\end{theorem}

\begin{theorem}[Elimination theorem of $APPTC$]
Let $p$ be a closed $APPTC$ term including the encapsulation operator $\partial_H$. Then there is a basic $APPTC$ term $q$ such that $APPTC\vdash p=q$.
\end{theorem}

\begin{theorem}[Congruence theorem of encapsulation operator $\partial_H$ with respect to probabilistic truly concurrent bisimulation equivalences]
Probabilistic truly concurrent bisimulation equivalences $\sim_{pp}$, $\sim_{ps}$, $\sim_{php}$ and $\sim_{phhp}$ are all congruences with respect to encapsulation operator $\partial_H$.
\end{theorem}

\begin{theorem}[Soundness of $APPTC$ modulo probabilistic truly concurrent bisimulation equivalences]
Let $x$ and $y$ be $APPTC$ terms including encapsulation operator $\partial_H$.

\begin{enumerate}
  \item If $APPTC\vdash x=y$, then $x\sim_{pp} y$;
  \item If $APPTC\vdash x=y$, then $x\sim_{ps} y$;
  \item If $APPTC\vdash x=y$, then $x\sim_{php} y$;
  \item If $APPTC\vdash x=y$, then $x\sim_{phhp} y$.
\end{enumerate}
\end{theorem}

\begin{theorem}[Completeness of $APPTC$ modulo probabilistic truly concurrent bisimulation equivalences]
Let $p$ and $q$ be closed $APPTC$ terms including encapsulation operator $\partial_H$.

\begin{enumerate}
  \item If $p\sim_{pp} q$ then $p=q$;
  \item If $p\sim_{ps} q$ then $p=q$;
  \item If $p\sim_{php} q$ then $p=q$;
  \item If $p\sim_{phhp} q$ then $p=q$.
\end{enumerate}
\end{theorem}

\subsubsection{Recursion}

\begin{definition}[Guarded recursive specification]
A recursive specification

$$X_1=t_1(X_1,\cdots,X_n)$$
$$...$$
$$X_n=t_n(X_1,\cdots,X_n)$$

is guarded if the right-hand sides of its recursive equations can be adapted to the form by applications of the axioms in $APPTC$ and replacing recursion variables by the right-hand
sides of their recursive equations,

$((a_{111}\leftmerge\cdots\leftmerge a_{11i_1})\cdot s_1(X_1,\cdots,X_n)+\cdots+(a_{1k1}\leftmerge\cdots\leftmerge a_{1ki_k})\cdot s_k(X_1,\cdots,X_n)+(b_{111}\leftmerge\cdots\leftmerge
b_{11j_1})+\cdots+(b_{11j_1}\leftmerge\cdots\leftmerge b_{1lj_l}))\boxplus_{\pi_1}\cdots\boxplus_{\pi_{m-1}}((a_{m11}\leftmerge\cdots\leftmerge a_{m1i_1})\cdot s_1(X_1,\cdots,X_n)+
\cdots+(a_{mk1}\leftmerge\cdots\leftmerge a_{mki_k})\cdot s_k(X_1,\cdots,X_n)+(b_{m11}\leftmerge\cdots\leftmerge b_{m1j_1})+\cdots+(b_{m1j_1}\leftmerge\cdots\leftmerge b_{mlj_l}))$

where $a_{111},\cdots,a_{11i_1},a_{1k1},\cdots,a_{1ki_k},b_{111},\cdots,b_{11j_1},b_{11j_1},\cdots,b_{1lj_l},\cdots, a_{m11},\cdots,a_{m1i_1},a_{mk1},\cdots,a_{mki_k},\\b_{m11},\cdots,
b_{m1j_1},b_{m1j_1},\cdots,b_{mlj_l}\in \mathbb{E}$, and the sum above is allowed to be empty, in which case it represents the deadlock $\delta$.
\end{definition}

\begin{definition}[Linear recursive specification]\label{LRS}
A recursive specification is linear if its recursive equations are of the form

$((a_{111}\leftmerge\cdots\leftmerge a_{11i_1})X_1+\cdots+(a_{1k1}\leftmerge\cdots\leftmerge a_{1ki_k})X_k+(b_{111}\leftmerge\cdots\leftmerge b_{11j_1})+\cdots+(b_{11j_1}\leftmerge\cdots
\leftmerge b_{1lj_l}))\boxplus_{\pi_1}\cdots\boxplus_{\pi_{m-1}}((a_{m11}\leftmerge\cdots\leftmerge a_{m1i_1})X_1+\cdots+(a_{mk1}\leftmerge\cdots\leftmerge a_{mki_k})X_k+
(b_{m11}\leftmerge\cdots\leftmerge b_{m1j_1})+\cdots+(b_{m1j_1}\leftmerge\cdots\leftmerge b_{mlj_l}))$

where $a_{111},\cdots,a_{11i_1},a_{1k1},\cdots,a_{1ki_k},b_{111},\cdots,b_{11j_1},b_{11j_1},\cdots,b_{1lj_l},\cdots,a_{m11},\cdots,a_{m1i_1},a_{mk1},\cdots,a_{mki_k},\\b_{m11},\cdots,
b_{m1j_1},b_{m1j_1},\cdots,b_{mlj_l}\in \mathbb{E}$, and the sum above is allowed to be empty, in which case it represents the deadlock $\delta$.
\end{definition}

We give the definition of PDFs of recursion in Table \ref{PDFGR22}.

\begin{center}
    \begin{table}
        $$\mu(\langle X|E\rangle,y)=\mu(\langle t_X|E\rangle,y)$$
        $$\mu(x,y)=0,\textrm{otherwise}$$
        \caption{PDF definitions of recursion}
        \label{PDFGR22}
    \end{table}
\end{center}

For a guarded recursive specifications $E$ with the form

$$X_1=t_1(X_1,\cdots,X_n)$$
$$\cdots$$
$$X_n=t_n(X_1,\cdots,X_n)$$

the behavior of the solution $\langle X_i|E\rangle$ for the recursion variable $X_i$ in $E$, where $i\in\{1,\cdots,n\}$, is exactly the behavior of their right-hand sides
$t_i(X_1,\cdots,X_n)$, which is captured by the two transition rules in Table \ref{TRForPGR22}.

\begin{center}
    \begin{table}
        $$\frac{t_i(\langle X_1|E\rangle,\cdots,\langle X_n|E\rangle)\rightsquigarrow y}{\langle X_i|E\rangle\rightsquigarrow y}$$
        $$\frac{t_i(\langle X_1|E\rangle,\cdots,\langle X_n|E\rangle)\xrightarrow{\{e_1,\cdots,e_k\}}\surd}{\langle X_i|E\rangle\xrightarrow{\{e_1,\cdots,e_k\}}\surd}$$
        $$\frac{t_i(\langle X_1|E\rangle,\cdots,\langle X_n|E\rangle)\xrightarrow{\{e_1,\cdots,e_k\}} y}{\langle X_i|E\rangle\xrightarrow{\{e_1,\cdots,e_k\}} y}$$
        \caption{Transition rules of guarded recursion}
        \label{TRForPGR22}
    \end{table}
\end{center}

\begin{theorem}[Conservitivity of $APPTC$ with guarded recursion]
$APPTC$ with guarded recursion is a conservative extension of $APPTC$.
\end{theorem}

\begin{theorem}[Congruence theorem of $APPTC$ with guarded recursion]
Probabilistic truly concurrent bisimulation equivalences $\sim_{pp}$, $\sim_{ps}$, $\sim_{php}$ and $\sim_{phhp}$ are all congruences with respect to $APPTC$ with guarded recursion.
\end{theorem}

\begin{theorem}[Elimination theorem of $APPTC$ with linear recursion]
Each process term in $APPTC$ with linear recursion is equal to a process term $\langle X_1|E\rangle$ with $E$ a linear recursive specification.
\end{theorem}

\begin{theorem}[Soundness of $APPTC$ with guarded recursion]
Let $x$ and $y$ be $APPTC$ with guarded recursion terms. If $APPTC\textrm{ with guarded recursion}\vdash x=y$, then
\begin{enumerate}
  \item $x\sim_{pp} y$;
  \item $x\sim_{ps} y$;
  \item $x\sim_{php} y$;
  \item $x\sim_{phhp} y$.
\end{enumerate}
\end{theorem}

\begin{theorem}[Completeness of $APPTC$ with linear recursion]
Let $p$ and $q$ be closed $APPTC$ with linear recursion terms, then,
\begin{enumerate}
  \item if $p\sim_{pp} q$ then $p=q$;
  \item if $p\sim_{ps} q$ then $p=q$;
  \item if $p\sim_{php} q$ then $p=q$;
  \item if $p\sim_{phhp} q$ then $p=q$.
\end{enumerate}
\end{theorem}

\subsubsection{Abstraction}

To abstract away from the internal implementations of a program, and verify that the program exhibits the desired external behaviors, the silent step $\tau$ and abstraction operator
$\tau_I$ are introduced, where $I\subseteq \mathbb{E}$ denotes the internal events. The silent step $\tau$ represents the internal events, when we consider the external behaviors of a
process, $\tau$ events can be removed, that is, $\tau$ events must keep silent. The transition rule of $\tau$ is shown in Table \ref{TRForPTau22}. In the following, let the atomic event
$e$ range over $\mathbb{E}\cup\{\delta\}\cup\{\tau\}$, and let the communication function $\gamma:\mathbb{E}\cup\{\tau\}\times \mathbb{E}\cup\{\tau\}\rightarrow \mathbb{E}\cup\{\delta\}$,
with each communication involved $\tau$ resulting into $\delta$.

\begin{center}
    \begin{table}
        $$\frac{}{\tau\rightsquigarrow\breve{\tau}}$$
        $$\frac{}{\tau\xrightarrow{\tau}\surd}$$
        \caption{Transition rules of the silent step}
        \label{TRForPTau22}
    \end{table}
\end{center}

The silent step $\tau$ as an atomic event, is introduced into $E$. Considering the recursive specification $X=\tau X$, $\tau s$, $\tau\tau s$, and $\tau\cdots s$ are all its solutions,
that is, the solutions make the existence of $\tau$-loops which cause unfairness. To prevent $\tau$-loops, we extend the definition of linear recursive specification
to the guarded one.

\begin{definition}[Guarded linear recursive specification]\label{GLRS22}
A recursive specification is linear if its recursive equations are of the form

$((a_{111}\parallel\cdots\parallel a_{11i_1})X_1+\cdots+(a_{1k1}\parallel\cdots\parallel a_{1ki_k})X_k+(b_{111}\parallel\cdots\parallel b_{11j_1})+\cdots+
(b_{11j_1}\parallel\cdots\parallel b_{1lj_l}))\boxplus_{\pi_1}\cdots\boxplus_{\pi_{m-1}}((a_{m11}\parallel\cdots\parallel a_{m1i_1})X_1+\cdots+(a_{mk1}\parallel\cdots\parallel a_{mki_k})X_k+
(b_{m11}\parallel\cdots\parallel b_{m1j_1})+\cdots+(b_{m1j_1}\parallel\cdots\parallel b_{mlj_l}))$

where $a_{111},\cdots,a_{11i_1},a_{1k1},\cdots,a_{1ki_k},b_{111},\cdots,b_{11j_1},b_{11j_1},\cdots,b_{1lj_l}\cdots\\
a_{m11},\cdots,a_{m1i_1},a_{mk1},\cdots,a_{mki_k},b_{m11},\cdots,b_{m1j_1},b_{m1j_1},\cdots,b_{mlj_l}\in \mathbb{E}\cup\{\tau\}$, and the sum above is allowed to be empty, in which
case it represents the deadlock $\delta$.

A linear recursive specification $E$ is guarded if there does not exist an infinite sequence of $\tau$-transitions $\langle X|E\rangle\rightsquigarrow\xrightarrow{\tau}
\langle X'|E\rangle\rightsquigarrow\xrightarrow{\tau}\langle X''|E\rangle\rightsquigarrow\xrightarrow{\tau}\cdots$.
\end{definition}

\begin{theorem}[Conservitivity of $APTC$ with silent step and guarded linear recursion]
$APTC$ with silent step and guarded linear recursion is a conservative extension of $APTC$ with linear recursion.
\end{theorem}

\begin{theorem}[Congruence theorem of $APTC$ with silent step and guarded linear recursion]
Probabilistic rooted branching truly concurrent bisimulation equivalences $\approx_{prbp}$, $\approx_{prbs}$, $\approx_{prbhp}$ and $\approx_{prbhhp}$ are all congruences with respect
to $APTC$ with silent step and guarded linear recursion.
\end{theorem}

We design the axioms for the silent step $\tau$ in Table \ref{AxiomsForPTau22}.

\begin{center}
\begin{table}
  \begin{tabular}{@{}ll@{}}
\hline No. &Axiom\\
  $B1$ & $(y=y+y,z=z+z)\quad x\cdot((y+\tau\cdot(y+z))\boxplus_{\pi}w)=x\cdot((y+z)\boxplus_{\pi}w)$\\
  $B2$ & $(y=y+y,z=z+z)\quad x\leftmerge((y+\tau\leftmerge(y+z))\boxplus_{\pi}w)=x\leftmerge((y+z)\boxplus_{\pi}w)$\\
\end{tabular}
\caption{Axioms of silent step}
\label{AxiomsForPTau22}
\end{table}
\end{center}

\begin{theorem}[Elimination theorem of $APTC$ with silent step and guarded linear recursion]\label{ETTau}
Each process term in $APTC$ with silent step and guarded linear recursion is equal to a process term $\langle X_1|E\rangle$ with $E$ a guarded linear recursive specification.
\end{theorem}

\begin{theorem}[Soundness of $APTC$ with silent step and guarded linear recursion]\label{SAPTCTAU}
Let $x$ and $y$ be $APTC$ with silent step and guarded linear recursion terms. If $APTC$ with silent step and guarded linear recursion $\vdash x=y$, then
\begin{enumerate}
  \item $x\approx_{prbp} y$;
  \item $x\approx_{prbs} y$;
  \item $x\approx_{prbhp} y$;
  \item $x\approx_{prbhhp}y$.
\end{enumerate}
\end{theorem}

\begin{theorem}[Completeness of $APTC$ with silent step and guarded linear recursion]\label{CAPTCTAU}
Let $p$ and $q$ be closed $APTC$ with silent step and guarded linear recursion terms, then,
\begin{enumerate}
  \item if $p\approx_{prbp} q$ then $p=q$;
  \item if $p\approx_{prbs} q$ then $p=q$;
  \item if $p\approx_{prbhp} q$ then $p=q$;
  \item if $p\approx_{prbhhp} q$ then $p=q$.
\end{enumerate}
\end{theorem}

The unary abstraction operator $\tau_I$ ($I\subseteq \mathbb{E}$) renames all atomic events in $I$ into $\tau$. $APTC$ with silent step and abstraction operator is called $APTC_{\tau}$.
The transition rules of operator $\tau_I$ are shown in Table \ref{TRForPAbstraction22}.

\begin{center}
    \begin{table}
        $$\frac{x\rightsquigarrow x'}{\tau_I(x)\rightsquigarrow\tau_I(x')}$$
        $$\frac{x\xrightarrow{e}\surd}{\tau_I(x)\xrightarrow{e}\surd}\quad e\notin I
        \quad\frac{x\xrightarrow{e}x'}{\tau_I(x)\xrightarrow{e}\tau_I(x')}\quad e\notin I$$

        $$\frac{x\xrightarrow{e}\surd}{\tau_I(x)\xrightarrow{\tau}\surd}\quad e\in I
        \quad\frac{x\xrightarrow{e}x'}{\tau_I(x)\xrightarrow{\tau}\tau_I(x')}\quad e\in I$$
        \caption{Transition rules of the abstraction operator}
        \label{TRForPAbstraction22}
    \end{table}
\end{center}

\begin{theorem}[Conservitivity of $APTC_{\tau}$ with guarded linear recursion]
$APTC_{\tau}$ with guarded linear recursion is a conservative extension of $APTC$ with silent step and guarded linear recursion.
\end{theorem}

\begin{theorem}[Congruence theorem of $APTC_{\tau}$ with guarded linear recursion]
Probabilistic rooted branching truly concurrent bisimulation equivalences $\approx_{prbp}$, $\approx_{prbs}$, $\approx_{prbhp}$ and $\approx_{prbhhp}$ are all congruences with respect
to $APTC_{\tau}$ with guarded linear recursion.
\end{theorem}

We design the axioms for the abstraction operator $\tau_I$ in Table \ref{AxiomsForPAbstraction22}.

\begin{center}
\begin{table}
  \begin{tabular}{@{}ll@{}}
\hline No. &Axiom\\
  $TI1$ & $e\notin I\quad \tau_I(e)=e$\\
  $TI2$ & $e\in I\quad \tau_I(e)=\tau$\\
  $TI3$ & $\tau_I(\delta)=\delta$\\
  $TI4$ & $\tau_I(x+y)=\tau_I(x)+\tau_I(y)$\\
  $PTI1$ & $\tau_I(x\boxplus_{\pi}y)=\tau_I(x)\boxplus_{\pi}\tau_I(y)$\\
  $TI5$ & $\tau_I(x\cdot y)=\tau_I(x)\cdot\tau_I(y)$\\
  $TI6$ & $\tau_I(x\leftmerge y)=\tau_I(x)\leftmerge\tau_I(y)$\\
\end{tabular}
\caption{Axioms of abstraction operator}
\label{AxiomsForPAbstraction22}
\end{table}
\end{center}

\begin{theorem}[Soundness of $APTC_{\tau}$ with guarded linear recursion]\label{SAPTCABS}
Let $x$ and $y$ be $APTC_{\tau}$ with guarded linear recursion terms. If $APTC_{\tau}$ with guarded linear recursion $\vdash x=y$, then
\begin{enumerate}
  \item $x\approx_{prbp} y$;
  \item $x\approx_{prbs} y$;
  \item $x\approx_{prbhp} y$;
  \item $x\approx_{prbhhp} y$.
\end{enumerate}
\end{theorem}

Though $\tau$-loops are prohibited in guarded linear recursive specifications in a specifiable way, they can be constructed using the abstraction operator, for example, there exist
$\tau$-loops in the process term $\tau_{\{a\}}(\langle X|X=aX\rangle)$. To avoid $\tau$-loops caused by $\tau_I$ and ensure fairness, we introduce the following recursive verification
rules as Table \ref{RVR22} shows, note that $i_1,\cdots, i_m,j_1,\cdots,j_n\in I\subseteq \mathbb{E}\setminus\{\tau\}$.

\begin{center}
\begin{table}
    $$VR_1\quad \frac{x=y+(i_1\leftmerge\cdots\leftmerge i_m)\cdot x, y=y+y}{\tau\cdot\tau_I(x)=\tau\cdot \tau_I(y)}$$
    $$VR_2\quad \frac{x=z\boxplus_{\pi}(u+(i_1\leftmerge\cdots\leftmerge i_m)\cdot x),z=z+u,z=z+z}{\tau\cdot\tau_I(x)=\tau\cdot\tau_I(z)}$$
    $$VR_3\quad \frac{x=z+(i_1\leftmerge\cdots\leftmerge i_m)\cdot y,y=z\boxplus_{\pi}(u+(j_1\leftmerge\cdots\leftmerge j_n)\cdot x), z=z+u,z=z+z}{\tau\cdot\tau_I(x)=\tau\cdot\tau_I(y')\textrm{ for }y'=z\boxplus_{\pi}(u+(i_1\leftmerge\cdots\leftmerge i_m)\cdot y')}$$
\caption{Recursive verification rules}
\label{RVR22}
\end{table}
\end{center}

\begin{theorem}[Soundness of $VR_1,VR_2,VR_3$]
$VR_1$, $VR_2$ and $VR_3$ are sound modulo probabilistic rooted branching truly concurrent bisimulation equivalences $\approx_{prbp}$, $\approx_{prbs}$, $\approx_{prbhp}$ and $\approx_{prbhhp}$.
\end{theorem}

\subsection{Operational Semantics for Quantum Computing}

In quantum processes, to avoid the abuse of quantum information which may violate the no-cloning theorem, a quantum configuration $\langle C,\varrho \rangle$
\cite{PSQP} \cite{QPA} \cite{QPA2} \cite{CQP} \cite{CQP2} \cite{qCCS} \cite{BQP} \cite{PSQP} \cite{SBQP} is usually consisted of a traditional configuration $C$ and state information $\varrho$ of
all (public) quantum information variables. Though quantum information variables are not explicitly defined and are hidden behind quantum operations or unitary operators, more importantly, the
state information $\varrho$ is the effects of execution of a series of quantum operations or unitary operators on involved quantum systems, the execution of a series of quantum operations
or unitary operators should not only obey the restrictions of the structure of the process terms, but also those of quantum mechanics principles. Through the state information
$\varrho$, we can check and observe the functions of quantum mechanics principles, such as quantum entanglement, quantum measurement, etc.

So, the operational semantics of quantum processes should be defined based on quantum process configuration $\langle C,\varrho\rangle$, in which $\varrho=\varsigma$ of two state
information $\varrho$ and $\varsigma$ means equality under the framework of quantum information and quantum computing, that is, these two quantum processes are in the same quantum
state.

\begin{definition}[FR pomset transitions and step]
Let $\mathcal{E}$ be a PES and let $C\in\mathcal{C}(\mathcal{E})$, and $\emptyset\neq X\subseteq \mathbb{E}$, if $C\cap X=\emptyset$ and $C'=C\cup X\in\mathcal{C}(\mathcal{E})$, then
$\langle C,\varrho\rangle\xrightarrow{X} \langle C',\varrho'\rangle$ is called a forward pomset transition from $\langle C,\varrho\rangle$ to $\langle C',\varrho'\rangle$ and
$\langle C',\varrho'\rangle\xtworightarrow{X[\mathcal{K}]} \langle C,\varrho\rangle$ is called a reverse pomset transition from $\langle C',\varrho'\rangle$ to $\langle C,\varrho\rangle$. When the events in
$X$ and $X[\mathcal{K}]$ are pairwise
concurrent, we say that $\langle C,\varrho\rangle\xrightarrow{X}\langle C',\varrho'\rangle$ is a forward step and $\langle C',\varrho'\rangle\xtworightarrow{X[\mathcal{K}]}\langle C,\varrho\rangle$ is a reverse step.
It is obvious that $\rightarrow^*\xrightarrow{X}\rightarrow^*=\xrightarrow{X}$ and
$\rightarrow^*\xrightarrow{e}\rightarrow^*=\xrightarrow{e}$ for any $e\in\mathbb{E}$ and $X\subseteq\mathbb{E}$.
\end{definition}

\begin{definition}[FR weak pomset transitions and weak step]
Let $\mathcal{E}$ be a PES and let $C\in\mathcal{C}(\mathcal{E})$, and $\emptyset\neq X\subseteq \hat{\mathbb{E}}$, if $C\cap X=\emptyset$ and
$\hat{C'}=\hat{C}\cup X\in\mathcal{C}(\mathcal{E})$, then $\langle C,\varrho\rangle\xRightarrow{X} \langle C',\varrho'\rangle$ is called a forward weak pomset transition from $\langle C,\varrho\rangle$ to
$\langle C',\varrho'\rangle$, where we define $\xRightarrow{e}\triangleq\xrightarrow{\tau^*}\xrightarrow{e}\xrightarrow{\tau^*}$. And $\langle C',\varrho'\rangle\xTworightarrow{X[\mathcal{K}]} \langle C,\varrho\rangle$
is called a reverse weak pomset transition from $\langle C',\varrho'\rangle$ to $\langle C,\varrho\rangle$. When the events in $X$ are pairwise concurrent, we say that
$\langle C,\varrho\rangle\xRightarrow{X}\langle C',\varrho'\rangle$ is a forward weak step, when the events in $X[\mathcal{K}]$ are pairwise concurrent, we say that
$\langle C',\varrho'\rangle\xTworightarrow{X}\langle C,\varrho\rangle$ is a reverse weak step.
\end{definition}

We will also suppose that all the PESs are image finite, that is, for any PES $\mathcal{E}$ and $C\in \mathcal{C}(\mathcal{E})$ and $a\in \Lambda$,
$\{e\in \mathbb{E}|\langle C,\varrho\rangle\xrightarrow{e} \langle C',\varrho'\rangle\wedge \lambda(e)=a\}$ and
$\{e\in\hat{\mathbb{E}}|\langle C,\varrho\rangle\xRightarrow{e} \langle C',\varrho'\rangle\wedge \lambda(e)=a\}$ and
$\{e\in \mathbb{E}|\langle C',\varrho'\rangle\xtworightarrow{e} \langle C,\varrho\rangle\wedge \lambda(e)=a\}$ and
$\{e\in\hat{\mathbb{E}}|\langle C',\varrho'\rangle\xTworightarrow{e} \langle C,\varrho\rangle\wedge \lambda(e)=a\}$ are finite.

\begin{definition}[FR pomset, step bisimulation]
Let $\mathcal{E}_1$, $\mathcal{E}_2$ be PESs. A FR pomset bisimulation is a relation $R\subseteq\langle\mathcal{C}(\mathcal{E}_1),S\rangle\times\langle\mathcal{C}(\mathcal{E}_2),S\rangle$,
such that (1) if $(\langle C_1,\varrho\rangle,\langle C_2,\varrho\rangle)\in R$, and $\langle C_1,\varrho\rangle\xrightarrow{X_1}\langle C_1',\varrho'\rangle$ then
$\langle C_2,\varrho\rangle\xrightarrow{X_2}\langle C_2',\varrho'\rangle$, with $X_1\subseteq \mathbb{E}_1$, $X_2\subseteq \mathbb{E}_2$, $X_1\sim X_2$ and
$(\langle C_1',\varrho'\rangle,\langle C_2',\varrho'\rangle)\in R$ for all $\varrho,\varrho'\in S$, and vice-versa;
(2) if $(\langle C_1,\varrho\rangle,\langle C_2,\varrho\rangle)\in R$, and $\langle C_1,\varrho\rangle\xtworightarrow{X_1[\mathcal{K}_1]}\langle C_1',\varrho'\rangle$ then
$\langle C_2,\varrho\rangle\xtworightarrow{X_2[\mathcal{K}_2]}\langle C_2',\varrho'\rangle$, with $X_1\subseteq \mathbb{E}_1$, $X_2\subseteq \mathbb{E}_2$, $X_1\sim X_2$ and
$(\langle C_1',\varrho'\rangle,\langle C_2',\varrho'\rangle)\in R$ for all $\varrho,\varrho'\in S$, and vice-versa. We say that $\mathcal{E}_1$, $\mathcal{E}_2$ are FR pomset bisimilar, written
$\mathcal{E}_1\sim_p^{fr}\mathcal{E}_2$, if there exists a FR pomset bisimulation $R$, such that $(\langle\emptyset,\emptyset\rangle,\langle\emptyset,\emptyset\rangle)\in R$. By replacing
FR pomset transitions with FR steps, we can get the definition of FR step bisimulation. When PESs $\mathcal{E}_1$ and $\mathcal{E}_2$ are FR step bisimilar, we write
$\mathcal{E}_1\sim_s^{fr}\mathcal{E}_2$.
\end{definition}

\begin{definition}[FR weak pomset, step bisimulation]
Let $\mathcal{E}_1$, $\mathcal{E}_2$ be PESs. A FR weak pomset bisimulation is a relation
$R\subseteq\langle\mathcal{C}(\mathcal{E}_1),S\rangle\times\langle\mathcal{C}(\mathcal{E}_2),S\rangle$, such that (1) if $(\langle C_1,\varrho\rangle,\langle C_2,\varrho\rangle)\in R$, and
$\langle C_1,\varrho\rangle\xRightarrow{X_1}\langle C_1',\varrho'\rangle$ then $\langle C_2,\varrho\rangle\xRightarrow{X_2}\langle C_2',\varrho'\rangle$, with $X_1\subseteq \hat{\mathbb{E}_1}$,
$X_2\subseteq \hat{\mathbb{E}_2}$, $X_1\sim X_2$ and $(\langle C_1',\varrho'\rangle,\langle C_2',\varrho'\rangle)\in R$ for all $\varrho,\varrho'\in S$, and vice-versa;
(2) if $(\langle C_1,\varrho\rangle,\langle C_2,\varrho\rangle)\in R$, and
$\langle C_1,\varrho\rangle\xTworightarrow{X_1[\mathcal{K}_1]}\langle C_1',\varrho'\rangle$ then $\langle C_2,\varrho\rangle\xTworightarrow{X_2[\mathcal{K}_2]}\langle C_2',\varrho'\rangle$, with $X_1\subseteq \hat{\mathbb{E}_1}$,
$X_2\subseteq \hat{\mathbb{E}_2}$, $X_1\sim X_2$ and $(\langle C_1',\varrho'\rangle,\langle C_2',\varrho'\rangle)\in R$ for all $\varrho,\varrho'\in S$, and vice-versa. We say that $\mathcal{E}_1$,
$\mathcal{E}_2$ are FR weak pomset bisimilar, written $\mathcal{E}_1\approx_p^{fr}\mathcal{E}_2$, if there exists a FR weak pomset bisimulation $R$, such that
$(\langle\emptyset,\emptyset\rangle,\langle\emptyset,\emptyset\rangle)\in R$. By replacing FR weak pomset transitions with FR weak steps, we can get the definition of FR weak step bisimulation.
When PESs $\mathcal{E}_1$ and $\mathcal{E}_2$ are FR weak step bisimilar, we write $\mathcal{E}_1\approx_s^{fr}\mathcal{E}_2$.
\end{definition}

\begin{definition}[Posetal product]
Given two PESs $\mathcal{E}_1$, $\mathcal{E}_2$, the posetal product of their configurations, denoted
$\langle\mathcal{C}(\mathcal{E}_1),S\rangle\overline{\times}\langle\mathcal{C}(\mathcal{E}_2),S\rangle$, is defined as

$$\{(\langle C_1,\varrho\rangle,f,\langle C_2,\varrho\rangle)|C_1\in\mathcal{C}(\mathcal{E}_1),C_2\in\mathcal{C}(\mathcal{E}_2),f:C_1\rightarrow C_2 \textrm{ isomorphism}\}.$$

A subset $R\subseteq\langle\mathcal{C}(\mathcal{E}_1),S\rangle\overline{\times}\langle\mathcal{C}(\mathcal{E}_2),S\rangle$ is called a posetal relation. We say that $R$ is downward
closed when for any
$(\langle C_1,\varrho\rangle,f,\langle C_2,\varrho\rangle),(\langle C_1',\varrho'\rangle,f',\langle C_2',\varrho'\rangle)\in \langle\mathcal{C}(\mathcal{E}_1),S\rangle\overline{\times}\langle\mathcal{C}(\mathcal{E}_2),S\rangle$,
if $(\langle C_1,\varrho\rangle,f,\langle C_2,\varrho\rangle)\subseteq (\langle C_1',\varrho'\rangle,f',\langle C_2',\varrho'\rangle)$ pointwise and $(\langle C_1',\varrho'\rangle,f',\langle C_2',\varrho'\rangle)\in R$,
then $(\langle C_1,\varrho\rangle,f,\langle C_2,\varrho\rangle)\in R$.

For $f:X_1\rightarrow X_2$, we define $f[x_1\mapsto x_2]:X_1\cup\{x_1\}\rightarrow X_2\cup\{x_2\}$, $z\in X_1\cup\{x_1\}$,(1)$f[x_1\mapsto x_2](z)=
x_2$,if $z=x_1$;(2)$f[x_1\mapsto x_2](z)=f(z)$, otherwise. Where $X_1\subseteq \mathbb{E}_1$, $X_2\subseteq \mathbb{E}_2$, $x_1\in \mathbb{E}_1$, $x_2\in \mathbb{E}_2$.
\end{definition}

\begin{definition}[Weakly posetal product]
Given two PESs $\mathcal{E}_1$, $\mathcal{E}_2$, the weakly posetal product of their configurations, denoted
$\langle\mathcal{C}(\mathcal{E}_1),S\rangle\overline{\times}\langle\mathcal{C}(\mathcal{E}_2),S\rangle$, is defined as

$$\{(\langle C_1,\varrho\rangle,f,\langle C_2,\varrho\rangle)|C_1\in\mathcal{C}(\mathcal{E}_1),C_2\in\mathcal{C}(\mathcal{E}_2),f:\hat{C_1}\rightarrow \hat{C_2} \textrm{ isomorphism}\}.$$

A subset $R\subseteq\langle\mathcal{C}(\mathcal{E}_1),S\rangle\overline{\times}\langle\mathcal{C}(\mathcal{E}_2),S\rangle$ is called a weakly posetal relation. We say that $R$ is
downward closed when for any
$(\langle C_1,\varrho\rangle,f,\langle C_2,\varrho\rangle),(\langle C_1',\varrho'\rangle,f,\langle C_2',\varrho'\rangle)\in \langle\mathcal{C}(\mathcal{E}_1),S\rangle\overline{\times}\langle\mathcal{C}(\mathcal{E}_2),S\rangle$,
if $(\langle C_1,\varrho\rangle,f,\langle C_2,\varrho\rangle)\subseteq (\langle C_1',\varrho'\rangle,f',\langle C_2',\varrho'\rangle)$ pointwise and $(\langle C_1',\varrho'\rangle,f',\langle C_2',\varrho'\rangle)\in R$,
then $(\langle C_1,\varrho\rangle,f,\langle C_2,\varrho\rangle)\in R$.

For $f:X_1\rightarrow X_2$, we define $f[x_1\mapsto x_2]:X_1\cup\{x_1\}\rightarrow X_2\cup\{x_2\}$, $z\in X_1\cup\{x_1\}$,(1)$f[x_1\mapsto x_2](z)=
x_2$,if $z=x_1$;(2)$f[x_1\mapsto x_2](z)=f(z)$, otherwise. Where $X_1\subseteq \hat{\mathbb{E}_1}$, $X_2\subseteq \hat{\mathbb{E}_2}$, $x_1\in \hat{\mathbb{E}}_1$,
$x_2\in \hat{\mathbb{E}}_2$. Also, we define $f(\tau^*)=f(\tau^*)$.
\end{definition}

\begin{definition}[FR (hereditary) history-preserving bisimulation]
A FR history-preserving (hp-) bisimulation is a posetal relation $R\subseteq\langle\mathcal{C}(\mathcal{E}_1),S\rangle\overline{\times}\langle\mathcal{C}(\mathcal{E}_2),S\rangle$ such
that (1) if $(\langle C_1,\varrho\rangle,f,\langle C_2,\varrho\rangle)\in R$, and $\langle C_1,\varrho\rangle\xrightarrow{e_1} \langle C_1',\varrho'\rangle$, then
$\langle C_2,\varrho\rangle\xrightarrow{e_2} \langle C_2',\varrho'\rangle$, with $(\langle C_1',\varrho'\rangle,f[e_1\mapsto e_2],\langle C_2',\varrho'\rangle)\in R$ for all $\varrho,\varrho'\in S$, and vice-versa;
(2) if $(\langle C_1,\varrho\rangle,f,\langle C_2,\varrho\rangle)\in R$, and $\langle C_1,\varrho\rangle\xtworightarrow{e_1[m]} \langle C_1',\varrho'\rangle$, then
$\langle C_2,\varrho\rangle\xtworightarrow{e_2[n]} \langle C_2',\varrho'\rangle$, with $(\langle C_1',\varrho'\rangle,f[e_1[m]\mapsto e_2[n],\langle C_2',\varrho'\rangle)\in R$ for all $\varrho,\varrho'\in S$, and vice-versa.
$\mathcal{E}_1,\mathcal{E}_2$ are FR history-preserving (hp-)bisimilar and are written $\mathcal{E}_1\sim_{hp}^{fr}\mathcal{E}_2$ if there exists a FR hp-bisimulation $R$ such that
$(\langle\emptyset,\emptyset\rangle,\emptyset,\langle\emptyset,\emptyset\rangle)\in R$.

A FR hereditary history-preserving (hhp-)bisimulation is a downward closed FR hp-bisimulation. $\mathcal{E}_1,\mathcal{E}_2$ are FR hereditary history-preserving (hhp-)bisimilar and
are written $\mathcal{E}_1\sim_{hhp}^{fr}\mathcal{E}_2$.
\end{definition}

\begin{definition}[FR weak (hereditary) history-preserving bisimulation]
A FR weak history-preserving (hp-) bisimulation is a weakly posetal relation
$R\subseteq\langle\mathcal{C}(\mathcal{E}_1),S\rangle\overline{\times}\langle\mathcal{C}(\mathcal{E}_2),S\rangle$ such that (1) if $(\langle C_1,\varrho\rangle,f,\langle C_2,\varrho\rangle)\in R$, and
$\langle C_1,\varrho\rangle\xRightarrow{e_1} \langle C_1',\varrho'\rangle$, then $\langle C_2,\varrho\rangle\xRightarrow{e_2} \langle C_2',\varrho'\rangle$, with $(\langle C_1',\varrho'\rangle,f[e_1\mapsto e_2],\langle C_2',\varrho'\rangle)\in R$
for all $\varrho,\varrho'\in S$, and vice-versa;
(2) if $(\langle C_1,\varrho\rangle,f,\langle C_2,\varrho\rangle)\in R$, and
$\langle C_1,\varrho\rangle\xTworightarrow{e_1[m]} \langle C_1',\varrho'\rangle$, then $\langle C_2,\varrho\rangle\xTworightarrow{e_2[n]} \langle C_2',\varrho'\rangle$, with
$(\langle C_1',\varrho'\rangle,f[e_1\mapsto e_2],\langle C_2',\varrho'\rangle)\in R$
for all $\varrho,\varrho'\in S$, and vice-versa. $\mathcal{E}_1,\mathcal{E}_2$ are FR weak history-preserving (hp-)bisimilar and are written $\mathcal{E}_1\approx_{hp}^{fr}\mathcal{E}_2$ if there exists
a FR weak hp-bisimulation $R$ such that $(\langle\emptyset,\emptyset\rangle,\emptyset,\langle\emptyset,\emptyset\rangle)\in R$.

A FR weakly hereditary history-preserving (hhp-)bisimulation is a downward closed FR weak hp-bisimulation. $\mathcal{E}_1,\mathcal{E}_2$ are FR weakly hereditary history-preserving
(hhp-)bisimilar and are written $\mathcal{E}_1\approx_{hhp}^{fr}\mathcal{E}_2$.
\end{definition}

\begin{definition}[FR Branching pomset, step bisimulation]
Assume a special termination predicate $\downarrow$, and let $\surd$ represent a state with $\surd\downarrow$. Let $\mathcal{E}_1$, $\mathcal{E}_2$ be PESs. A FR branching pomset
bisimulation is a relation $R\subseteq\langle\mathcal{C}(\mathcal{E}_1),S\rangle\times\langle\mathcal{C}(\mathcal{E}_2),S\rangle$, such that:
 \begin{enumerate}
   \item if $(\langle C_1,\varrho\rangle,\langle C_2,\varrho\rangle)\in R$, and $\langle C_1,\varrho\rangle\xrightarrow{X}\langle C_1',\varrho'\rangle$ then
   \begin{itemize}
     \item either $X\equiv \tau^*$, and $(\langle C_1',\varrho'\rangle,\langle C_2,\varrho\rangle)\in R$ with $\varrho'\in \tau(\varrho)$;
     \item or there is a sequence of (zero or more) $\tau$-transitions $\langle C_2,\varrho\rangle\xrightarrow{\tau^*} \langle C_2^0,\varrho^0\rangle$, such that
     $(\langle C_1,\varrho\rangle,\langle C_2^0,\varrho^0\rangle)\in R$ and $\langle C_2^0,\varrho^0\rangle\xRightarrow{X}\langle C_2',\varrho'\rangle$ with
     $(\langle C_1',\varrho'\rangle,\langle C_2',\varrho'\rangle)\in R$;
   \end{itemize}
   \item if $(\langle C_1,\varrho\rangle,\langle C_2,\varrho\rangle)\in R$, and $\langle C_2,\varrho\rangle\xrightarrow{X}\langle C_2',\varrho'\rangle$ then
   \begin{itemize}
     \item either $X\equiv \tau^*$, and $(\langle C_1,\varrho\rangle,\langle C_2',\varrho'\rangle)\in R$;
     \item or there is a sequence of (zero or more) $\tau$-transitions $\langle C_1,\varrho\rangle\xrightarrow{\tau^*} \langle C_1^0,\varrho^0\rangle$, such that $(\langle C_1^0,\varrho^0\rangle,\langle C_2,\varrho\rangle)\in R$ and $\langle C_1^0,\varrho^0\rangle\xRightarrow{X}\langle C_1',\varrho'\rangle$ with $(\langle C_1',\varrho'\rangle,\langle C_2',\varrho'\rangle)\in R$;
   \end{itemize}
   \item if $(\langle C_1,\varrho\rangle,\langle C_2,\varrho\rangle)\in R$ and $\langle C_1,\varrho\rangle\downarrow$, then there is a sequence of (zero or more) $\tau$-transitions
   $\langle C_2,\varrho\rangle\xrightarrow{\tau^*}\langle C_2^0,\varrho^0\rangle$ such that $(\langle C_1,\varrho\rangle,\langle C_2^0,\varrho^0\rangle)\in R$ and
   $\langle C_2^0,\varrho^0\rangle\downarrow$;
   \item if $(\langle C_1,\varrho\rangle,\langle C_2,\varrho\rangle)\in R$ and $\langle C_2,\varrho\rangle\downarrow$, then there is a sequence of (zero or more) $\tau$-transitions
   $\langle C_1,\varrho\rangle\xrightarrow{\tau^*}\langle C_1^0,\varrho^0\rangle$ such that $(\langle C_1^0,\varrho^0\rangle,\langle C_2,\varrho\rangle)\in R$ and
   $\langle C_1^0,\varrho^0\rangle\downarrow$;
   \item if $(\langle C_1,\varrho\rangle,\langle C_2,\varrho\rangle)\in R$, and $\langle C_1,\varrho\rangle\xtworightarrow{X[\mathcal{K}]}\langle C_1',\varrho'\rangle$ then
   \begin{itemize}
     \item either $X[\mathcal{K}]\equiv \tau^*$, and $(\langle C_1',\varrho'\rangle,\langle C_2,\varrho\rangle)\in R$ with $\varrho'\in \tau(\varrho)$;
     \item or there is a sequence of (zero or more) $\tau$-transitions $\langle C_2,\varrho\rangle\xtworightarrow{\tau^*} \langle C_2^0,\varrho^0\rangle$, such that
     $(\langle C_1,\varrho\rangle,\langle C_2^0,\varrho^0\rangle)\in R$ and $\langle C_2^0,\varrho^0\rangle\xTworightarrow{X[\mathcal{K}]}\langle C_2',\varrho'\rangle$ with
     $(\langle C_1',\varrho'\rangle,\langle C_2',\varrho'\rangle)\in R$;
   \end{itemize}
   \item if $(\langle C_1,\varrho\rangle,\langle C_2,\varrho\rangle)\in R$, and $\langle C_2,\varrho\rangle\xtworightarrow{X}\langle C_2',\varrho'\rangle$ then
   \begin{itemize}
     \item either $X[\mathcal{K}]\equiv \tau^*$, and $(\langle C_1,\varrho\rangle,\langle C_2',\varrho'\rangle)\in R$;
     \item or there is a sequence of (zero or more) $\tau$-transitions $\langle C_1,\varrho\rangle\xtworightarrow{\tau^*} \langle C_1^0,\varrho^0\rangle$, such that
     $(\langle C_1^0,\varrho^0\rangle,\langle C_2,\varrho\rangle)\in R$ and $\langle C_1^0,\varrho^0\rangle\xTworightarrow{X}\langle C_1',\varrho'\rangle$ with
     $(\langle C_1',\varrho'\rangle,\langle C_2',\varrho'\rangle)\in R$;
   \item if $(\langle C_1,\varrho\rangle,\langle C_2,\varrho\rangle)\in R$ and $\langle C_1,\varrho\rangle\downarrow$, then there is a sequence of (zero or more) $\tau$-transitions
   $\langle C_2,\varrho\rangle\xtworightarrow{\tau^*}\langle C_2^0,\varrho^0\rangle$ such that $(\langle C_1,\varrho\rangle,\langle C_2^0,\varrho^0\rangle)\in R$ and
   $\langle C_2^0,\varrho^0\rangle\downarrow$;
   \item if $(\langle C_1,\varrho\rangle,\langle C_2,\varrho\rangle)\in R$ and $\langle C_2,\varrho\rangle\downarrow$, then there is a sequence of (zero or more) $\tau$-transitions
   $\langle C_1,\varrho\rangle\xtworightarrow{\tau^*}\langle C_1^0,\varrho^0\rangle$ such that $(\langle C_1^0,\varrho^0\rangle,\langle C_2,\varrho\rangle)\in R$ and
   $\langle C_1^0,\varrho^0\rangle\downarrow$;
   \end{itemize}
 \end{enumerate}

We say that $\mathcal{E}_1$, $\mathcal{E}_2$ are FR branching pomset bisimilar, written $\mathcal{E}_1\approx_{bp}^{fr}\mathcal{E}_2$, if there exists a FR branching pomset bisimulation $R$, such
that $(\langle\emptyset,\emptyset\rangle,\langle\emptyset,\emptyset\rangle)\in R$.

By replacing FR pomset transitions with FR steps, we can get the definition of FR branching step bisimulation. When PESs $\mathcal{E}_1$ and $\mathcal{E}_2$ are FR branching step bisimilar, we
write $\mathcal{E}_1\approx_{bs}^{fr}\mathcal{E}_2$.
\end{definition}

\begin{definition}[FR rooted branching pomset, step bisimulation]
Assume a special termination predicate $\downarrow$, and let $\surd$ represent a state with $\surd\downarrow$. Let $\mathcal{E}_1$, $\mathcal{E}_2$ be PESs. A FR rooted branching pomset bisimulation is a relation $R\subseteq\langle\mathcal{C}(\mathcal{E}_1),S\rangle\times\langle\mathcal{C}(\mathcal{E}_2),S\rangle$, such that:
 \begin{enumerate}
   \item if $(\langle C_1,\varrho\rangle,\langle C_2,\varrho\rangle)\in R$, and $\langle C_1,\varrho\rangle\xrightarrow{X}\langle C_1',\varrho'\rangle$ then
   $\langle C_2,\varrho\rangle\xrightarrow{X}\langle C_2',\varrho'\rangle$ with $\langle C_1',\varrho'\rangle\approx_{bp}^{fr}\langle C_2',\varrho'\rangle$;
   \item if $(\langle C_1,\varrho\rangle,\langle C_2,\varrho\rangle)\in R$, and $\langle C_2,\varrho\rangle\xrightarrow{X}\langle C_2',\varrho'\rangle$ then
   $\langle C_1,\varrho\rangle\xrightarrow{X}\langle C_1',\varrho'\rangle$ with $\langle C_1',\varrho'\rangle\approx_{bp}^{fr}\langle C_2',\varrho'\rangle$;
   \item if $(\langle C_1,\varrho\rangle,\langle C_2,\varrho\rangle)\in R$, and $\langle C_1,\varrho\rangle\xtworightarrow{X[\mathcal{K}]}\langle C_1',\varrho'\rangle$ then
   $\langle C_2,\varrho\rangle\xtworightarrow{X[\mathcal{K}]}\langle C_2',\varrho'\rangle$ with $\langle C_1',\varrho'\rangle\approx_{bp}^{fr}\langle C_2',\varrho'\rangle$;
   \item if $(\langle C_1,\varrho\rangle,\langle C_2,\varrho\rangle)\in R$, and $\langle C_2,\varrho\rangle\xtworightarrow{X[\mathcal{K}]}\langle C_2',\varrho'\rangle$ then
   $\langle C_1,\varrho\rangle\xtworightarrow{X[\mathcal{K}]}\langle C_1',\varrho'\rangle$ with $\langle C_1',\varrho'\rangle\approx_{bp}^{fr}\langle C_2',\varrho'\rangle$;
   \item if $(\langle C_1,\varrho\rangle,\langle C_2,\varrho\rangle)\in R$ and $\langle C_1,\varrho\rangle\downarrow$, then $\langle C_2,\varrho\rangle\downarrow$;
   \item if $(\langle C_1,\varrho\rangle,\langle C_2,\varrho\rangle)\in R$ and $\langle C_2,\varrho\rangle\downarrow$, then $\langle C_1,\varrho\rangle\downarrow$.
 \end{enumerate}

We say that $\mathcal{E}_1$, $\mathcal{E}_2$ are FR rooted branching pomset bisimilar, written $\mathcal{E}_1\approx_{rbp}^{fr}\mathcal{E}_2$, if there exists a FR rooted branching pomset
bisimulation $R$, such that $(\langle\emptyset,\emptyset\rangle,\langle\emptyset,\emptyset\rangle)\in R$.

By replacing FR pomset transitions with FR steps, we can get the definition of FR rooted branching step bisimulation. When PESs $\mathcal{E}_1$ and $\mathcal{E}_2$ are FR rooted branching step
bisimilar, we write $\mathcal{E}_1\approx_{rbs}^{fr}\mathcal{E}_2$.
\end{definition}

\begin{definition}[FR branching (hereditary) history-preserving bisimulation]\label{BHHPBG}
Assume a special termination predicate $\downarrow$, and let $\surd$ represent a state with $\surd\downarrow$. A FR branching history-preserving (hp-) bisimulation is a weakly posetal
relation $R\subseteq\langle\mathcal{C}(\mathcal{E}_1),S\rangle\overline{\times}\langle\mathcal{C}(\mathcal{E}_2),S\rangle$ such that:

 \begin{enumerate}
   \item if $(\langle C_1,\varrho\rangle,f,\langle C_2,\varrho\rangle)\in R$, and $\langle C_1,\varrho\rangle\xrightarrow{e_1}\langle C_1',\varrho'\rangle$ then
   \begin{itemize}
     \item either $e_1\equiv \tau$, and $(\langle C_1',\varrho'\rangle,f[e_1\mapsto \tau^{e_1}],\langle C_2,\varrho\rangle)\in R$;
     \item or there is a sequence of (zero or more) $\tau$-transitions $\langle C_2,\varrho\rangle\xrightarrow{\tau^*} \langle C_2^0,\varrho^0\rangle$, such that
     $(\langle C_1,\varrho\rangle,f,\langle C_2^0,\varrho^0\rangle)\in R$ and $\langle C_2^0,\varrho^0\rangle\xrightarrow{e_2}\langle C_2',\varrho'\rangle$ with
     $(\langle C_1',\varrho'\rangle,f[e_1\mapsto e_2],\langle C_2',\varrho'\rangle)\in R$;
   \end{itemize}
   \item if $(\langle C_1,\varrho\rangle,f,\langle C_2,\varrho\rangle)\in R$, and $\langle C_2,\varrho\rangle\xrightarrow{e_2}\langle C_2',\varrho'\rangle$ then
   \begin{itemize}
     \item either $e_2\equiv \tau$, and $(\langle C_1,\varrho\rangle,f[e_2\mapsto \tau^{e_2}],\langle C_2',\varrho'\rangle)\in R$;
     \item or there is a sequence of (zero or more) $\tau$-transitions $\langle C_1,\varrho\rangle\xrightarrow{\tau^*} \langle C_1^0,\varrho^0\rangle$, such that
     $(\langle C_1^0,\varrho^0\rangle,f,\langle C_2,\varrho\rangle)\in R$ and $\langle C_1^0,\varrho^0\rangle\xrightarrow{e_1}\langle C_1',\varrho'\rangle$ with
     $(\langle C_1',\varrho'\rangle,f[e_2\mapsto e_1],\langle C_2',\varrho'\rangle)\in R$;
   \end{itemize}
   \item if $(\langle C_1,\varrho\rangle,f,\langle C_2,\varrho\rangle)\in R$ and $\langle C_1,\varrho\rangle\downarrow$, then there is a sequence of (zero or more)
   $\tau$-transitions $\langle C_2,\varrho\rangle\xrightarrow{\tau^*}\langle C_2^0,\varrho^0\rangle$ such that $(\langle C_1,\varrho\rangle,f,\langle C_2^0,\varrho^0\rangle)\in R$
   and    $\langle C_2^0,\varrho^0\rangle\downarrow$;
   \item if $(\langle C_1,\varrho\rangle,f,\langle C_2,\varrho\rangle)\in R$ and $\langle C_2,\varrho\rangle\downarrow$, then there is a sequence of (zero or more) $\tau$-transitions
   $\langle C_1,\varrho\rangle\xrightarrow{\tau^*}\langle C_1^0,\varrho^0\rangle$ such that $(\langle C_1^0,\varrho^0\rangle,f,\langle C_2,\varrho\rangle)\in R$ and
   $\langle C_1^0,\varrho^0\rangle\downarrow$;
   \item if $(\langle C_1,\varrho\rangle,f,\langle C_2,\varrho\rangle)\in R$, and $\langle C_1,\varrho\rangle\xtworightarrow{e_1[m]}\langle C_1',\varrho'\rangle$ then
   \begin{itemize}
     \item either $e_1[m]\equiv \tau$, and $(\langle C_1',\varrho'\rangle,f[e_1\mapsto \tau^{e_1[m]}],\langle C_2,\varrho\rangle)\in R$;
     \item or there is a sequence of (zero or more) $\tau$-transitions $\langle C_2,\varrho\rangle\xtworightarrow{\tau^*} \langle C_2^0,\varrho^0\rangle$, such that
     $(\langle C_1,\varrho\rangle,f,\langle C_2^0,\varrho^0\rangle)\in R$ and $\langle C_2^0,\varrho^0\rangle\xtworightarrow{e_2[n]}\langle C_2',\varrho'\rangle$ with
     $(\langle C_1',\varrho'\rangle,f[e_1[m]\mapsto e_2[n]],\langle C_2',\varrho'\rangle)\in R$;
   \end{itemize}
   \item if $(\langle C_1,\varrho\rangle,f,\langle C_2,\varrho\rangle)\in R$, and $\langle C_2,\varrho\rangle\xtworightarrow{e_2[n]}\langle C_2',\varrho'\rangle$ then
   \begin{itemize}
     \item either $e_2[n]\equiv \tau$, and $(\langle C_1,\varrho\rangle,f[e_2[n]\mapsto \tau^{e_2}],\langle C_2',\varrho'\rangle)\in R$;
     \item or there is a sequence of (zero or more) $\tau$-transitions $\langle C_1,\varrho\rangle\xtworightarrow{\tau^*} \langle C_1^0,\varrho^0\rangle$, such that
     $(\langle C_1^0,\varrho^0\rangle,f,\langle C_2,\varrho\rangle)\in R$ and $\langle C_1^0,\varrho^0\rangle\xtworightarrow{e_1[m]}\langle C_1',\varrho'\rangle$ with
     $(\langle C_1',\varrho'\rangle,f[e_2[n]\mapsto e_1[m]],\langle C_2',\varrho'\rangle)\in R$;
   \end{itemize}
   \item if $(\langle C_1,\varrho\rangle,f,\langle C_2,\varrho\rangle)\in R$ and $\langle C_1,\varrho\rangle\downarrow$, then there is a sequence of (zero or more)
   $\tau$-transitions $\langle C_2,\varrho\rangle\xtworightarrow{\tau^*}\langle C_2^0,\varrho^0\rangle$ such that $(\langle C_1,\varrho\rangle,f,\langle C_2^0,\varrho^0\rangle)\in R$
   and    $\langle C_2^0,\varrho^0\rangle\downarrow$;
   \item if $(\langle C_1,\varrho\rangle,f,\langle C_2,\varrho\rangle)\in R$ and $\langle C_2,\varrho\rangle\downarrow$, then there is a sequence of (zero or more) $\tau$-transitions
   $\langle C_1,\varrho\rangle\xtworightarrow{\tau^*}\langle C_1^0,\varrho^0\rangle$ such that $(\langle C_1^0,\varrho^0\rangle,f,\langle C_2,\varrho\rangle)\in R$ and
   $\langle C_1^0,\varrho^0\rangle\downarrow$.
 \end{enumerate}

$\mathcal{E}_1,\mathcal{E}_2$ are FR branching history-preserving (hp-)bisimilar and are written $\mathcal{E}_1\approx_{bhp}^{fr}\mathcal{E}_2$ if there exists a FR branching hp-bisimulation $R$
such that $(\langle\emptyset,\emptyset\rangle,\emptyset,\langle\emptyset,\emptyset\rangle)\in R$.

A FR branching hereditary history-preserving (hhp-)bisimulation is a downward closed FR branching hp-bisimulation. $\mathcal{E}_1,\mathcal{E}_2$ are FR branching hereditary history-preserving
(hhp-)bisimilar and are written $\mathcal{E}_1\approx_{bhhp}^{fr}\mathcal{E}_2$.
\end{definition}

\begin{definition}[FR rooted branching (hereditary) history-preserving bisimulation]
Assume a special termination predicate $\downarrow$, and let $\surd$ represent a state with $\surd\downarrow$. A FR rooted branching history-preserving (hp-) bisimulation is a weakly
posetal relation $R\subseteq\langle\mathcal{C}(\mathcal{E}_1),S\rangle\overline{\times}\langle\mathcal{C}(\mathcal{E}_2),S\rangle$ such that:

 \begin{enumerate}
   \item if $(\langle C_1,\varrho\rangle,f,\langle C_2,\varrho\rangle)\in R$, and $\langle C_1,\varrho\rangle\xrightarrow{e_1}\langle C_1',\varrho'\rangle$, then
   $\langle C_2,\varrho\rangle\xrightarrow{e_2}\langle C_2',\varrho'\rangle$ with $\langle C_1',\varrho'\rangle\approx_{bhp}^{fr}\langle C_2',\varrho'\rangle$;
   \item if $(\langle C_1,\varrho\rangle,f,\langle C_2,\varrho\rangle)\in R$, and $\langle C_2,\varrho\rangle\xrightarrow{e_2}\langle C_2',\varrho'\rangle$, then
   $\langle C_1,\varrho\rangle\xrightarrow{e_1}\langle C_1',\varrho'\rangle$ with $\langle C_1',\varrho'\rangle\approx_{bhp}^{fr}\langle C_2',\varrho'\rangle$;
   \item if $(\langle C_1,\varrho\rangle,f,\langle C_2,\varrho\rangle)\in R$, and $\langle C_1,\varrho\rangle\xtworightarrow{e_1[m]}\langle C_1',\varrho'\rangle$, then
   $\langle C_2,\varrho\rangle\xtworightarrow{e_2[n]}\langle C_2',\varrho'\rangle$ with $\langle C_1',\varrho'\rangle\approx_{bhp}^{fr}\langle C_2',\varrho'\rangle$;
   \item if $(\langle C_1,\varrho\rangle,f,\langle C_2,\varrho\rangle)\in R$, and $\langle C_2,\varrho\rangle\xtworightarrow{e_2[n]}\langle C_2',\varrho'\rangle$, then
   $\langle C_1,\varrho\rangle\xtworightarrow{e_1[m]}\langle C_1',\varrho'\rangle$ with $\langle C_1',\varrho'\rangle\approx_{bhp}^{fr}\langle C_2',\varrho'\rangle$;
   \item if $(\langle C_1,\varrho\rangle,f,\langle C_2,\varrho\rangle)\in R$ and $\langle C_1,\varrho\rangle\downarrow$, then $\langle C_2,\varrho\rangle\downarrow$;
   \item if $(\langle C_1,\varrho\rangle,f,\langle C_2,\varrho\rangle)\in R$ and $\langle C_2,\varrho\rangle\downarrow$, then $\langle C_1,\varrho\rangle\downarrow$.
 \end{enumerate}

$\mathcal{E}_1,\mathcal{E}_2$ are FR rooted branching history-preserving (hp-)bisimilar and are written $\mathcal{E}_1\approx_{rbhp}^{fr}\mathcal{E}_2$ if there exists a FR rooted branching
hp-bisimulation $R$ such that $(\langle\emptyset,\emptyset\rangle,\emptyset,\langle\emptyset,\emptyset\rangle)\in R$.

A FR rooted branching hereditary history-preserving (hhp-)bisimulation is a downward closed FR rooted branching hp-bisimulation. $\mathcal{E}_1,\mathcal{E}_2$ are FR rooted branching hereditary
history-preserving (hhp-)bisimilar and are written $\mathcal{E}_1\approx_{rbhhp}^{fr}\mathcal{E}_2$.
\end{definition}

\begin{definition}[Probabilistic transitions]
Let $\mathcal{E}$ be a PES and let $C\in\mathcal{C}(\mathcal{E})$, the transition $\langle C,\varrho\rangle\xrsquigarrow{\pi} \langle C^{\pi},\varrho\rangle$ is called a probabilistic
transition
from $\langle C,\varrho\rangle$ to $\langle C^{\pi},\varrho\rangle$.
\end{definition}

\begin{definition}[FR probabilistic pomset, step bisimulation]\label{PSBG}
Let $\mathcal{E}_1$, $\mathcal{E}_2$ be PESs. A FR probabilistic pomset bisimulation is a relation $R\subseteq\langle\mathcal{C}(\mathcal{E}_1),S\rangle\times\langle\mathcal{C}(\mathcal{E}_2),S\rangle$,
such that (1) if $(\langle C_1,\varrho\rangle,\langle C_2,\varrho\rangle)\in R$, and $\langle C_1,\varrho\rangle\xrightarrow{X_1}\langle C_1',\varrho'\rangle$ then
$\langle C_2,\varrho\rangle\xrightarrow{X_2}\langle C_2',\varrho'\rangle$, with $X_1\subseteq \mathbb{E}_1$, $X_2\subseteq \mathbb{E}_2$, $X_1\sim X_2$ and
$(\langle C_1',\varrho'\rangle,\langle C_2',\varrho'\rangle)\in R$ for all $\varrho,\varrho'\in S$, and vice-versa;
(2) if $(\langle C_1,\varrho\rangle,\langle C_2,\varrho\rangle)\in R$, and $\langle C_1,\varrho\rangle\xtworightarrow{X_1[\mathcal{K}_1]}\langle C_1',\varrho'\rangle$ then
$\langle C_2,\varrho\rangle\xrightarrow{X_2[\mathcal{K}_2]}\langle C_2',\varrho'\rangle$, with $X_1\subseteq \mathbb{E}_1$, $X_2\subseteq \mathbb{E}_2$, $X_1\sim X_2$ and
$(\langle C_1',\varrho'\rangle,\langle C_2',\varrho'\rangle)\in R$ for all $\varrho,\varrho'\in S$, and vice-versa;
(3) if $(\langle C_1,\varrho\rangle,\langle C_2,\varrho\rangle)\in R$, and $\langle C_1,\varrho\rangle\xrsquigarrow{\pi}\langle C_1^{\pi},\varrho\rangle$
then $\langle C_2,\varrho\rangle\xrsquigarrow{\pi}\langle C_2^{\pi},\varrho\rangle$ and $(\langle C_1^{\pi},\varrho\rangle,\langle C_2^{\pi},\varrho\rangle)\in R$, and vice-versa; (4) if $(\langle C_1,\varrho\rangle,\langle C_2,\varrho\rangle)\in R$,
then $\mu(C_1,C)=\mu(C_2,C)$ for each $C\in\mathcal{C}(\mathcal{E})/R$; (5) $[\surd]_R=\{\surd\}$. We say that $\mathcal{E}_1$, $\mathcal{E}_2$ are FR probabilistic pomset bisimilar, written
$\mathcal{E}_1\sim_{pp}^{fr}\mathcal{E}_2$, if there exists a probabilistic pomset bisimulation $R$, such that $(\langle\emptyset,\emptyset\rangle,\langle\emptyset,\emptyset\rangle)\in R$.
By replacing FR probabilistic pomset transitions with FR probabilistic steps, we can get the definition of FR probabilistic step bisimulation. When PESs $\mathcal{E}_1$ and $\mathcal{E}_2$ are FR
probabilistic step bisimilar, we write $\mathcal{E}_1\sim_{ps}^{fr}\mathcal{E}_2$.
\end{definition}

\begin{definition}[FR weakly probabilistic pomset, step bisimulation]
Let $\mathcal{E}_1$, $\mathcal{E}_2$ be PESs. A FR weakly probabilistic pomset bisimulation is a relation $R\subseteq\langle\mathcal{C}(\mathcal{E}_1),S\rangle\times\langle\mathcal{C}(\mathcal{E}_2),S\rangle$,
such that (1) if $(\langle C_1,\varrho\rangle,\langle C_2,\varrho\rangle)\in R$, and $\langle C_1,\varrho\rangle\xRightarrow{X_1}\langle C_1',\varrho'\rangle$ then
$\langle C_2,\varrho\rangle\xRightarrow{X_2}\langle C_2',\varrho'\rangle$, with $X_1\subseteq \hat{\mathbb{E}_1}$, $X_2\subseteq \hat{\mathbb{E}_2}$, $X_1\sim X_2$ and
$(\langle C_1',\varrho'\rangle,\langle C_2',\varrho'\rangle)\in R$ for all $\varrho,\varrho'\in S$, and vice-versa;
(2) if $(\langle C_1,\varrho\rangle,\langle C_2,\varrho\rangle)\in R$, and $\langle C_1,\varrho\rangle\xTworightarrow{X_1[\mathcal{K}_1]}\langle C_1',\varrho'\rangle$ then
$\langle C_2,\varrho\rangle\xTworightarrow{X_2[\mathcal{K}_2]}\langle C_2',\varrho'\rangle$, with $X_1\subseteq \hat{\mathbb{E}_1}$, $X_2\subseteq \hat{\mathbb{E}_2}$, $X_1\sim X_2$ and
$(\langle C_1',\varrho'\rangle,\langle C_2',\varrho'\rangle)\in R$ for all $\varrho,\varrho'\in S$, and vice-versa;
(3) if $(\langle C_1,\varrho\rangle,\langle C_2,\varrho\rangle)\in R$, and $\langle C_1,\varrho\rangle\xrsquigarrow{\pi}\langle C_1^{\pi},\varrho\rangle$
then $\langle C_2,\varrho\rangle\xrsquigarrow{\pi}\langle C_2^{\pi},\varrho\rangle$ and $(\langle C_1^{\pi},\varrho\rangle,\langle C_2^{\pi},\varrho\rangle)\in R$, and vice-versa; (4) if $(\langle C_1,\varrho\rangle,\langle C_2,\varrho\rangle)\in R$,
then $\mu(C_1,C)=\mu(C_2,C)$ for each $C\in\mathcal{C}(\mathcal{E})/R$; (5) $[\surd]_R=\{\surd\}$. We say that $\mathcal{E}_1$, $\mathcal{E}_2$ are FR weakly probabilistic pomset bisimilar,
written $\mathcal{E}_1\approx_{pp}^{fr}\mathcal{E}_2$, if there exists a FR weakly probabilistic pomset bisimulation $R$, such that
$(\langle\emptyset,\emptyset\rangle,\langle\emptyset,\emptyset\rangle)\in R$. By replacing FR weakly probabilistic pomset transitions with FR weakly probabilistic steps, we can get the
definition of FR weakly probabilistic step bisimulation. When PESs $\mathcal{E}_1$ and $\mathcal{E}_2$ are FR weakly probabilistic step bisimilar, we write
$\mathcal{E}_1\approx_{ps}^{FR}\mathcal{E}_2$.
\end{definition}

\begin{definition}[Posetal product]
Given two PESs $\mathcal{E}_1$, $\mathcal{E}_2$, the posetal product of their configurations, denoted
$\langle\mathcal{C}(\mathcal{E}_1),S\rangle\overline{\times}\langle\mathcal{C}(\mathcal{E}_2),S\rangle$, is defined as

$$\{(\langle C_1,\varrho\rangle,f,\langle C_2,\varrho\rangle)|C_1\in\mathcal{C}(\mathcal{E}_1),C_2\in\mathcal{C}(\mathcal{E}_2),f:C_1\rightarrow C_2 \textrm{ isomorphism}\}.$$

A subset $R\subseteq\langle\mathcal{C}(\mathcal{E}_1),S\rangle\overline{\times}\langle\mathcal{C}(\mathcal{E}_2),S\rangle$ is called a posetal relation. We say that $R$ is downward
closed when for any $(\langle C_1,\varrho\rangle,f,\langle C_2,\varrho\rangle),(\langle C_1',\varrho'\rangle,f',\langle C_2',\varrho'\rangle)\in \langle\mathcal{C}(\mathcal{E}_1),S\rangle\overline{\times}\langle\mathcal{C}(\mathcal{E}_2),S\rangle$,
if $(\langle C_1,\varrho\rangle,f,\langle C_2,\varrho\rangle)\subseteq (\langle C_1',\varrho'\rangle,f',\langle C_2',\varrho'\rangle)$ pointwise and
$(\langle C_1',\varrho'\rangle,f',\langle C_2',\varrho'\rangle)\in R$, then $(\langle C_1,\varrho\rangle,f,\langle C_2,\varrho\rangle)\in R$.

For $f:X_1\rightarrow X_2$, we define $f[x_1\mapsto x_2]:X_1\cup\{x_1\}\rightarrow X_2\cup\{x_2\}$, $z\in X_1\cup\{x_1\}$,(1)$f[x_1\mapsto x_2](z)=
x_2$,if $z=x_1$;(2)$f[x_1\mapsto x_2](z)=f(z)$, otherwise. Where $X_1\subseteq \mathbb{E}_1$, $X_2\subseteq \mathbb{E}_2$, $x_1\in \mathbb{E}_1$, $x_2\in \mathbb{E}_2$.
\end{definition}

\begin{definition}[Weakly posetal product]
Given two PESs $\mathcal{E}_1$, $\mathcal{E}_2$, the weakly posetal product of their configurations, denoted
$\langle\mathcal{C}(\mathcal{E}_1),S\rangle\overline{\times}\langle\mathcal{C}(\mathcal{E}_2),S\rangle$, is defined as

$$\{(\langle C_1,\varrho\rangle,f,\langle C_2,\varrho\rangle)|C_1\in\mathcal{C}(\mathcal{E}_1),C_2\in\mathcal{C}(\mathcal{E}_2),f:\hat{C_1}\rightarrow \hat{C_2} \textrm{ isomorphism}\}.$$

A subset $R\subseteq\langle\mathcal{C}(\mathcal{E}_1),S\rangle\overline{\times}\langle\mathcal{C}(\mathcal{E}_2),S\rangle$ is called a weakly posetal relation. We say that $R$ is
downward closed when for any $(\langle C_1,\varrho\rangle,f,\langle C_2,\varrho\rangle),(\langle C_1',\varrho'\rangle,f,\langle C_2',\varrho'\rangle)\in \langle\mathcal{C}(\mathcal{E}_1),S\rangle\overline{\times}\langle\mathcal{C}(\mathcal{E}_2),S\rangle$,
if $(\langle C_1,\varrho\rangle,f,\langle C_2,\varrho\rangle)\subseteq (\langle C_1',\varrho'\rangle,f',\langle C_2',\varrho'\rangle)$ pointwise and
$(\langle C_1',\varrho'\rangle,f',\langle C_2',\varrho'\rangle)\in R$, then $(\langle C_1,\varrho\rangle,f,\langle C_2,\varrho\rangle)\in R$.

For $f:X_1\rightarrow X_2$, we define $f[x_1\mapsto x_2]:X_1\cup\{x_1\}\rightarrow X_2\cup\{x_2\}$, $z\in X_1\cup\{x_1\}$,(1)$f[x_1\mapsto x_2](z)=
x_2$,if $z=x_1$;(2)$f[x_1\mapsto x_2](z)=f(z)$, otherwise. Where $X_1\subseteq \hat{\mathbb{E}_1}$, $X_2\subseteq \hat{\mathbb{E}_2}$, $x_1\in \hat{\mathbb{E}}_1$,
$x_2\in \hat{\mathbb{E}}_2$. Also, we define $f(\tau^*)=f(\tau^*)$.
\end{definition}

\begin{definition}[FR probabilistic (hereditary) history-preserving bisimulation]
A FR probabilistic history-preserving (hp-) bisimulation is a posetal relation
$R\subseteq\langle\mathcal{C}(\mathcal{E}_1),S\rangle\overline{\times}\langle\mathcal{C}(\mathcal{E}_2),S\rangle$ such that (1) if $(\langle C_1,\varrho\rangle,f,\langle C_2,\varrho\rangle)\in R$,
and $\langle C_1,\varrho\rangle\xrightarrow{e_1} \langle C_1',\varrho'\rangle$, then $\langle C_2,\varrho\rangle\xrightarrow{e_2} \langle C_2',\varrho'\rangle$, with
$(\langle C_1',\varrho'\rangle,f[e_1\mapsto e_2],\langle C_2',\varrho'\rangle)\in R$ for all $\varrho,\varrho'\in S$, and vice-versa;
(2) if $(\langle C_1,\varrho\rangle,f,\langle C_2,\varrho\rangle)\in R$,
and $\langle C_1,\varrho\rangle\xtworightarrow{e_1[m]} \langle C_1',\varrho'\rangle$, then $\langle C_2,\varrho\rangle\xtworightarrow{e_2[n]} \langle C_2',\varrho'\rangle$, with
$(\langle C_1',\varrho'\rangle,f[e_1[m]\mapsto e_2[n]],\langle C_2',\varrho'\rangle)\in R$ for all $\varrho,\varrho'\in S$, and vice-versa;
(3) if $(\langle C_1,\varrho\rangle,f,\langle C_2,\varrho\rangle)\in R$, and
$\langle C_1,\varrho\rangle\xrsquigarrow{\pi}\langle C_1^{\pi},\varrho\rangle$ then $\langle C_2,\varrho\rangle\xrsquigarrow{\pi}\langle C_2^{\pi},\varrho\rangle$ and $(\langle C_1^{\pi},\varrho\rangle,f,\langle C_2^{\pi},\varrho\rangle)\in R$,
and vice-versa; (4) if $(C_1,f,C_2)\in R$, then $\mu(C_1,C)=\mu(C_2,C)$ for each $C\in\mathcal{C}(\mathcal{E})/R$; (5) $[\surd]_R=\{\surd\}$. $\mathcal{E}_1,\mathcal{E}_2$ are
probabilistic history-preserving (hp-)bisimilar and are written $\mathcal{E}_1\sim_{php}\mathcal{E}_2$ if there exists a probabilistic hp-bisimulation $R$ such that
$(\langle\emptyset,\emptyset\rangle,\emptyset,\langle\emptyset,\emptyset\rangle)\in R$.

A FR probabilistic hereditary history-preserving (hhp-)bisimulation is a downward closed FR probabilistic hp-bisimulation. $\mathcal{E}_1,\mathcal{E}_2$ are FR probabilistic hereditary
history-preserving (hhp-)bisimilar and are written $\mathcal{E}_1\sim_{phhp}^{fr}\mathcal{E}_2$.
\end{definition}

\begin{definition}[FR weakly probabilistic (hereditary) history-preserving bisimulation]
A FR weakly probabilistic history-preserving (hp-) bisimulation is a weakly posetal relation
$R\subseteq\langle\mathcal{C}(\mathcal{E}_1),S\rangle\overline{\times}\langle\mathcal{C}(\mathcal{E}_2),S\rangle$ such that (1) if $(\langle C_1,\varrho\rangle,f,\langle C_2,\varrho\rangle)\in R$,
and $\langle C_1,\varrho\rangle\xRightarrow{e_1} \langle C_1',\varrho'\rangle$, then $\langle C_2,\varrho\rangle\xRightarrow{e_2} \langle C_2',\varrho'\rangle$, with
$(\langle C_1',\varrho'\rangle,f[e_1\mapsto e_2],\langle C_2',\varrho'\rangle)\in R$ for all $\varrho,\varrho'\in S$, and vice-versa;
(2) if $(\langle C_1,\varrho\rangle,f,\langle C_2,\varrho\rangle)\in R$,
and $\langle C_1,\varrho\rangle\xTworightarrow{e_1[m]} \langle C_1',\varrho'\rangle$, then $\langle C_2,\varrho\rangle\xTworightarrow{e_2[n]} \langle C_2',\varrho'\rangle$, with
$(\langle C_1',\varrho'\rangle,f[e_1[m]\mapsto e_2[n]],\langle C_2',\varrho'\rangle)\in R$ for all $\varrho,\varrho'\in S$, and vice-versa;
(3) if $(\langle C_1,\varrho\rangle,f,\langle C_2,\varrho\rangle)\in R$, and
$\langle C_1,\varrho\rangle\xrsquigarrow{\pi}\langle C_1^{\pi},\varrho\rangle$ then $\langle C_2,\varrho\rangle\xrsquigarrow{\pi}\langle C_2^{\pi},\varrho\rangle$ and
$(\langle C_1^{\pi},\varrho\rangle,f,\langle C_2^{\pi},\varrho\rangle)\in R$, and vice-versa; (4) if $(C_1,f,C_2)\in R$, then $\mu(C_1,C)=\mu(C_2,C)$ for each $C\in\mathcal{C}(\mathcal{E})/R$;
(5) $[\surd]_R=\{\surd\}$. $\mathcal{E}_1,\mathcal{E}_2$ are FR weakly probabilistic history-preserving (hp-)bisimilar and are written $\mathcal{E}_1\approx_{php}^{fr}\mathcal{E}_2$ if there
exists a FR weakly probabilistic hp-bisimulation $R$ such that $(\langle\emptyset,\emptyset\rangle,\emptyset,\langle\emptyset,\emptyset\rangle)\in R$.

A FR weakly probabilistic hereditary history-preserving (hhp-)bisimulation is a downward closed FR weakly probabilistic hp-bisimulation. $\mathcal{E}_1,\mathcal{E}_2$ are FR weakly
probabilistic hereditary history-preserving (hhp-)bisimilar and are written $\mathcal{E}_1\approx_{phhp}^{fr}\mathcal{E}_2$.
\end{definition}

\begin{definition}[FR probabilistic branching pomset, step bisimulation]
Assume a special termination predicate $\downarrow$, and let $\surd$ represent a state with $\surd\downarrow$. Let $\mathcal{E}_1$, $\mathcal{E}_2$ be PESs. A FR probabilistic branching
pomset bisimulation is a relation $R\subseteq\langle\mathcal{C}(\mathcal{E}_1),S\rangle\times\langle\mathcal{C}(\mathcal{E}_2),S\rangle$, such that:

 \begin{enumerate}
   \item if $(\langle C_1,\varrho\rangle,\langle C_2,\varrho\rangle)\in R$, and $\langle C_1,\varrho\rangle\xrightarrow{X}\langle C_1',\varrho'\rangle$ then
   \begin{itemize}
     \item either $X\equiv \tau^*$, and $(\langle C_1',\varrho'\rangle,\langle C_2,\varrho\rangle)\in R$ with $\varrho'\in \tau(\varrho)$;
     \item or there is a sequence of (zero or more) probabilistic transitions and $\tau$-transitions $\langle C_2,\varrho\rangle\rightsquigarrow^*\xrightarrow{\tau^*} \langle C_2^0,\varrho^0\rangle$, such that
     $(\langle C_1,\varrho\rangle,\langle C_2^0,\varrho^0\rangle)\in R$ and $\langle C_2^0,\varrho^0\rangle\xRightarrow{X}\langle C_2',\varrho'\rangle$ with
     $(\langle C_1',\varrho'\rangle,\langle C_2',\varrho'\rangle)\in R$;
   \end{itemize}
   \item if $(\langle C_1,\varrho\rangle,\langle C_2,\varrho\rangle)\in R$, and $\langle C_2,\varrho\rangle\xrightarrow{X}\langle C_2',\varrho'\rangle$ then
   \begin{itemize}
     \item either $X\equiv \tau^*$, and $(\langle C_1,\varrho\rangle,\langle C_2',\varrho'\rangle)\in R$;
     \item or there is a sequence of (zero or more) probabilistic transitions and $\tau$-transitions $\langle C_1,\varrho\rangle\rightsquigarrow^*\xrightarrow{\tau^*} \langle C_1^0,\varrho^0\rangle$, such that
     $(\langle C_1^0,\varrho^0\rangle,\langle C_2,\varrho\rangle)\in R$ and $\langle C_1^0,\varrho^0\rangle\xRightarrow{X}\langle C_1',\varrho'\rangle$ with
     $(\langle C_1',\varrho'\rangle,\langle C_2',\varrho'\rangle)\in R$;
   \end{itemize}
   \item if $(\langle C_1,\varrho\rangle,\langle C_2,\varrho\rangle)\in R$ and $\langle C_1,\varrho\rangle\downarrow$, then there is a sequence of (zero or more) probabilistic transitions and $\tau$-transitions
   $\langle C_2,\varrho\rangle\rightsquigarrow^*\xrightarrow{\tau^*}\langle C_2^0,\varrho^0\rangle$ such that $(\langle C_1,\varrho\rangle,\langle C_2^0,\varrho^0\rangle)\in R$ and
   $\langle C_2^0,\varrho^0\rangle\downarrow$;
   \item if $(\langle C_1,\varrho\rangle,\langle C_2,\varrho\rangle)\in R$ and $\langle C_2,\varrho\rangle\downarrow$, then there is a sequence of (zero or more) probabilistic transitions and $\tau$-transitions
   $\langle C_1,\varrho\rangle\rightsquigarrow^*\xrightarrow{\tau^*}\langle C_1^0,\varrho^0\rangle$ such that $(\langle C_1^0,\varrho^0\rangle,\langle C_2,\varrho\rangle)\in R$ and
   $\langle C_1^0,\varrho^0\rangle\downarrow$;
   \item if $(\langle C_1,\varrho\rangle,\langle C_2,\varrho\rangle)\in R$, and $\langle C_1,\varrho\rangle\xtworightarrow{X[\mathcal{K}]}\langle C_1',\varrho'\rangle$ then
   \begin{itemize}
     \item either $X[\mathcal{K}]\equiv \tau^*$, and $(\langle C_1',\varrho'\rangle,\langle C_2,\varrho\rangle)\in R$ with $\varrho'\in \tau(\varrho)$;
     \item or there is a sequence of (zero or more) probabilistic transitions and $\tau$-transitions $\langle C_2,\varrho\rangle\rightsquigarrow^*\xtworightarrow{\tau^*} \langle C_2^0,\varrho^0\rangle$, such that
     $(\langle C_1,\varrho\rangle,\langle C_2^0,\varrho^0\rangle)\in R$ and $\langle C_2^0,\varrho^0\rangle\xTworightarrow{X[\mathcal{K}]}\langle C_2',\varrho'\rangle$ with
     $(\langle C_1',\varrho'\rangle,\langle C_2',\varrho'\rangle)\in R$;
   \end{itemize}
   \item if $(\langle C_1,\varrho\rangle,\langle C_2,\varrho\rangle)\in R$, and $\langle C_2,\varrho\rangle\xtworightarrow{X[\mathcal{K}]}\langle C_2',\varrho'\rangle$ then
   \begin{itemize}
     \item either $X[\mathcal{K}]\equiv \tau^*$, and $(\langle C_1,\varrho\rangle,\langle C_2',\varrho'\rangle)\in R$;
     \item or there is a sequence of (zero or more) probabilistic transitions and $\tau$-transitions $\langle C_1,\varrho\rangle\rightsquigarrow^*\xtworightarrow{\tau^*} \langle C_1^0,\varrho^0\rangle$, such that
     $(\langle C_1^0,\varrho^0\rangle,\langle C_2,\varrho\rangle)\in R$ and $\langle C_1^0,\varrho^0\rangle\xTworightarrow{X[\mathcal{K}]}\langle C_1',\varrho'\rangle$ with
     $(\langle C_1',\varrho'\rangle,\langle C_2',\varrho'\rangle)\in R$;
   \end{itemize}
   \item if $(\langle C_1,\varrho\rangle,\langle C_2,\varrho\rangle)\in R$ and $\langle C_1,\varrho\rangle\downarrow$, then there is a sequence of (zero or more) probabilistic transitions and $\tau$-transitions
   $\langle C_2,\varrho\rangle\rightsquigarrow^*\xtworightarrow{\tau^*}\langle C_2^0,\varrho^0\rangle$ such that $(\langle C_1,\varrho\rangle,\langle C_2^0,\varrho^0\rangle)\in R$ and
   $\langle C_2^0,\varrho^0\rangle\downarrow$;
   \item if $(\langle C_1,\varrho\rangle,\langle C_2,\varrho\rangle)\in R$ and $\langle C_2,\varrho\rangle\downarrow$, then there is a sequence of (zero or more) probabilistic transitions and $\tau$-transitions
   $\langle C_1,\varrho\rangle\rightsquigarrow^*\xtworightarrow{\tau^*}\langle C_1^0,\varrho^0\rangle$ such that $(\langle C_1^0,\varrho^0\rangle,\langle C_2,\varrho\rangle)\in R$ and
   $\langle C_1^0,\varrho^0\rangle\downarrow$;
   \item if $(C_1,C_2)\in R$,then $\mu(C_1,C)=\mu(C_2,C)$ for each $C\in\mathcal{C}(\mathcal{E})/R$;
   \item $[\surd]_R=\{\surd\}$.
 \end{enumerate}

We say that $\mathcal{E}_1$, $\mathcal{E}_2$ are FR probabilistic branching pomset bisimilar, written $\mathcal{E}_1\approx_{pbp}^{fr}\mathcal{E}_2$, if there exists a FR probabilistic branching
pomset bisimulation $R$, such that $(\langle\emptyset,\emptyset\rangle,\langle\emptyset,\emptyset\rangle)\in R$.

By replacing FR probabilistic pomset transitions with steps, we can get the definition of FR probabilistic branching step bisimulation. When PESs $\mathcal{E}_1$ and $\mathcal{E}_2$ are
FR probabilistic branching step bisimilar, we write $\mathcal{E}_1\approx_{pbs}^{fr}\mathcal{E}_2$.
\end{definition}

\begin{definition}[FR probabilistic rooted branching pomset, step bisimulation]
Assume a special termination predicate $\downarrow$, and let $\surd$ represent a state with $\surd\downarrow$. Let $\mathcal{E}_1$, $\mathcal{E}_2$ be PESs. A FR probabilistic rooted
branching pomset bisimulation is a relation $R\subseteq\langle\mathcal{C}(\mathcal{E}_1),S\rangle\times\langle\mathcal{C}(\mathcal{E}_2),S\rangle$, such that:

 \begin{enumerate}
   \item if $(\langle C_1,\varrho\rangle,\langle C_2,\varrho\rangle)\in R$, and $\langle C_1,\varrho\rangle\rightsquigarrow\xrightarrow{X}\langle C_1',\varrho'\rangle$ then
   $\langle C_2,\varrho\rangle\rightsquigarrow\xrightarrow{X}\langle C_2',\varrho'\rangle$ with $\langle C_1',\varrho'\rangle\approx_{pbp}^{fr}\langle C_2',\varrho'\rangle$;
   \item if $(\langle C_1,\varrho\rangle,\langle C_2,\varrho\rangle)\in R$, and $\langle C_2,\varrho\rangle\rightsquigarrow\xrightarrow{X}\langle C_2',\varrho'\rangle$ then
   $\langle C_1,\varrho\rangle\rightsquigarrow\xrightarrow{X}\langle C_1',\varrho'\rangle$ with $\langle C_1',\varrho'\rangle\approx_{pbp}^{fr}\langle C_2',\varrho'\rangle$;
   \item if $(\langle C_1,\varrho\rangle,\langle C_2,\varrho\rangle)\in R$, and $\langle C_1,\varrho\rangle\rightsquigarrow\xtworightarrow{X[\mathcal{K}]}\langle C_1',\varrho'\rangle$ then
   $\langle C_2,\varrho\rangle\rightsquigarrow\xtworightarrow{X[\mathcal{K}]}\langle C_2',\varrho'\rangle$ with $\langle C_1',\varrho'\rangle\approx_{pbp}^{fr}\langle C_2',\varrho'\rangle$;
   \item if $(\langle C_1,\varrho\rangle,\langle C_2,\varrho\rangle)\in R$, and $\langle C_2,\varrho\rangle\rightsquigarrow\xtworightarrow{X[\mathcal{K}]}\langle C_2',\varrho'\rangle$ then
   $\langle C_1,\varrho\rangle\rightsquigarrow\xtworightarrow{X[\mathcal{K}]}\langle C_1',\varrho'\rangle$ with $\langle C_1',\varrho'\rangle\approx_{pbp}^{fr}\langle C_2',\varrho'\rangle$;
   \item if $(\langle C_1,\varrho\rangle,\langle C_2,\varrho\rangle)\in R$ and $\langle C_1,\varrho\rangle\downarrow$, then $\langle C_2,\varrho\rangle\downarrow$;
   \item if $(\langle C_1,\varrho\rangle,\langle C_2,\varrho\rangle)\in R$ and $\langle C_2,\varrho\rangle\downarrow$, then $\langle C_1,\varrho\rangle\downarrow$.
 \end{enumerate}

We say that $\mathcal{E}_1$, $\mathcal{E}_2$ are FR probabilistic rooted branching pomset bisimilar, written $\mathcal{E}_1\approx_{prbp}\mathcal{E}_2$, if there exists a FR probabilistic
rooted branching pomset bisimulation $R$, such that $(\langle\emptyset,\emptyset\rangle,\langle\emptyset,\emptyset\rangle)\in R$.

By replacing FR pomset transitions with steps, we can get the definition of FR probabilistic rooted branching step bisimulation. When PESs $\mathcal{E}_1$ and $\mathcal{E}_2$ are FR probabilistic
rooted branching step bisimilar, we write $\mathcal{E}_1\approx_{prbs}^{fr}\mathcal{E}_2$.
\end{definition}

\begin{definition}[FR probabilistic branching (hereditary) history-preserving bisimulation]
Assume a special termination predicate $\downarrow$, and let $\surd$ represent a state with $\surd\downarrow$. A FR probabilistic branching history-preserving (hp-) bisimulation is a
weakly posetal relation $R\subseteq\langle\mathcal{C}(\mathcal{E}_1),S\rangle\overline{\times}\langle\mathcal{C}(\mathcal{E}_2),S\rangle$ such that:

 \begin{enumerate}
   \item if $(\langle C_1,\varrho\rangle,f,\langle C_2,\varrho\rangle)\in R$, and $\langle C_1,\varrho\rangle\xrightarrow{e_1}\langle C_1',\varrho'\rangle$ then
   \begin{itemize}
     \item either $e_1\equiv \tau$, and $(\langle C_1',\varrho'\rangle,f[e_1\mapsto \tau],\langle C_2,\varrho\rangle)\in R$;
     \item or there is a sequence of (zero or more) probabilistic transitions and $\tau$-transitions $\langle C_2,\varrho\rangle\rightsquigarrow^*\xrightarrow{\tau^*} \langle C_2^0,\varrho^0\rangle$, such that
     $(\langle C_1,\varrho\rangle,f,\langle C_2^0,\varrho^0\rangle)\in R$ and $\langle C_2^0,\varrho^0\rangle\xrightarrow{e_2}\langle C_2',\varrho'\rangle$ with
     $(\langle C_1',\varrho'\rangle,f[e_1\mapsto e_2],\langle C_2',\varrho'\rangle)\in R$;
   \end{itemize}
   \item if $(\langle C_1,\varrho\rangle,f,\langle C_2,\varrho\rangle)\in R$, and $\langle C_2,\varrho\rangle\xrightarrow{e_2}\langle C_2',\varrho'\rangle$ then
   \begin{itemize}
     \item either $e_2\equiv \tau$, and $(\langle C_1,\varrho\rangle,f[e_2\mapsto \tau],\langle C_2',\varrho'\rangle)\in R$;
     \item or there is a sequence of (zero or more) probabilistic transitions and $\tau$-transitions $\langle C_1,\varrho\rangle\rightsquigarrow^*\xrightarrow{\tau^*} \langle C_1^0,\varrho^0\rangle$, such that
     $(\langle C_1^0,\varrho^0\rangle,f,\langle C_2,\varrho\rangle)\in R$ and $\langle C_1^0,\varrho^0\rangle\xrightarrow{e_1}\langle C_1',\varrho'\rangle$ with
     $(\langle C_1',\varrho'\rangle,f[e_2\mapsto e_1],\langle C_2',\varrho'\rangle)\in R$;
   \end{itemize}
   \item if $(\langle C_1,\varrho\rangle,f,\langle C_2,\varrho\rangle)\in R$ and $\langle C_1,\varrho\rangle\downarrow$, then there is a sequence of (zero or more) probabilistic transitions and $\tau$-transitions
   $\langle C_2,\varrho\rangle\rightsquigarrow^*\xrightarrow{\tau^*}\langle C_2^0,\varrho^0\rangle$ such that $(\langle C_1,\varrho\rangle,f,\langle C_2^0,\varrho^0\rangle)\in R$ and
   $\langle C_2^0,\varrho^0\rangle\downarrow$;
   \item if $(\langle C_1,\varrho\rangle,f,\langle C_2,\varrho\rangle)\in R$ and $\langle C_2,\varrho\rangle\downarrow$, then there is a sequence of (zero or more) probabilistic transitions and $\tau$-transitions
   $\langle C_1,\varrho\rangle\rightsquigarrow^*\xrightarrow{\tau^*}\langle C_1^0,\varrho^0\rangle$ such that $(\langle C_1^0,\varrho^0\rangle,f,\langle C_2,\varrho\rangle)\in R$ and
   $\langle C_1^0,\varrho^0\rangle\downarrow$;
   \item if $(\langle C_1,\varrho\rangle,f,\langle C_2,\varrho\rangle)\in R$, and $\langle C_1,\varrho\rangle\xtworightarrow{e_1[m]}\langle C_1',\varrho'\rangle$ then
   \begin{itemize}
     \item either $e_1[m]\equiv \tau$, and $(\langle C_1',\varrho'\rangle,f[e_1[m]\mapsto \tau],\langle C_2,\varrho\rangle)\in R$;
     \item or there is a sequence of (zero or more) probabilistic transitions and $\tau$-transitions $\langle C_2,\varrho\rangle\rightsquigarrow^*\xtworightarrow{\tau^*} \langle C_2^0,\varrho^0\rangle$, such that
     $(\langle C_1,\varrho\rangle,f,\langle C_2^0,\varrho^0\rangle)\in R$ and $\langle C_2^0,\varrho^0\rangle\xtworightarrow{e_2[n]}\langle C_2',\varrho'\rangle$ with
     $(\langle C_1',\varrho'\rangle,f[e_1[m]\mapsto e_2[n]],\langle C_2',\varrho'\rangle)\in R$;
   \end{itemize}
   \item if $(\langle C_1,\varrho\rangle,f,\langle C_2,\varrho\rangle)\in R$, and $\langle C_2,\varrho\rangle\xtworightarrow{e_2[n]}\langle C_2',\varrho'\rangle$ then
   \begin{itemize}
     \item either $e_2[n]\equiv \tau$, and $(\langle C_1,\varrho\rangle,f[e_2\mapsto \tau],\langle C_2',\varrho'\rangle)\in R$;
     \item or there is a sequence of (zero or more) probabilistic transitions and $\tau$-transitions $\langle C_1,\varrho\rangle\rightsquigarrow^*\xtworightarrow{\tau^*} \langle C_1^0,\varrho^0\rangle$, such that
     $(\langle C_1^0,\varrho^0\rangle,f,\langle C_2,\varrho\rangle)\in R$ and $\langle C_1^0,\varrho^0\rangle\xtworightarrow{e_1[m]}\langle C_1',\varrho'\rangle$ with
     $(\langle C_1',\varrho'\rangle,f[e_2[n]\mapsto e_1[m]],\langle C_2',\varrho'\rangle)\in R$;
   \end{itemize}
   \item if $(\langle C_1,\varrho\rangle,f,\langle C_2,\varrho\rangle)\in R$ and $\langle C_1,\varrho\rangle\downarrow$, then there is a sequence of (zero or more) probabilistic transitions and $\tau$-transitions
   $\langle C_2,\varrho\rangle\rightsquigarrow^*\xtworightarrow{\tau^*}\langle C_2^0,\varrho^0\rangle$ such that $(\langle C_1,\varrho\rangle,f,\langle C_2^0,\varrho^0\rangle)\in R$ and
   $\langle C_2^0,\varrho^0\rangle\downarrow$;
   \item if $(\langle C_1,\varrho\rangle,f,\langle C_2,\varrho\rangle)\in R$ and $\langle C_2,\varrho\rangle\downarrow$, then there is a sequence of (zero or more) probabilistic transitions and $\tau$-transitions
   $\langle C_1,\varrho\rangle\rightsquigarrow^*\xtworightarrow{\tau^*}\langle C_1^0,\varrho^0\rangle$ such that $(\langle C_1^0,\varrho^0\rangle,f,\langle C_2,\varrho\rangle)\in R$ and
   $\langle C_1^0,\varrho^0\rangle\downarrow$;
   \item if $(C_1,C_2)\in R$,then $\mu(C_1,C)=\mu(C_2,C)$ for each $C\in\mathcal{C}(\mathcal{E})/R$;
   \item $[\surd]_R=\{\surd\}$.
 \end{enumerate}

$\mathcal{E}_1,\mathcal{E}_2$ are FR probabilistic branching history-preserving (hp-)bisimilar and are written $\mathcal{E}_1\approx_{pbhp}^{fr}\mathcal{E}_2$ if there exists a FR probabilistic
branching hp-bisimulation $R$ such that $(\langle\emptyset,\emptyset\rangle,\emptyset,\langle\emptyset,\emptyset\rangle)\in R$.

A FR probabilistic branching hereditary history-preserving (hhp-)bisimulation is a downward closed FR probabilistic branching hp-bisimulation. $\mathcal{E}_1,\mathcal{E}_2$ are FR probabilistic
branching hereditary history-preserving (hhp-)bisimilar and are written $\mathcal{E}_1\approx_{pbhhp}^{fr}\mathcal{E}_2$.
\end{definition}

\begin{definition}[FR probabilistic rooted branching (hereditary) history-preserving bisimulation]
Assume a special termination predicate $\downarrow$, and let $\surd$ represent a state with $\surd\downarrow$. A FR probabilistic rooted branching history-preserving (hp-) bisimulation is
a weakly posetal relation $R\subseteq\langle\mathcal{C}(\mathcal{E}_1),S\rangle\overline{\times}\langle\mathcal{C}(\mathcal{E}_2),S\rangle$ such that:

 \begin{enumerate}
   \item if $(\langle C_1,\varrho\rangle,f,\langle C_2,\varrho\rangle)\in R$, and $\langle C_1,\varrho\rangle\rightsquigarrow\xrightarrow{e_1}\langle C_1',\varrho'\rangle$, then
   $\langle C_2,\varrho\rangle\rightsquigarrow\xrightarrow{e_2}\langle C_2',\varrho'\rangle$ with $\langle C_1',\varrho'\rangle\approx_{pbhp}^{fr}\langle C_2',\varrho'\rangle$;
   \item if $(\langle C_1,\varrho\rangle,f,\langle C_2,\varrho\rangle)\in R$, and $\langle C_2,\varrho\rangle\rightsquigarrow\xrightarrow{e_2}\langle C_2',\varrho'\rangle$, then
   $\langle C_1,\varrho\rangle\rightsquigarrow\xrightarrow{e_1}\langle C_1',\varrho'\rangle$ with $\langle C_1',\varrho'\rangle\approx_{pbhp}^{fr}\langle C_2',\varrho'\rangle$;
   \item if $(\langle C_1,\varrho\rangle,f,\langle C_2,\varrho\rangle)\in R$, and $\langle C_1,\varrho\rangle\rightsquigarrow\xtworightarrow{e_1[m]}\langle C_1',\varrho'\rangle$, then
   $\langle C_2,\varrho\rangle\rightsquigarrow\xtworightarrow{e_2[n]}\langle C_2',\varrho'\rangle$ with $\langle C_1',\varrho'\rangle\approx_{pbhp}^{fr}\langle C_2',\varrho'\rangle$;
   \item if $(\langle C_1,\varrho\rangle,f,\langle C_2,\varrho\rangle)\in R$, and $\langle C_2,\varrho\rangle\rightsquigarrow\xtworightarrow{e_2[n]}\langle C_2',\varrho'\rangle$, then
   $\langle C_1,\varrho\rangle\rightsquigarrow\xtworightarrow{e_1[m]}\langle C_1',\varrho'\rangle$ with $\langle C_1',\varrho'\rangle\approx_{pbhp}^{fr}\langle C_2',\varrho'\rangle$;
   \item if $(\langle C_1,\varrho\rangle,f,\langle C_2,\varrho\rangle)\in R$ and $\langle C_1,\varrho\rangle\downarrow$, then $\langle C_2,\varrho\rangle\downarrow$;
   \item if $(\langle C_1,\varrho\rangle,f,\langle C_2,\varrho\rangle)\in R$ and $\langle C_2,\varrho\rangle\downarrow$, then $\langle C_1,\varrho\rangle\downarrow$.
 \end{enumerate}

$\mathcal{E}_1,\mathcal{E}_2$ are FR probabilistic rooted branching history-preserving (hp-)bisimilar and are written $\mathcal{E}_1\approx_{prbhp}^{fr}\mathcal{E}_2$ if there exists a FR probabilistic
rooted branching hp-bisimulation $R$ such that $(\langle\emptyset,\emptyset\rangle,\emptyset,\langle\emptyset,\emptyset\rangle)\in R$.

A FR probabilistic rooted branching hereditary history-preserving (hhp-)bisimulation is a downward closed FR probabilistic rooted branching hp-bisimulation. $\mathcal{E}_1,\mathcal{E}_2$ are FR
probabilistic rooted branching hereditary history-preserving (hhp-)bisimilar and are written $\mathcal{E}_1\approx_{prbhhp}^{fr}\mathcal{E}_2$.
\end{definition}

\newpage\section{APRTC for Open Quantum Systems}\label{qaprtc}

In this chapter, we introduce APRTC for open quantum systems, including BARTC for open quantum systems abbreviated qBARTC in section \ref{qbartc}, APRTC for open quantum systems
abbreviated qAPRTC in section \ref{qaprtc2}, recursion in section \ref{qorec}, abstraction in section \ref{qoabs}, quantum entanglement in section \ref{qe1} and unification of quantum
and classical computing for open quantum systems in section \ref{uni1}.

Note that, in open quantum systems, quantum operations denoted $\mathbb{E}$ are the atomic actions (events), and a quantum operation $e\in\mathbb{E}$.

\subsection{BARTC for Open Quantum Systems}\label{qbartc}

In the following, let $e_1, e_2, e_1', e_2'\in \mathbb{E}$, and let variables $x,y,z$ range over the set of terms for true concurrency, $p,q,s$ range over the set of closed terms. The
predicate $Std(x)$ denotes that $x$ contains only standard events (no histories of events) and $NStd(x)$ means that $x$ only contains histories of events. The set of axioms of qBARTC
consists of the laws given in Table \ref{AxiomsForqBARTC5}.

\begin{center}
    \begin{table}
        \begin{tabular}{@{}ll@{}}
            \hline No. &Axiom\\
            $A1$ & $x+ y = y+ x$\\
            $A2$ & $(x+ y)+ z = x+ (y+ z)$\\
            $A3$ & $x+ x = x$\\
            $A41$ & $(x+ y)\cdot z = x\cdot z + y\cdot z\quad (Std(x),Std(y), Std(z))$\\
            $A42$ & $x\cdot (y+z) = x\cdot y + x\cdot z\quad (NStd(x),NStd(y), NStd(z))$\\
            $A5$ & $(x\cdot y)\cdot z = x\cdot(y\cdot z)$\\
        \end{tabular}
        \caption{Axioms of qBARTC}
        \label{AxiomsForqBARTC5}
    \end{table}
\end{center}

\begin{definition}[Basic terms of qBARTC]\label{BTqBARTC5}
The set of basic terms of qBARTC, $\mathcal{B}(qBARTC)$, is inductively defined as follows:
\begin{enumerate}
  \item $\mathbb{E}\subset\mathcal{B}(qBARTC)$;
  \item if $e\in \mathbb{E}, t\in\mathcal{B}(qBARTC)$ then $e\cdot t\in\mathcal{B}(qBARTC)$;
  \item if $e[m]\in \mathbb{E}, t\in\mathcal{B}(qBARTC)$ then $t\cdot e[m]\in\mathcal{B}(qBARTC)$;
  \item if $t,s\in\mathcal{B}(qBARTC)$ then $t+ s\in\mathcal{B}(qBARTC)$.
\end{enumerate}
\end{definition}

\begin{theorem}[Elimination theorem of qBARTC]\label{ETqBARTC5}
Let $p$ be a closed qBARTC term. Then there is a basic qBARTC term $q$ such that $qBARTC\vdash p=q$.
\end{theorem}

\begin{proof}
The same as that of $BARTC$, we omit the proof, please refer to \cite{APRTC} for details.
\end{proof}

In this subsection, we will define a term-deduction system which gives the operational semantics of qBARTC. We give the forward operational transition rules of operators $\cdot$ and $+$
as Table \ref{SETRForqBARTC5} shows, and the reverse rules of operators $\cdot$ and $+$ as Table \ref{RSETRForqBARTC5} shows. And the predicate $\xrightarrow{e}e[m]$ represents
successful forward termination after forward execution of the event $e$, the predicate $\xtworightarrow{e[m]}e$ represents successful reverse termination after reverse execution of
the event history $e[m]$.

\begin{center}
    \begin{table}
        $$\frac{}{\langle e,\varrho\rangle\xrightarrow{e}\langle e[m],\varrho'\rangle}$$
        $$\frac{\langle x,\varrho\rangle\xrightarrow{e}\langle e[m],\varrho'\rangle}{\langle x+ y,\varrho\rangle\xrightarrow{e}\langle e[m],\varrho'\rangle}
        \quad\frac{\langle x,\varrho\rangle\xrightarrow{e}\langle x',\varrho'\rangle}{\langle x+ y,\varrho\rangle\xrightarrow{e}\langle x',\varrho'\rangle}
        \quad\frac{\langle y,\varrho\rangle\xrightarrow{e}\langle e[m],\varrho'\rangle}{\langle x+ y,\varrho\rangle\xrightarrow{e}\langle e[m],\varrho'\rangle}
        \quad\frac{\langle y,\varrho\rangle\xrightarrow{e}\langle y',\varrho'\rangle}{\langle x+ y,\varrho\rangle\xrightarrow{e}\langle y',\varrho'\rangle}$$
        $$\frac{\langle x,\varrho\rangle\xrightarrow{e}\langle e[m],\varrho'\rangle}{\langle x\cdot y,\varrho\rangle\xrightarrow{e} \langle e[m]\cdot y,\varrho'\rangle}
        \quad\frac{\langle x,\varrho\rangle\xrightarrow{e}\langle x',\varrho'\rangle}{\langle x\cdot y,\varrho\rangle\xrightarrow{e}\langle x'\cdot y,\varrho'\rangle}$$
        \caption{Forward single event transition rules of qBARTC}
        \label{SETRForqBARTC5}
    \end{table}
\end{center}

\begin{center}
    \begin{table}
        $$\frac{}{\langle e[m],\varrho\rangle\xtworightarrow{e[m]}\langle e,\varrho'\rangle}$$
        $$\frac{\langle x,\varrho\rangle\xtworightarrow{e[m]}\langle e,\varrho'\rangle}{\langle x+ y,\varrho\rangle\xtworightarrow{e[m]}\langle e,\varrho'\rangle}
        \quad\frac{\langle x,\varrho\rangle\xtworightarrow{e[m]}\langle x',\varrho'\rangle}{\langle x+ y,\varrho\rangle\xtworightarrow{e[m]}\langle x',\varrho'\rangle}
        \quad\frac{\langle y,\varrho\rangle\xtworightarrow{e[m]}\langle e,\varrho'\rangle}{\langle x+ y,\varrho\rangle\xtworightarrow{e[m]}\langle e,\varrho'\rangle}
        \quad\frac{\langle y,\varrho\rangle\xtworightarrow{e[m]}\langle y',\varrho'\rangle}{\langle x+ y,\varrho\rangle\xtworightarrow{e[m]}\langle y',\varrho'\rangle}$$
        $$\frac{\langle y,\varrho\rangle\xtworightarrow{e[m]}\langle e,\varrho'\rangle}{\langle x\cdot y,\varrho\rangle\xtworightarrow{e[m]} \langle x\cdot e,\varrho'\rangle}
        \quad\frac{\langle y,\varrho\rangle\xtworightarrow{e[m]}\langle y',\varrho'\rangle}{\langle x\cdot y,\varrho\rangle\xtworightarrow{e[m]}\langle x\cdot y',\varrho'\rangle}$$
        \caption{Reverse single event transition rules of qBARTC}
        \label{RSETRForqBARTC5}
    \end{table}
\end{center}

The forward pomset transition rules are shown in Table \ref{PTRForqBARTC5}, and reverse pomset transition rules are shown in Table \ref{RPTRForqBARTC5}, different to single event
transition rules, the pomset transition rules are labeled by pomsets, which are defined by causality $\cdot$ and conflict $+$.

\begin{center}
    \begin{table}
        $$\frac{}{\langle X,\varrho\rangle\xrightarrow{X}\langle X[\mathcal{K}],\varrho'\rangle}$$
        $$\frac{\langle x,\varrho\rangle\xrightarrow{X}\langle X[\mathcal{K}],\varrho'\rangle}{\langle x+ y,\varrho\rangle\xrightarrow{X}\langle X[\mathcal{K}],\varrho'\rangle}(X\subseteq x)
        \quad\frac{\langle x,\varrho\rangle\xrightarrow{X}\langle x',\varrho'\rangle}{\langle x+ y,\varrho\rangle\xrightarrow{X}\langle x',\varrho'\rangle}(X\subseteq x)$$
        $$\frac{\langle y,\varrho\rangle\xrightarrow{X}\langle X[\mathcal{K}],\varrho'\rangle}{\langle x+ y,\varrho\rangle\xrightarrow{X}\langle X[\mathcal{K}],\varrho'\rangle}(X\subseteq x)
        \quad\frac{\langle y,\varrho\rangle\xrightarrow{X}\langle y',\varrho'\rangle}{\langle x+ y,\varrho\rangle\xrightarrow{X}\langle y',\varrho'\rangle}(X\subseteq x)$$
        $$\frac{\langle x,\varrho\rangle\xrightarrow{X}\langle X[\mathcal{K}],\varrho'\rangle}{\langle x\cdot y,\varrho\rangle\xrightarrow{X} \langle X[\mathcal{K}]\cdot y,\varrho'\rangle}(X\subseteq x)
        \quad\frac{\langle x,\varrho\rangle\xrightarrow{X}\langle x',\varrho'\rangle}{\langle x\cdot y,\varrho\rangle\xrightarrow{X}\langle x'\cdot y,\varrho'\rangle}(X\subseteq x)$$
        \caption{Forward pomset transition rules of qBARTC}
        \label{PTRForqBARTC5}
    \end{table}
\end{center}

\begin{center}
    \begin{table}
        $$\frac{}{\langle X[\mathcal{K}],\varrho\rangle\xtworightarrow{X[\mathcal{K}]}\langle X,\varrho'\rangle}$$
        $$\frac{\langle x,\varrho\rangle\xtworightarrow{X[\mathcal{K}]}\langle X,\varrho'\rangle}{\langle x+ y,\varrho\rangle\xtworightarrow{X[\mathcal{K}]}\langle X,\varrho'rangle} (X\subseteq x)
        \quad\frac{\langle x,\varrho\rangle\xtworightarrow{X[\mathcal{K}]}\langle x',\varrho'\rangle}{\langle x+ y,\varrho\rangle\xtworightarrow{X[\mathcal{K}]}\langle x',\varrho'\rangle} (X\subseteq x)$$
        $$\frac{\langle y,\varrho\rangle\xtworightarrow{Y[\mathcal{K}]}\langle Y,\varrho'\rangle}{\langle x+ y,\varrho\rangle\xtworightarrow{Y[\mathcal{K}]}\langle Y,\varrho'\rangle} (Y\subseteq y)
        \quad\frac{\langle y,\varrho\rangle\xtworightarrow{Y[\mathcal{K}]}\langle y',\varrho'\rangle}{\langle x+ y,\varrho\rangle\xtworightarrow{Y[\mathcal{K}]}\langle y',\varrho'\rangle}(Y\subseteq y)$$
        $$\frac{\langle y,\varrho\rangle\xtworightarrow{Y[\mathcal{K}]}\langle Y,\varrho'\rangle}{\langle x\cdot y,\varrho\rangle\xtworightarrow{Y[\mathcal{K}]} \langle x\cdot Y,\varrho'\rangle} (Y\subseteq y)
        \quad\frac{\langle y,\varrho\rangle\xtworightarrow{Y[\mathcal{K}]}\langle y',\varrho'\rangle}{\langle x\cdot y,\varrho\rangle\xtworightarrow{Y[\mathcal{K}]}\langle x\cdot y',\varrho'\rangle} (Y\subseteq y)$$
        \caption{Reverse pomset transition rules of qBARTC}
        \label{RPTRForqBARTC5}
    \end{table}
\end{center}

\begin{theorem}[Congruence of qBARTC with respect to FR truly concurrent bisimulation equivalences]
FR truly concurrent bisimulation equivalence $\sim_p^{fr}$, $\sim_s^{fr}$, $\sim_{hp}^{fr}$, $\sim_{hhp}^{fr}$ are all congruences with respect to qBARTC.
\end{theorem}

\begin{proof}
It is obvious that FR truly concurrent bisimulations $\sim_{p}^{fr}$, $\sim_s^{fr}$, $\sim_{hp}^{fr}$ and $\sim_{hhp}^{fr}$ are all equivalent relations with respect to $qBARTC$. So, it is sufficient to prove
that FR truly concurrent bisimulations $\sim_{p}^{fr}$, $\sim_s^{fr}$, $\sim_{hp}^{fr}$ and $\sim_{hhp}^{fr}$ are preserved for $\cdot$ and $+$ according to the transition rules in Table \ref{SETRForqBARTC5},
\ref{RSETRForqBARTC5}, \ref{PTRForqBARTC5}, and \ref{RPTRForqBARTC5},
that is, if $x\sim_{p}^{fr}x'$ and $y\sim_{p}^{fr}y'$, then $x+ y\sim_{p}^{fr}x'+ y'$ and $x\cdot y\sim_{p}^{fr}x'\cdot y'$; if $x\sim_{s}^{fr}x'$ and $y\sim_{s}^{fr}y'$, then $x+ y\sim_{s}^{fr}x'+ y'$ and $x\cdot y\sim_{s}^{fr}x'\cdot y'$;
if $x\sim_{hp}^{fr}x'$ and $y\sim_{hp}^{fr}y'$, then$x+ y\sim_{hp}^{fr}x'+ y'$ and  $x\cdot y\sim_{hp}^{fr}x'\cdot y'$; and if $x\sim_{hhp}^{fr}x'$ and $y\sim_{hhp}^{fr}y'$, then $x+ y\sim_{hhp}^{fr}x'+ y'$ and $x\cdot y\sim_{hhp}^{fr}x'\cdot y'$.
The proof is quit trivial, and we leave the proof as an exercise for the readers.
\end{proof}

\begin{theorem}[Soundness of qBARTC modulo FR truly concurrent bisimulation equivalence]
Let $x$ and $y$ be qBARTC terms.

\begin{enumerate}
  \item If $qBARTC\vdash x=y$, then $x\sim_p^{fr} y$;
  \item If $qBARTC\vdash x=y$, then $x\sim_s^{fr} y$;
  \item If $qBARTC\vdash x=y$, then $x\sim_{hp}^{fr} y$;
  \item If $qBARTC\vdash x=y$, then $x\sim_{hhp}^{fr} y$.
\end{enumerate}
\end{theorem}

\begin{proof}
(1) Since FR pomset bisimulation $\sim_{p}^{fr}$ is both an equivalent and a congruent relation, we only need to check if each axiom in Table \ref{AxiomsForqBARTC5} is sound
modulo FR pomset bisimulation equivalence. We leave the proof as an exercise for the readers.

(2) Since FR step bisimulation $\sim_{s}^{fr}$ is both an equivalent and a congruent relation, we only need to check if each axiom in Table \ref{AxiomsForqBARTC5} is sound modulo
FR step bisimulation equivalence. We leave the proof as an exercise for the readers.

(3) Since FR hp-bisimulation $\sim_{hp}^{fr}$ is both an equivalent and a congruent relation, we only need to check if each axiom in Table \ref{AxiomsForqBARTC5} is sound modulo
FR hp-bisimulation equivalence. We leave the proof as an exercise for the readers.

(4) Since FR hhp-bisimulation $\sim_{hhp}^{fr}$ is both an equivalent and a congruent relation, we only need to check if each axiom in Table \ref{AxiomsForqBARTC5} is sound modulo
FR hhp-bisimulation equivalence. We leave the proof as an exercise for the readers.
\end{proof}

\begin{theorem}[Completeness of qBARTC modulo FR truly concurrent bisimulation equivalence]
Let $p$ and $q$ be closed qBARTC terms, then,

\begin{enumerate}
  \item if $p\sim_p^{fr} q$ then $p=q$;
  \item if $p\sim_s^{fr} q$ then $p=q$;
  \item if $p\sim_{hp}^{fr} q$ then $p=q$;
  \item if $p\sim_{hhp}^{fr} q$ then $p=q$.
\end{enumerate}
\end{theorem}

\begin{proof}
According to the definition of FR truly concurrent bisimulation equivalences $\sim_{p}^{fr}$, $\sim_{s}^{fr}$, $\sim_{hp}^{fr}$ and $\sim_{hhp}^{fr}$, $p\sim_{p}^{fr}q$, $p\sim_{s}^{fr}q$, $p\sim_{hp}^{fr}q$ and $p\sim_{hhp}^{fr}q$ implies
both the bisimilarities between $p$ and $q$, and also the in the same quantum states. According to the completeness of BARTC (please refer to \cite{APRTC} for details), we can get the
completeness of qBARTC.
\end{proof}

\subsection{APRTC for Open Quantum Systems}\label{qaprtc2}

The forward transition rules for parallelism $\parallel$ are shown in Table \ref{TRForqParallel5}, and the reverse transition rules for $\parallel$ are shown in Table
\ref{RTRForqParallel5}.

\begin{center}
    \begin{table}
        $$\frac{\langle x,\varrho\rangle\xrightarrow{e_1}\langle e_1[m],\varrho'\rangle\quad \langle y,\varrho\rangle\xrightarrow{e_2}\langle e_2[m],\varrho''\rangle}{\langle x\parallel y,\varrho\rangle\xrightarrow{\{e_1,e_2\}}\langle e_1[m]\parallel e_2[m],\varrho'\cup\varrho''\rangle}
        \quad\frac{\langle x,\varrho\rangle\xrightarrow{e_1}\langle x',\varrho'\rangle\quad \langle y,\varrho\rangle\xrightarrow{e_2}\langle e_2[m],\varrho''\rangle}{\langle x\parallel y,\varrho\rangle\xrightarrow{\{e_1,e_2\}}\langle x'\parallel e_2[m],\varrho'\cup\varrho''\rangle}$$
        $$\frac{\langle x,\varrho\rangle\xrightarrow{e_1}\langle e_1[m],\varrho'\rangle\quad \langle y,\varrho\rangle\xrightarrow{e_2}\langle y',\varrho''\rangle}{\langle x\parallel y,\varrho\rangle\xrightarrow{\{e_1,e_2\}}\langle e_1[m]\parallel y',\varrho'\cup\varrho''\rangle}
        \quad\frac{\langle x,\varrho\rangle\xrightarrow{e_1}\langle x',\varrho'\rangle\quad \langle y,\varrho\rangle\xrightarrow{e_2}\langle y',\varrho''\rangle}{\langle x\parallel y,\varrho\rangle\xrightarrow{\{e_1,e_2\}}\langle x'\between y',\varrho'\cup\varrho''\rangle}$$
        $$\frac{\langle x,\varrho\rangle\xrightarrow{e_1}\langle e_1[m],\varrho'\rangle\quad \langle y,\varrho\rangle\xrightarrow{e_2}\langle e_2[m],\varrho''\rangle \quad(e_1\leq e_2)}{\langle x\leftmerge y,\varrho\rangle\xrightarrow{\{e_1,e_2\}}\langle e_1[m]\leftmerge e_2[m],\varrho'\cup\varrho''\rangle}
        \quad\frac{\langle x,\varrho\rangle\xrightarrow{e_1}\langle x',\varrho'\rangle\quad \langle y,\varrho\rangle\xrightarrow{e_2}\langle e_2[m],\varrho''\rangle \quad(e_1\leq e_2)}{\langle x\leftmerge y,\varrho\rangle\xrightarrow{\{e_1,e_2\}}\langle x'\leftmerge e_2[m],\varrho'\cup\varrho''\rangle}$$
        $$\frac{\langle x,\varrho\rangle\xrightarrow{e_1}\langle e_1[m],\varrho'\rangle\quad \langle y,\varrho\rangle\xrightarrow{e_2}\langle y',\varrho''\rangle \quad(e_1\leq e_2)}{\langle x\leftmerge y,\varrho\rangle\xrightarrow{\{e_1,e_2\}}\langle e_1[m]\leftmerge y',\varrho'\cup\varrho''\rangle}
        \quad\frac{\langle x,\varrho\rangle\xrightarrow{e_1}\langle x',\varrho'\rangle\quad \langle y,\varrho\rangle\xrightarrow{e_2}\langle y',\varrho''\rangle \quad(e_1\leq e_2)}{\langle x\leftmerge y,\varrho\rangle\xrightarrow{\{e_1,e_2\}}\langle x'\between y',\varrho'\cup\varrho''\rangle}$$
        \caption{Forward transition rules of parallel operator $\parallel$}
        \label{TRForqParallel5}
    \end{table}
\end{center}

\begin{center}
    \begin{table}
        $$\frac{\langle x,\varrho\rangle\xtworightarrow{e_1[m]}\langle e_1,\varrho'\rangle\quad \langle y,\varrho\rangle\xtworightarrow{e_2[m]}\langle e_2,\varrho''\rangle}{\langle x\parallel y,\varrho\rangle\xtworightarrow{\{e_1[m],e_2[m]\}}\langle e_1\parallel e_2,\varrho'\cup\varrho''\rangle}
        \quad\frac{\langle x,\varrho\rangle\xtworightarrow{e_1[m]}\langle x',\varrho'\rangle\quad \langle y,\varrho\rangle\xtworightarrow{e_2[m]}\langle e_2,\varrho''\rangle}{\langle x\parallel y,\varrho\rangle\xtworightarrow{\{e_1[m],e_2[m]\}}\langle x'\parallel e_2,\varrho'\cup\varrho''\rangle}$$
        $$\frac{\langle x,\varrho\rangle\xtworightarrow{e_1[m]}\langle e_1,\varrho'\rangle\quad \langle y,\varrho\rangle\xtworightarrow{e_2[m]}\langle y',\varrho''\rangle}{\langle x\parallel y,\varrho\rangle\xtworightarrow{\{e_1[m],e_2[m]\}}\langle e_1\parallel y',\varrho'\cup\varrho''\rangle}
        \quad\frac{\langle x,\varrho\rangle\xtworightarrow{e_1[m]}\langle x',\varrho'\rangle\quad \langle y,\varrho\rangle\xtworightarrow{e_2[m]}\langle y',\varrho''\rangle}{\langle x\parallel y,\varrho\rangle\xtworightarrow{\{e_1[m],e_2[m]\}}\langle x'\between y',\varrho'\cup\varrho''\rangle}$$
        $$\frac{\langle x,\varrho\rangle\xtworightarrow{e_1[m]}\langle e_1,\varrho'\rangle\quad \langle y,\varrho\rangle\xtworightarrow{e_2[m]}\langle e_2,\varrho''\rangle \quad(e_1\leq e_2)}{\langle x\leftmerge y,\varrho\rangle\xtworightarrow{\{e_1[m],e_2[m]\}}\langle e_1\leftmerge e_2,\varrho'\cup\varrho''\rangle}
        \quad\frac{\langle x,\varrho\rangle\xtworightarrow{e_1[m]}\langle x',\varrho'\rangle\quad \langle y,\varrho\rangle\xtworightarrow{e_2[m]}\langle e_2,\varrho''\rangle \quad(e_1\leq e_2)}{\langle x\leftmerge y,\varrho\rangle\xtworightarrow{\{e_1[m],e_2[m]\}}\langle x'\leftmerge e_2,\varrho'\cup\varrho''\rangle}$$
        $$\frac{\langle x,\varrho\rangle\xtworightarrow{e_1[m]}\langle e_1,\varrho'\rangle\quad \langle y,\varrho\rangle\xtworightarrow{e_2[m]}\langle y',\varrho''\rangle \quad(e_1\leq e_2)}{\langle x\leftmerge y,\varrho\rangle\xtworightarrow{\{e_1[m],e_2[m]\}}\langle e_1\leftmerge y',\varrho'\cup\varrho''\rangle}
        \quad\frac{\langle x,\varrho\rangle\xtworightarrow{e_1[m]}\langle x',\varrho'\rangle\quad \langle y,\varrho\rangle\xtworightarrow{e_2[m]}\langle y',\varrho''\rangle \quad(e_1\leq e_2)}{\langle x\leftmerge y,\varrho\rangle\xtworightarrow{\{e_1[m],e_2[m]\}}\langle x'\between y',\varrho'\cup\varrho''\rangle}$$
        \caption{Reverse transition rules of parallel operator $\parallel$}
        \label{RTRForqParallel5}
    \end{table}
\end{center}

The forward and reverse transition rules of communication $\mid$ are shown in Table \ref{TRForqCommunication5} and Table \ref{RTRForqCommunication5}.

\begin{center}
    \begin{table}
        $$\frac{\langle x,\varrho\rangle\xrightarrow{e_1}\langle e_1[m],\varrho'\rangle\quad \langle y,\varrho\rangle\xrightarrow{e_2}e_2[m]}{\langle x\mid y,\varrho\rangle\xrightarrow{\gamma(e_1,e_2)}\langle\gamma(e_1,e_2)[m],\varrho'\cup\varrho''\rangle}
        \quad\frac{\langle x,\varrho\rangle\xrightarrow{e_1}\langle x',\varrho'\rangle\quad \langle y,\varrho\rangle\xrightarrow{e_2}\langle e_2[m],\varrho''\rangle}{\langle x\mid y,\varrho\rangle\xrightarrow{\gamma(e_1,e_2)}\langle\gamma(e_1,e_2)[m]\cdot x',\varrho'\cup\varrho''\rangle}$$
        $$\frac{\langle x,\varrho\rangle\xrightarrow{e_1}\langle e_1[m],\varrho'\rangle\quad \langle y,\varrho\rangle\xrightarrow{e_2}\langle y',\varrho''\rangle}{\langle x\mid y,\varrho\rangle\xrightarrow{\gamma(e_1,e_2)}\langle \gamma(e_1,e_2)[m]\cdot y',\varrho'\cup\varrho''\rangle}
        \quad\frac{\langle x,\varrho\rangle\xrightarrow{e_1}\langle x',\varrho'\rangle\quad \langle y,\varrho\rangle\xrightarrow{e_2}\langle y',\varrho''\rangle}{\langle x\mid y,\varrho\rangle\xrightarrow{\gamma(e_1,e_2)}\langle\gamma(e_1,e_2)[m]\cdot x'\between y',\varrho'\cup\varrho''\rangle}$$
        \caption{Forward transition rules of communication operator $\mid$}
        \label{TRForqCommunication5}
    \end{table}
\end{center}

\begin{center}
    \begin{table}
        $$\frac{\langle x,\varrho\rangle\xtworightarrow{e_1[m]}\langle e_1,\varrho'\rangle\quad \langle y,\varrho\rangle\xtworightarrow{e_2[m]}\langle e_2,\varrho''\rangle}{\langle x\mid y,\varrho\rangle\xtworightarrow{\gamma(e_1,e_2)[m]}\langle\gamma(e_1,e_2),\varrho'\cup\varrho''\rangle}
        \quad\frac{\langle x,\varrho\rangle\xtworightarrow{e_1[m]}\langle x',\varrho'\rangle\quad \langle y,\varrho\rangle\xtworightarrow{e_2[m]}\langle e_2,\varrho''\rangle}{\langle x\mid y,\varrho\rangle\xtworightarrow{\gamma(e_1,e_2)[m]}\langle\gamma(e_1,e_2)\cdot x',\varrho'\cup\varrho''\rangle}$$
        $$\frac{\langle x,\varrho\rangle\xtworightarrow{e_1[m]}\langle e_1,\varrho'\rangle\quad \langle y,\varrho\rangle\xtworightarrow{e_2[m]}\langle y',\varrho''\rangle}{\langle x\mid y,\varrho\rangle\xtworightarrow{\gamma(e_1,e_2)[m]}\langle\gamma(e_1,e_2)\cdot y',\varrho'\cup\varrho''\rangle}
        \quad\frac{\langle x,\varrho\rangle\xtworightarrow{e_1[m]}\langle x',\varrho'\rangle\quad \langle y,\varrho\rangle\xtworightarrow{e_2[m]}\langle y',\varrho''\rangle}{\langle x\mid y,\varrho\rangle\xtworightarrow{\gamma(e_1,e_2)[m]}\langle\gamma(e_1,e_2)\cdot x'\between y',\varrho'\cup\varrho''\rangle}$$
        \caption{Reverse transition rules of communication operator $\mid$}
        \label{RTRForqCommunication5}
    \end{table}
\end{center}

The conflict elimination is also captured by two auxiliary operators, the unary conflict elimination operator $\Theta$ and the binary unless operator $\triangleleft$. The forward and
reverse transition rules for $\Theta$ and $\triangleleft$ are expressed by ten transition rules in Table \ref{TRForqConflict5} and Table \ref{RTRForqConflict5}.

\begin{center}
    \begin{table}
        $$\frac{\langle x,\varrho\rangle\xrightarrow{e_1}\langle e_1[m],\varrho'\rangle\quad (\sharp(e_1,e_2))}{\langle\Theta(x),\varrho\rangle\xrightarrow{e_1}\langle e_1[m],\varrho'\rangle}
        \quad\frac{\langle x,\varrho\rangle\xrightarrow{e_2}\langle e_2[n],\varrho'\rangle\quad (\sharp(e_1,e_2))}{\langle\Theta(x),\varrho\rangle\xrightarrow{e_2}\langle e_2[n],\varrho'\rangle}$$
        $$\frac{\langle x,\varrho\rangle\xrightarrow{e_1}\langle x',\varrho'\rangle\quad (\sharp(e_1,e_2))}{\langle\Theta(x),\varrho\rangle\xrightarrow{e_1}\langle\Theta(x'),\varrho'\rangle}
        \quad\frac{\langle x,\varrho\rangle\xrightarrow{e_2}\langle x',\varrho'\rangle\quad (\sharp(e_1,e_2))}{\langle \Theta(x),\varrho\rangle\xrightarrow{e_2}\langle\Theta(x'),\varrho'\rangle}$$
        $$\frac{\langle x,\varrho\rangle\xrightarrow{e_1}\langle e_1[m],\varrho'\rangle \quad \langle y,\varrho\rangle\nrightarrow^{e_2}\quad (\sharp(e_1,e_2))}{\langle x\triangleleft y,\varrho\rangle\xrightarrow{\tau}\langle\surd,\tau(\varrho')\rangle}
        \quad\frac{\langle x,\varrho\rangle\xrightarrow{e_1}\langle x',\varrho'\rangle \quad \langle y,\varrho\rangle\nrightarrow^{e_2}\quad (\sharp(e_1,e_2))}{\langle x\triangleleft y,\varrho\rangle\xrightarrow{\tau}\langle x',\tau(\varrho')\rangle}$$
        $$\frac{\langle x,\varrho\rangle\xrightarrow{e_1}\langle e_1[m],\varrho'\rangle \quad \langle y,\varrho\rangle\nrightarrow^{e_3}\quad (\sharp(e_1,e_2),e_2\leq e_3)}{\langle x\triangleleft y,\varrho\rangle\xrightarrow{e_1}\langle e_1[m],\varrho'\rangle}
        \quad\frac{\langle x,\varrho\rangle\xrightarrow{e_1}\langle x',\varrho'\rangle \quad \langle y,\varrho\rangle\nrightarrow^{e_3}\quad (\sharp(e_1,e_2),e_2\leq e_3)}{\langle x\triangleleft y,\varrho\rangle\xrightarrow{e_1}\langle x',\varrho'\rangle}$$
        $$\frac{\langle x,\varrho\rangle\xrightarrow{e_3}\langle e_3[l],\varrho'\rangle \quad \langle y,\varrho\rangle\nrightarrow^{e_2}\quad (\sharp(e_1,e_2),e_1\leq e_3)}{\langle x\triangleleft y,\varrho\rangle\xrightarrow{\tau}\langle\surd,\tau(\varrho')\rangle}
        \quad\frac{\langle x,\varrho\rangle\xrightarrow{e_3}\langle x',\varrho'\rangle \quad \langle y,\varrho\rangle\nrightarrow^{e_2}\quad (\sharp(e_1,e_2),e_1\leq e_3)}{\langle x\triangleleft y,\varrho\rangle\xrightarrow{\tau}\langle x',\tau(\varrho')\rangle}$$
        \caption{Forward transition rules of conflict elimination}
        \label{TRForqConflict5}
    \end{table}
\end{center}

\begin{center}
    \begin{table}
        $$\frac{\langle x,\varrho\rangle\xtworightarrow{e_1[m]}\langle e_1,\varrho'\rangle\quad (\sharp(e_1,e_2))}{\langle\Theta(x),\varrho\rangle\xtworightarrow{e_1[m]}\langle e_1,\varrho'\rangle}
        \quad\frac{\langle x,\varrho\rangle\xtworightarrow{e_2[n]}\langle e_2,\varrho'\rangle\quad (\sharp(e_1,e_2))}{\langle\Theta(x),\varrho\rangle\xtworightarrow{e_2[n]}\langle e_2,\varrho'\rangle}$$
        $$\frac{\langle x,\varrho\rangle\xtworightarrow{e_1[m]}\langle x',\varrho'\rangle\quad (\sharp(e_1,e_2))}{\langle\Theta(x),\varrho\rangle\xtworightarrow{e_1[m]}\langle \Theta(x'),\varrho'\rangle}
        \quad\frac{\langle x,\varrho\rangle\xtworightarrow{e_2[n]}\langle x',\varrho'\rangle\quad (\sharp(e_1,e_2))}{\langle\Theta(x),\varrho\rangle\xtworightarrow{e_2[n]}\langle\Theta(x'),\varrho'\rangle}$$
        $$\frac{\langle x,\varrho\rangle\xtworightarrow{e_1[m]}\langle e_1,\varrho'\rangle \quad \langle y,\varrho\rangle\xntworightarrow{e_2[n]}\quad (\sharp(e_1,e_2))}{\langle x\triangleleft y,\varrho\rangle\xtworightarrow{\tau}\langle\surd,\tau(\varrho')\rangle}
        \quad\frac{\langle x,\varrho\rangle\xtworightarrow{e_1[m]}\langle x',\varrho'\rangle \quad \langle y,\varrho\rangle\xntworightarrow{e_2[n]}\quad (\sharp(e_1,e_2))}{\langle x\triangleleft y,\varrho\rangle\xtworightarrow{\tau}\langle x',\tau(\varrho')\rangle}$$
        $$\frac{\langle x,\varrho\rangle\xtworightarrow{e_1[m]}\langle e_1,\varrho'\rangle \quad \langle y,\varrho\rangle\xntworightarrow{e_3[l]}\quad (\sharp(e_1,e_2),e_2\geq e_3)}{\langle x\triangleleft y,\varrho\rangle\xtworightarrow{e_1[m]}\langle e_1,\varrho'\rangle}
        \quad\frac{\langle x,\varrho\rangle\xtworightarrow{e_1[m]}x' \quad \langle y,\varrho\rangle\xntworightarrow{e_3[l]}\quad (\sharp(e_1,e_2),e_2\geq e_3)}{\langle x\triangleleft y,\varrho\rangle\xtworightarrow{e_1[m]}\langle x',\varrho'\rangle}$$
        $$\frac{\langle x,\varrho\rangle\xtworightarrow{e_3[l]}e_3 \quad \langle y,\varrho\rangle\xntworightarrow{e_2[n]}\quad (\sharp(e_1,e_2),e_1\geq e_3)}{\langle x\triangleleft y,\varrho\rangle\xtworightarrow{\tau}\langle\surd,\tau(\varrho')\rangle}
        \quad\frac{\langle x,\varrho\rangle\xtworightarrow{e_3[l]}x' \quad \langle y,\varrho\rangle\xntworightarrow{e_2[n]}\quad (\sharp(e_1,e_2),e_1\geq e_3)}{\langle x\triangleleft y,\varrho\rangle\xtworightarrow{\tau}\langle x',\tau(\varrho')\rangle}$$
        \caption{Reverse transition rules of conflict elimination}
        \label{RTRForqConflict5}
    \end{table}
\end{center}

\begin{theorem}[Congruence theorem of qAPRTC]
FR truly concurrent bisimulation equivalences $\sim_{p}^{fr}$, $\sim_s^{fr}$, $\sim_{hp}^{fr}$ and $\sim_{hhp}^{fr}$ are all congruences with respect to qAPRTC.
\end{theorem}

\begin{proof}
It is obvious that FR truly concurrent bisimulations $\sim_{p}^{fr}$, $\sim_s^{fr}$, $\sim_{hp}^{fr}$ and $\sim_{hhp}^{fr}$ are all equivalent relations with respect to $qAPRTC$. So, it is sufficient to prove
that truly concurrent bisimulations $\sim_{p}^{fr}$, $\sim_s^{fr}$, $\sim_{hp}^{fr}$ and $\sim_{hhp}^{fr}$ are preserved for $\between$, $\parallel$, $\leftmerge$, $\mid$, $\Theta$, $\triangleleft$ and $\partial_H$
according to the transition rules in Table \ref{TRForqParallel5}, \ref{RTRForqParallel5}, \ref{TRForqCommunication5}, \ref{RTRForqCommunication5}, \ref{TRForqConflict5} and \ref{RTRForqConflict5},
that is, if $x\sim_{p}^{fr}x'$ and $y\sim_{p}^{fr}y'$, then $x\between y\sim_{p}^{fr}x'\between y'$, $x\parallel y\sim_{p}^{fr}x'\parallel y'$,
$x\leftmerge y\sim_{p}^{fr}x'\leftmerge y'$, $x\mid y\sim_{p}^{fr}x'\mid y'$, $\Theta(x)\sim_{p}^{fr}\Theta(x')$, $x\triangleleft y\sim_{p}^{fr}x'\triangleleft y'$; if $x\sim_{s}^{fr}x'$ and $y\sim_{s}^{fr}y'$,
then $x\between y\sim_{s}^{fr}x'\between y'$, $x\parallel y\sim_{s}^{fr}x'\parallel y'$,
$x\leftmerge y\sim_{s}^{fr}x'\leftmerge y'$, $x\mid y\sim_{s}^{fr}x'\mid y'$, $\Theta(x)\sim_{s}^{fr}\Theta(x')$, $x\triangleleft y\sim_{s}^{fr}x'\triangleleft y'$;
if $x\sim_{hp}^{fr}x'$ and $y\sim_{hp}^{fr}y'$, then $x\between y\sim_{hp}^{fr}x'\between y'$, $x\parallel y\sim_{hp}^{fr}x'\parallel y'$,
$x\leftmerge y\sim_{hp}^{fr}x'\leftmerge y'$, $x\mid y\sim_{hp}^{fr}x'\mid y'$, $\Theta(x)\sim_{hp}^{fr}\Theta(x')$, $x\triangleleft y\sim_{hp}^{fr}x'\triangleleft y'$; and if $x\sim_{hhp}^{fr}x'$ and $y\sim_{hhp}^{fr}y'$,
then $x\between y\sim_{hhp}^{fr}x'\between y'$, $x\parallel y\sim_{hhp}^{fr}x'\parallel y'$,
$x\leftmerge y\sim_{hhp}^{fr}x'\leftmerge y'$, $x\mid y\sim_{hhp}^{fr}x'\mid y'$, $\Theta(x)\sim_{hhp}^{fr}\Theta(x')$, $x\triangleleft y\sim_{hhp}^{fr}x'\triangleleft y'$.
The proof is quit trivial, and we leave the proof as an exercise for the readers.
\end{proof}

\begin{definition}[Basic terms of qAPRTC]\label{BTAPTC5}
The set of basic terms of qAPRTC, $\mathcal{B}(qAPRTC)$, is inductively defined as follows:
\begin{enumerate}
  \item $\mathbb{E}\subset\mathcal{B}(qAPRTC)$;
  \item if $e\in \mathbb{E}, t\in\mathcal{B}(qAPRTC)$ then $e\cdot t\in\mathcal{B}(qAPRTC)$;
  \item if $e[m]\in \mathbb{E}, t\in\mathcal{B}(qAPRTC)$ then $t\cdot e[m]\in\mathcal{B}(qAPRTC)$;
  \item if $t,s\in\mathcal{B}(qAPRTC)$ then $t+ s\in\mathcal{B}(qAPRTC)$;
  \item if $t,s\in\mathcal{B}(qAPRTC)$ then $t\leftmerge s\in\mathcal{B}(qAPRTC)$.
\end{enumerate}
\end{definition}

We design the axioms of parallelism in Table \ref{AxiomsForqParallelism5}, including algebraic laws for parallel operator $\parallel$, communication operator $\mid$, conflict
elimination operator $\Theta$ and unless operator $\triangleleft$, and also the whole parallel operator $\between$. Since the communication between two communicating events in
different parallel branches may cause deadlock (a state of inactivity), which is caused by mismatch of two communicating events or the imperfectness of the communication channel.
We introduce a new constant $\delta$ to denote the deadlock, and let the atomic event $e\in \mathbb{E}\cup\{\delta\}$.

\begin{center}
    \begin{table}
        \begin{tabular}{@{}ll@{}}
            \hline No. &Axiom\\
            $A6$ & $x+ \delta = x$\\
            $A7$ & $\delta\cdot x =\delta(\textrm{Std(x)})$\\
            $RA7$ & $x\cdot \delta =\delta(\textrm{NStd(x)})$\\
            $P1$ & $x\between y = x\parallel y + x\mid y$\\
            $P2$ & $x\parallel y = y \parallel x$\\
            $P3$ & $(x\parallel y)\parallel z = x\parallel (y\parallel z)$\\
            $P4$ & $x\parallel y=x\leftmerge y+y\leftmerge x$\\
            $P5$ & $(e_1\leq e_2)\quad e_1\leftmerge (e_2\cdot y) = (e_1\leftmerge e_2)\cdot y$\\
            $RP5$ & $(e_1[m]\leq e_2[m])\quad e_1[m]\leftmerge (y\cdot e_2[m]) = y\cdot(e_1[m]\leftmerge e_2[m])$\\
            $P6$ & $(e_1\leq e_2)\quad (e_1\cdot x)\leftmerge e_2 = (e_1\leftmerge e_2)\cdot x$\\
            $RP6$ & $(e_1[m]\leq e_2[m])\quad (x\cdot e_1[m])\leftmerge e_2[m] = x\cdot(e_1[m]\leftmerge e_2[m])$\\
            $P7$ & $(e_1\leq e_2)\quad (e_1\cdot x)\leftmerge (e_2\cdot y) = (e_1\leftmerge e_2)\cdot (x\between y)$\\
            $RP7$ & $(e_1[m]\leq e_2[m])\quad(x\cdot e_1[m])\leftmerge (y\cdot e_2[m]) = (x\between y)\cdot(e_1[m]\leftmerge e_2[m])$\\
            $P8$ & $(x+ y)\leftmerge z = (x\leftmerge z)+ (y\leftmerge z)(\textrm{Std(x)})$\\
            $RP8$ & $x\leftmerge (y+ z) = (x\leftmerge y)+ (x\leftmerge z)(\textrm{NStd(x)})$\\
            $P9$ & $\delta\leftmerge x = \delta(\textrm{Std(x)},\textrm{Std(y)},\textrm{Std(z)})$\\
            $RP9$ & $x\leftmerge \delta = \delta(\textrm{NStd(x)},\textrm{NStd(y)},\textrm{NStd(z)})$\\
            $C1$ & $e_1\mid e_2 = \gamma(e_1,e_2)$\\
            $RC1$ & $e_1[m]\mid e_2[m] = \gamma(e_1,e_2)[m]$\\
            $C2$ & $e_1\mid (e_2\cdot y) = \gamma(e_1,e_2)\cdot y$\\
            $RC2$ & $e_1[m]\mid (y \cdot e_2[m]) =y\cdot \gamma(e_1,e_2)[m]$\\
            $C3$ & $(e_1\cdot x)\mid e_2 = \gamma(e_1,e_2)\cdot x$\\
            $RC3$ & $(x \cdot e_1[m])\mid e_2[m] =x\cdot \gamma(e_1,e_2)[m]$\\
            $C4$ & $(e_1\cdot x)\mid (e_2\cdot y) = \gamma(e_1,e_2)\cdot (x\between y)$\\
            $RC4$ & $(x \cdot e_1[m])\mid (y \cdot e_2[m]) =(x\between y)\cdot \gamma(e_1,e_2)[m]$\\
            $C5$ & $(x+ y)\mid z = (x\mid z) + (y\mid z)$\\
            $C6$ & $x\mid (y+ z) = (x\mid y)+ (x\mid z)$\\
            $C7$ & $\delta\mid x = \delta$\\
            $C8$ & $x\mid\delta = \delta$\\
            $CE1$ & $\Theta(e) = e$\\
            $RCE1$ & $\Theta(e[m]) = e[m]$\\
            $CE2$ & $\Theta(\delta) = \delta$\\
            $CE3$ & $\Theta(x+ y) = \Theta(x)\triangleleft y + \Theta(y)\triangleleft x$\\
            $CE4$ & $\Theta(x\cdot y)=\Theta(x)\cdot\Theta(y)$\\
            $CE5$ & $\Theta(x\parallel y) = ((\Theta(x)\triangleleft y)\parallel y)+ ((\Theta(y)\triangleleft x)\parallel x)$\\
            $CE6$ & $\Theta(x\mid y) = ((\Theta(x)\triangleleft y)\mid y)+ ((\Theta(y)\triangleleft x)\mid x)$\\
        \end{tabular}
        \caption{Axioms of parallelism}
        \label{AxiomsForqParallelism5}
    \end{table}
\end{center}

\begin{center}
    \begin{table}
        \begin{tabular}{@{}ll@{}}
            \hline No. &Axiom\\
            $U1$ & $(\sharp(e_1,e_2))\quad e_1\triangleleft e_2 = \tau$\\
            $RU1$ & $(\sharp(e_1[m],e_2[n]))\quad e_1[m]\triangleleft e_2[n] = \tau$\\
            $U2$ & $(\sharp(e_1,e_2),e_2\leq e_3)\quad e_1\triangleleft e_3 = e_1$\\
            $RU2$ & $(\sharp(e_1[m],e_2[n]),e_2[n]\geq e_3[l])\quad e_1[m]\triangleleft e_3[l] = e_1[m]$\\
            $U3$ & $(\sharp(e_1,e_2),e_2\leq e_3)\quad e3\triangleleft e_1 = \tau$\\
            $RU3$ & $(\sharp(e_1[m],e_2[n]),e_2[n]\geq e_3[l])\quad e3[l]\triangleleft e_1[m] = \tau$\\
            $U4$ & $e\triangleleft \delta = e$\\
            $U5$ & $\delta \triangleleft e = \delta$\\
            $U6$ & $(x+ y)\triangleleft z = (x\triangleleft z)+ (y\triangleleft z)$\\
            $U7$ & $(x\cdot y)\triangleleft z = (x\triangleleft z)\cdot (y\triangleleft z)$\\
            $U8$ & $(x\parallel y)\triangleleft z = (x\triangleleft z)\parallel (y\triangleleft z)$\\
            $U9$ & $(x\mid y)\triangleleft z = (x\triangleleft z)\mid (y\triangleleft z)$\\
            $U10$ & $x\triangleleft (y+ z) = (x\triangleleft y)\triangleleft z$\\
            $U11$ & $x\triangleleft (y\cdot z)=(x\triangleleft y)\triangleleft z$\\
            $U12$ & $x\triangleleft (y\parallel z) = (x\triangleleft y)\triangleleft z$\\
            $U13$ & $x\triangleleft (y\mid z) = (x\triangleleft y)\triangleleft z$\\
        \end{tabular}
        \caption{Axioms of parallelism (continuing)}
        \label{AxiomsForqParallelism52}
    \end{table}
\end{center}

Based on the definition of basic terms for qAPRTC and axioms of parallelism (see Table \ref{AxiomsForqParallelism5}), we can prove the elimination theorem
of parallelism.

\begin{theorem}[Elimination theorem of FR parallelism]\label{ETParallelism5}
Let $p$ be a closed qAPRTC term. Then there is a basic qAPRTC term $q$ such that $qAPRTC\vdash p=q$.
\end{theorem}

\begin{proof}
The same as that of $APRTC$, we omit the proof, please refer to \cite{APRTC} for details.
\end{proof}

\begin{theorem}[Generalization of the algebra for parallelism with respect to qBARTC]
The algebra for parallelism is a generalization of qBARTC.
\end{theorem}

\begin{proof}
It follows from the following three facts.

\begin{enumerate}
  \item The transition rules of $qBARTC$ in are all source-dependent;
  \item The sources of the transition rules of the algebra for parallelism contain an occurrence of $\between$, or $\parallel$, or $\leftmerge$, or $\mid$, or $\Theta$, or $\triangleleft$;
  \item The transition rules of the algebra for parallelism are all source-dependent.
\end{enumerate}
\end{proof}

\begin{theorem}[Soundness of parallelism modulo FR truly concurrent bisimulation equivalence]
Let $x$ and $y$ be qAPRTC terms.

\begin{enumerate}
  \item If $qAPRTC\vdash x=y$, then $x\sim_{p}^{fr} y$;
  \item If $qAPRTC\vdash x=y$, then $x\sim_{s}^{fr} y$;
  \item If $qAPRTC\vdash x=y$, then $x\sim_{hp}^{fr} y$;
  \item If $qAPRTC\vdash x=y$, then $x\sim_{hhp}^{fr} y$.
\end{enumerate}
\end{theorem}

\begin{proof}
(1) Since FR pomset bisimulation $\sim_{p}^{fr}$ is both an equivalent and a congruent relation, we only need to check if each axiom in Table \ref{AxiomsForqParallelism5} is sound
modulo FR pomset bisimulation equivalence. We leave the proof as an exercise for the readers.

(2) Since FR step bisimulation $\sim_{s}^{fr}$ is both an equivalent and a congruent relation, we only need to check if each axiom in Table \ref{AxiomsForqParallelism5} is sound modulo
FR step bisimulation equivalence. We leave the proof as an exercise for the readers.

(3) Since FR hp-bisimulation $\sim_{hp}^{fr}$ is both an equivalent and a congruent relation, we only need to check if each axiom in Table \ref{AxiomsForqParallelism5} is sound modulo
FR hp-bisimulation equivalence. We leave the proof as an exercise for the readers.

(4) Since FR hhp-bisimulation $\sim_{hhp}^{fr}$ is both an equivalent and a congruent relation, we only need to check if each axiom in Table \ref{AxiomsForqParallelism5} is sound modulo
FR hhp-bisimulation equivalence. We leave the proof as an exercise for the readers.
\end{proof}

\begin{theorem}[Completeness of parallelism modulo FR truly concurrent bisimulation equivalence]
Let $p$ and $q$ be closed qAPRTC terms, then,

\begin{enumerate}
  \item if $p\sim_{p}^{fr} q$ then $p=q$;
  \item if $p\sim_{s}^{fr} q$ then $p=q$;
  \item if $p\sim_{hp}^{fr} q$ then $p=q$;
  \item if $p\sim_{hhp}^{fr} q$ then $p=q$.
\end{enumerate}
\end{theorem}

\begin{proof}
According to the definition of FR truly concurrent bisimulation equivalences $\sim_{p}^{fr}$, $\sim_{s}^{fr}$, $\sim_{hp}^{fr}$ and $\sim_{hhp}^{fr}$, $p\sim_{p}^{fr}q$, $p\sim_{s}^{fr}q$,
$p\sim_{hp}^{fr}q$ and $p\sim_{hhp}^{fr}q$ implies
both the bisimilarities between $p$ and $q$, and also the in the same quantum states. According to the completeness of APRTC (please refer to \cite{APRTC} for details), we can get the
completeness of qAPRTC.
\end{proof}

The mismatch of two communicating events in different parallel branches can cause deadlock, so the deadlocks in the concurrent processes should be eliminated. Like $APRTC$ \cite{APRTC},
we also introduce the unary encapsulation operator $\partial_H$ for set $H$ of atomic events, which renames all atomic events in $H$ into $\delta$. The whole algebra including
parallelism for true concurrency in the above subsections, deadlock $\delta$ and encapsulation operator $\partial_H$, is called Reversible Algebra for Parallelism in True Concurrency,
abbreviated qAPRTC.

The forward transition rules of encapsulation operator $\partial_H$ are shown in Table \ref{TRForqEncapsulation5}, and the reverse transition rules of encapsulation operator
$\partial_H$ are shown in Table \ref{RTRForqEncapsulation5}.

\begin{center}
    \begin{table}
        $$\frac{\langle x,\varrho\rangle\xrightarrow{e}\langle e[m],\varrho'\rangle}{\langle\partial_H(x),\varrho\rangle\xrightarrow{e}\langle\partial_H(e[m]),\varrho'\rangle}\quad (e\notin H)
        \quad\frac{\langle x,\varrho\rangle\xrightarrow{e}\langle x',\varrho'\rangle}{\langle\partial_H(x),\varrho\rangle\xrightarrow{e}\langle\partial_H(x'),\varrho'\rangle}\quad(e\notin H)$$
        \caption{Forward transition rules of encapsulation operator $\partial_H$}
        \label{TRForqEncapsulation5}
    \end{table}
\end{center}

\begin{center}
    \begin{table}
        $$\frac{\langle x,\varrho\rangle\xtworightarrow{e[m]}\langle e,\varrho'\rangle}{\langle\partial_H(x),\varrho\rangle\xtworightarrow{e[m]}\langle e,\varrho'\rangle}\quad (e\notin H)
        \quad\frac{\langle x,\varrho\rangle\xtworightarrow{e}\langle x',\varrho'\rangle}{\langle \partial_H(x),\varrho\rangle\xtworightarrow{e}\langle\partial_H(x'),\varrho'\rangle}\quad(e\notin H)$$
        \caption{Reverse transition rules of encapsulation operator $\partial_H$}
        \label{RTRForqEncapsulation5}
    \end{table}
\end{center}

Based on the transition rules for encapsulation operator $\partial_H$ in Table \ref{TRForqEncapsulation5} and Table \ref{RTRForqEncapsulation5}, we design the axioms as Table
\ref{AxiomsForqEncapsulation5} shows.

\begin{center}
    \begin{table}
        \begin{tabular}{@{}ll@{}}
            \hline No. &Axiom\\
            $D1$ & $e\notin H\quad\partial_H(e) = e$\\
            $RD1$ & $e\notin H\quad\partial_H(e[m]) = e[m]$\\
            $D2$ & $e\in H\quad \partial_H(e) = \delta$\\
            $RD2$ & $e\in H\quad \partial_H(e[m]) = \delta$\\
            $D3$ & $\partial_H(\delta) = \delta$\\
            $D4$ & $\partial_H(x+ y) = \partial_H(x)+\partial_H(y)$\\
            $D5$ & $\partial_H(x\cdot y) = \partial_H(x)\cdot\partial_H(y)$\\
            $D6$ & $\partial_H(x\leftmerge y) = \partial_H(x)\leftmerge\partial_H(y)$\\
        \end{tabular}
        \caption{Axioms of encapsulation operator}
        \label{AxiomsForqEncapsulation5}
    \end{table}
\end{center}

\begin{theorem}[Conservativity of qAPRTC with respect to the algebra for parallelism]
qAPRTC is a conservative extension of the algebra for parallelism.
\end{theorem}

\begin{proof}
It follows from the following three facts.

\begin{enumerate}
  \item The transition rules of the algebra for parallelism in are all source-dependent;
  \item The sources of the transition rules of $qAPRTC$ contain an occurrence of $\partial_H$;
  \item The transition rules of $qAPRTC$ are all source-dependent.
\end{enumerate}
\end{proof}

\begin{theorem}[Congruence theorem of encapsulation operator $\partial_H$]
Truly concurrent bisimulation equivalences $\sim_p^{fr}$, $\sim_s^{fr}$, $\sim_{hp}^{fr}$ and $\sim_{hhp}^{fr}$ are all congruences with respect to encapsulation operator $\partial_H$.
\end{theorem}

\begin{proof}
It is obvious that FR truly concurrent bisimulations $\sim_{p}^{fr}$, $\sim_s^{fr}$, $\sim_{hp}^{fr}$ and $\sim_{hhp}^{fr}$ are all equivalent relations with respect to $qBARTC$. So, it is sufficient to prove
that FR truly concurrent bisimulations $\sim_{p}^{fr}$, $\sim_s^{fr}$, $\sim_{hp}^{fr}$ and $\sim_{hhp}^{fr}$ are preserved for $\partial_H$ according to the transition rules in Table \ref{TRForqEncapsulation5},
\ref{RTRForqEncapsulation5},
that is, if $x\sim_{p}^{fr}x'$, then $\partial_H(x)\sim_{p}^{fr}\partial_H(x')$; if $x\sim_{s}^{fr}x'$, then $\partial_H(x)\sim_{s}^{fr}\partial_H(x')$;
if $x\sim_{hp}^{fr}x'$, then $\partial_H(x)\sim_{hp}^{fr}\partial_H(x')$; and if $x\sim_{hhp}^{fr}x'$, then $\partial_H(x)\sim_{hhp}^{fr}\partial_H(x')$.
The proof is quit trivial, and we leave the proof as an exercise for the readers.
\end{proof}

\begin{theorem}[Elimination theorem of qAPRTC]
Let $p$ be a closed qAPRTC term including the encapsulation operator $\partial_H$. Then there is a basic qAPRTC term $q$ such that $qAPRTC\vdash p=q$.
\end{theorem}

\begin{proof}
The same as that of $APRTC$, we omit the proof, please refer to \cite{APRTC} for details.
\end{proof}

\begin{theorem}[Soundness of qAPRTC modulo FR truly concurrent bisimulation equivalence]
Let $x$ and $y$ be qAPRTC terms including encapsulation operator $\partial_H$.

\begin{enumerate}
  \item If $qAPRTC\vdash x=y$, then $x\sim_{p}^{fr} y$;
  \item If $qAPRTC\vdash x=y$, then $x\sim_{s}^{fr} y$;
  \item If $qAPRTC\vdash x=y$, then $x\sim_{hp}^{fr} y$;
  \item If $qAPRTC\vdash x=y$, then $x\sim_{hhp}^{fr} y$.
\end{enumerate}
\end{theorem}

\begin{proof}
(1) Since FR pomset bisimulation $\sim_{p}^{fr}$ is both an equivalent and a congruent relation, we only need to check if each axiom in Table \ref{AxiomsForqEncapsulation5} is sound
modulo FR pomset bisimulation equivalence. We leave the proof as an exercise for the readers.

(2) Since FR step bisimulation $\sim_{s}^{fr}$ is both an equivalent and a congruent relation, we only need to check if each axiom in Table \ref{AxiomsForqEncapsulation5} is sound modulo
FR step bisimulation equivalence. We leave the proof as an exercise for the readers.

(3) Since FR hp-bisimulation $\sim_{hp}^{fr}$ is both an equivalent and a congruent relation, we only need to check if each axiom in Table \ref{AxiomsForqEncapsulation5} is sound modulo
FR hp-bisimulation equivalence. We leave the proof as an exercise for the readers.

(4) Since FR hhp-bisimulation $\sim_{hhp}^{fr}$ is both an equivalent and a congruent relation, we only need to check if each axiom in Table \ref{AxiomsForqEncapsulation5} is sound modulo
FR hhp-bisimulation equivalence. We leave the proof as an exercise for the readers.
\end{proof}

\begin{theorem}[Completeness of qAPRTC modulo FR truly concurrent bisimulation equivalence]
Let $p$ and $q$ be closed qAPRTC terms including encapsulation operator $\partial_H$, then,

\begin{enumerate}
  \item if $p\sim_{p}^{fr} q$ then $p=q$;
  \item if $p\sim_{s}^{fr} q$ then $p=q$;
  \item if $p\sim_{hp}^{fr} q$ then $p=q$;
  \item if $p\sim_{hhp}^{fr} q$ then $p=q$.
\end{enumerate}
\end{theorem}

\begin{proof}
According to the definition of FR truly concurrent bisimulation equivalences $\sim_{p}^{fr}$, $\sim_{s}^{fr}$, $\sim_{hp}^{fr}$ and $\sim_{hhp}^{fr}$, $p\sim_{p}^{fr}q$, $p\sim_{s}^{fr}q$,
$p\sim_{hp}^{fr}q$ and $p\sim_{hhp}^{fr}q$ implies
both the bisimilarities between $p$ and $q$, and also the in the same quantum states. According to the completeness of APRTC (please refer to \cite{APRTC} for details), we can get the
completeness of qAPRTC.
\end{proof}

\subsection{Recursion}\label{qorec}

\begin{definition}[Recursive specification]
A recursive specification is a finite set of recursive equations

$$X_1=t_1(X_1,\cdots,X_n)$$
$$\cdots$$
$$X_n=t_n(X_1,\cdots,X_n)$$

where the left-hand sides of $X_i$ are called recursion variables, and the right-hand sides $t_i(X_1,\cdots,X_n)$ are process terms in $qAPRTC$ with possible occurrences of the recursion
variables $X_1,\cdots,X_n$.
\end{definition}

\begin{definition}[Solution]
Processes $p_1,\cdots,p_n$ are a solution for a recursive specification $\{X_i=t_i(X_1,\cdots,X_n)|i\in\{1,\cdots,n\}\}$ (with respect to truly concurrent bisimulation equivalences
$\sim_s^{fr}$($\sim_p^{fr}$, $\sim_{hp}^{fr}$, $\sim_{hhp}^{fr}$)) if $p_i\sim_s^{fr} (\sim_p^{fr}, \sim_{hp}^{fr},\sim{hhp}^{fr})t_i(p_1,\cdots,p_n)$ for $i\in\{1,\cdots,n\}$.
\end{definition}

\begin{definition}[Guarded recursive specification]
A recursive specification

$$X_1=t_1(X_1,\cdots,X_n)$$
$$...$$
$$X_n=t_n(X_1,\cdots,X_n)$$

is guarded if the right-hand sides of its recursive equations can be adapted to the form by applications of the axioms in $qAPRTC$ and replacing recursion variables by the right-hand
sides of their recursive equations,

$$(a_{11}\leftmerge\cdots\leftmerge a_{1i_1})\cdot s_1(X_1,\cdots,X_n)+\cdots+(a_{k1}\leftmerge\cdots\leftmerge a_{ki_k})\cdot s_k(X_1,\cdots,X_n)+(b_{11}\leftmerge\cdots\leftmerge b_{1j_1})+\cdots+(b_{1j_1}\leftmerge\cdots\leftmerge b_{lj_l})$$

where $a_{11},\cdots,a_{1i_1},a_{k1},\cdots,a_{ki_k},b_{11},\cdots,b_{1j_1},b_{1j_1},\cdots,b_{lj_l}\in \mathbb{E}$, and the sum above is allowed to be empty, in which case it
represents the deadlock $\delta$.
\end{definition}

\begin{definition}[Linear recursive specification]
A recursive specification is linear if its recursive equations are of the form

$$(a_{11}\leftmerge\cdots\leftmerge a_{1i_1})X_1+\cdots+(a_{k1}\leftmerge\cdots\leftmerge a_{ki_k})X_k+(b_{11}\leftmerge\cdots\leftmerge b_{1j_1})+\cdots+(b_{1j_1}\leftmerge\cdots\leftmerge b_{lj_l})$$

where $a_{11},\cdots,a_{1i_1},a_{k1},\cdots,a_{ki_k},b_{11},\cdots,b_{1j_1},b_{1j_1},\cdots,b_{lj_l}\in \mathbb{E}$, and the sum above is allowed to be empty, in which case it
represents the deadlock $\delta$.
\end{definition}

\begin{center}
    \begin{table}
        $$\frac{\langle t_i(\langle X_1|E\rangle,\cdots,\langle X_n|E\rangle),\varrho\rangle\xrightarrow{\{e_1,\cdots,e_k\}}\langle e_1[m]\leftmerge\cdots\leftmerge e_k[m],\varrho'\rangle}{\langle\langle X_i|E\rangle,\varrho\rangle\xrightarrow{\{e_1,\cdots,e_k\}}\langle e_1[m]\leftmerge\cdots\leftmerge e_k[m],\varrho'\rangle}$$
        $$\frac{\langle t_i(\langle X_1|E\rangle,\cdots,\langle X_n|E\rangle),\varrho\rangle\xrightarrow{\{e_1,\cdots,e_k\}} \langle y,\varrho'\rangle}{\langle\langle X_i|E\rangle,\varrho\rangle\xrightarrow{\{e_1,\cdots,e_k\}} \langle y,\varrho'\rangle}$$
        $$\frac{\langle t_i(\langle X_1|E\rangle,\cdots,\langle X_n|E\rangle),\varrho\rangle\xtworightarrow{\{e_1[m],\cdots,e_k[m]\}}\langle e_1\leftmerge\cdots\leftmerge e_k,\varrho'\rangle}{\langle\langle X_i|E\rangle,\varrho\rangle\xtworightarrow{\{e_1[m],\cdots,e_k[m]\}}\langle e_1\leftmerge\cdots\leftmerge e_k,\varrho'\rangle}$$
        $$\frac{\langle t_i(\langle X_1|E\rangle,\cdots,\langle X_n|E\rangle),\varrho\rangle\xtworightarrow{\{e_1[m],\cdots,e_k[m]\}} \langle y,\varrho'\rangle}{\langle\langle X_i|E\rangle,\varrho\rangle\xtworightarrow{\{e_1[m],\cdots,e_k[m]\}} \langle y,\varrho'\rangle}$$
        \caption{Transition rules of guarded recursion}
        \label{TRForGR}
    \end{table}
\end{center}

The $RDP$ (Recursive Definition Principle) and the $RSP$ (Recursive Specification Principle) are shown in Table \ref{RDPRSP}.

\begin{center}
\begin{table}
  \begin{tabular}{@{}ll@{}}
\hline No. &Axiom\\
  $RDP$ & $\langle X_i|E\rangle = t_i(\langle X_1|E,\cdots,X_n|E\rangle)\quad (i\in\{1,\cdots,n\})$\\
  $RSP$ & if $y_i=t_i(y_1,\cdots,y_n)$ for $i\in\{1,\cdots,n\}$, then $y_i=\langle X_i|E\rangle \quad(i\in\{1,\cdots,n\})$\\
\end{tabular}
\caption{Recursive definition and specification principle}
\label{RDPRSP}
\end{table}
\end{center}

\begin{theorem}[Conservitivity of qAPRTC with guarded recursion]
qAPRTC with guarded recursion is a conservative extension of qAPRTC.
\end{theorem}

\begin{proof}
It follows from the following three facts.

\begin{enumerate}
  \item The transition rules of $qAPRTC$ in are all source-dependent;
  \item The sources of the transition rules $qAPRTC$ with guarded recursion contain only one constant;
  \item The transition rules of $qAPRTC$ with guarded recursion are all source-dependent.
\end{enumerate}

So, $qAPRTC$ with guarded recursion is a conservative extension of $qAPRTC$, as desired.
\end{proof}

\begin{theorem}[Congruence theorem of qAPRTC with guarded recursion]
Truly concurrent bisimulation equivalences $\sim_p^{fr}$, $\sim_s^{fr}$ and $\sim_{hp}^{fr}$ are all congruences with respect to qAPRTC with guarded recursion.
\end{theorem}

\begin{proof}
It follows the following two facts:
\begin{enumerate}
  \item in a guarded recursive specification, right-hand sides of its recursive equations can be adapted to the form by applications of the axioms in $qAPRTC$ and replacing recursion
  variables by the right-hand sides of their recursive equations;
  \item truly concurrent bisimulation equivalences $\sim_{p}^{fr}$, $\sim_{s}^{fr}$, $\sim_{hp}^{fr}$ and $\sim_{hhp}^{fr}$ are all congruences with respect to all operators of $qAPRTC$.
\end{enumerate}
\end{proof}

\begin{theorem}[Elimination theorem of qAPRTC with linear recursion]\label{ETRecursion5}
Each process term in qAPRTC with linear recursion is equal to a process term $\langle X_1|E\rangle$ with $E$ a linear recursive specification.
\end{theorem}

\begin{proof}
The same as that of $APRTC$ with linear recursion, we omit the proof, please refer to \cite{APRTC} for details.
\end{proof}

\begin{theorem}[Soundness of qAPRTC with guarded recursion]\label{SqAPRTCR5}
Let $x$ and $y$ be qAPRTC with guarded recursion terms. If $qAPRTC\textrm{ with guarded recursion}\vdash x=y$, then
\begin{enumerate}
  \item $x\sim_{s}^{fr} y$;
  \item $x\sim_p^{fr} y$;
  \item $x\sim_{hp}^{fr} y$;
  \item $x\sim_{hhp}^{fr} y$.
\end{enumerate}
\end{theorem}

\begin{proof}
(1) Since FR pomset bisimulation $\sim_{p}^{fr}$ is both an equivalent and a congruent relation, we only need to check if each axiom in Table \ref{RDPRSP} is sound
modulo FR pomset bisimulation equivalence. We leave the proof as an exercise for the readers.

(2) Since FR step bisimulation $\sim_{s}^{fr}$ is both an equivalent and a congruent relation, we only need to check if each axiom in Table \ref{RDPRSP} is sound modulo FR
step bisimulation equivalence. We leave the proof as an exercise for the readers.

(3) Since FR hp-bisimulation $\sim_{hp}^{fr}$ is both an equivalent and a congruent relation, we only need to check if each axiom in Table \ref{RDPRSP} is sound modulo FR
hp-bisimulation equivalence. We leave the proof as an exercise for the readers.

(4) Since FR hhp-bisimulation $\sim_{hhp}^{fr}$ is both an equivalent and a congruent relation, we only need to check if each axiom in Table \ref{RDPRSP} is sound modulo FR
hhp-bisimulation equivalence. We leave the proof as an exercise for the readers.
\end{proof}

\begin{theorem}[Completeness of qAPRTC with linear recursion]\label{CqAPRTCR5}
Let $p$ and $q$ be closed qAPRTC with linear recursion terms, then,
\begin{enumerate}
  \item if $p\sim_{s}^{fr} q$ then $p=q$;
  \item if $p\sim_p^{fr} q$ then $p=q$;
  \item if $p\sim_{hp}^{fr} q$ then $p=q$;
  \item if $p\sim_{hhp}^{fr} q$ then $p=q$.
\end{enumerate}
\end{theorem}

\begin{proof}
According to the definition of truly concurrent bisimulation equivalences $\sim_{p}^{fr}$, $\sim_{s}^{fr}$, $\sim_{hp}^{fr}$ and $\sim_{hhp}^{fr}$, $p\sim_{p}^{fr}q$, $p\sim_{s}^{fr}q$,
$p\sim_{hp}^{fr}q$ and $p\sim_{hhp}^{fr}q$ implies
both the bisimilarities between $p$ and $q$, and also the in the same quantum states. According to the completeness of APRTC with linear recursion (please refer to \cite{APRTC} for details), we can get the
completeness of qAPRTC with linear recursion.
\end{proof}

\subsection{Abstraction}\label{qoabs}

\begin{definition}[Guarded linear recursive specification]
A recursive specification is linear if its recursive equations are of the form

$$(a_{11}\leftmerge\cdots\leftmerge a_{1i_1})X_1+\cdots+(a_{k1}\leftmerge\cdots\leftmerge a_{ki_k})X_k+(b_{11}\leftmerge\cdots\leftmerge b_{1j_1})+\cdots+(b_{1j_1}\leftmerge\cdots\leftmerge b_{lj_l})$$

where $a_{11},\cdots,a_{1i_1},a_{k1},\cdots,a_{ki_k},b_{11},\cdots,b_{1j_1},b_{1j_1},\cdots,b_{lj_l}\in \mathbb{E}\cup\{\tau\}$, and the sum above is allowed to be empty, in which case
it represents the deadlock $\delta$.

A linear recursive specification $E$ is guarded if there does not exist an infinite sequence of $\tau$-transitions
$\langle X|E\rangle\xrightarrow{\tau}\langle X'|E\rangle\xrightarrow{\tau}\langle X''|E\rangle\xrightarrow{\tau}\cdots$.
\end{definition}

To abstract away from the internal implementations of a program, and verify that the program exhibits the desired external behaviors, the silent step $\tau$ and abstraction operator
$\tau_I$ are introduced, where $I\subseteq \mathbb{E}$ denotes the internal events. The transition rule of $\tau$ is shown in Table \ref{TRForqTau5}. In the following, let the atomic
event $e$ range over $\mathbb{E}\cup\{\delta\}\cup\{\tau\}$, and let the communication function $\gamma:\mathbb{E}\cup\{\tau\}\times \mathbb{E}\cup\{\tau\}\rightarrow
\mathbb{E}\cup\{\delta\}$, with each communication involved $\tau$ resulting in $\delta$.

\begin{center}
    \begin{table}
        $$\frac{}{\langle\tau,\varrho\rangle\xrightarrow{\tau}\langle\surd,\tau(\varrho)\rangle}$$
        $$\frac{}{\langle\tau,\varrho\rangle\xtworightarrow{\tau}\langle\surd,\tau(\varrho)\rangle}$$
        \caption{Transition rule of the silent step}
        \label{TRForqTau5}
    \end{table}
\end{center}

\begin{theorem}[Conservitivity of $RAPTC$ with guarded linear recursion and silent step]
$RAPTC$ with guarded linear recursion and silent step is a conservative extension of $RAPTC$ with guarded linear recursion.
\end{theorem}

\begin{proof}
Since the transition rules of $qAPRTC$ with silent step and guarded linear recursion are source-dependent, and the transition rules for $\tau$ in Table
\ref{TRForqTau5} contain only a fresh constant $\tau$ in their source, so the transition rules of $qAPTC$ with silent step and guarded linear recursion is a conservative extension
of those of $qAPRTC$ with guarded linear recursion.
\end{proof}

\begin{theorem}[Congruence theorem of $RAPTC$ with guarded linear recursion and silent step]
FR Rooted branching truly concurrent bisimulation equivalences $\approx_{rbp}^{fr}$, $\approx_{rbs}^{fr}$ and $\approx_{rbhp}^{fr}$ are all congruences with respect to $RAPTC$ with guarded linear recursion and silent step.
\end{theorem}

\begin{proof}
It follows the following three facts:
\begin{enumerate}
  \item in a guarded linear recursive specification, right-hand sides of its recursive equations can be adapted to the form by applications of the axioms in $qAPRTC$ and replacing
  recursion variables by the right-hand sides of their recursive equations;
  \item FR truly concurrent bisimulation equivalences $\sim_{p}^{fr}$, $\sim_{s}^{fr}$, $\sim_{hp}^{fr}$ and $\sim_{hhp}^{fr}$ are all congruences with respect to all operators of
  $qAPRTC$, while FR truly concurrent bisimulation equivalences $\sim_{p}^{fr}$, $\sim_{s}^{fr}$, $\sim_{hp}^{fr}$ and $\sim_{hhp}^{fr}$ imply the corresponding FR rooted
  branching truly concurrent bisimulations $\approx_{rbp}^{fr}$, $\approx_{rbs}^{fr}$, $\approx_{rbhp}^{fr}$ and $\approx_{rbhhp}^{fr}$, so FR rooted branching truly concurrent
  bisimulations $\approx_{rbp}^{fr}$, $\approx_{rbs}^{fr}$, $\approx_{rbhp}^{fr}$ and $\approx_{rbhhp}^{fr}$ are all congruences with respect to all operators of $qAPRTC$;
  \item While $\mathbb{E}$ is extended to $\mathbb{E}\cup\{\tau\}$, it can be proved that FR rooted branching truly concurrent
  bisimulations $\approx_{rbp}^{fr}$, $\approx_{rbs}^{fr}$, $\approx_{rbhp}^{fr}$ and $\approx_{rbhhp}^{fr}$ are all congruences with respect to all operators of $qAPRTC$, we omit it.
\end{enumerate}
\end{proof}

We design the axioms for the silent step $\tau$ in Table \ref{AxiomsForqTau5}.

\begin{center}
\begin{table}
  \begin{tabular}{@{}ll@{}}
\hline No. &Axiom\\
  $B1$ & $e\cdot\tau=e$\\
  $RB1$ & $\tau\cdot e[m]=e[m]$\\
  $B2$ & $e\cdot(\tau\cdot(x+y)+x)=e\cdot(x+y)$\\
  $RB2$ & $((x+y)\cdot\tau+x)\cdot e[m]=(x+y)\cdot e[m]$\\
  $B3$ & $x\leftmerge\tau=x(\textrm{Std(x)})$\\
  $RB3$ & $\tau\leftmerge x=x(\textrm{NStd(x)})$\\
\end{tabular}
\caption{Axioms of silent step}
\label{AxiomsForqTau5}
\end{table}
\end{center}

\begin{theorem}[Elimination theorem of qAPRTC with silent step and guarded linear recursion]\label{ETTau5}
Each process term in qAPRTC with silent step and guarded linear recursion is equal to a process term $\langle X_1|E\rangle$ with $E$ a guarded linear recursive specification.
\end{theorem}

\begin{proof}
The same as that of $APRTC$ with silent step and guarded linear recursion, we omit the proof, please refer to \cite{APRTC} for details.
\end{proof}

\begin{theorem}[Soundness of qAPRTC with silent step and guarded linear recursion]\label{SqAPRTCTAU5}
Let $x$ and $y$ be qAPRTC with silent step and guarded linear recursion terms. If qAPRTC with silent step and guarded linear recursion $\vdash x=y$, then
\begin{enumerate}
  \item $x\approx_{rbs}^{fr} y$;
  \item $x\approx_{rbp}^{fr} y$;
  \item $x\approx_{rbhp}^{fr} y$;
  \item $x\approx_{rbhhp}^{fr} y$.
\end{enumerate}
\end{theorem}

\begin{proof}
(1) Since FR rooted branching pomset bisimulation $\approx_{rbp}^{fr}$ is both an equivalent and a congruent relation with respect to $qAPRTC$ with silent step and guarded
linear recursion, we only need to check if each axiom in Table \ref{AxiomsForqTau5} is sound modulo FR rooted branching pomset bisimulation $\approx_{rbp}^{fr}$. We leave them as
exercises to the readers.

(2) Since FR rooted branching step bisimulation $\approx_{rbs}^{fr}$ is both an equivalent and a congruent relation with respect to $qAPRTC$ with silent step and guarded
linear recursion, we only need to check if each axiom in Table \ref{AxiomsForqTau5} is sound modulo FR rooted branching step bisimulation $\approx_{rbs}^{fr}$. We leave them
as exercises to the readers.

(3) Since FR rooted branching hp-bisimulation $\approx_{rbhp}^{fr}$ is both an equivalent and a congruent relation with respect to $qAPRTC$ with silent step and guarded linear
recursion, we only need to check if each axiom in Table \ref{AxiomsForqTau5} is sound modulo FR rooted branching hp-bisimulation $\approx_{rbhp}^{fr}$. We leave them as exercises
to the readers.

(4) Since FR rooted branching hhp-bisimulation $\approx_{rbhhp}^{fr}$ is both an equivalent and a congruent relation with respect to $qAPRTC$ with silent step and guarded linear
recursion, we only need to check if each axiom in Table \ref{AxiomsForqTau5} is sound modulo FR rooted branching hhp-bisimulation $\approx_{rbhhp}^{fr}$. We leave them as exercises
to the readers.
\end{proof}

\begin{theorem}[Completeness of qAPRTC with silent step and guarded linear recursion]
Let $p$ and $q$ be closed qAPRTC with silent step and guarded linear recursion terms, then,
\begin{enumerate}
  \item if $p\approx_{rbs}^{fr} q$ then $p=q$;
  \item if $p\approx_{rbp}^{fr} q$ then $p=q$;
  \item if $p\approx_{rbhp}^{fr} q$ then $p=q$;
  \item if $p\approx_{rbhhp}^{fr} q$ then $p=q$.
\end{enumerate}
\end{theorem}

\begin{proof}
According to the definition of FR truly concurrent rooted branching bisimulation equivalences $\approx_{rbp}^{fr}$, $\approx_{rbs}^{fr}$, $\approx_{rbhp}^{fr}$ and $\approx_{rbhhp}^{fr}$,
$p\approx_{rbp}^{fr}q$, $p\approx_{rbs}^{fr}q$, $p\approx_{rbhp}^{fr}q$ and $p\approx_{rbhhp}^{fr}q$ implies
both the rooted branching bisimilarities between $p$ and $q$, and also the in the same quantum states. According to the completeness of APRTC with silent step and guarded linear recursion (please refer to \cite{ATC} for details), we can get the
completeness of qAPRTC with silent step and guarded linear recursion.
\end{proof}

The unary abstraction operator $\tau_I$ ($I\subseteq \mathbb{E}$) renames all atomic events in $I$ into $\tau$. qAPRTC with silent step and abstraction operator is called
$qAPRTC_{\tau}$. The transition rules of operator $\tau_I$ are shown in Table \ref{TRForqAbstraction5}.

\begin{center}
    \begin{table}
        $$\frac{\langle x,\varrho\rangle\xrightarrow{e}\langle e[m],\varrho'\rangle}{\langle \tau_I(x),\varrho\rangle\xrightarrow{e}\langle e[m],\varrho'\rangle}\quad e\notin I
        \quad\frac{\langle x,\varrho\rangle\xrightarrow{e}\langle x',\varrho'\rangle}{\langle\tau_I(x),\varrho\rangle\xrightarrow{e}\langle \tau_I(x'),\varrho'\rangle}\quad e\notin I$$

        $$\frac{\langle x,\varrho\rangle\xrightarrow{e}\langle\surd,\tau(\varrho)\rangle}{\langle\tau_I(x),\varrho\rangle\xrightarrow{\tau}\langle\surd,\tau(\varrho)\rangle}\quad e\in I
        \quad\frac{\langle x,\varrho\rangle\xrightarrow{e}\langle x',\tau(\varrho)\rangle}{\langle\tau_I(x),\varrho\rangle\xrightarrow{\tau}\langle\tau_I(x'),\tau(\varrho)\rangle}\quad e\in I$$

        $$\frac{\langle x,\varrho\rangle\xtworightarrow{e[m]}\langle e,\varrho'\rangle}{\langle\tau_I(x),\varrho\rangle\xtworightarrow{e[m]}\langle e,\varrho'\rangle}\quad e[m]\notin I
        \quad\frac{\langle x,\varrho\rangle\xtworightarrow{e[m]}\langle x',\varrho\rangle}{\langle\tau_I(x),\varrho\rangle\xtworightarrow{e[m]}\langle\tau_I(x'),\varrho'\rangle}\quad e[m]\notin I$$

        $$\frac{\langle x,\varrho\rangle\xtworightarrow{e[m]}\langle\surd,\tau(\varrho)\rangle}{\langle\tau_I(x),\varrho\rangle\xtworightarrow{\tau}\langle\surd,\tau(\varrho)\rangle}\quad e[m]\in I
        \quad\frac{\langle x,\varrho\rangle\xtworightarrow{e[m]}\langle x',\tau(\varrho)\rangle}{\langle\tau_I(x),\varrho\rangle\xtworightarrow{\tau}\langle\tau_I(x'),\tau(\varrho)\rangle}\quad e[m]\in I$$
        \caption{Transition rule of the abstraction operator}
        \label{TRForqAbstraction5}
    \end{table}
\end{center}

\begin{theorem}[Conservitivity of $qAPRTC_{\tau}$ with guarded linear recursion]
$qAPRTC_{\tau}$ with guarded linear recursion is a conservative extension of qAPRTC with silent step and guarded linear recursion.
\end{theorem}

\begin{proof}
Since the transition rules of $qAPRTC$ with silent step and guarded linear recursion are source-dependent, and the transition rules for abstraction operator in Table
\ref{TRForqAbstraction5} contain only a fresh operator $\tau_I$ in their source, so the transition rules of $qAPRTC_{\tau}$ with guarded linear recursion is a conservative extension
of those of $qAPRTC$ with silent step and guarded linear recursion.
\end{proof}

\begin{theorem}[Congruence theorem of $qAPRTC_{\tau}$ with guarded linear recursion]
Rooted branching FR truly concurrent bisimulation equivalences $\approx_{rbp}^{fr}$, $\approx_{rbs}^{fr}$, $\approx_{rbhp}^{fr}$ and $\approx_{rbhhp}^{fr}$ are all congruences with respect to $qAPRTC_{\tau}$.
\end{theorem}

\begin{proof}
(1) It is easy to see that FR rooted branching pomset bisimulation is an equivalent relation on $qAPRTC_{\tau}$ with guarded linear recursion terms, we only need to
prove that $\approx_{rbp}^{fr}$ is preserved by the operator $\tau_I$. It is trivial and we leave the proof as an exercise for the readers.

(2) It is easy to see that FR rooted branching step bisimulation is an equivalent relation on $qAPRTC_{\tau}$ with guarded linear recursion terms, we only need to
prove that $\approx_{rbs}^{fr}$ is preserved by the operator $\tau_I$. It is trivial and we leave the proof as an exercise for the readers.

(3) It is easy to see that FR rooted branching hp-bisimulation is an equivalent relation on $qAPRTC_{\tau}$ with guarded linear recursion terms, we only need to
prove that $\approx_{rbhp}^{fr}$ is preserved by the operator $\tau_I$. It is trivial and we leave the proof as an exercise for the readers.

(4) It is easy to see that FR rooted branching hhp-bisimulation is an equivalent relation on $qAPRTC_{\tau}$ with guarded linear recursion terms, we only need to
prove that $\approx_{rbhhp}^{fr}$ is preserved by the operator $\tau_I$. It is trivial and we leave the proof as an exercise for the readers.
\end{proof}

We design the axioms for the abstraction operator $\tau_I$ in Table \ref{AxiomsForqAbstraction5}.

\begin{center}
\begin{table}
  \begin{tabular}{@{}ll@{}}
\hline No. &Axiom\\
  $TI1$ & $e\notin I\quad \tau_I(e)=e$\\
  $RTI1$ & $e[m]\notin I\quad \tau_I(e[m])=e[m]$\\
  $TI2$ & $e\in I\quad \tau_I(e)=\tau$\\
  $RTI2$ & $e[m]\in I\quad \tau_I(e[m])=\tau$\\
  $TI3$ & $\tau_I(\delta)=\delta$\\
  $TI4$ & $\tau_I(x+y)=\tau_I(x)+\tau_I(y)$\\
  $TI5$ & $\tau_I(x\cdot y)=\tau_I(x)\cdot\tau_I(y)$\\
  $TI6$ & $\tau_I(x\leftmerge y)=\tau_I(x)\leftmerge\tau_I(y)$\\
\end{tabular}
\caption{Axioms of abstraction operator}
\label{AxiomsForqAbstraction5}
\end{table}
\end{center}

\begin{theorem}[Soundness of $qAPRTC_{\tau}$ with guarded linear recursion]\label{SqAPRTCABS5}
Let $x$ and $y$ be $qAPRTC_{\tau}$ with guarded linear recursion terms. If $qAPRTC_{\tau}$ with guarded linear recursion $\vdash x=y$, then
\begin{enumerate}
  \item $x\approx_{rbs}^{fr} y$;
  \item $x\approx_{rbp}^{fr} y$;
  \item $x\approx_{rbhp}^{fr} y$;
  \item $x\approx_{rbhp}^{fr} y$.
\end{enumerate}
\end{theorem}

\begin{proof}
(1) Since FR rooted branching step bisimulation $\approx_{rbs}^{fr}$ is both an equivalent and a congruent relation with respect to $APPTC_{\tau}$ with guarded linear
recursion, we only need to check if each axiom in Table \ref{AxiomsForqAbstraction5} is sound modulo FR rooted branching step bisimulation $\approx_{rbs}^{fr}$. We leave them as
exercises to the readers.

(2) Since FR rooted branching pomset bisimulation $\approx_{rbp}^{fr}$ is both an equivalent and a congruent relation with respect to $APPTC_{\tau}$ with guarded linear
recursion, we only need to check if each axiom in Table \ref{AxiomsForqAbstraction5} is sound modulo FR rooted branching pomset bisimulation $\approx_{rbp}^{fr}$. We leave them
as exercises to the readers.

(3) Since FR rooted branching hp-bisimulation $\approx_{rbhp}^{fr}$ is both an equivalent and a congruent relation with respect to $APPTC_{\tau}$ with guarded linear
recursion, we only need to check if each axiom in Table \ref{AxiomsForqAbstraction5} is sound modulo FR rooted branching hp-bisimulation $\approx_{rbhp}^{fr}$. We leave them as
exercises to the readers.

(4) Since FR rooted branching hhp-bisimulation $\approx_{rbhhp}^{fr}$ is both an equivalent and a congruent relation with respect to $APPTC_{\tau}$ with guarded linear
recursion, we only need to check if each axiom in Table \ref{AxiomsForqAbstraction5} is sound modulo FR rooted branching hhp-bisimulation $\approx_{rbhhp}^{fr}$. We leave them as
exercises to the readers.
\end{proof}

\begin{definition}[Cluster]\label{CLUSTER5}
Let $E$ be a guarded linear recursive specification, and $I\subseteq \mathbb{E}$. Two recursion variable $X$ and $Y$ in $E$ are in the same cluster for $I$ iff there exist sequences of transitions $\langle X|E\rangle\xrightarrow{\{b_{11},\cdots, b_{1i}\}}\cdots\xrightarrow{\{b_{m1},\cdots, b_{mi}\}}\langle Y|E\rangle$ and $\langle Y|E\rangle\xrightarrow{\{c_{11},\cdots, c_{1j}\}}\cdots\xrightarrow{\{c_{n1},\cdots, c_{nj}\}}\langle X|E\rangle$, or $\langle X|E\rangle\xtworightarrow{\{b_{11}[m],\cdots, b_{1i}[m]\}}\cdots\xtworightarrow{\{b_{m1}[m],\cdots, b_{mi}[m]\}}\langle Y|E\rangle$ and $\langle Y|E\rangle\xtworightarrow{\{c_{11}[n],\cdots, c_{1j}[n]\}}\cdots\xtworightarrow{\{c_{n1}[n],\cdots, c_{nj}[n]\}}\langle X|E\rangle$, where $b_{11},\cdots,b_{mi},c_{11},\cdots,c_{nj}, b_{11}[m],\cdots,b_{mi}[m],c_{11}[n],\cdots,c_{nj}[n]\in I\cup\{\tau\}$.

$a_1\parallel\cdots\parallel a_k$, or $(a_1\parallel\cdots\parallel a_k) X$, or $a_1[m]\parallel\cdots\parallel a_k[m]$, or $X (a_1[m]\parallel\cdots\parallel a_k[m])$ is an exit for the cluster $C$ iff: (1) $a_1\parallel\cdots\parallel a_k$, or $(a_1\parallel\cdots\parallel a_k) X$, or $a_1[m]\parallel\cdots\parallel a_k[m]$, or $X (a_1[m]\parallel\cdots\parallel a_k[m])$ is a summand at the right-hand side of the recursive equation for a recursion variable in $C$, and (2) in the case of $(a_1\parallel\cdots\parallel a_k) X$, and $X (a_1[m]\parallel\cdots\parallel a_k[m])$ either $a_l, a_l[m]\notin I\cup\{\tau\}(l\in\{1,2,\cdots,k\})$ or $X\notin C$.
\end{definition}

\begin{center}
\begin{table}
  \begin{tabular}{@{}ll@{}}
\hline No. &Axiom\\
  CFAR & If $X$ is in a cluster for $I$ with exits \\
           & $\{(a_{11}\leftmerge\cdots\leftmerge a_{1i})Y_1,\cdots,(a_{m1}\leftmerge\cdots\leftmerge a_{mi})Y_m, b_{11}\leftmerge\cdots\leftmerge b_{1j},\cdots,b_{n1}\leftmerge\cdots\leftmerge b_{nj}\}$, \\
           & then $\tau\cdot\tau_I(\langle X|E\rangle)=$\\
           & $\tau\cdot\tau_I((a_{11}\leftmerge\cdots\leftmerge a_{1i})\langle Y_1|E\rangle+\cdots+(a_{m1}\leftmerge\cdots\leftmerge a_{mi})\langle Y_m|E\rangle+b_{11}\leftmerge\cdots\leftmerge b_{1j}+\cdots+b_{n1}\leftmerge\cdots\leftmerge b_{nj})$\\
           & Or exists,\\
           & $\{Y_1(a_{11}[m]\leftmerge\cdots\leftmerge a_{1i}[m1]),\cdots,Y_m(a_{m1}[mm]\leftmerge\cdots\leftmerge a_{mi}[mm]),$\\
           & $b_{11}[n1]\leftmerge\cdots\leftmerge b_{1j}[n1],\cdots,b_{n1}[nn]\leftmerge\cdots\leftmerge b_{nj}[nn]\}$, \\
           & then $\tau_I(\langle X|E\rangle)\cdot\tau=$\\
           & $\tau_I(\langle Y_1|E\rangle(a_{11}[m1]\leftmerge\cdots\leftmerge a_{1i}[m1])+\cdots+\langle Y_m|E\rangle(a_{m1}[mm]\leftmerge\cdots\leftmerge a_{mi}[mm])$\\
           & $+b_{11}[n1]\leftmerge\cdots\leftmerge b_{1j}[n1]+\cdots+b_{n1}[nn]\leftmerge\cdots\leftmerge b_{nj}[nn])\cdot\tau$\\
\end{tabular}
\caption{Cluster fair abstraction rule}
\label{qCFAR5}
\end{table}
\end{center}

\begin{theorem}[Soundness of CFAR]\label{SCFAR5}
CFAR is sound modulo rooted branching FR truly concurrent bisimulation equivalences $\approx_{rbs}^{fr}$, $\approx_{rbp}^{fr}$, $\approx_{rbhp}^{fr}$ and $\approx_{rbhhp}^{fr}$.
\end{theorem}

\begin{proof}
(1) Since FR rooted branching step bisimulation $\approx_{rbs}^{fr}$ is both an equivalent and a congruent relation with respect to $APPTC_{\tau}$ with guarded linear
recursion, we only need to check if each axiom in Table \ref{qCFAR5} is sound modulo FR rooted branching step bisimulation $\approx_{rbs}^{fr}$. We leave them as
exercises to the readers.

(2) Since FR rooted branching pomset bisimulation $\approx_{rbp}^{fr}$ is both an equivalent and a congruent relation with respect to $APPTC_{\tau}$ with guarded linear
recursion, we only need to check if each axiom in Table \ref{qCFAR5} is sound modulo FR rooted branching pomset bisimulation $\approx_{rbp}^{fr}$. We leave them
as exercises to the readers.

(3) Since FR rooted branching hp-bisimulation $\approx_{rbhp}^{fr}$ is both an equivalent and a congruent relation with respect to $APPTC_{\tau}$ with guarded linear
recursion, we only need to check if each axiom in Table \ref{qCFAR5} is sound modulo FR rooted branching hp-bisimulation $\approx_{rbhp}^{fr}$. We leave them as
exercises to the readers.

(4) Since FR rooted branching hhp-bisimulation $\approx_{rbhhp}^{fr}$ is both an equivalent and a congruent relation with respect to $APPTC_{\tau}$ with guarded linear
recursion, we only need to check if each axiom in Table \ref{qCFAR5} is sound modulo FR rooted branching hhp-bisimulation $\approx_{rbhhp}^{fr}$. We leave them as
exercises to the readers.
\end{proof}

\begin{theorem}[Completeness of $qAPRTC_{\tau}$ with guarded linear recursion and CFAR]\label{CCFAR5}
Let $p$ and $q$ be closed $qAPRTC_{\tau}$ with guarded linear recursion and CFAR terms, then,
\begin{enumerate}
  \item if $p\approx_{rbs}^{fr} q$ then $p=q$;
  \item if $p\approx_{rbp}^{fr} q$ then $p=q$;
  \item if $p\approx_{rbhp}^{fr} q$ then $p=q$;
  \item if $p\approx_{rbhhp}^{fr} q$ then $p=q$.
\end{enumerate}
\end{theorem}

\begin{proof}
According to the definition of FR truly concurrent rooted branching bisimulation equivalences $\approx_{rbp}^{fr}$, $\approx_{rbs}^{fr}$, $\approx_{rbhp}^{fr}$ and $\approx_{rbhhp}^{fr}$,
$p\approx_{rbp}^{fr}q$, $p\approx_{rbs}^{fr}q$, $p\approx_{rbhp}^{fr}q$ and $p\approx_{rbhhp}^{fr}q$ implies
both the rooted branching bisimilarities between $p$ and $q$, and also the in the same quantum states. According to the completeness of $APRTC_{\tau}$ guarded linear recursion (please refer to \cite{ATC} for details), we can get the
completeness of $qAPRTC_{\tau}$ with guarded linear recursion.
\end{proof}

\subsection{Quantum Entanglement}\label{qe1}

If two quantum variables are entangled, then a quantum operation performed on one variable, then state of the other quantum variable is also changed. So, the entangled states must be
all the inner variables or all the public variables. We will introduced a mechanism to explicitly define quantum entanglement in open quantum systems.
A new constant called shadow constant denoted $\circledS^e_i$ corresponding to a specific quantum operation, $e$ is added to $qAPTC$.
If there are $n$ quantum variables entangled, they maybe be distributed in different quantum systems, with a quantum operation performed on one variable, there should be one
$\circledS^e_i$ ($1\leq i\leq n-1$) executed on each variable in the other $n-1$ variables. Thus, distributed variables are all hidden behind actions.
In the following, we let $\circledS\in \mathbb{E}$.

The axiom system of the shadow constant $\circledS$ is shown in Table \ref{AxiomsForQE1}.

\begin{center}
\begin{table}
  \begin{tabular}{@{}ll@{}}
\hline No. &Axiom\\
  $SC1$ & $\circledS\cdot x = x$ \\
  $SC2$ & $x\cdot\circledS = x$\\
  $SC3$ & $e\leftmerge\circledS^e=e$\\
  $SC4$ & $\circledS^e\leftmerge e=e$\\
  $SC5$ & $e\leftmerge(\circledS^e\cdot y) = e\cdot y$\\
  $SC6$ & $\circledS^e\leftmerge(e\cdot y) = e\cdot y$\\
  $SC7$ & $(e\cdot x)\leftmerge\circledS^e = e\cdot x$\\
  $SC8$ & $(\circledS^e\cdot x)\leftmerge e = e\cdot x$\\
  $SC9$ & $(e\cdot x)\leftmerge(\circledS^e\cdot y) = e\cdot (x\between y)$\\
  $SC10$ & $(\circledS^e\cdot x)\leftmerge(e\cdot y) = e\cdot (x\between y)$\\
  $RSC3$ & $e[m]\leftmerge\circledS^e[m]=e[m]$\\
  $RSC4$ & $\circledS^e[m]\leftmerge e[m]=e[m]$\\
  $RSC5$ & $e[m]\leftmerge(y\cdot\circledS^e[m]) = y\cdot e[m]$\\
  $RSC6$ & $\circledS^e[m]\leftmerge(y\cdot e[m]) =y\cdot e[m]$\\
  $RSC7$ & $(x\cdot e[m])\leftmerge\circledS^e[m] =x\cdot e[m]$\\
  $RSC8$ & $(x\cdot \circledS^e[m])\leftmerge e[m] =x\cdot e[m]$\\
  $RSC9$ & $(x\cdot e[m])\leftmerge(y\cdot \circledS^e[m]) =(x\between y)\cdot e[m]$\\
  $RSC10$ & $(x\cdot \circledS^e[m])\leftmerge(y \cdot e[m]) =(x\between y)\cdot e[m]$\\
\end{tabular}
\caption{Axioms of quantum entanglement}
\label{AxiomsForQE1}
\end{table}
\end{center}

The transition rules of constant $\circledS$ are as Table \ref{TRForENT1} shows.

\begin{center}
    \begin{table}
        $$\frac{}{\langle\circledS,\varrho\rangle\rightarrow\langle\surd,\varrho\rangle}$$
        $$\frac{\langle x, \varrho\rangle\xrightarrow{e}\langle x',\varrho'\rangle\quad \langle y, \varrho'\rangle\xrightarrow{\circledS^e}\langle y',\varrho'\rangle}{\langle x\leftmerge y,\varrho\rangle\xrightarrow{e}\langle x'\between y', \varrho'\rangle}$$
        $$\frac{\langle x, \varrho\rangle\xrightarrow{e}\langle e[m],\varrho'\rangle\quad \langle y, \varrho'\rangle\xrightarrow{\circledS^e}\langle y',\varrho'\rangle}{\langle x\leftmerge y,\varrho\rangle\xrightarrow{e}\langle e[m]\leftmerge y', \varrho'\rangle}$$
        $$\frac{\langle x, \varrho'\rangle\xrightarrow{\circledS^e}\langle \surd,\varrho'\rangle\quad \langle y, \varrho\rangle\xrightarrow{e}\langle y',\varrho'\rangle}{\langle x\leftmerge y,\varrho\rangle\xrightarrow{e}\langle y', \varrho'\rangle}$$
        $$\frac{\langle x, \varrho\rangle\xrightarrow{e}\langle e[m],\varrho'\rangle\quad \langle y, \varrho'\rangle\xrightarrow{\circledS^e}\langle\surd,\varrho'\rangle}{\langle x\leftmerge y,\varrho\rangle\xrightarrow{e}\langle e[m], \varrho'\rangle}$$
        $$\frac{}{\langle\circledS[m],\varrho\rangle\xtworightarrow{ }\langle\surd,\varrho\rangle}$$
        $$\frac{\langle x, \varrho\rangle\xtworightarrow{e[m]}\langle x',\varrho'\rangle\quad \langle y, \varrho'\rangle\xtworightarrow{\circledS^e[m]}\langle y',\varrho'\rangle}{\langle x\leftmerge y,\varrho\rangle\xtworightarrow{e[m]}\langle x'\between y', \varrho'\rangle}$$
        $$\frac{\langle x, \varrho\rangle\xtworightarrow{e[m]}\langle e,\varrho'\rangle\quad \langle y, \varrho'\rangle\xtworightarrow{\circledS^e[m]}\langle y',\varrho'\rangle}{\langle x\leftmerge y,\varrho\rangle\xtworightarrow{e[m]}\langle e\leftmerge y', \varrho'\rangle}$$
        $$\frac{\langle x, \varrho'\rangle\xtworightarrow{\circledS^e}\langle \surd,\varrho'\rangle\quad \langle y, \varrho\rangle\xtworightarrow{e[m]}\langle y',\varrho'\rangle}{\langle x\leftmerge y,\varrho\rangle\xtworightarrow{e[m]}\langle y', \varrho'\rangle}$$
        $$\frac{\langle x, \varrho\rangle\xtworightarrow{e[m]}\langle e,\varrho'\rangle\quad \langle y, \varrho'\rangle\xtworightarrow{\circledS^e[m]}\langle\surd,\varrho'\rangle}{\langle x\leftmerge y,\varrho\rangle\xtworightarrow{e[m]}\langle e, \varrho'\rangle}$$
        \caption{Transition rules of constant $\circledS$}
        \label{TRForENT1}
    \end{table}
\end{center}

\begin{theorem}[Elimination theorem of $qAPRTC_{\tau}$ with guarded linear recursion and shadow constant]
Let $p$ be a closed $qAPRTC_{\tau}$ with guarded linear recursion and shadow constant term. Then there is a closed $qAPRTC$ term such that $qAPRTC_{\tau}$ with guarded linear recursion
and shadow constant$\vdash p=q$.
\end{theorem}

\begin{proof}
We leave the proof to the readers as an excise.
\end{proof}

\begin{theorem}[Conservitivity of $qAPTC_{\tau}$ with guarded linear recursion and shadow constant]
$qAPRTC_{\tau}$ with guarded linear recursion and shadow constant is a conservative extension of $qAPRTC_{\tau}$ with guarded linear recursion.
\end{theorem}

\begin{proof}
We leave the proof to the readers as an excise.
\end{proof}

\begin{theorem}[Congruence theorem of $qAPRTC_{\tau}$ with guarded linear recursion and shadow constant]
FR rooted branching truly concurrent bisimulation equivalences $\approx_{rbp}^{fr}$, $\approx_{rbs}^{fr}$, $\approx_{rbhp}^{fr}$ and $\approx_{rbhhp}^{fr}$ are all congruences with respect to $qAPRTC_{\tau}$
with guarded linear recursion and shadow constant.
\end{theorem}

\begin{proof}
We leave the proof to the readers as an excise.
\end{proof}

\begin{theorem}[Soundness of $qAPRTC_{\tau}$ with guarded linear recursion and shadow constant]
Let $p$ and $q$ be closed $qAPRTC_{\tau}$ with guarded linear recursion and shadow constant terms. If $qAPRTC_{\tau}$ with guarded linear recursion and shadow constant$\vdash x=y$, then

\begin{enumerate}
  \item $x\approx_{rbs}^{fr} y$;
  \item $x\approx_{rbp}^{fr} y$;
  \item $x\approx_{rbhp}^{fr} y$;
  \item $x\approx_{rbhhp}^{fr} y$.
\end{enumerate}
\end{theorem}

\begin{proof}
We leave the proof to the readers as an excise.
\end{proof}

\begin{theorem}[Completeness of $qAPRTC_{\tau}$ with guarded linear recursion and shadow constant]
Let $p$ and $q$ are closed $qAPRTC_{\tau}$ with guarded linear recursion and shadow constant terms, then,

\begin{enumerate}
  \item if $p\approx_{rbs}^{fr} q$ then $p=q$;
  \item if $p\approx_{rbp}^{fr} q$ then $p=q$;
  \item if $p\approx_{rbhp}^{fr} q$ then $p=q$;
  \item if $p\approx_{rbhhp}^{fr} q$ then $p=q$.
\end{enumerate}
\end{theorem}

\begin{proof}
We leave the proof to the readers as an excise.
\end{proof}

\subsection{Unification of Quantum and Classical Computing for Open Quantum Systems}\label{uni1}

We give the transition rules under quantum configuration for traditional atomic actions (events) $e'\in\mathbb{E}$ as Table \ref{TRForBPA3} shows.

\begin{center}
    \begin{table}
        $$\frac{}{\langle e',\varrho\rangle\xrightarrow{e'}\langle e'[m],\varrho\rangle}$$
        $$\frac{\langle x,\varrho\rangle\xrightarrow{e'}\langle e[m],\varrho\rangle}{\langle x+ y,\varrho\rangle\xrightarrow{e'}\langle e'[m],\varrho\rangle}$$
        $$\frac{\langle x,\varrho\rangle\xrightarrow{e'}\langle x',\varrho\rangle}{\langle x+ y,\varrho\rangle\xrightarrow{e'}\langle x',\varrho\rangle}$$
        $$\frac{\langle y,\varrho\rangle\xrightarrow{e'}\langle e'[m],\varrho\rangle}{\langle x+ y,\varrho\rangle\xrightarrow{e'}\langle e'[m],\varrho\rangle}$$
        $$\frac{\langle y,\varrho\rangle\xrightarrow{e'}\langle y',\varrho\rangle}{\langle x+ y,\varrho\rangle\xrightarrow{e'}\langle y',\varrho\rangle}$$
        $$\frac{\langle x,\varrho\rangle\xrightarrow{e'}\langle e'[m],\varrho\rangle}{\langle x\cdot y,\varrho\rangle\xrightarrow{e'} \langle e'[m]\cdot y,\varrho\rangle}$$
        $$\frac{\langle x,\varrho\rangle\xrightarrow{e'}\langle x',\varrho\rangle}{\langle x\cdot y,\varrho\rangle\xrightarrow{e'}\langle x'\cdot y,\varrho\rangle}$$

        $$\frac{}{\langle e'[m],\varrho\rangle\xtworightarrow{e'[m]}\langle e',\varrho\rangle}$$
        $$\frac{\langle x,\varrho\rangle\xtworightarrow{e'[m]}\langle e',\varrho\rangle}{\langle x+ y,\varrho\rangle\xtworightarrow{e'[m]}\langle e',\varrho\rangle}$$
        $$\frac{\langle x,\varrho\rangle\xtworightarrow{e'[m]}\langle x',\varrho\rangle}{\langle x+ y,\varrho\rangle\xtworightarrow{e'[m]}\langle x',\varrho\rangle}$$
        $$\frac{\langle y,\varrho\rangle\xtworightarrow{e'[m]}\langle e',\varrho\rangle}{\langle x+ y,\varrho\rangle\xtworightarrow{e'[m]}\langle e',\varrho\rangle}$$
        $$\frac{\langle y,\varrho\rangle\xtworightarrow{e'[m]}\langle y',\varrho\rangle}{\langle x+ y,\varrho\rangle\xtworightarrow{e'[m]}\langle y',\varrho\rangle}$$
        $$\frac{\langle y,\varrho\rangle\xtworightarrow{e'[m]}\langle e',\varrho\rangle}{\langle x\cdot y,\varrho\rangle\xtworightarrow{e'[m]} \langle x\cdot e',\varrho\rangle}$$
        $$\frac{\langle y,\varrho\rangle\xtworightarrow{e'[m]}\langle y',\varrho\rangle}{\langle x\cdot y,\varrho\rangle\xtworightarrow{e'[m]}\langle x\cdot y',\varrho\rangle}$$

        \caption{Transition rules of BATC under quantum configuration}
        \label{TRForBPA3}
    \end{table}
\end{center}

And the axioms for traditional actions are the same as those of qBARTC. And it is natural can be extended to qAPRTC, recursion and abstraction. So, quantum and classical computing
are unified under the framework of qAPRTC for open quantum systems.

\newpage\section{Applications of qAPRTC}\label{aqaprtc}

Quantum and classical computing in open systems are unified with qAPRTC, which have the same equational logic and the same quantum configuration based operational semantics.
The unification can be used widely in verification for the behaviors of quantum and classical computing mixed systems. In this chapter, we show its usage in verification of the
quantum communication protocols.

\subsection{Verification of BB84 Protocol}\label{VBB844}

\begin{figure}
  \centering
  \includegraphics{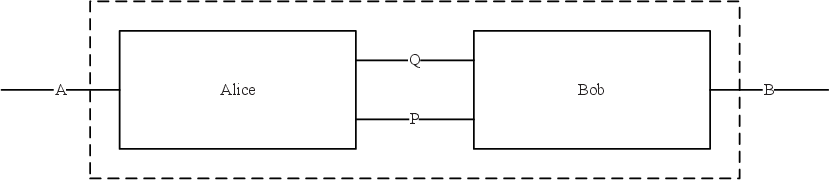}
  \caption{The BB84 protocol.}
  \label{BB844}
\end{figure}

The BB84 protocol is used to create a private key between two parities, Alice and Bob. Firstly, we introduce the basic BB84 protocol briefly, which is illustrated in Figure \ref{BB844}.

\begin{enumerate}
  \item Alice create two string of bits with size $n$ randomly, denoted as $B_a$ and $K_a$.
  \item Alice generates a string of qubits $q$ with size $n$, and the $i$th qubit in $q$ is $|x_y\rangle$, where $x$ is the $i$th bit of $B_a$ and $y$ is the $i$th bit of $K_a$.
  \item Alice sends $q$ to Bob through a quantum channel $Q$ between Alice and Bob.
  \item Bob receives $q$ and randomly generates a string of bits $B_b$ with size $n$.
  \item Bob measures each qubit of $q$ according to a basis by bits of $B_b$. And the measurement results would be $K_b$, which is also with size $n$.
  \item Bob sends his measurement bases $B_b$ to Alice through a public channel $P$.
  \item Once receiving $B_b$, Alice sends her bases $B_a$ to Bob through channel $P$, and Bob receives $B_a$.
  \item Alice and Bob determine that at which position the bit strings $B_a$ and $B_b$ are equal, and they discard the mismatched bits of $B_a$ and $B_b$. Then the remaining bits of $K_a$ and $K_b$, denoted as $K_a'$ and $K_b'$ with $K_{a,b}=K_a'=K_b'$.
\end{enumerate}

We re-introduce the basic BB84 protocol in an abstract way with more technical details as Figure \ref{BB844} illustrates.

Now, we assume a special measurement operation $Rand[q;B_a]$ which create a string of $n$ random bits $B_a$ from the $q$ quantum system, and the same as $Rand[q;K_a]$, $Rand[q';B_b]$. $M[q;K_b]$ denotes the Bob's measurement operation of $q$. The generation of $n$ qubits $q$ through two quantum operations $Set_{K_a}[q]$ and $H_{B_a}[q]$. Alice sends $q$ to Bob through the quantum channel $Q$ by quantum communicating action $send_{Q}(q)$ and Bob receives $q$ through $Q$ by quantum communicating action $receive_{Q}(q)$. Bob sends $B_b$ to Alice through the public channel $P$ by classical communicating action $send_{P}(B_b)$ and Alice receives $B_b$ through channel $P$ by classical communicating action $receive_{P}(B_b)$, and the same as $send_{P}(B_a)$ and $receive_{P}(B_a)$. Alice and Bob generate the private key $K_{a,b}$ by a classical comparison action $cmp(K_{a,b},K_a,K_b,B_a,B_b)$. Let Alice and Bob be a system $AB$ and let interactions between Alice and Bob be internal actions. $AB$ receives external input $D_i$ through channel $A$ by communicating action $receive_A(D_i)$ and sends results $D_o$ through channel $B$ by communicating action $send_B(D_o)$.

Then the state transition of Alice can be described by qACP as follows.

\begin{eqnarray}
&&A=\sum_{D_i\in \Delta_i}receive_A(D_i)\cdot A_1\nonumber\\
&&A_1=Rand[q;B_a]\cdot A_2\nonumber\\
&&A_2=Rand[q;K_a]\cdot A_3\nonumber\\
&&A_3=Set_{K_a}[q]\cdot A_4\nonumber\\
&&A_4=H_{B_a}[q]\cdot A_5\nonumber\\
&&A_5=send_Q(q)\cdot A_6\nonumber\\
&&A_6=receive_P(B_b)\cdot A_7\nonumber\\
&&A_7=send_P(B_a)\cdot A_8\nonumber\\
&&A_8=cmp(K_{a,b},K_a,K_b,B_a,B_b)\cdot A\nonumber
\end{eqnarray}

where $\Delta_i$ is the collection of the input data.

And the state transition of Bob can be described by qACP as follows.

\begin{eqnarray}
&&B=receive_Q(q)\cdot B_1\nonumber\\
&&B_1=Rand[q';B_b]\cdot B_2\nonumber\\
&&B_2=M[q;K_b]\cdot B_3\nonumber\\
&&B_3=send_P(B_b)\cdot B_4\nonumber\\
&&B_4=receive_P(B_a)\cdot B_5\nonumber\\
&&B_5=cmp(K_{a,b},K_a,K_b,B_a,B_b)\cdot B_6\nonumber\\
&&B_6=\sum_{D_o\in\Delta_o}send_B(D_o)\cdot B\nonumber
\end{eqnarray}

where $\Delta_o$ is the collection of the output data.

The send action and receive action of the same data through the same channel can communicate each other, otherwise, a deadlock $\delta$ will be caused. We define the following communication functions.

\begin{eqnarray}
&&\gamma(send_Q(q),receive_Q(q))\triangleq c_Q(q)\nonumber\\
&&\gamma(send_P(B_b),receive_P(B_b))\triangleq c_P(B_b)\nonumber\\
&&\gamma(send_P(B_a),receive_P(B_a))\triangleq c_P(B_a)\nonumber
\end{eqnarray}

Let $A$ and $B$ in parallel, then the system $AB$ can be represented by the following process term.

$$\tau_I(\partial_H(\Theta(A\between B)))$$

where $H=\{send_Q(q),receive_Q(q),send_P(B_b),receive_P(B_b),send_P(B_a),receive_P(B_a)\}$ and $I=\{Rand[q;B_a], Rand[q;K_a], Set_{K_a}[q], H_{B_a}[q], Rand[q';B_b], M[q;K_b], c_Q(q), c_P(B_b),\\ c_P(B_a), cmp(K_{a,b},K_a,K_b,B_a,B_b)\}$.

Then we get the following conclusion.

\begin{theorem}
The basic BB84 protocol $\tau_I(\partial_H(\Theta(A\between B)))$ exhibits desired external behaviors.
\end{theorem}

\begin{proof}
We can get $\tau_I(\partial_H(\Theta(A\between B)))=\sum_{D_i\in \Delta_i}\sum_{D_o\in\Delta_o}receive_A(D_i)\leftmerge send_B(D_o)\leftmerge \tau_I(\partial_H(\Theta(A\between B)))$. So, the basic
BB84 protocol $\tau_I(\partial_H(\Theta(A\between B)))$ exhibits desired external behaviors.
\end{proof}

\subsection{Verification of E91 Protocol}\label{VE914}

With support of Entanglement merge $\between$, now, qACP can be used to verify quantum protocols utilizing entanglement. The E91 protocol\cite{E91} is the first quantum protocol which utilizes entanglement and mixes quantum and classical information. In this section, we take an example of verification for the E91 protocol.

\begin{figure}
  \centering
  \includegraphics{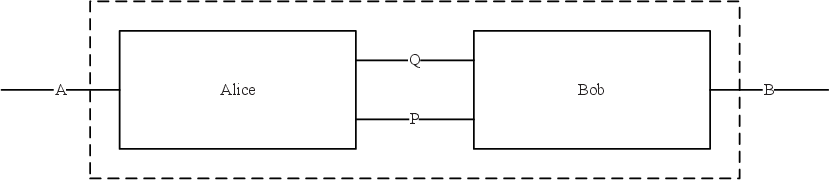}
  \caption{The E91 protocol.}
  \label{E914}
\end{figure}

The E91 protocol is used to create a private key between two parities, Alice and Bob. Firstly, we introduce the basic E91 protocol briefly, which is illustrated in Figure \ref{E914}.

\begin{enumerate}
  \item Alice generates a string of EPR pairs $q$ with size $n$, i.e., $2n$ particles, and sends a string of qubits $q_b$ from each EPR pair with $n$ to Bob through a quantum channel $Q$, remains the other string of qubits $q_a$ from each pair with size $n$.
  \item Alice create two string of bits with size $n$ randomly, denoted as $B_a$ and $K_a$.
  \item Bob receives $q_b$ and randomly generates a string of bits $B_b$ with size $n$.
  \item Alice measures each qubit of $q_a$ according to a basis by bits of $B_a$. And the measurement results would be $K_a$, which is also with size $n$.
  \item Bob measures each qubit of $q_b$ according to a basis by bits of $B_b$. And the measurement results would be $K_b$, which is also with size $n$.
  \item Bob sends his measurement bases $B_b$ to Alice through a public channel $P$.
  \item Once receiving $B_b$, Alice sends her bases $B_a$ to Bob through channel $P$, and Bob receives $B_a$.
  \item Alice and Bob determine that at which position the bit strings $B_a$ and $B_b$ are equal, and they discard the mismatched bits of $B_a$ and $B_b$. Then the remaining bits of $K_a$ and $K_b$, denoted as $K_a'$ and $K_b'$ with $K_{a,b}=K_a'=K_b'$.
\end{enumerate}

We re-introduce the basic E91 protocol in an abstract way with more technical details as Figure \ref{E914} illustrates.

Now, $M[q_a;K_a]$ denotes the Alice's measurement operation of $q_a$, and $\circledS_{M[q_a;K_a]}$ denotes the responding shadow constant; $M[q_b;K_b]$ denotes the Bob's measurement operation of $q_b$, and $\circledS_{M[q_b;K_b]}$ denotes the responding shadow constant. Alice sends $q_b$ to Bob through the quantum channel $Q$ by quantum communicating action $send_{Q}(q_b)$ and Bob receives $q_b$ through $Q$ by quantum communicating action $receive_{Q}(q_b)$. Bob sends $B_b$ to Alice through the public channel $P$ by classical communicating action $send_{P}(B_b)$ and Alice receives $B_b$ through channel $P$ by classical communicating action $receive_{P}(B_b)$, and the same as $send_{P}(B_a)$ and $receive_{P}(B_a)$. Alice and Bob generate the private key $K_{a,b}$ by a classical comparison action $cmp(K_{a,b},K_a,K_b,B_a,B_b)$. Let Alice and Bob be a system $AB$ and let interactions between Alice and Bob be internal actions. $AB$ receives external input $D_i$ through channel $A$ by communicating action $receive_A(D_i)$ and sends results $D_o$ through channel $B$ by communicating action $send_B(D_o)$.

Then the state transition of Alice can be described by qACP as follows.

\begin{eqnarray}
&&A=\sum_{D_i\in \Delta_i}receive_A(D_i)\cdot A_1\nonumber\\
&&A_1=send_Q(q_b)\cdot A_2\nonumber\\
&&A_2=M[q_a;K_a]\cdot A_3\nonumber\\
&&A_3=\circledS_{M[q_b;K_b]}\cdot A_4\nonumber\\
&&A_4=receive_P(B_b)\cdot A_5\nonumber\\
&&A_5=send_P(B_a)\cdot A_6\nonumber\\
&&A_6=cmp(K_{a,b},K_a,K_b,B_a,B_b)\cdot A\nonumber
\end{eqnarray}

where $\Delta_i$ is the collection of the input data.

And the state transition of Bob can be described by qACP as follows.

\begin{eqnarray}
&&B=receive_Q(q_b)\cdot B_1\nonumber\\
&&B_1=\circledS_{M[q_a;K_a]}\cdot B_2\nonumber\\
&&B_2=M[q_b;K_b]\cdot B_3\nonumber\\
&&B_3=send_P(B_b)\cdot B_4\nonumber\\
&&B_4=receive_P(B_a)\cdot B_5\nonumber\\
&&B_5=cmp(K_{a,b},K_a,K_b,B_a,B_b)\cdot B_6\nonumber\\
&&B_6=\sum_{D_o\in\Delta_o}send_B(D_o)\cdot B\nonumber
\end{eqnarray}

where $\Delta_o$ is the collection of the output data.

The send action and receive action of the same data through the same channel can communicate each other, otherwise, a deadlock $\delta$ will be caused. The quantum operation and its shadow constant pair will lead entanglement occur, otherwise, a deadlock $\delta$ will occur. We define the following communication functions.

\begin{eqnarray}
&&\gamma(send_Q(q_b),receive_Q(q_b))\triangleq c_Q(q_b)\nonumber\\
&&\gamma(send_P(B_b),receive_P(B_b))\triangleq c_P(B_b)\nonumber\\
&&\gamma(send_P(B_a),receive_P(B_a))\triangleq c_P(B_a)\nonumber
\end{eqnarray}

Let $A$ and $B$ in parallel, then the system $AB$ can be represented by the following process term.

$$\tau_I(\partial_H(\Theta(A\between B)))$$

where $H=\{send_Q(q_b),receive_Q(q_b),send_P(B_b),receive_P(B_b),send_P(B_a),receive_P(B_a),\\ M[q_a;K_a], \circledS_{M[q_a;K_a]}, M[q_b;K_b], \circledS_{M[q_b;K_b]}\}$ and $I=\{c_Q(q_b), c_P(B_b), c_P(B_a), M[q_a;K_a], M[q_b;K_b],\\ cmp(K_{a,b},K_a,K_b,B_a,B_b)\}$.

Then we get the following conclusion.

\begin{theorem}
The basic E91 protocol $\tau_I(\partial_H(A\parallel B))$ exhibits desired external behaviors.
\end{theorem}

\begin{proof}
We can get $\tau_I(\partial_H(\Theta(A\between B)))=\sum_{D_i\in \Delta_i}\sum_{D_o\in\Delta_o}receive_A(D_i)\leftmerge send_B(D_o)\leftmerge \tau_I(\partial_H(\Theta(A\between B)))$. 
So, the basic E91 protocol $\tau_I(\partial_H(\Theta(A\between B)))$ exhibits desired external behaviors.
\end{proof}

\subsection{Verification of B92 Protocol}\label{VB924}

The famous B92 protocol\cite{B92} is a quantum key distribution protocol, in which quantum information and classical information are mixed. We take an example of the B92 protocol to illustrate the usage of qACP in verification of quantum protocols.

\begin{figure}
  \centering
  \includegraphics{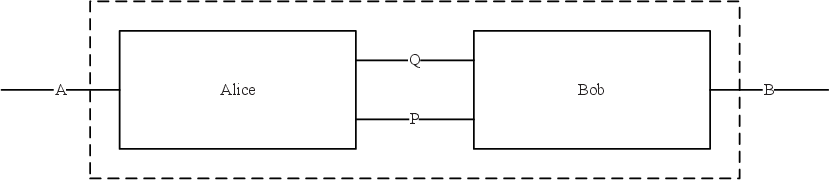}
  \caption{The B92 protocol.}
  \label{B924}
\end{figure}

The B92 protocol is used to create a private key between two parities, Alice and Bob. B92 is a protocol of quantum key distribution (QKD) which uses polarized photons as information carriers. Firstly, we introduce the basic B92 protocol briefly, which is illustrated in Figure \ref{B924}.

\begin{enumerate}
  \item Alice create a string of bits with size $n$ randomly, denoted as $A$.
  \item Alice generates a string of qubits $q$ with size $n$, carried by polarized photons. If $A_i=0$, the ith qubit is $|0\rangle$; else if $A_i=1$, the ith qubit is $|+\rangle$.
  \item Alice sends $q$ to Bob through a quantum channel $Q$ between Alice and Bob.
  \item Bob receives $q$ and randomly generates a string of bits $B$ with size $n$.
  \item If $B_i=0$, Bob chooses the basis $\oplus$; else if $B_i=1$, Bob chooses the basis $\otimes$. Bob measures each qubit of $q$ according to the above basses. And Bob builds a String of bits $T$, if the measurement produces $|0\rangle$ or $|+\rangle$, then $T_i=0$; else if the measurement produces $|1\rangle$ or $|-\rangle$, then $T_i=1$, which is also with size $n$.
  \item Bob sends $T$ to Alice through a public channel $P$.
  \item Alice and Bob determine that at which position the bit strings $A$ and $B$ are remained for which $T_i=1$. In absence of Eve, $A_i=1-B_i$, a shared raw key $K_{a,b}$ is formed by $A_i$.
\end{enumerate}

We re-introduce the basic B92 protocol in an abstract way with more technical details as Figure \ref{B924} illustrates.

Now, we assume a special measurement operation $Rand[q;A]$ which create a string of $n$ random bits $A$ from the $q$ quantum system, and the same as $Rand[q';B]$. $M[q;T]$ denotes the Bob's measurement operation of $q$. The generation of $n$ qubits $q$ through a quantum operation $Set_{A}[q]$. Alice sends $q$ to Bob through the quantum channel $Q$ by quantum communicating action $send_{Q}(q)$ and Bob receives $q$ through $Q$ by quantum communicating action $receive_{Q}(q)$. Bob sends $T$ to Alice through the public channel $P$ by classical communicating action $send_{P}(T)$ and Alice receives $T$ through channel $P$ by classical communicating action $receive_{P}(T)$. Alice and Bob generate the private key $K_{a,b}$ by a classical comparison action $cmp(K_{a,b},T,A,B)$. Let Alice and Bob be a system $AB$ and let interactions between Alice and Bob be internal actions. $AB$ receives external input $D_i$ through channel $A$ by communicating action $receive_A(D_i)$ and sends results $D_o$ through channel $B$ by communicating action $send_B(D_o)$.

Then the state transition of Alice can be described by qACP as follows.

\begin{eqnarray}
&&A=\sum_{D_i\in \Delta_i}receive_A(D_i)\cdot A_1\nonumber\\
&&A_1=Rand[q;A]\cdot A_2\nonumber\\
&&A_2=Set_{A}[q]\cdot A_3\nonumber\\
&&A_3=send_Q(q)\cdot A_4\nonumber\\
&&A_4=receive_P(T)\cdot A_5\nonumber\\
&&A_5=cmp(K_{a,b},T,A,B)\cdot A\nonumber
\end{eqnarray}

where $\Delta_i$ is the collection of the input data.

And the state transition of Bob can be described by qACP as follows.

\begin{eqnarray}
&&B=receive_Q(q)\cdot B_1\nonumber\\
&&B_1=Rand[q';B]\cdot B_2\nonumber\\
&&B_2=M[q;T]\cdot B_3\nonumber\\
&&B_3=send_P(T)\cdot B_4\nonumber\\
&&B_4=cmp(K_{a,b},T,A,B)\cdot B_5\nonumber\\
&&B_5=\sum_{D_o\in\Delta_o}send_B(D_o)\cdot B\nonumber
\end{eqnarray}

where $\Delta_o$ is the collection of the output data.

The send action and receive action of the same data through the same channel can communicate each other, otherwise, a deadlock $\delta$ will be caused. We define the following communication functions.

\begin{eqnarray}
&&\gamma(send_Q(q),receive_Q(q))\triangleq c_Q(q)\nonumber\\
&&\gamma(send_P(T),receive_P(T))\triangleq c_P(T)\nonumber
\end{eqnarray}

Let $A$ and $B$ in parallel, then the system $AB$ can be represented by the following process term.

$$\tau_I(\partial_H(\Theta(A\between B)))$$

where $H=\{send_Q(q),receive_Q(q),send_P(T),receive_P(T)\}$ and $I=\{Rand[q;A], Set_{A}[q], Rand[q';B], \\ M[q;T], c_Q(q), c_P(T), cmp(K_{a,b},T,A,B)\}$.

Then we get the following conclusion.

\begin{theorem}
The basic B92 protocol $\tau_I(\partial_H(A\parallel B))$ exhibits desired external behaviors.
\end{theorem}

\begin{proof}
We can get $\tau_I(\partial_H(\Theta(A\between B)))=\sum_{D_i\in \Delta_i}\sum_{D_o\in\Delta_o}receive_A(D_i)\leftmerge send_B(D_o)\leftmerge \tau_I(\partial_H(\Theta(A\between B)))$. 
So, the basic B92 protocol $\tau_I(\partial_H(\Theta(A\between B)))$ exhibits desired external behaviors.
\end{proof}

\subsection{Verification of DPS Protocol}\label{VDPS4}

The famous DPS protocol\cite{DPS} is a quantum key distribution protocol, in which quantum information and classical information are mixed. We take an example of the DPS protocol to illustrate the usage of qACP in verification of quantum protocols.

\begin{figure}
  \centering
  \includegraphics{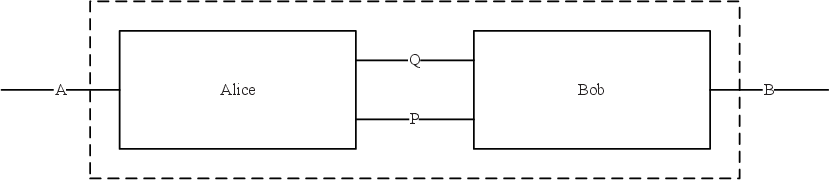}
  \caption{The DPS protocol.}
  \label{DPS4}
\end{figure}

The DPS protocol is used to create a private key between two parities, Alice and Bob. DPS is a protocol of quantum key distribution (QKD) which uses pulses of a photon which has nonorthogonal four states. Firstly, we introduce the basic DPS protocol briefly, which is illustrated in Figure \ref{DPS4}.

\begin{enumerate}
  \item Alice generates a string of qubits $q$ with size $n$, carried by a series of single photons possily at four time instances.
  \item Alice sends $q$ to Bob through a quantum channel $Q$ between Alice and Bob.
  \item Bob receives $q$ by detectors clicking at the second or third time instance, and records the time into $T$ with size $n$ and which detector clicks into $D$ with size $n$.
  \item Bob sends $T$ to Alice through a public channel $P$.
  \item Alice receives $T$. From $T$ and her modulation data, Alice knows which detector clicked in Bob's site, i.e. $D$.
  \item Alice and Bob have an identical bit string, provided that the first detector click represents "0" and the other detector represents "1", then a shared raw key $K_{a,b}$ is formed.
\end{enumerate}

We re-introduce the basic DPS protocol in an abstract way with more technical details as Figure \ref{DPS4} illustrates.

Now, we assume $M[q;T]$ denotes the Bob's measurement operation of $q$. The generation of $n$ qubits $q$ through a quantum operation $Set_{A}[q]$. Alice sends $q$ to Bob through the quantum channel $Q$ by quantum communicating action $send_{Q}(q)$ and Bob receives $q$ through $Q$ by quantum communicating action $receive_{Q}(q)$. Bob sends $T$ to Alice through the public channel $P$ by classical communicating action $send_{P}(T)$ and Alice receives $T$ through channel $P$ by classical communicating action $receive_{P}(T)$. Alice and Bob generate the private key $K_{a,b}$ by a classical comparison action $cmp(K_{a,b},D)$. Let Alice and Bob be a system $AB$ and let interactions between Alice and Bob be internal actions. $AB$ receives external input $D_i$ through channel $A$ by communicating action $receive_A(D_i)$ and sends results $D_o$ through channel $B$ by communicating action $send_B(D_o)$.

Then the state transition of Alice can be described by qACP as follows.

\begin{eqnarray}
&&A=\sum_{D_i\in \Delta_i}receive_A(D_i)\cdot A_1\nonumber\\
&&A_1=Set_{A}[q]\cdot A_2\nonumber\\
&&A_2=send_Q(q)\cdot A_3\nonumber\\
&&A_3=receive_P(T)\cdot A_4\nonumber\\
&&A_4=cmp(K_{a,b},D)\cdot A\nonumber
\end{eqnarray}

where $\Delta_i$ is the collection of the input data.

And the state transition of Bob can be described by qACP as follows.

\begin{eqnarray}
&&B=receive_Q(q)\cdot B_1\nonumber\\
&&B_1=M[q;T]\cdot B_2\nonumber\\
&&B_2=send_P(T)\cdot B_3\nonumber\\
&&B_3=cmp(K_{a,b},D)\cdot B_4\nonumber\\
&&B_4=\sum_{D_o\in\Delta_o}send_B(D_o)\cdot B\nonumber
\end{eqnarray}

where $\Delta_o$ is the collection of the output data.

The send action and receive action of the same data through the same channel can communicate each other, otherwise, a deadlock $\delta$ will be caused. We define the following communication functions.

\begin{eqnarray}
&&\gamma(send_Q(q),receive_Q(q))\triangleq c_Q(q)\nonumber\\
&&\gamma(send_P(T),receive_P(T))\triangleq c_P(T)\nonumber\\
\end{eqnarray}

Let $A$ and $B$ in parallel, then the system $AB$ can be represented by the following process term.

$$\tau_I(\partial_H(\Theta(A\between B)))$$

where $H=\{send_Q(q),receive_Q(q),send_P(T),receive_P(T)\}$

and $I=\{Set_{A}[q], M[q;T], c_Q(q), c_P(T), cmp(K_{a,b},D)\}$.

Then we get the following conclusion.

\begin{theorem}
The basic DPS protocol $\tau_I(\partial_H(\Theta(A\between B)))$ exhibits desired external behaviors.
\end{theorem}

\begin{proof}
We can get $\tau_I(\partial_H(\Theta(A\between B)))=\sum_{D_i\in \Delta_i}\sum_{D_o\in\Delta_o}receive_A(D_i)\leftmerge send_B(D_o)\leftmerge \tau_I(\partial_H(\Theta(A\between B)))$. 
So, the basic DPS protocol $\tau_I(\partial_H(\Theta(A\between B)))$ exhibits desired external behaviors.
\end{proof}

\subsection{Verification of BBM92 Protocol}\label{VBBM924}

The famous BBM92 protocol\cite{BBM92} is a quantum key distribution protocol, in which quantum information and classical information are mixed. We take an example of the BBM92 protocol to illustrate the usage of qACP in verification of quantum protocols.

\begin{figure}
  \centering
  \includegraphics{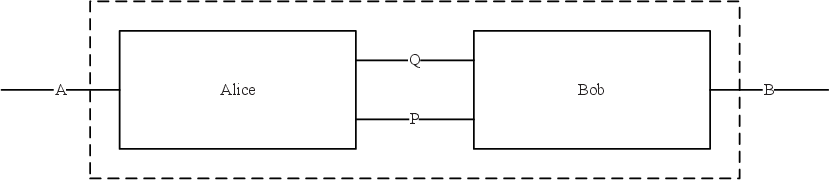}
  \caption{The BBM92 protocol.}
  \label{BBM924}
\end{figure}

The BBM92 protocol is used to create a private key between two parities, Alice and Bob. BBM92 is a protocol of quantum key distribution (QKD) which uses EPR pairs as information carriers. Firstly, we introduce the basic BBM92 protocol briefly, which is illustrated in Figure \ref{BBM924}.

\begin{enumerate}
  \item Alice generates a string of EPR pairs $q$ with size $n$, i.e., $2n$ particles, and sends a string of qubits $q_b$ from each EPR pair with $n$ to Bob through a quantum channel $Q$, remains the other string of qubits $q_a$ from each pair with size $n$.
  \item Alice create a string of bits with size $n$ randomly, denoted as $B_a$.
  \item Bob receives $q_b$ and randomly generates a string of bits $B_b$ with size $n$.
  \item Alice measures each qubit of $q_a$ according to bits of $B_a$, if $B_{a_i}=0$, then uses $x$ axis ($\rightarrow$); else if $B_{a_i}=1$, then uses $z$ axis ($\uparrow$).
  \item Bob measures each qubit of $q_b$ according to bits of $B_b$, if $B_{b_i}=0$, then uses $x$ axis ($\rightarrow$); else if $B_{b_i}=1$, then uses $z$ axis ($\uparrow$).
  \item Bob sends his measurement axis choices $B_b$ to Alice through a public channel $P$.
  \item Once receiving $B_b$, Alice sends her axis choices $B_a$ to Bob through channel $P$, and Bob receives $B_a$.
  \item Alice and Bob agree to discard all instances in which they happened to measure along different axes, as well as instances in which measurements fails because of imperfect quantum efficiency of the detectors. Then the remaining instances can be used to generate a private key $K_{a,b}$.
\end{enumerate}

We re-introduce the basic BBM92 protocol in an abstract way with more technical details as Figure \ref{BBM924} illustrates.

Now, $M[q_a;B_a]$ denotes the Alice's measurement operation of $q_a$, and $\circledS_{M[q_a;B_a]}$ denotes the responding shadow constant; $M[q_b;B_b]$ denotes the Bob's measurement operation of $q_b$, and $\circledS_{M[q_b;B_b]}$ denotes the responding shadow constant. Alice sends $q_b$ to Bob through the quantum channel $Q$ by quantum communicating action $send_{Q}(q_b)$ and Bob receives $q_b$ through $Q$ by quantum communicating action $receive_{Q}(q_b)$. Bob sends $B_b$ to Alice through the public channel $P$ by classical communicating action $send_{P}(B_b)$ and Alice receives $B_b$ through channel $P$ by classical communicating action $receive_{P}(B_b)$, and the same as $send_{P}(B_a)$ and $receive_{P}(B_a)$. Alice and Bob generate the private key $K_{a,b}$ by a classical comparison action $cmp(K_{a,b},B_a,B_b)$. Let Alice and Bob be a system $AB$ and let interactions between Alice and Bob be internal actions. $AB$ receives external input $D_i$ through channel $A$ by communicating action $receive_A(D_i)$ and sends results $D_o$ through channel $B$ by communicating action $send_B(D_o)$.

Then the state transition of Alice can be described by qACP as follows.

\begin{eqnarray}
&&A=\sum_{D_i\in \Delta_i}receive_A(D_i)\cdot A_1\nonumber\\
&&A_1=send_Q(q_b)\cdot A_2\nonumber\\
&&A_2=M[q_a;B_a]\cdot A_3\nonumber\\
&&A_3=\circledS_{M[q_b;B_b]}\cdot A_4\nonumber\\
&&A_4=receive_P(B_b)\cdot A_5\nonumber\\
&&A_5=send_P(B_a)\cdot A_6\nonumber\\
&&A_6=cmp(K_{a,b},B_a,B_b)\cdot A\nonumber
\end{eqnarray}

where $\Delta_i$ is the collection of the input data.

And the state transition of Bob can be described by qACP as follows.

\begin{eqnarray}
&&B=receive_Q(q_b)\cdot B_1\nonumber\\
&&B_1=\circledS_{M[q_a;B_a]}\cdot B_2\nonumber\\
&&B_2=M[q_b;B_b]\cdot B_3\nonumber\\
&&B_3=send_P(B_b)\cdot B_4\nonumber\\
&&B_4=receive_P(B_a)\cdot B_5\nonumber\\
&&B_5=cmp(K_{a,b},B_a,B_b)\cdot B_6\nonumber\\
&&B_6=\sum_{D_o\in\Delta_o}send_B(D_o)\cdot B\nonumber
\end{eqnarray}

where $\Delta_o$ is the collection of the output data.

The send action and receive action of the same data through the same channel can communicate each other, otherwise, a deadlock $\delta$ will be caused. The quantum operation and its shadow constant pair will lead entanglement occur, otherwise, a deadlock $\delta$ will occur. We define the following communication functions.

\begin{eqnarray}
&&\gamma(send_Q(q_b),receive_Q(q_b))\triangleq c_Q(q_b)\nonumber\\
&&\gamma(send_P(B_b),receive_P(B_b))\triangleq c_P(B_b)\nonumber\\
&&\gamma(send_P(B_a),receive_P(B_a))\triangleq c_P(B_a)\nonumber
\end{eqnarray}

Let $A$ and $B$ in parallel, then the system $AB$ can be represented by the following process term.

$$\tau_I(\partial_H(\Theta(A\between B)))$$

where $H=\{send_Q(q_b),receive_Q(q_b),send_P(B_b),receive_P(B_b),send_P(B_a),receive_P(B_a),\\ M[q_a;B_a], \circledS_{M[q_a;B_a]}, M[q_b;B_b], \circledS_{M[q_b;B_b]}\}$ and $I=\{c_Q(q_b), c_P(B_b), c_P(B_a), M[q_a;B_a], M[q_b;B_b],\\ cmp(K_{a,b},B_a,B_b)\}$.

Then we get the following conclusion.

\begin{theorem}
The basic BBM92 protocol $\tau_I(\partial_H(\Theta(A\between B)))$ exhibits desired external behaviors.
\end{theorem}

\begin{proof}
We can get $\tau_I(\partial_H(\Theta(A\between B)))=\sum_{D_i\in \Delta_i}\sum_{D_o\in\Delta_o}receive_A(D_i)\leftmerge send_B(D_o)\leftmerge \tau_I(\partial_H(\Theta(A\between B)))$. 
So, the basic BBM92 protocol $\tau_I(\partial_H(\Theta(A\between B)))$ exhibits desired external behaviors.
\end{proof}

\subsection{Verification of SARG04 Protocol}\label{VSARG044}

The famous SARG04 protocol\cite{SARG04} is a quantum key distribution protocol, in which quantum information and classical information are mixed. We take an example of the SARG04 protocol to illustrate the usage of qACP in verification of quantum protocols.

\begin{figure}
  \centering
  \includegraphics{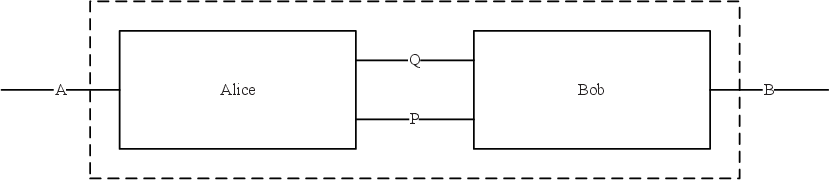}
  \caption{The SARG04 protocol.}
  \label{SARG044}
\end{figure}

The SARG04 protocol is used to create a private key between two parities, Alice and Bob. SARG04 is a protocol of quantum key distribution (QKD) which refines the BB84 protocol against PNS (Photon Number Splitting) attacks. The main innovations are encoding bits in nonorthogonal states and the classical sifting procedure. Firstly, we introduce the basic SARG04 protocol briefly, which is illustrated in Figure \ref{SARG044}.

\begin{enumerate}
  \item Alice create a string of bits with size $n$ randomly, denoted as $K_a$.
  \item Alice generates a string of qubits $q$ with size $n$, and the $i$th qubit of $q$ has four nonorthogonal states, it is $|\pm x\rangle$ if $K_a=0$; it is $|\pm z\rangle$ if $K_a=1$. And she records the corresponding one of the four pairs of nonorthogonal states into $B_a$ with size $2n$.
  \item Alice sends $q$ to Bob through a quantum channel $Q$ between Alice and Bob.
  \item Alice sends $B_a$ through a public channel $P$.
  \item Bob measures each qubit of $q$ $\sigma_x$ or $\sigma_z$. And he records the unambiguous discriminations into $K_b$ with a raw size $n/4$, and the unambiguous discrimination information into $B_b$ with size $n$.
  \item Bob sends $B_b$ to Alice through the public channel $P$.
  \item Alice and Bob determine that at which position the bit should be remained. Then the remaining bits of $K_a$ and $K_b$ is the private key $K_{a,b}$.
\end{enumerate}

We re-introduce the basic SARG04 protocol in an abstract way with more technical details as Figure \ref{SARG044} illustrates.

Now, we assume a special measurement operation $Rand[q;K_a]$ which create a string of $n$ random bits $K_a$ from the $q$ quantum system. $M[q;K_b]$ denotes the Bob's measurement operation of $q$. The generation of $n$ qubits $q$ through a quantum operation $Set_{K_a}[q]$. Alice sends $q$ to Bob through the quantum channel $Q$ by quantum communicating action $send_{Q}(q)$ and Bob receives $q$ through $Q$ by quantum communicating action $receive_{Q}(q)$. Bob sends $B_b$ to Alice through the public channel $P$ by classical communicating action $send_{P}(B_b)$ and Alice receives $B_b$ through channel $P$ by classical communicating action $receive_{P}(B_b)$, and the same as $send_{P}(B_a)$ and $receive_{P}(B_a)$. Alice and Bob generate the private key $K_{a,b}$ by a classical comparison action $cmp(K_{a,b},K_a,K_b,B_a,B_b)$. Let Alice and Bob be a system $AB$ and let interactions between Alice and Bob be internal actions. $AB$ receives external input $D_i$ through channel $A$ by communicating action $receive_A(D_i)$ and sends results $D_o$ through channel $B$ by communicating action $send_B(D_o)$.

Then the state transition of Alice can be described by qACP as follows.

\begin{eqnarray}
&&A=\sum_{D_i\in \Delta_i}receive_A(D_i)\cdot A_1\nonumber\\
&&A_1=Rand[q;K_a]\cdot A_2\nonumber\\
&&A_2=Set_{K_a}[q]\cdot A_3\nonumber\\
&&A_3=send_Q(q)\cdot A_4\nonumber\\
&&A_4=send_P(B_a)\cdot A_5\nonumber\\
&&A_5=receive_P(B_b)\cdot A_6\nonumber\\
&&A_6=cmp(K_{a,b},K_a,K_b,B_a,B_b)\cdot A\nonumber
\end{eqnarray}

where $\Delta_i$ is the collection of the input data.

And the state transition of Bob can be described by qACP as follows.

\begin{eqnarray}
&&B=receive_Q(q)\cdot B_1\nonumber\\
&&B_1=receive_P(B_a)\cdot B_2\nonumber\\
&&B_2=M[q;K_b]\cdot B_3\nonumber\\
&&B_3=send_P(B_b)\cdot B_4\nonumber\\
&&B_4=cmp(K_{a,b},K_a,K_b,B_a,B_b)\cdot B_5\nonumber\\
&&B_5=\sum_{D_o\in\Delta_o}send_B(D_o)\cdot B\nonumber
\end{eqnarray}

where $\Delta_o$ is the collection of the output data.

The send action and receive action of the same data through the same channel can communicate each other, otherwise, a deadlock $\delta$ will be caused. We define the following communication functions.

\begin{eqnarray}
&&\gamma(send_Q(q),receive_Q(q))\triangleq c_Q(q)\nonumber\\
&&\gamma(send_P(B_b),receive_P(B_b))\triangleq c_P(B_b)\nonumber\\
&&\gamma(send_P(B_a),receive_P(B_a))\triangleq c_P(B_a)\nonumber
\end{eqnarray}

Let $A$ and $B$ in parallel, then the system $AB$ can be represented by the following process term.

$$\tau_I(\partial_H(\Theta(A\between B)))$$

where $H=\{send_Q(q),receive_Q(q),send_P(B_b),receive_P(B_b),send_P(B_a),receive_P(B_a)\}$ and $I=\{Rand[q;K_a], Set_{K_a}[q], M[q;K_b], c_Q(q), c_P(B_b),\\ c_P(B_a), cmp(K_{a,b},K_a,K_b,B_a,B_b)\}$.

Then we get the following conclusion.

\begin{theorem}
The basic SARG04 protocol $\tau_I(\partial_H(\Theta(A\between B)))$ exhibits desired external behaviors.
\end{theorem}

\begin{proof}
We can get $\tau_I(\partial_H(\Theta(A\between B)))=\sum_{D_i\in \Delta_i}\sum_{D_o\in\Delta_o}receive_A(D_i)\leftmerge send_B(D_o)\leftmerge \tau_I(\partial_H(\Theta(A\between B)))$. 
So, the basic SARG04 protocol $\tau_I(\partial_H(\Theta(A\between B)))$ exhibits desired external behaviors.
\end{proof}

\subsection{Verification of COW Protocol}\label{VCOW4}

The famous COW protocol\cite{COW} is a quantum key distribution protocol, in which quantum information and classical information are mixed. We take an example of the COW protocol to illustrate the usage of qACP in verification of quantum protocols.

\begin{figure}
  \centering
  \includegraphics{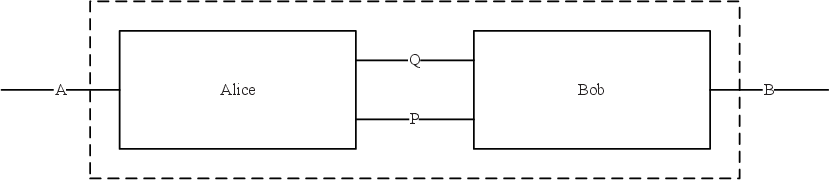}
  \caption{The COW protocol.}
  \label{COW4}
\end{figure}

The COW protocol is used to create a private key between two parities, Alice and Bob. COW is a protocol of quantum key distribution (QKD) which is practical. Firstly, we introduce the basic COW protocol briefly, which is illustrated in Figure \ref{COW4}.

\begin{enumerate}
  \item Alice generates a string of qubits $q$ with size $n$, and the $i$th qubit of $q$ is "0" with probability $\frac{1-f}{2}$, "1" with probability $\frac{1-f}{2}$ and the decoy sequence with probability $f$.
  \item Alice sends $q$ to Bob through a quantum channel $Q$ between Alice and Bob.
  \item Alice sends $A$ of the items corresponding to a decoy sequence through a public channel $P$.
  \item Bob removes all the detections at times $2A-1$ and $2A$ from his raw key and looks whether detector $D_{2M}$ has ever fired at time $2A$.
  \item Bob sends $B$ of the times $2A+1$ in which he had a detector in $D_{2M}$ to Alice through the public channel $P$.
  \item Alice receives $B$ and verifies if some of these items corresponding to a bit sequence "1,0".
  \item Bob sends $C$ of the items that he has detected through the public channel $P$.
  \item Alice and Bob run error correction and privacy amplification on these bits, and the private key $K_{a,b}$ is established.
\end{enumerate}

We re-introduce the basic COW protocol in an abstract way with more technical details as Figure \ref{COW4} illustrates.

Now, we assume The generation of $n$ qubits $q$ through a quantum operation $Set[q]$. $M[q]$ denotes the Bob's measurement operation of $q$.  Alice sends $q$ to Bob through the quantum channel $Q$ by quantum communicating action $send_{Q}(q)$ and Bob receives $q$ through $Q$ by quantum communicating action $receive_{Q}(q)$. Alice sends $A$ to Alice through the public channel $P$ by classical communicating action $send_{P}(A)$ and Alice receives $A$ through channel $P$ by classical communicating action $receive_{P}(A)$, and the same as $send_{P}(B)$ and $receive_{P}(B)$, and $send_{P}(C)$ and $receive_{P}(C)$. Alice and Bob generate the private key $K_{a,b}$ by a classical comparison action $cmp(K_{a,b})$. Let Alice and Bob be a system $AB$ and let interactions between Alice and Bob be internal actions. $AB$ receives external input $D_i$ through channel $A$ by communicating action $receive_A(D_i)$ and sends results $D_o$ through channel $B$ by communicating action $send_B(D_o)$.

Then the state transition of Alice can be described by qACP as follows.

\begin{eqnarray}
&&A=\sum_{D_i\in \Delta_i}receive_A(D_i)\cdot A_1\nonumber\\
&&A_1=Set[q]\cdot A_2\nonumber\\
&&A_2=send_Q(q)\cdot A_3\nonumber\\
&&A_3=send_P(A)\cdot A_4\nonumber\\
&&A_4=receive_P(B)\cdot A_5\nonumber\\
&&A_5=receive_P(C)\cdot A_6\nonumber\\
&&A_6=cmp(K_{a,b})\cdot A\nonumber
\end{eqnarray}

where $\Delta_i$ is the collection of the input data.

And the state transition of Bob can be described by qACP as follows.

\begin{eqnarray}
&&B=receive_Q(q)\cdot B_1\nonumber\\
&&B_1=receive_P(A)\cdot B_2\nonumber\\
&&B_2=M[q]\cdot B_3\nonumber\\
&&B_3=send_P(B)\cdot B_4\nonumber\\
&&B_4=send_P(C)\cdot B_5\nonumber\\
&&B_5=cmp(K_{a,b})\cdot B_6\nonumber\\
&&B_6=\sum_{D_o\in\Delta_o}send_B(D_o)\cdot B\nonumber
\end{eqnarray}

where $\Delta_o$ is the collection of the output data.

The send action and receive action of the same data through the same channel can communicate each other, otherwise, a deadlock $\delta$ will be caused. We define the following communication functions.

\begin{eqnarray}
&&\gamma(send_Q(q),receive_Q(q))\triangleq c_Q(q)\nonumber\\
&&\gamma(send_P(A),receive_P(A))\triangleq c_P(A)\nonumber\\
&&\gamma(send_P(B),receive_P(B))\triangleq c_P(B)\nonumber\\
&&\gamma(send_P(C),receive_P(C))\triangleq c_P(C)\nonumber
\end{eqnarray}

Let $A$ and $B$ in parallel, then the system $AB$ can be represented by the following process term.

$$\tau_I(\partial_H(\Theta(A\between B)))$$

where $H=\{send_Q(q),receive_Q(q),send_P(A),receive_P(A),send_P(B),receive_P(B),\\send_P(C),receive_P(C)\}$ and $I=\{Set[q], M[q], c_Q(q), c_P(A),\\ c_P(B),c_P(C), cmp(K_{a,b})\}$.

Then we get the following conclusion.

\begin{theorem}
The basic COW protocol $\tau_I(\partial_H(\Theta(A\between B)))$ exhibits desired external behaviors.
\end{theorem}

\begin{proof}
We can get $\tau_I(\partial_H(\Theta(A\between B)))=\sum_{D_i\in \Delta_i}\sum_{D_o\in\Delta_o}receive_A(D_i)\leftmerge send_B(D_o)\leftmerge \tau_I(\partial_H(\Theta(A\between B)))$. 
So, the basic COW protocol $\tau_I(\partial_H(\Theta(A\between B)))$ exhibits desired external behaviors.
\end{proof}

\subsection{Verification of SSP Protocol}\label{VSSP4}

The famous SSP protocol\cite{SSP} is a quantum key distribution protocol, in which quantum information and classical information are mixed. We take an example of the SSP protocol to illustrate the usage of qACP in verification of quantum protocols.

\begin{figure}
  \centering
  \includegraphics{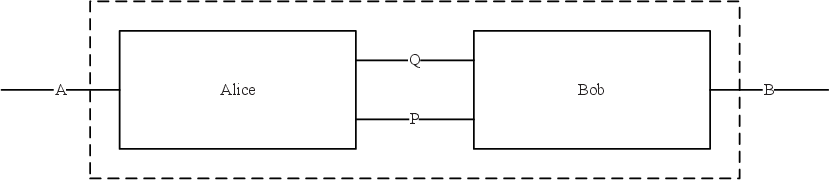}
  \caption{The SSP protocol.}
  \label{SSP4}
\end{figure}

The SSP protocol is used to create a private key between two parities, Alice and Bob. SSP is a protocol of quantum key distribution (QKD) which uses six states. Firstly, we introduce the basic SSP protocol briefly, which is illustrated in Figure \ref{SSP4}.

\begin{enumerate}
  \item Alice create two string of bits with size $n$ randomly, denoted as $B_a$ and $K_a$.
  \item Alice generates a string of qubits $q$ with size $n$, and the $i$th qubit in $q$ is one of the six states $\pm x$, $\pm y$ and $\pm z$.
  \item Alice sends $q$ to Bob through a quantum channel $Q$ between Alice and Bob.
  \item Bob receives $q$ and randomly generates a string of bits $B_b$ with size $n$.
  \item Bob measures each qubit of $q$ according to a basis by bits of $B_b$, i.e., $x$, $y$ or $z$ basis. And the measurement results would be $K_b$, which is also with size $n$.
  \item Bob sends his measurement bases $B_b$ to Alice through a public channel $P$.
  \item Once receiving $B_b$, Alice sends her bases $B_a$ to Bob through channel $P$, and Bob receives $B_a$.
  \item Alice and Bob determine that at which position the bit strings $B_a$ and $B_b$ are equal, and they discard the mismatched bits of $B_a$ and $B_b$. Then the remaining bits of $K_a$ and $K_b$, denoted as $K_a'$ and $K_b'$ with $K_{a,b}=K_a'=K_b'$.
\end{enumerate}

We re-introduce the basic SSP protocol in an abstract way with more technical details as Figure \ref{SSP4} illustrates.

Now, we assume a special measurement operation $Rand[q;B_a]$ which create a string of $n$ random bits $B_a$ from the $q$ quantum system, and the same as $Rand[q;K_a]$, $Rand[q';B_b]$. $M[q;K_b]$ denotes the Bob's measurement operation of $q$. The generation of $n$ qubits $q$ through two quantum operations $Set_{K_a}[q]$ and $H_{B_a}[q]$. Alice sends $q$ to Bob through the quantum channel $Q$ by quantum communicating action $send_{Q}(q)$ and Bob receives $q$ through $Q$ by quantum communicating action $receive_{Q}(q)$. Bob sends $B_b$ to Alice through the public channel $P$ by classical communicating action $send_{P}(B_b)$ and Alice receives $B_b$ through channel $P$ by classical communicating action $receive_{P}(B_b)$, and the same as $send_{P}(B_a)$ and $receive_{P}(B_a)$. Alice and Bob generate the private key $K_{a,b}$ by a classical comparison action $cmp(K_{a,b},K_a,K_b,B_a,B_b)$. Let Alice and Bob be a system $AB$ and let interactions between Alice and Bob be internal actions. $AB$ receives external input $D_i$ through channel $A$ by communicating action $receive_A(D_i)$ and sends results $D_o$ through channel $B$ by communicating action $send_B(D_o)$.

Then the state transition of Alice can be described by qACP as follows.

\begin{eqnarray}
&&A=\sum_{D_i\in \Delta_i}receive_A(D_i)\cdot A_1\nonumber\\
&&A_1=Rand[q;B_a]\cdot A_2\nonumber\\
&&A_2=Rand[q;K_a]\cdot A_3\nonumber\\
&&A_3=Set_{K_a}[q]\cdot A_4\nonumber\\
&&A_4=H_{B_a}[q]\cdot A_5\nonumber\\
&&A_5=send_Q(q)\cdot A_6\nonumber\\
&&A_6=receive_P(B_b)\cdot A_7\nonumber\\
&&A_7=send_P(B_a)\cdot A_8\nonumber\\
&&A_8=cmp(K_{a,b},K_a,K_b,B_a,B_b)\cdot A\nonumber
\end{eqnarray}

where $\Delta_i$ is the collection of the input data.

And the state transition of Bob can be described by qACP as follows.

\begin{eqnarray}
&&B=receive_Q(q)\cdot B_1\nonumber\\
&&B_1=Rand[q';B_b]\cdot B_2\nonumber\\
&&B_2=M[q;K_b]\cdot B_3\nonumber\\
&&B_3=send_P(B_b)\cdot B_4\nonumber\\
&&B_4=receive_P(B_a)\cdot B_5\nonumber\\
&&B_5=cmp(K_{a,b},K_a,K_b,B_a,B_b)\cdot B_6\nonumber\\
&&B_6=\sum_{D_o\in\Delta_o}send_B(D_o)\cdot B\nonumber
\end{eqnarray}

where $\Delta_o$ is the collection of the output data.

The send action and receive action of the same data through the same channel can communicate each other, otherwise, a deadlock $\delta$ will be caused. We define the following communication functions.

\begin{eqnarray}
&&\gamma(send_Q(q),receive_Q(q))\triangleq c_Q(q)\nonumber\\
&&\gamma(send_P(B_b),receive_P(B_b))\triangleq c_P(B_b)\nonumber\\
&&\gamma(send_P(B_a),receive_P(B_a))\triangleq c_P(B_a)\nonumber
\end{eqnarray}

Let $A$ and $B$ in parallel, then the system $AB$ can be represented by the following process term.

$$\tau_I(\partial_H(\Theta(A\between B)))$$

where $H=\{send_Q(q),receive_Q(q),send_P(B_b),receive_P(B_b),send_P(B_a),receive_P(B_a)\}$ and $I=\{Rand[q;B_a], Rand[q;K_a], Set_{K_a}[q], H_{B_a}[q], Rand[q';B_b], M[q;K_b], c_Q(q), c_P(B_b),\\ c_P(B_a), cmp(K_{a,b},K_a,K_b,B_a,B_b)\}$.

Then we get the following conclusion.

\begin{theorem}
The basic SSP protocol $\tau_I(\partial_H(\Theta(A\between B)))$ exhibits desired external behaviors.
\end{theorem}

\begin{proof}
We can get $\tau_I(\partial_H(\Theta(A\between B)))=\sum_{D_i\in \Delta_i}\sum_{D_o\in\Delta_o}receive_A(D_i)\leftmerge send_B(D_o)\leftmerge \tau_I(\partial_H(\Theta(A\between B)))$. 
So, the basic SSP protocol $\tau_I(\partial_H(\Theta(A\between B)))$ exhibits desired external behaviors.
\end{proof}

\subsection{Verification of S09 Protocol}\label{VS094}

The famous S09 protocol\cite{S09} is a quantum key distribution protocol, in which quantum information and classical information are mixed. We take an example of the S09 protocol to illustrate the usage of qACP in verification of quantum protocols.

\begin{figure}
  \centering
  \includegraphics{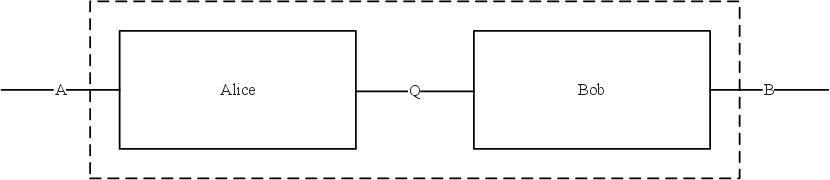}
  \caption{The S09 protocol.}
  \label{S094}
\end{figure}

The S09 protocol is used to create a private key between two parities, Alice and Bob, by use of pure quantum information. Firstly, we introduce the basic S09 protocol briefly, which is illustrated in Figure \ref{S094}.

\begin{enumerate}
  \item Alice create two string of bits with size $n$ randomly, denoted as $B_a$ and $K_a$.
  \item Alice generates a string of qubits $q$ with size $n$, and the $i$th qubit in $q$ is $|x_y\rangle$, where $x$ is the $i$th bit of $B_a$ and $y$ is the $i$th bit of $K_a$.
  \item Alice sends $q$ to Bob through a quantum channel $Q$ between Alice and Bob.
  \item Bob receives $q$ and randomly generates a string of bits $B_b$ with size $n$.
  \item Bob measures each qubit of $q$ according to a basis by bits of $B_b$. After the measurement, the state of $q$ evolves into $q'$.
  \item Bob sends $q'$ to Alice through the quantum channel $Q$.
  \item Alice measures each qubit of $q'$ to generate a string $C$.
  \item Alice sums $C_i\oplus B_{a_i}$ to get the private key $K_{a,b}=B_b$.
\end{enumerate}

We re-introduce the basic S09 protocol in an abstract way with more technical details as Figure \ref{S094} illustrates.

Now, we assume a special measurement operation $Rand[q;B_a]$ which create a string of $n$ random bits $B_a$ from the $q$ quantum system, and the same as $Rand[q;K_a]$, $Rand[q';B_b]$. $M[q;B_b]$ denotes the Bob's measurement operation of $q$, and the same as $M[q';C]$. The generation of $n$ qubits $q$ through two quantum operations $Set_{K_a}[q]$ and $H_{B_a}[q]$. Alice sends $q$ to Bob through the quantum channel $Q$ by quantum communicating action $send_{Q}(q)$ and Bob receives $q$ through $Q$ by quantum communicating action $receive_{Q}(q)$, and the same as $send_{Q}(q')$ and $receive_{Q}(q')$. Alice and Bob generate the private key $K_{a,b}$ by a classical comparison action $cmp(K_{a,b},B_b)$. We omit the sum classical $\oplus$ actions without of loss of generality. Let Alice and Bob be a system $AB$ and let interactions between Alice and Bob be internal actions. $AB$ receives external input $D_i$ through channel $A$ by communicating action $receive_A(D_i)$ and sends results $D_o$ through channel $B$ by communicating action $send_B(D_o)$.

Then the state transition of Alice can be described by qACP as follows.

\begin{eqnarray}
&&A=\sum_{D_i\in \Delta_i}receive_A(D_i)\cdot A_1\nonumber\\
&&A_1=Rand[q;B_a]\cdot A_2\nonumber\\
&&A_2=Rand[q;K_a]\cdot A_3\nonumber\\
&&A_3=Set_{K_a}[q]\cdot A_4\nonumber\\
&&A_4=H_{B_a}[q]\cdot A_5\nonumber\\
&&A_5=send_Q(q)\cdot A_6\nonumber\\
&&A_6=receive_Q(q')\cdot A_{7}\nonumber\\
&&A_7=M[q';C]\cdot A_8\nonumber\\
&&A_{8}=cmp(K_{a,b},B_b)\cdot A\nonumber
\end{eqnarray}

where $\Delta_i$ is the collection of the input data.

And the state transition of Bob can be described by qACP as follows.

\begin{eqnarray}
&&B=receive_Q(q)\cdot B_1\nonumber\\
&&B_1=Rand[q';B_b]\cdot B_2\nonumber\\
&&B_2=M[q;B_b]\cdot B_3\nonumber\\
&&B_3=send_Q(q')\cdot B_4\nonumber\\
&&B_4=cmp(K_{a,b},B_b)\cdot B_{5}\nonumber\\
&&B_{5}=\sum_{D_o\in\Delta_o}send_B(D_o)\cdot B\nonumber
\end{eqnarray}

where $\Delta_o$ is the collection of the output data.

The send action and receive action of the same data through the same channel can communicate each other, otherwise, a deadlock $\delta$ will be caused. We define the following communication functions.

\begin{eqnarray}
&&\gamma(send_Q(q),receive_Q(q))\triangleq c_Q(q)\nonumber\\
&&\gamma(send_Q(q'),receive_Q(q'))\triangleq c_Q(q')\nonumber
\end{eqnarray}

Let $A$ and $B$ in parallel, then the system $AB$ can be represented by the following process term.

$$\tau_I(\partial_H(\Theta(A\between B)))$$

where $H=\{send_Q(q),receive_Q(q),send_Q(q'),receive_Q(q')\}$ and $I=\{Rand[q;B_a], Rand[q;K_a], Set_{K_a}[q], \\ H_{B_a}[q], Rand[q';B_b], M[q;K_b], M[q';C], c_Q(q), c_Q(q'), cmp(K_{a,b},B_b)\}$.

Then we get the following conclusion.

\begin{theorem}
The basic S09 protocol $\tau_I(\partial_H(\Theta(A\between B)))$ exhibits desired external behaviors.
\end{theorem}

\begin{proof}
We can get $\tau_I(\partial_H(\Theta(A\between B)))=\sum_{D_i\in \Delta_i}\sum_{D_o\in\Delta_o}receive_A(D_i)\leftmerge send_B(D_o)\leftmerge \tau_I(\partial_H(\Theta(A\between B)))$. 
So, the basic S09 protocol $\tau_I(\partial_H(\Theta(A\between B)))$ exhibits desired external behaviors.
\end{proof}

\subsection{Verification of KMB09 Protocol}\label{VKMB094}

The famous KMB09 protocol\cite{KMB09} is a quantum key distribution protocol, in which quantum information and classical information are mixed. We take an example of the KMB09 protocol to illustrate the usage of qACP in verification of quantum protocols.

\begin{figure}
  \centering
  \includegraphics{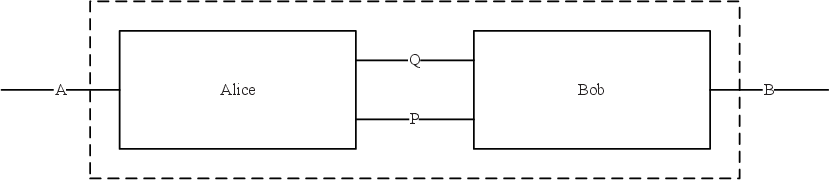}
  \caption{The KMB09 protocol.}
  \label{KMB094}
\end{figure}

The KMB09 protocol is used to create a private key between two parities, Alice and Bob. KMB09 is a protocol of quantum key distribution (QKD) which refines the BB84 protocol against PNS (Photon Number Splitting) attacks. The main innovations are encoding bits in nonorthogonal states and the classical sifting procedure. Firstly, we introduce the basic KMB09 protocol briefly, which is illustrated in Figure \ref{KMB094}.

\begin{enumerate}
  \item Alice create a string of bits with size $n$ randomly, denoted as $K_a$, and randomly assigns each bit value a random index $i=1,2,...,N$ into $B_a$.
  \item Alice generates a string of qubits $q$ with size $n$, accordingly either in $|e_i\rangle$ or $|f_i\rangle$.
  \item Alice sends $q$ to Bob through a quantum channel $Q$ between Alice and Bob.
  \item Alice sends $B_a$ through a public channel $P$.
  \item Bob measures each qubit of $q$ by randomly switching the measurement basis between $e$ and $f$. And he records the unambiguous discriminations into $K_b$, and the unambiguous discrimination information into $B_b$.
  \item Bob sends $B_b$ to Alice through the public channel $P$.
  \item Alice and Bob determine that at which position the bit should be remained. Then the remaining bits of $K_a$ and $K_b$ is the private key $K_{a,b}$.
\end{enumerate}

We re-introduce the basic KMB09 protocol in an abstract way with more technical details as Figure \ref{KMB094} illustrates.

Now, we assume a special measurement operation $Rand[q;K_a]$ which create a string of $n$ random bits $K_a$ from the $q$ quantum system. $M[q;K_b]$ denotes the Bob's measurement operation of $q$. The generation of $n$ qubits $q$ through a quantum operation $Set_{K_a}[q]$. Alice sends $q$ to Bob through the quantum channel $Q$ by quantum communicating action $send_{Q}(q)$ and Bob receives $q$ through $Q$ by quantum communicating action $receive_{Q}(q)$. Bob sends $B_b$ to Alice through the public channel $P$ by classical communicating action $send_{P}(B_b)$ and Alice receives $B_b$ through channel $P$ by classical communicating action $receive_{P}(B_b)$, and the same as $send_{P}(B_a)$ and $receive_{P}(B_a)$. Alice and Bob generate the private key $K_{a,b}$ by a classical comparison action $cmp(K_{a,b},K_a,K_b,B_a,B_b)$. Let Alice and Bob be a system $AB$ and let interactions between Alice and Bob be internal actions. $AB$ receives external input $D_i$ through channel $A$ by communicating action $receive_A(D_i)$ and sends results $D_o$ through channel $B$ by communicating action $send_B(D_o)$.

Then the state transition of Alice can be described by qACP as follows.

\begin{eqnarray}
&&A=\sum_{D_i\in \Delta_i}receive_A(D_i)\cdot A_1\nonumber\\
&&A_1=Rand[q;K_a]\cdot A_2\nonumber\\
&&A_2=Set_{K_a}[q]\cdot A_3\nonumber\\
&&A_3=send_Q(q)\cdot A_4\nonumber\\
&&A_4=send_P(B_a)\cdot A_5\nonumber\\
&&A_5=receive_P(B_b)\cdot A_6\nonumber\\
&&A_6=cmp(K_{a,b},K_a,K_b,B_a,B_b)\cdot A\nonumber
\end{eqnarray}

where $\Delta_i$ is the collection of the input data.

And the state transition of Bob can be described by qACP as follows.

\begin{eqnarray}
&&B=receive_Q(q)\cdot B_1\nonumber\\
&&B_1=receive_P(B_a)\cdot B_2\nonumber\\
&&B_2=M[q;K_b]\cdot B_3\nonumber\\
&&B_3=send_P(B_b)\cdot B_4\nonumber\\
&&B_4=cmp(K_{a,b},K_a,K_b,B_a,B_b)\cdot B_5\nonumber\\
&&B_5=\sum_{D_o\in\Delta_o}send_B(D_o)\cdot B\nonumber
\end{eqnarray}

where $\Delta_o$ is the collection of the output data.

The send action and receive action of the same data through the same channel can communicate each other, otherwise, a deadlock $\delta$ will be caused. We define the following communication functions.

\begin{eqnarray}
&&\gamma(send_Q(q),receive_Q(q))\triangleq c_Q(q)\nonumber\\
&&\gamma(send_P(B_b),receive_P(B_b))\triangleq c_P(B_b)\nonumber\\
&&\gamma(send_P(B_a),receive_P(B_a))\triangleq c_P(B_a)\nonumber
\end{eqnarray}

Let $A$ and $B$ in parallel, then the system $AB$ can be represented by the following process term.

$$\tau_I(\partial_H(\Theta(A\between B)))$$

where $H=\{send_Q(q),receive_Q(q),send_P(B_b),receive_P(B_b),send_P(B_a),receive_P(B_a)\}$ and $I=\{Rand[q;K_a], Set_{K_a}[q], M[q;K_b], c_Q(q), c_P(B_b),\\ c_P(B_a), cmp(K_{a,b},K_a,K_b,B_a,B_b)\}$.

Then we get the following conclusion.

\begin{theorem}
The basic KMB09 protocol $\tau_I(\partial_H(\Theta(A\between B)))$ exhibits desired external behaviors.
\end{theorem}

\begin{proof}
We can get $\tau_I(\partial_H(\Theta(A\between B)))=\sum_{D_i\in \Delta_i}\sum_{D_o\in\Delta_o}receive_A(D_i)\leftmerge send_B(D_o)\leftmerge \tau_I(\partial_H(\Theta(A\between B)))$. 
So, the basic KMB09 protocol $\tau_I(\partial_H(\Theta(A\between B)))$ exhibits desired external behaviors.
\end{proof}

\subsection{Verification of S13 Protocol}\label{VS134}

The famous S13 protocol\cite{S13} is a quantum key distribution protocol, in which quantum information and classical information are mixed. We take an example of the S13 protocol to illustrate the usage of qACP in verification of quantum protocols.

\begin{figure}
  \centering
  \includegraphics{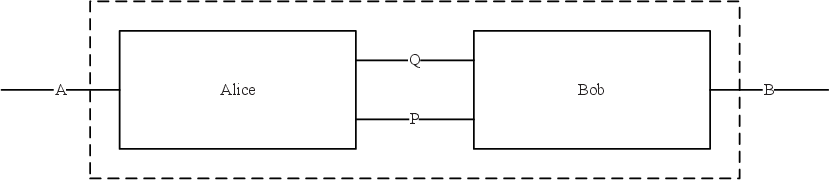}
  \caption{The S13 protocol.}
  \label{S134}
\end{figure}

The S13 protocol is used to create a private key between two parities, Alice and Bob. Firstly, we introduce the basic S13 protocol briefly, which is illustrated in Figure \ref{S134}.

\begin{enumerate}
  \item Alice create two string of bits with size $n$ randomly, denoted as $B_a$ and $K_a$.
  \item Alice generates a string of qubits $q$ with size $n$, and the $i$th qubit in $q$ is $|x_y\rangle$, where $x$ is the $i$th bit of $B_a$ and $y$ is the $i$th bit of $K_a$.
  \item Alice sends $q$ to Bob through a quantum channel $Q$ between Alice and Bob.
  \item Bob receives $q$ and randomly generates a string of bits $B_b$ with size $n$.
  \item Bob measures each qubit of $q$ according to a basis by bits of $B_b$. And the measurement results would be $K_b$, which is also with size $n$.
  \item Alice sends a random binary string $C$ to Bob through the public channel $P$.
  \item Alice sums $B_{a_i}\oplus C_i$ to obtain $T$ and generates other random string of binary values $J$. From the elements occupying a concrete position, $i$, of the preceding strings, Alice get the new states of $q'$, and sends it to Bob through the quantum channel $Q$.
  \item Bob sums $1\oplus B_{b_i}$ to obtain the string of binary basis $N$ and measures $q'$ according to these bases, and generating $D$.
  \item Alice sums $K_{a_i}\oplus J_i$ to obtain the binary string $Y$ and sends it to Bob through the public channel $P$.
  \item Bob encrypts $B_b$ to obtain $U$ and sends to Alice through the public channel $P$.
  \item Alice decrypts $U$ to obtain $B_b$. She sums $B_{a_i}\oplus B_{b_i}$ to obtain $L$ and sends $L$ to Bob through the public channel $P$.
  \item Bob sums $B_{b_i}\oplus L_i$ to get the private key $K_{a,b}$.
\end{enumerate}

We re-introduce the basic S13 protocol in an abstract way with more technical details as Figure \ref{S134} illustrates.

Now, we assume a special measurement operation $Rand[q;B_a]$ which create a string of $n$ random bits $B_a$ from the $q$ quantum system, and the same as $Rand[q;K_a]$, $Rand[q';B_b]$. $M[q;K_b]$ denotes the Bob's measurement operation of $q$, and the same as $M[q';D]$. The generation of $n$ qubits $q$ through two quantum operations $Set_{K_a}[q]$ and $H_{B_a}[q]$, and the same as $Set_{T}[q']$. Alice sends $q$ to Bob through the quantum channel $Q$ by quantum communicating action $send_{Q}(q)$ and Bob receives $q$ through $Q$ by quantum communicating action $receive_{Q}(q)$, and the same as $send_{Q}(q')$ and $receive_{Q}(q')$. Bob sends $B_b$ to Alice through the public channel $P$ by classical communicating action $send_{P}(B_b)$ and Alice receives $B_b$ through channel $P$ by classical communicating action $receive_{P}(B_b)$, and the same as $send_{P}(B_a)$ and $receive_{P}(B_a)$, $send_{P}(C)$ and $receive_{P}(C)$, $send_{P}(Y)$ and $receive_{P}(Y)$, $send_{P}(U)$ and $receive_{P}(U)$, $send_{P}(L)$ and $receive_{P}(L)$. Alice and Bob generate the private key $K_{a,b}$ by a classical comparison action $cmp(K_{a,b},K_a,K_b,B_a,B_b)$. We omit the sum classical $\oplus$ actions without of loss of generality. Let Alice and Bob be a system $AB$ and let interactions between Alice and Bob be internal actions. $AB$ receives external input $D_i$ through channel $A$ by communicating action $receive_A(D_i)$ and sends results $D_o$ through channel $B$ by communicating action $send_B(D_o)$.

Then the state transition of Alice can be described by qACP as follows.

\begin{eqnarray}
&&A=\sum_{D_i\in \Delta_i}receive_A(D_i)\cdot A_1\nonumber\\
&&A_1=Rand[q;B_a]\cdot A_2\nonumber\\
&&A_2=Rand[q;K_a]\cdot A_3\nonumber\\
&&A_3=Set_{K_a}[q]\cdot A_4\nonumber\\
&&A_4=H_{B_a}[q]\cdot A_5\nonumber\\
&&A_5=send_Q(q)\cdot A_6\nonumber\\
&&A_6=send_P(C)\cdot A_7\nonumber\\
&&A_7=send_Q(q')\cdot A_8\nonumber\\
&&A_8=send_P(Y)\cdot A_9\nonumber\\
&&A_9=receive_P(U)\cdot A_{10}\nonumber\\
&&A_{10}=send_P(L)\cdot A_{11}\nonumber\\
&&A_{11}=cmp(K_{a,b},K_a,K_b,B_a,B_b)\cdot A\nonumber
\end{eqnarray}

where $\Delta_i$ is the collection of the input data.

And the state transition of Bob can be described by qACP as follows.

\begin{eqnarray}
&&B=receive_Q(q)\cdot B_1\nonumber\\
&&B_1=Rand[q';B_b]\cdot B_2\nonumber\\
&&B_2=M[q;K_b]\cdot B_3\nonumber\\
&&B_3=receive_P(C)\cdot B_4\nonumber\\
&&B_4=receive_Q(q')\cdot B_5\nonumber\\
&&B_5=M[q';D]\cdot B_6\nonumber\\
&&B_6=receive_P(Y)\cdot B_7\nonumber\\
&&B_7=send_P(U)\cdot B_8\nonumber\\
&&B_8=receive_P(L)\cdot B_9\nonumber\\
&&B_9=cmp(K_{a,b},K_a,K_b,B_a,B_b)\cdot B_{10}\nonumber\\
&&B_{10}=\sum_{D_o\in\Delta_o}send_B(D_o)\cdot B\nonumber
\end{eqnarray}

where $\Delta_o$ is the collection of the output data.

The send action and receive action of the same data through the same channel can communicate each other, otherwise, a deadlock $\delta$ will be caused. We define the following communication functions.

\begin{eqnarray}
&&\gamma(send_Q(q),receive_Q(q))\triangleq c_Q(q)\nonumber\\
&&\gamma(send_Q(q'),receive_Q(q'))\triangleq c_Q(q')\nonumber\\
&&\gamma(send_P(C),receive_P(C))\triangleq c_P(C)\nonumber\\
&&\gamma(send_P(Y),receive_P(Y))\triangleq c_P(Y)\nonumber\\
&&\gamma(send_P(U),receive_P(U))\triangleq c_P(U)\nonumber\\
&&\gamma(send_P(L),receive_P(L))\triangleq c_P(L)\nonumber
\end{eqnarray}

Let $A$ and $B$ in parallel, then the system $AB$ can be represented by the following process term.

$$\tau_I(\partial_H(\Theta(A\between B)))$$

where $H=\{send_Q(q),receive_Q(q),send_Q(q'),receive_Q(q'),send_P(C),\\receive_P(C),send_P(Y),receive_P(Y),send_P(U),receive_P(U),send_P(L),receive_P(L)\}$ and $I=\{Rand[q;B_a], Rand[q;K_a],\\ Set_{K_a}[q], H_{B_a}[q], Rand[q';B_b], M[q;K_b], M[q';D], c_Q(q), c_P(C),\\c_Q(q'), c_P(Y), c_P(U), c_P(L), cmp(K_{a,b},K_a,K_b,B_a,B_b)\}$.

Then we get the following conclusion.

\begin{theorem}
The basic S13 protocol $\tau_I(\partial_H(\Theta(A\between B)))$ exhibits desired external behaviors.
\end{theorem}

\begin{proof}
We can get $\tau_I(\partial_H(\Theta(A\between B)))=\sum_{D_i\in \Delta_i}\sum_{D_o\in\Delta_o}receive_A(D_i)\leftmerge send_B(D_o)\leftmerge \tau_I(\partial_H(\Theta(A\between B)))$. 
So, the basic S13 protocol $\tau_I(\partial_H(\Theta(A\between B)))$ exhibits desired external behaviors.
\end{proof}

\newpage\section{Algebra for Parallelism in Reversible Probabilistic True Concurrency}\label{aprptc2}

The theory $APRPTC$ has four modules: $BARPTC$ , $APRPTC$, recursion and abstraction.

This chapter is organized as follows. We introduce $BARPTC$ in section \ref{barptc}, $APRPTC$ in section \ref{aprptc}, recursion in section \ref{rprec}, and abstraction in section
\ref{rpabs}.

\subsection{Basic Algebra for Reversible Probabilistic True Concurrency}{\label{barptc}}

In this section, we will discuss the algebraic laws for prime event structure $\mathcal{E}$ under the reversible computational framework, exactly for causality $\leq$, conflict $\sharp$ and probabilistic conflict $\sharp_{\pi}$.
We will follow the conventions of process algebra, using $\cdot$ instead of $\leq$, $+$ instead of $\sharp$ and $\boxplus_{\pi}$ instead of $\sharp_{\pi}$. The resulted algebra is called
Basic Algebra for Reversible Probabilistic True Concurrency, abbreviated $BARPTC$.

\subsubsection{Axiom System of $BARPTC$}

In the following, let $e_1, e_2, e_1', e_2'\in \mathbb{E}$, and let variables $x,y,z$ range over the set of terms for true concurrency, $p,q,s$ range over the set of closed terms. The
set of axioms of $BARPTC$ consists of the laws given in Table \ref{AxiomsForBARPTC}.

\begin{center}
    \begin{table}
        \begin{tabular}{@{}ll@{}}
            \hline No. &Axiom\\
            $A1$ & $x+ y = y+ x$\\
            $A2$ & $(x+ y)+ z = x+ (y+ z)$\\
            $A3$ & $e+ e = e$\\
            $A41$ & $(x+ y)\cdot z = x\cdot z + y\cdot z\quad (Std(x),Std(y), Std(z))$\\
            $A42$ & $x\cdot (y+z) = x\cdot y + x\cdot z\quad (NStd(x),NStd(y), NStd(z))$\\
            $A5$ & $(x\cdot y)\cdot z = x\cdot(y\cdot z)$\\
            $A6$ & $x+ \delta = x$\\
            $A7$ & $\delta\cdot x =\delta({Std(x)})$\\
            $RA7$ & $x\cdot\delta =\delta({NStd(x)})$\\
            $PA1$ & $x\boxplus_{\pi} y=y\boxplus_{1-\pi} x$\\
            $PA2$ & $x\boxplus_{\pi}(y\boxplus_{\rho} z)=(x\boxplus_{\frac{\pi}{\pi+\rho-\pi\rho}}y)\boxplus_{\pi+\rho-\pi\rho} z$\\
            $PA3$ & $x\boxplus_{\pi}x=x$\\
            $PA41$ & $(x\boxplus_{\pi}y)\cdot z=x\cdot z\boxplus_{\pi}y\cdot z\quad (Std(x),Std(y), Std(z))$\\
            $PA42$ & $x\cdot (y\boxplus_{\pi} z)=x\cdot y\boxplus_{\pi}x\cdot z\quad (NStd(x),NStd(y), NStd(z))$\\
            $PA5$ & $(x\boxplus_{\pi}y)+z=(x+z)\boxplus_{\pi}(y+z)$\\
        \end{tabular}
        \caption{Axioms of $BARPTC$}
        \label{AxiomsForBARPTC}
    \end{table}
\end{center}


\subsubsection{Properties of $BARPTC$}

\begin{definition}[Basic terms of $BARPTC$]\label{BTBARPTC}
The set of basic terms of $BARPTC$, $\mathcal{B}(BARPTC)$, is inductively defined as follows:

\begin{enumerate}
  \item $\mathbb{E}\subset\mathcal{B}(BARPTC)$;
  \item if $e\in \mathbb{E}, t\in\mathcal{B}(BARPTC)$ then $e\cdot t\in\mathcal{B}(BARPTC)$;
  \item if $e[m]\in \mathbb{E}, t\in\mathcal{B}(BARPTC)$ then $t\cdot e[m]\in\mathcal{B}(BARPTC)$;
  \item if $t,s\in\mathcal{B}(BARPTC)$ then $t+ s\in\mathcal{B}(BARPTC)$;
  \item if $t,s\in\mathcal{B}(BARPTC)$ then $t\boxplus_{\pi} s\in\mathcal{B}(BARPTC)$.
\end{enumerate}
\end{definition}

\begin{theorem}[Elimination theorem of $BARPTC$]\label{ETBARPTC}
Let $p$ be a closed $BARPTC$ term. Then there is a basic $BARPTC$ term $q$ such that $BARPTC\vdash p=q$.
\end{theorem}

\begin{proof}
(1) Firstly, suppose that the following ordering on the signature of $BARPTC$ is defined: $\cdot > +>\boxplus_{\pi}$ and the symbol $\cdot$ is given the lexicographical status for the
first argument, then for each rewrite rule $p\rightarrow q$ in Table \ref{TRSForBARPTC} relation $p>_{lpo} q$ can easily be proved. We obtain that the term rewrite system shown in
Table \ref{TRSForBARPTC} is strongly normalizing, for it has finitely many rewriting rules, and $>$ is a well-founded ordering on the signature of $BARPTC$, and if $s>_{lpo} t$,
for each rewriting rule $s\rightarrow t$ is in Table \ref{TRSForBARPTC} (see Theorem \ref{SN}).

\begin{center}
    \begin{table}
        \begin{tabular}{@{}ll@{}}
            \hline No. &Rewriting Rule\\
            $RA3$ & $x+ x \rightarrow x$\\
            $RA41$ & $(x+ y)\cdot z \rightarrow x\cdot z + y\cdot z\quad (Std(x),Std(y), Std(z))$\\
            $RA42$ & $x\cdot (y+z) \rightarrow x\cdot y + x\cdot z\quad (NStd(x),NStd(y), NStd(z))$\\
            $RA5$ & $(x\cdot y)\cdot z \rightarrow x\cdot(y\cdot z)$\\
            $RA6$ & $x+ \delta \rightarrow x$\\
            $RA7$ & $\delta\cdot x \rightarrow\delta({Std(x)})$\\
            $RRA7$ & $x\cdot\delta \rightarrow\delta({NStd(x)})$\\
            $RPA1$ & $x\boxplus_{\pi} y\rightarrow y\boxplus_{1-\pi} x$\\
            $RPA2$ & $x\boxplus_{\pi}(y\boxplus_{\rho} z)\rightarrow (x\boxplus_{\frac{\pi}{\pi+\rho-\pi\rho}}y)\boxplus_{\pi+\rho-\pi\rho} z$\\
            $RPA3$ & $x\boxplus_{\pi}x\rightarrow x$\\
            $RPA41$ & $(x\boxplus_{\pi}y)\cdot z\rightarrow x\cdot z\boxplus_{\pi}y\cdot z\quad (Std(x),Std(y), Std(z))$\\
            $RPA42$ & $x\cdot (y\boxplus_{\pi} z)\rightarrow x\cdot y\boxplus_{\pi}x\cdot z\quad (NStd(x),NStd(y), NStd(z))$\\
            $RPA5$ & $(x\boxplus_{\pi}y)+z\rightarrow (x+z)\boxplus_{\pi}(y+z)$\\
        \end{tabular}
        \caption{Term rewrite system of $BARPTC$}
        \label{TRSForBARPTC}
    \end{table}
\end{center}

(2) Then we prove that the normal forms of closed $BARPTC$ terms are basic $BARPTC$ terms.

Suppose that $p$ is a normal form of some closed $BARPTC$ term and suppose that $p$ is not a basic term. Let $p'$ denote the smallest sub-term of $p$ which is not a basic term. It
implies that each sub-term of $p'$ is a basic term. Then we prove that $p$ is not a term in normal form. It is sufficient to induct on the structure of $p'$:

\begin{itemize}
  \item Case $p'\equiv e, e\in \mathbb{E}$. $p'$ is a basic term, which contradicts the assumption that $p'$ is not a basic term, so this case should not occur.
  \item Case $p'\equiv e[m], e[m]\in \mathbb{E}$. $p'$ is a basic term, which contradicts the assumption that $p'$ is not a basic term, so this case should not occur.
  \item Case $p'\equiv p_1\cdot p_2$. By induction on the structure of the basic term $p_1$:
      \begin{itemize}
        \item Subcase $p_1\in \mathbb{E}$. $p'$ would be a basic term, which contradicts the assumption that $p'$ is not a basic term;
        \item Subcase $p_1\equiv e\cdot p_1'$. $RA5$ rewriting rule can be applied. So $p$ is not a normal form;
        \item Subcase $p_1\equiv p_1'\cdot e[m]$. $RA5$ rewriting rule can be applied. So $p$ is not a normal form;
        \item Subcase $p_1\equiv p_1'+ p_1''$. $RA41$ and $RA42$ rewriting rules can be applied. So $p$ is not a normal form;
        \item Subcase $p_1\equiv p_1'\boxplus_{\pi} p_1''$. $RPA41$ and $RPA42$ rewriting rules can be applied. So $p$ is not a normal form.
      \end{itemize}
  \item Case $p'\equiv p_1+ p_2$. By induction on the structure of the basic terms both $p_1$ and $p_2$, all subcases will lead to that $p'$ would be a basic term, which contradicts
  the assumption that $p'$ is not a basic term;
  \item Case $p'\equiv p_1\boxplus_{\pi} p_2$. By induction on the structure of the basic terms both $p_1$ and $p_2$, all subcases will lead to that $p'$ would be a basic term, which contradicts
  the assumption that $p'$ is not a basic term.
\end{itemize}
\end{proof}

\subsubsection{Structured Operational Semantics of $BARPTC$}

In this subsection, we will define a term-deduction system which gives the operational semantics of $BARPTC$. Like the way in \cite{PPA}, we also introduce the counterpart $\breve{e}$
of the event $e$, and also the set $\breve{\mathbb{E}}=\{\breve{e}|e\in\mathbb{E}\}$.

Firstly, we give the definition of PDFs in Table \ref{PDFBARPTC}.

\begin{center}
    \begin{table}
        $$\mu(e,\breve{e})=1$$
        $$\mu(x\cdot y, x'\cdot y)=\mu(x,x')$$
        $$\mu(x+y,x'+y')=\mu(x,x')\cdot \mu(y,y')$$
        $$\mu(x\boxplus_{\pi}y,z)=\pi\mu(x,z)+(1-\pi)\mu(y,z)$$
        $$\mu(x,y)=0,\textrm{otherwise}$$
        \caption{PDF definitions of $BARPTC$}
        \label{PDFBARPTC}
    \end{table}
\end{center}

We give the operational transition rules for operators $\cdot$, $+$ and
$\boxplus_{\pi}$ as Table \ref{SETRForBARPTC} shows. And the predicate $\xrightarrow{e}\surd$ represents successful termination after execution of the event $e$.

\begin{center}
    \begin{table}
        $$\frac{}{e\rightsquigarrow\breve{e}}$$
        $$\frac{x\rightsquigarrow x'}{x\cdot y\rightsquigarrow x'\cdot y}$$
        $$\frac{y\rightsquigarrow y'}{x\cdot y\rightsquigarrow x\cdot y'}$$
        $$\frac{x\rightsquigarrow x'\quad y\rightsquigarrow y'}{x+y\rightsquigarrow x'+y'}$$
        $$\frac{x\rightsquigarrow x'}{x\boxplus_{\pi}y\rightsquigarrow x'}\quad \frac{y\rightsquigarrow y'}{x\boxplus_{\pi}y\rightsquigarrow y'}$$
        $$\frac{}{\breve{e}\xrightarrow{e}\breve{e}[m]}$$
        $$\frac{x\xrightarrow{e}\breve{e}[m]}{x+ y\xrightarrow{e}\breve{e}[m]} \quad\frac{x\xrightarrow{e}x'}{x+ y\xrightarrow{e}x'} \quad\frac{y\xrightarrow{e}\breve{e}[m]}{x+ y\xrightarrow{e}\breve{e}[m]} \quad\frac{y\xrightarrow{e}y'}{x+ y\xrightarrow{e}y'}$$
        $$\frac{x\xrightarrow{e}\breve{e}[m]}{x\cdot y\xrightarrow{e} \breve{e}[m]\cdot y} \quad\frac{x\xrightarrow{e}x'}{x\cdot y\xrightarrow{e}x'\cdot y}$$

        $$\frac{}{\breve{e}[m]\xtworightarrow{{e}[m]}\breve{e}}$$
        $$\frac{x\xtworightarrow{{e}[m]}\breve{e}}{x+ y\xtworightarrow{{e}[m]}\breve{e}} \quad\frac{x\xtworightarrow{{e}[m]}x'}{x+ y\xtworightarrow{{e}[m]}x'} \quad\frac{y\xtworightarrow{{e}[m]}\breve{e}}{x+ y\xtworightarrow{{e}[m]}\breve{e}} \quad\frac{y\xtworightarrow{{e}[m]}y'}{x+ y\xtworightarrow{{e}[m]}y'}$$
        $$\frac{y\xtworightarrow{{e}[m]}\breve{e}}{x\cdot y\xtworightarrow{{e}[m]} x\cdot \breve{e}} \quad\frac{y\xtworightarrow{{e}[m]}y'}{x\cdot y\xtworightarrow{{e}[m]}x\cdot y'}$$
        \caption{Single event transition rules of $BARPTC$}
        \label{SETRForBARPTC}
    \end{table}
\end{center}

The pomset transition rules are shown in Table \ref{PTRForBARPTC}, different to single event transition rules in Table \ref{SETRForBARPTC}, the pomset transition rules are labeled by
pomsets, which are defined by causality $\cdot$, conflict $+$ and $\boxplus_{\pi}$.

\begin{center}
    \begin{table}
        $$\frac{}{X\rightsquigarrow\breve{X}}$$
        $$\frac{x\rightsquigarrow x'}{x\cdot y\rightsquigarrow x'\cdot y}$$
        $$\frac{y\rightsquigarrow y'}{x\cdot y\rightsquigarrow x\cdot y'}$$
        $$\frac{x\rightsquigarrow x'\quad y\rightsquigarrow y'}{x+y\rightsquigarrow x'+y'}$$
        $$\frac{x\rightsquigarrow x'}{x\boxplus_{\pi}y\rightsquigarrow x'}\quad \frac{y\rightsquigarrow y'}{x\boxplus_{\pi}y\rightsquigarrow y'}$$
        $$\frac{}{\breve{X}\xrightarrow{X}\breve{X}[\mathcal{K}]}$$
        $$\frac{x\xrightarrow{X}\breve{X}[\mathcal{K}]}{x+ y\xrightarrow{X}\breve{X}[\mathcal{K}]} (X\subseteq x)\quad\frac{x\xrightarrow{X}x'}{x+ y\xrightarrow{X}x'} (X\subseteq x) \quad\frac{y\xrightarrow{Y}\breve{Y}[\mathcal{L}]}{x+ y\xrightarrow{Y}\breve{Y}[\mathcal{L}]} (Y\subseteq y)\quad\frac{y\xrightarrow{Y}y'}{x+ y\xrightarrow{Y}y'}(Y\subseteq y)$$
        $$\frac{x\xrightarrow{X}\breve{X}[\mathcal{K}]}{x\cdot y\xrightarrow{X} \breve{X}[\mathcal{K}]\cdot y} (X\subseteq x)\quad\frac{x\xrightarrow{X}x'}{x\cdot y\xrightarrow{X}x'\cdot y} (X\subseteq x)$$

        $$\frac{}{\breve{X}[\mathcal{K}]\xtworightarrow{{X}[\mathcal{K}]}\breve{X}}$$
        $$\frac{x\xtworightarrow{{X}[\mathcal{K}]}\breve{X}}{x+ y\xtworightarrow{{X}[\mathcal{K}]}\breve{X}} (X\subseteq x)\quad\frac{x\xtworightarrow{{X}[\mathcal{K}]}x'}{x+ y\xtworightarrow{{X}[\mathcal{K}]}x'} (X\subseteq x) \quad\frac{y\xtworightarrow{{Y}[\mathcal{L}]}\breve{Y}}{x+ y\xtworightarrow{{Y}[\mathcal{L}]}\breve{Y}} (Y\subseteq y)\quad\frac{y\xtworightarrow{{Y}[\mathcal{L}]}y'}{x+ y\xtworightarrow{{Y}[\mathcal{L}]}y'}(Y\subseteq y)$$
        $$\frac{y\xtworightarrow{{Y}[\mathcal{L}]}\breve{Y}}{x\cdot y\xtworightarrow{{Y}[\mathcal{L}]} x\cdot \breve{Y}} (Y\subseteq y)\quad\frac{y\xtworightarrow{{X}[\mathcal{L}]}y'}{x\cdot y\xtworightarrow{{Y}[\mathcal{L}]}x\cdot y'} (Y\subseteq y)$$
        \caption{Pomset transition rules of $BARPTC$}
        \label{PTRForBARPTC}
    \end{table}
\end{center}

\begin{theorem}[Congruence of $BARPTC$ with respect to FR probabilistic pomset bisimulation equivalence]
FR probabilistic pomset bisimulation equivalence $\sim_{pp}^{fr}$ is a congruence with respect to $BARPTC$.
\end{theorem}

\begin{proof}
It is easy to see that FR probabilistic pomset bisimulation is an equivalent relation on $BARPTC$ terms, we only need to prove that $\sim_{pp}^{fr}$ is preserved by the operators $\cdot$, $+$ and $\boxplus_{\pi}$.
That is, if $x\sim_{pp}^{fr} x'$ and $y\sim_{pp}^{fr}y'$, we need to prove that $x\cdot y\sim_{pp}^{fr}x'\cdot y'$, $x+ y\sim_{pp}^{fr}x'+ y'$ and $x\boxplus_{\pi} y\sim_{pp}^{fr}x'\boxplus_{\pi} y'$. The proof
is quite trivial and we omit it.
\end{proof}

\begin{theorem}[Soundness of $BARPTC$ modulo probabilistic pomset bisimulation equivalence]\label{SBARPTCPBE}
Let $x$ and $y$ be $BARPTC$ terms. If $BARPTC\vdash x=y$, then $x\sim_{pp}^{fr} y$.
\end{theorem}

\begin{proof}
Since probabilistic pomset bisimulation $\sim_{pp}^{fr}$ is both an equivalent and a congruent relation, we only need to check if each axiom in Table \ref{AxiomsForBARPTC} is sound modulo
FR probabilistic pomset bisimulation equivalence. The proof is quite trivial and we omit it.
\end{proof}

\begin{theorem}[Completeness of $BARPTC$ modulo FR probabilistic pomset bisimulation equivalence]\label{CBARPTCPBE}
Let $p$ and $q$ be closed $BARPTC$ terms, if $p\sim_{pp}^{fr} q$ then $p=q$.
\end{theorem}

\begin{proof}
Firstly, by the elimination theorem of $BARPTC$, we know that for each closed $BARPTC$ term $p$, there exists a closed basic $BARPTC$ term $p'$, such that $BARPTC\vdash p=p'$, so, we only
need to consider closed basic $BARPTC$ terms.

The basic terms (see Definition \ref{BTBARPTC}) modulo associativity and commutativity (AC) of conflict $+$ (defined by axioms $A1$ and $A2$ in Table \ref{AxiomsForBARPTC}), and this
equivalence is denoted by $=_{AC}$. Then, each equivalence class $s$ modulo AC of $+$ has the following normal form

$$s_1\boxplus_{\pi_1}\cdots\boxplus_{\pi_{k-1}} s_k$$

with each $s_i$ has the following form

$$t_1+\cdots+ t_l$$

with each $t_j$ either an atomic event or of the form $u_1\cdot u_2$, and each $t_j$ is called the summand of $s$.

Now, we prove that for normal forms $n$ and $n'$, if $n\sim_{pp}^{fr} n'$ then $n=_{AC}n'$. It is sufficient to induct on the sizes of $n$ and $n'$.

\begin{itemize}
  \item Consider a summand $e$ of $n$. Then $n\rightsquigarrow\breve{e}\xrightarrow{e}\breve{e}[m]$, so $n\sim_{pp}^{fr} n'$ implies $n'\rightsquigarrow\breve{e}\xrightarrow{e}\breve{e}[m]$, meaning that
  $n'$ also contains the summand $e$.
  \item Consider a summand $e[m]$ of $n$. Then $n\rightsquigarrow\breve{e}[m]\xtworightarrow{e[m]}\breve{e}$, so $n\sim_{pp}^{fr} n'$ implies $n'\rightsquigarrow\breve{e}\xtworightarrow{e[m]}\breve{e}$, meaning that
  $n'$ also contains the summand $e[m]$.
  \item Consider a summand $t_1\cdot t_2$ of $n$. Then $n\rightsquigarrow\breve{t_1}\xrightarrow{t_1}\breve{t_1}[m]\cdot t_2$, so $n\sim_{pp}^{fr} n'$ implies $n'\rightsquigarrow\breve{t_1}\xrightarrow{t_1}\breve{t_1}[m]\cdot t_2'$
  with $t_2\sim_{pp}^{fr} t_2'$, meaning that $n'$ contains a summand $t_1\cdot t_2'$. Since $t_2$ and $t_2'$ are normal forms and have sizes smaller than $n$ and $n'$, by
  the induction hypotheses $t_2\sim_{pp}^{fr} t_2'$ implies $t_2=_{AC} t_2'$.
  \item Consider a summand $t_1\cdot t_2[m]$ of $n$. Then $n\rightsquigarrow\breve{t_2}[m]\xtworightarrow{t_2[m]}t_1\cdot \breve{t_2}$, so $n\sim_{pp}^{fr} n'$ implies $n'\rightsquigarrow\breve{t_2}[m]\xtworightarrow{t_2[m]}t_1'\cdot \breve{t_2}$
  with $t_1\sim_{pp}^{fr} t_1'$, meaning that $n'$ contains a summand $t_1'\cdot t_2[m]$. Since $t_1$ and $t_1'$ are normal forms and have sizes smaller than $n$ and $n'$, by
  the induction hypotheses $t_1\sim_{pp}^{fr} t_1'$ implies $t_1=_{AC} t_1'$.
\end{itemize}

So, we get $n=_{AC} n'$.

Finally, let $s$ and $t$ be basic terms, and $s\sim_{pp}^{fr} t$, there are normal forms $n$ and $n'$, such that $s=n$ and $t=n'$. The soundness theorem of $BARPTC$ modulo FR probabilistic
pomset bisimulation equivalence (see Theorem \ref{SBARPTCPBE}) yields $s\sim_{pp}^{fr} n$ and $t\sim_{pp}^{fr} n'$, so $n\sim_{pp}^{fr} s\sim_{pp}^{fr} t\sim_{pp}^{fr} n'$. Since if $n\sim_{pp}^{fr} n'$ then
$n=_{AC}n'$, $s=n=_{AC}n'=t$, as desired.
\end{proof}

The step transition rules are similar in Table \ref{PTRForBARPTC}, different to pomset transition rules, the step transition rules are labeled by steps, in which every event is pairwise
concurrent.

\begin{theorem}[Congruence of $BARPTC$ with respect to FR probabilistic step bisimulation equivalence]
FR probabilistic probabilistic step bisimulation equivalence $\sim_{ps}^{fr}$ is a congruence with respect to $BARPTC$.
\end{theorem}

\begin{proof}
It is easy to see that FR probabilistic step bisimulation is an equivalent relation on $BARPTC$ terms, we only need to prove that $\sim_{ps}^{fr}$ is preserved by the operators $\cdot$, $+$ and
$\boxplus_{\pi}$.
That is, if $x\sim_{ps}^{fr} x'$ and $y\sim_{ps}^{fr}y'$, we need to prove that $x\cdot y\sim_{ps}^{fr}x'\cdot y'$, $x+ y\sim_{ps}^{fr}x'+ y'$ and $x\boxplus_{\pi} y\sim_{ps}^{fr}x'\boxplus_{\pi} y'$. The proof
is quite trivial and we omit it
\end{proof}

\begin{theorem}[Soundness of $BARPTC$ modulo FR probabilistic step bisimulation equivalence]\label{SBARPTCSBE}
Let $x$ and $y$ be $BARPTC$ terms. If $BARPTC\vdash x=y$, then $x\sim_{ps}^{fr} y$.
\end{theorem}

\begin{proof}
Since FR probabilistic step bisimulation $\sim_{ps}^{fr}$ is both an equivalent and a congruent relation, we only need to check if each axiom in Table \ref{AxiomsForBARPTC} is sound modulo
FR probabilistic step bisimulation equivalence. The proof is quite trivial and we omit it.
\end{proof}

\begin{theorem}[Completeness of $BARPTC$ modulo FR probabilistic step bisimulation equivalence]\label{CBARPTCSBE}
Let $p$ and $q$ be closed $BARPTC$ terms, if $p\sim_{ps}^{fr} q$ then $p=q$.
\end{theorem}

\begin{proof}
Firstly, by the elimination theorem of $BARPTC$, we know that for each closed $BARPTC$ term $p$, there exists a closed basic $BARPTC$ term $p'$, such that $BARPTC\vdash p=p'$, so, we only
need to consider closed basic $BARPTC$ terms.

The basic terms (see Definition \ref{BTBARPTC}) modulo associativity and commutativity (AC) of conflict $+$ (defined by axioms $A1$ and $A2$ in Table \ref{AxiomsForBARPTC}), and this
equivalence is denoted by $=_{AC}$. Then, each equivalence class $s$ modulo AC of $+$ has the following normal form

$$s_1\boxplus_{\pi_1}\cdots\boxplus_{\pi_{k-1}} s_k$$

with each $s_i$ has the following form

$$t_1+\cdots+ t_l$$

with each $t_j$ either an atomic event or of the form $u_1\cdot u_2$, and each $t_j$ is called the summand of $s$.

Now, we prove that for normal forms $n$ and $n'$, if $n\sim_{ps}^{fr} n'$ then $n=_{AC}n'$. It is sufficient to induct on the sizes of $n$ and $n'$.

\begin{itemize}
  \item Consider a summand $e$ of $n$. Then $n\rightsquigarrow\breve{e}\xrightarrow{e}\breve{e}[m]$, so $n\sim_{ps}^{fr} n'$ implies $n'\rightsquigarrow\breve{e}\xrightarrow{e}\breve{e}[m]$, meaning that
  $n'$ also contains the summand $e$.
  \item Consider a summand $e[m]$ of $n$. Then $n\rightsquigarrow\breve{e}[m]\xtworightarrow{e[m]}\breve{e}$, so $n\sim_{ps}^{fr} n'$ implies $n'\rightsquigarrow\breve{e}\xtworightarrow{e[m]}\breve{e}$, meaning that
  $n'$ also contains the summand $e[m]$.
  \item Consider a summand $t_1\cdot t_2$ of $n$. Then $n\rightsquigarrow\breve{t_1}\xrightarrow{t_1}\breve{t_1}[m]\cdot t_2$, so $n\sim_{ps}^{fr} n'$ implies $n'\rightsquigarrow\breve{t_1}\xrightarrow{t_1}\breve{t_1}[m]\cdot t_2'$
  with $t_2\sim_{ps}^{fr} t_2'$, meaning that $n'$ contains a summand $t_1\cdot t_2'$. Since $t_2$ and $t_2'$ are normal forms and have sizes smaller than $n$ and $n'$, by
  the induction hypotheses $t_2\sim_{ps}^{fr} t_2'$ implies $t_2=_{AC} t_2'$.
  \item Consider a summand $t_1\cdot t_2[m]$ of $n$. Then $n\rightsquigarrow\breve{t_2}[m]\xtworightarrow{t_2[m]}t_1\cdot \breve{t_2}$, so $n\sim_{ps}^{fr} n'$ implies $n'\rightsquigarrow\breve{t_2}[m]\xtworightarrow{t_2[m]}t_1'\cdot \breve{t_2}$
  with $t_1\sim_{ps}^{fr} t_1'$, meaning that $n'$ contains a summand $t_1'\cdot t_2[m]$. Since $t_1$ and $t_1'$ are normal forms and have sizes smaller than $n$ and $n'$, by
  the induction hypotheses $t_1\sim_{ps}^{fr} t_1'$ implies $t_1=_{AC} t_1'$.
\end{itemize}

So, we get $n=_{AC} n'$.

Finally, let $s$ and $t$ be basic terms, and $s\sim_{ps}^{fr} t$, there are normal forms $n$ and $n'$, such that $s=n$ and $t=n'$. The soundness theorem of $BARPTC$ modulo FR probabilistic
pomset bisimulation equivalence (see Theorem \ref{SBARPTCPBE}) yields $s\sim_{ps}^{fr} n$ and $t\sim_{ps}^{fr} n'$, so $n\sim_{ps}^{fr} s\sim_{ps}^{fr} t\sim_{ps}^{fr} n'$. Since if $n\sim_{ps}^{fr} n'$ then
$n=_{AC}n'$, $s=n=_{AC}n'=t$, as desired.
\end{proof}

The transition rules for (hereditary) hp-bisimulation of $BARPTC$ are same as single event transition rules in Table \ref{SETRForBARPTC}.

\begin{theorem}[Congruence of $BARPTC$ with respect to FR probabilistic hp-bisimulation equivalence]
FR probabilistic hp-bisimulation equivalence $\sim_{php}$ is a congruence with respect to $BARPTC$.
\end{theorem}

\begin{proof}
It is easy to see that FR probabilistic history-preserving bisimulation is an equivalent relation on $BARPTC$ terms, we only need to prove that $\sim_{php}^{fr}$ is preserved by the operators $\cdot$, $+$
and $\boxplus_{\pi}$.
That is, if $x\sim_{php}^{fr} x'$ and $y\sim_{php}^{fr}y'$, we need to prove that $x\cdot y\sim_{php}^{fr}x'\cdot y'$, $x+ y\sim_{php}^{fr}x'+ y'$ and $x\boxplus_{\pi} y\sim_{php}^{fr}x'\boxplus_{\pi} y'$. The proof
is quite trivial and we omit it.
\end{proof}

\begin{theorem}[Soundness of $BARPTC$ modulo FR probabilistic hp-bisimulation equivalence]\label{SBARPTCHPBE}
Let $x$ and $y$ be $BARPTC$ terms. If $BARPTC\vdash x=y$, then $x\sim_{php}^{fr} y$.
\end{theorem}

\begin{proof}
Since FR probabilistic hp-bisimulation $\sim_{php}^{fr}$ is both an equivalent and a congruent relation, we only need to check if each axiom in Table \ref{AxiomsForBARPTC} is sound modulo
FR probabilistic hp-bisimulation equivalence. The proof is quite trivial and we omit it.
\end{proof}

\begin{theorem}[Completeness of $BARPTC$ modulo FR probabilistic hp-bisimulation equivalence]\label{CBARPTCHPBE}
Let $p$ and $q$ be closed $BARPTC$ terms, if $p\sim_{php}^{fr} q$ then $p=q$.
\end{theorem}

\begin{proof}
Firstly, by the elimination theorem of $BARPTC$, we know that for each closed $BARPTC$ term $p$, there exists a closed basic $BARPTC$ term $p'$, such that $BARPTC\vdash p=p'$, so, we only
need to consider closed basic $BARPTC$ terms.

The basic terms (see Definition \ref{BTBARPTC}) modulo associativity and commutativity (AC) of conflict $+$ (defined by axioms $A1$ and $A2$ in Table \ref{AxiomsForBARPTC}), and this
equivalence is denoted by $=_{AC}$. Then, each equivalence class $s$ modulo AC of $+$ has the following normal form

$$s_1\boxplus_{\pi_1}\cdots\boxplus_{\pi_{k-1}} s_k$$

with each $s_i$ has the following form

$$t_1+\cdots+ t_l$$

with each $t_j$ either an atomic event or of the form $u_1\cdot u_2$, and each $t_j$ is called the summand of $s$.

Now, we prove that for normal forms $n$ and $n'$, if $n\sim_{php}^{fr} n'$ then $n=_{AC}n'$. It is sufficient to induct on the sizes of $n$ and $n'$.

\begin{itemize}
  \item Consider a summand $e$ of $n$. Then $n\rightsquigarrow\breve{e}\xrightarrow{e}\breve{e}[m]$, so $n\sim_{php}^{fr} n'$ implies $n'\rightsquigarrow\breve{e}\xrightarrow{e}\breve{e}[m]$, meaning that
  $n'$ also contains the summand $e$.
  \item Consider a summand $e[m]$ of $n$. Then $n\rightsquigarrow\breve{e}[m]\xtworightarrow{e[m]}\breve{e}$, so $n\sim_{php}^{fr} n'$ implies $n'\rightsquigarrow\breve{e}\xtworightarrow{e[m]}\breve{e}$, meaning that
  $n'$ also contains the summand $e[m]$.
  \item Consider a summand $t_1\cdot t_2$ of $n$. Then $n\rightsquigarrow\breve{t_1}\xrightarrow{t_1}\breve{t_1}[m]\cdot t_2$, so $n\sim_{php}^{fr} n'$ implies $n'\rightsquigarrow\breve{t_1}\xrightarrow{t_1}\breve{t_1}[m]\cdot t_2'$
  with $t_2\sim_{php}^{fr} t_2'$, meaning that $n'$ contains a summand $t_1\cdot t_2'$. Since $t_2$ and $t_2'$ are normal forms and have sizes smaller than $n$ and $n'$, by
  the induction hypotheses $t_2\sim_{php}^{fr} t_2'$ implies $t_2=_{AC} t_2'$.
  \item Consider a summand $t_1\cdot t_2[m]$ of $n$. Then $n\rightsquigarrow\breve{t_2}[m]\xtworightarrow{t_2[m]}t_1\cdot \breve{t_2}$, so $n\sim_{php}^{fr} n'$ implies $n'\rightsquigarrow\breve{t_2}[m]\xtworightarrow{t_2[m]}t_1'\cdot \breve{t_2}$
  with $t_1\sim_{php}^{fr} t_1'$, meaning that $n'$ contains a summand $t_1'\cdot t_2[m]$. Since $t_1$ and $t_1'$ are normal forms and have sizes smaller than $n$ and $n'$, by
  the induction hypotheses $t_1\sim_{php}^{fr} t_1'$ implies $t_1=_{AC} t_1'$.
\end{itemize}

So, we get $n=_{AC} n'$.

Finally, let $s$ and $t$ be basic terms, and $s\sim_{php}^{fr} t$, there are normal forms $n$ and $n'$, such that $s=n$ and $t=n'$. The soundness theorem of $BARPTC$ modulo FR probabilistic
pomset bisimulation equivalence (see Theorem \ref{SBARPTCPBE}) yields $s\sim_{php}^{fr} n$ and $t\sim_{php}^{fr} n'$, so $n\sim_{php}^{fr} s\sim_{php}^{fr} t\sim_{php}^{fr} n'$. Since if $n\sim_{php}^{fr} n'$ then
$n=_{AC}n'$, $s=n=_{AC}n'=t$, as desired.
\end{proof}

\begin{theorem}[Congruence of $BARPTC$ with respect to FR probabilistic hhp-bisimulation equivalence]
FR probabilistic hhp-bisimulation equivalence $\sim_{phhp}^{fr}$ is a congruence with respect to $BARPTC$.
\end{theorem}

\begin{proof}
It is easy to see that FR probabilistic hhp-bisimulation is an equivalent relation on $BARPTC$ terms, we only need to prove that $\sim_{phhp}^{fr}$ is preserved by the operators $\cdot$, $+$ and
$\boxplus_{\pi}$.
That is, if $x\sim_{phhp}^{fr} x'$ and $y\sim_{phhp}^{fr}y'$, we need to prove that $x\cdot y\sim_{phhp}^{fr}x'\cdot y'$, $x+ y\sim_{phhp}^{fr}x'+ y'$ and $x\boxplus_{\pi} y\sim_{phhp}^{fr}x'\boxplus_{\pi} y'$. The proof
is quite trivial and we omit it.
\end{proof}

\begin{theorem}[Soundness of $BARPTC$ modulo FR probabilistic hhp-bisimulation equivalence]\label{SBARPTCHHPBE}
Let $x$ and $y$ be $BARPTC$ terms. If $BARPTC\vdash x=y$, then $x\sim_{phhp}^{fr} y$.
\end{theorem}

\begin{proof}
Since FR probabilistic hhp-bisimulation $\sim_{phhp}^{fr}$ is both an equivalent and a congruent relation, we only need to check if each axiom in Table \ref{AxiomsForBARPTC} is sound modulo
FR probabilistic hhp-bisimulation equivalence. It is quite trivial and we omit it.
\end{proof}

\begin{theorem}[Completeness of $BARPTC$ modulo FR probabilistic hhp-bisimulation equivalence]\label{CBARPTCHHPBE}
Let $p$ and $q$ be closed $BARPTC$ terms, if $p\sim_{phhp}^{fr} q$ then $p=q$.
\end{theorem}

\begin{proof}
Firstly, by the elimination theorem of $BARPTC$, we know that for each closed $BARPTC$ term $p$, there exists a closed basic $BARPTC$ term $p'$, such that $BARPTC\vdash p=p'$, so, we only
need to consider closed basic $BARPTC$ terms.

The basic terms (see Definition \ref{BTBARPTC}) modulo associativity and commutativity (AC) of conflict $+$ (defined by axioms $A1$ and $A2$ in Table \ref{AxiomsForBARPTC}), and this
equivalence is denoted by $=_{AC}$. Then, each equivalence class $s$ modulo AC of $+$ has the following normal form

$$s_1\boxplus_{\pi_1}\cdots\boxplus_{\pi_{k-1}} s_k$$

with each $s_i$ has the following form

$$t_1+\cdots+ t_l$$

with each $t_j$ either an atomic event or of the form $u_1\cdot u_2$, and each $t_j$ is called the summand of $s$.

Now, we prove that for normal forms $n$ and $n'$, if $n\sim_{phhp}^{fr} n'$ then $n=_{AC}n'$. It is sufficient to induct on the sizes of $n$ and $n'$.

\begin{itemize}
  \item Consider a summand $e$ of $n$. Then $n\rightsquigarrow\breve{e}\xrightarrow{e}\breve{e}[m]$, so $n\sim_{phhp}^{fr} n'$ implies $n'\rightsquigarrow\breve{e}\xrightarrow{e}\breve{e}[m]$, meaning that
  $n'$ also contains the summand $e$.
  \item Consider a summand $e[m]$ of $n$. Then $n\rightsquigarrow\breve{e}[m]\xtworightarrow{e[m]}\breve{e}$, so $n\sim_{phhp}^{fr} n'$ implies $n'\rightsquigarrow\breve{e}\xtworightarrow{e[m]}\breve{e}$, meaning that
  $n'$ also contains the summand $e[m]$.
  \item Consider a summand $t_1\cdot t_2$ of $n$. Then $n\rightsquigarrow\breve{t_1}\xrightarrow{t_1}\breve{t_1}[m]\cdot t_2$, so $n\sim_{phhp}^{fr} n'$ implies $n'\rightsquigarrow\breve{t_1}\xrightarrow{t_1}\breve{t_1}[m]\cdot t_2'$
  with $t_2\sim_{phhp}^{fr} t_2'$, meaning that $n'$ contains a summand $t_1\cdot t_2'$. Since $t_2$ and $t_2'$ are normal forms and have sizes smaller than $n$ and $n'$, by
  the induction hypotheses $t_2\sim_{phhp}^{fr} t_2'$ implies $t_2=_{AC} t_2'$.
  \item Consider a summand $t_1\cdot t_2[m]$ of $n$. Then $n\rightsquigarrow\breve{t_2}[m]\xtworightarrow{t_2[m]}t_1\cdot \breve{t_2}$, so $n\sim_{phhp}^{fr} n'$ implies $n'\rightsquigarrow\breve{t_2}[m]\xtworightarrow{t_2[m]}t_1'\cdot \breve{t_2}$
  with $t_1\sim_{phhp}^{fr} t_1'$, meaning that $n'$ contains a summand $t_1'\cdot t_2[m]$. Since $t_1$ and $t_1'$ are normal forms and have sizes smaller than $n$ and $n'$, by
  the induction hypotheses $t_1\sim_{phhp}^{fr} t_1'$ implies $t_1=_{AC} t_1'$.
\end{itemize}

So, we get $n=_{AC} n'$.

Finally, let $s$ and $t$ be basic terms, and $s\sim_{phhp}^{fr} t$, there are normal forms $n$ and $n'$, such that $s=n$ and $t=n'$. The soundness theorem of $BARPTC$ modulo FR probabilistic
pomset bisimulation equivalence (see Theorem \ref{SBARPTCPBE}) yields $s\sim_{phhp}^{fr} n$ and $t\sim_{phhp}^{fr} n'$, so $n\sim_{phhp}^{fr} s\sim_{phhp}^{fr} t\sim_{phhp}^{fr} n'$. Since if $n\sim_{phhp}^{fr} n'$ then
$n=_{AC}n'$, $s=n=_{AC}n'=t$, as desired.
\end{proof}

\subsection{Algebra for Parallelism in Reversible Probabilistic True Concurrency}\label{aprptc}

In this section, we will discuss parallelism in reversible probabilistic true concurrency. The resulted algebra is called Algebra for Parallelism in reversible probabilistic True
Concurrency, abbreviated $APRPTC$.

\subsubsection{Axiom System of Parallelism}

We design the axioms of parallelism in Table \ref{AxiomsForRPParallelism}, including algebraic laws for parallel operator $\parallel$, communication operator $\mid$, conflict elimination
operator $\Theta$ and unless operator $\triangleleft$, and also the whole parallel operator $\between$. Since the communication between two communicating events in different parallel
branches may cause deadlock (a state of inactivity), which is caused by mismatch of two communicating events or the imperfectness of the communication channel. We introduce a new
constant $\delta$ to denote the deadlock, and let the atomic event $e\in \mathbb{E}\cup\{\delta\}$.

\begin{center}
    \begin{table}
        \begin{tabular}{@{}ll@{}}
            \hline No. &Axiom\\
            $P1$ & $(x+x=x,y+y=y)\quad x\between y = x\parallel y + x\mid y$\\
            $P2$ & $x\parallel y = y \parallel x$\\
            $P3$ & $(x\parallel y)\parallel z = x\parallel (y\parallel z)$\\
            $P4$ & $(x+x=x,y+y=y)\quad x\parallel y = x\leftmerge y + y\leftmerge x$\\
            $P5$ & $(e_1\leq e_2)\quad e_1\leftmerge (e_2\cdot y) = (e_1\leftmerge e_2)\cdot y$\\
            $RP5$ & $(e_1[m]\leq e_2[m])\quad e_1[m]\leftmerge (y\cdot e_2[m]) = y\cdot(e_1[m]\leftmerge e_2[m])$\\
            $P6$ & $(e_1\leq e_2)\quad (e_1\cdot x)\leftmerge e_2 = (e_1\leftmerge e_2)\cdot x$\\
            $RP6$ & $(e_1[m]\leq e_2[m])\quad (x\cdot e_1[m])\leftmerge e_2[m] = x\cdot(e_1[m]\leftmerge e_2[m])$\\
            $P7$ & $(e_1\leq e_2)\quad (e_1\cdot x)\leftmerge (e_2\cdot y) = (e_1\leftmerge e_2)\cdot (x\between y)$\\
            $RP7$ & $(e_1[m]\leq e_2[m])\quad(x\cdot e_1[m])\leftmerge (y\cdot e_2[m]) = (x\between y)\cdot(e_1[m]\leftmerge e_2[m])$\\
            $P8$ & $(x+ y)\leftmerge z = (x\leftmerge z)+ (y\leftmerge z)(\textrm{Std(x)},\textrm{Std(y)},\textrm{Std(z)})$\\
            $RP8$ & $x\leftmerge (y+ z) = (x\leftmerge y)+ (x\leftmerge z)(\textrm{NStd(x)},\textrm{NStd(y)},\textrm{NStd(z)})$\\
            $P9$ & $\delta\leftmerge x = \delta(\textrm{Std(x)})$\\
            $RP9$ & $x\leftmerge \delta = \delta(\textrm{NStd(x)})$\\
            $C1$ & $e_1\mid e_2 = \gamma(e_1,e_2)$\\
            $RC1$ & $e_1[m]\mid e_2[m] = \gamma(e_1,e_2)[m]$\\
            $C2$ & $e_1\mid (e_2\cdot y) = \gamma(e_1,e_2)\cdot y$\\
            $RC2$ & $e_1[m]\mid (y \cdot e_2[m]) =y\cdot \gamma(e_1,e_2)[m]$\\
            $C3$ & $(e_1\cdot x)\mid e_2 = \gamma(e_1,e_2)\cdot x$\\
            $RC3$ & $(x \cdot e_1[m])\mid e_2[m] =x\cdot \gamma(e_1,e_2)[m]$\\
            $C4$ & $(e_1\cdot x)\mid (e_2\cdot y) = \gamma(e_1,e_2)\cdot (x\between y)$\\
            $RC4$ & $(x \cdot e_1[m])\mid (y \cdot e_2[m]) =(x\between y)\cdot \gamma(e_1,e_2)[m]$\\
            $C5$ & $(x+ y)\mid z = (x\mid z) + (y\mid z)$\\
            $C6$ & $x\mid (y+ z) = (x\mid y)+ (x\mid z)$\\
            $C7$ & $\delta\mid x = \delta$\\
            $C8$ & $x\mid\delta = \delta$\\
            $PM1$ & $x\parallel (y\boxplus_{\pi} z)=(x\parallel y)\boxplus_{\pi}(x\parallel z)$\\
            $PM2$ & $(x\boxplus_{\pi} y)\parallel z=(x\parallel z)\boxplus_{\pi}(y\parallel z)$\\
            $PM3$ & $x\mid (y\boxplus_{\pi} z)=(x\mid y)\boxplus_{\pi}(x\mid z)$\\
            $PM4$ & $(x\boxplus_{\pi} y)\mid z=(x\mid z)\boxplus_{\pi}(y\mid z)$\\
            $CE1$ & $\Theta(e) = e$\\
            $RCE1$ & $\Theta(e[m]) = e[m]$\\
            $CE2$ & $\Theta(\delta) = \delta$\\
            $CE3$ & $\Theta(x+ y) = \Theta(x)\triangleleft y + \Theta(y)\triangleleft x$\\
            $PCE1$ & $\Theta(x\boxplus_{\pi} y) = \Theta(x)\triangleleft y \boxplus_{\pi} \Theta(y)\triangleleft x$\\
            $CE4$ & $\Theta(x\cdot y)=\Theta(x)\cdot\Theta(y)$\\
            $CE5$ & $\Theta(x\leftmerge y) = ((\Theta(x)\triangleleft y)\leftmerge y)+ ((\Theta(y)\triangleleft x)\leftmerge x)$\\
            $CE6$ & $\Theta(x\mid y) = ((\Theta(x)\triangleleft y)\mid y)+ ((\Theta(y)\triangleleft x)\mid x)$\\
        \end{tabular}
        \caption{Axioms of parallelism}
        \label{AxiomsForRPParallelism}
    \end{table}
\end{center}

\begin{center}
    \begin{table}
        \begin{tabular}{@{}ll@{}}
            \hline No. &Axiom\\
            $U1$ & $(\sharp(e_1,e_2))\quad e_1\triangleleft e_2 = \tau$\\
            $RU1$ & $(\sharp(e_1[m],e_2[n]))\quad e_1[m]\triangleleft e_2[n] = \tau$\\
            $U2$ & $(\sharp(e_1,e_2),e_2\leq e_3)\quad e_1\triangleleft e_3 = e_1$\\
            $RU2$ & $(\sharp(e_1[m],e_2[n]),e_2[n]\geq e_3[l])\quad e_1[m]\triangleleft e_3[l] = e_1[m]$\\
            $U3$ & $(\sharp(e_1,e_2),e_2\leq e_3)\quad e3\triangleleft e_1 = \tau$\\
            $RU3$ & $(\sharp(e_1[m],e_2[n]),e_2[n]\geq e_3[l])\quad e3[l]\triangleleft e_1[m] = \tau$\\
            $PU1$ & $(\sharp_{\pi}(e_1,e_2))\quad e_1\triangleleft e_2 = \tau$\\
            $RPU1$ & $(\sharp_{\pi}(e_1[m],e_2[n]))\quad e_1[m]\triangleleft e_2[n] = \tau$\\
            $PU2$ & $(\sharp_{\pi}(e_1,e_2),e_2\leq e_3)\quad e_1\triangleleft e_3 = e_1$\\
            $RPU2$ & $(\sharp_{\pi}(e_1[m],e_2[n]),e_2[n]\leq e_3[l])\quad e_1[m]\triangleleft e_3[l] = e_1[m]$\\
            $PU3$ & $(\sharp_{\pi}(e_1,e_2),e_2\leq e_3)\quad e_3\triangleleft e_1 = \tau$\\
            $RPU3$ & $(\sharp_{\pi}(e_1[m],e_2[n]),e_2[n]\leq e_3[l])\quad e_3[l]\triangleleft e_1[m] = \tau$\\
            $U4$ & $e\triangleleft \delta = e$\\
            $RU4$ & $e[m]\triangleleft \delta = \delta$\\
            $U5$ & $\delta \triangleleft e = \delta$\\
            $RU5$ & $\delta \triangleleft e[m] = e[m]$\\
            $U6$ & $(x+ y)\triangleleft z = (x\triangleleft z)+ (y\triangleleft z)$\\
            $PU4$ & $(x\boxplus_{\pi} y)\triangleleft z = (x\triangleleft z)\boxplus_{\pi} (y\triangleleft z)$\\
            $U7$ & $(x\cdot y)\triangleleft z = (x\triangleleft z)\cdot (y\triangleleft z)$\\
            $U8$ & $(x\leftmerge y)\triangleleft z = (x\triangleleft z)\leftmerge (y\triangleleft z)$\\
            $U9$ & $(x\mid y)\triangleleft z = (x\triangleleft z)\mid (y\triangleleft z)$\\
            $U10$ & $x\triangleleft (y+ z) = (x\triangleleft y)\triangleleft z$\\
            $PU5$ & $x\triangleleft (y\boxplus_{\pi} z) = (x\triangleleft y)\triangleleft z$\\
            $U11$ & $x\triangleleft (y\cdot z)=(x\triangleleft y)\triangleleft z$\\
            $U12$ & $x\triangleleft (y\leftmerge z) = (x\triangleleft y)\triangleleft z$\\
            $U13$ & $x\triangleleft (y\mid z) = (x\triangleleft y)\triangleleft z$\\
        \end{tabular}
        \caption{Axioms of parallelism (continuing)}
        \label{AxiomsForRPParallelism2}
    \end{table}
\end{center}

\begin{definition}[Basic terms of $APRPTC$]\label{BTAPRPTC}
The set of basic terms of $APRPTC$, $\mathcal{B}(APRPTC)$, is inductively defined as follows:
\begin{enumerate}
  \item $\mathbb{E}\subset\mathcal{B}(APRPTC)$;
  \item if $e\in \mathbb{E}, t\in\mathcal{B}(APRPTC)$ then $e\cdot t\in\mathcal{B}(APRPTC)$;
  \item if $e[m]\in \mathbb{E}, t\in\mathcal{B}(APRPTC)$ then $t\cdot e[m]\in\mathcal{B}(APRPTC)$;
  \item if $t,s\in\mathcal{B}(APRPTC)$ then $t+ s\in\mathcal{B}(APRPTC)$;
  \item if $t,s\in\mathcal{B}(APRPTC)$ then $t\boxplus_{\pi} s\in\mathcal{B}(APRPTC)$;
  \item if $t,s\in\mathcal{B}(APRPTC)$ then $t\leftmerge s\in\mathcal{B}(APRPTC)$.
\end{enumerate}
\end{definition}

Based on the definition of basic terms for $APRPTC$ (see Definition \ref{BTAPRPTC}) and axioms of parallelism (see Table \ref{AxiomsForRPParallelism}), we can prove the elimination
theorem of parallelism.

\begin{theorem}[Elimination theorem of parallelism]\label{ETPParallelism}
Let $p$ be a closed $APRPTC$ term. Then there is a basic $APRPTC$ term $q$ such that $APRPTC\vdash p=q$.
\end{theorem}

\begin{proof}
(1) Firstly, suppose that the following ordering on the signature of $APRPTC$ is defined: $\leftmerge > \cdot > +>\boxplus_{\pi}$ and the symbol $\parallel$ is given the lexicographical
status for the first argument, then for each rewrite rule $p\rightarrow q$ in Table \ref{TRSForAPRPTC} relation $p>_{lpo} q$ can easily be proved. We obtain that the term rewrite system
shown in Table \ref{TRSForAPRPTC} is strongly normalizing, for it has finitely many rewriting rules, and $>$ is a well-founded ordering on the signature of $APRPTC$, and if $s>_{lpo} t$,
for each rewriting rule $s\rightarrow t$ is in Table \ref{TRSForAPRPTC} (see Theorem \ref{SN}).

\begin{center}
    \begin{table}
        \begin{tabular}{@{}ll@{}}
            \hline No. &Rewriting Rule\\
            $RP1$ & $(x+x=x,y+y=y)\quad x\between y \rightarrow x\parallel y + x\mid y$\\
            $RP2$ & $x\parallel y \rightarrow y \parallel x$\\
            $RP3$ & $(x\parallel y)\parallel z \rightarrow x\parallel (y\parallel z)$\\
            $RP4$ & $(x+x=x,y+y=y)\quad x\parallel y \rightarrow x\leftmerge y + y\leftmerge x$\\
            $RP5$ & $(e_1\leq e_2)\quad e_1\leftmerge (e_2\cdot y) \rightarrow (e_1\leftmerge e_2)\cdot y$\\
            $RRP5$ & $(e_1[m]\leq e_2[m])\quad e_1[m]\leftmerge (y\cdot e_2[m]) \rightarrow y\cdot(e_1[m]\leftmerge e_2[m])$\\
            $RP6$ & $(e_1\leq e_2)\quad (e_1\cdot x)\leftmerge e_2 \rightarrow (e_1\leftmerge e_2)\cdot x$\\
            $RRP6$ & $(e_1[m]\leq e_2[m])\quad (x\cdot e_1[m])\leftmerge e_2[m] \rightarrow x\cdot(e_1[m]\leftmerge e_2[m])$\\
            $RP7$ & $(e_1\leq e_2)\quad (e_1\cdot x)\leftmerge (e_2\cdot y) \rightarrow (e_1\leftmerge e_2)\cdot (x\between y)$\\
            $RRP7$ & $(e_1[m]\leq e_2[m])\quad(x\cdot e_1[m])\leftmerge (y\cdot e_2[m]) \rightarrow (x\between y)\cdot(e_1[m]\leftmerge e_2[m])$\\
            $RP8$ & $(x+ y)\leftmerge z \rightarrow (x\leftmerge z)+ (y\leftmerge z)(\textrm{Std(x)},\textrm{Std(y)},\textrm{Std(z)})$\\
            $RRP8$ & $x\leftmerge (y+ z) \rightarrow (x\leftmerge y)+ (x\leftmerge z)(\textrm{NStd(x)},\textrm{NStd(y)},\textrm{NStd(z)})$\\
            $RP9$ & $\delta\leftmerge x \rightarrow \delta(\textrm{Std(x)})$\\
            $RRP9$ & $x\leftmerge \delta \rightarrow \delta(\textrm{NStd(x)})$\\
            $RC1$ & $e_1\mid e_2 \rightarrow \gamma(e_1,e_2)$\\
            $RRC1$ & $e_1[m]\mid e_2[m] \rightarrow \gamma(e_1,e_2)[m]$\\
            $RC2$ & $e_1\mid (e_2\cdot y) \rightarrow \gamma(e_1,e_2)\cdot y$\\
            $RRC2$ & $e_1[m]\mid (y \cdot e_2[m]) \rightarrow y\cdot \gamma(e_1,e_2)[m]$\\
            $RC3$ & $(e_1\cdot x)\mid e_2 \rightarrow \gamma(e_1,e_2)\cdot x$\\
            $RRC3$ & $(x \cdot e_1[m])\mid e_2[m] \rightarrow x\cdot \gamma(e_1,e_2)[m]$\\
            $RC4$ & $(e_1\cdot x)\mid (e_2\cdot y) \rightarrow \gamma(e_1,e_2)\cdot (x\between y)$\\
            $RRC4$ & $(x \cdot e_1[m])\mid (y \cdot e_2[m]) \rightarrow(x\between y)\cdot \gamma(e_1,e_2)[m]$\\
            $RC5$ & $(x+ y)\mid z \rightarrow (x\mid z) + (y\mid z)$\\
            $RC6$ & $x\mid (y+ z) \rightarrow (x\mid y)+ (x\mid z)$\\
            $RC7$ & $\delta\mid x \rightarrow \delta$\\
            $RC8$ & $x\mid\delta \rightarrow \delta$\\
            $RPM1$ & $x\parallel (y\boxplus_{\pi} z)\rightarrow(x\parallel y)\boxplus_{\pi}(x\parallel z)$\\
            $RPM2$ & $(x\boxplus_{\pi} y)\parallel z\rightarrow(x\parallel z)\boxplus_{\pi}(y\parallel z)$\\
            $RPM3$ & $x\mid (y\boxplus_{\pi} z)\rightarrow(x\mid y)\boxplus_{\pi}(x\mid z)$\\
            $RPM4$ & $(x\boxplus_{\pi} y)\mid z\rightarrow(x\mid z)\boxplus_{\pi}(y\mid z)$\\
            $RCE1$ & $\Theta(e) \rightarrow e$\\
            $RRCE1$ & $\Theta(e[m]) \rightarrow e[m]$\\
            $RCE2$ & $\Theta(\delta) \rightarrow \delta$\\
            $RCE3$ & $\Theta(x+ y) \rightarrow \Theta(x)\triangleleft y + \Theta(y)\triangleleft x$\\
            $RPCE1$ & $\Theta(x\boxplus_{\pi} y) \rightarrow \Theta(x)\triangleleft y \boxplus_{\pi} \Theta(y)\triangleleft x$\\
            $RCE4$ & $\Theta(x\cdot y)\rightarrow\Theta(x)\cdot\Theta(y)$\\
            $RCE5$ & $\Theta(x\leftmerge y) \rightarrow ((\Theta(x)\triangleleft y)\leftmerge y)+ ((\Theta(y)\triangleleft x)\leftmerge x)$\\
            $RCE6$ & $\Theta(x\mid y) \rightarrow ((\Theta(x)\triangleleft y)\mid y)+ ((\Theta(y)\triangleleft x)\mid x)$\\
        \end{tabular}
        \caption{Term rewrite system of $APRPTC$}
        \label{TRSForAPRPTC}
    \end{table}
\end{center}

\begin{center}
    \begin{table}
        \begin{tabular}{@{}ll@{}}
            \hline No. &Rewriting Rule\\
            $RU1$ & $(\sharp(e_1,e_2))\quad e_1\triangleleft e_2 \rightarrow \tau$\\
            $RRU1$ & $(\sharp(e_1[m],e_2[n]))\quad e_1[m]\triangleleft e_2[n] \rightarrow \tau$\\
            $RU2$ & $(\sharp(e_1,e_2),e_2\leq e_3)\quad e_1\triangleleft e_3 \rightarrow e_1$\\
            $RRU2$ & $(\sharp(e_1[m],e_2[n]),e_2[n]\geq e_3[l])\quad e_1[m]\triangleleft e_3[l] \rightarrow e_1[m]$\\
            $RU3$ & $(\sharp(e_1,e_2),e_2\leq e_3)\quad e3\triangleleft e_1 \rightarrow \tau$\\
            $RRU3$ & $(\sharp(e_1[m],e_2[n]),e_2[n]\geq e_3[l])\quad e3[l]\triangleleft e_1[m] \rightarrow \tau$\\
            $RPU1$ & $(\sharp_{\pi}(e_1,e_2))\quad e_1\triangleleft e_2 \rightarrow \tau$\\
            $RRPU1$ & $(\sharp_{\pi}(e_1[m],e_2[n]))\quad e_1[m]\triangleleft e_2[n] \rightarrow \tau$\\
            $RPU2$ & $(\sharp_{\pi}(e_1,e_2),e_2\leq e_3)\quad e_1\triangleleft e_3 \rightarrow e_1$\\
            $RRPU2$ & $(\sharp_{\pi}(e_1[m],e_2[n]),e_2[n]\leq e_3[l])\quad e_1[m]\triangleleft e_3[l] \rightarrow e_1[m]$\\
            $RPU3$ & $(\sharp_{\pi}(e_1,e_2),e_2\leq e_3)\quad e_3\triangleleft e_1 \rightarrow \tau$\\
            $RRPU3$ & $(\sharp_{\pi}(e_1[m],e_2[n]),e_2[n]\leq e_3[l])\quad e_3[l]\triangleleft e_1[m] \rightarrow \tau$\\
            $RU4$ & $e\triangleleft \delta \rightarrow e$\\
            $RRU4$ & $e[m]\triangleleft \delta \rightarrow e[m]$\\
            $RU5$ & $\delta \triangleleft e \rightarrow \delta$\\
            $RRU5$ & $\delta \triangleleft e[m] \rightarrow \delta$\\
            $RU6$ & $(x+ y)\triangleleft z \rightarrow (x\triangleleft z)+ (y\triangleleft z)$\\
            $RPU4$ & $(x\boxplus_{\pi} y)\triangleleft z \rightarrow (x\triangleleft z)\boxplus_{\pi} (y\triangleleft z)$\\
            $RU7$ & $(x\cdot y)\triangleleft z \rightarrow (x\triangleleft z)\cdot (y\triangleleft z)$\\
            $RU8$ & $(x\leftmerge y)\triangleleft z \rightarrow (x\triangleleft z)\leftmerge (y\triangleleft z)$\\
            $RU9$ & $(x\mid y)\triangleleft z \rightarrow (x\triangleleft z)\mid (y\triangleleft z)$\\
            $RU10$ & $x\triangleleft (y+ z) \rightarrow (x\triangleleft y)\triangleleft z$\\
            $RPU5$ & $x\triangleleft (y\boxplus_{\pi} z) \rightarrow (x\triangleleft y)\triangleleft z$\\
            $RU11$ & $x\triangleleft (y\cdot z)\rightarrow(x\triangleleft y)\triangleleft z$\\
            $RU12$ & $x\triangleleft (y\leftmerge z) \rightarrow (x\triangleleft y)\triangleleft z$\\
            $RU13$ & $x\triangleleft (y\mid z) \rightarrow (x\triangleleft y)\triangleleft z$\\
        \end{tabular}
        \caption{Term rewrite system of $APRPTC$ (continuing)}
        \label{TRSForAPRPTC2}
    \end{table}
\end{center}

(2) Then we prove that the normal forms of closed $APRPTC$ terms are basic $APRPTC$ terms.

Suppose that $p$ is a normal form of some closed $APRPTC$ term and suppose that $p$ is not a basic $APRPTC$ term. Let $p'$ denote the smallest sub-term of $p$ which is not a basic $APRPTC$ term. It implies that each sub-term of $p'$ is a basic $APRPTC$ term. Then we prove that $p$ is not a term in normal form. It is sufficient to induct on the structure of $p'$:

\begin{itemize}
  \item Case $p'\equiv e, e\in \mathbb{E}$. $p'$ is a basic $APRPTC$ term, which contradicts the assumption that $p'$ is not a basic $APRPTC$ term, so this case should not occur.
  \item Case $p'\equiv e[m], e[m]\in \mathbb{E}$. $p'$ is a basic $APRPTC$ term, which contradicts the assumption that $p'$ is not a basic $APRPTC$ term, so this case should not occur.
  \item Case $p'\equiv p_1\cdot p_2$. By induction on the structure of the basic $APRPTC$ term $p_1$:
      \begin{itemize}
        \item Subcase $p_1\in \mathbb{E}$. $p'$ would be a basic $APRPTC$ term, which contradicts the assumption that $p'$ is not a basic $APRPTC$ term;
        \item Subcase $p_1\equiv e\cdot p_1'$. $RA5$ rewriting rule in Table \ref{TRSForBARPTC} can be applied. So $p$ is not a normal form;
        \item Subcase $p_1\equiv p_1'\cdot e[m]$. $RA5$ rewriting rule in Table \ref{TRSForBARPTC} can be applied. So $p$ is not a normal form;
        \item Subcase $p_1\equiv p_1'+ p_1''$. $RA41$ and $RA42$ rewriting rules in Table \ref{TRSForBARPTC} can be applied. So $p$ is not a normal form;
        \item Subcase $p_1\equiv p_1'\boxplus_{\pi} p_1''$. $RPA41$, $RPA42$ and $RPA5$ rewriting rules in Table \ref{TRSForBARPTC} can be applied. So $p$ is not a normal form;
        \item Subcase $p_1\equiv p_1'\leftmerge p_1''$. $p'$ would be a basic $APRPTC$ term, which contradicts the assumption that $p'$ is not a basic $APRPTC$ term;
        \item Subcase $p_1\equiv p_1'\mid p_1''$. $RC2$ and $RRC2$ rewrite rule in Table \ref{TRSForAPRPTC} can be applied. So $p$ is not a normal form;
        \item Subcase $p_1\equiv \Theta(p_1')$. $RCE1$, $RRCE1$ and $RCE2$, $RRCE2$ rewrite rules in Table \ref{TRSForAPRPTC} can be applied. So $p$ is not a normal form.
      \end{itemize}
  \item Case $p'\equiv p_1+ p_2$. By induction on the structure of the basic $APRPTC$ terms both $p_1$ and $p_2$, all subcases will lead to that $p'$ would be a basic $APRPTC$ term,
  which contradicts the assumption that $p'$ is not a basic $APRPTC$ term.
  \item Case $p'\equiv p_1\boxplus_{\pi} p_2$. By induction on the structure of the basic $APRPTC$ terms both $p_1$ and $p_2$, all subcases will lead to that $p'$ would be a basic $APRPTC$ term,
  which contradicts the assumption that $p'$ is not a basic $APRPTC$ term.
  \item Case $p'\equiv p_1\leftmerge p_2$. By induction on the structure of the basic $APRPTC$ terms both $p_1$ and $p_2$, all subcases will lead to that $p'$ would be a basic $APRPTC$
  term, which contradicts the assumption that $p'$ is not a basic $APRPTC$ term.
  \item Case $p'\equiv p_1\mid p_2$. By induction on the structure of the basic $APRPTC$ terms both $p_1$ and $p_2$, all subcases will lead to that $p'$ would be a basic $APRPTC$ term,
  which contradicts the assumption that $p'$ is not a basic $APRPTC$ term.
  \item Case $p'\equiv \Theta(p_1)$. By induction on the structure of the basic $APRPTC$ term $p_1$, $RCE19-RCE24$ rewrite rules in Table \ref{TRSForAPRPTC} can be applied. So $p$ is not
  a normal form.
  \item Case $p'\equiv p_1\triangleleft p_2$. By induction on the structure of the basic $APRPTC$ terms both $p_1$ and $p_2$, all subcases will lead to that $p'$ would be a basic
  $APRPTC$ term, which contradicts the assumption that $p'$ is not a basic $APRPTC$ term.
\end{itemize}
\end{proof}

\subsubsection{Structured Operational Semantics of Parallelism}

Firstly, we give the definition of PDFs in Table \ref{PDFAPRPTC}.

\begin{center}
    \begin{table}
        $$\mu(\delta,\breve{\delta})=1$$
        $$\mu(x\between y,x'\parallel y'+x'\mid y')=\mu(x,x')\cdot\mu(y,y')$$
        $$\mu(x\parallel y,x'\leftmerge y+y'\leftmerge x)=\mu(x,x')\cdot \mu(y,y')$$
        $$\mu(x\leftmerge y, x'\leftmerge y)=\mu(x,x')$$
        $$\mu(x\mid y,x'\mid y')=\mu(x,x')\cdot \mu(y,y')$$
        $$\mu(\Theta(x),\Theta(x'))=\mu(x,x')$$
        $$\mu(x\triangleleft y, x'\triangleleft y)=\mu(x,x')$$
        $$\mu(x,y)=0,\textrm{otherwise}$$
        \caption{PDF definitions of $APRPTC$}
        \label{PDFAPRPTC}
    \end{table}
\end{center}

We give the transition rules of APTC in Table \ref{TRForAPRPTC1}, \ref{TRForAPRPTC}, it is suitable for all truly concurrent behavioral equivalence, including probabilistic pomset bisimulation,
probabilistic step bisimulation, probabilistic hp-bisimulation and probabilistic hhp-bisimulation.

\begin{center}
    \begin{table}
        $$\frac{x\rightsquigarrow x'\quad y\rightsquigarrow y'}{x\between y\rightsquigarrow x'\parallel y'+x'\mid y'}$$
        $$\frac{x\rightsquigarrow x'\quad y\rightsquigarrow y'}{x\parallel y\rightsquigarrow x'\leftmerge y+y'\leftmerge x}$$
        $$\frac{x\rightsquigarrow x'}{x\leftmerge y\rightsquigarrow x'\leftmerge y}$$
        $$\frac{x\rightsquigarrow x'\quad y\rightsquigarrow y'}{x\mid y\rightsquigarrow x'\mid y'}$$
        $$\frac{x\rightsquigarrow x'}{\Theta(x)\rightsquigarrow \Theta(x')}$$
        $$\frac{x\rightsquigarrow x'}{x\triangleleft y\rightsquigarrow x'\triangleleft y}$$
        \caption{Probabilistic transition rules of APRPTC}
        \label{TRForAPRPTC1}
    \end{table}
\end{center}

\begin{center}
    \begin{table}
        $$\frac{x\xrightarrow{e_1}e_1[m]\quad y\xrightarrow{e_2}e_2[m]}{x\parallel y\xrightarrow{\{e_1,e_2\}}e_1[m]\parallel e_2[m]}
        \quad\frac{x\xrightarrow{e_1}x'\quad y\xrightarrow{e_2}e_2[m]}{x\parallel y\xrightarrow{\{e_1,e_2\}}x'\parallel e_2[m]}$$
        $$\frac{x\xrightarrow{e_1}e_1[m]\quad y\xrightarrow{e_2}y'}{x\parallel y\xrightarrow{\{e_1,e_2\}}e_1[m]\parallel y'}
        \quad\frac{x\xrightarrow{e_1}x'\quad y\xrightarrow{e_2}y'}{x\parallel y\xrightarrow{\{e_1,e_2\}}x'\between y'}$$
        $$\frac{x\xrightarrow{e_1}e_1[m]\quad y\xrightarrow{e_2}e_2[m] \quad(e_1\leq e_2)}{x\leftmerge y\xrightarrow{\{e_1,e_2\}}e_1[m]\leftmerge e_2[m]}
        \quad\frac{x\xrightarrow{e_1}x'\quad y\xrightarrow{e_2}e_2[m] \quad(e_1\leq e_2)}{x\leftmerge y\xrightarrow{\{e_1,e_2\}}x'\leftmerge e_2[m]}$$
        $$\frac{x\xrightarrow{e_1}e_1[m]\quad y\xrightarrow{e_2}y' \quad(e_1\leq e_2)}{x\leftmerge y\xrightarrow{\{e_1,e_2\}}e_1[m]\leftmerge y'}
        \quad\frac{x\xrightarrow{e_1}x'\quad y\xrightarrow{e_2}y' \quad(e_1\leq e_2)}{x\leftmerge y\xrightarrow{\{e_1,e_2\}}x'\between y'}$$
        $$\frac{x\xrightarrow{e_1}e_1[m]\quad y\xrightarrow{e_2}e_2[m]}{x\mid y\xrightarrow{\gamma(e_1,e_2)}\gamma(e_1,e_2)[m]}
        \quad\frac{x\xrightarrow{e_1}x'\quad y\xrightarrow{e_2}e_2[m]}{x\mid y\xrightarrow{\gamma(e_1,e_2)}\gamma(e_1,e_2)[m]\cdot x'}$$
        $$\frac{x\xrightarrow{e_1}e_1[m]\quad y\xrightarrow{e_2}y'}{x\mid y\xrightarrow{\gamma(e_1,e_2)}\gamma(e_1,e_2)[m]\cdot y'}
        \quad\frac{x\xrightarrow{e_1}x'\quad y\xrightarrow{e_2}y'}{x\mid y\xrightarrow{\gamma(e_1,e_2)}\gamma(e_1,e_2)[m]\cdot x'\between y'}$$
        $$\frac{x\xrightarrow{e_1}e_1[m]\quad (\sharp(e_1,e_2))}{\Theta(x)\xrightarrow{e_1}e_1[m]} \quad\frac{x\xrightarrow{e_2}e_2[n]\quad (\sharp(e_1,e_2))}{\Theta(x)\xrightarrow{e_2}e_2[n]}$$
        $$\frac{x\xrightarrow{e_1}x'\quad (\sharp(e_1,e_2))}{\Theta(x)\xrightarrow{e_1}\Theta(x')} \quad\frac{x\xrightarrow{e_2}x'\quad (\sharp(e_1,e_2))}{\Theta(x)\xrightarrow{e_2}\Theta(x')}$$
        $$\frac{x\xrightarrow{e_1}e_1[m] \quad y\nrightarrow^{e_2}\quad (\sharp(e_1,e_2))}{x\triangleleft y\xrightarrow{\tau}\surd}
        \quad\frac{x\xrightarrow{e_1}x' \quad y\nrightarrow^{e_2}\quad (\sharp(e_1,e_2))}{x\triangleleft y\xrightarrow{\tau}x'}$$
        $$\frac{x\xrightarrow{e_1}e_1[m] \quad y\nrightarrow^{e_3}\quad (\sharp(e_1,e_2),e_2\leq e_3)}{x\triangleleft y\xrightarrow{e_1}e_1[m]}
        \quad\frac{x\xrightarrow{e_1}x' \quad y\nrightarrow^{e_3}\quad (\sharp(e_1,e_2),e_2\leq e_3)}{x\triangleleft y\xrightarrow{e_1}x'}$$
        $$\frac{x\xrightarrow{e_3}e_3[l] \quad y\nrightarrow^{e_2}\quad (\sharp(e_1,e_2),e_1\leq e_3)}{x\triangleleft y\xrightarrow{\tau}\surd}
        \quad\frac{x\xrightarrow{e_3}x' \quad y\nrightarrow^{e_2}\quad (\sharp(e_1,e_2),e_1\leq e_3)}{x\triangleleft y\xrightarrow{\tau}x'}$$
%
        \caption{Forward action transition rules of APRPTC}
        \label{TRForAPRPTC}
    \end{table}
\end{center}

\begin{center}
    \begin{table}
        $$\frac{x\xrightarrow{e_1}e_1[m]\quad (\sharp_{\pi}(e_1,e_2))}{\Theta(x)\xrightarrow{e_1}e_1[m]} \quad\frac{x\xrightarrow{e_2}e_2[n]\quad (\sharp_{\pi}(e_1,e_2))}{\Theta(x)\xrightarrow{e_2}e_2[n]}$$
        $$\frac{x\xrightarrow{e_1}x'\quad (\sharp_{\pi}(e_1,e_2))}{\Theta(x)\xrightarrow{e_1}\Theta(x')} \quad\frac{x\xrightarrow{e_2}x'\quad (\sharp_{\pi}(e_1,e_2))}{\Theta(x)\xrightarrow{e_2}\Theta(x')}$$
        $$\frac{x\xrightarrow{e_1}e_1[m] \quad y\nrightarrow^{e_2}\quad (\sharp_{\pi}(e_1,e_2))}{x\triangleleft y\xrightarrow{\tau}\surd}
        \quad\frac{x\xrightarrow{e_1}x' \quad y\nrightarrow^{e_2}\quad (\sharp_{\pi}(e_1,e_2))}{x\triangleleft y\xrightarrow{\tau}x'}$$
        $$\frac{x\xrightarrow{e_1}e_1[m] \quad y\nrightarrow^{e_3}\quad (\sharp_{\pi}(e_1,e_2),e_2\leq e_3)}{x\triangleleft y\xrightarrow{e_1}e_1[m]}
        \quad\frac{x\xrightarrow{e_1}x' \quad y\nrightarrow^{e_3}\quad (\sharp_{\pi}(e_1,e_2),e_2\leq e_3)}{x\triangleleft y\xrightarrow{e_1}x'}$$
        $$\frac{x\xrightarrow{e_3}e_3[l] \quad y\nrightarrow^{e_2}\quad (\sharp_{\pi}(e_1,e_2),e_1\leq e_3)}{x\triangleleft y\xrightarrow{\tau}\surd}
        \quad\frac{x\xrightarrow{e_3}x' \quad y\nrightarrow^{e_2}\quad (\sharp_{\pi}(e_1,e_2),e_1\leq e_3)}{x\triangleleft y\xrightarrow{\tau}x'}$$
        \caption{Forward action transition rules of APRPTC (continuing)}
        \label{TRForAPRPTC2}
    \end{table}
\end{center}

\begin{center}
    \begin{table}
        $$\frac{x\xtworightarrow{e_1[m]}e_1\quad y\xtworightarrow{e_2[m]}e_2}{x\parallel y\xtworightarrow{\{e_1[m],e_2[m]\}}e_1\parallel e_2}
        \quad\frac{x\xtworightarrow{e_1[m]}x'\quad y\xtworightarrow{e_2[m]}e_2}{x\parallel y\xtworightarrow{\{e_1[m],e_2[m]\}}x'\parallel e_2}$$
        $$\frac{x\xtworightarrow{e_1[m]}e_1\quad y\xtworightarrow{e_2[m]}y'}{x\parallel y\xtworightarrow{\{e_1[m],e_2[m]\}}e_1\parallel y'}
        \quad\frac{x\xtworightarrow{e_1[m]}x'\quad y\xtworightarrow{e_2[m]}y'}{x\parallel y\xtworightarrow{\{e_1[m],e_2[m]\}}x'\between y'}$$
        $$\frac{x\xtworightarrow{e_1[m]}e_1\quad y\xtworightarrow{e_2[m]}e_2 \quad(e_1\leq e_2)}{x\leftmerge y\xtworightarrow{\{e_1[m],e_2[m]\}}e_1\leftmerge e_2}
        \quad\frac{x\xtworightarrow{e_1[m]}x'\quad y\xtworightarrow{e_2[m]}e_2 \quad(e_1\leq e_2)}{x\leftmerge y\xtworightarrow{\{e_1[m],e_2[m]\}}x'\leftmerge e_2}$$
        $$\frac{x\xtworightarrow{e_1[m]}e_1\quad y\xtworightarrow{e_2[m]}y' \quad(e_1\leq e_2)}{x\leftmerge y\xtworightarrow{\{e_1[m],e_2[m]\}}e_1\leftmerge y'}
        \quad\frac{x\xtworightarrow{e_1[m]}x'\quad y\xtworightarrow{e_2[m]}y' \quad(e_1\leq e_2)}{x\leftmerge y\xtworightarrow{\{e_1[m],e_2[m]\}}x'\between y'}$$
        $$\frac{x\xtworightarrow{e_1[m]}e_1\quad y\xtworightarrow{e_2[m]}e_2}{x\mid y\xtworightarrow{\gamma(e_1,e_2)[m]}\gamma(e_1,e_2)} \quad\frac{x\xtworightarrow{e_1[m]}x'\quad y\xtworightarrow{e_2[m]}e_2}{x\mid y\xtworightarrow{\gamma(e_1,e_2)[m]}\gamma(e_1,e_2)\cdot x'}$$
        $$\frac{x\xtworightarrow{e_1[m]}e_1\quad y\xtworightarrow{e_2[m]}y'}{x\mid y\xtworightarrow{\gamma(e_1,e_2)[m]}\gamma(e_1,e_2)\cdot y'} \quad\frac{x\xtworightarrow{e_1[m]}x'\quad y\xtworightarrow{e_2[m]}y'}{x\mid y\xtworightarrow{\gamma(e_1,e_2)[m]}\gamma(e_1,e_2)\cdot x'\between y'}$$
        $$\frac{x\xtworightarrow{e_1[m]}e_1\quad (\sharp(e_1,e_2))}{\Theta(x)\xtworightarrow{e_1[m]}e_1} \quad\frac{x\xtworightarrow{e_2[n]}e_2\quad (\sharp(e_1,e_2))}{\Theta(x)\xtworightarrow{e_2[n]}e_2}$$
        $$\frac{x\xtworightarrow{e_1[m]}x'\quad (\sharp(e_1,e_2))}{\Theta(x)\xtworightarrow{e_1[m]}\Theta(x')} \quad\frac{x\xtworightarrow{e_2[n]}x'\quad (\sharp(e_1,e_2))}{\Theta(x)\xtworightarrow{e_2[n]}\Theta(x')}$$
        $$\frac{x\xtworightarrow{e_1[m]}e_1 \quad y\xntworightarrow{e_2[n]}\quad (\sharp(e_1,e_2))}{x\triangleleft y\xtworightarrow{\tau}\surd}
        \quad\frac{x\xtworightarrow{e_1[m]}x' \quad y\xntworightarrow{e_2[n]}\quad (\sharp(e_1,e_2))}{x\triangleleft y\xtworightarrow{\tau}x'}$$
        $$\frac{x\xtworightarrow{e_1[m]}e_1 \quad y\xntworightarrow{e_3[l]}\quad (\sharp(e_1,e_2),e_2\geq e_3)}{x\triangleleft y\xtworightarrow{e_1[m]}e_1}
        \quad\frac{x\xtworightarrow{e_1[m]}x' \quad y\xntworightarrow{e_3[l]}\quad (\sharp(e_1,e_2),e_2\geq e_3)}{x\triangleleft y\xtworightarrow{e_1[m]}x'}$$
        $$\frac{x\xtworightarrow{e_3[l]}e_3 \quad y\xntworightarrow{e_2[n]}\quad (\sharp(e_1,e_2),e_1\geq e_3)}{x\triangleleft y\xtworightarrow{\tau}\surd}
        \quad\frac{x\xtworightarrow{e_3[l]}x' \quad y\xntworightarrow{e_2[n]}\quad (\sharp(e_1,e_2),e_1\geq e_3)}{x\triangleleft y\xtworightarrow{\tau}x'}$$
%
        \caption{Reverse action transition rules of APRPTC}
        \label{RTRForAPRPTC}
    \end{table}
\end{center}

\begin{center}
    \begin{table}
        $$\frac{x\xtworightarrow{e_1[m]}e_1\quad (\sharp_{\pi}(e_1,e_2))}{\Theta(x)\xtworightarrow{e_1[m]}e_1} \quad\frac{x\xtworightarrow{e_2[n]}e_2\quad (\sharp_{\pi}(e_1,e_2))}{\Theta(x)\xtworightarrow{e_2[n]}e_2}$$
        $$\frac{x\xtworightarrow{e_1[m]}x'\quad (\sharp_{\pi}(e_1,e_2))}{\Theta(x)\xtworightarrow{e_1[m]}\Theta(x')} \quad\frac{x\xtworightarrow{e_2[n]}x'\quad (\sharp_{\pi}(e_1,e_2))}{\Theta(x)\xtworightarrow{e_2[n]}\Theta(x')}$$
        $$\frac{x\xtworightarrow{e_1[m]}e_1 \quad y\xntworightarrow{e_2[n]}\quad (\sharp_{\pi}(e_1,e_2))}{x\triangleleft y\xtworightarrow{\tau}\surd}
        \quad\frac{x\xtworightarrow{e_1[m]}x' \quad y\xntworightarrow{e_2[n]}\quad (\sharp_{\pi}(e_1,e_2))}{x\triangleleft y\xtworightarrow{\tau}x'}$$
        $$\frac{x\xtworightarrow{e_1[m]}e_1 \quad y\xntworightarrow{e_3[l]}\quad (\sharp_{\pi}(e_1,e_2),e_2\geq e_3)}{x\triangleleft y\xtworightarrow{e_1[m]}e_1}
        \quad\frac{x\xtworightarrow{e_1[m]}x' \quad y\xntworightarrow{e_3[l]}\quad (\sharp_{\pi}(e_1,e_2),e_2\geq e_3)}{x\triangleleft y\xtworightarrow{e_1[m]}x'}$$
        $$\frac{x\xtworightarrow{e_3[l]}e_3 \quad y\xntworightarrow{e_2[n]}\quad (\sharp_{\pi}(e_1,e_2),e_1\geq e_3)}{x\triangleleft y\xtworightarrow{\tau}\surd}
        \quad\frac{x\xtworightarrow{e_3[l]}x' \quad y\xntworightarrow{e_2[n]}\quad (\sharp_{\pi}(e_1,e_2),e_1\geq e_3)}{x\triangleleft y\xtworightarrow{\tau}x'}$$
        \caption{Reverse action transition rules of APRPTC}
        \label{RTRForAPRPTC2}
    \end{table}
\end{center}

\begin{theorem}[Generalization of the algebra for parallelism with respect to $BARPTC$]
The algebra for parallelism is a generalization of $BARPTC$.
\end{theorem}

\begin{proof}
It follows from the following three facts.

\begin{enumerate}
  \item The transition rules of $BARPTC$ in section \ref{barptc} are all source-dependent;
  \item The sources of the transition rules for the algebra for parallelism contain an occurrence of $\between$, or $\parallel$, or $\leftmerge$, or $\mid$, or $\Theta$, or $\triangleleft$;
  \item The transition rules of $APRPTC$ are all source-dependent.
\end{enumerate}

So, the algebra for parallelism is a generalization of $BARPTC$, that is, $BARPTC$ is an embedding of the algebra for parallelism, as desired.
\end{proof}

\begin{theorem}[Congruence of $APRPTC$ with respect to FR probabilistic pomset bisimulation equivalence]
FR probabilistic pomset bisimulation equivalence $\sim_{pp}^{fr}$ is a congruence with respect to $APRPTC$.
\end{theorem}

\begin{proof}
It is easy to see that FR probabilistic pomset bisimulation is an equivalent relation on $APRPTC$ terms, we only need to prove that $\sim_{pp}^{fr}$ is preserved by the operators
$\between$, $\parallel$, $\leftmerge$, $\mid$, $\Theta$ and $\triangleleft$.
That is, if $x\sim_{pp}^{fr} x'$ and $y\sim_{pp}^{fr}y'$, we need to prove that $x\between y\sim_{pp}^{fr}x'\between y'$, $x\parallel y\sim_{pp}^{fr}x'\parallel y'$, $x\leftmerge y\sim_{pp}^{fr}x'\leftmerge y'$,
$x\mid y\sim_{pp}^{fr}x'\mid y'$, $x\triangleleft y\sim_{pp}^{fr}x'\triangleleft y'$, and $\Theta(x)\sim_{pp}^{fr}\Theta(x')$. The proof is quite trivial and we omit it.
\end{proof}

\begin{theorem}[Soundness of parallelism modulo FR probabilistic pomset bisimulation equivalence]\label{SPPPBE}
Let $x$ and $y$ be $APRPTC$ terms. If $APRPTC\vdash x=y$, then $x\sim_{pp}^{fr} y$.
\end{theorem}

\begin{proof}
Since FR probabilistic pomset bisimulation $\sim_{pp}^{fr}$ is both an equivalent and a congruent relation with respect to the operators $\between$, $\parallel$, $\leftmerge$, $\mid$, $\Theta$ and
$\triangleleft$, we only need to check if each axiom in Table \ref{AxiomsForRPParallelism} is sound modulo FR probabilistic pomset bisimulation equivalence. The proof is quite trivial, and we omit it.
\end{proof}

\begin{theorem}[Completeness of parallelism modulo probabilistic pomset bisimulation equivalence]\label{CPPPBE}
Let $p$ and $q$ be closed $APRPTC$ terms, if $p\sim_{pp}^{fr} q$ then $p=q$.
\end{theorem}

\begin{proof}
Firstly, by the elimination theorem of $APRPTC$ (see Theorem \ref{ETPParallelism}), we know that for each closed $APRPTC$ term $p$, there exists a closed basic $APRPTC$ term $p'$,
such that $APRPTC\vdash p=p'$, so, we only need to consider closed basic $APRPTC$ terms.

The basic terms (see Definition \ref{BTAPRPTC}) modulo associativity and commutativity (AC) of conflict $+$ (defined by axioms $A1$ and $A2$ in Table \ref{AxiomsForBARPTC} and these
equivalences is denoted by $=_{AC}$. Then, each equivalence class $s$ modulo AC of $+$ has the following normal form

$$s_1\boxplus_{\pi_1}\cdots\boxplus_{\pi_{k-1}} s_k$$

with each $s_i$ has the following form

$$t_1+\cdots+ t_l$$

with each $t_j$ either an atomic event or of the form

$$u_1\cdot\cdots\cdot u_m$$

with each $u_l$ either an atomic event or of the form

$$v_1\leftmerge\cdots\leftmerge v_m$$

with each $v_m$ an atomic event, and each $t_j$ is called the summand of $s$.

Now, we prove that for normal forms $n$ and $n'$, if $n\sim_{pp}^{fr} n'$ then $n=_{AC}n'$. It is sufficient to induct on the sizes of $n$ and $n'$.

\begin{itemize}
  \item Consider a summand $e$ of $n$. Then $n\rightsquigarrow\breve{e}\xrightarrow{e}\breve{e}[m]$, so $n\sim_{pp}^{fr} n'$ implies $n'\rightsquigarrow\breve{e}\xrightarrow{e}\breve{e}[m]$, meaning that
  $n'$ also contains the summand $e$.
  \item Consider a summand $e[m]$ of $n$. Then $n\rightsquigarrow\breve{e}[m]\xtworightarrow{e[m]}\breve{e}$, so $n\sim_{pp}^{fr} n'$ implies $n'\rightsquigarrow\breve{e}\xtworightarrow{e[m]}\breve{e}$, meaning that
  $n'$ also contains the summand $e[m]$.
  \item Consider a summand $t_1\cdot t_2$ of $n$,
  \begin{itemize}
    \item if $t_1\equiv e'$, then $n\rightsquigarrow\breve{e'}\xrightarrow{e'}\breve{e'}[m]\cdot t_2$, so $n\sim_{pp}^{fr} n'$ implies $n'\rightsquigarrow\breve{e'}\xrightarrow{e'}\breve{e'}[m]\cdot t_2'$ with $t_2\sim_{pp}^{fr} t_2'$,
    meaning that $n'$ contains a summand $e'\cdot t_2'$. Since $t_2$ and $t_2'$ are normal forms and have sizes smaller than $n$ and $n'$, by the induction hypotheses if
    $t_2\sim_{pp}^{fr} t_2'$ then $t_2=_{AC} t_2'$;
    \item if $t_1\equiv e_1\leftmerge\cdots\leftmerge e_m$, then $n\rightsquigarrow\breve{e_1}\leftmerge\cdots\leftmerge\breve{e_m}\xrightarrow{\{e_1,\cdots,e_m\}}\breve{e_1}[m]\leftmerge\cdots\leftmerge \breve{e_m}[m]\cdot t_2$,
    so $n\sim_{pp}^{fr} n'$ implies $n'\rightsquigarrow\breve{e_1}\leftmerge\cdots\leftmerge\breve{e_m}\xrightarrow{\{e_1,\cdots,e_m\}}\breve{e_1}[m]\leftmerge\cdots\leftmerge \breve{e_m}[m]\cdot t_2'$ with $t_2\sim_{pp}^{fr} t_2'$, meaning that $n'$
    contains a summand $(e_1\leftmerge\cdots\leftmerge e_m)\cdot t_2'$. Since $t_2$ and $t_2'$ are normal forms and have sizes smaller than $n$ and $n'$, by the induction hypotheses
    if $t_2\sim_{pp}^{fr} t_2'$ then $t_2=_{AC} t_2'$.
  \end{itemize}
  \item Consider a summand $t_1\cdot t_2[m]$ of $n$,
  \begin{itemize}
    \item if $t_2\equiv e'[m]$, then $n\rightsquigarrow\breve{e'}[m]\xtworightarrow{e'[m]} t_1\cdot\breve{e'}$, so $n\sim_{pp}^{fr} n'$ implies $n'\rightsquigarrow\breve{e'}[m]\xtworightarrow{e'[m]} t_1'\cdot\breve{e'}$ with $t_1\sim_{pp}^{fr} t_1'$,
    meaning that $n'$ contains a summand $ t_1'\cdot e'[m]$. Since $t_1$ and $t_1'$ are normal forms and have sizes smaller than $n$ and $n'$, by the induction hypotheses if
    $t_1\sim_{pp}^{fr} t_1'$ then $t_1=_{AC} t_1'$;
    \item if $t_2\equiv e_1[m]\leftmerge\cdots\leftmerge e_m[m]$, then $n\rightsquigarrow\breve{e_1}[m]\leftmerge\cdots\leftmerge\breve{e_m}[m]\xtworightarrow{\{e_1[m],\cdots,e_m[m]\}} t_1\cdot\breve{e_1}\leftmerge\cdots\leftmerge \breve{e_m}$,
    so $n\sim_{pp}^{fr} n'$ implies $n'\rightsquigarrow\breve{e_1}[m]\leftmerge\cdots\leftmerge\breve{e_m}[m]\xtworightarrow{\{e_1[m],\cdots,e_m[m]\}} t_1'\cdot\breve{e_1}\leftmerge\cdots\leftmerge \breve{e_m}$ with $t_1\sim_{pp}^{fr} t_1'$, meaning that $n'$
    contains a summand $ t_1'\cdot(e_1[m]\leftmerge\cdots\leftmerge e_m[m])$. Since $t_1$ and $t_1'$ are normal forms and have sizes smaller than $n$ and $n'$, by the induction hypotheses
    if $t_1\sim_{pp}^{fr} t_1'$ then $t_1=_{AC} t_1'$.
  \end{itemize}
\end{itemize}

So, we get $n=_{AC} n'$.

Finally, let $s$ and $t$ be basic $APRPTC$ terms, and $s\sim_{pp}^{fr} t$, there are normal forms $n$ and $n'$, such that $s=n$ and $t=n'$. The soundness theorem of parallelism modulo
FR probabilistic pomset bisimulation equivalence yields $s\sim_{pp}^{fr} n$ and $t\sim_{pp}^{fr} n'$, so $n\sim_{pp}^{fr} s\sim_{pp}^{fr} t\sim_{pp}^{fr} n'$. Since if $n\sim_{pp}^{fr} n'$
then $n=_{AC}n'$, $s=n=_{AC}n'=t$, as desired.
\end{proof}

\begin{theorem}[Congruence of $APRPTC$ with respect to FR probabilistic step bisimulation equivalence]
FR probabilistic step bisimulation equivalence $\sim_{ps}^{fr}$ is a congruence with respect to $APRPTC$.
\end{theorem}

\begin{proof}
It is easy to see that FR probabilistic step bisimulation is an equivalent relation on $APRPTC$ terms, we only need to prove that $\sim_{ps}^{fr}$ is preserved by the operators
$\between$, $\parallel$, $\leftmerge$, $\mid$, $\Theta$ and $\triangleleft$.
That is, if $x\sim_{ps}^{fr} x'$ and $y\sim_{ps}^{fr}y'$, we need to prove that $x\between y\sim_{ps}^{fr}x'\between y'$, $x\parallel y\sim_{ps}^{fr}x'\parallel y'$, $x\leftmerge y\sim_{ps}^{fr}x'\leftmerge y'$,
$x\mid y\sim_{ps}^{fr}x'\mid y'$, $x\triangleleft y\sim_{ps}^{fr}x'\triangleleft y'$, and $\Theta(x)\sim_{ps}^{fr}\Theta(x')$. The proof is quite trivial and we omit it.
\end{proof}

\begin{theorem}[Soundness of parallelism modulo FR probabilistic step bisimulation equivalence]\label{SPPSBE}
Let $x$ and $y$ be $APRPTC$ terms. If $APRPTC\vdash x=y$, then $x\sim_{ps}^{fr} y$.
\end{theorem}

\begin{proof}
Since FR probabilistic step bisimulation $\sim_{ps}^{fr}$ is both an equivalent and a congruent relation with respect to the operators $\between$, $\parallel$, $\leftmerge$, $\mid$, $\Theta$ and
$\triangleleft$, we only need to check if each axiom in Table \ref{AxiomsForRPParallelism} is sound modulo FR probabilistic step bisimulation equivalence. The proof is quite trivial, and we omit it.
\end{proof}

\begin{theorem}[Completeness of parallelism modulo FR probabilistic step bisimulation equivalence]\label{CPPSBE}
Let $p$ and $q$ be closed $APRPTC$ terms, if $p\sim_{ps}^{fr} q$ then $p=q$.
\end{theorem}

\begin{proof}
Firstly, by the elimination theorem of $APRPTC$ (see Theorem \ref{ETPParallelism}), we know that for each closed $APRPTC$ term $p$, there exists a closed basic $APRPTC$ term $p'$,
such that $APRPTC\vdash p=p'$, so, we only need to consider closed basic $APRPTC$ terms.

The basic terms (see Definition \ref{BTAPRPTC}) modulo associativity and commutativity (AC) of conflict $+$ (defined by axioms $A1$ and $A2$ in Table \ref{AxiomsForBARPTC} and these
equivalences is denoted by $=_{AC}$. Then, each equivalence class $s$ modulo AC of $+$ has the following normal form

$$s_1\boxplus_{\pi_1}\cdots\boxplus_{\pi_{k-1}} s_k$$

with each $s_i$ has the following form

$$t_1+\cdots+ t_l$$

with each $t_j$ either an atomic event or of the form

$$u_1\cdot\cdots\cdot u_m$$

with each $u_l$ either an atomic event or of the form

$$v_1\leftmerge\cdots\leftmerge v_m$$

with each $v_m$ an atomic event, and each $t_j$ is called the summand of $s$.

Now, we prove that for normal forms $n$ and $n'$, if $n\sim_{ps}^{fr} n'$ then $n=_{AC}n'$. It is sufficient to induct on the sizes of $n$ and $n'$.

\begin{itemize}
  \item Consider a summand $e$ of $n$. Then $n\rightsquigarrow\breve{e}\xrightarrow{e}\breve{e}[m]$, so $n\sim_{ps}^{fr} n'$ implies $n'\rightsquigarrow\breve{e}\xrightarrow{e}\breve{e}[m]$, meaning that
  $n'$ also contains the summand $e$.
  \item Consider a summand $e[m]$ of $n$. Then $n\rightsquigarrow\breve{e}[m]\xtworightarrow{e[m]}\breve{e}$, so $n\sim_{ps}^{fr} n'$ implies $n'\rightsquigarrow\breve{e}\xtworightarrow{e[m]}\breve{e}$, meaning that
  $n'$ also contains the summand $e[m]$.
  \item Consider a summand $t_1\cdot t_2$ of $n$,
  \begin{itemize}
    \item if $t_1\equiv e'$, then $n\rightsquigarrow\breve{e'}\xrightarrow{e'}\breve{e'}[m]\cdot t_2$, so $n\sim_{ps}^{fr} n'$ implies $n'\rightsquigarrow\breve{e'}\xrightarrow{e'}\breve{e'}[m]\cdot t_2'$ with $t_2\sim_{ps}^{fr} t_2'$,
    meaning that $n'$ contains a summand $e'\cdot t_2'$. Since $t_2$ and $t_2'$ are normal forms and have sizes smaller than $n$ and $n'$, by the induction hypotheses if
    $t_2\sim_{ps}^{fr} t_2'$ then $t_2=_{AC} t_2'$;
    \item if $t_1\equiv e_1\leftmerge\cdots\leftmerge e_m$, then $n\rightsquigarrow\breve{e_1}\leftmerge\cdots\leftmerge\breve{e_m}\xrightarrow{\{e_1,\cdots,e_m\}}\breve{e_1}[m]\leftmerge\cdots\leftmerge \breve{e_m}[m]\cdot t_2$,
    so $n\sim_{ps}^{fr} n'$ implies $n'\rightsquigarrow\breve{e_1}\leftmerge\cdots\leftmerge\breve{e_m}\xrightarrow{\{e_1,\cdots,e_m\}}\breve{e_1}[m]\leftmerge\cdots\leftmerge \breve{e_m}[m]\cdot t_2'$ with $t_2\sim_{ps}^{fr} t_2'$, meaning that $n'$
    contains a summand $(e_1\leftmerge\cdots\leftmerge e_m)\cdot t_2'$. Since $t_2$ and $t_2'$ are normal forms and have sizes smaller than $n$ and $n'$, by the induction hypotheses
    if $t_2\sim_{ps}^{fr} t_2'$ then $t_2=_{AC} t_2'$.
  \end{itemize}
  \item Consider a summand $t_1\cdot t_2[m]$ of $n$,
  \begin{itemize}
    \item if $t_2\equiv e'[m]$, then $n\rightsquigarrow\breve{e'}[m]\xtworightarrow{e'[m]} t_1\cdot\breve{e'}$, so $n\sim_{ps}^{fr} n'$ implies $n'\rightsquigarrow\breve{e'}[m]\xtworightarrow{e'[m]} t_1'\cdot\breve{e'}$ with $t_1\sim_{ps}^{fr} t_1'$,
    meaning that $n'$ contains a summand $ t_1'\cdot e'[m]$. Since $t_1$ and $t_1'$ are normal forms and have sizes smaller than $n$ and $n'$, by the induction hypotheses if
    $t_1\sim_{ps}^{fr} t_1'$ then $t_1=_{AC} t_1'$;
    \item if $t_2\equiv e_1[m]\leftmerge\cdots\leftmerge e_m[m]$, then $n\rightsquigarrow\breve{e_1}[m]\leftmerge\cdots\leftmerge\breve{e_m}[m]\xtworightarrow{\{e_1[m],\cdots,e_m[m]\}} t_1\cdot\breve{e_1}\leftmerge\cdots\leftmerge \breve{e_m}$,
    so $n\sim_{ps}^{fr} n'$ implies $n'\rightsquigarrow\breve{e_1}[m]\leftmerge\cdots\leftmerge\breve{e_m}[m]\xtworightarrow{\{e_1[m],\cdots,e_m[m]\}} t_1'\cdot\breve{e_1}\leftmerge\cdots\leftmerge \breve{e_m}$ with $t_1\sim_{ps}^{fr} t_1'$, meaning that $n'$
    contains a summand $ t_1'\cdot(e_1[m]\leftmerge\cdots\leftmerge e_m[m])$. Since $t_1$ and $t_1'$ are normal forms and have sizes smaller than $n$ and $n'$, by the induction hypotheses
    if $t_1\sim_{ps}^{fr} t_1'$ then $t_1=_{AC} t_1'$.
  \end{itemize}
\end{itemize}

So, we get $n=_{AC} n'$.

Finally, let $s$ and $t$ be basic $APRPTC$ terms, and $s\sim_{ps}^{fr} t$, there are normal forms $n$ and $n'$, such that $s=n$ and $t=n'$. The soundness theorem of parallelism modulo
FR probabilistic step bisimulation equivalence yields $s\sim_{ps}^{fr} n$ and $t\sim_{ps}^{fr} n'$, so $n\sim_{ps}^{fr} s\sim_{ps}^{fr} t\sim_{ps}^{fr} n'$. Since if $n\sim_{ps}^{fr} n'$
then $n=_{AC}n'$, $s=n=_{AC}n'=t$, as desired.
\end{proof}

\begin{theorem}[Congruence of $APRPTC$ with respect to FR probabilistic hp-bisimulation equivalence]
FR probabilistic hp-bisimulation equivalence $\sim_{php}^{fr}$ is a congruence with respect to $APRPTC$.
\end{theorem}

\begin{proof}
It is easy to see that FR probabilistic hp-bisimulation is an equivalent relation on $APRPTC$ terms, we only need to prove that $\sim_{php}^{fr}$ is preserved by the operators
$\between$, $\parallel$, $\leftmerge$, $\mid$, $\Theta$ and $\triangleleft$.
That is, if $x\sim_{php}^{fr} x'$ and $y\sim_{php}^{fr}y'$, we need to prove that $x\between y\sim_{php}^{fr}x'\between y'$, $x\parallel y\sim_{php}^{fr}x'\parallel y'$, $x\leftmerge y\sim_{php}^{fr}x'\leftmerge y'$,
$x\mid y\sim_{php}^{fr}x'\mid y'$, $x\triangleleft y\sim_{php}^{fr}x'\triangleleft y'$, and $\Theta(x)\sim_{php}^{fr}\Theta(x')$. The proof is quite trivial and we omit it.
\end{proof}

\begin{theorem}[Soundness of parallelism modulo FR probabilistic hp-bisimulation equivalence]\label{SPPHPBE}
Let $x$ and $y$ be $APRPTC$ terms. If $APRPTC\vdash x=y$, then $x\sim_{php}^{fr} y$.
\end{theorem}

\begin{proof}
Since FR probabilistic hp-bisimulation $\sim_{php}^{fr}$ is both an equivalent and a congruent relation with respect to the operators $\between$, $\parallel$, $\leftmerge$, $\mid$, $\Theta$ and
$\triangleleft$, we only need to check if each axiom in Table \ref{AxiomsForRPParallelism} is sound modulo FR probabilistic hp-bisimulation equivalence. The proof is quite trivial, and we omit it.
\end{proof}

\begin{theorem}[Completeness of parallelism modulo FR probabilistic hp-bisimulation equivalence]\label{CPPHPBE}
Let $p$ and $q$ be closed $APRPTC$ terms, if $p\sim_{php}^{fr} q$ then $p=q$.
\end{theorem}

\begin{proof}
Firstly, by the elimination theorem of $APRPTC$ (see Theorem \ref{ETPParallelism}), we know that for each closed $APRPTC$ term $p$, there exists a closed basic $APRPTC$ term $p'$,
such that $APRPTC\vdash p=p'$, so, we only need to consider closed basic $APRPTC$ terms.

The basic terms (see Definition \ref{BTAPRPTC}) modulo associativity and commutativity (AC) of conflict $+$ (defined by axioms $A1$ and $A2$ in Table \ref{AxiomsForBARPTC} and these
equivalences is denoted by $=_{AC}$. Then, each equivalence class $s$ modulo AC of $+$ has the following normal form

$$s_1\boxplus_{\pi_1}\cdots\boxplus_{\pi_{k-1}} s_k$$

with each $s_i$ has the following form

$$t_1+\cdots+ t_l$$

with each $t_j$ either an atomic event or of the form

$$u_1\cdot\cdots\cdot u_m$$

with each $u_l$ either an atomic event or of the form

$$v_1\leftmerge\cdots\leftmerge v_m$$

with each $v_m$ an atomic event, and each $t_j$ is called the summand of $s$.

Now, we prove that for normal forms $n$ and $n'$, if $n\sim_{php}^{fr} n'$ then $n=_{AC}n'$. It is sufficient to induct on the sizes of $n$ and $n'$.

\begin{itemize}
  \item Consider a summand $e$ of $n$. Then $n\rightsquigarrow\breve{e}\xrightarrow{e}\breve{e}[m]$, so $n\sim_{php}^{fr} n'$ implies $n'\rightsquigarrow\breve{e}\xrightarrow{e}\breve{e}[m]$, meaning that
  $n'$ also contains the summand $e$.
  \item Consider a summand $e[m]$ of $n$. Then $n\rightsquigarrow\breve{e}[m]\xtworightarrow{e[m]}\breve{e}$, so $n\sim_{php}^{fr} n'$ implies $n'\rightsquigarrow\breve{e}\xtworightarrow{e[m]}\breve{e}$, meaning that
  $n'$ also contains the summand $e[m]$.
  \item Consider a summand $t_1\cdot t_2$ of $n$,
  \begin{itemize}
    \item if $t_1\equiv e'$, then $n\rightsquigarrow\breve{e'}\xrightarrow{e'}\breve{e'}[m]\cdot t_2$, so $n\sim_{php}^{fr} n'$ implies $n'\rightsquigarrow\breve{e'}\xrightarrow{e'}\breve{e'}[m]\cdot t_2'$ with $t_2\sim_{php}^{fr} t_2'$,
    meaning that $n'$ contains a summand $e'\cdot t_2'$. Since $t_2$ and $t_2'$ are normal forms and have sizes smaller than $n$ and $n'$, by the induction hypotheses if
    $t_2\sim_{php}^{fr} t_2'$ then $t_2=_{AC} t_2'$;
    \item if $t_1\equiv e_1\leftmerge\cdots\leftmerge e_m$, then $n\rightsquigarrow\breve{e_1}\leftmerge\cdots\leftmerge\breve{e_m}\xrightarrow{\{e_1,\cdots,e_m\}}\breve{e_1}[m]\leftmerge\cdots\leftmerge \breve{e_m}[m]\cdot t_2$,
    so $n\sim_{php}^{fr} n'$ implies $n'\rightsquigarrow\breve{e_1}\leftmerge\cdots\leftmerge\breve{e_m}\xrightarrow{\{e_1,\cdots,e_m\}}\breve{e_1}[m]\leftmerge\cdots\leftmerge \breve{e_m}[m]\cdot t_2'$ with $t_2\sim_{php}^{fr} t_2'$, meaning that $n'$
    contains a summand $(e_1\leftmerge\cdots\leftmerge e_m)\cdot t_2'$. Since $t_2$ and $t_2'$ are normal forms and have sizes smaller than $n$ and $n'$, by the induction hypotheses
    if $t_2\sim_{php}^{fr} t_2'$ then $t_2=_{AC} t_2'$.
  \end{itemize}
  \item Consider a summand $t_1\cdot t_2[m]$ of $n$,
  \begin{itemize}
    \item if $t_2\equiv e'[m]$, then $n\rightsquigarrow\breve{e'}[m]\xtworightarrow{e'[m]} t_1\cdot\breve{e'}$, so $n\sim_{php}^{fr} n'$ implies $n'\rightsquigarrow\breve{e'}[m]\xtworightarrow{e'[m]} t_1'\cdot\breve{e'}$ with $t_1\sim_{php}^{fr} t_1'$,
    meaning that $n'$ contains a summand $ t_1'\cdot e'[m]$. Since $t_1$ and $t_1'$ are normal forms and have sizes smaller than $n$ and $n'$, by the induction hypotheses if
    $t_1\sim_{php}^{fr} t_1'$ then $t_1=_{AC} t_1'$;
    \item if $t_2\equiv e_1[m]\leftmerge\cdots\leftmerge e_m[m]$, then $n\rightsquigarrow\breve{e_1}[m]\leftmerge\cdots\leftmerge\breve{e_m}[m]\xtworightarrow{\{e_1[m],\cdots,e_m[m]\}} t_1\cdot\breve{e_1}\leftmerge\cdots\leftmerge \breve{e_m}$,
    so $n\sim_{php}^{fr} n'$ implies $n'\rightsquigarrow\breve{e_1}[m]\leftmerge\cdots\leftmerge\breve{e_m}[m]\xtworightarrow{\{e_1[m],\cdots,e_m[m]\}} t_1'\cdot\breve{e_1}\leftmerge\cdots\leftmerge \breve{e_m}$ with $t_1\sim_{php}^{fr} t_1'$, meaning that $n'$
    contains a summand $ t_1'\cdot(e_1[m]\leftmerge\cdots\leftmerge e_m[m])$. Since $t_1$ and $t_1'$ are normal forms and have sizes smaller than $n$ and $n'$, by the induction hypotheses
    if $t_1\sim_{php}^{fr} t_1'$ then $t_1=_{AC} t_1'$.
  \end{itemize}
\end{itemize}

So, we get $n=_{AC} n'$.

Finally, let $s$ and $t$ be basic $APRPTC$ terms, and $s\sim_{php}^{fr} t$, there are normal forms $n$ and $n'$, such that $s=n$ and $t=n'$. The soundness theorem of parallelism modulo
FR probabilistic hp-bisimulation equivalence yields $s\sim_{php}^{fr} n$ and $t\sim_{php}^{fr} n'$, so $n\sim_{php}^{fr} s\sim_{php}^{fr} t\sim_{php}^{fr} n'$. Since if $n\sim_{php}^{fr} n'$
then $n=_{AC}n'$, $s=n=_{AC}n'=t$, as desired.
\end{proof}

\begin{theorem}[Congruence of $APRPTC$ with respect to FR probabilistic hhp-bisimulation equivalence]
FR probabilistic hhp-bisimulation equivalence $\sim_{phhp}^{fr}$ is a congruence with respect to $APRPTC$.
\end{theorem}

\begin{proof}
It is easy to see that FR probabilistic hhp-bisimulation is an equivalent relation on $APRPTC$ terms, we only need to prove that $\sim_{phhp}^{fr}$ is preserved by the operators
$\between$, $\parallel$, $\leftmerge$, $\mid$, $\Theta$ and $\triangleleft$.
That is, if $x\sim_{phhp}^{fr} x'$ and $y\sim_{phhp}^{fr}y'$, we need to prove that $x\between y\sim_{phhp}^{fr}x'\between y'$, $x\parallel y\sim_{phhp}^{fr}x'\parallel y'$, $x\leftmerge y\sim_{phhp}^{fr}x'\leftmerge y'$,
$x\mid y\sim_{phhp}^{fr}x'\mid y'$, $x\triangleleft y\sim_{phhp}^{fr}x'\triangleleft y'$, and $\Theta(x)\sim_{phhp}^{fr}\Theta(x')$. The proof is quite trivial and we omit it.
\end{proof}

\begin{theorem}[Soundness of parallelism modulo FR probabilistic hhp-bisimulation equivalence]\label{SPPHHPBE}
Let $x$ and $y$ be $APRPTC$ terms. If $APRPTC\vdash x=y$, then $x\sim_{phhp}^{fr} y$.
\end{theorem}

\begin{proof}
Since FR probabilistic hhp-bisimulation $\sim_{phhp}^{fr}$ is both an equivalent and a congruent relation with respect to the operators $\between$, $\parallel$, $\leftmerge$, $\mid$, $\Theta$ and
$\triangleleft$, we only need to check if each axiom in Table \ref{AxiomsForRPParallelism} is sound modulo FR probabilistic hhp-bisimulation equivalence. The proof is quite trivial, and we omit it.
\end{proof}

\begin{theorem}[Completeness of parallelism modulo FR probabilistic hhp-bisimulation equivalence]\label{CPPHHPBE}
Let $p$ and $q$ be closed $APRPTC$ terms, if $p\sim_{phhp}^{fr} q$ then $p=q$.
\end{theorem}

\begin{proof}
Firstly, by the elimination theorem of $APRPTC$ (see Theorem \ref{ETPParallelism}), we know that for each closed $APRPTC$ term $p$, there exists a closed basic $APRPTC$ term $p'$,
such that $APRPTC\vdash p=p'$, so, we only need to consider closed basic $APRPTC$ terms.

The basic terms (see Definition \ref{BTAPRPTC}) modulo associativity and commutativity (AC) of conflict $+$ (defined by axioms $A1$ and $A2$ in Table \ref{AxiomsForBARPTC} and these
equivalences is denoted by $=_{AC}$. Then, each equivalence class $s$ modulo AC of $+$ has the following normal form

$$s_1\boxplus_{\pi_1}\cdots\boxplus_{\pi_{k-1}} s_k$$

with each $s_i$ has the following form

$$t_1+\cdots+ t_l$$

with each $t_j$ either an atomic event or of the form

$$u_1\cdot\cdots\cdot u_m$$

with each $u_l$ either an atomic event or of the form

$$v_1\leftmerge\cdots\leftmerge v_m$$

with each $v_m$ an atomic event, and each $t_j$ is called the summand of $s$.

Now, we prove that for normal forms $n$ and $n'$, if $n\sim_{phhp}^{fr} n'$ then $n=_{AC}n'$. It is sufficient to induct on the sizes of $n$ and $n'$.

\begin{itemize}
  \item Consider a summand $e$ of $n$. Then $n\rightsquigarrow\breve{e}\xrightarrow{e}\breve{e}[m]$, so $n\sim_{phhp}^{fr} n'$ implies $n'\rightsquigarrow\breve{e}\xrightarrow{e}\breve{e}[m]$, meaning that
  $n'$ also contains the summand $e$.
  \item Consider a summand $e[m]$ of $n$. Then $n\rightsquigarrow\breve{e}[m]\xtworightarrow{e[m]}\breve{e}$, so $n\sim_{phhp}^{fr} n'$ implies $n'\rightsquigarrow\breve{e}\xtworightarrow{e[m]}\breve{e}$, meaning that
  $n'$ also contains the summand $e[m]$.
  \item Consider a summand $t_1\cdot t_2$ of $n$,
  \begin{itemize}
    \item if $t_1\equiv e'$, then $n\rightsquigarrow\breve{e'}\xrightarrow{e'}\breve{e'}[m]\cdot t_2$, so $n\sim_{phhp}^{fr} n'$ implies $n'\rightsquigarrow\breve{e'}\xrightarrow{e'}\breve{e'}[m]\cdot t_2'$ with $t_2\sim_{phhp}^{fr} t_2'$,
    meaning that $n'$ contains a summand $e'\cdot t_2'$. Since $t_2$ and $t_2'$ are normal forms and have sizes smaller than $n$ and $n'$, by the induction hypotheses if
    $t_2\sim_{phhp}^{fr} t_2'$ then $t_2=_{AC} t_2'$;
    \item if $t_1\equiv e_1\leftmerge\cdots\leftmerge e_m$, then $n\rightsquigarrow\breve{e_1}\leftmerge\cdots\leftmerge\breve{e_m}\xrightarrow{\{e_1,\cdots,e_m\}}\breve{e_1}[m]\leftmerge\cdots\leftmerge \breve{e_m}[m]\cdot t_2$,
    so $n\sim_{phhp}^{fr} n'$ implies $n'\rightsquigarrow\breve{e_1}\leftmerge\cdots\leftmerge\breve{e_m}\xrightarrow{\{e_1,\cdots,e_m\}}\breve{e_1}[m]\leftmerge\cdots\leftmerge \breve{e_m}[m]\cdot t_2'$ with $t_2\sim_{phhp}^{fr} t_2'$, meaning that $n'$
    contains a summand $(e_1\leftmerge\cdots\leftmerge e_m)\cdot t_2'$. Since $t_2$ and $t_2'$ are normal forms and have sizes smaller than $n$ and $n'$, by the induction hypotheses
    if $t_2\sim_{phhp}^{fr} t_2'$ then $t_2=_{AC} t_2'$.
  \end{itemize}
  \item Consider a summand $t_1\cdot t_2[m]$ of $n$,
  \begin{itemize}
    \item if $t_2\equiv e'[m]$, then $n\rightsquigarrow\breve{e'}[m]\xtworightarrow{e'[m]} t_1\cdot\breve{e'}$, so $n\sim_{phhp}^{fr} n'$ implies $n'\rightsquigarrow\breve{e'}[m]\xtworightarrow{e'[m]} t_1'\cdot\breve{e'}$ with $t_1\sim_{phhp}^{fr} t_1'$,
    meaning that $n'$ contains a summand $ t_1'\cdot e'[m]$. Since $t_1$ and $t_1'$ are normal forms and have sizes smaller than $n$ and $n'$, by the induction hypotheses if
    $t_1\sim_{phhp}^{fr} t_1'$ then $t_1=_{AC} t_1'$;
    \item if $t_2\equiv e_1[m]\leftmerge\cdots\leftmerge e_m[m]$, then $n\rightsquigarrow\breve{e_1}[m]\leftmerge\cdots\leftmerge\breve{e_m}[m]\xtworightarrow{\{e_1[m],\cdots,e_m[m]\}} t_1\cdot\breve{e_1}\leftmerge\cdots\leftmerge \breve{e_m}$,
    so $n\sim_{phhp}^{fr} n'$ implies $n'\rightsquigarrow\breve{e_1}[m]\leftmerge\cdots\leftmerge\breve{e_m}[m]\xtworightarrow{\{e_1[m],\cdots,e_m[m]\}} t_1'\cdot\breve{e_1}\leftmerge\cdots\leftmerge \breve{e_m}$ with $t_1\sim_{phhp}^{fr} t_1'$, meaning that $n'$
    contains a summand $ t_1'\cdot(e_1[m]\leftmerge\cdots\leftmerge e_m[m])$. Since $t_1$ and $t_1'$ are normal forms and have sizes smaller than $n$ and $n'$, by the induction hypotheses
    if $t_1\sim_{phhp}^{fr} t_1'$ then $t_1=_{AC} t_1'$.
  \end{itemize}
\end{itemize}

So, we get $n=_{AC} n'$.

Finally, let $s$ and $t$ be basic $APRPTC$ terms, and $s\sim_{phhp}^{fr} t$, there are normal forms $n$ and $n'$, such that $s=n$ and $t=n'$. The soundness theorem of parallelism modulo
FR probabilistic hhp-bisimulation equivalence yields $s\sim_{phhp}^{fr} n$ and $t\sim_{phhp}^{fr} n'$, so $n\sim_{phhp}^{fr} s\sim_{phhp}^{fr} t\sim_{phhp}^{fr} n'$. Since if $n\sim_{phhp}^{fr} n'$
then $n=_{AC}n'$, $s=n=_{AC}n'=t$, as desired.
\end{proof}

\subsubsection{Encapsulation}

The mismatch of two communicating events in different parallel branches can cause deadlock, so the deadlocks in the concurrent processes should be eliminated. Like $ACP$ \cite{ACP}, we
also introduce the unary encapsulation operator $\partial_H$ for set $H$ of atomic events, which renames all atomic events in $H$ into $\delta$. The whole algebra including parallelism
for true concurrency in the above subsections, deadlock $\delta$ and encapsulation operator $\partial_H$, is also called Algebra for Parallelism in Reversible Probabilistic True Concurrency, abbreviated $APRPTC$.

Firstly, we give the definition of PDFs in Table \ref{PDFEncap}.

\begin{center}
    \begin{table}
        $$\mu(\partial_H(x),\partial_H(x'))=\mu(x,x')$$
        $$\mu(x,y)=0,\textrm{otherwise}$$
        \caption{PDF definitions of $\partial_H$}
        \label{PDFEncap}
    \end{table}
\end{center}

The transition rules of encapsulation operator $\partial_H$ are shown in Table \ref{TRForPEncapsulation}.

\begin{center}
    \begin{table}
        $$\frac{x\rightsquigarrow x'}{\partial_H(x)\rightsquigarrow\partial_H(x')}$$
        $$\frac{x\xrightarrow{e}e[m]}{\partial_H(x)\xrightarrow{e}\partial_H(e[m])}\quad (e\notin H)
        \quad\frac{x\xrightarrow{e}x'}{\partial_H(x)\xrightarrow{e}\partial_H(x')}\quad(e\notin H)$$
        $$\frac{x\xtworightarrow{e[m]}e}{\partial_H(x)\xtworightarrow{e[m]}e}\quad (e\notin H)
        \quad\frac{x\xtworightarrow{e}x'}{\partial_H(x)\xtworightarrow{e}\partial_H(x')}\quad(e\notin H)$$
        \caption{Transition rules of encapsulation operator $\partial_H$}
        \label{TRForPEncapsulation}
    \end{table}
\end{center}

Based on the transition rules for encapsulation operator $\partial_H$ in Table \ref{TRForPEncapsulation}, we design the axioms as Table \ref{AxiomsForPEncapsulation} shows.

\begin{center}
    \begin{table}
        \begin{tabular}{@{}ll@{}}
            \hline No. &Axiom\\
            $D1$ & $e\notin H\quad\partial_H(e) = e$\\
            $RD1$ & $e\notin H\quad\partial_H(e[m]) = e[m]$\\
            $D2$ & $e\in H\quad \partial_H(e) = \delta$\\
            $RD2$ & $e\in H\quad \partial_H(e[m]) = \delta$\\
            $D3$ & $\partial_H(\delta) = \delta$\\
            $D4$ & $\partial_H(x+ y) = \partial_H(x)+\partial_H(y)$\\
            $D5$ & $\partial_H(x\cdot y) = \partial_H(x)\cdot\partial_H(y)$\\
            $D6$ & $\partial_H(x\leftmerge y) = \partial_H(x)\leftmerge\partial_H(y)$\\
            $PD1$ & $\partial_H(x\boxplus_{\pi}y)=\partial_H(x)\boxplus_{\pi}\partial_H(y)$\\
        \end{tabular}
        \caption{Axioms of encapsulation operator}
        \label{AxiomsForPEncapsulation}
    \end{table}
\end{center}


\begin{theorem}[Conservativity of $APRPTC$ with respect to the algebra for parallelism]
$APRPTC$ is a conservative extension of the algebra for parallelism.
\end{theorem}

\begin{proof}
It follows from the following two facts (see Theorem \ref{TCE}).

\begin{enumerate}
  \item The transition rules of the algebra for parallelism in the above subsections are all source-dependent;
  \item The sources of the transition rules for the encapsulation operator contain an occurrence of $\partial_H$.
\end{enumerate}

So, $APRPTC$ is a conservative extension of the algebra for parallelism, as desired.
\end{proof}

\begin{theorem}[Elimination theorem of $APRPTC$]\label{ETPEncapsulation}
Let $p$ be a closed $APRPTC$ term including the encapsulation operator $\partial_H$. Then there is a basic $APRPTC$ term $q$ such that $APRPTC\vdash p=q$.
\end{theorem}

\begin{proof}
(1) Firstly, suppose that the following ordering on the signature of $APRPTC$ is defined: $\leftmerge > \cdot > +>\boxplus_{\pi}$ and the symbol $\leftmerge$ is given the
lexicographical status for the first argument, then for each rewrite rule $p\rightarrow q$ in Table \ref{TRSForPEncapsulation} relation $p>_{lpo} q$ can easily be proved. We obtain
that the term rewrite system shown in Table \ref{TRSForPEncapsulation} is strongly normalizing, for it has finitely many rewriting rules, and $>$ is a well-founded ordering on the
signature of $APRPTC$, and if $s>_{lpo} t$, for each rewriting rule $s\rightarrow t$ is in Table \ref{TRSForPEncapsulation} (see Theorem \ref{SN}).

\begin{center}
    \begin{table}
        \begin{tabular}{@{}ll@{}}
            \hline No. &Rewriting Rule\\
            $RD1$ & $e\notin H\quad\partial_H(e) \rightarrow e$\\
            $RRD1$ & $e\notin H\quad\partial_H(e[m]) \rightarrow e[m]$\\
            $RD2$ & $e\in H\quad \partial_H(e) \rightarrow \delta$\\
            $RRD2$ & $e\in H\quad \partial_H(e[m]) \rightarrow \delta$\\
            $RD3$ & $\partial_H(\delta) \rightarrow \delta$\\
            $RD4$ & $\partial_H(x+ y) \rightarrow \partial_H(x)+\partial_H(y)$\\
            $RD5$ & $\partial_H(x\cdot y) \rightarrow \partial_H(x)\cdot\partial_H(y)$\\
            $RD6$ & $\partial_H(x\leftmerge y) \rightarrow \partial_H(x)\leftmerge\partial_H(y)$\\
            $RPD1$ & $\partial_H(x\boxplus_{\pi}y)\rightarrow\partial_H(x)\boxplus_{\pi}\partial_H(y)$\\
        \end{tabular}
        \caption{Term rewrite system of encapsulation operator $\partial_H$}
        \label{TRSForPEncapsulation}
    \end{table}
\end{center}

(2) Then we prove that the normal forms of closed $APRPTC$ terms including encapsulation operator $\partial_H$ are basic $APRPTC$ terms.

Suppose that $p$ is a normal form of some closed $APRPTC$ term and suppose that $p$ is not a basic $APRPTC$ term. Let $p'$ denote the smallest sub-term of $p$ which is not a
basic $APRPTC$ term. It implies that each sub-term of $p'$ is a basic $APRPTC$ term. Then we prove that $p$ is not a term in normal form. It is sufficient to induct on the structure of
$p'$, following from Theorem \ref{ETPParallelism}, we only prove the new case $p'\equiv \partial_H(p_1)$:

\begin{itemize}
  \item Case $p_1\equiv e$. The transition rules $RD1$ or $RD2$ can be applied, so $p$ is not a normal form;
  \item Case $p_1\equiv e[m]$. The transition rules $RRD1$ or $RRD2$ can be applied, so $p$ is not a normal form;
  \item Case $p_1\equiv \delta$. The transition rules $RD3$ can be applied, so $p$ is not a normal form;
  \item Case $p_1\equiv p_1'+ p_1''$. The transition rules $RD4$ can be applied, so $p$ is not a normal form;
  \item Case $p_1\equiv p_1'\cdot p_1''$. The transition rules $RD5$ can be applied, so $p$ is not a normal form;
  \item Case $p_1\equiv p_1'\leftmerge p_1''$. The transition rules $RD6$ can be applied, so $p$ is not a normal form;
  \item Case $p_1\equiv p_1'\boxplus_{\pi} p_1''$. The transition rules $RPD1$ can be applied, so $p$ is not a normal form.
\end{itemize}
\end{proof}

\begin{theorem}[Congruence theorem of encapsulation operator $\partial_H$ with respect to FR probabilistic pomset bisimulation equivalence]
FR probabilistic pomset bisimulation equivalence $\sim_{pp}^{fr}$ is a congruence with respect to encapsulation operator $\partial_H$.
\end{theorem}

\begin{proof}
It is easy to see that FR probabilistic pomset bisimulation is an equivalent relation on $APRPTC$ terms, we only need to prove that $\sim_{pp}^{fr}$ is preserved by the operators $\partial_H$.
That is, if $x\sim_{pp}^{fr} x'$, we need to prove that $\partial_H(x)\sim_{pp}^{fr}\partial_H(x')$. The proof is quite trivial and we omit it.
\end{proof}

\begin{theorem}[Soundness of $APRPTC$ modulo FR probabilistic pomset bisimulation equivalence]\label{SAPRPTCPPBE}
Let $x$ and $y$ be $APRPTC$ terms including encapsulation operator $\partial_H$. If $APRPTC\vdash x=y$, then $x\sim_{pp}^{fr} y$.
\end{theorem}

\begin{proof}
Since FR probabilistic pomset bisimulation $\sim_{pp}^{fr}$ is both an equivalent and a congruent relation with respect to the operator $\partial_H$, we only need to check if each axiom in
Table \ref{AxiomsForPEncapsulation} is sound modulo FR probabilistic pomset bisimulation equivalence. The proof is quite trivial and we omit it.
\end{proof}

\begin{theorem}[Completeness of $APRPTC$ modulo FR probabilistic pomset bisimulation equivalence]\label{CAPRPTCPPBE}
Let $p$ and $q$ be closed $APRPTC$ terms including encapsulation operator $\partial_H$, if $p\sim_{pp}^{fr} q$ then $p=q$.
\end{theorem}

\begin{proof}
Firstly, by the elimination theorem of $APRPTC$ (see Theorem \ref{ETPEncapsulation}), we know that the normal form of $APRPTC$ does not contain $\partial_H$, and for each closed $APRPTC$
term $p$, there exists a closed basic $APRPTC$ term $p'$, such that $APRPTC\vdash p=p'$, so, we only need to consider closed basic $APRPTC$ terms.

Similarly to Theorem \ref{CPPPBE}, we can prove that for normal forms $n$ and $n'$, if $n\sim_{pp}^{fr} n'$ then $n=_{AC}n'$.

Finally, let $s$ and $t$ be basic $APRPTC$ terms, and $s\sim_{pp}^{fr} t$, there are normal forms $n$ and $n'$, such that $s=n$ and $t=n'$. The soundness theorem of $APRPTC$ modulo FR probabilistic
pomset bisimulation equivalence (see Theorem \ref{SAPRPTCPPBE}) yields $s\sim_{pp}^{fr} n$ and $t\sim_{pp}^{fr} n'$, so $n\sim_{pp}^{fr} s\sim_{pp}^{fr} t\sim_{pp}^{fr} n'$. Since if $n\sim_{pp}^{fr} n'$ then
$n=_{AC}n'$, $s=n=_{AC}n'=t$, as desired.
\end{proof}

\begin{theorem}[Congruence theorem of encapsulation operator $\partial_H$ with respect to FR probabilistic step bisimulation equivalence]
FR probabilistic step bisimulation equivalence $\sim_{ps}^{fr}$ is a congruence with respect to encapsulation operator $\partial_H$.
\end{theorem}

\begin{proof}
It is easy to see that FR probabilistic step bisimulation is an equivalent relation on $APRPTC$ terms, we only need to prove that $\sim_{ps}^{fr}$ is preserved by the operators $\partial_H$.
That is, if $x\sim_{ps}^{fr} x'$, we need to prove that $\partial_H(x)\sim_{ps}^{fr}\partial_H(x')$. The proof is quite trivial and we omit it.
\end{proof}

\begin{theorem}[Soundness of $APRPTC$ modulo FR probabilistic step bisimulation equivalence]\label{SAPRPTCPSBE}
Let $x$ and $y$ be $APRPTC$ terms including encapsulation operator $\partial_H$. If $APRPTC\vdash x=y$, then $x\sim_{ps}^{fr} y$.
\end{theorem}

\begin{proof}
Since FR probabilistic step bisimulation $\sim_{ps}^{fr}$ is both an equivalent and a congruent relation with respect to the operator $\partial_H$, we only need to check if each axiom in
Table \ref{AxiomsForPEncapsulation} is sound modulo FR probabilistic step bisimulation equivalence. The proof is quite trivial and we omit it.
\end{proof}

\begin{theorem}[Completeness of $APRPTC$ modulo FR probabilistic step bisimulation equivalence]\label{CAPRPTCPSBE}
Let $p$ and $q$ be closed $APRPTC$ terms including encapsulation operator $\partial_H$, if $p\sim_{ps}^{fr} q$ then $p=q$.
\end{theorem}

\begin{proof}
Firstly, by the elimination theorem of $APRPTC$ (see Theorem \ref{ETPEncapsulation}), we know that the normal form of $APRPTC$ does not contain $\partial_H$, and for each closed $APRPTC$
term $p$, there exists a closed basic $APRPTC$ term $p'$, such that $APRPTC\vdash p=p'$, so, we only need to consider closed basic $APRPTC$ terms.

Similarly to Theorem \ref{CPPSBE}, we can prove that for normal forms $n$ and $n'$, if $n\sim_{ps}^{fr} n'$ then $n=_{AC}n'$.

Finally, let $s$ and $t$ be basic $APRPTC$ terms, and $s\sim_{ps}^{fr} t$, there are normal forms $n$ and $n'$, such that $s=n$ and $t=n'$. The soundness theorem of $APRPTC$ modulo FR probabilistic
step bisimulation equivalence (see Theorem \ref{SAPRPTCPSBE}) yields $s\sim_{ps}^{fr} n$ and $t\sim_{ps}^{fr} n'$, so $n\sim_{ps}^{fr} s\sim_{ps}^{fr} t\sim_{ps}^{fr} n'$. Since if $n\sim_{ps}^{fr} n'$ then
$n=_{AC}n'$, $s=n=_{AC}n'=t$, as desired.
\end{proof}

\begin{theorem}[Congruence theorem of encapsulation operator $\partial_H$ with respect to FR probabilistic hp-bisimulation equivalence]
FR probabilistic hp-bisimulation equivalence $\sim_{php}^{fr}$ is a congruence with respect to encapsulation operator $\partial_H$.
\end{theorem}

\begin{proof}
It is easy to see that FR probabilistic hp-bisimulation is an equivalent relation on $APRPTC$ terms, we only need to prove that $\sim_{php}^{fr}$ is preserved by the operators $\partial_H$.
That is, if $x\sim_{php}^{fr} x'$, we need to prove that $\partial_H(x)\sim_{php}^{fr}\partial_H(x')$. The proof is quite trivial and we omit it.
\end{proof}

\begin{theorem}[Soundness of $APRPTC$ modulo FR probabilistic hp-bisimulation equivalence]\label{SAPRPTCPHPBE}
Let $x$ and $y$ be $APRPTC$ terms including encapsulation operator $\partial_H$. If $APRPTC\vdash x=y$, then $x\sim_{php}^{fr} y$.
\end{theorem}

\begin{proof}
Since FR probabilistic hp-bisimulation $\sim_{php}^{fr}$ is both an equivalent and a congruent relation with respect to the operator $\partial_H$, we only need to check if each axiom in
Table \ref{AxiomsForPEncapsulation} is sound modulo FR probabilistic hp-bisimulation equivalence. The proof is quite trivial and we omit it.
\end{proof}

\begin{theorem}[Completeness of $APRPTC$ modulo FR probabilistic hp-bisimulation equivalence]\label{CAPRPTCPHPBE}
Let $p$ and $q$ be closed $APRPTC$ terms including encapsulation operator $\partial_H$, if $p\sim_{php}^{fr} q$ then $p=q$.
\end{theorem}

\begin{proof}
Firstly, by the elimination theorem of $APRPTC$ (see Theorem \ref{ETPEncapsulation}), we know that the normal form of $APRPTC$ does not contain $\partial_H$, and for each closed $APRPTC$
term $p$, there exists a closed basic $APRPTC$ term $p'$, such that $APRPTC\vdash p=p'$, so, we only need to consider closed basic $APRPTC$ terms.

Similarly to Theorem \ref{CPPHPBE}, we can prove that for normal forms $n$ and $n'$, if $n\sim_{php}^{fr} n'$ then $n=_{AC}n'$.

Finally, let $s$ and $t$ be basic $APRPTC$ terms, and $s\sim_{php}^{fr} t$, there are normal forms $n$ and $n'$, such that $s=n$ and $t=n'$. The soundness theorem of $APRPTC$ modulo FR probabilistic
hp-bisimulation equivalence (see Theorem \ref{SAPRPTCPHPBE}) yields $s\sim_{php}^{fr} n$ and $t\sim_{php}^{fr} n'$, so $n\sim_{php}^{fr} s\sim_{php}^{fr} t\sim_{php}^{fr} n'$. Since if $n\sim_{php}^{fr} n'$ then
$n=_{AC}n'$, $s=n=_{AC}n'=t$, as desired.
\end{proof}

\begin{theorem}[Congruence theorem of encapsulation operator $\partial_H$ with respect to FR probabilistic hhp-bisimulation equivalence]
FR probabilistic hhp-bisimulation equivalence $\sim_{phhp}^{fr}$ is a congruence with respect to encapsulation operator $\partial_H$.
\end{theorem}

\begin{proof}
It is easy to see that FR probabilistic hhp-bisimulation is an equivalent relation on $APRPTC$ terms, we only need to prove that $\sim_{phhp}^{fr}$ is preserved by the operators $\partial_H$.
That is, if $x\sim_{phhp}^{fr} x'$, we need to prove that $\partial_H(x)\sim_{phhp}^{fr}\partial_H(x')$. The proof is quite trivial and we omit it.
\end{proof}

\begin{theorem}[Soundness of $APRPTC$ modulo FR probabilistic hhp-bisimulation equivalence]\label{SAPRPTCPHHPBE}
Let $x$ and $y$ be $APRPTC$ terms including encapsulation operator $\partial_H$. If $APRPTC\vdash x=y$, then $x\sim_{phhp}^{fr} y$.
\end{theorem}

\begin{proof}
Since FR probabilistic hhp-bisimulation $\sim_{phhp}^{fr}$ is both an equivalent and a congruent relation with respect to the operator $\partial_H$, we only need to check if each axiom in
Table \ref{AxiomsForPEncapsulation} is sound modulo FR probabilistic hhp-bisimulation equivalence. The proof is quite trivial and we omit it.
\end{proof}

\begin{theorem}[Completeness of $APRPTC$ modulo FR probabilistic hhp-bisimulation equivalence]\label{CAPRPTCPHHPBE}
Let $p$ and $q$ be closed $APRPTC$ terms including encapsulation operator $\partial_H$, if $p\sim_{phhp}^{fr} q$ then $p=q$.
\end{theorem}

\begin{proof}
Firstly, by the elimination theorem of $APRPTC$ (see Theorem \ref{ETPEncapsulation}), we know that the normal form of $APRPTC$ does not contain $\partial_H$, and for each closed $APRPTC$
term $p$, there exists a closed basic $APRPTC$ term $p'$, such that $APRPTC\vdash p=p'$, so, we only need to consider closed basic $APRPTC$ terms.

Similarly to Theorem \ref{CPPHHPBE}, we can prove that for normal forms $n$ and $n'$, if $n\sim_{phhp}^{fr} n'$ then $n=_{AC}n'$.

Finally, let $s$ and $t$ be basic $APRPTC$ terms, and $s\sim_{phhp}^{fr} t$, there are normal forms $n$ and $n'$, such that $s=n$ and $t=n'$. The soundness theorem of $APRPTC$ modulo FR probabilistic
hhp-bisimulation equivalence (see Theorem \ref{SAPRPTCPHHPBE}) yields $s\sim_{phhp}^{fr} n$ and $t\sim_{phhp}^{fr} n'$, so $n\sim_{phhp}^{fr} s\sim_{phhp}^{fr} t\sim_{phhp}^{fr} n'$. Since if $n\sim_{phhp}^{fr} n'$ then
$n=_{AC}n'$, $s=n=_{AC}n'=t$, as desired.
\end{proof}

\subsection{Recursion}\label{rprec}

In this section, we introduce recursion to capture infinite processes based on $APRPTC$. Since in $APRPTC$, there are four basic operators $\cdot$, $+$, $\boxplus_{\pi}$ and $\leftmerge$,
the recursion must be adapted this situation to include $\boxplus_{\pi}$ and $\leftmerge$.

In the following, $E,F,G$ are recursion specifications, $X,Y,Z$ are recursive variables.

\subsubsection{Guarded Recursive Specifications}

\begin{definition}[Guarded recursive specification]
A recursive specification

$$X_1=t_1(X_1,\cdots,X_n)$$
$$...$$
$$X_n=t_n(X_1,\cdots,X_n)$$

is guarded if the right-hand sides of its recursive equations can be adapted to the form by applications of the axioms in $APRPTC$ and replacing recursion variables by the right-hand
sides of their recursive equations,

$((a_{111}\leftmerge\cdots\leftmerge a_{11i_1})\cdot s_1(X_1,\cdots,X_n)+\cdots+(a_{1k1}\leftmerge\cdots\leftmerge a_{1ki_k})\cdot s_k(X_1,\cdots,X_n)+(b_{111}\leftmerge\cdots\leftmerge
b_{11j_1})+\cdots+(b_{11j_1}\leftmerge\cdots\leftmerge b_{1lj_l}))\boxplus_{\pi_1}\cdots\boxplus_{\pi_{m-1}}((a_{m11}\leftmerge\cdots\leftmerge a_{m1i_1})\cdot s_1(X_1,\cdots,X_n)+
\cdots+(a_{mk1}\leftmerge\cdots\leftmerge a_{mki_k})\cdot s_k(X_1,\cdots,X_n)+(b_{m11}\leftmerge\cdots\leftmerge b_{m1j_1})+\cdots+(b_{m1j_1}\leftmerge\cdots\leftmerge b_{mlj_l}))$

where $a_{111},\cdots,a_{11i_1},a_{1k1},\cdots,a_{1ki_k},b_{111},\cdots,b_{11j_1},b_{11j_1},\cdots,b_{1lj_l},\cdots, a_{m11},\cdots,a_{m1i_1},a_{mk1},\cdots,a_{mki_k},\\b_{m11},\cdots,
b_{m1j_1},b_{m1j_1},\cdots,b_{mlj_l}\in \mathbb{E}$, and the sum above is allowed to be empty, in which case it represents the deadlock $\delta$.
\end{definition}

\begin{definition}[Linear recursive specification]\label{LRS}
A recursive specification is linear if its recursive equations are of the form

$((a_{111}\leftmerge\cdots\leftmerge a_{11i_1})X_1+\cdots+(a_{1k1}\leftmerge\cdots\leftmerge a_{1ki_k})X_k+(b_{111}\leftmerge\cdots\leftmerge b_{11j_1})+\cdots+(b_{11j_1}\leftmerge\cdots
\leftmerge b_{1lj_l}))\boxplus_{\pi_1}\cdots\boxplus_{\pi_{m-1}}((a_{m11}\leftmerge\cdots\leftmerge a_{m1i_1})X_1+\cdots+(a_{mk1}\leftmerge\cdots\leftmerge a_{mki_k})X_k+
(b_{m11}\leftmerge\cdots\leftmerge b_{m1j_1})+\cdots+(b_{m1j_1}\leftmerge\cdots\leftmerge b_{mlj_l}))$

where $a_{111},\cdots,a_{11i_1},a_{1k1},\cdots,a_{1ki_k},b_{111},\cdots,b_{11j_1},b_{11j_1},\cdots,b_{1lj_l},\cdots,a_{m11},\cdots,a_{m1i_1},a_{mk1},\cdots,a_{mki_k},\\b_{m11},\cdots,
b_{m1j_1},b_{m1j_1},\cdots,b_{mlj_l}\in \mathbb{E}$, and the sum above is allowed to be empty, in which case it represents the deadlock $\delta$.
\end{definition}

Firstly, we give the definition of PDFs in Table \ref{PDFGR}.

\begin{center}
    \begin{table}
        $$\mu(\langle X|E\rangle,y)=\mu(\langle t_X|E\rangle,y)$$
        $$\mu(x,y)=0,\textrm{otherwise}$$
        \caption{PDF definitions of recursion}
        \label{PDFGR}
    \end{table}
\end{center}

For a guarded recursive specifications $E$ with the form

$$X_1=t_1(X_1,\cdots,X_n)$$
$$\cdots$$
$$X_n=t_n(X_1,\cdots,X_n)$$

the behavior of the solution $\langle X_i|E\rangle$ for the recursion variable $X_i$ in $E$, where $i\in\{1,\cdots,n\}$, is exactly the behavior of their right-hand sides
$t_i(X_1,\cdots,X_n)$, which is captured by the two transition rules in Table \ref{TRForPGR}.

\begin{center}
    \begin{table}
        $$\frac{t_i(\langle X_1|E\rangle,\cdots,\langle X_n|E\rangle)\rightsquigarrow y}{\langle X_i|E\rangle\rightsquigarrow y}$$
        $$\frac{t_i(\langle X_1|E\rangle,\cdots,\langle X_n|E\rangle)\xrightarrow{\{e_1,\cdots,e_k\}}e_1[m]\leftmerge\cdots\leftmerge e_k[m]}{\langle X_i|E\rangle\xrightarrow{\{e_1,\cdots,e_k\}}e_1[m]\leftmerge\cdots\leftmerge e_k[m]}$$
        $$\frac{t_i(\langle X_1|E\rangle,\cdots,\langle X_n|E\rangle)\xrightarrow{\{e_1,\cdots,e_k\}} y}{\langle X_i|E\rangle\xrightarrow{\{e_1,\cdots,e_k\}} y}$$
        $$\frac{t_i(\langle X_1|E\rangle,\cdots,\langle X_n|E\rangle)\xtworightarrow{\{e_1[m],\cdots,e_k[m]\}}e_1\leftmerge\cdots\leftmerge e_k}{\langle X_i|E\rangle\xtworightarrow{\{e_1[m],\cdots,e_k[m]\}}e_1\leftmerge\cdots\leftmerge e_k}$$
        $$\frac{t_i(\langle X_1|E\rangle,\cdots,\langle X_n|E\rangle)\xtworightarrow{\{e_1[m],\cdots,e_k[m]\}} y}{\langle X_i|E\rangle\xtworightarrow{\{e_1[m],\cdots,e_k[m]\}} y}$$
        \caption{Transition rules of guarded recursion}
        \label{TRForPGR}
    \end{table}
\end{center}

\begin{theorem}[Conservitivity of $APRPTC$ with guarded recursion]
$APRPTC$ with guarded recursion is a conservative extension of $APRPTC$.
\end{theorem}

\begin{proof}
Since the transition rules of $APRPTC$ are source-dependent, and the transition rules for guarded recursion in Table \ref{TRForPGR} contain only a fresh constant in their source, so the
transition rules of $APRPTC$ with guarded recursion are a conservative extension of those of $APRPTC$.
\end{proof}

\begin{theorem}[Congruence theorem of $APRPTC$ with guarded recursion]
FR probabilistic truly concurrent bisimulation equivalences $\sim_{pp}^{fr}$, $\sim_{ps}^{fr}$, $\sim_{php}^{fr}$ and $\sim_{phhp}^{fr}$ are all congruences with respect to $APRPTC$ with guarded recursion.
\end{theorem}

\begin{proof}
It follows the following two facts:
\begin{enumerate}
  \item in a guarded recursive specification, right-hand sides of its recursive equations can be adapted to the form by applications of the axioms in $APRPTC$ and replacing recursion
  variables by the right-hand sides of their recursive equations;
  \item FR truly concurrent bisimulation equivalences $\sim_{pp}^{fr}$, $\sim_{ps}^{fr}$, $\sim_{php}^{fr}$ and $\sim_{phhp}^{fr}$ are all congruences with respect to all operators of $APRPTC$.
\end{enumerate}
\end{proof}

\subsubsection{Recursive Definition and Specification Principles}

The $RDP$ (Recursive Definition Principle) and the $RSP$ (Recursive Specification Principle) are shown in Table \ref{PRDPRSP}.

\begin{center}
\begin{table}
  \begin{tabular}{@{}ll@{}}
\hline No. &Axiom\\
  $RDP$ & $\langle X_i|E\rangle = t_i(\langle X_1|E\rangle,\cdots,\langle X_n|E\rangle)\quad (i\in\{1,\cdots,n\})$\\
  $RSP$ & if $y_i=t_i(y_1,\cdots,y_n)$ for $i\in\{1,\cdots,n\}$, then $y_i=\langle X_i|E\rangle \quad(i\in\{1,\cdots,n\})$\\
\end{tabular}
\caption{Recursive definition and specification principle}
\label{PRDPRSP}
\end{table}
\end{center}

$RDP$ follows immediately from the two transition rules for guarded recursion, which express that $\langle X_i|E\rangle$ and $t_i(\langle X_1|E\rangle,\cdots,\langle X_n|E\rangle)$
have the same initial transitions for $i\in\{1,\cdots,n\}$. $RSP$ follows from the fact that guarded recursive specifications have only one solution.

\begin{theorem}[Elimination theorem of $APRPTC$ with linear recursion]\label{ETPRecursion}
Each process term in $APRPTC$ with linear recursion is equal to a process term $\langle X_1|E\rangle$ with $E$ a linear recursive specification.
\end{theorem}

\begin{proof}
By applying structural induction with respect to term size, each process term $t_1$ in $APRPTC$ with linear recursion generates a process can be expressed in the form of equations

$t_i=((a_{1i11}\leftmerge\cdots\leftmerge a_{1i1i_1})t_{i1}+\cdots+(a_{1ik_i1}\leftmerge\cdots\leftmerge a_{1ik_ii_k})t_{ik_i}+(b_{1i11}\leftmerge\cdots\leftmerge b_{1i1i_1})+\cdots+
(b_{1il_i1}\leftmerge\cdots\leftmerge b_{1il_ii_l}))\boxplus_{\pi_1}\cdots\boxplus_{\pi_{m-1}}((a_{mi11}\leftmerge\cdots\leftmerge a_{mi1i_1})t_{i1}+\cdots+(a_{mik_i1}\leftmerge\cdots
\leftmerge a_{mik_ii_k})t_{ik_i}+(b_{mi11}\leftmerge\cdots\leftmerge b_{mi1i_1})+\cdots+(b_{mil_i1}\leftmerge\cdots\leftmerge b_{mil_ii_l}))$

for $i\in\{1,\cdots,n\}$. Let the linear recursive specification $E$ consist of the recursive equations

$X_i=((a_{1i11}\leftmerge\cdots\leftmerge a_{1i1i_1})X_{i1}+\cdots+(a_{1ik_i1}\leftmerge\cdots\leftmerge a_{1ik_ii_k})X_{ik_i}+(b_{1i11}\leftmerge\cdots\leftmerge b_{1i1i_1})+\cdots+
(b_{1il_i1}\leftmerge\cdots\leftmerge b_{1il_ii_l}))\boxplus_{\pi_1}\cdots\boxplus_{\pi_{m-1}}((a_{mi11}\leftmerge\cdots\leftmerge a_{mi1i_1})X_{i1}+\cdots+(a_{mik_i1}\leftmerge\cdots
\leftmerge a_{mik_ii_k})X_{ik_i}+(b_{mi11}\leftmerge\cdots\leftmerge b_{mi1i_1})+\cdots+(b_{mil_i1}\leftmerge\cdots\leftmerge b_{mil_ii_l}))$

for $i\in\{1,\cdots,n\}$. Replacing $X_i$ by $t_i$ for $i\in\{1,\cdots,n\}$ is a solution for $E$, $RSP$ yields $t_1=\langle X_1|E\rangle$.
\end{proof}

\begin{theorem}[Soundness of $APRPTC$ with guarded recursion]\label{SAPRPTCR}
Let $x$ and $y$ be $APRPTC$ with guarded recursion terms. If $APRPTC\textrm{ with guarded recursion}\vdash x=y$, then

\begin{enumerate}
  \item $x\sim_{pp}^{fr} y$;
  \item $x\sim_{ps}^{fr} y$;
  \item $x\sim_{php}^{fr} y$;
  \item $x\sim_{phhp}^{fr} y$.
\end{enumerate}
\end{theorem}

\begin{proof}
Since $\sim_{pp}^{fr}$, $\sim_{ps}^{fr}$, $\sim_{php}^{fr}$ and $\sim_{phhp}^{fr}$ are all both an equivalent and a congruent relation with respect to $APRPTC$ with guarded recursion, we only need to
check if each axiom in Table \ref{PRDPRSP} is sound modulo $\sim_{pp}^{fr}$, $\sim_{ps}^{fr}$, $\sim_{php}^{fr}$ and $\sim_{phhp}^{fr}$. The proof is quite trivial and we omit it.
\end{proof}

\begin{theorem}[Completeness of $APRPTC$ with linear recursion]\label{CAPRPTCR}
Let $p$ and $q$ be closed $APRPTC$ with linear recursion terms, then,

\begin{enumerate}
  \item if $p\sim_{pp}^{fr} q$ then $p=q$;
  \item if $p\sim_{ps}^{fr} q$ then $p=q$;
  \item if $p\sim_{php}^{fr} q$ then $p=q$;
  \item if $p\sim_{phhp}^{fr} q$ then $p=q$.
\end{enumerate}
\end{theorem}

\begin{proof}
Firstly, by the elimination theorem of $APRPTC$ with guarded recursion (see Theorem \ref{ETPRecursion}), we know that each process term in $APRPTC$ with linear recursion is equal to a
process term $\langle X_1|E\rangle$ with $E$ a linear recursive specification.

It remains to prove the following cases.

(1) If $\langle X_1|E_1\rangle \sim_{pp}^{fr} \langle Y_1|E_2\rangle$ for linear recursive specification $E_1$ and $E_2$, then $\langle X_1|E_1\rangle = \langle Y_1|E_2\rangle$.

Let $E_1$ consist of recursive equations $X=t_X$ for $X\in \mathcal{X}$ and $E_2$
consists of recursion equations $Y=t_Y$ for $Y\in\mathcal{Y}$. Let the linear recursive specification $E$ consist of recursion equations $Z_{XY}=t_{XY}$, and
$\langle X|E_1\rangle\sim_{pp}^{fr}\langle Y|E_2\rangle$, and $t_{XY}$ consists of the following summands:

\begin{enumerate}
  \item $t_{XY}$ contains a summand $(a_1\leftmerge\cdots\leftmerge a_m)Z_{X'Y'}$ iff $t_X$ contains the summand $(a_1\leftmerge\cdots\leftmerge a_m)X'$ and $t_Y$ contains the
  summand $(a_1\leftmerge\cdots\leftmerge a_m)Y'$ such that $\langle X'|E_1\rangle\sim_{pp}^{fr}\langle Y'|E_2\rangle$;
  \item $t_{XY}$ contains a summand $b_1\leftmerge\cdots\leftmerge b_n$ iff $t_X$ contains the summand $b_1\leftmerge\cdots\leftmerge b_n$ and $t_Y$ contains the summand
  $b_1\leftmerge\cdots\leftmerge b_n$.
\end{enumerate}

Let $\sigma$ map recursion variable $X$ in $E_1$ to $\langle X|E_1\rangle$, and let $\psi$ map recursion variable $Z_{XY}$ in $E$ to $\langle X|E_1\rangle$. So,
$\sigma((a_1\leftmerge\cdots\leftmerge a_m)X')\equiv(a_1\leftmerge\cdots\leftmerge a_m)\langle X'|E_1\rangle\equiv\psi((a_1\leftmerge\cdots\leftmerge a_m)Z_{X'Y'})$,
so by $RDP$, we get $\langle X|E_1\rangle=\sigma(t_X)=\psi(t_{XY})$. Then by $RSP$, $\langle X|E_1\rangle=\langle Z_{XY}|E\rangle$, particularly,
$\langle X_1|E_1\rangle=\langle Z_{X_1Y_1}|E\rangle$. Similarly, we can obtain $\langle Y_1|E_2\rangle=\langle Z_{X_1Y_1}|E\rangle$. Finally,
$\langle X_1|E_1\rangle=\langle Z_{X_1Y_1}|E\rangle=\langle Y_1|E_2\rangle$, as desired.

Similarly, we can prove the case of reverse transitions, we omit it.

(2) If $\langle X_1|E_1\rangle \sim_{ps}^{fr} \langle Y_1|E_2\rangle$ for linear recursive specification $E_1$ and $E_2$, then $\langle X_1|E_1\rangle = \langle Y_1|E_2\rangle$.

It can be proven similarly to (1), we omit it.

(3) If $\langle X_1|E_1\rangle \sim_{php}^{fr} \langle Y_1|E_2\rangle$ for linear recursive specification $E_1$ and $E_2$, then $\langle X_1|E_1\rangle = \langle Y_1|E_2\rangle$.

It can be proven similarly to (1), we omit it.

(4) If $\langle X_1|E_1\rangle \sim_{phhp}^{fr} \langle Y_1|E_2\rangle$ for linear recursive specification $E_1$ and $E_2$, then $\langle X_1|E_1\rangle = \langle Y_1|E_2\rangle$.

It can be proven similarly to (1), we omit it.
\end{proof}

\subsubsection{Approximation Induction Principle}

In this subsection, we introduce approximation induction principle ($AIP$) and try to explain that $AIP$ is still valid in reversible probabilistic true concurrency. $AIP$ can be used to try and
equate FR probabilistic truly concurrent bisimilar guarded recursive specifications. $AIP$ says that if two process terms are FR probabilistic truly concurrent bisimilar up to any finite depth,
then they are FR probabilistic truly concurrent bisimilar.

Also, we need the auxiliary unary projection operator $\Pi_n$ for $n\in\mathbb{N}$ and $\mathbb{N}\triangleq\{0,1,2,\cdots\}$.

Firstly, we give the definition of PDFs in Table \ref{PDFAIP}.

\begin{center}
    \begin{table}
        $$\mu(\Pi_n(x),\Pi_n(x'))=\mu(x,x')\quad n\geq 1$$
        $$\mu(x,y)=0,\textrm{otherwise}$$
        \caption{PDF definitions of approximation induction principle}
        \label{PDFAIP}
    \end{table}
\end{center}

The transition rules of $\Pi_n$ are expressed in Table \ref{TRForPProjection}.

\begin{center}
    \begin{table}
        $$\frac{x\rightsquigarrow x'}{\Pi_n(x)\rightsquigarrow\Pi_n(x')}$$
        $$\frac{x\xrightarrow{\{e_1,\cdots,e_k\}}e_1[m]\leftmerge\cdots\leftmerge e_k[m]}{\Pi_{n+1}(x)\xrightarrow{\{e_1,\cdots,e_k\}}e_1[m]\leftmerge\cdots\leftmerge e_k[m]}
        \quad\frac{x\xrightarrow{\{e_1,\cdots,e_k\}}x'}{\Pi_{n+1}(x)\xrightarrow{\{e_1,\cdots,e_k\}}\Pi_n(x')}$$
        $$\frac{x\xtworightarrow{\{e_1[m],\cdots,e_k[m]\}}e_1\leftmerge\cdots\leftmerge e_k]}{\Pi_{n+1}(x)\xtworightarrow{\{e_1[m],\cdots,e_k[m]\}}e_1\leftmerge\cdots\leftmerge e_k]}
        \quad\frac{x\xtworightarrow{\{e_1[m],\cdots,e_k[m]\}}x'}{\Pi_{n+1}(x)\xtworightarrow{\{e_1[m],\cdots,e_k[m]\}}\Pi_n(x')}$$
        \caption{Transition rules of encapsulation operator $\partial_H$}
        \label{TRForPProjection}
    \end{table}
\end{center}

Based on the transition rules for projection operator $\Pi_n$ in Table \ref{TRForPProjection}, we design the axioms as Table \ref{AxiomsForPProjection} shows.

\begin{center}
    \begin{table}
        \begin{tabular}{@{}ll@{}}
            \hline No. &Axiom\\
            $PR1$ & $\Pi_n(x+y)=\Pi_n(x)+\Pi_n(y)$\\
            $PPR1$ & $\Pi_n(x\boxplus_{\rho} y)=\Pi_n(x)\boxplus_{\rho}\Pi_n(y)$\\
            $PR2$ & $\Pi_n(x\leftmerge y)=\Pi_n(x)\leftmerge \Pi_n(y)$\\
            $PR3$ & $\Pi_{n+1}(e_1\leftmerge\cdots\leftmerge e_k)=e_1\leftmerge\cdots\leftmerge e_k$\\
            $RPR3$ & $\Pi_{n+1}(e_1[m]\leftmerge\cdots\leftmerge e_k[m])=e_1[m]\leftmerge\cdots\leftmerge e_k[m]$\\
            $PR4$ & $\Pi_{n+1}((e_1\leftmerge\cdots\leftmerge e_k)\cdot x)=(e_1\leftmerge\cdots\leftmerge e_k)\cdot\Pi_n(x)$\\
            $RPR4$ & $\Pi_{n+1}( x\cdot(e_1[m]\leftmerge\cdots\leftmerge e_k[m]))=\Pi_n(x)\cdot(e_1[m]\leftmerge\cdots\leftmerge e_k[m])$\\
            $PR5$ & $\Pi_0(x)=\delta$\\
            $PR6$ & $\Pi_n(\delta)=\delta$\\
        \end{tabular}
        \caption{Axioms of projection operator}
        \label{AxiomsForPProjection}
    \end{table}
\end{center}


\begin{theorem}[Conservativity of $APRPTC$ with projection operator and guarded recursion]
$APRPTC$  with projection operator and guarded recursion is a conservative extension of $APRPTC$ with guarded recursion.
\end{theorem}

\begin{proof}
It follows from the following two facts (see Theorem \ref{TCE}).

\begin{enumerate}
  \item The transition rules of $APRPTC$ with guarded recursion are all source-dependent;
  \item The sources of the transition rules for the projection operator contain an occurrence of $\Pi_n$.
\end{enumerate}
\end{proof}

\begin{theorem}[Congruence theorem of projection operator $\Pi_n$]
FR probabilistic truly concurrent bisimulation equivalences $\sim_{pp}^{fr}$, $\sim_{ps}^{fr}$, $\sim_{php}^{fr}$ and $\sim_{phhp}^{fr}$ are all congruences with respect to projection
operator $\Pi_n$.
\end{theorem}

\begin{proof}
It is easy to see that $\sim_{pp}^{fr}$, $\sim_{ps}^{fr}$, $\sim_{php}^{fr}$ and $\sim_{phhp}^{fr}$ are all an equivalent relation with respect to projection operator $\Pi_n$, we only need to prove that
$\sim_{pp}^{fr}$, $\sim_{ps}^{fr}$, $\sim_{php}^{fr}$ and $\sim_{phhp}^{fr}$ are preserved by the operators $\Pi_n$.
That is, if $x\sim_{pp}^{fr} x'$, $x\sim_{ps}^{fr} x'$, $x\sim_{php}^{fr} x'$ and $x\sim_{phhp}^{fr} x'$, we need to prove that $\Pi_n(x)\sim_{pp}^{fr} \Pi_n(x')$, $\Pi_n(x)\sim_{ps}^{fr} \Pi_n(x')$,
$\Pi_n(x)\sim_{php}^{fr} \Pi_n(x')$ and $\Pi_n(x)\sim_{phhp}^{fr} \Pi_n(x')$. The proof is quite trivial and we omit it.
\end{proof}

\begin{theorem}[Elimination theorem of $APRPTC$ with linear recursion and projection operator]\label{ETPProjection}
Each process term in $APRPTC$ with linear recursion and projection operator is equal to a process term $\langle X_1|E\rangle$ with $E$ a linear recursive specification.
\end{theorem}

\begin{proof}
By applying structural induction with respect to term size, each process term $t_1$ in $APRPTC$ with linear recursion and projection operator $\Pi_n$ generates a process can be
expressed in the form of equations

$t_i=((a_{1i11}\leftmerge\cdots\leftmerge a_{1i1i_1})t_{i1}+\cdots+(a_{1ik_i1}\leftmerge\cdots\leftmerge a_{1ik_ii_k})t_{ik_i}+(b_{1i11}\leftmerge\cdots\leftmerge b_{1i1i_1})+\cdots+
(b_{1il_i1}\leftmerge\cdots\leftmerge b_{1il_ii_l}))\boxplus_{\pi_1}\cdots\boxplus_{\pi_{m-1}}((a_{mi11}\leftmerge\cdots\leftmerge a_{mi1i_1})t_{i1}+\cdots+(a_{mik_i1}\leftmerge\cdots
\leftmerge a_{mik_ii_k})t_{ik_i}+(b_{mi11}\leftmerge\cdots\leftmerge b_{mi1i_1})+\cdots+(b_{mil_i1}\leftmerge\cdots\leftmerge b_{mil_ii_l}))$

for $i\in\{1,\cdots,n\}$. Let the linear recursive specification $E$ consist of the recursive equations

$X_i=((a_{1i11}\leftmerge\cdots\leftmerge a_{1i1i_1})X_{i1}+\cdots+(a_{1ik_i1}\leftmerge\cdots\leftmerge a_{1ik_ii_k})X_{ik_i}+(b_{1i11}\leftmerge\cdots\leftmerge b_{1i1i_1})+\cdots+
(b_{1il_i1}\leftmerge\cdots\leftmerge b_{1il_ii_l}))\boxplus_{\pi_1}\cdots\boxplus_{\pi_{m-1}}((a_{mi11}\leftmerge\cdots\leftmerge a_{mi1i_1})X_{i1}+\cdots+(a_{mik_i1}\leftmerge\cdots
\leftmerge a_{mik_ii_k})X_{ik_i}+(b_{mi11}\leftmerge\cdots\leftmerge b_{mi1i_1})+\cdots+(b_{mil_i1}\leftmerge\cdots\leftmerge b_{mil_ii_l}))$

for $i\in\{1,\cdots,n\}$. Replacing $X_i$ by $t_i$ for $i\in\{1,\cdots,n\}$ is a solution for $E$, $RSP$ yields $t_1=\langle X_1|E\rangle$.

That is, in $E$, there is not the occurrence of projection operator $\Pi_n$.
\end{proof}

\begin{theorem}[Soundness of $APRPTC$ with projection operator and guarded recursion]\label{SAPRPTCR}
Let $x$ and $y$ be $APRPTC$ with projection operator and guarded recursion terms. If $APRPTC$ with projection operator and guarded recursion $\vdash x=y$, then
\begin{enumerate}
  \item $x\sim_{pp}^{fr} y$;
  \item $x\sim_{ps}^{fr} y$;
  \item $x\sim_{php}^{fr} y$;
  \item $x\sim_{phhp}^{fr} y$.
\end{enumerate}
\end{theorem}

\begin{proof}
Since $\sim_{pp}^{fr}$, $\sim_{ps}^{fr}$, $\sim_{php}^{fr}$ and $\sim_{phhp}^{fr}$ are all both an equivalent and a congruent relation with respect to $APRPTC$ with guarded recursion, we only need to
check if each axiom in Table \ref{TRForPProjection} is sound modulo $\sim_{pp}^{fr}$, $\sim_{ps}^{fr}$, $\sim_{php}^{fr}$ and $\sim_{phhp}^{fr}$. The proof is quite trivial and we omit it.
\end{proof}

Then $AIP$ is given in Table \ref{PAIP}.

\begin{center}
    \begin{table}
        \begin{tabular}{@{}ll@{}}
            \hline No. &Axiom\\
            $AIP$ & if $\Pi_n(x)=\Pi_n(y)$ for $n\in\mathbb{N}$, then $x=y$\\
        \end{tabular}
        \caption{$AIP$}
        \label{PAIP}
    \end{table}
\end{center}

\begin{theorem}[Soundness of $AIP$]\label{SPAIP}
Let $x$ and $y$ be $APRPTC$ with projection operator and guarded recursion terms.

\begin{enumerate}
  \item If $\Pi_n(x)\sim_{pp}^{fr}\Pi_n(y)$ for $n\in\mathbb{N}$, then $x\sim_{pp}^{fr} y$;
  \item If $\Pi_n(x)\sim_{ps}^{fr}\Pi_n(y)$ for $n\in\mathbb{N}$, then $x\sim_{ps}^{fr} y$;
  \item If $\Pi_n(x)\sim_{php}^{fr}\Pi_n(y)$ for $n\in\mathbb{N}$, then $x\sim_{php}^{fr} y$;
  \item If $\Pi_n(x)\sim_{phhp}^{fr}\Pi_n(y)$ for $n\in\mathbb{N}$, then $x\sim_{phhp}^{fr} y$.
\end{enumerate}
\end{theorem}

\begin{proof}
(1) If $\Pi_n(x)\sim_{pp}^{fr}\Pi_n(y)$ for $n\in\mathbb{N}$, then $x\sim_{pp}^{fr} y$.

Since $\sim_{pp}^{fr}$ is both an equivalent and a congruent relation with respect to $APRPTC$ with guarded recursion and projection operator, we only need to check if $AIP$ in Table
\ref{PAIP} is sound modulo $\sim_{pp}^{fr}$.

Let $p,p_0$ and $q,q_0$ be closed $APRPTC$ with projection operator and guarded recursion terms such that $\Pi_n(p_0)\sim_{pp}^{fr}\Pi_n(q_0)$ for $n\in\mathbb{N}$. We define a relation
$R$ such that $p R q$ iff $\Pi_n(p)\sim_{pp}^{fr} \Pi_n(q)$. Obviously, $p_0 R q_0$, next, we prove that $R\in\sim_{pp}^{fr}$.

Let $p R q$ and $p\rightsquigarrow\xrightarrow{\{e_1,\cdots,e_k\}}\surd$, then $\Pi_1(p)\rightsquigarrow\xrightarrow{\{e_1,\cdots,e_k\}}\surd$, $\Pi_1(p)\sim_{pp}^{fr}\Pi_1(q)$ yields
$\Pi_1(q)\rightsquigarrow\xrightarrow{\{e_1,\cdots,e_k\}}\surd$. Similarly, $q\rightsquigarrow\xrightarrow{\{e_1,\cdots,e_k\}}\surd$ implies
$p\rightsquigarrow\xrightarrow{\{e_1,\cdots,e_k\}}\surd$.

Let $p R q$ and $p\rightsquigarrow\xrightarrow{\{e_1,\cdots,e_k\}}p'$. We define the set of process terms

$$S_n\triangleq\{q'|q\rightsquigarrow\xrightarrow{\{e_1,\cdots,e_k\}}q'\textrm{ and }\Pi_n(p')\sim_{pp}^{fr}\Pi_n(q')\}$$

\begin{enumerate}
  \item Since $\Pi_{n+1}(p)\sim_{pp}^{fr}\Pi_{n+1}(q)$ and $\Pi_{n+1}(p)\rightsquigarrow\xrightarrow{\{e_1,\cdots,e_k\}}\Pi_n(p')$, there exist $q'$ such that
  $\Pi_{n+1}(q)\rightsquigarrow\xrightarrow{\{e_1,\cdots,e_k\}}\Pi_n(q')$ and $\Pi_{n}(p')\sim_{pp}^{fr}\Pi_{n}(q')$. So, $S_n$ is not empty.
  \item There are only finitely many $q'$ such that $q\rightsquigarrow\xrightarrow{\{e_1,\cdots,e_k\}}q'$, so, $S_n$ is finite.
  \item $\Pi_{n+1}(p)\sim_{pp}^{fr}\Pi_{n+1}(q)$ implies $\Pi_{n}(p')\sim_{pp}^{fr}\Pi_{n}(q')$, so $S_n\supseteq S_{n+1}$.
\end{enumerate}

So, $S_n$ has a non-empty intersection, and let $q'$ be in this intersection, then $q\rightsquigarrow\xrightarrow{\{e_1,\cdots,e_k\}}q'$ and $\Pi_n(p')\sim_{pp}^{fr}\Pi_n(q')$,
so $p' R q'$. Similarly, let $p\mathcal{q}q$, we can obtain $q\rightsquigarrow\xrightarrow{\{e_1,\cdots,e_k\}}q'$ implies $p\rightsquigarrow\xrightarrow{\{e_1,\cdots,e_k\}}p'$
such that $p' R q'$.

Similarly, we can prove the case of reverse transitions, we omit it.

Finally, $R\in\sim_{pp}^{fr}$ and $p_0\sim_{pp}^{fr} q_0$, as desired.

(2) If $\Pi_n(x)\sim_{ps}^{fr}\Pi_n(y)$ for $n\in\mathbb{N}$, then $x\sim_{ps}^{fr} y$.

It can be proven similarly to (1).

(3) If $\Pi_n(x)\sim_{php}^{fr}\Pi_n(y)$ for $n\in\mathbb{N}$, then $x\sim_{php}^{fr} y$.

It can be proven similarly to (1).

(4) If $\Pi_n(x)\sim_{phhp}^{fr}\Pi_n(y)$ for $n\in\mathbb{N}$, then $x\sim_{phhp}^{fr} y$.

It can be proven similarly to (1).
\end{proof}

\begin{theorem}[Completeness of $AIP$]\label{CPAIP}
Let $p$ and $q$ be closed $APRPTC$ with linear recursion and projection operator terms, then,
\begin{enumerate}
  \item if $p\sim_{pp}^{fr} q$ then $\Pi_n(p)=\Pi_n(q)$;
  \item if $p\sim_{ps}^{fr} q$ then $\Pi_n(p)=\Pi_n(q)$;
  \item if $p\sim_{php}^{fr} q$ then $\Pi_n(p)=\Pi_n(q)$;
  \item if $p\sim_{phhp}^{fr} q$ then $\Pi_n(p)=\Pi_n(q)$.
\end{enumerate}
\end{theorem}

\begin{proof}
Firstly, by the elimination theorem of $APRPTC$ with guarded recursion and projection operator (see Theorem \ref{ETPProjection}), we know that each process term in $APRPTC$ with linear
recursion and projection operator is equal to a process term $\langle X_1|E\rangle$ with $E$ a linear recursive specification:

$X_i=((a_{1i11}\leftmerge\cdots\leftmerge a_{1i1i_1})X_{i1}+\cdots+(a_{1ik_i1}\leftmerge\cdots\leftmerge a_{1ik_ii_k})X_{ik_i}+(b_{1i11}\leftmerge\cdots\leftmerge b_{1i1i_1})+\cdots+
(b_{1il_i1}\leftmerge\cdots\leftmerge b_{1il_ii_l}))\boxplus_{\pi_1}\cdots\boxplus_{\pi_{m-1}}((a_{mi11}\leftmerge\cdots\leftmerge a_{mi1i_1})X_{i1}+\cdots+(a_{mik_i1}\leftmerge\cdots
\leftmerge a_{mik_ii_k})X_{ik_i}+(b_{mi11}\leftmerge\cdots\leftmerge b_{mi1i_1})+\cdots+(b_{mil_i1}\leftmerge\cdots\leftmerge b_{mil_ii_l}))$

for $i\in\{1,\cdots,n\}$.

It remains to prove the following cases.

(1) if $p\sim_{pp}^{fr} q$ then $\Pi_n(p)=\Pi_n(q)$.

Let $p\sim_{pp}^{fr} q$, and fix an $n\in\mathbb{N}$, there are $p',q'$ in basic $APRPTC$ terms such that $p'=\Pi_n(p)$ and $q'=\Pi_n(q)$. Since $\sim_{pp}^{fr}$ is a congruence with respect to
$APRPTC$, if $p\sim_{pp}^{fr} q$ then $\Pi_n(p)\sim_{pp}^{fr}\Pi_n(q)$. The soundness theorem yields $p'\sim_{pp}^{fr}\Pi_n(p)\sim_{pp}^{fr}\Pi_n(q)\sim_{pp}^{fr} q'$. Finally, the completeness of $APRPTC$
modulo $\sim_{pp}^{fr}$ (see Theorem \ref{SAPRPTCPPBE}) ensures $p'=q'$, and $\Pi_n(p)=p'=q'=\Pi_n(q)$, as desired.

(2) if $p\sim_{ps}^{fr} q$ then $\Pi_n(p)=\Pi_n(q)$.

Let $p\sim_{ps}^{fr} q$, and fix an $n\in\mathbb{N}$, there are $p',q'$ in basic $APRPTC$ terms such that $p'=\Pi_n(p)$ and $q'=\Pi_n(q)$. Since $\sim_{ps}^{fr}$ is a congruence with respect to
$APRPTC$, if $p\sim_{ps}^{fr} q$ then $\Pi_n(p)\sim_{ps}^{fr}\Pi_n(q)$. The soundness theorem yields $p'\sim_{ps}^{fr}\Pi_n(p)\sim_{ps}^{fr}\Pi_n(q)\sim_{ps}^{fr} q'$. Finally, the completeness of $APRPTC$
modulo $\sim_{ps}^{fr}$ (see Theorem \ref{SAPRPTCPSBE}) ensures $p'=q'$, and $\Pi_n(p)=p'=q'=\Pi_n(q)$, as desired.

(3) if $p\sim_{php}^{fr} q$ then $\Pi_n(p)=\Pi_n(q)$.

Let $p\sim_{php}^{fr} q$, and fix an $n\in\mathbb{N}$, there are $p',q'$ in basic $APRPTC$ terms such that $p'=\Pi_n(p)$ and $q'=\Pi_n(q)$. Since $\sim_{php}^{fr}$ is a congruence with respect to
$APRPTC$, if $p\sim_{php}^{fr} q$ then $\Pi_n(p)\sim_{php}^{fr}\Pi_n(q)$. The soundness theorem yields $p'\sim_{php}^{fr}\Pi_n(p)\sim_{php}^{fr}\Pi_n(q)\sim_{php}^{fr} q'$. Finally, the completeness of $APRPTC$
modulo $\sim_{php}^{fr}$ (see Theorem \ref{SAPRPTCPHPBE}) ensures $p'=q'$, and $\Pi_n(p)=p'=q'=\Pi_n(q)$, as desired.

(4) if $p\sim_{phhp}^{fr} q$ then $\Pi_n(p)=\Pi_n(q)$.

Let $p\sim_{phhp}^{fr} q$, and fix an $n\in\mathbb{N}$, there are $p',q'$ in basic $APRPTC$ terms such that $p'=\Pi_n(p)$ and $q'=\Pi_n(q)$. Since $\sim_{phhp}^{fr}$ is a congruence with respect to
$APRPTC$, if $p\sim_{phhp}^{fr} q$ then $\Pi_n(p)\sim_{phhp}^{fr}\Pi_n(q)$. The soundness theorem yields $p'\sim_{phhp}^{fr}\Pi_n(p)\sim_{phhp}^{fr}\Pi_n(q)\sim_{phhp}^{fr} q'$. Finally, the completeness of $APRPTC$
modulo $\sim_{phhp}^{fr}$ (see Theorem \ref{SAPRPTCPHHPBE}) ensures $p'=q'$, and $\Pi_n(p)=p'=q'=\Pi_n(q)$, as desired.

Similarly, we can prove the case of reverse transitions, we omit it.
\end{proof}

\subsection{Abstraction}\label{rpabs}

To abstract away from the internal implementations of a program, and verify that the program exhibits the desired external behaviors, the silent step $\tau$ and abstraction operator
$\tau_I$ are introduced, where $I\subseteq \mathbb{E}$ denotes the internal events. The silent step $\tau$ represents the internal events, when we consider the external behaviors of a
process, $\tau$ events can be removed, that is, $\tau$ events must keep silent. The transition rule of $\tau$ is shown in Table \ref{TRForPTau}. In the following, let the atomic event
$e$ range over $\mathbb{E}\cup\{\delta\}\cup\{\tau\}$, and let the communication function $\gamma:\mathbb{E}\cup\{\tau\}\times \mathbb{E}\cup\{\tau\}\rightarrow \mathbb{E}\cup\{\delta\}$,
with each communication involved $\tau$ resulting into $\delta$.

\begin{center}
    \begin{table}
        $$\frac{}{\tau\rightsquigarrow\breve{\tau}}$$
        $$\frac{}{\tau\xrightarrow{\tau}\surd}$$
         $$\frac{}{\tau\xtworightarrow{\tau}\surd}$$
        \caption{Transition rules of the silent step}
        \label{TRForPTau}
    \end{table}
\end{center}

In this section, we try to find the algebraic laws of $\tau$ and $\tau_I$ in probabilistic true concurrency.

\subsubsection{Guarded Linear Recursion}

The silent step $\tau$ as an atomic event, is introduced into $E$. Considering the recursive specification $X=\tau X$, $\tau s$, $\tau\tau s$, and $\tau\cdots s$ are all its solutions,
that is, the solutions make the existence of $\tau$-loops which cause unfairness. To prevent $\tau$-loops, we extend the definition of linear recursive specification
to the guarded one.

\begin{definition}[Guarded linear recursive specification]\label{GLRS}
A recursive specification is linear if its recursive equations are of the form

$((a_{111}\parallel\cdots\parallel a_{11i_1})X_1+\cdots+(a_{1k1}\parallel\cdots\parallel a_{1ki_k})X_k+(b_{111}\parallel\cdots\parallel b_{11j_1})+\cdots+
(b_{11j_1}\parallel\cdots\parallel b_{1lj_l}))\boxplus_{\pi_1}\cdots\boxplus_{\pi_{m-1}}((a_{m11}\parallel\cdots\parallel a_{m1i_1})X_1+\cdots+(a_{mk1}\parallel\cdots\parallel a_{mki_k})X_k+
(b_{m11}\parallel\cdots\parallel b_{m1j_1})+\cdots+(b_{m1j_1}\parallel\cdots\parallel b_{mlj_l}))$

where $a_{111},\cdots,a_{11i_1},a_{1k1},\cdots,a_{1ki_k},b_{111},\cdots,b_{11j_1},b_{11j_1},\cdots,b_{1lj_l}\cdots\\
a_{m11},\cdots,a_{m1i_1},a_{mk1},\cdots,a_{mki_k},b_{m11},\cdots,b_{m1j_1},b_{m1j_1},\cdots,b_{mlj_l}\in \mathbb{E}\cup\{\tau\}$, and the sum above is allowed to be empty, in which
case it represents the deadlock $\delta$.

A linear recursive specification $E$ is guarded if there does not exist an infinite sequence of $\tau$-transitions $\langle X|E\rangle\rightsquigarrow\xrightarrow{\tau}
\langle X'|E\rangle\rightsquigarrow\xrightarrow{\tau}\langle X''|E\rangle\rightsquigarrow\xrightarrow{\tau}\cdots$.
\end{definition}

\begin{theorem}[Conservitivity of $APRPTC$ with silent step and guarded linear recursion]
$APRPTC$ with silent step and guarded linear recursion is a conservative extension of $APRPTC$ with linear recursion.
\end{theorem}

\begin{proof}
Since the transition rules of $APRPTC$ with linear recursion are source-dependent, and the transition rules for silent step in Table \ref{TRForPTau} contain only a fresh constant $\tau$
in their source, so the transition rules of $APRPTC$ with silent step and guarded linear recursion is a conservative extension of those of $APRPTC$ with linear recursion.
\end{proof}

\begin{theorem}[Congruence theorem of $APRPTC$ with silent step and guarded linear recursion]
FR probabilistic rooted branching truly concurrent bisimulation equivalences $\approx_{prbp}^{fr}$, $\approx_{prbs}^{fr}$, $\approx_{prbhp}^{fr}$ and $\approx_{prbhhp}^{fr}$ are all congruences with respect
to $APRPTC$ with silent step and guarded linear recursion.
\end{theorem}

\begin{proof}
It follows the following three facts:
\begin{enumerate}
  \item in a guarded linear recursive specification, right-hand sides of its recursive equations can be adapted to the form by applications of the axioms in $APRPTC$ and replacing
  recursion variables by the right-hand sides of their recursive equations;
  \item FR probabilistic truly concurrent bisimulation equivalences $\sim_{pp}^{fr}$, $\sim_{ps}^{fr}$, $\sim_{php}^{fr}$ and $\sim_{phhp}^{fr}$ are all congruences with respect to all operators of $APRPTC$,
  while FR probabilistic truly concurrent bisimulation equivalences $\sim_{pp}^{fr}$, $\sim_{ps}^{fr}$, $\sim_{php}^{fr}$ and $\sim_{phhp}^{fr}$ imply the corresponding FR probabilistic rooted branching truly
  concurrent bisimulations $\approx_{prbp}^{fr}$, $\approx_{prbs}^{fr}$, $\approx_{prbhp}^{fr}$ and $\approx_{prbhhp}^{fr}$, so FR probabilistic rooted branching truly concurrent bisimulations $\approx_{prbp}^{fr}$,
  $\approx_{prbs}^{fr}$, $\approx_{prbhp}^{fr}$ and $\approx_{prbhhp}^{fr}$ are all congruences with respect to all operators of $APRPTC$;
  \item While $\mathbb{E}$ is extended to $\mathbb{E}\cup\{\tau\}$, it can be proved that FR probabilistic rooted branching truly concurrent bisimulations $\approx_{prbp}^{fr}$, $\approx_{prbs}^{fr}$,
  $\approx_{prbhp}^{fr}$ and $\approx_{prbhhp}^{fr}$ are all congruences with respect to all operators of $APRPTC$, we omit it.
\end{enumerate}
\end{proof}

\subsubsection{Algebraic Laws for the Silent Step}

We design the axioms for the silent step $\tau$ in Table \ref{AxiomsForPTau}.

\begin{center}
\begin{table}
  \begin{tabular}{@{}ll@{}}
\hline No. &Axiom\\
  $B1$ & $(y=y+y,z=z+z,(Std(x),Std(y),Std(z),Std(w)))$\\
  &$x\cdot((y+\tau\cdot(y+z))\boxplus_{\pi}w)=x\cdot((y+z)\boxplus_{\pi}w)$\\
  $RB1$ & $(y=y+y,z=z+z,(NStd(x),NStd(y),NStd(z),NStd(w)))$\\
  &$((y+(y+z)\cdot\tau)\boxplus_{\pi}w)\cdot x=((y+z)\boxplus_{\pi}w)\cdot x$\\
  $B2$ & $(y=y+y,z=z+z,(Std(x),Std(y),Std(z),Std(w)))$\\
  &$x\leftmerge((y+\tau\leftmerge(y+z))\boxplus_{\pi}w)=x\leftmerge((y+z)\boxplus_{\pi}w)$\\
  $RB2$ & $(y=y+y,z=z+z,(NStd(x),NStd(y),NStd(z),NStd(w)))$\\
  &$((y+(y+z)\leftmerge\tau)\boxplus_{\pi}w)\leftmerge x=((y+z)\boxplus_{\pi}w)\leftmerge x$\\
\end{tabular}
\caption{Axioms of silent step}
\label{AxiomsForPTau}
\end{table}
\end{center}


\begin{theorem}[Elimination theorem of $APRPTC$ with silent step and guarded linear recursion]\label{ETTau}
Each process term in $APRPTC$ with silent step and guarded linear recursion is equal to a process term $\langle X_1|E\rangle$ with $E$ a guarded linear recursive specification.
\end{theorem}

\begin{proof}
By applying structural induction with respect to term size, each process term $t_1$ in $APRPTC$ with silent step and guarded linear recursion generates a process can be expressed in the
form of equations

$t_i=((a_{1i11}\leftmerge\cdots\leftmerge a_{1i1i_1})t_{i1}+\cdots+(a_{1ik_i1}\leftmerge\cdots\leftmerge a_{1ik_ii_k})t_{ik_i}+(b_{1i11}\leftmerge\cdots\leftmerge b_{1i1i_1})+\cdots+
(b_{1il_i1}\leftmerge\cdots\leftmerge b_{1il_ii_l}))\boxplus_{\pi_1}\cdots\boxplus_{\pi_{m-1}}((a_{mi11}\leftmerge\cdots\leftmerge a_{mi1i_1})t_{i1}+\cdots+(a_{mik_i1}\leftmerge\cdots
\leftmerge a_{mik_ii_k})t_{ik_i}+(b_{mi11}\leftmerge\cdots\leftmerge b_{mi1i_1})+\cdots+(b_{mil_i1}\leftmerge\cdots\leftmerge b_{m1il_ii_l}))$

for $i\in\{1,\cdots,n\}$. Let the linear recursive specification $E$ consist of the recursive equations

$X_i=((a_{1i11}\leftmerge\cdots\leftmerge a_{1i1i_1})X_{i1}+\cdots+(a_{1ik_i1}\leftmerge\cdots\leftmerge a_{1ik_ii_k})X_{ik_i}+(b_{1i11}\leftmerge\cdots\leftmerge b_{1i1i_1})+\cdots+
(b_{1il_i1}\leftmerge\cdots\leftmerge b_{1il_ii_l}))\boxplus_{\pi_1}\cdots\boxplus_{\pi_{m-1}}((a_{mi11}\leftmerge\cdots\leftmerge a_{mi1i_1})X_{i1}+\cdots+(a_{mik_i1}\leftmerge\cdots
\leftmerge a_{mik_ii_k})X_{ik_i}+(b_{mi11}\leftmerge\cdots\leftmerge b_{mi1i_1})+\cdots+(b_{mil_i1}\leftmerge\cdots\leftmerge b_{mil_ii_l}))$

for $i\in\{1,\cdots,n\}$. Replacing $X_i$ by $t_i$ for $i\in\{1,\cdots,n\}$ is a solution for $E$, $RSP$ yields $t_1=\langle X_1|E\rangle$.
\end{proof}

\begin{theorem}[Soundness of $APRPTC$ with silent step and guarded linear recursion]\label{SAPRPTCTAU}
Let $x$ and $y$ be $APRPTC$ with silent step and guarded linear recursion terms. If $APRPTC$ with silent step and guarded linear recursion $\vdash x=y$, then
\begin{enumerate}
  \item $x\approx_{prbp}^{fr} y$;
  \item $x\approx_{prbs}^{fr} y$;
  \item $x\approx_{prbhp}^{fr} y$;
  \item $x\approx_{prbhhp}^{fr}y$.
\end{enumerate}
\end{theorem}

\begin{proof}
Since FR probabilistic truly concurrent rooted branching bisimulation $\approx_{prbp}^{fr}$, $\approx_{prbs}^{fr}$, $\approx_{prbhp}^{fr}$ and $\approx_{prbhhp}^{fr}$ are all both an equivalent and a
congruent relation with respect to $APRPTC$ with silent step and guarded linear recursion, we only need to check if each axiom in Table \ref{AxiomsForPTau} is sound modulo
FR probabilistic truly concurrent rooted branching bisimulation $\approx_{prbp}^{fr}$, $\approx_{prbs}^{fr}$, $\approx_{prbhp}^{fr}$ and $\approx_{prbhhp}^{fr}$. The proof is quite trivial and we omit it.
\end{proof}

\begin{theorem}[Completeness of $APRPTC$ with silent step and guarded linear recursion]\label{CAPRPTCTAU}
Let $p$ and $q$ be closed $APRPTC$ with silent step and guarded linear recursion terms, then,
\begin{enumerate}
  \item if $p\approx_{prbp}^{fr} q$ then $p=q$;
  \item if $p\approx_{prbs}^{fr} q$ then $p=q$;
  \item if $p\approx_{prbhp}^{fr} q$ then $p=q$;
  \item if $p\approx_{prbhhp}^{fr} q$ then $p=q$.
\end{enumerate}
\end{theorem}

\begin{proof}
Firstly, by the elimination theorem of $APRPTC$ with silent step and guarded linear recursion (see Theorem \ref{ETTau}), we know that each process term in $APRPTC$ with silent step and guarded linear recursion is equal to a process term $\langle X_1|E\rangle$ with $E$ a guarded linear recursive specification.

It remains to prove the following cases.

(1) If $\langle X_1|E_1\rangle \approx_{prbp}^{fr} \langle Y_1|E_2\rangle$ for guarded linear recursive specification $E_1$ and $E_2$, then $\langle X_1|E_1\rangle = \langle Y_1|E_2\rangle$.

Firstly, the recursive equation $W=\tau+\cdots+\tau$ with $W\nequiv X_1$ in $E_1$ and $E_2$, can be removed, and the corresponding summands $aW$ are replaced by $a$, to get $E_1'$ and
$E_2'$, by use of the axioms $RDP$, $A3$ and $B1-2$, and $\langle X|E_1\rangle = \langle X|E_1'\rangle$, $\langle Y|E_2\rangle = \langle Y|E_2'\rangle$.

Let $E_1$ consists of recursive equations $X=t_X$ for $X\in \mathcal{X}$ and $E_2$
consists of recursion equations $Y=t_Y$ for $Y\in\mathcal{Y}$, and are not the form $\tau+\cdots+\tau$. Let the guarded linear recursive specification $E$ consists of recursion
equations $Z_{XY}=t_{XY}$, and $\langle X|E_1\rangle\approx_{prbp}^{fr}\langle Y|E_2\rangle$, and $t_{XY}$ consists of the following summands:

\begin{enumerate}
  \item $t_{XY}$ contains a summand $(a_1\leftmerge\cdots\leftmerge a_m)Z_{X'Y'}$ iff $t_X$ contains the summand $(a_1\leftmerge\cdots\leftmerge a_m)X'$ and $t_Y$ contains the summand
  $(a_1\leftmerge\cdots\leftmerge a_m)Y'$ such that $\langle X'|E_1\rangle\approx_{prbp}^{fr}\langle Y'|E_2\rangle$;
  \item $t_{XY}$ contains a summand $b_1\leftmerge\cdots\leftmerge b_n$ iff $t_X$ contains the summand $b_1\leftmerge\cdots\leftmerge b_n$ and $t_Y$ contains the summand
  $b_1\leftmerge\cdots\leftmerge b_n$;
  \item $t_{XY}$ contains a summand $\tau Z_{X'Y}$ iff $XY\nequiv X_1Y_1$, $t_X$ contains the summand $\tau X'$, and $\langle X'|E_1\rangle\approx_{prbp}^{fr}\langle Y|E_2\rangle$;
  \item $t_{XY}$ contains a summand $\tau Z_{XY'}$ iff $XY\nequiv X_1Y_1$, $t_Y$ contains the summand $\tau Y'$, and $\langle X|E_1\rangle\approx_{prbp}^{fr}\langle Y'|E_2\rangle$.
\end{enumerate}

Since $E_1$ and $E_2$ are guarded, $E$ is guarded. Constructing the process term $u_{XY}$ consist of the following summands:

\begin{enumerate}
  \item $u_{XY}$ contains a summand $(a_1\leftmerge\cdots\leftmerge a_m)\langle X'|E_1\rangle$ iff $t_X$ contains the summand $(a_1\leftmerge\cdots\leftmerge a_m)X'$ and $t_Y$ contains
  the summand $(a_1\leftmerge\cdots\leftmerge a_m)Y'$ such that $\langle X'|E_1\rangle\approx_{prbp}^{fr}\langle Y'|E_2\rangle$;
  \item $u_{XY}$ contains a summand $b_1\leftmerge\cdots\leftmerge b_n$ iff $t_X$ contains the summand $b_1\leftmerge\cdots\leftmerge b_n$ and $t_Y$ contains the summand
  $b_1\leftmerge\cdots\leftmerge b_n$;
  \item $u_{XY}$ contains a summand $\tau \langle X'|E_1\rangle$ iff $XY\nequiv X_1Y_1$, $t_X$ contains the summand $\tau X'$, and
  $\langle X'|E_1\rangle\approx_{rbs}\langle Y|E_2\rangle$.
\end{enumerate}

Let the process term $s_{XY}$ be defined as follows:

\begin{enumerate}
  \item $s_{XY}\triangleq\tau\langle X|E_1\rangle + u_{XY}$ iff $XY\nequiv X_1Y_1$, $t_Y$ contains the summand $\tau Y'$, and $\langle X|E_1\rangle\approx_{prbp}^{fr}\langle Y'|E_2\rangle$;
  \item $s_{XY}\triangleq\langle X|E_1\rangle$, otherwise.
\end{enumerate}

So, $\langle X|E_1\rangle=\langle X|E_1\rangle+u_{XY}$, and $(a_1\leftmerge\cdots\leftmerge a_m)(\tau\langle X|E_1\rangle+u_{XY})
=(a_1\leftmerge\cdots\leftmerge a_m)((\tau\langle X|E_1\rangle+u_{XY})+u_{XY})=(a_1\leftmerge\cdots\leftmerge a_m)(\langle X|E_1\rangle+u_{XY})
=(a_1\leftmerge\cdots\leftmerge a_m)\langle X|E_1\rangle$, hence, $(a_1\leftmerge\cdots\leftmerge a_m)s_{XY}=(a_1\leftmerge\cdots\leftmerge a_m)\langle X|E_1\rangle$.

Let $\sigma$ map recursion variable $X$ in $E_1$ to $\langle X|E_1\rangle$, and let $\psi$ map recursion variable $Z_{XY}$ in $E$ to $s_{XY}$. It is sufficient to prove
$s_{XY}=\psi(t_{XY})$ for recursion variables $Z_{XY}$ in $E$. Either $XY\equiv X_1Y_1$ or $XY\nequiv X_1Y_1$, we all can get $s_{XY}=\psi(t_{XY})$.
So, $s_{XY}=\langle Z_{XY}|E\rangle$ for recursive variables $Z_{XY}$ in $E$ is a solution for $E$. Then by $RSP$, particularly, $\langle X_1|E_1\rangle=\langle Z_{X_1Y_1}|E\rangle$.
Similarly, we can obtain $\langle Y_1|E_2\rangle=\langle Z_{X_1Y_1}|E\rangle$. Finally, $\langle X_1|E_1\rangle=\langle Z_{X_1Y_1}|E\rangle=\langle Y_1|E_2\rangle$, as desired.

Similarly, we can prove the case of reverse transitions, we omit it.

(2) If $\langle X_1|E_1\rangle \approx_{prbs}^{fr} \langle Y_1|E_2\rangle$ for guarded linear recursive specification $E_1$ and $E_2$, then $\langle X_1|E_1\rangle = \langle Y_1|E_2\rangle$.

It can be proven similarly to (1), we omit it.

(3) If $\langle X_1|E_1\rangle \approx_{prbhp}^{fr} \langle Y_1|E_2\rangle$ for guarded linear recursive specification $E_1$ and $E_2$, then $\langle X_1|E_1\rangle = \langle Y_1|E_2\rangle$.

It can be proven similarly to (1), we omit it.

(4) If $\langle X_1|E_1\rangle \approx_{prbhhp}^{fr} \langle Y_1|E_2\rangle$ for guarded linear recursive specification $E_1$ and $E_2$, then $\langle X_1|E_1\rangle = \langle Y_1|E_2\rangle$.

It can be proven similarly to (1), we omit it.
\end{proof}

\subsubsection{Abstraction}

The unary abstraction operator $\tau_I$ ($I\subseteq \mathbb{E}$) renames all atomic events in $I$ into $\tau$. $APRPTC$ with silent step and abstraction operator is called $APRPTC_{\tau}$.
The transition rules of operator $\tau_I$ are shown in Table \ref{TRForPAbstraction}.

\begin{center}
    \begin{table}
        $$\frac{x\rightsquigarrow x'}{\tau_I(x)\rightsquigarrow\tau_I(x')}$$
        $$\frac{x\xrightarrow{e}\surd}{\tau_I(x)\xrightarrow{e}\surd}\quad e\notin I
        \quad\frac{x\xrightarrow{e}x'}{\tau_I(x)\xrightarrow{e}\tau_I(x')}\quad e\notin I$$

        $$\frac{x\xrightarrow{e}\surd}{\tau_I(x)\xrightarrow{\tau}\surd}\quad e\in I
        \quad\frac{x\xrightarrow{e}x'}{\tau_I(x)\xrightarrow{\tau}\tau_I(x')}\quad e\in I$$

        $$\frac{x\xtworightarrow{e[m]}e}{\tau_I(x)\xtworightarrow{e[m]}e}\quad e[m]\notin I
        \quad\frac{x\xtworightarrow{e[m]}x'}{\tau_I(x)\xtworightarrow{e[m]}\tau_I(x')}\quad e[m]\notin I$$

        $$\frac{x\xtworightarrow{e[m]}\surd}{\tau_I(x)\xtworightarrow{\tau}\surd}\quad e[m]\in I
        \quad\frac{x\xtworightarrow{e[m]}x'}{\tau_I(x)\xtworightarrow{\tau}\tau_I(x')}\quad e[m]\in I$$
        \caption{Transition rules of the abstraction operator}
        \label{TRForPAbstraction}
    \end{table}
\end{center}

\begin{theorem}[Conservitivity of $APRPTC_{\tau}$ with guarded linear recursion]
$APRPTC_{\tau}$ with guarded linear recursion is a conservative extension of $APRPTC$ with silent step and guarded linear recursion.
\end{theorem}

\begin{proof}
Since the transition rules of $APRPTC$ with silent step and guarded linear recursion are source-dependent, and the transition rules for abstraction operator in Table
\ref{TRForPAbstraction} contain only a fresh operator $\tau_I$ in their source, so the transition rules of $APRPTC_{\tau}$ with guarded linear recursion is a conservative extension of
those of $APRPTC$ with silent step and guarded linear recursion.
\end{proof}

\begin{theorem}[Congruence theorem of $APRPTC_{\tau}$ with guarded linear recursion]
FR probabilistic rooted branching truly concurrent bisimulation equivalences $\approx_{prbp}^{fr}$, $\approx_{prbs}^{fr}$, $\approx_{prbhp}^{fr}$ and $\approx_{prbhhp}^{fr}$ are all congruences with respect
to $APRPTC_{\tau}$ with guarded linear recursion.
\end{theorem}

\begin{proof}
It is easy to see that FR probabilistic rooted branching truly concurrent bisimulations $\approx_{prbp}^{fr}$, $\approx_{prbs}^{fr}$, $\approx_{prbhp}^{fr}$ and $\approx_{prbhhp}^{fr}$ are all equivalent
relations on $APRPTC$ terms, we only need to prove that $\approx_{prbp}^{fr}$, $\approx_{prbs}^{fr}$, $\approx_{prbhp}^{fr}$ and $\approx_{prbhhp}^{fr}$ are preserved by the operators $\tau_I$.
That is, if $x\approx_{prbp}^{fr} x'$, $x\approx_{prbs}^{fr} x'$, $x\approx_{prbhp}^{fr} x'$ and $x\approx_{prbhhp}^{fr} x'$, we need to prove that $\tau_I(x)\approx_{prbp}^{fr}\tau_I(x')$,
$\tau_I(x)\approx_{prbs}^{fr}\tau_I(x')$, $\tau_I(x)\approx_{prbhp}^{fr}\tau_I(x')$ and $\tau_I(x)\approx_{prbhhp}^{fr}\tau_I(x')$. The proof is quite trivial and we omit it.
\end{proof}

We design the axioms for the abstraction operator $\tau_I$ in Table \ref{AxiomsForPAbstraction}.

\begin{center}
\begin{table}
  \begin{tabular}{@{}ll@{}}
\hline No. &Axiom\\
  $TI1$ & $e\notin I\quad \tau_I(e)=e$\\
  $RTI1$ & $e[m]\notin I\quad \tau_I(e[m])=e[m]$\\
  $TI2$ & $e\in I\quad \tau_I(e)=\tau$\\
  $RTI2$ & $e[m]\in I\quad \tau_I(e[m])=\tau$\\
  $TI3$ & $\tau_I(\delta)=\delta$\\
  $TI4$ & $\tau_I(x+y)=\tau_I(x)+\tau_I(y)$\\
  $PTI1$ & $\tau_I(x\boxplus_{\pi}y)=\tau_I(x)\boxplus_{\pi}\tau_I(y)$\\
  $TI5$ & $\tau_I(x\cdot y)=\tau_I(x)\cdot\tau_I(y)$\\
  $TI6$ & $\tau_I(x\leftmerge y)=\tau_I(x)\leftmerge\tau_I(y)$\\
\end{tabular}
\caption{Axioms of abstraction operator}
\label{AxiomsForPAbstraction}
\end{table}
\end{center}


\begin{theorem}[Soundness of $APRPTC_{\tau}$ with guarded linear recursion]\label{SAPRPTCABS}
Let $x$ and $y$ be $APRPTC_{\tau}$ with guarded linear recursion terms. If $APRPTC_{\tau}$ with guarded linear recursion $\vdash x=y$, then
\begin{enumerate}
  \item $x\approx_{prbp}^{fr} y$;
  \item $x\approx_{prbs}^{fr} y$;
  \item $x\approx_{prbhp}^{fr} y$;
  \item $x\approx_{prbhhp}^{fr} y$.
\end{enumerate}
\end{theorem}

\begin{proof}
Since FR probabilistic rooted branching step bisimulations $\approx_{prbp}^{fr}$, $\approx_{prbs}^{fr}$, $\approx_{prbhp}^{fr}$ and $\approx_{prbhhp}^{fr}$ are all both equivalent and congruent relations
with respect to $APRPTC_{\tau}$ with guarded linear recursion, we only need to check if each axiom in Table \ref{AxiomsForPAbstraction} is sound modulo $\approx_{prbp}^{fr}$, $\approx_{prbs}^{fr}$, $\approx_{prbhp}^{fr}$ and $\approx_{prbhhp}^{fr}$.
The proof is quite trivial and we omit it.
\end{proof}

Though $\tau$-loops are prohibited in guarded linear recursive specifications in a specifiable way, they can be constructed using the abstraction operator, for example, there exist
$\tau$-loops in the process term $\tau_{\{a\}}(\langle X|X=aX\rangle)$. To avoid $\tau$-loops caused by $\tau_I$ and ensure fairness, we introduce the following recursive verification
rules as Table \ref{RVR5} shows, note that $i_1,\cdots, i_m,j_1,\cdots,j_n\in I\subseteq \mathbb{E}\setminus\{\tau\}$.

\begin{center}
\begin{table}
    $$VR_1\quad\frac{x=y+(i_1\leftmerge\cdots\leftmerge i_m)\cdot x, y=y+y,(Std(x),Std(y))}{\tau\cdot\tau_I(x)=\tau\cdot \tau_I(y)}$$
    $$RVR_1\quad\frac{x=y+ x\cdot(i_1[n]\leftmerge\cdots\leftmerge i_m[n]), y=y+y,(NStd(x),NStd(y))}{\tau_I(x)\cdot\tau= \tau_I(y)\cdot\tau}$$
    $$VR_2\quad \frac{x=z\boxplus_{\pi}(u+(i_1\leftmerge\cdots\leftmerge i_m)\cdot x),z=z+u,z=z+z,(Std(x),Std(z),Std(u))}{\tau\cdot\tau_I(x)=\tau\cdot\tau_I(z)}$$
    $$RVR_2\quad \frac{x=z\boxplus_{\pi}(u+ x\cdot(i_1[n]\leftmerge\cdots\leftmerge i_m[n])),z=z+u,z=z+z,(NStd(x),NStd(z),NStd(u))}{\tau_I(x)\cdot\tau=\tau_I(z)\cdot\tau}$$
    $$VR_3\quad \frac{x=z+(i_1\leftmerge\cdots\leftmerge i_m)\cdot y,y=z\boxplus_{\pi}(u+(j_1\leftmerge\cdots\leftmerge j_n)\cdot x), z=z+u,z=z+z}{\tau\cdot\tau_I(x)=\tau\cdot\tau_I(y')\textrm{ for }y'=z\boxplus_{\pi}(u+(i_1\leftmerge\cdots\leftmerge i_m)\cdot y')}$$
    $(Std(x),Std(y),Std(z),Std(u))$
    $$RVR_3\quad \frac{x=z+ y\cdot(i_1[k]\leftmerge\cdots\leftmerge i_m[k]),y=z\boxplus_{\pi}(u+ x\cdot(j_1[l]\leftmerge\cdots\leftmerge j_n[l])), z=z+u,z=z+z}{\tau_I(x)\cdot\tau=\tau_I(y')\cdot\tau\textrm{ for }y'=z\boxplus_{\pi}(u+ y'\cdot(i_1[k]\leftmerge\cdots\leftmerge i_m[k]))}$$
    $(Std(x),Std(y),Std(z),Std(u))$
\caption{Recursive verification rules}
\label{RVR5}
\end{table}
\end{center}

\begin{theorem}[Soundness of $VR_1,VR_2,VR_3$]
$VR_1$, $VR_2$ and $VR_3$ are sound modulo FR probabilistic rooted branching truly concurrent bisimulation equivalences $\approx_{prbp}^{fr}$, $\approx_{prbs}^{fr}$, $\approx_{prbhp}^{fr}$ and $\approx_{prbhhp}^{fr}$.
\end{theorem}

\newpage\section{APRPTC for Closed Quantum Systems}\label{qaprptc2}

The theory $APRPTC$ for closed quantum systems abbreviated qAPRPTC has four modules: $qBARPTC$ , $qAPRPTC$, recursion and abstraction.

This chapter is organized as follows. We introduce $qBARPTC$ in section \ref{qbarptc}, $APRPTC$ in section \ref{qaprptc}, recursion in section \ref{qcrec}, and abstraction in section
\ref{qcabs}. And we introduce quantum measurement in section \ref{qm}, quantum entanglement in section \ref{qe2}, and unification of quantum and classical computing in section \ref{uni2}.

Note that, for a closed quantum system, the unitary operators are the atomic actions (events) and let unitary operators into $\mathbb{E}$. And for the existence of quantum measurement,
the probabilism is unavoidable.

\subsection{$BARPTC$ for Closed Quantum Systems}{\label{qbarptc}}

In this subsection, we will discuss $qBARPTC$. Let $\mathbb{E}$ be the set of atomic events (actions, unitary operators).

In the following, let $e_1, e_2, e_1', e_2'\in \mathbb{E}$, and let variables $x,y,z$ range over the set of terms for true concurrency, $p,q$ range over the set of
closed terms.

The set of axioms of $BARPTC$ consists of the laws given in Table \ref{AxiomsForqBARPTC}.

\begin{center}
    \begin{table}
        \begin{tabular}{@{}ll@{}}
            \hline No. &Axiom\\
            $A1$ & $x+ y = y+ x$\\
            $A2$ & $(x+ y)+ z = x+ (y+ z)$\\
            $A3$ & $e+ e = e$\\
            $A41$ & $(x+ y)\cdot z = x\cdot z + y\cdot z\quad (Std(x),Std(y), Std(z))$\\
            $A42$ & $x\cdot (y+z) = x\cdot y + x\cdot z\quad (NStd(x),NStd(y), NStd(z))$\\
            $A5$ & $(x\cdot y)\cdot z = x\cdot(y\cdot z)$\\
            $A6$ & $x+\delta = x$\\
            $A7$ & $\delta\cdot x =\delta({Std(x)})$\\
            $RA7$ & $x\cdot\delta =\delta({NStd(x)})$\\
            $PA1$ & $x\boxplus_{\pi} y=y\boxplus_{1-\pi} x$\\
            $PA2$ & $x\boxplus_{\pi}(y\boxplus_{\rho} z)=(x\boxplus_{\frac{\pi}{\pi+\rho-\pi\rho}}y)\boxplus_{\pi+\rho-\pi\rho} z$\\
            $PA3$ & $x\boxplus_{\pi}x=x$\\
            $PA41$ & $(x\boxplus_{\pi}y)\cdot z=x\cdot z\boxplus_{\pi}y\cdot z\quad (Std(x),Std(y), Std(z))$\\
            $PA42$ & $x\cdot (y\boxplus_{\pi} z)=x\cdot y\boxplus_{\pi}x\cdot z\quad (NStd(x),NStd(y), NStd(z))$\\
            $PA5$ & $(x\boxplus_{\pi}y)+z=(x+z)\boxplus_{\pi}(y+z)$\\
        \end{tabular}
        \caption{Axioms of $qBARPTC$}
        \label{AxiomsForqBARPTC}
    \end{table}
\end{center}

\begin{definition}[Basic terms of $qBARPTC$]\label{BTBATCG}
The set of basic terms of $qBARPTC$, $\mathcal{B}(qBARPTC)$, is inductively defined as follows:

\begin{enumerate}
  \item $\mathbb{E}\subset\mathcal{B}(qBARPTC)$;
  \item if $e\in \mathbb{E}, t\in\mathcal{B}(qBARPTC)$ then $e\cdot t\in\mathcal{B}(qBARPTC)$;
  \item if $e[m]\in \mathbb{E}, t\in\mathcal{B}(qBARPTC)$ then $t\cdot e[m]\in\mathcal{B}(qBARPTC)$;
  \item if $t,t'\in\mathcal{B}(qBARPTC)$ then $t+ t'\in\mathcal{B}(qBARPTC)$;
  \item if $t,t'\in\mathcal{B}(qBARPTC)$ then $t\boxplus_{\pi} t'\in\mathcal{B}(qBARPTC)$.
\end{enumerate}
\end{definition}

\begin{theorem}[Elimination theorem of $qBARPTC$]\label{ETBATCG}
Let $p$ be a closed $qBARPTC$ term. Then there is a basic $qBARPTC$ term $q$ such that $qBARPTC\vdash p=q$.
\end{theorem}

\begin{proof}
The same as that of $BARPTC$, we omit the proof, please refer to chapter \ref{aprptc2} for details.
\end{proof}

We will define a term-deduction system which gives the operational semantics of $qBARPTC$. We give the operational transition rules for
atomic event $e\in\mathbb{E}$, operators $\cdot$ and $+$ as Table \ref{SETRForqBARPTC} shows. And the predicate $\xrightarrow{e}\surd$ represents successful termination after execution
of the event $e$.

\begin{center}
    \begin{table}
        $$\frac{}{\langle e,\varrho\rangle\rightsquigarrow\langle\breve{e},\varrho\rangle}$$
        $$\frac{\langle x,\varrho\rangle\rightsquigarrow \langle x',\varrho\rangle}{\langle x\cdot y,\varrho\rangle\rightsquigarrow \langle x'\cdot y,\varrho\rangle}$$
        $$\frac{\langle x,\varrho\rangle\rightsquigarrow \langle x',\varrho\rangle\quad \langle y,\varrho\rangle\rightsquigarrow \langle y',\varrho\rangle}{\langle x+y,\varrho\rangle\rightsquigarrow \langle x'+y',\varrho\rangle}$$
        $$\frac{\langle x,\varrho\rangle\rightsquigarrow \langle x',\varrho\rangle}{\langle x\boxplus_{\pi}y,\varrho\rangle\rightsquigarrow \langle x',\varrho\rangle}\quad \frac{\langle y,\varrho\rangle\rightsquigarrow \langle y',\varrho\rangle}{\langle x\boxplus_{\pi}y,\varrho\rangle\rightsquigarrow \langle y',\varrho\rangle}$$

        $$\frac{\langle x,\varrho\rangle\xrightarrow{e}\langle e[m],\varrho'\rangle}{\langle x+ y,\varrho\rangle\xrightarrow{e}\langle e[m],\varrho'\rangle}
        \quad\frac{\langle x,\varrho\rangle\xrightarrow{e}\langle x',\varrho'\rangle}{\langle x+ y,\varrho\rangle\xrightarrow{e}\langle x',\varrho'\rangle}
        \quad\frac{\langle y,\varrho\rangle\xrightarrow{e}\langle e[m],\varrho'\rangle}{\langle x+ y,\varrho\rangle\xrightarrow{e}\langle e[m],\varrho'\rangle}
        \quad\frac{\langle y,\varrho\rangle\xrightarrow{e}\langle y',\varrho'\rangle}{\langle x+ y,\varrho\rangle\xrightarrow{e}\langle y',\varrho'\rangle}$$
        $$\frac{\langle x,\varrho\rangle\xrightarrow{e}\langle e[m],\varrho'\rangle}{\langle x\cdot y,\varrho\rangle\xrightarrow{e} \langle e[m]\cdot y,\varrho'\rangle}
        \quad\frac{\langle x,\varrho\rangle\xrightarrow{e}\langle x',\varrho'\rangle}{\langle x\cdot y,\varrho\rangle\xrightarrow{e}\langle x'\cdot y,\varrho'\rangle}$$
        $$\frac{}{\langle e[m],\varrho\rangle\xtworightarrow{e[m]}\langle e,\varrho'\rangle}$$
        $$\frac{\langle x,\varrho\rangle\xtworightarrow{e[m]}\langle e,\varrho'\rangle}{\langle x+ y,\varrho\rangle\xtworightarrow{e[m]}\langle e,\varrho'\rangle}
        \quad\frac{\langle x,\varrho\rangle\xtworightarrow{e[m]}\langle x',\varrho'\rangle}{\langle x+ y,\varrho\rangle\xtworightarrow{e[m]}\langle x',\varrho'\rangle}
        \quad\frac{\langle y,\varrho\rangle\xtworightarrow{e[m]}\langle e,\varrho'\rangle}{\langle x+ y,\varrho\rangle\xtworightarrow{e[m]}\langle e,\varrho'\rangle}
        \quad\frac{\langle y,\varrho\rangle\xtworightarrow{e[m]}\langle y',\varrho'\rangle}{\langle x+ y,\varrho\rangle\xtworightarrow{e[m]}\langle y',\varrho'\rangle}$$
        $$\frac{\langle y,\varrho\rangle\xtworightarrow{e[m]}\langle e,\varrho'\rangle}{\langle x\cdot y,\varrho\rangle\xtworightarrow{e[m]} \langle x\cdot e,\varrho'\rangle}
        \quad\frac{\langle y,\varrho\rangle\xtworightarrow{e[m]}\langle y',\varrho'\rangle}{\langle x\cdot y,\varrho\rangle\xtworightarrow{e[m]}\langle x\cdot y',\varrho'\rangle}$$
        \caption{Single event transition rules of $qBARPTC$}
        \label{SETRForqBARPTC}
    \end{table}
\end{center}

Note that, we replace the single atomic event $e\in\mathbb{E}$ by $X\subseteq\mathbb{E}$, we can obtain the pomset transition rules of $qBARPTC$, and omit them.

\begin{theorem}[Congruence of $qBARPTC$ with respect to probabilistic truly concurrent bisimulation equivalences]
(1) FR probabilistic pomset bisimulation equivalence $\sim_{pp}^{fr}$ is a congruence with respect to $qBARPTC$.

(2) FR probabilistic step bisimulation equivalence $\sim_{ps}^{fr}$ is a congruence with respect to $qBARPTC$.

(3) FR probabilistic hp-bisimulation equivalence $\sim_{php}^{fr}$ is a congruence with respect to $qBARPTC$.

(4) FR probabilistic hhp-bisimulation equivalence $\sim_{phhp}^{fr}$ is a congruence with respect to $qBARPTC$.
\end{theorem}

\begin{proof}
(1) It is easy to see that FR probabilistic pomset bisimulation is an equivalent relation on $qBARPTC$ terms, we only need to prove that $\sim_{pp}^{fr}$ is preserved by the operators $\cdot$
, $+$ and $\boxplus_{\pi}$. It is trivial and we leave the proof as an exercise for the readers.

(2) It is easy to see that FR probabilistic step bisimulation is an equivalent relation on $qBARPTC$ terms, we only need to prove that $\sim_{ps}^{fr}$ is preserved by the operators $\cdot$,
$+$ and $\boxplus_{\pi}$. It is trivial and we leave the proof as an exercise for the readers.

(3) It is easy to see that FR probabilistic hp-bisimulation is an equivalent relation on $qBARPTC$ terms, we only need to prove that $\sim_{php}^{fr}$ is preserved by the operators $\cdot$,
$+$, and $\boxplus_{\pi}$. It is trivial and we leave the proof as an exercise for the readers.

(4) It is easy to see that FR probabilistic hhp-bisimulation is an equivalent relation on $qBARPTC$ terms, we only need to prove that $\sim_{phhp}^{fr}$ is preserved by the operators $\cdot$,
$+$, and $\boxplus_{\pi}$. It is trivial and we leave the proof as an exercise for the readers.
\end{proof}

\begin{theorem}[Soundness of $qBARPTC$ modulo FR probabilistic truly concurrent bisimulation equivalences]
(1) Let $x$ and $y$ be $qBARPTC$ terms. If $BATC\vdash x=y$, then $x\sim_{pp}^{fr} y$.

(2) Let $x$ and $y$ be $qBARPTC$ terms. If $BATC\vdash x=y$, then $x\sim_{ps}^{fr} y$.

(3) Let $x$ and $y$ be $qBARPTC$ terms. If $BATC\vdash x=y$, then $x\sim_{php}^{fr} y$.

(4) Let $x$ and $y$ be $qBARPTC$ terms. If $BATC\vdash x=y$, then $x\sim_{phhp}^{fr} y$.
\end{theorem}

\begin{proof}
(1) Since FR probabilistic pomset bisimulation $\sim_{pp}^{fr}$ is both an equivalent and a congruent relation, we only need to check if each axiom in Table \ref{AxiomsForqBARPTC} is sound
modulo FR probabilistic pomset bisimulation equivalence. We leave the proof as an exercise for the readers.

(2) Since FR probabilistic step bisimulation $\sim_{ps}^{fr}$ is both an equivalent and a congruent relation, we only need to check if each axiom in Table \ref{AxiomsForqBARPTC} is sound modulo
FR probabilistic step bisimulation equivalence. We leave the proof as an exercise for the readers.

(3) Since FR probabilistic hp-bisimulation $\sim_{php}^{fr}$ is both an equivalent and a congruent relation, we only need to check if each axiom in Table \ref{AxiomsForqBARPTC} is sound modulo
FR probabilistic hp-bisimulation equivalence. We leave the proof as an exercise for the readers.

(4) Since FR probabilistic hhp-bisimulation $\sim_{phhp}^{fr}$ is both an equivalent and a congruent relation, we only need to check if each axiom in Table \ref{AxiomsForqBARPTC} is sound modulo
FR probabilistic hhp-bisimulation equivalence. We leave the proof as an exercise for the readers.
\end{proof}

\begin{theorem}[Completeness of $qBARPTC$ modulo FR probabilistic truly concurrent bisimulation equivalences]\label{CBATCG}
(1) Let $p$ and $q$ be closed $qBARPTC$ terms, if $p\sim_{pp}^{fr} q$ then $p=q$.

(2) Let $p$ and $q$ be closed $qBARPTC$ terms, if $p\sim_{ps}^{fr} q$ then $p=q$.

(3) Let $p$ and $q$ be closed $qBARPTC$ terms, if $p\sim_{php}^{fr} q$ then $p=q$.

(4) Let $p$ and $q$ be closed $qBARPTC$ terms, if $p\sim_{phhp}^{fr} q$ then $p=q$.
\end{theorem}

\begin{proof}
According to the definition of FR probabilistic truly concurrent bisimulation equivalences $\sim_{pp}^{fr}$, $\sim_{ps}^{fr}$, $\sim_{php}^{fr}$ and $\sim_{phhp}^{fr}$, $p\sim_{pp}^{fr}q$, $p\sim_{ps}^{fr}q$, $p\sim_{php}^{fr}q$ and $p\sim_{phhp}^{fr}q$ implies
both the bisimilarities between $p$ and $q$, and also the in the same quantum states. According to the completeness of BARPTC (please refer to chapter \ref{aprptc2} for details), we can get the
completeness of qBARPTC.
\end{proof}

\subsection{$APRPTC$ for Closed Quantum Systems}{\label{qaprptc}}

In this subsection, we will extend $APRPTC$ for closed quantum systems, which is abbreviated $qAPRPTC$.

The set of axioms of $qAPRPTC$ including axioms of $qBARPTC$ in Table \ref{AxiomsForqBARPTC} and the axioms are shown in Table \ref{AxiomsForqAPRPTC}.

\begin{center}
    \begin{table}
        \begin{tabular}{@{}ll@{}}
            \hline No. &Axiom\\
            $P1$ & $(x+x=x,y+y=y)\quad x\between y = x\parallel y + x\mid y$\\
            $P2$ & $x\parallel y = y \parallel x$\\
            $P3$ & $(x\parallel y)\parallel z = x\parallel (y\parallel z)$\\
            $P4$ & $(x+x=x,y+y=y)\quad x\parallel y = x\leftmerge y + y\leftmerge x$\\
            $P5$ & $(e_1\leq e_2)\quad e_1\leftmerge (e_2\cdot y) = (e_1\leftmerge e_2)\cdot y$\\
            $RP5$ & $(e_1[m]\leq e_2[m])\quad e_1[m]\leftmerge (y\cdot e_2[m]) = y\cdot(e_1[m]\leftmerge e_2[m])$\\
            $P6$ & $(e_1\leq e_2)\quad (e_1\cdot x)\leftmerge e_2 = (e_1\leftmerge e_2)\cdot x$\\
            $RP6$ & $(e_1[m]\leq e_2[m])\quad (x\cdot e_1[m])\leftmerge e_2[m] = x\cdot(e_1[m]\leftmerge e_2[m])$\\
            $P7$ & $(e_1\leq e_2)\quad (e_1\cdot x)\leftmerge (e_2\cdot y) = (e_1\leftmerge e_2)\cdot (x\between y)$\\
            $RP7$ & $(e_1[m]\leq e_2[m])\quad(x\cdot e_1[m])\leftmerge (y\cdot e_2[m]) = (x\between y)\cdot(e_1[m]\leftmerge e_2[m])$\\
            $P8$ & $(x+ y)\leftmerge z = (x\leftmerge z)+ (y\leftmerge z)(\textrm{Std(x)},\textrm{Std(y)},\textrm{Std(z)})$\\
            $RP8$ & $x\leftmerge (y+ z) = (x\leftmerge y)+ (x\leftmerge z)(\textrm{NStd(x)},\textrm{NStd(y)},\textrm{NStd(z)})$\\
            $P9$ & $\delta\leftmerge x = \delta(\textrm{Std(x)})$\\
            $RP9$ & $x\leftmerge \delta = \delta(\textrm{NStd(x)})$\\
            $C1$ & $e_1\mid e_2 = \gamma(e_1,e_2)$\\
            $RC1$ & $e_1[m]\mid e_2[m] = \gamma(e_1,e_2)[m]$\\
            $C2$ & $e_1\mid (e_2\cdot y) = \gamma(e_1,e_2)\cdot y$\\
            $RC2$ & $e_1[m]\mid (y \cdot e_2[m]) =y\cdot \gamma(e_1,e_2)[m]$\\
            $C3$ & $(e_1\cdot x)\mid e_2 = \gamma(e_1,e_2)\cdot x$\\
            $RC3$ & $(x \cdot e_1[m])\mid e_2[m] =x\cdot \gamma(e_1,e_2)[m]$\\
            $C4$ & $(e_1\cdot x)\mid (e_2\cdot y) = \gamma(e_1,e_2)\cdot (x\between y)$\\
            $RC4$ & $(x \cdot e_1[m])\mid (y \cdot e_2[m]) =(x\between y)\cdot \gamma(e_1,e_2)[m]$\\
            $C5$ & $(x+ y)\mid z = (x\mid z) + (y\mid z)$\\
            $C6$ & $x\mid (y+ z) = (x\mid y)+ (x\mid z)$\\
            $C7$ & $\delta\mid x = \delta$\\
            $C8$ & $x\mid\delta = \delta$\\
            $PM1$ & $x\parallel (y\boxplus_{\pi} z)=(x\parallel y)\boxplus_{\pi}(x\parallel z)$\\
            $PM2$ & $(x\boxplus_{\pi} y)\parallel z=(x\parallel z)\boxplus_{\pi}(y\parallel z)$\\
            $PM3$ & $x\mid (y\boxplus_{\pi} z)=(x\mid y)\boxplus_{\pi}(x\mid z)$\\
            $PM4$ & $(x\boxplus_{\pi} y)\mid z=(x\mid z)\boxplus_{\pi}(y\mid z)$\\
            $CE1$ & $\Theta(e) = e$\\
            $RCE1$ & $\Theta(e[m]) = e[m]$\\
            $CE2$ & $\Theta(\delta) = \delta$\\
            $CE3$ & $\Theta(x+ y) = \Theta(x)\triangleleft y + \Theta(y)\triangleleft x$\\
            $PCE1$ & $\Theta(x\boxplus_{\pi} y) = \Theta(x)\triangleleft y \boxplus_{\pi} \Theta(y)\triangleleft x$\\
            $CE4$ & $\Theta(x\cdot y)=\Theta(x)\cdot\Theta(y)$\\
            $CE5$ & $\Theta(x\leftmerge y) = ((\Theta(x)\triangleleft y)\leftmerge y)+ ((\Theta(y)\triangleleft x)\leftmerge x)$\\
            $CE6$ & $\Theta(x\mid y) = ((\Theta(x)\triangleleft y)\mid y)+ ((\Theta(y)\triangleleft x)\mid x)$\\
        \end{tabular}
        \caption{Axioms of $qAPRPTC$}
        \label{AxiomsForqAPRPTC}
    \end{table}
\end{center}

\begin{center}
    \begin{table}
        \begin{tabular}{@{}ll@{}}
            \hline No. &Axiom\\
            $U1$ & $(\sharp(e_1,e_2))\quad e_1\triangleleft e_2 = \tau$\\
            $RU1$ & $(\sharp(e_1[m],e_2[n]))\quad e_1[m]\triangleleft e_2[n] = \tau$\\
            $U2$ & $(\sharp(e_1,e_2),e_2\leq e_3)\quad e_1\triangleleft e_3 = e_1$\\
            $RU2$ & $(\sharp(e_1[m],e_2[n]),e_2[n]\geq e_3[l])\quad e_1[m]\triangleleft e_3[l] = e_1[m]$\\
            $U3$ & $(\sharp(e_1,e_2),e_2\leq e_3)\quad e3\triangleleft e_1 = \tau$\\
            $RU3$ & $(\sharp(e_1[m],e_2[n]),e_2[n]\geq e_3[l])\quad e3[l]\triangleleft e_1[m] = \tau$\\
            $PU1$ & $(\sharp_{\pi}(e_1,e_2))\quad e_1\triangleleft e_2 = \tau$\\
            $RPU1$ & $(\sharp_{\pi}(e_1[m],e_2[n]))\quad e_1[m]\triangleleft e_2[n] = \tau$\\
            $PU2$ & $(\sharp_{\pi}(e_1,e_2),e_2\leq e_3)\quad e_1\triangleleft e_3 = e_1$\\
            $RPU2$ & $(\sharp_{\pi}(e_1[m],e_2[n]),e_2[n]\leq e_3[l])\quad e_1[m]\triangleleft e_3[l] = e_1[m]$\\
            $PU3$ & $(\sharp_{\pi}(e_1,e_2),e_2\leq e_3)\quad e_3\triangleleft e_1 = \tau$\\
            $RPU3$ & $(\sharp_{\pi}(e_1[m],e_2[n]),e_2[n]\leq e_3[l])\quad e_3[l]\triangleleft e_1[m] = \tau$\\
            $U4$ & $e\triangleleft \delta = e$\\
            $RU4$ & $e[m]\triangleleft \delta = \delta$\\
            $U5$ & $\delta \triangleleft e = \delta$\\
            $RU5$ & $\delta \triangleleft e[m] = e[m]$\\
            $U6$ & $(x+ y)\triangleleft z = (x\triangleleft z)+ (y\triangleleft z)$\\
            $PU4$ & $(x\boxplus_{\pi} y)\triangleleft z = (x\triangleleft z)\boxplus_{\pi} (y\triangleleft z)$\\
            $U7$ & $(x\cdot y)\triangleleft z = (x\triangleleft z)\cdot (y\triangleleft z)$\\
            $U8$ & $(x\leftmerge y)\triangleleft z = (x\triangleleft z)\leftmerge (y\triangleleft z)$\\
            $U9$ & $(x\mid y)\triangleleft z = (x\triangleleft z)\mid (y\triangleleft z)$\\
            $U10$ & $x\triangleleft (y+ z) = (x\triangleleft y)\triangleleft z$\\
            $PU5$ & $x\triangleleft (y\boxplus_{\pi} z) = (x\triangleleft y)\triangleleft z$\\
            $U11$ & $x\triangleleft (y\cdot z)=(x\triangleleft y)\triangleleft z$\\
            $U12$ & $x\triangleleft (y\leftmerge z) = (x\triangleleft y)\triangleleft z$\\
            $U13$ & $x\triangleleft (y\mid z) = (x\triangleleft y)\triangleleft z$\\
            $D1$ & $e\notin H\quad\partial_H(e) = e$\\
            $RD1$ & $e\notin H\quad\partial_H(e[m]) = e[m]$\\
            $D2$ & $e\in H\quad \partial_H(e) = \delta$\\
            $RD2$ & $e\in H\quad \partial_H(e[m]) = \delta$\\
            $D3$ & $\partial_H(\delta) = \delta$\\
            $D4$ & $\partial_H(x+ y) = \partial_H(x)+\partial_H(y)$\\
            $D5$ & $\partial_H(x\cdot y) = \partial_H(x)\cdot\partial_H(y)$\\
            $D6$ & $\partial_H(x\leftmerge y) = \partial_H(x)\leftmerge\partial_H(y)$\\
            $PD1$ & $\partial_H(x\boxplus_{\pi}y)=\partial_H(x)\boxplus_{\pi}\partial_H(y)$\\
        \end{tabular}
        \caption{Axioms of $qAPRPTC$ (continuing)}
        \label{AxiomsForqAPRPTC2}
    \end{table}
\end{center}

\begin{definition}[Basic terms of $qAPRPTC$]\label{BTqAPRPTCG}
The set of basic terms of $qAPRPTC$, $\mathcal{B}(qAPRPTC)$, is inductively defined as follows:

\begin{enumerate}
    \item $\mathbb{E}\subset\mathcal{B}(qAPRPTC)$;
    \item if $e\in \mathbb{E}, t\in\mathcal{B}(qAPRPTC)$ then $e\cdot t\in\mathcal{B}(qAPRPTC)$;
    \item if $e[m]\in \mathbb{E}, t\in\mathcal{B}(APRPTC)$ then $t\cdot e[m]\in\mathcal{B}(APRPTC)$;
    \item if $t,t'\in\mathcal{B}(qAPRPTC)$ then $t+ t'\in\mathcal{B}(qAPRPTC)$;
    \item if $t,t'\in\mathcal{B}(qAPRPTC)$ then $t\boxplus_{\pi} t'\in\mathcal{B}(qAPRPTC)$
    \item if $t,t'\in\mathcal{B}(qAPRPTC)$ then $t\leftmerge t'\in\mathcal{B}(qAPRPTC)$.
\end{enumerate}
\end{definition}

Based on the definition of basic terms for $qAPRPTC$ (see Definition \ref{BTqAPRPTCG}) and axioms of $qAPRPTC$, we can prove the elimination theorem of $qAPRPTC$.

\begin{theorem}[Elimination theorem of $qAPRPTC$]\label{ETqAPRPTCG}
Let $p$ be a closed $qAPRPTC$ term. Then there is a basic $qAPRPTC$ term $q$ such that $qAPRPTC\vdash p=q$.
\end{theorem}

\begin{proof}
The same as that of $APRPTC$, we omit the proof, please refer to chapter \ref{aprptc2} for details.
\end{proof}

We will define a term-deduction system which gives the operational semantics of $qAPRPTC$. Two atomic events $e_1$ and $e_2$ are in race condition, which are denoted $e_1\% e_2$.

\begin{center}
    \begin{table}
        $$\frac{x\rightsquigarrow x'\quad y\rightsquigarrow y'}{x\between y\rightsquigarrow x'\parallel y'+x'\mid y'}$$
        $$\frac{x\rightsquigarrow x'\quad y\rightsquigarrow y'}{x\parallel y\rightsquigarrow x'\leftmerge y+y'\leftmerge x}$$
        $$\frac{x\rightsquigarrow x'}{x\leftmerge y\rightsquigarrow x'\leftmerge y}$$
        $$\frac{x\rightsquigarrow x'\quad y\rightsquigarrow y'}{x\mid y\rightsquigarrow x'\mid y'}$$
        $$\frac{x\rightsquigarrow x'}{\Theta(x)\rightsquigarrow \Theta(x')}$$
        $$\frac{x\rightsquigarrow x'}{x\triangleleft y\rightsquigarrow x'\triangleleft y}$$
        \caption{Probabilistic transition rules of $qAPRPTC$}
        \label{TRForAPRPTCG1}
    \end{table}
\end{center}

\begin{center}
    \begin{table}
        $$\frac{\langle x,\varrho\rangle\xrightarrow{e_1}\langle e_1[m],\varrho'\rangle\quad \langle y,\varrho\rangle\xrightarrow{e_2}\langle e_2[m],\varrho''\rangle}{\langle x\parallel y,\varrho\rangle\xrightarrow{\{e_1,e_2\}}\langle e_1[m]\parallel e_2[m],\varrho'\cup\varrho''\rangle}
        \quad\frac{\langle x,\varrho\rangle\xrightarrow{e_1}\langle x',\varrho'\rangle\quad \langle y,\varrho\rangle\xrightarrow{e_2}\langle e_2[m],\varrho''\rangle}{\langle x\parallel y,\varrho\rangle\xrightarrow{\{e_1,e_2\}}\langle x'\parallel e_2[m],\varrho'\cup\varrho''\rangle}$$
        $$\frac{\langle x,\varrho\rangle\xrightarrow{e_1}\langle e_1[m],\varrho'\rangle\quad \langle y,\varrho\rangle\xrightarrow{e_2}\langle y',\varrho''\rangle}{\langle x\parallel y,\varrho\rangle\xrightarrow{\{e_1,e_2\}}\langle e_1[m]\parallel y',\varrho'\cup\varrho''\rangle}
        \quad\frac{\langle x,\varrho\rangle\xrightarrow{e_1}\langle x',\varrho'\rangle\quad \langle y,\varrho\rangle\xrightarrow{e_2}\langle y',\varrho''\rangle}{\langle x\parallel y,\varrho\rangle\xrightarrow{\{e_1,e_2\}}\langle x'\between y',\varrho'\cup\varrho''\rangle}$$
        $$\frac{\langle x,\varrho\rangle\xrightarrow{e_1}\langle e_1[m],\varrho'\rangle\quad \langle y,\varrho\rangle\xrightarrow{e_2}\langle e_2[m],\varrho''\rangle \quad(e_1\leq e_2)}{\langle x\leftmerge y,\varrho\rangle\xrightarrow{\{e_1,e_2\}}\langle e_1[m]\leftmerge e_2[m],\varrho'\cup\varrho''\rangle}
        \quad\frac{\langle x,\varrho\rangle\xrightarrow{e_1}\langle x',\varrho'\rangle\quad \langle y,\varrho\rangle\xrightarrow{e_2}\langle e_2[m],\varrho''\rangle \quad(e_1\leq e_2)}{\langle x\leftmerge y,\varrho\rangle\xrightarrow{\{e_1,e_2\}}\langle x'\leftmerge e_2[m],\varrho'\cup\varrho''\rangle}$$
        $$\frac{\langle x,\varrho\rangle\xrightarrow{e_1}\langle e_1[m],\varrho'\rangle\quad \langle y,\varrho\rangle\xrightarrow{e_2}\langle y',\varrho''\rangle \quad(e_1\leq e_2)}{\langle x\leftmerge y,\varrho\rangle\xrightarrow{\{e_1,e_2\}}\langle e_1[m]\leftmerge y',\varrho'\cup\varrho''\rangle}
        \quad\frac{\langle x,\varrho\rangle\xrightarrow{e_1}\langle x',\varrho'\rangle\quad \langle y,\varrho\rangle\xrightarrow{e_2}\langle y',\varrho''\rangle \quad(e_1\leq e_2)}{\langle x\leftmerge y,\varrho\rangle\xrightarrow{\{e_1,e_2\}}\langle x'\between y',\varrho'\cup\varrho''\rangle}$$
        $$\frac{\langle x,\varrho\rangle\xrightarrow{e_1}\langle e_1[m],\varrho'\rangle\quad \langle y,\varrho\rangle\xrightarrow{e_2}e_2[m]}{\langle x\mid y,\varrho\rangle\xrightarrow{\gamma(e_1,e_2)}\langle\gamma(e_1,e_2)[m],\varrho'\cup\varrho''\rangle}
        \quad\frac{\langle x,\varrho\rangle\xrightarrow{e_1}\langle x',\varrho'\rangle\quad \langle y,\varrho\rangle\xrightarrow{e_2}\langle e_2[m],\varrho''\rangle}{\langle x\mid y,\varrho\rangle\xrightarrow{\gamma(e_1,e_2)}\langle\gamma(e_1,e_2)[m]\cdot x',\varrho'\cup\varrho''\rangle}$$
        $$\frac{\langle x,\varrho\rangle\xrightarrow{e_1}\langle e_1[m],\varrho'\rangle\quad \langle y,\varrho\rangle\xrightarrow{e_2}\langle y',\varrho''\rangle}{\langle x\mid y,\varrho\rangle\xrightarrow{\gamma(e_1,e_2)}\langle \gamma(e_1,e_2)[m]\cdot y',\varrho'\cup\varrho''\rangle}
        \quad\frac{\langle x,\varrho\rangle\xrightarrow{e_1}\langle x',\varrho'\rangle\quad \langle y,\varrho\rangle\xrightarrow{e_2}\langle y',\varrho''\rangle}{\langle x\mid y,\varrho\rangle\xrightarrow{\gamma(e_1,e_2)}\langle\gamma(e_1,e_2)[m]\cdot x'\between y',\varrho'\cup\varrho''\rangle}$$
        $$\frac{\langle x,\varrho\rangle\xrightarrow{e_1}\langle e_1[m],\varrho'\rangle\quad (\sharp(e_1,e_2))}{\langle\Theta(x),\varrho\rangle\xrightarrow{e_1}\langle e_1[m],\varrho'\rangle}
        \quad\frac{\langle x,\varrho\rangle\xrightarrow{e_2}\langle e_2[n],\varrho'\rangle\quad (\sharp(e_1,e_2))}{\langle\Theta(x),\varrho\rangle\xrightarrow{e_2}\langle e_2[n],\varrho'\rangle}$$
        $$\frac{\langle x,\varrho\rangle\xrightarrow{e_1}\langle x',\varrho'\rangle\quad (\sharp(e_1,e_2))}{\langle\Theta(x),\varrho\rangle\xrightarrow{e_1}\langle\Theta(x'),\varrho'\rangle}
        \quad\frac{\langle x,\varrho\rangle\xrightarrow{e_2}\langle x',\varrho'\rangle\quad (\sharp(e_1,e_2))}{\langle \Theta(x),\varrho\rangle\xrightarrow{e_2}\langle\Theta(x'),\varrho'\rangle}$$
        $$\frac{\langle x,\varrho\rangle\xrightarrow{e_1}\langle e_1[m],\varrho'\rangle \quad \langle y,\varrho\rangle\nrightarrow^{e_2}\quad (\sharp(e_1,e_2))}{\langle x\triangleleft y,\varrho\rangle\xrightarrow{\tau}\langle\surd,\tau(\varrho')\rangle}
        \quad\frac{\langle x,\varrho\rangle\xrightarrow{e_1}\langle x',\varrho'\rangle \quad \langle y,\varrho\rangle\nrightarrow^{e_2}\quad (\sharp(e_1,e_2))}{\langle x\triangleleft y,\varrho\rangle\xrightarrow{\tau}\langle x',\tau(\varrho')\rangle}$$
        $$\frac{\langle x,\varrho\rangle\xrightarrow{e_1}\langle e_1[m],\varrho'\rangle \quad \langle y,\varrho\rangle\nrightarrow^{e_3}\quad (\sharp(e_1,e_2),e_2\leq e_3)}{\langle x\triangleleft y,\varrho\rangle\xrightarrow{e_1}\langle e_1[m],\varrho'\rangle}
        \quad\frac{\langle x,\varrho\rangle\xrightarrow{e_1}\langle x',\varrho'\rangle \quad \langle y,\varrho\rangle\nrightarrow^{e_3}\quad (\sharp(e_1,e_2),e_2\leq e_3)}{\langle x\triangleleft y,\varrho\rangle\xrightarrow{e_1}\langle x',\varrho'\rangle}$$
        $$\frac{\langle x,\varrho\rangle\xrightarrow{e_3}\langle e_3[l],\varrho'\rangle \quad \langle y,\varrho\rangle\nrightarrow^{e_2}\quad (\sharp(e_1,e_2),e_1\leq e_3)}{\langle x\triangleleft y,\varrho\rangle\xrightarrow{\tau}\langle\surd,\tau(\varrho')\rangle}
        \quad\frac{\langle x,\varrho\rangle\xrightarrow{e_3}\langle x',\varrho'\rangle \quad \langle y,\varrho\rangle\nrightarrow^{e_2}\quad (\sharp(e_1,e_2),e_1\leq e_3)}{\langle x\triangleleft y,\varrho\rangle\xrightarrow{\tau}\langle x',\tau(\varrho')\rangle}$$

        \caption{Forward action transition rules of $qAPRPTC$}
        \label{TRForqAPRPTCG}
    \end{table}
\end{center}

\begin{center}
    \begin{table}
        $$\frac{\langle x,\varrho\rangle\xrightarrow{e_1}\langle e_1[m],\varrho'\rangle\quad (\sharp_{\pi}(e_1,e_2))}{\langle\Theta(x),\varrho\rangle\xrightarrow{e_1}\langle e_1[m],\varrho'\rangle}
        \quad\frac{\langle x,\varrho\rangle\xrightarrow{e_2}\langle e_2[n],\varrho'\rangle\quad (\sharp_{\pi}(e_1,e_2))}{\langle\Theta(x),\varrho\rangle\xrightarrow{e_2}\langle e_2[n],\varrho'\rangle}$$
        $$\frac{\langle x,\varrho\rangle\xrightarrow{e_1}\langle x',\varrho'\rangle\quad (\sharp_{\pi}(e_1,e_2))}{\langle\Theta(x),\varrho\rangle\xrightarrow{e_1}\langle\Theta(x'),\varrho'\rangle}
        \quad\frac{\langle x,\varrho\rangle\xrightarrow{e_2}\langle x',\varrho'\rangle\quad (\sharp_{\pi}(e_1,e_2))}{\langle \Theta(x),\varrho\rangle\xrightarrow{e_2}\langle\Theta(x'),\varrho'\rangle}$$
        $$\frac{\langle x,\varrho\rangle\xrightarrow{e_1}\langle e_1[m],\varrho'\rangle \quad \langle y,\varrho\rangle\nrightarrow^{e_2}\quad (\sharp_{\pi}(e_1,e_2))}{\langle x\triangleleft y,\varrho\rangle\xrightarrow{\tau}\langle\surd,\tau(\varrho')\rangle}
        \quad\frac{\langle x,\varrho\rangle\xrightarrow{e_1}\langle x',\varrho'\rangle \quad \langle y,\varrho\rangle\nrightarrow^{e_2}\quad (\sharp_{\pi}(e_1,e_2))}{\langle x\triangleleft y,\varrho\rangle\xrightarrow{\tau}\langle x',\tau(\varrho')\rangle}$$
        $$\frac{\langle x,\varrho\rangle\xrightarrow{e_1}\langle e_1[m],\varrho'\rangle \quad \langle y,\varrho\rangle\nrightarrow^{e_3}\quad (\sharp_{\pi}(e_1,e_2),e_2\leq e_3)}{\langle x\triangleleft y,\varrho\rangle\xrightarrow{e_1}\langle e_1[m],\varrho'\rangle}
        \quad\frac{\langle x,\varrho\rangle\xrightarrow{e_1}\langle x',\varrho'\rangle \quad \langle y,\varrho\rangle\nrightarrow^{e_3}\quad (\sharp_{\pi}(e_1,e_2),e_2\leq e_3)}{\langle x\triangleleft y,\varrho\rangle\xrightarrow{e_1}\langle x',\varrho'\rangle}$$
        $$\frac{\langle x,\varrho\rangle\xrightarrow{e_3}\langle e_3[l],\varrho'\rangle \quad \langle y,\varrho\rangle\nrightarrow^{e_2}\quad (\sharp_{\pi}(e_1,e_2),e_1\leq e_3)}{\langle x\triangleleft y,\varrho\rangle\xrightarrow{\tau}\langle\surd,\tau(\varrho')\rangle}
        \quad\frac{\langle x,\varrho\rangle\xrightarrow{e_3}\langle x',\varrho'\rangle \quad \langle y,\varrho\rangle\nrightarrow^{e_2}\quad (\sharp_{\pi}(e_1,e_2),e_1\leq e_3)}{\langle x\triangleleft y,\varrho\rangle\xrightarrow{\tau}\langle x',\tau(\varrho')\rangle}$$
        $$\frac{\langle x\varrho\rangle\xrightarrow{e}\langle e[m],\varrho'\rangle}{\langle\partial_H(x),\varrho\rangle\xrightarrow{e}\langle\partial_H(e[m]),\varrho'\rangle}\quad (e\notin H)
        \quad\frac{\langle x,\varrho\rangle\xrightarrow{e}\angle x',\varrho'\rangle}{\langle\partial_H(x),\varrho\rangle\xrightarrow{e}\langle\partial_H(x'),\varrho'\rangle}\quad(e\notin H)$$
        \caption{Forward action transition rules of $qAPRPTC$ (continuing)}
        \label{TRForqAPRPTCG2}
    \end{table}
\end{center}

\begin{center}
    \begin{table}
        $$\frac{\langle x,\varrho\rangle\xtworightarrow{e_1[m]}\langle e_1,\varrho'\rangle\quad \langle y,\varrho\rangle\xtworightarrow{e_2[m]}\langle e_2,\varrho''\rangle}{\langle x\parallel y,\varrho\rangle\xtworightarrow{\{e_1[m],e_2[m]\}}\langle e_1\parallel e_2,\varrho'\cup\varrho''\rangle}
        \quad\frac{\langle x,\varrho\rangle\xtworightarrow{e_1[m]}\langle x',\varrho'\rangle\quad \langle y,\varrho\rangle\xtworightarrow{e_2[m]}\langle e_2,\varrho''\rangle}{\langle x\parallel y,\varrho\rangle\xtworightarrow{\{e_1[m],e_2[m]\}}\langle x'\parallel e_2,\varrho'\cup\varrho''\rangle}$$
        $$\frac{\langle x,\varrho\rangle\xtworightarrow{e_1[m]}\langle e_1,\varrho'\rangle\quad \langle y,\varrho\rangle\xtworightarrow{e_2[m]}\langle y',\varrho''\rangle}{\langle x\parallel y,\varrho\rangle\xtworightarrow{\{e_1[m],e_2[m]\}}\langle e_1\parallel y',\varrho'\cup\varrho''\rangle}
        \quad\frac{\langle x,\varrho\rangle\xtworightarrow{e_1[m]}\langle x',\varrho'\rangle\quad \langle y,\varrho\rangle\xtworightarrow{e_2[m]}\langle y',\varrho''\rangle}{\langle x\parallel y,\varrho\rangle\xtworightarrow{\{e_1[m],e_2[m]\}}\langle x'\between y',\varrho'\cup\varrho''\rangle}$$
        $$\frac{\langle x,\varrho\rangle\xtworightarrow{e_1[m]}\langle e_1,\varrho'\rangle\quad \langle y,\varrho\rangle\xtworightarrow{e_2[m]}\langle e_2,\varrho''\rangle \quad(e_1\leq e_2)}{\langle x\leftmerge y,\varrho\rangle\xtworightarrow{\{e_1[m],e_2[m]\}}\langle e_1\leftmerge e_2,\varrho'\cup\varrho''\rangle}
        \quad\frac{\langle x,\varrho\rangle\xtworightarrow{e_1[m]}\langle x',\varrho'\rangle\quad \langle y,\varrho\rangle\xtworightarrow{e_2[m]}\langle e_2,\varrho''\rangle \quad(e_1\leq e_2)}{\langle x\leftmerge y,\varrho\rangle\xtworightarrow{\{e_1[m],e_2[m]\}}\langle x'\leftmerge e_2,\varrho'\cup\varrho''\rangle}$$
        $$\frac{\langle x,\varrho\rangle\xtworightarrow{e_1[m]}\langle e_1,\varrho'\rangle\quad \langle y,\varrho\rangle\xtworightarrow{e_2[m]}\langle y',\varrho''\rangle \quad(e_1\leq e_2)}{\langle x\leftmerge y,\varrho\rangle\xtworightarrow{\{e_1[m],e_2[m]\}}\langle e_1\leftmerge y',\varrho'\cup\varrho''\rangle}
        \quad\frac{\langle x,\varrho\rangle\xtworightarrow{e_1[m]}\langle x',\varrho'\rangle\quad \langle y,\varrho\rangle\xtworightarrow{e_2[m]}\langle y',\varrho''\rangle \quad(e_1\leq e_2)}{\langle x\leftmerge y,\varrho\rangle\xtworightarrow{\{e_1[m],e_2[m]\}}\langle x'\between y',\varrho'\cup\varrho''\rangle}$$
        $$\frac{\langle x,\varrho\rangle\xtworightarrow{e_1[m]}\langle e_1,\varrho'\rangle\quad \langle y,\varrho\rangle\xtworightarrow{e_2[m]}\langle e_2,\varrho''\rangle}{\langle x\mid y,\varrho\rangle\xtworightarrow{\gamma(e_1,e_2)[m]}\langle\gamma(e_1,e_2),\varrho'\cup\varrho''\rangle}
        \quad\frac{\langle x,\varrho\rangle\xtworightarrow{e_1[m]}\langle x',\varrho'\rangle\quad \langle y,\varrho\rangle\xtworightarrow{e_2[m]}\langle e_2,\varrho''\rangle}{\langle x\mid y,\varrho\rangle\xtworightarrow{\gamma(e_1,e_2)[m]}\langle\gamma(e_1,e_2)\cdot x',\varrho'\cup\varrho''\rangle}$$
        $$\frac{\langle x,\varrho\rangle\xtworightarrow{e_1[m]}\langle e_1,\varrho'\rangle\quad \langle y,\varrho\rangle\xtworightarrow{e_2[m]}\langle y',\varrho''\rangle}{\langle x\mid y,\varrho\rangle\xtworightarrow{\gamma(e_1,e_2)[m]}\langle\gamma(e_1,e_2)\cdot y',\varrho'\cup\varrho''\rangle}
        \quad\frac{\langle x,\varrho\rangle\xtworightarrow{e_1[m]}\langle x',\varrho'\rangle\quad \langle y,\varrho\rangle\xtworightarrow{e_2[m]}\langle y',\varrho''\rangle}{\langle x\mid y,\varrho\rangle\xtworightarrow{\gamma(e_1,e_2)[m]}\langle\gamma(e_1,e_2)\cdot x'\between y',\varrho'\cup\varrho''\rangle}$$
        $$\frac{\langle x,\varrho\rangle\xtworightarrow{e_1[m]}\langle e_1,\varrho'\rangle\quad (\sharp(e_1,e_2))}{\langle\Theta(x),\varrho\rangle\xtworightarrow{e_1[m]}\langle e_1,\varrho'\rangle}
        \quad\frac{\langle x,\varrho\rangle\xtworightarrow{e_2[n]}\langle e_2,\varrho'\rangle\quad (\sharp(e_1,e_2))}{\langle\Theta(x),\varrho\rangle\xtworightarrow{e_2[n]}\langle e_2,\varrho'\rangle}$$
        $$\frac{\langle x,\varrho\rangle\xtworightarrow{e_1[m]}\langle x',\varrho'\rangle\quad (\sharp(e_1,e_2))}{\langle\Theta(x),\varrho\rangle\xtworightarrow{e_1[m]}\langle \Theta(x'),\varrho'\rangle}
        \quad\frac{\langle x,\varrho\rangle\xtworightarrow{e_2[n]}\langle x',\varrho'\rangle\quad (\sharp(e_1,e_2))}{\langle\Theta(x),\varrho\rangle\xtworightarrow{e_2[n]}\langle\Theta(x'),\varrho'\rangle}$$
        $$\frac{\langle x,\varrho\rangle\xtworightarrow{e_1[m]}\langle e_1,\varrho'\rangle \quad \langle y,\varrho\rangle\xntworightarrow{e_2[n]}\quad (\sharp(e_1,e_2))}{\langle x\triangleleft y,\varrho\rangle\xtworightarrow{\tau}\langle\surd,\tau(\varrho')\rangle}
        \quad\frac{\langle x,\varrho\rangle\xtworightarrow{e_1[m]}\langle x',\varrho'\rangle \quad \langle y,\varrho\rangle\xntworightarrow{e_2[n]}\quad (\sharp(e_1,e_2))}{\langle x\triangleleft y,\varrho\rangle\xtworightarrow{\tau}\langle x',\tau(\varrho')\rangle}$$
        $$\frac{\langle x,\varrho\rangle\xtworightarrow{e_1[m]}\langle e_1,\varrho'\rangle \quad \langle y,\varrho\rangle\xntworightarrow{e_3[l]}\quad (\sharp(e_1,e_2),e_2\geq e_3)}{\langle x\triangleleft y,\varrho\rangle\xtworightarrow{e_1[m]}\langle e_1,\varrho'\rangle}
        \quad\frac{\langle x,\varrho\rangle\xtworightarrow{e_1[m]}x' \quad \langle y,\varrho\rangle\xntworightarrow{e_3[l]}\quad (\sharp(e_1,e_2),e_2\geq e_3)}{\langle x\triangleleft y,\varrho\rangle\xtworightarrow{e_1[m]}\langle x',\varrho'\rangle}$$
        $$\frac{\langle x,\varrho\rangle\xtworightarrow{e_3[l]}e_3 \quad \langle y,\varrho\rangle\xntworightarrow{e_2[n]}\quad (\sharp(e_1,e_2),e_1\geq e_3)}{\langle x\triangleleft y,\varrho\rangle\xtworightarrow{\tau}\langle\surd,\tau(\varrho')\rangle}
        \quad\frac{\langle x,\varrho\rangle\xtworightarrow{e_3[l]}x' \quad \langle y,\varrho\rangle\xntworightarrow{e_2[n]}\quad (\sharp(e_1,e_2),e_1\geq e_3)}{\langle x\triangleleft y,\varrho\rangle\xtworightarrow{\tau}\langle x',\tau(\varrho')\rangle}$$

        \caption{Reverse action transition rules of $qAPRPTC$ (continuing)}
        \label{RTRForqAPRPTCG}
    \end{table}
\end{center}

\begin{center}
    \begin{table}
        $$\frac{\langle x,\varrho\rangle\xtworightarrow{e_1[m]}\langle e_1,\varrho'\rangle\quad (\sharp_{\pi}(e_1,e_2))}{\langle\Theta(x),\varrho\rangle\xtworightarrow{e_1[m]}\langle e_1,\varrho'\rangle}
        \quad\frac{\langle x,\varrho\rangle\xtworightarrow{e_2[n]}\langle e_2,\varrho'\rangle\quad (\sharp_{\pi}(e_1,e_2))}{\langle\Theta(x),\varrho\rangle\xtworightarrow{e_2[n]}\langle e_2,\varrho'\rangle}$$
        $$\frac{\langle x,\varrho\rangle\xtworightarrow{e_1[m]}\langle x',\varrho'\rangle\quad (\sharp_{\pi}(e_1,e_2))}{\langle\Theta(x),\varrho\rangle\xtworightarrow{e_1[m]}\langle \Theta(x'),\varrho'\rangle}
        \quad\frac{\langle x,\varrho\rangle\xtworightarrow{e_2[n]}\langle x',\varrho'\rangle\quad (\sharp_{\pi}(e_1,e_2))}{\langle\Theta(x),\varrho\rangle\xtworightarrow{e_2[n]}\langle\Theta(x'),\varrho'\rangle}$$
        $$\frac{\langle x,\varrho\rangle\xtworightarrow{e_1[m]}\langle e_1,\varrho'\rangle \quad \langle y,\varrho\rangle\xntworightarrow{e_2[n]}\quad (\sharp_{\pi}(e_1,e_2))}{\langle x\triangleleft y,\varrho\rangle\xtworightarrow{\tau}\langle\surd,\tau(\varrho')\rangle}
        \quad\frac{\langle x,\varrho\rangle\xtworightarrow{e_1[m]}\langle x',\varrho'\rangle \quad \langle y,\varrho\rangle\xntworightarrow{e_2[n]}\quad (\sharp_{\pi}(e_1,e_2))}{\langle x\triangleleft y,\varrho\rangle\xtworightarrow{\tau}\langle x',\tau(\varrho')\rangle}$$
        $$\frac{\langle x,\varrho\rangle\xtworightarrow{e_1[m]}\langle e_1,\varrho'\rangle \quad \langle y,\varrho\rangle\xntworightarrow{e_3[l]}\quad (\sharp_{\pi}(e_1,e_2),e_2\geq e_3)}{\langle x\triangleleft y,\varrho\rangle\xtworightarrow{e_1[m]}\langle e_1,\varrho'\rangle}
        \quad\frac{\langle x,\varrho\rangle\xtworightarrow{e_1[m]}x' \quad \langle y,\varrho\rangle\xntworightarrow{e_3[l]}\quad (\sharp_{\pi}(e_1,e_2),e_2\geq e_3)}{\langle x\triangleleft y,\varrho\rangle\xtworightarrow{e_1[m]}\langle x',\varrho'\rangle}$$
        $$\frac{\langle x,\varrho\rangle\xtworightarrow{e_3[l]}e_3 \quad \langle y,\varrho\rangle\xntworightarrow{e_2[n]}\quad (\sharp_{\pi}(e_1,e_2),e_1\geq e_3)}{\langle x\triangleleft y,\varrho\rangle\xtworightarrow{\tau}\langle\surd,\tau(\varrho')\rangle}
        \quad\frac{\langle x,\varrho\rangle\xtworightarrow{e_3[l]}x' \quad \langle y,\varrho\rangle\xntworightarrow{e_2[n]}\quad (\sharp_{\pi}(e_1,e_2),e_1\geq e_3)}{\langle x\triangleleft y,\varrho\rangle\xtworightarrow{\tau}\langle x',\tau(\varrho')\rangle}$$
        $$\frac{\langle x,\varrho\rangle\xtworightarrow{e[m]}\langle e,\varrho'\rangle}{\langle\partial_H(x),\varrho\rangle\xtworightarrow{e[m]}\langle e,\varrho'\rangle}\quad (e\notin H)
        \quad\frac{\langle x,\varrho\rangle\xtworightarrow{e}\langle x',\varrho'\rangle}{\langle \partial_H(x),\varrho\rangle\xtworightarrow{e}\langle\partial_H(x'),\varrho'\rangle}\quad(e\notin H)$$
        \caption{Reverse action transition rules of $qAPRPTC$ (continuing)}
        \label{RTRForqAPRPTCG2}
    \end{table}
\end{center}

\begin{theorem}[Generalization of $qAPRPTC$ with respect to $qBARPTC$]
$qAPRPTC$ is a generalization of $qBARPTC$.
\end{theorem}

\begin{proof}
It follows from the following three facts.

\begin{enumerate}
  \item The transition rules of $qBARPTC$ in section \ref{qbarptc} are all source-dependent;
  \item The sources of the transition rules $qAPRPTC$ contain an occurrence of $\between$, or $\parallel$, or $\leftmerge$, or $\mid$, or $\Theta$, or $\triangleleft$;
  \item The transition rules of $qAPRPTC$ are all source-dependent.
\end{enumerate}

So, $qAPRPTC$ is a generalization of $qBARPTC$, that is, $qBARPTC$ is an embedding of $qAPRPTC$, as desired.
\end{proof}

\begin{theorem}[Congruence of $qAPRPTC$ with respect to probabilistic truly concurrent bisimulation equivalences]\label{CqAPRPTCG}
(1) FR probabilistic pomset bisimulation equivalence $\sim_{pp}^{fr}$ is a congruence with respect to $qAPRPTC$.

(2) FR probabilistic step bisimulation equivalence $\sim_{ps}^{fr}$ is a congruence with respect to $qAPRPTC$.

(3) FR probabilistic hp-bisimulation equivalence $\sim_{php}^{fr}$ is a congruence with respect to $qAPRPTC$.

(4) FR probabilistic hhp-bisimulation equivalence $\sim_{phhp}^{fr}$ is a congruence with respect to $qAPRPTC$.
\end{theorem}

\begin{proof}
(1) It is easy to see that FR probabilistic pomset bisimulation is an equivalent relation on $qAPRPTC$ terms, we only need to prove that $\sim_{pp}^{fr}$ is preserved by the operators
$\parallel$, $\leftmerge$, $\mid$, $\Theta$, $\triangleleft$, $\partial_H$. It is trivial and we leave the proof as an exercise for the readers.

(2) It is easy to see that FR probabilistic step bisimulation is an equivalent relation on $qAPRPTC$ terms, we only need to prove that $\sim_{ps}^{fr}$ is preserved by the operators
$\parallel$, $\leftmerge$, $\mid$, $\Theta$, $\triangleleft$, $\partial_H$. It is trivial and we leave the proof as an exercise for the readers.

(3) It is easy to see that FR probabilistic hp-bisimulation is an equivalent relation on $qAPRPTC$ terms, we only need to prove that $\sim_{php}^{fr}$ is preserved by the operators
$\parallel$, $\leftmerge$, $\mid$, $\Theta$, $\triangleleft$, $\partial_H$. It is trivial and we leave the proof as an exercise for the readers.

(4) It is easy to see that FR probabilistic hhp-bisimulation is an equivalent relation on $qAPRPTC$ terms, we only need to prove that $\sim_{phhp}^{fr}$ is preserved by the operators
$\parallel$, $\leftmerge$, $\mid$, $\Theta$, $\triangleleft$, $\partial_H$. It is trivial and we leave the proof as an exercise for the readers.
\end{proof}

\begin{theorem}[Soundness of $qAPRPTC$ modulo FR probabilistic truly concurrent bisimulation equivalences]\label{SqAPRPTCG}
(1) Let $x$ and $y$ be $qAPRPTC$ terms. If $qAPRPTC\vdash x=y$, then $x\sim_{pp}^{fr} y$.

(2) Let $x$ and $y$ be $qAPRPTC$ terms. If $qAPRPTC\vdash x=y$, then $x\sim_{ps}^{fr} y$.

(3) Let $x$ and $y$ be $qAPRPTC$ terms. If $qAPRPTC\vdash x=y$, then $x\sim_{php}^{fr} y$;

(3) Let $x$ and $y$ be $qAPRPTC$ terms. If $qAPRPTC\vdash x=y$, then $x\sim_{phhp}^{fr} y$.
\end{theorem}

\begin{proof}
(1) Since FR probabilistic pomset bisimulation $\sim_{pp}^{fr}$ is both an equivalent and a congruent relation, we only need to check if each axiom in Table \ref{AxiomsForqAPRPTC} is sound modulo
FR probabilistic pomset bisimulation equivalence. We leave the proof as an exercise for the readers.

(2) Since FR probabilistic step bisimulation $\sim_{ps}^{fr}$ is both an equivalent and a congruent relation, we only need to check if each axiom in Table \ref{AxiomsForqAPRPTC} is sound modulo
FR probabilistic step bisimulation equivalence. We leave the proof as an exercise for the readers.

(3) Since FR probabilistic hp-bisimulation $\sim_{php}^{fr}$ is both an equivalent and a congruent relation, we only need to check if each axiom in Table \ref{AxiomsForqAPRPTC} is sound modulo
FR probabilistic hp-bisimulation equivalence. We leave the proof as an exercise for the readers.

(4) Since FR probabilistic hhp-bisimulation $\sim_{phhp}^{fr}$ is both an equivalent and a congruent relation, we only need to check if each axiom in Table \ref{AxiomsForqAPRPTC} is sound modulo
FR probabilistic hhp-bisimulation equivalence. We leave the proof as an exercise for the readers.
\end{proof}

\begin{theorem}[Completeness of $qAPRPTC$ modulo FR probabilistic truly concurrent bisimulation equivalences]\label{CqAPRPTCG}
(1) Let $p$ and $q$ be closed $qAPRPTC$ terms, if $p\sim_{pp}^{fr} q$ then $p=q$.

(2) Let $p$ and $q$ be closed $qAPRPTC$ terms, if $p\sim_{ps}^{fr} q$ then $p=q$.

(3) Let $p$ and $q$ be closed $qAPRPTC$ terms, if $p\sim_{php}^{fr} q$ then $p=q$.

(3) Let $p$ and $q$ be closed $qAPRPTC$ terms, if $p\sim_{phhp}^{fr} q$ then $p=q$.
\end{theorem}

\begin{proof}
According to the definition of FR probabilistic truly concurrent bisimulation equivalences $\sim_{pp}^{fr}$, $\sim_{ps}^{fr}$, $\sim_{php}^{fr}$ and $\sim_{phhp}^{fr}$, $p\sim_{pp}^{fr}q$, $p\sim_{ps}^{fr}q$, $p\sim_{php}^{fr}q$ and $p\sim_{phhp}^{fr}q$ implies
both the bisimilarities between $p$ and $q$, and also the in the same quantum states. According to the completeness of APRPTC (please refer to chapter \ref{aprptc2} for details), we can get the
completeness of qAPRPTC.
\end{proof}

\subsection{Recursion}{\label{qcrec}}

In this subsection, we introduce recursion to capture infinite processes based on $qAPRPTC$. In the following, $E,F,G$ are recursion specifications, $X,Y,Z$ are recursive variables.

\begin{definition}[Guarded recursive specification]
A recursive specification

$$X_1=t_1(X_1,\cdots,X_n)$$
$$...$$
$$X_n=t_n(X_1,\cdots,X_n)$$

is guarded if the right-hand sides of its recursive equations can be adapted to the form by applications of the axioms in $qAPRPTC$ and replacing recursion variables by the right-hand
sides of their recursive equations,

$((a_{111}\leftmerge\cdots\leftmerge a_{11i_1})\cdot s_1(X_1,\cdots,X_n)+\cdots+(a_{1k1}\leftmerge\cdots\leftmerge a_{1ki_k})\cdot s_k(X_1,\cdots,X_n)+(b_{111}\leftmerge\cdots\leftmerge
b_{11j_1})+\cdots+(b_{11j_1}\leftmerge\cdots\leftmerge b_{1lj_l}))\boxplus_{\pi_1}\cdots\boxplus_{\pi_{m-1}}((a_{m11}\leftmerge\cdots\leftmerge a_{m1i_1})\cdot s_1(X_1,\cdots,X_n)+
\cdots+(a_{mk1}\leftmerge\cdots\leftmerge a_{mki_k})\cdot s_k(X_1,\cdots,X_n)+(b_{m11}\leftmerge\cdots\leftmerge b_{m1j_1})+\cdots+(b_{m1j_1}\leftmerge\cdots\leftmerge b_{mlj_l}))$

where $a_{111},\cdots,a_{11i_1},a_{1k1},\cdots,a_{1ki_k},b_{111},\cdots,b_{11j_1},b_{11j_1},\cdots,b_{1lj_l},\cdots, a_{m11},\cdots,a_{m1i_1},a_{1k1},\cdots,a_{mki_k},\\b_{111},\cdots,
b_{m1j_1},b_{m1j_1},\cdots,b_{mlj_l}\in \mathbb{E}$, and the sum above is allowed to be empty, in which case it represents the deadlock $\delta$. And there does not exist an infinite
sequence of $\epsilon$-transitions $\langle X|E\rangle\rightarrow\langle X'|E\rangle\rightarrow\langle X''|E\rangle\rightarrow\cdots$.
\end{definition}

\begin{center}
    \begin{table}
        $$\frac{\langle t_i(\langle X_1|E\rangle,\cdots,\langle X_n|E\rangle),\varrho\rangle\rightsquigarrow \langle y,\varrho\rangle}{\langle\langle X_i|E\rangle,\varrho\rangle\rightsquigarrow \langle y,\varrho\rangle}$$
        $$\frac{\langle t_i(\langle X_1|E\rangle,\cdots,\langle X_n|E\rangle),\varrho\rangle\xrightarrow{\{e_1,\cdots,e_k\}}\langle e_1[m]\leftmerge\cdots\leftmerge e_k[m],\varrho'\rangle}{\langle\langle X_i|E\rangle,\varrho\rangle\xrightarrow{\{e_1,\cdots,e_k\}}\langle e_1[m]\leftmerge\cdots\leftmerge e_k[m],\varrho'\rangle}$$
        $$\frac{\langle t_i(\langle X_1|E\rangle,\cdots,\langle X_n|E\rangle),\varrho\rangle\xrightarrow{\{e_1,\cdots,e_k\}} \langle y,\varrho'\rangle}{\langle\langle X_i|E\rangle,\varrho\rangle\xrightarrow{\{e_1,\cdots,e_k\}} \langle y,\varrho'\rangle}$$
        $$\frac{\langle t_i(\langle X_1|E\rangle,\cdots,\langle X_n|E\rangle),\varrho\rangle\xtworightarrow{\{e_1[m],\cdots,e_k[m]\}}\langle e_1\leftmerge\cdots\leftmerge e_k,\varrho'\rangle}{\langle\langle X_i|E\rangle,\varrho\rangle\xtworightarrow{\{e_1[m],\cdots,e_k[m]\}}\langle e_1\leftmerge\cdots\leftmerge e_k,\varrho'\rangle}$$
        $$\frac{\langle t_i(\langle X_1|E\rangle,\cdots,\langle X_n|E\rangle),\varrho\rangle\xtworightarrow{\{e_1[m],\cdots,e_k[m]\}} \langle y,\varrho'\rangle}{\langle\langle X_i|E\rangle,\varrho\rangle\xtworightarrow{\{e_1[m],\cdots,e_k[m]\}} \langle y,\varrho'\rangle}$$
        \caption{Transition rules of guarded recursion}
        \label{TRForGRG}
    \end{table}
\end{center}

\begin{theorem}[Conservitivity of $qAPRPTC$ with guarded recursion]
$qAPRPTC$ with guarded recursion is a conservative extension of $qAPRPTC$.
\end{theorem}

\begin{proof}
Since the transition rules of $qAPRPTC$ are source-dependent, and the transition rules for guarded recursion in Table \ref{TRForGRG} contain only a fresh constant in their source, so
the transition rules of $qAPRPTC$ with guarded recursion are a conservative extension of those of $qAPRPTC$.
\end{proof}

\begin{theorem}[Congruence theorem of $qAPRPTC$ with guarded recursion]
FR probabilistic truly concurrent bisimulation equivalences $\sim_{pp}^{fr}$, $\sim_{p}$, $\sim_{php}^{fr}$ and $\sim_{phhp}^{fr}$ are all congruences with respect to $qAPRPTC$ with guarded recursion.
\end{theorem}

\begin{proof}
It follows the following two facts:
\begin{enumerate}
  \item in a guarded recursive specification, right-hand sides of its recursive equations can be adapted to the form by applications of the axioms in $qAPRPTC$ and replacing recursion
  variables by the right-hand sides of their recursive equations;
  \item FR probabilistic truly concurrent bisimulation equivalences $\sim_{pp}^{fr}$, $\sim_{ps}^{fr}$, $\sim_{php}^{fr}$ and $\sim_{phhp}^{fr}$ are all congruences with respect to all operators of $qAPRPTC$.
\end{enumerate}
\end{proof}

\begin{theorem}[Elimination theorem of $qAPRPTC$ with linear recursion]\label{ETRecursionG}
Each process term in $qAPRPTC$ with linear recursion is equal to a process term $\langle X_1|E\rangle$ with $E$ a linear recursive specification.
\end{theorem}

\begin{proof}
The same as that of $APRPTC$, we omit the proof, please refer to chapter \ref{aprptc2} for details.
\end{proof}

\begin{theorem}[Soundness of $qAPRPTC$ with guarded recursion]\label{SqAPRPTC_GRG}
Let $x$ and $y$ be $qAPRPTC$ with guarded recursion terms. If $qAPRPTC\textrm{ with guarded recursion}\vdash x=y$, then

(1) $x\sim_{ps}^{fr} y$.

(2) $x\sim_{pp}^{fr} y$.

(3) $x\sim_{php}^{fr} y$.

(4) $x\sim_{phhp}^{fr} y$.
\end{theorem}

\begin{proof}
(1) Since FR probabilistic step bisimulation $\sim_{ps}^{fr}$ is both an equivalent and a congruent relation with respect to $qAPRPTC$ with guarded recursion, we only need to check if each
axiom in Table \ref{RDPRSP} is sound modulo FR probabilistic step bisimulation equivalence. We leave them as exercises to the readers.

(2) Since FR probabilistic pomset bisimulation $\sim_{pp}^{fr}$ is both an equivalent and a congruent relation with respect to the guarded recursion, we only need to check if each axiom in
Table \ref{RDPRSP} is sound modulo FR probabilistic pomset bisimulation equivalence. We leave them as exercises to the readers.

(3) Since FR probabilistic hp-bisimulation $\sim_{php}^{fr}$ is both an equivalent and a congruent relation with respect to guarded recursion, we only need to check if each axiom in Table
\ref{RDPRSP} is sound modulo FR probabilistic hp-bisimulation equivalence. We leave them as exercises to the readers.

(4) Since FR probabilistic hhp-bisimulation $\sim_{phhp}^{fr}$ is both an equivalent and a congruent relation with respect to guarded recursion, we only need to check if each axiom in Table
\ref{RDPRSP} is sound modulo FR probabilistic hhp-bisimulation equivalence. We leave them as exercises to the readers.
\end{proof}

\begin{theorem}[Completeness of $qAPRPTC$ with linear recursion]\label{CqAPRPTC_GRG}
Let $p$ and $q$ be closed $qAPRPTC$ with linear recursion terms, then,

(1) if $p\sim_{ps}^{fr} q$ then $p=q$.

(2) if $p\sim_{pp}^{fr} q$ then $p=q$.

(3) if $p\sim_{php}^{fr} q$ then $p=q$.

(4) if $p\sim_{phhp}^{fr} q$ then $p=q$.
\end{theorem}

\begin{proof}
According to the definition of FR probabilistic truly concurrent bisimulation equivalences $\sim_{pp}^{fr}$, $\sim_{ps}^{fr}$, $\sim_{php}^{fr}$ and $\sim_{phhp}^{fr}$, $p\sim_{pp}^{fr}q$, $p\sim_{ps}^{fr}q$, $p\sim_{php}^{fr}q$ and $p\sim_{phhp}^{fr}q$ implies
both the bisimilarities between $p$ and $q$, and also the in the same quantum states. According to the completeness of APRPTC with linear recursion (please refer to chapter \ref{aprptc2} for details), we can get the
completeness of qAPRPTC with linear recursion.
\end{proof}

\subsection{Abstraction}{\label{qcabs}}

To abstract away from the internal implementations of a program, and verify that the program exhibits the desired external behaviors, the silent step $\tau$ and abstraction operator
$\tau_I$ are introduced, where $I\subseteq \mathbb{E}\cup G_{at}$ denotes the internal events or guards. The silent step $\tau$ represents the internal events or guards, when we
consider the external behaviors of a process, $\tau$ steps can be removed, that is, $\tau$ steps must keep silent. The transition rule of $\tau$ is shown in Table \ref{TRForqTau2}. In
the following, let the atomic event $e$ range over $\mathbb{E}\cup\{\epsilon\}\cup\{\delta\}\cup\{\tau\}$, and $\phi$ range over $G\cup \{\tau\}$, and let the communication function
$\gamma:\mathbb{E}\cup\{\tau\}\times \mathbb{E}\cup\{\tau\}\rightarrow \mathbb{E}\cup\{\delta\}$, with each communication involved $\tau$ resulting in $\delta$. We use $\tau(\varrho)$ to
denote $effect(\tau,\varrho)$, for the fact that $\tau$ only change the state of internal data environment, that is, for the external data environments, $\varrho=\tau(\varrho)$.

\begin{center}
    \begin{table}
        $$\frac{}{\tau\rightsquigarrow\breve{\tau}}$$
        $$\frac{}{\langle\tau,\varrho\rangle\xrightarrow{\tau}\langle\surd,\tau(\varrho)\rangle}$$
        $$\frac{}{\langle\tau,\varrho\rangle\xtworightarrow{\tau}\langle\surd,\tau(\varrho)\rangle}$$
        \caption{Transition rule of the silent step}
        \label{TRForqTau2}
    \end{table}
\end{center}

\begin{theorem}[Conservitivity of $qAPRPTC$ with silent step and guarded linear recursion]
$qAPRPTC$ with silent step and guarded linear recursion is a conservative extension of $qAPRPTC$ with linear recursion.
\end{theorem}

\begin{proof}
Since the transition rules of $qAPRPTC$ with linear recursion are source-dependent, and the transition rules for silent step in Table \ref{TRForqTau2} contain only a fresh constant
$\tau$ in their source, so the transition rules of $qAPRPTC$ with silent step and guarded linear recursion is a conservative extension of those of $qAPRPTC$ with linear recursion.
\end{proof}

\begin{theorem}[Congruence theorem of $qAPRPTC$ with silent step and guarded linear recursion]
FR probabilistic rooted branching truly concurrent bisimulation equivalences $\approx_{prbp}^{fr}$, $\approx_{prbs}^{fr}$, $\approx_{prbhp}^{fr}$ and $\approx_{rbhhp}$ are all congruences with respect
to $qAPRPTC$ with silent step and guarded linear recursion.
\end{theorem}

\begin{proof}
It follows the following three facts:
\begin{enumerate}
  \item in a guarded linear recursive specification, right-hand sides of its recursive equations can be adapted to the form by applications of the axioms in $qAPRPTC$ and replacing
  recursion variables by the right-hand sides of their recursive equations;
  \item FR probabilistic truly concurrent bisimulation equivalences $\sim_{pp}^{fr}$, $\sim_{ps}^{fr}$, $\sim_{php}^{fr}$ and $\sim_{phhp}^{fr}$ are all congruences with respect to all operators of
  $qAPRPTC$, while FR probabilistic truly concurrent bisimulation equivalences $\sim_{pp}^{fr}$, $\sim_{ps}^{fr}$, $\sim_{php}^{fr}$ and $\sim_{phhp}^{fr}$ imply the corresponding FR probabilistic rooted
  branching truly concurrent bisimulations $\approx_{prbp}^{fr}$, $\approx_{prbs}^{fr}$, $\approx_{prbhp}^{fr}$ and $\approx_{prbhhp}^{fr}$, so probabilistic rooted branching truly concurrent
  bisimulations $\approx_{prbp}^{fr}$, $\approx_{prbs}^{fr}$, $\approx_{prbhp}^{fr}$ and $\approx_{prbhhp}^{fr}$ are all congruences with respect to all operators of $qAPRPTC$;
  \item While $\mathbb{E}$ is extended to $\mathbb{E}\cup\{\tau\}$, and $G$ is extended to $G\cup\{\tau\}$, it can be proved that FR probabilistic rooted branching truly concurrent
  bisimulations $\approx_{prbp}^{fr}$, $\approx_{prbs}^{fr}$, $\approx_{prbhp}^{fr}$ and $\approx_{prbhhp}^{fr}$ are all congruences with respect to all operators of $qAPRPTC$, we omit it.
\end{enumerate}
\end{proof}

We design the axioms for the silent step $\tau$ in Table \ref{AxiomsForqTau2}.

\begin{center}
\begin{table}
  \begin{tabular}{@{}ll@{}}
  \hline No. &Axiom\\
  $B1$ & $(y=y+y,z=z+z,(Std(x),Std(y),Std(z),Std(w)))$\\
  &$x\cdot((y+\tau\cdot(y+z))\boxplus_{\pi}w)=x\cdot((y+z)\boxplus_{\pi}w)$\\
  $RB1$ & $(y=y+y,z=z+z,(NStd(x),NStd(y),NStd(z),NStd(w)))$\\
  &$((y+(y+z)\cdot\tau)\boxplus_{\pi}w)\cdot x=((y+z)\boxplus_{\pi}w)\cdot x$\\
  $B2$ & $(y=y+y,z=z+z,(Std(x),Std(y),Std(z),Std(w)))$\\
  &$x\leftmerge((y+\tau\leftmerge(y+z))\boxplus_{\pi}w)=x\leftmerge((y+z)\boxplus_{\pi}w)$\\
  $RB2$ & $(y=y+y,z=z+z,(NStd(x),NStd(y),NStd(z),NStd(w)))$\\
  &$((y+(y+z)\leftmerge\tau)\boxplus_{\pi}w)\leftmerge x=((y+z)\boxplus_{\pi}w)\leftmerge x$\\
\end{tabular}
\caption{Axioms of silent step}
\label{AxiomsForqTau2}
\end{table}
\end{center}

\begin{theorem}[Elimination theorem of $qAPRPTC$ with silent step and guarded linear recursion]\label{ETTauG}
Each process term in $qAPRPTC$ with silent step and guarded linear recursion is equal to a process term $\langle X_1|E\rangle$ with $E$ a guarded linear recursive specification.
\end{theorem}

\begin{proof}
The same as that of $APRPTC$, we omit the proof, please refer to chapter \ref{aprptc2} for details.
\end{proof}

\begin{theorem}[Soundness of $qAPRPTC$ with silent step and guarded linear recursion]\label{SqAPRPTC_GTAUG}
Let $x$ and $y$ be $qAPRPTC$ with silent step and guarded linear recursion terms. If $qAPRPTC$ with silent step and guarded linear recursion $\vdash x=y$, then

(1) $x\approx_{prbs}^{fr} y$.

(2) $x\approx_{prbp}^{fr} y$.

(3) $x\approx_{prbhp}^{fr} y$.

(4) $x\approx_{prbhhp}^{fr} y$.
\end{theorem}

\begin{proof}
(1) Since FR probabilistic rooted branching step bisimulation $\approx_{prbs}^{fr}$ is both an equivalent and a congruent relation with respect to $qAPRPTC$ with silent step and guarded
linear recursion, we only need to check if each axiom in Table \ref{AxiomsForqTau2} is sound modulo FR probabilistic rooted branching step bisimulation $\approx_{prbs}^{fr}$. We leave them as
exercises to the readers.

(2) Since FR probabilistic rooted branching pomset bisimulation $\approx_{prbp}^{fr}$ is both an equivalent and a congruent relation with respect to $qAPRPTC$ with silent step and guarded
linear recursion, we only need to check if each axiom in Table \ref{AxiomsForqTau2} is sound modulo FR probabilistic rooted branching pomset bisimulation $\approx_{prbp}^{fr}$. We leave them
as exercises to the readers.

(3) Since FR probabilistic rooted branching hp-bisimulation $\approx_{prbhp}^{fr}$ is both an equivalent and a congruent relation with respect to $qAPRPTC$ with silent step and guarded linear
recursion, we only need to check if each axiom in Table \ref{AxiomsForqTau2} is sound modulo FR probabilistic rooted branching hp-bisimulation $\approx_{prbhp}^{fr}$. We leave them as exercises
to the readers.

(4) Since FR probabilistic rooted branching hhp-bisimulation $\approx_{prbhhp}^{fr}$ is both an equivalent and a congruent relation with respect to $qAPRPTC$ with silent step and guarded linear
recursion, we only need to check if each axiom in Table \ref{AxiomsForqTau2} is sound modulo FR probabilistic rooted branching hhp-bisimulation $\approx_{prbhhp}^{fr}$. We leave them as exercises
to the readers.
\end{proof}

\begin{theorem}[Completeness of $qAPRPTC$ with silent step and guarded linear recursion]\label{CqAPRPTC_GTAUG}
Let $p$ and $q$ be closed $qAPRPTC$ with silent step and guarded linear recursion terms, then,

(1) if $p\approx_{prbs}^{fr} q$ then $p=q$.

(2) if $p\approx_{prbp}^{fr} q$ then $p=q$.

(3) if $p\approx_{prbhp}^{fr} q$ then $p=q$.

(3) if $p\approx_{prbhhp}^{fr} q$ then $p=q$.
\end{theorem}

\begin{proof}
According to the definition of FR probabilistic rooted branching truly concurrent bisimulation equivalences $\approx_{prbp}^{fr}$, $\approx_{prbs}^{fr}$, $\approx_{prbhp}^{fr}$ and $\approx_{prbhhp}^{fr}$, and $\approx_{prbp}^{fr}$, $\approx_{prbs}^{fr}$, $\approx_{prbhp}^{fr}$ and $\approx_{prbhhp}^{fr}$ implies
both the bisimilarities between $p$ and $q$, and also the in the same quantum states. According to the completeness of APRPTC with silent step and guarded linear recursion (please refer to chapter \ref{aprptc2} for details), we can get the
completeness of qAPRPTC with silent step and guarded linear recursion.
\end{proof}

The unary abstraction operator $\tau_I$ ($I\subseteq \mathbb{E}\cup G_{at}$) renames all atomic events or atomic guards in $I$ into $\tau$. $qAPRPTC$ with silent step and abstraction
operator is called $qAPRPTC_{\tau}$. The transition rules of operator $\tau_I$ are shown in Table \ref{TRForqAbstraction2}.

\begin{center}
    \begin{table}
        $$\frac{\langle x,\varrho\rangle\rightsquigarrow \langle x',\varrho\rangle}{\langle \tau_I(x),\varrho\rangle\rightsquigarrow\langle\tau_I(x'),\varrho\rangle}$$
        $$\frac{\langle x,\varrho\rangle\xrightarrow{e}\langle e[m],\varrho'\rangle}{\langle \tau_I(x),\varrho\rangle\xrightarrow{e}\langle e[m],\varrho'\rangle}\quad e\notin I
        \quad\frac{\langle x,\varrho\rangle\xrightarrow{e}\langle x',\varrho'\rangle}{\langle\tau_I(x),\varrho\rangle\xrightarrow{e}\langle \tau_I(x'),\varrho'\rangle}\quad e\notin I$$

        $$\frac{\langle x,\varrho\rangle\xrightarrow{e}\langle\surd,\tau(\varrho)\rangle}{\langle\tau_I(x),\varrho\rangle\xrightarrow{\tau}\langle\surd,\tau(\varrho)\rangle}\quad e\in I
        \quad\frac{\langle x,\varrho\rangle\xrightarrow{e}\langle x',\tau(\varrho)\rangle}{\langle\tau_I(x),\varrho\rangle\xrightarrow{\tau}\langle\tau_I(x'),\tau(\varrho)\rangle}\quad e\in I$$

        $$\frac{\langle x,\varrho\rangle\xtworightarrow{e[m]}\langle e,\varrho'\rangle}{\langle\tau_I(x),\varrho\rangle\xtworightarrow{e[m]}\langle e,\varrho'\rangle}\quad e[m]\notin I
        \quad\frac{\langle x,\varrho\rangle\xtworightarrow{e[m]}\langle x',\varrho\rangle}{\langle\tau_I(x),\varrho\rangle\xtworightarrow{e[m]}\langle\tau_I(x'),\varrho'\rangle}\quad e[m]\notin I$$

        $$\frac{\langle x,\varrho\rangle\xtworightarrow{e[m]}\langle\surd,\tau(\varrho)\rangle}{\langle\tau_I(x),\varrho\rangle\xtworightarrow{\tau}\langle\surd,\tau(\varrho)\rangle}\quad e[m]\in I
        \quad\frac{\langle x,\varrho\rangle\xtworightarrow{e[m]}\langle x',\tau(\varrho)\rangle}{\langle\tau_I(x),\varrho\rangle\xtworightarrow{\tau}\langle\tau_I(x'),\tau(\varrho)\rangle}\quad e[m]\in I$$
        \caption{Transition rule of the abstraction operator}
        \label{TRForqAbstraction2}
    \end{table}
\end{center}

\begin{theorem}[Conservitivity of $qAPRPTC_{\tau}$ with guarded linear recursion]
$qAPRPTC_{\tau}$ with guarded linear recursion is a conservative extension of $qAPRPTC$ with silent step and guarded linear recursion.
\end{theorem}

\begin{proof}
Since the transition rules of $qAPRPTC$ with silent step and guarded linear recursion are source-dependent, and the transition rules for abstraction operator in Table
\ref{TRForqAbstraction2} contain only a fresh operator $\tau_I$ in their source, so the transition rules of $qAPRPTC_{\tau}$ with guarded linear recursion is a conservative extension
of those of $qAPRPTC$ with silent step and guarded linear recursion.
\end{proof}

\begin{theorem}[Congruence theorem of $qAPRPTC_{\tau}$ with guarded linear recursion]
FR probabilistic rooted branching truly concurrent bisimulation equivalences $\approx_{prbp}^{fr}$, $\approx_{prbs}^{fr}$, $\approx_{prbhp}^{fr}$ and $\approx_{prbhhp}^{fr}$ are all congruences with respect
to $qAPRPTC_{\tau}$ with guarded linear recursion.
\end{theorem}

\begin{proof}
(1) It is easy to see that FR probabilistic rooted branching pomset bisimulation is an equivalent relation on $qAPRPTC_{\tau}$ with guarded linear recursion terms, we only need to
prove that $\approx_{prbp}^{fr}$ is preserved by the operators $\tau_I$. It is trivial and we leave the proof as an exercise for the readers.

(2) It is easy to see that FR probabilistic rooted branching step bisimulation is an equivalent relation on $qAPRPTC_{\tau}$ with guarded linear recursion terms, we only need to
prove that $\approx_{prbs}^{fr}$ is preserved by the operators $\tau_I$. It is trivial and we leave the proof as an exercise for the readers.

(3) It is easy to see that FR probabilistic rooted branching hp-bisimulation is an equivalent relation on $qAPRPTC_{\tau}$ with guarded linear recursion terms, we only need to
prove that $\approx_{prbhp}^{fr}$ is preserved by the operators $\tau_I$. It is trivial and we leave the proof as an exercise for the readers.

(4) It is easy to see that FR probabilistic rooted branching hhp-bisimulation is an equivalent relation on $qAPRPTC_{\tau}$ with guarded linear recursion terms, we only need to
prove that $\approx_{prbhhp}^{fr}$ is preserved by the operators $\tau_I$. It is trivial and we leave the proof as an exercise for the readers.
\end{proof}

We design the axioms for the abstraction operator $\tau_I$ in Table \ref{AxiomsForqAbstraction2}.

\begin{center}
\begin{table}
  \begin{tabular}{@{}ll@{}}
\hline No. &Axiom\\
  $TI1$ & $e\notin I\quad \tau_I(e)=e$\\
  $RTI1$ & $e[m]\notin I\quad \tau_I(e[m])=e[m]$\\
  $TI2$ & $e\in I\quad \tau_I(e)=\tau$\\
  $RTI2$ & $e[m]\in I\quad \tau_I(e[m])=\tau$\\
  $TI3$ & $\tau_I(\delta)=\delta$\\
  $TI4$ & $\tau_I(x+y)=\tau_I(x)+\tau_I(y)$\\
  $PTI1$ & $\tau_I(x\boxplus_{\pi}y)=\tau_I(x)\boxplus_{\pi}\tau_I(y)$\\
  $TI5$ & $\tau_I(x\cdot y)=\tau_I(x)\cdot\tau_I(y)$\\
  $TI6$ & $\tau_I(x\leftmerge y)=\tau_I(x)\leftmerge\tau_I(y)$\\
\end{tabular}
\caption{Axioms of abstraction operator}
\label{AxiomsForqAbstraction2}
\end{table}
\end{center}

\begin{theorem}[Soundness of $qAPRPTC_{\tau}$ with guarded linear recursion]
Let $x$ and $y$ be $qAPRPTC_{\tau}$ with guarded linear recursion terms. If $qAPRPTC_{\tau}$ with guarded linear recursion $\vdash x=y$, then

(1) $x\approx_{prbs}^{fr} y$.

(2) $x\approx_{prbp}^{fr} y$.

(3) $x\approx_{prbhp}^{fr} y$.

(4) $x\approx_{prbhhp}^{fr} y$.
\end{theorem}

\begin{proof}
(1) Since FR probabilistic rooted branching step bisimulation $\approx_{prbs}^{fr}$ is both an equivalent and a congruent relation with respect to $qAPRPTC_{\tau}$ with guarded linear
recursion, we only need to check if each axiom in Table \ref{AxiomsForqAbstraction2} is sound modulo FR probabilistic rooted branching step bisimulation $\approx_{prbs}^{fr}$. We leave them as
exercises to the readers.

(2) Since FR probabilistic rooted branching pomset bisimulation $\approx_{prbp}^{fr}$ is both an equivalent and a congruent relation with respect to $qAPRPTC_{\tau}$ with guarded linear
recursion, we only need to check if each axiom in Table \ref{AxiomsForqAbstraction2} is sound modulo FR probabilistic rooted branching pomset bisimulation $\approx_{prbp}^{fr}$. We leave them
as exercises to the readers.

(3) Since FR probabilistic rooted branching hp-bisimulation $\approx_{prbhp}^{fr}$ is both an equivalent and a congruent relation with respect to $qAPRPTC_{\tau}$ with guarded linear
recursion, we only need to check if each axiom in Table \ref{AxiomsForqAbstraction2} is sound modulo FR probabilistic rooted branching hp-bisimulation $\approx_{prbhp}^{fr}$. We leave them as
exercises to the readers.

(4) Since FR probabilistic rooted branching hhp-bisimulation $\approx_{prbhhp}^{fr}$ is both an equivalent and a congruent relation with respect to $qAPRPTC_{\tau}$ with guarded linear
recursion, we only need to check if each axiom in Table \ref{AxiomsForqAbstraction2} is sound modulo FR probabilistic rooted branching hhp-bisimulation $\approx_{prbhhp}^{fr}$. We leave them as
exercises to the readers.
\end{proof}

Though $\tau$-loops are prohibited in guarded linear recursive specifications in a specifiable way, they can be constructed using the abstraction operator, for example, there exist
$\tau$-loops in the process term $\tau_{\{a\}}(\langle X|X=aX\rangle)$. To avoid $\tau$-loops caused by $\tau_I$ and ensure fairness, we introduce the following recursive verification
rules as Table \ref{RVR} shows, note that $i_1,\cdots, i_m,j_1,\cdots,j_n\in I\subseteq \mathbb{E}\setminus\{\tau\}$.

\begin{center}
\begin{table}
    $$VR_1\quad\frac{x=y+(i_1\leftmerge\cdots\leftmerge i_m)\cdot x, y=y+y,(Std(x),Std(y))}{\tau\cdot\tau_I(x)=\tau\cdot \tau_I(y)}$$
    $$RVR_1\quad\frac{x=y+ x\cdot(i_1[n]\leftmerge\cdots\leftmerge i_m[n]), y=y+y,(NStd(x),NStd(y))}{\tau_I(x)\cdot\tau= \tau_I(y)\cdot\tau}$$
    $$VR_2\quad \frac{x=z\boxplus_{\pi}(u+(i_1\leftmerge\cdots\leftmerge i_m)\cdot x),z=z+u,z=z+z,(Std(x),Std(z),Std(u))}{\tau\cdot\tau_I(x)=\tau\cdot\tau_I(z)}$$
    $$RVR_2\quad \frac{x=z\boxplus_{\pi}(u+ x\cdot(i_1[n]\leftmerge\cdots\leftmerge i_m[n])),z=z+u,z=z+z,(NStd(x),NStd(z),NStd(u))}{\tau_I(x)\cdot\tau=\tau_I(z)\cdot\tau}$$
    $$VR_3\quad \frac{x=z+(i_1\leftmerge\cdots\leftmerge i_m)\cdot y,y=z\boxplus_{\pi}(u+(j_1\leftmerge\cdots\leftmerge j_n)\cdot x), z=z+u,z=z+z}{\tau\cdot\tau_I(x)=\tau\cdot\tau_I(y')\textrm{ for }y'=z\boxplus_{\pi}(u+(i_1\leftmerge\cdots\leftmerge i_m)\cdot y')}$$
    $(Std(x),Std(y),Std(z),Std(u))$
    $$RVR_3\quad \frac{x=z+ y\cdot(i_1[k]\leftmerge\cdots\leftmerge i_m[k]),y=z\boxplus_{\pi}(u+ x\cdot(j_1[l]\leftmerge\cdots\leftmerge j_n[l])), z=z+u,z=z+z}{\tau_I(x)\cdot\tau=\tau_I(y')\cdot\tau\textrm{ for }y'=z\boxplus_{\pi}(u+ y'\cdot(i_1[k]\leftmerge\cdots\leftmerge i_m[k]))}$$
    $(Std(x),Std(y),Std(z),Std(u))$
\caption{Recursive verification rules}
\label{RVR}
\end{table}
\end{center}

\begin{theorem}[Soundness of $VR_1,VR_2,VR_3$]
$VR_1$, $VR_2$ and $VR_3$ are sound modulo probabilistic rooted branching truly concurrent bisimulation equivalences $\approx_{prbp}^{fr}$, $\approx_{prbs}^{fr}$, $\approx_{prbhp}^{fr}$ and $\approx_{prbhhp}^{fr}$.
\end{theorem}

\subsection{Quantum Measurement}\label{qm}

In closed quantum systems, there is another basic quantum operation -- quantum measurement, besides the unitary operator. Quantum measurements have a probabilistic nature.

There is a concrete but non-trivial problem in modeling quantum measurement.

Let the following process term represent quantum measurement during modeling phase,

$$\beta_1\cdot t_1\boxplus_{\pi_1}\beta_2\cdot t_2\boxplus_{\pi_2}\cdots\boxplus_{\pi_{i-1}}\beta_i\cdot t_i$$

where $\sum_i \pi_i=1$, $t_i\in\mathcal{B}(qBARPTC)$, $\beta$ denotes a quantum measurement, and $\beta=\sum_i\lambda_i \beta_i$, $\beta_i$ denotes the projection performed on the quantum
system $\varrho$, $\pi_i=Tr(\beta_i\varrho)$, $\varrho_i=\beta_i\varrho \beta_i/\pi_i$.

The above term means that, firstly, we choose a projection $\beta_i$ in a quantum measurement $\beta=\sum_i\lambda_i\beta_i$ probabilistically, then, we execute (perform) the
projection $\beta_i$ on the closed quantum system. This also adheres to the intuition on quantum mechanics.

We define $B$ as the collection of all projections of all quantum measurements, and make the collection of atomic actions in $PQRA$ be $\mathbb{E}=\mathbb{E}\cup B$. We see that a
projection $\beta_i\in B$ has the almost same semantics as a unitary operator $\alpha\in A$. So, we add the following (probabilistic and action)
transition rules into those of $PQRA$.

$$\frac{}{\langle\beta_i,\varrho\rangle\rightsquigarrow\langle\breve{\beta_i},\varrho\rangle}$$

$$\frac{}{\langle\breve{\beta_i},\varrho\rangle\xrightarrow{\beta_i}\langle\surd,\varrho'\rangle}$$

Until now, $qAPRPTC$ works again. The two main quantum operations in a closed quantum system -- the unitary operator and the quantum measurement, are fully modeled in probabilistic
process algebra.

\subsection{Quantum Entanglement}\label{qe2}

As in section \ref{qe1}, The axiom system of the shadow constant $\circledS$ is shown in Table \ref{AxiomsForQE2}.

\begin{center}
\begin{table}
  \begin{tabular}{@{}ll@{}}
\hline No. &Axiom\\
  $SC1$ & $\circledS\cdot x = x$ \\
  $SC2$ & $x\cdot\circledS = x$\\
  $SC3$ & $e\leftmerge\circledS^e=e$\\
  $SC4$ & $\circledS^e\leftmerge e=e$\\
  $SC5$ & $e\leftmerge(\circledS^e\cdot y) = e\cdot y$\\
  $SC6$ & $\circledS^e\leftmerge(e\cdot y) = e\cdot y$\\
  $SC7$ & $(e\cdot x)\leftmerge\circledS^e = e\cdot x$\\
  $SC8$ & $(\circledS^e\cdot x)\leftmerge e = e\cdot x$\\
  $SC9$ & $(e\cdot x)\leftmerge(\circledS^e\cdot y) = e\cdot (x\between y)$\\
  $SC10$ & $(\circledS^e\cdot x)\leftmerge(e\cdot y) = e\cdot (x\between y)$\\
  $RSC3$ & $e[m]\leftmerge\circledS^e[m]=e[m]$\\
  $RSC4$ & $\circledS^e[m]\leftmerge e[m]=e[m]$\\
  $RSC5$ & $e[m]\leftmerge(y\cdot\circledS^e[m]) = y\cdot e[m]$\\
  $RSC6$ & $\circledS^e[m]\leftmerge(y\cdot e[m]) =y\cdot e[m]$\\
  $RSC7$ & $(x\cdot e[m])\leftmerge\circledS^e[m] =x\cdot e[m]$\\
  $RSC8$ & $(x\cdot \circledS^e[m])\leftmerge e[m] =x\cdot e[m]$\\
  $RSC9$ & $(x\cdot e[m])\leftmerge(y\cdot \circledS^e[m]) =(x\between y)\cdot e[m]$\\
  $RSC10$ & $(x\cdot \circledS^e[m])\leftmerge(y \cdot e[m]) =(x\between y)\cdot e[m]$\\
\end{tabular}
\caption{Axioms of quantum entanglement}
\label{AxiomsForQE2}
\end{table}
\end{center}

The transition rules of constant $\circledS$ are as Table \ref{TRForENT2} shows.

\begin{center}
    \begin{table}
        $$\frac{}{\langle\circledS,\varrho\rangle\rightsquigarrow\langle\breve{\circledS},\varrho\rangle}$$
        $$\frac{}{\langle\breve{\circledS},\varrho\rangle\rightarrow\langle\surd,\varrho\rangle}$$
        $$\frac{\langle x, \varrho\rangle\xrightarrow{e}\langle x',\varrho'\rangle\quad \langle y, \varrho'\rangle\xrightarrow{\circledS^e}\langle y',\varrho'\rangle}{\langle x\leftmerge y,\varrho\rangle\xrightarrow{e}\langle x'\between y', \varrho'\rangle}$$
        $$\frac{\langle x, \varrho\rangle\xrightarrow{e}\langle e[m],\varrho'\rangle\quad \langle y, \varrho'\rangle\xrightarrow{\circledS^e}\langle y',\varrho'\rangle}{\langle x\leftmerge y,\varrho\rangle\xrightarrow{e}\langle e[m]\leftmerge y', \varrho'\rangle}$$
        $$\frac{\langle x, \varrho'\rangle\xrightarrow{\circledS^e}\langle \surd,\varrho'\rangle\quad \langle y, \varrho\rangle\xrightarrow{e}\langle y',\varrho'\rangle}{\langle x\leftmerge y,\varrho\rangle\xrightarrow{e}\langle y', \varrho'\rangle}$$
        $$\frac{\langle x, \varrho\rangle\xrightarrow{e}\langle e[m],\varrho'\rangle\quad \langle y, \varrho'\rangle\xrightarrow{\circledS^e}\langle\surd,\varrho'\rangle}{\langle x\leftmerge y,\varrho\rangle\xrightarrow{e}\langle e[m], \varrho'\rangle}$$
        $$\frac{}{\langle\circledS[m],\varrho\rangle\xtworightarrow{ }\langle\surd,\varrho\rangle}$$
        $$\frac{\langle x, \varrho\rangle\xtworightarrow{e[m]}\langle x',\varrho'\rangle\quad \langle y, \varrho'\rangle\xtworightarrow{\circledS^e[m]}\langle y',\varrho'\rangle}{\langle x\leftmerge y,\varrho\rangle\xtworightarrow{e[m]}\langle x'\between y', \varrho'\rangle}$$
        $$\frac{\langle x, \varrho\rangle\xtworightarrow{e[m]}\langle e,\varrho'\rangle\quad \langle y, \varrho'\rangle\xtworightarrow{\circledS^e[m]}\langle y',\varrho'\rangle}{\langle x\leftmerge y,\varrho\rangle\xtworightarrow{e[m]}\langle e\leftmerge y', \varrho'\rangle}$$
        $$\frac{\langle x, \varrho'\rangle\xtworightarrow{\circledS^e}\langle \surd,\varrho'\rangle\quad \langle y, \varrho\rangle\xtworightarrow{e[m]}\langle y',\varrho'\rangle}{\langle x\leftmerge y,\varrho\rangle\xtworightarrow{e[m]}\langle y', \varrho'\rangle}$$
        $$\frac{\langle x, \varrho\rangle\xtworightarrow{e[m]}\langle e,\varrho'\rangle\quad \langle y, \varrho'\rangle\xtworightarrow{\circledS^e[m]}\langle\surd,\varrho'\rangle}{\langle x\leftmerge y,\varrho\rangle\xtworightarrow{e[m]}\langle e, \varrho'\rangle}$$
        \caption{Transition rules of constant $\circledS$}
        \label{TRForENT2}
    \end{table}
\end{center}

\begin{theorem}[Elimination theorem of $qAPRPTC_{\tau}$ with guarded linear recursion and shadow constant]
Let $p$ be a closed $qAPRPTC_{\tau}$ with guarded linear recursion and shadow constant term. Then there is a closed $qAPRPTC$ term such that $qAPRPTC_{\tau}$ with guarded linear recursion and shadow constant$\vdash p=q$.
\end{theorem}

\begin{proof}
We leave the proof to the readers as an excise.
\end{proof}

\begin{theorem}[Conservitivity of $qAPRPTC_{\tau}$ with guarded linear recursion and shadow constant]
$qAPRPTC_{\tau}$ with guarded linear recursion and shadow constant is a conservative extension of $qAPRPTC_{\tau}$ with guarded linear recursion.
\end{theorem}

\begin{proof}
We leave the proof to the readers as an excise.
\end{proof}

\begin{theorem}[Congruence theorem of $qAPRPTC_{\tau}$ with guarded linear recursion and shadow constant]
FR probabilistic rooted branching truly concurrent bisimulation equivalences $\approx_{prbp}^{fr}$, $\approx_{prbs}^{fr}$, $\approx_{prbhp}^{fr}$ and $\approx_{prbhhp}^{fr}$ are all congruences with respect
to $qAPRPTC_{\tau}$ with guarded linear recursion and shadow constant.
\end{theorem}

\begin{proof}
We leave the proof to the readers as an excise.
\end{proof}

\begin{theorem}[Soundness of $qAPRPTC_{\tau}$ with guarded linear recursion and shadow constant]
Let $p$ and $q$ be closed $qAPRPTC_{\tau}$ with guarded linear recursion and shadow constant terms. If $qAPRPTC_{\tau}$ with guarded linear recursion and shadow constant$\vdash x=y$, then

\begin{enumerate}
  \item $x\approx_{prbs}^{fr} y$;
  \item $x\approx_{prbp}^{fr} y$;
  \item $x\approx_{prbhp}^{fr} y$;
  \item $x\approx_{prbhhp}^{fr} y$.
\end{enumerate}
\end{theorem}

\begin{proof}
We leave the proof to the readers as an excise.
\end{proof}

\begin{theorem}[Completeness of $qAPRPTC_{\tau}$ with guarded linear recursion and shadow constant]
Let $p$ and $q$ are closed $qAPRPTC_{\tau}$ with guarded linear recursion and shadow constant terms, then,

\begin{enumerate}
  \item if $p\approx_{prbs}^{fr} q$ then $p=q$;
  \item if $p\approx_{prbp}^{fr} q$ then $p=q$;
  \item if $p\approx_{prbhp}^{fr} q$ then $p=q$;
  \item if $p\approx_{prbhhp}^{fr} q$ then $p=q$.
\end{enumerate}
\end{theorem}

\begin{proof}
We leave the proof to the readers as an excise.
\end{proof}

\subsection{Unification of Quantum and Classical Computing for Closed Quantum Systems}\label{uni2}

We give the transition rules under quantum configuration for traditional atomic actions (events) $e'\in\mathbb{E}$ as Table \ref{TRForBPA5} shows.

\begin{center}
    \begin{table}
        $$\frac{}{\langle e,\varrho\rangle\rightsquigarrow\langle\breve{e},\varrho\rangle}$$
        $$\frac{\langle x,\varrho\rangle\rightsquigarrow \langle x',\varrho\rangle}{\langle x\cdot y,\varrho\rangle\rightsquigarrow \langle x'\cdot y,\varrho\rangle}$$
        $$\frac{\langle x,\varrho\rangle\rightsquigarrow \langle x',\varrho\rangle\quad \langle y,\varrho\rangle\rightsquigarrow \langle y',\varrho\rangle}{\langle x+y,\varrho\rangle\rightsquigarrow \langle x'+y',\varrho\rangle}$$
        $$\frac{\langle x,\varrho\rangle\rightsquigarrow \langle x',\varrho\rangle}{\langle x\boxplus_{\pi}y,\varrho\rangle\rightsquigarrow \langle x',\varrho\rangle}\quad \frac{\langle y,\varrho\rangle\rightsquigarrow \langle y',\varrho\rangle}{\langle x\boxplus_{\pi}y,\varrho\rangle\rightsquigarrow \langle y',\varrho\rangle}$$

        $$\frac{\langle x,\varrho\rangle\xrightarrow{e}\langle e[m],\varrho\rangle}{\langle x+ y,\varrho\rangle\xrightarrow{e}\langle e[m],\varrho\rangle}
        \quad\frac{\langle x,\varrho\rangle\xrightarrow{e}\langle x',\varrho\rangle}{\langle x+ y,\varrho\rangle\xrightarrow{e}\langle x',\varrho\rangle}
        \quad\frac{\langle y,\varrho\rangle\xrightarrow{e}\langle e[m],\varrho\rangle}{\langle x+ y,\varrho\rangle\xrightarrow{e}\langle e[m],\varrho\rangle}
        \quad\frac{\langle y,\varrho\rangle\xrightarrow{e}\langle y',\varrho\rangle}{\langle x+ y,\varrho\rangle\xrightarrow{e}\langle y',\varrho\rangle}$$
        $$\frac{\langle x,\varrho\rangle\xrightarrow{e}\langle e[m],\varrho\rangle}{\langle x\cdot y,\varrho\rangle\xrightarrow{e} \langle e[m]\cdot y,\varrho\rangle}
        \quad\frac{\langle x,\varrho\rangle\xrightarrow{e}\langle x',\varrho\rangle}{\langle x\cdot y,\varrho\rangle\xrightarrow{e}\langle x'\cdot y,\varrho\rangle}$$
        $$\frac{}{\langle e[m],\varrho\rangle\xtworightarrow{e[m]}\langle e,\varrho\rangle}$$
        $$\frac{\langle x,\varrho\rangle\xtworightarrow{e[m]}\langle e,\varrho\rangle}{\langle x+ y,\varrho\rangle\xtworightarrow{e[m]}\langle e,\varrho\rangle}
        \quad\frac{\langle x,\varrho\rangle\xtworightarrow{e[m]}\langle x',\varrho\rangle}{\langle x+ y,\varrho\rangle\xtworightarrow{e[m]}\langle x',\varrho\rangle}
        \quad\frac{\langle y,\varrho\rangle\xtworightarrow{e[m]}\langle e,\varrho\rangle}{\langle x+ y,\varrho\rangle\xtworightarrow{e[m]}\langle e,\varrho\rangle}
        \quad\frac{\langle y,\varrho\rangle\xtworightarrow{e[m]}\langle y',\varrho\rangle}{\langle x+ y,\varrho\rangle\xtworightarrow{e[m]}\langle y',\varrho\rangle}$$
        $$\frac{\langle y,\varrho\rangle\xtworightarrow{e[m]}\langle e,\varrho\rangle}{\langle x\cdot y,\varrho\rangle\xtworightarrow{e[m]} \langle x\cdot e,\varrho\rangle}
        \quad\frac{\langle y,\varrho\rangle\xtworightarrow{e[m]}\langle y',\varrho\rangle}{\langle x\cdot y,\varrho\rangle\xtworightarrow{e[m]}\langle x\cdot y',\varrho\rangle}$$
        \caption{Transition rules of BARPTC under quantum configuration}
        \label{TRForBPA5}
    \end{table}
\end{center}

And the axioms for traditional actions are the same as those of qBARPTC. And it is natural can be extended to qAPRPTC, recursion and abstraction. So, quantum and classical computing
are unified under the framework of qAPRPTC for closed quantum systems.

\newpage\section{Applications of qAPRPTC}\label{aqaprptc}

Quantum and classical computing in closed systems are unified with qAPRPTC, which have the same equational logic and the same quantum configuration based operational semantics.
The unification can be used widely in verification for the behaviors of quantum and classical computing mixed systems. In this chapter, we show its usage in verification of the
quantum communication protocols.

\subsection{Verification of Quantum Teleportation Protocol}\label{VQT6}

\begin{figure}
  \centering
  \includegraphics{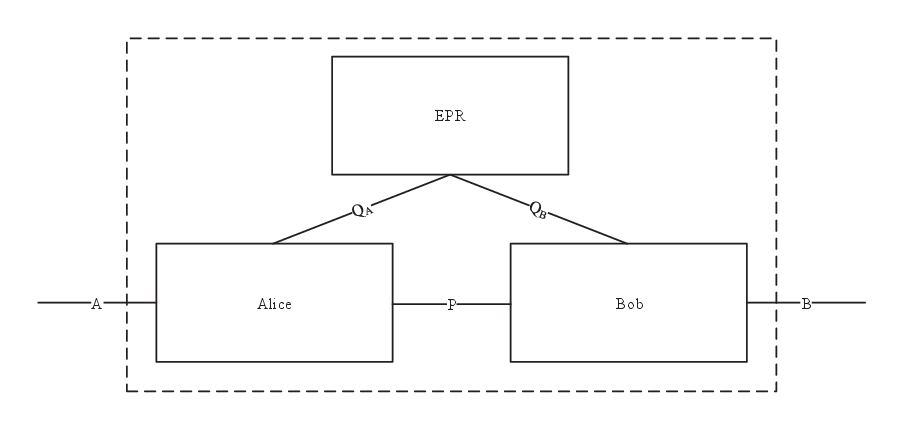}
  \caption{Quantum teleportation protocol.}
  \label{QT}
\end{figure}

Quantum teleportation \cite{QT} is a famous quantum protocol in quantum information theory to teleport an unknown quantum state by sending only classical information, provided that the sender and the receiver, Alice and Bob, shared an entangled state in advance. Firstly, we introduce the basic quantum teleportation protocol briefly, which is illustrated in Figure \ref{QT}. In this section, we show how to process quantum entanglement in an implicit way.

\begin{enumerate}
  \item EPR generates 2-qubits entangled EPR pair $q=q_1\otimes q_2$, and he sends $q_1$ to Alice through quantum channel $Q_A$ and $q_2$ to Bob through quantum channel $Q_B$;
  \item Alice receives $q_1$, after some preparations, she measures on $q_1$, and sends the measurement results $x$ to Bob through classical channel $P$;
  \item Bob receives $q_2$ from EPR, and also the classical information $x$ from Alice. According to $x$, he chooses specific Pauli transformation on $q_2$.
\end{enumerate}

We re-introduce the basic quantum teleportation protocol in an abstract way with more technical details as Figure \ref{QT} illustrates.

Now, we assume the generation of 2-qubits $q$ through two unitary operators $Set[q]$ and $H[q]$. EPR sends $q_1$ to Alice through the quantum channel $Q_A$ by quantum communicating action $send_{Q_A}(q_1)$ and Alice receives $q_1$ through $Q_A$ by quantum communicating action $receive_{Q_A}(q_1)$. Similarly, for Bob, those are $send_{Q_B}(q_2)$ and $receive_{Q_B}(q_2)$. After Alice receives $q_1$, she does some preparations, including a unitary transformation $CNOT$ and a Hadamard transformation $H$, then Alice do measurement $M=\sum^3_{i=0}M_i$, and sends measurement results $x$ to Bob through the public classical channel $P$ by classical communicating action $send_{P}(x)$, and Bob receives $x$ through channel $P$ by classical communicating action $receive_{P}(x)$. According to $x$, Bob performs specific Pauli transformations $\sigma_x$ on $q_2$. Let Alice, Bob and EPR be a system $ABE$ and let interactions between Alice, Bob and EPR be internal actions. $ABE$ receives external input $D_i$ through channel $A$ by communicating action $receive_A(D_i)$ and sends results $D_o$ through channel $B$ by communicating action $send_B(D_o)$. Note that the entangled EPR pair $q=q_1\otimes q_2$ is within $ABE$, so quantum entanglement can be processed implicitly.

Then the state transitions of EPR can be described by PQRA as follows.

\begin{eqnarray}
&&E=Set[q]\cdot E_1\nonumber\\
&&E_1=H[q]\cdot E_2\nonumber\\
&&E_2=send_{Q_A}(q_1)\cdot E_3\nonumber\\
&&E_3=send_{Q_B}(q_2)\cdot E\nonumber
\end{eqnarray}

And the state transitions of Alice can be described by PQRA as follows.

\begin{eqnarray}
&&A=\sum_{D_i\in \Delta_i}receive_A(D_i)\cdot A_1\nonumber\\
&&A_1=receive_{Q_A}(q_1)\cdot A_2\nonumber\\
&&A_2=CNOT\cdot A_3\nonumber\\
&&A_3=H\cdot A_4\nonumber\\
&&A_4=(M_0\cdot send_P(0)\boxplus_{\frac{1}{4}}M_1\cdot send_P(1)\boxplus_{\frac{1}{4}}M_2\cdot send_P(2)\boxplus_{\frac{1}{4}}M_3\cdot send_P(3))\cdot A\nonumber
\end{eqnarray}

where $\Delta_i$ is the collection of the input data.

And the state transitions of Bob can be described by PQRA as follows.

\begin{eqnarray}
&&B=receive_{Q_B}(q_2)\cdot B_1\nonumber\\
&&B_1=(receive_P(0)\cdot\sigma_0\boxplus_{\frac{1}{4}}receive_P(1)\cdot\sigma_1\boxplus_{\frac{1}{4}}receive_P(2) \cdot\sigma_2\boxplus_{\frac{1}{4}}receive_P(3)\cdot\sigma_3)\cdot B_2\nonumber\\
&&B_2=\sum_{D_o\in\Delta_o}send_B(D_o)\cdot B\nonumber
\end{eqnarray}

where $\Delta_o$ is the collection of the output data.

The send action and receive action of the same data through the same channel can communicate each other, otherwise, a deadlock $\delta$ will be caused. We define the following communication functions.

\begin{eqnarray}
&&\gamma(send_{Q_A}(q_1),receive_{Q_A}(q_1))\triangleq c_{Q_A}(q_1)\nonumber\\
&&\gamma(send_{Q_B}(q_2),receive_{Q_B}(q_2))\triangleq c_{Q_B}(q_2)\nonumber\\
&&\gamma(send_P(0),receive_P(0))\triangleq c_P(0)\nonumber\\
&&\gamma(send_P(1),receive_P(1))\triangleq c_P(1)\nonumber\\
&&\gamma(send_P(2),receive_P(2))\triangleq c_P(2)\nonumber\\
&&\gamma(send_P(3),receive_P(3))\triangleq c_P(3)\nonumber
\end{eqnarray}

Let $A$, $B$ and $E$ in parallel, then the system $ABE$ can be represented by the following process term.

$$\tau_I(\partial_H(\Theta(A\between B\between E)))$$

where $H=\{send_{Q_A}(q_1), receive_{Q_A}(q_1), send_{Q_B}(q_2), receive_{Q_B}(q_2),\\
send_P(0), receive_P(0), send_P(1), receive_P(1),\\
send_P(2), receive_P(2), send_P(3), receive_P(3)\}$ and $I=\{Set[q], H[q], CNOT, H, M_0, M_1,\\ M_2, M_3, \sigma_0, \sigma_1, \sigma_2, \sigma_3, \\ c_{Q_A}(q_1), c_{Q_B}(q_2), c_P(0), c_P(1), c_P(2), c_P(3)\}$.

Then we get the following conclusion.

\begin{theorem}
The basic quantum teleportation protocol $\tau_I(\partial_H(\Theta(A\between B\between E)))$ exhibits desired external behaviors.
\end{theorem}

\begin{proof}
We can get $\tau_I(\partial_H(\Theta(A\between B\between E)))=\sum_{D_i\in \Delta_i}\sum_{D_o\in\Delta_o}receive_A(D_i)\leftmerge send_B(D_o)\leftmerge
\tau_I(\partial_H(\Theta(A\between B\between E)))$. So, the basic quantum teleportation protocol $\tau_I(\partial_H(\Theta(A\between B\between E)))$ exhibits desired external behaviors.
\end{proof}

\subsection{Verification of BB84 Protocol}\label{VBB86}

\begin{figure}
  \centering
  \includegraphics{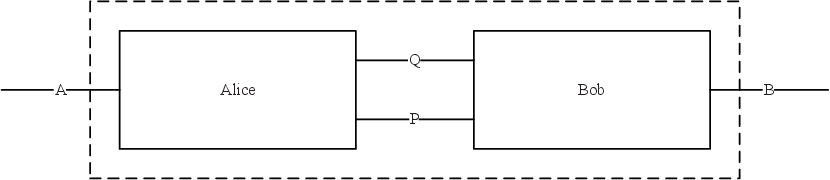}
  \caption{BB84 protocol.}
  \label{BB84}
\end{figure}

The BB84 protocol \cite{BB84} is used to create a private key between two parities, Alice and Bob. Firstly, we introduce the basic BB84 protocol briefly, which is illustrated in Figure \ref{BB84}.

\begin{enumerate}
  \item Alice create two string of bits with size $n$ randomly, denoted as $B_a$ and $K_a$;
  \item Alice generates a string of qubits $q$ with size $n$, and the $i$th qubit in $q$ is $|x_y\rangle$, where $x$ is the $i$th bit of $B_a$ and $y$ is the $i$th bit of $K_a$;
  \item Alice sends $q$ to Bob through a quantum channel $Q$ between Alice and Bob;
  \item Bob receives $q$ and randomly generates a string of bits $B_b$ with size $n$;
  \item Bob measures each qubit of $q$ according to a basis by bits of $B_b$. And the measurement results would be $K_b$, which is also with size $n$;
  \item Bob sends his measurement bases $B_b$ to Alice through a public channel $P$;
  \item Once receiving $B_b$, Alice sends her bases $B_a$ to Bob through channel $P$, and Bob receives $B_a$;
  \item Alice and Bob determine that at which position the bit strings $B_a$ and $B_b$ are equal, and they discard the mismatched bits of $B_a$ and $B_b$. Then the remaining bits of $K_a$ and $K_b$, denoted as $K_a'$ and $K_b'$ with $K_{a,b}=K_a'=K_b'$.
\end{enumerate}

We re-introduce the basic BB84 protocol in an abstract way with more technical details as Figure \ref{BB84} illustrates.

Now, we assume a special measurement operation $Rand[q;B_a]=\sum^{2n-1}_{i=0}Rand[q;B_a]_i$ which create a string of $n$ random bits $B_a$ from the $q$ quantum system, and the same as $Rand[q;K_a]=\sum^{2n-1}_{i=0}Rand[q;K_a]_i$, $Rand[q';B_b]=\sum^{2n-1}_{i=0}Rand[q';B_b]_i$. $M[q;K_b]=\sum^{2n-1}_{i=0}M[q;K_b]_i$ denotes the Bob's measurement on $q$. The generation of $n$ qubits $q$ through two unitary operators $Set_{K_a}[q]$ and $H_{B_a}[q]$. Alice sends $q$ to Bob through the quantum channel $Q$ by quantum communicating action $send_{Q}(q)$ and Bob receives $q$ through $Q$ by quantum communicating action $receive_{Q}(q)$. Bob sends $B_b$ to Alice through the public classical channel $P$ by classical communicating action $send_{P}(B_b)$ and Alice receives $B_b$ through channel $P$ by classical communicating action $receive_{P}(B_b)$, and the same as $send_{P}(B_a)$ and $receive_{P}(B_a)$. Alice and Bob generate the private key $K_{a,b}$ by a classical comparison action $cmp(K_{a,b},K_a,K_b,B_a,B_b)$. Let Alice and Bob be a system $AB$ and let interactions between Alice and Bob be internal actions. $AB$ receives external input $D_i$ through channel $A$ by communicating action $receive_A(D_i)$ and sends results $D_o$ through channel $B$ by communicating action $send_B(D_o)$.

Then the state transitions of Alice can be described by PQRA as follows.

\begin{eqnarray}
&&A=\sum_{D_i\in \Delta_i}receive_A(D_i)\cdot A_1\nonumber\\
&&A_1=\boxplus_{\frac{1}{2n},i=0}^{2n-1}Rand[q;B_a]_i\cdot A_2\nonumber\\
&&A_2=\boxplus_{\frac{1}{2n},i=0}^{2n-1}Rand[q;K_a]_i\cdot A_3\nonumber\\
&&A_3=Set_{K_a}[q]\cdot A_4\nonumber\\
&&A_4=H_{B_a}[q]\cdot A_5\nonumber\\
&&A_5=send_Q(q)\cdot A_6\nonumber\\
&&A_6=receive_P(B_b)\cdot A_7\nonumber\\
&&A_7=send_P(B_a)\cdot A_8\nonumber\\
&&A_8=cmp(K_{a,b},K_a,K_b,B_a,B_b)\cdot A\nonumber
\end{eqnarray}

where $\Delta_i$ is the collection of the input data.

And the state transitions of Bob can be described by PQRA as follows.

\begin{eqnarray}
&&B=receive_Q(q)\cdot B_1\nonumber\\
&&B_1=\boxplus_{\frac{1}{2n},i=0}^{2n-1}Rand[q';B_b]_i\cdot B_2\nonumber\\
&&B_2=\boxplus_{\frac{1}{2n},i=0}^{2n-1}M[q;K_b]_i\cdot B_3\nonumber\\
&&B_3=send_P(B_b)\cdot B_4\nonumber\\
&&B_4=receive_P(B_a)\cdot B_5\nonumber\\
&&B_5=cmp(K_{a,b},K_a,K_b,B_a,B_b)\cdot B_6\nonumber\\
&&B_6=\sum_{D_o\in\Delta_o}send_B(D_o)\cdot B\nonumber
\end{eqnarray}

where $\Delta_o$ is the collection of the output data.

The send action and receive action of the same data through the same channel can communicate each other, otherwise, a deadlock $\delta$ will be caused. We define the following communication functions.

\begin{eqnarray}
&&\gamma(send_Q(q),receive_Q(q))\triangleq c_Q(q)\nonumber\\
&&\gamma(send_P(B_b),receive_P(B_b))\triangleq c_P(B_b)\nonumber\\
&&\gamma(send_P(B_a),receive_P(B_a))\triangleq c_P(B_a)\nonumber
\end{eqnarray}

Let $A$ and $B$ in parallel, then the system $AB$ can be represented by the following process term.

$$\tau_I(\partial_H(\Theta(A\between B)))$$

where $H=\{send_Q(q),receive_Q(q),send_P(B_b),receive_P(B_b),send_P(B_a),receive_P(B_a)\}$ and $I=\{Rand[q;B_a]_i, Rand[q;K_a]_i, Set_{K_a}[q], H_{B_a}[q], Rand[q';B_b]_i, M[q;K_b]_i, \\c_Q(q), c_P(B_b), c_P(B_a), cmp(K_{a,b},K_a,K_b,B_a,B_b)\}$.

Then we get the following conclusion.

\begin{theorem}
The basic BB84 protocol $\tau_I(\partial_H(\Theta(A\between B)))$ exhibits desired external behaviors.
\end{theorem}

\begin{proof}
We can get $\tau_I(\partial_H(\Theta(A\between B)))=\sum_{D_i\in \Delta_i}\sum_{D_o\in\Delta_o}receive_A(D_i)\leftmerge send_B(D_o)\leftmerge \tau_I(\partial_H(\Theta(A\between B)))$.
So, the basic BB84 protocol $\tau_I(\partial_H(\Theta(A\between B)))$ exhibits desired external behaviors.
\end{proof}

\subsection{Verification of E91 Protocol}\label{VE916}

\begin{figure}
  \centering
  \includegraphics{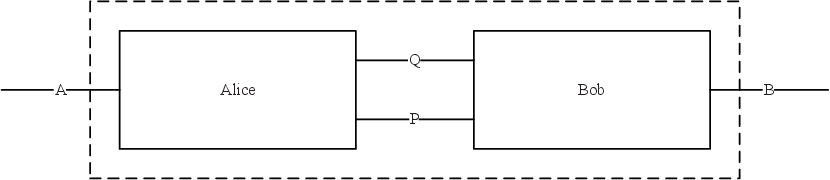}
  \caption{E91 protocol.}
  \label{E91}
\end{figure}

With support of Entanglement merge $\between$, PQRA can be used to verify quantum protocols utilizing entanglement explicitly. E91 protocol\cite{E91} is the first quantum protocol which utilizes entanglement. E91 protocol is used to create a private key between two parities, Alice and Bob. Firstly, we introduce the basic E91 protocol briefly, which is illustrated in Figure \ref{E91}.

\begin{enumerate}
  \item Alice generates a string of EPR pairs $q$ with size $n$, i.e., $2n$ particles, and sends a string of qubits $q_b$ from each EPR pair with $n$ to Bob through a quantum channel $Q$, remains the other string of qubits $q_a$ from each pair with size $n$;
  \item Alice create two string of bits with size $n$ randomly, denoted as $B_a$ and $K_a$;
  \item Bob receives $q_b$ and randomly generates a string of bits $B_b$ with size $n$;
  \item Alice measures each qubit of $q_a$ according to a basis by bits of $B_a$. And the measurement results would be $K_a$, which is also with size $n$;
  \item Bob measures each qubit of $q_b$ according to a basis by bits of $B_b$. And the measurement results would be $K_b$, which is also with size $n$;
  \item Bob sends his measurement bases $B_b$ to Alice through a public channel $P$;
  \item Once receiving $B_b$, Alice sends her bases $B_a$ to Bob through channel $P$, and Bob receives $B_a$;
  \item Alice and Bob determine that at which position the bit strings $B_a$ and $B_b$ are equal, and they discard the mismatched bits of $B_a$ and $B_b$. Then the remaining bits of $K_a$ and $K_b$, denoted as $K_a'$ and $K_b'$ with $K_{a,b}=K_a'=K_b'$.
\end{enumerate}

We re-introduce the basic E91 protocol in an abstract way with more technical details as Figure \ref{E91} illustrates.

Now, $M[q_a;K_a]=\sum_{i=0}^{2n-1}M[q_a;K_a]_i$ denotes the Alice's measurement operation of $q_a$, and $\circledS_{M[q_a;K_a]}=\sum_{i=0}^{2n-1}\circledS_{M[q_a;K_a]_i}$ denotes the responding shadow constant; $M[q_b;K_b]=\sum_{i=0}^{2n-1}M[q_b;K_b]_i$ denotes the Bob's measurement operation of $q_b$, and $\circledS_{M[q_b;K_b]}=\sum_{i=0}^{2n-1}\circledS_{M[q_b;K_b]_i}$ denotes the responding shadow constant. Alice sends $q_b$ to Bob through the quantum channel $Q$ by quantum communicating action $send_{Q}(q_b)$ and Bob receives $q_b$ through $Q$ by quantum communicating action $receive_{Q}(q_b)$. Bob sends $B_b$ to Alice through the public channel $P$ by classical communicating action $send_{P}(B_b)$ and Alice receives $B_b$ through channel $P$ by classical communicating action $receive_{P}(B_b)$, and the same as $send_{P}(B_a)$ and $receive_{P}(B_a)$. Alice and Bob generate the private key $K_{a,b}$ by a classical comparison action $cmp(K_{a,b},K_a,K_b,B_a,B_b)$. Let Alice and Bob be a system $AB$ and let interactions between Alice and Bob be internal actions. $AB$ receives external input $D_i$ through channel $A$ by communicating action $receive_A(D_i)$ and sends results $D_o$ through channel $B$ by communicating action $send_B(D_o)$.

Then the state transitions of Alice can be described by PQRA as follows.

\begin{eqnarray}
&&A=\sum_{D_i\in \Delta_i}receive_A(D_i)\cdot A_1\nonumber\\
&&A_1=send_Q(q_b)\cdot A_2\nonumber\\
&&A_2=\boxplus_{\frac{1}{2n},i=0}^{2n-1}M[q_a;K_a]_i\cdot A_3\nonumber\\
&&A_3=\boxplus_{\frac{1}{2n},i=0}^{2n-1}\circledS_{M[q_b;K_b]_i}\cdot A_4\nonumber\\
&&A_4=receive_P(B_b)\cdot A_5\nonumber\\
&&A_5=send_P(B_a)\cdot A_6\nonumber\\
&&A_6=cmp(K_{a,b},K_a,K_b,B_a,B_b)\cdot A\nonumber
\end{eqnarray}

where $\Delta_i$ is the collection of the input data.

And the state transitions of Bob can be described by PQRA as follows.

\begin{eqnarray}
&&B=receive_Q(q_b)\cdot B_1\nonumber\\
&&B_1=\boxplus_{\frac{1}{2n},i=0}^{2n-1}\circledS_{M[q_a;K_a]_i}\cdot B_2\nonumber\\
&&B_2=\boxplus_{\frac{1}{2n},i=0}^{2n-1}M[q_b;K_b]_i\cdot B_3\nonumber\\
&&B_3=send_P(B_b)\cdot B_4\nonumber\\
&&B_4=receive_P(B_a)\cdot B_5\nonumber\\
&&B_5=cmp(K_{a,b},K_a,K_b,B_a,B_b)\cdot B_6\nonumber\\
&&B_6=\sum_{D_o\in\Delta_o}send_B(D_o)\cdot B\nonumber
\end{eqnarray}

where $\Delta_o$ is the collection of the output data.

The send action and receive action of the same data through the same channel can communicate each other, otherwise, a deadlock $\delta$ will be caused. The quantum operation and its shadow constant pair will lead entanglement occur, otherwise, a deadlock $\delta$ will occur. We define the following communication functions.

\begin{eqnarray}
&&\gamma(send_Q(q_b),receive_Q(q_b))\triangleq c_Q(q_b)\nonumber\\
&&\gamma(send_P(B_b),receive_P(B_b))\triangleq c_P(B_b)\nonumber\\
&&\gamma(send_P(B_a),receive_P(B_a))\triangleq c_P(B_a)\nonumber
\end{eqnarray}

Let $A$ and $B$ in parallel, then the system $AB$ can be represented by the following process term.

$$\tau_I(\partial_H(\Theta(A\between B)))$$

where $H=\{send_Q(q_b),receive_Q(q_b),send_P(B_b),receive_P(B_b),send_P(B_a),receive_P(B_a),\\ M[q_a;K_a]_i, \circledS_{M[q_a;K_a]_i}, M[q_b;K_b]_i, \circledS_{M[q_b;K_b]_i}\}$ and $I=\{c_Q(q_b), c_P(B_b), c_P(B_a), M[q_a;K_a], M[q_b;K_b],\\ cmp(K_{a,b},K_a,K_b,B_a,B_b)\}$.

Then we get the following conclusion.

\begin{theorem}
The basic E91 protocol $\tau_I(\partial_H(\Theta(A\between B)))$ exhibits desired external behaviors.
\end{theorem}

\begin{proof}
We can get $\tau_I(\partial_H(\Theta(A\between B)))=\sum_{D_i\in \Delta_i}\sum_{D_o\in\Delta_o}receive_A(D_i)\leftmerge send_B(D_o)\leftmerge \tau_I(\partial_H(\Theta(A\between B)))$.
So, the basic E91 protocol $\tau_I(\partial_H(\Theta(A\between B)))$ exhibits desired external behaviors.
\end{proof}

\subsection{Verification of B92 Protocol}\label{VB926}

The famous B92 protocol\cite{B92} is a quantum key distribution protocol, in which quantum information and classical information are mixed. We take an example of the B92 protocol to illustrate the usage of probabilistic quantum process algebra in verification of quantum protocols.

\begin{figure}
  \centering
  \includegraphics{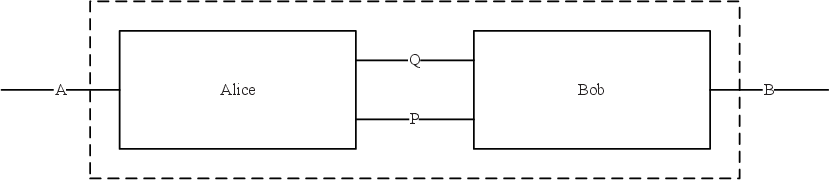}
  \caption{The B92 protocol.}
  \label{B92}
\end{figure}

The B92 protocol is used to create a private key between two parities, Alice and Bob. B92 is a protocol of quantum key distribution (QKD) which uses polarized photons as information carriers. Firstly, we introduce the basic B92 protocol briefly, which is illustrated in Figure \ref{B92}.

\begin{enumerate}
  \item Alice create a string of bits with size $n$ randomly, denoted as $A$.
  \item Alice generates a string of qubits $q$ with size $n$, carried by polarized photons. If $A_i=0$, the ith qubit is $|0\rangle$; else if $A_i=1$, the ith qubit is $|+\rangle$.
  \item Alice sends $q$ to Bob through a quantum channel $Q$ between Alice and Bob.
  \item Bob receives $q$ and randomly generates a string of bits $B$ with size $n$.
  \item If $B_i=0$, Bob chooses the basis $\oplus$; else if $B_i=1$, Bob chooses the basis $\otimes$. Bob measures each qubit of $q$ according to the above basses. And Bob builds a String of bits $T$, if the measurement produces $|0\rangle$ or $|+\rangle$, then $T_i=0$; else if the measurement produces $|1\rangle$ or $|-\rangle$, then $T_i=1$, which is also with size $n$.
  \item Bob sends $T$ to Alice through a public channel $P$.
  \item Alice and Bob determine that at which position the bit strings $A$ and $B$ are remained for which $T_i=1$. In absence of Eve, $A_i=1-B_i$, a shared raw key $K_{a,b}$ is formed by $A_i$.
\end{enumerate}

We re-introduce the basic B92 protocol in an abstract way with more technical details as Figure \ref{B92} illustrates.

Now, we assume a special measurement operation $Rand[q;A]=\sum^{2n-1}_{i=0}Rand[q;A]_i$ which create a string of $n$ random bits $A$ from the $q$ quantum system, and the same as $Rand[q';B]=\sum^{2n-1}_{i=0}Rand[q';B]_i$. $M[q;T]=\sum^{2n-1}_{i=0}M[q;T]_i$ denotes the Bob's measurement operation of $q$. The generation of $n$ qubits $q$ through a unitary operator $Set_{A}[q]$. Alice sends $q$ to Bob through the quantum channel $Q$ by quantum communicating action $send_{Q}(q)$ and Bob receives $q$ through $Q$ by quantum communicating action $receive_{Q}(q)$. Bob sends $T$ to Alice through the public channel $P$ by classical communicating action $send_{P}(T)$ and Alice receives $T$ through channel $P$ by classical communicating action $receive_{P}(T)$. Alice and Bob generate the private key $K_{a,b}$ by a classical comparison action $cmp(K_{a,b},T,A,B)$. Let Alice and Bob be a system $AB$ and let interactions between Alice and Bob be internal actions. $AB$ receives external input $D_i$ through channel $A$ by communicating action $receive_A(D_i)$ and sends results $D_o$ through channel $B$ by communicating action $send_B(D_o)$.

Then the state transition of Alice can be described by probabilistic quantum process algebra as follows.

\begin{eqnarray}
&&A=\sum_{D_i\in \Delta_i}receive_A(D_i)\cdot A_1\nonumber\\
&&A_1=\boxplus_{\frac{1}{2n},i=0}^{2n-1}Rand[q;A]_i\cdot A_2\nonumber\\
&&A_2=Set_{A}[q]\cdot A_3\nonumber\\
&&A_3=send_Q(q)\cdot A_4\nonumber\\
&&A_4=receive_P(T)\cdot A_5\nonumber\\
&&A_5=cmp(K_{a,b},T,A,B)\cdot A\nonumber
\end{eqnarray}

where $\Delta_i$ is the collection of the input data.

And the state transition of Bob can be described by probabilistic quantum process algebra as follows.

\begin{eqnarray}
&&B=receive_Q(q)\cdot B_1\nonumber\\
&&B_1=\boxplus_{\frac{1}{2n},i=0}^{2n-1}Rand[q';B]_i\cdot B_2\nonumber\\
&&B_2=\boxplus_{\frac{1}{2n},i=0}^{2n-1}M[q;T]_i\cdot B_3\nonumber\\
&&B_3=send_P(T)\cdot B_4\nonumber\\
&&B_4=cmp(K_{a,b},T,A,B)\cdot B_5\nonumber\\
&&B_5=\sum_{D_o\in\Delta_o}send_B(D_o)\cdot B\nonumber
\end{eqnarray}

where $\Delta_o$ is the collection of the output data.

The send action and receive action of the same data through the same channel can communicate each other, otherwise, a deadlock $\delta$ will be caused. We define the following communication functions.

\begin{eqnarray}
&&\gamma(send_Q(q),receive_Q(q))\triangleq c_Q(q)\nonumber\\
&&\gamma(send_P(T),receive_P(T))\triangleq c_P(T)\nonumber\\
\end{eqnarray}

Let $A$ and $B$ in parallel, then the system $AB$ can be represented by the following process term.

$$\tau_I(\partial_H(\Theta(A\between B)))$$

where $H=\{send_Q(q),receive_Q(q),send_P(T),receive_P(T)\}$ and $I=\{\boxplus_{\frac{1}{2n},i=0}^{2n-1}Rand[q;A]_i, \\Set_{A}[q], \boxplus_{\frac{1}{2n},i=0}^{2n-1}Rand[q';B]_i, \boxplus_{\frac{1}{2n},i=0}^{2n-1}M[q;T]_i, c_Q(q), c_P(T), cmp(K_{a,b},T,A,B)\}$.

Then we get the following conclusion.

\begin{theorem}
The basic B92 protocol $\tau_I(\partial_H(\Theta(A\between B)))$ exhibits desired external behaviors.
\end{theorem}

\begin{proof}
We can get $\tau_I(\partial_H(\Theta(A\between B)))=\sum_{D_i\in \Delta_i}\sum_{D_o\in\Delta_o}receive_A(D_i)\leftmerge send_B(D_o)\leftmerge \tau_I(\partial_H(\Theta(A\between B)))$.
So, the basic B92 protocol $\tau_I(\partial_H(\Theta(A\between B)))$ exhibits desired external behaviors.
\end{proof}

\subsection{Verification of DPS Protocol}\label{VDPS6}

The famous DPS protocol\cite{DPS} is a quantum key distribution protocol, in which quantum information and classical information are mixed. We take an example of the DPS protocol to illustrate the usage of probabilistic quantum process algebra in verification of quantum protocols.

\begin{figure}
  \centering
  \includegraphics{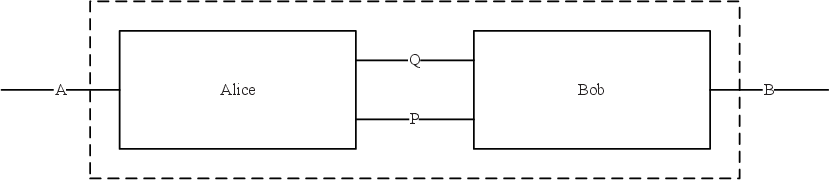}
  \caption{The DPS protocol.}
  \label{DPS}
\end{figure}

The DPS protocol is used to create a private key between two parities, Alice and Bob. DPS is a protocol of quantum key distribution (QKD) which uses pulses of a photon which has nonorthogonal four states. Firstly, we introduce the basic DPS protocol briefly, which is illustrated in Figure \ref{DPS}.

\begin{enumerate}
  \item Alice generates a string of qubits $q$ with size $n$, carried by a series of single photons possily at four time instances.
  \item Alice sends $q$ to Bob through a quantum channel $Q$ between Alice and Bob.
  \item Bob receives $q$ by detectors clicking at the second or third time instance, and records the time into $T$ with size $n$ and which detector clicks into $D$ with size $n$.
  \item Bob sends $T$ to Alice through a public channel $P$.
  \item Alice receives $T$. From $T$ and her modulation data, Alice knows which detector clicked in Bob's site, i.e. $D$.
  \item Alice and Bob have an identical bit string, provided that the first detector click represents "0" and the other detector represents "1", then a shared raw key $K_{a,b}$ is formed.
\end{enumerate}

We re-introduce the basic DPS protocol in an abstract way with more technical details as Figure \ref{DPS} illustrates.

Now, we assume $M[q;T]=\sum^{2n-1}_{i=0}M[q;T]_i$ denotes the Bob's measurement operation of $q$. The generation of $n$ qubits $q$ through a unitary operator $Set_{A}[q]$. Alice sends $q$ to Bob through the quantum channel $Q$ by quantum communicating action $send_{Q}(q)$ and Bob receives $q$ through $Q$ by quantum communicating action $receive_{Q}(q)$. Bob sends $T$ to Alice through the public channel $P$ by classical communicating action $send_{P}(T)$ and Alice receives $T$ through channel $P$ by classical communicating action $receive_{P}(T)$. Alice and Bob generate the private key $K_{a,b}$ by a classical comparison action $cmp(K_{a,b},D)$. Let Alice and Bob be a system $AB$ and let interactions between Alice and Bob be internal actions. $AB$ receives external input $D_i$ through channel $A$ by communicating action $receive_A(D_i)$ and sends results $D_o$ through channel $B$ by communicating action $send_B(D_o)$.

Then the state transition of Alice can be described by probabilistic quantum process algebra as follows.

\begin{eqnarray}
&&A=\sum_{D_i\in \Delta_i}receive_A(D_i)\cdot A_1\nonumber\\
&&A_1=Set_{A}[q]\cdot A_2\nonumber\\
&&A_2=send_Q(q)\cdot A_3\nonumber\\
&&A_3=receive_P(T)\cdot A_4\nonumber\\
&&A_4=cmp(K_{a,b},D)\cdot A\nonumber
\end{eqnarray}

where $\Delta_i$ is the collection of the input data.

And the state transition of Bob can be described by probabilistic quantum process algebra as follows.

\begin{eqnarray}
&&B=receive_Q(q)\cdot B_1\nonumber\\
&&B_1=\boxplus_{\frac{1}{2n},i=0}^{2n-1}M[q;T]_i\cdot B_2\nonumber\\
&&B_2=send_P(T)\cdot B_3\nonumber\\
&&B_3=cmp(K_{a,b},D)\cdot B_4\nonumber\\
&&B_4=\sum_{D_o\in\Delta_o}send_B(D_o)\cdot B\nonumber
\end{eqnarray}

where $\Delta_o$ is the collection of the output data.

The send action and receive action of the same data through the same channel can communicate each other, otherwise, a deadlock $\delta$ will be caused. We define the following communication functions.

\begin{eqnarray}
&&\gamma(send_Q(q),receive_Q(q))\triangleq c_Q(q)\nonumber\\
&&\gamma(send_P(T),receive_P(T))\triangleq c_P(T)\nonumber\\
\end{eqnarray}

Let $A$ and $B$ in parallel, then the system $AB$ can be represented by the following process term.

$$\tau_I(\partial_H(\Theta(A\between B)))$$

where $H=\{send_Q(q),receive_Q(q),send_P(T),receive_P(T)\}$ and $I=\{Set_{A}[q], \\ \boxplus_{\frac{1}{2n},i=0}^{2n-1}M[q;T]_i, c_Q(q), c_P(T), cmp(K_{a,b},D)\}$.

Then we get the following conclusion.

\begin{theorem}
The basic DPS protocol $\tau_I(\partial_H(\Theta(A\between B)))$ exhibits desired external behaviors.
\end{theorem}

\begin{proof}
We can get $\tau_I(\partial_H(\Theta(A\between B)))=\sum_{D_i\in \Delta_i}\sum_{D_o\in\Delta_o}receive_A(D_i)\leftmerge send_B(D_o)\leftmerge \tau_I(\partial_H(\Theta(A\between B)))$.
So, the basic DPS protocol $\tau_I(\partial_H(\Theta(A\between B)))$ exhibits desired external behaviors.
\end{proof}

\subsection{Verification of BBM92 Protocol}\label{VBBM926}

The famous BBM92 protocol\cite{BBM92} is a quantum key distribution protocol, in which quantum information and classical information are mixed. We take an example of the BBM92 protocol to illustrate the usage of probabilistic quantum process algebra in verification of quantum protocols.

\begin{figure}
  \centering
  \includegraphics{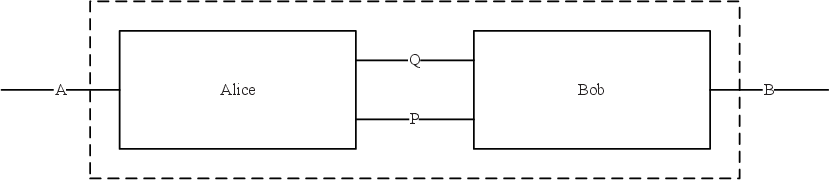}
  \caption{The BBM92 protocol.}
  \label{BBM92}
\end{figure}

The BBM92 protocol is used to create a private key between two parities, Alice and Bob. BBM92 is a protocol of quantum key distribution (QKD) which uses EPR pairs as information carriers. Firstly, we introduce the basic BBM92 protocol briefly, which is illustrated in Figure \ref{BBM92}.

\begin{enumerate}
  \item Alice generates a string of EPR pairs $q$ with size $n$, i.e., $2n$ particles, and sends a string of qubits $q_b$ from each EPR pair with $n$ to Bob through a quantum channel $Q$, remains the other string of qubits $q_a$ from each pair with size $n$.
  \item Alice create a string of bits with size $n$ randomly, denoted as $B_a$.
  \item Bob receives $q_b$ and randomly generates a string of bits $B_b$ with size $n$.
  \item Alice measures each qubit of $q_a$ according to bits of $B_a$, if $B_{a_i}=0$, then uses $x$ axis ($\rightarrow$); else if $B_{a_i}=1$, then uses $z$ axis ($\uparrow$).
  \item Bob measures each qubit of $q_b$ according to bits of $B_b$, if $B_{b_i}=0$, then uses $x$ axis ($\rightarrow$); else if $B_{b_i}=1$, then uses $z$ axis ($\uparrow$).
  \item Bob sends his measurement axis choices $B_b$ to Alice through a public channel $P$.
  \item Once receiving $B_b$, Alice sends her axis choices $B_a$ to Bob through channel $P$, and Bob receives $B_a$.
  \item Alice and Bob agree to discard all instances in which they happened to measure along different axes, as well as instances in which measurements fails because of imperfect quantum efficiency of the detectors. Then the remaining instances can be used to generate a private key $K_{a,b}$.
\end{enumerate}

We re-introduce the basic BBM92 protocol in an abstract way with more technical details as Figure \ref{BBM92} illustrates.

Now, $M[q_a;B_a]=\sum_{i=0}^{2n-1}M[q_a;K_a]_i$ denotes the Alice's measurement operation of $q_a$, and $\circledS_{M[q_a;B_a]}=\sum_{i=0}^{2n-1}\circledS_{M[q_a;B_a]_i}$ denotes the responding shadow constant; $M[q_b;B_b]=\sum_{i=0}^{2n-1}M[q_b;B_b]_i$ denotes the Bob's measurement operation of $q_b$, and $\circledS_{M[q_b;B_b]}=\sum_{i=0}^{2n-1}\circledS_{M[q_b;B_n]_i}$ denotes the responding shadow constant. Alice sends $q_b$ to Bob through the quantum channel $Q$ by quantum communicating action $send_{Q}(q_b)$ and Bob receives $q_b$ through $Q$ by quantum communicating action $receive_{Q}(q_b)$. Bob sends $B_b$ to Alice through the public channel $P$ by classical communicating action $send_{P}(B_b)$ and Alice receives $B_b$ through channel $P$ by classical communicating action $receive_{P}(B_b)$, and the same as $send_{P}(B_a)$ and $receive_{P}(B_a)$. Alice and Bob generate the private key $K_{a,b}$ by a classical comparison action $cmp(K_{a,b},B_a,B_b)$. Let Alice and Bob be a system $AB$ and let interactions between Alice and Bob be internal actions. $AB$ receives external input $D_i$ through channel $A$ by communicating action $receive_A(D_i)$ and sends results $D_o$ through channel $B$ by communicating action $send_B(D_o)$.

Then the state transition of Alice can be described by probabilistic quantum process algebra as follows.

\begin{eqnarray}
&&A=\sum_{D_i\in \Delta_i}receive_A(D_i)\cdot A_1\nonumber\\
&&A_1=send_Q(q_b)\cdot A_2\nonumber\\
&&A_2=\boxplus_{\frac{1}{2n},i=0}^{2n-1}M[q_a;B_a]_i\cdot A_3\nonumber\\
&&A_3=\boxplus_{\frac{1}{2n},i=0}^{2n-1}\circledS_{M[q_b;B_b]_i}\cdot A_4\nonumber\\
&&A_4=receive_P(B_b)\cdot A_5\nonumber\\
&&A_5=send_P(B_a)\cdot A_6\nonumber\\
&&A_6=cmp(K_{a,b},B_a,B_b)\cdot A\nonumber
\end{eqnarray}

where $\Delta_i$ is the collection of the input data.

And the state transition of Bob can be described by probabilistic quantum process algebra as follows.

\begin{eqnarray}
&&B=receive_Q(q_b)\cdot B_1\nonumber\\
&&B_1=\boxplus_{\frac{1}{2n},i=0}^{2n-1}\circledS_{M[q_a;B_a]_i}\cdot B_2\nonumber\\
&&B_2=\boxplus_{\frac{1}{2n},i=0}^{2n-1}M[q_b;B_b]_i\cdot B_3\nonumber\\
&&B_3=send_P(B_b)\cdot B_4\nonumber\\
&&B_4=receive_P(B_a)\cdot B_5\nonumber\\
&&B_5=cmp(K_{a,b},B_a,B_b)\cdot B_6\nonumber\\
&&B_6=\sum_{D_o\in\Delta_o}send_B(D_o)\cdot B\nonumber
\end{eqnarray}

where $\Delta_o$ is the collection of the output data.

The send action and receive action of the same data through the same channel can communicate each other, otherwise, a deadlock $\delta$ will be caused. The quantum measurement and its shadow constant pair will lead entanglement occur, otherwise, a deadlock $\delta$ will occur. We define the following communication functions.

\begin{eqnarray}
&&\gamma(send_Q(q_b),receive_Q(q_b))\triangleq c_Q(q_b)\nonumber\\
&&\gamma(send_P(B_b),receive_P(B_b))\triangleq c_P(B_b)\nonumber\\
&&\gamma(send_P(B_a),receive_P(B_a))\triangleq c_P(B_a)\nonumber
\end{eqnarray}

Let $A$ and $B$ in parallel, then the system $AB$ can be represented by the following process term.

$$\tau_I(\partial_H(\Theta(A\between B)))$$

where $H=\{send_Q(q_b),receive_Q(q_b),send_P(B_b),receive_P(B_b),send_P(B_a),receive_P(B_a),\\ \boxplus_{\frac{1}{2n},i=0}^{2n-1}M[q_a;B_a]_i, \boxplus_{\frac{1}{2n},i=0}^{2n-1}\circledS_{M[q_a;B_a]_i}, \boxplus_{\frac{1}{2n},i=0}^{2n-1}M[q_b;B_b]_i, \boxplus_{\frac{1}{2n},i=0}^{2n-1}\circledS_{M[q_b;B_b]_i}\}$

and $I=\{c_Q(q_b), c_P(B_b), c_P(B_a), M[q_a;B_a], M[q_b;B_b],\\ cmp(K_{a,b},B_a,B_b)\}$.

Then we get the following conclusion.

\begin{theorem}
The basic BBM92 protocol $\tau_I(\partial_H(\Theta(A\between B)))$ exhibits desired external behaviors.
\end{theorem}

\begin{proof}
We can get $\tau_I(\partial_H(\Theta(A\between B)))=\sum_{D_i\in \Delta_i}\sum_{D_o\in\Delta_o}receive_A(D_i)\leftmerge send_B(D_o)\leftmerge \tau_I(\partial_H(\Theta(A\between B)))$.
So, the basic BBM92 protocol $\tau_I(\partial_H(\Theta(A\between B)))$ exhibits desired external behaviors.
\end{proof}

\subsection{Verification of SARG04 Protocol}\label{VSARG046}

The famous SARG04 protocol\cite{SARG04} is a quantum key distribution protocol, in which quantum information and classical information are mixed. We take an example of the SARG04 protocol to illustrate the usage of probabilistic quantum process algebra in verification of quantum protocols.

\begin{figure}
  \centering
  \includegraphics{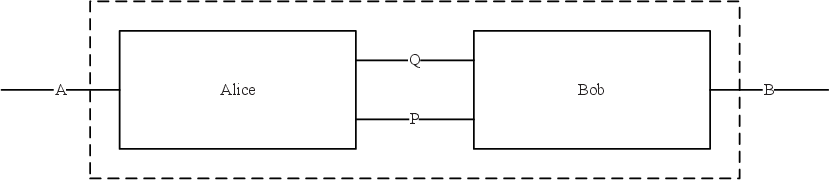}
  \caption{The SARG04 protocol.}
  \label{SARG04}
\end{figure}

The SARG04 protocol is used to create a private key between two parities, Alice and Bob. SARG04 is a protocol of quantum key distribution (QKD) which refines the BB84 protocol against PNS (Photon Number Splitting) attacks. The main innovations are encoding bits in nonorthogonal states and the classical sifting procedure. Firstly, we introduce the basic SARG04 protocol briefly, which is illustrated in Figure \ref{SARG04}.

\begin{enumerate}
  \item Alice create a string of bits with size $n$ randomly, denoted as $K_a$.
  \item Alice generates a string of qubits $q$ with size $n$, and the $i$th qubit of $q$ has four nonorthogonal states, it is $|\pm x\rangle$ if $K_a=0$; it is $|\pm z\rangle$ if $K_a=1$. And she records the corresponding one of the four pairs of nonorthogonal states into $B_a$ with size $2n$.
  \item Alice sends $q$ to Bob through a quantum channel $Q$ between Alice and Bob.
  \item Alice sends $B_a$ through a public channel $P$.
  \item Bob measures each qubit of $q$ $\sigma_x$ or $\sigma_z$. And he records the unambiguous discriminations into $K_b$ with a raw size $n/4$, and the unambiguous discrimination information into $B_b$ with size $n$.
  \item Bob sends $B_b$ to Alice through the public channel $P$.
  \item Alice and Bob determine that at which position the bit should be remained. Then the remaining bits of $K_a$ and $K_b$ is the private key $K_{a,b}$.
\end{enumerate}

We re-introduce the basic SARG04 protocol in an abstract way with more technical details as Figure \ref{SARG04} illustrates.

Now, we assume a special measurement operation $Rand[q;K_a]=\sum^{2n-1}_{i=0}Rand[q;K_a]_i$ which create a string of $n$ random bits $K_a$ from the $q$ quantum system. $M[q;K_b]=\sum^{2n-1}_{i=0}M[q;K_b]_i$ denotes the Bob's measurement operation of $q$. The generation of $n$ qubits $q$ through a unitary operator $Set_{K_a}[q]$. Alice sends $q$ to Bob through the quantum channel $Q$ by quantum communicating action $send_{Q}(q)$ and Bob receives $q$ through $Q$ by quantum communicating action $receive_{Q}(q)$. Bob sends $B_b$ to Alice through the public channel $P$ by classical communicating action $send_{P}(B_b)$ and Alice receives $B_b$ through channel $P$ by classical communicating action $receive_{P}(B_b)$, and the same as $send_{P}(B_a)$ and $receive_{P}(B_a)$. Alice and Bob generate the private key $K_{a,b}$ by a classical comparison action $cmp(K_{a,b},K_a,K_b,B_a,B_b)$. Let Alice and Bob be a system $AB$ and let interactions between Alice and Bob be internal actions. $AB$ receives external input $D_i$ through channel $A$ by communicating action $receive_A(D_i)$ and sends results $D_o$ through channel $B$ by communicating action $send_B(D_o)$.

Then the state transition of Alice can be described by probabilistic quantum process algebra as follows.

\begin{eqnarray}
&&A=\sum_{D_i\in \Delta_i}receive_A(D_i)\cdot A_1\nonumber\\
&&A_1=\boxplus_{\frac{1}{2n},i=0}^{2n-1}Rand[q;K_a]_i\cdot A_2\nonumber\\
&&A_2=Set_{K_a}[q]\cdot A_3\nonumber\\
&&A_3=send_Q(q)\cdot A_4\nonumber\\
&&A_4=send_P(B_a)\cdot A_5\nonumber\\
&&A_5=receive_P(B_b)\cdot A_6\nonumber\\
&&A_6=cmp(K_{a,b},K_a,K_b,B_a,B_b)\cdot A\nonumber
\end{eqnarray}

where $\Delta_i$ is the collection of the input data.

And the state transition of Bob can be described by probabilistic quantum process algebra as follows.

\begin{eqnarray}
&&B=receive_Q(q)\cdot B_1\nonumber\\
&&B_1=receive_P(B_a)\cdot B_2\nonumber\\
&&B_2=\boxplus_{\frac{1}{2n},i=0}^{2n-1}M[q;K_b]_i\cdot B_3\nonumber\\
&&B_3=send_P(B_b)\cdot B_4\nonumber\\
&&B_4=cmp(K_{a,b},K_a,K_b,B_a,B_b)\cdot B_5\nonumber\\
&&B_5=\sum_{D_o\in\Delta_o}send_B(D_o)\cdot B\nonumber
\end{eqnarray}

where $\Delta_o$ is the collection of the output data.

The send action and receive action of the same data through the same channel can communicate each other, otherwise, a deadlock $\delta$ will be caused. We define the following communication functions.

\begin{eqnarray}
&&\gamma(send_Q(q),receive_Q(q))\triangleq c_Q(q)\nonumber\\
&&\gamma(send_P(B_b),receive_P(B_b))\triangleq c_P(B_b)\nonumber\\
&&\gamma(send_P(B_a),receive_P(B_a))\triangleq c_P(B_a)\nonumber
\end{eqnarray}

Let $A$ and $B$ in parallel, then the system $AB$ can be represented by the following process term.

$$\tau_I(\partial_H(\Theta(A\between B)))$$

where $H=\{send_Q(q),receive_Q(q),send_P(B_b),receive_P(B_b),send_P(B_a),receive_P(B_a)\}$ and $I=\{\boxplus_{\frac{1}{2n},i=0}^{2n-1}Rand[q;K_a]_i, Set_{K_a}[q], \boxplus_{\frac{1}{2n},i=0}^{2n-1}M[q;K_b]_i, c_Q(q), c_P(B_b),\\ c_P(B_a), cmp(K_{a,b},K_a,K_b,B_a,B_b)\}$.

Then we get the following conclusion.

\begin{theorem}
The basic SARG04 protocol $\tau_I(\partial_H(\Theta(A\between B)))$ exhibits desired external behaviors.
\end{theorem}

\begin{proof}
We can get $\tau_I(\partial_H(\Theta(A\between B)))=\sum_{D_i\in \Delta_i}\sum_{D_o\in\Delta_o}receive_A(D_i)\leftmerge send_B(D_o)\leftmerge \tau_I(\partial_H(\Theta(A\between B)))$.
So, the basic SARG04 protocol $\tau_I(\partial_H(\Theta(A\between B)))$ exhibits desired external behaviors.
\end{proof}

\subsection{Verification of COW Protocol}\label{VCOW6}

The famous COW protocol\cite{COW} is a quantum key distribution protocol, in which quantum information and classical information are mixed. We take an example of the COW protocol to illustrate the usage of probabilistic quantum process algebra in verification of quantum protocols.

\begin{figure}
  \centering
  \includegraphics{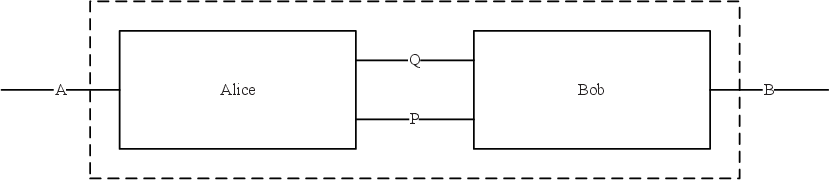}
  \caption{The COW protocol.}
  \label{COW}
\end{figure}

The COW protocol is used to create a private key between two parities, Alice and Bob. COW is a protocol of quantum key distribution (QKD) which is practical. Firstly, we introduce the basic COW protocol briefly, which is illustrated in Figure \ref{COW}.

\begin{enumerate}
  \item Alice generates a string of qubits $q$ with size $n$, and the $i$th qubit of $q$ is "0" with probability $\frac{1-f}{2}$, "1" with probability $\frac{1-f}{2}$ and the decoy sequence with probability $f$.
  \item Alice sends $q$ to Bob through a quantum channel $Q$ between Alice and Bob.
  \item Alice sends $A$ of the items corresponding to a decoy sequence through a public channel $P$.
  \item Bob removes all the detections at times $2A-1$ and $2A$ from his raw key and looks whether detector $D_{2M}$ has ever fired at time $2A$.
  \item Bob sends $B$ of the times $2A+1$ in which he had a detector in $D_{2M}$ to Alice through the public channel $P$.
  \item Alice receives $B$ and verifies if some of these items corresponding to a bit sequence "1,0".
  \item Bob sends $C$ of the items that he has detected through the public channel $P$.
  \item Alice and Bob run error correction and privacy amplification on these bits, and the private key $K_{a,b}$ is established.
\end{enumerate}

We re-introduce the basic COW protocol in an abstract way with more technical details as Figure \ref{COW} illustrates.

Now, we assume The generation of $n$ qubits $q$ through a unitary operator $Set[q]$. $M[q]=\sum^{2n-1}_{i=0}M[q]_i$ denotes the Bob's measurement operation of $q$.  Alice sends $q$ to Bob through the quantum channel $Q$ by quantum communicating action $send_{Q}(q)$ and Bob receives $q$ through $Q$ by quantum communicating action $receive_{Q}(q)$. Alice sends $A$ to Alice through the public channel $P$ by classical communicating action $send_{P}(A)$ and Alice receives $A$ through channel $P$ by classical communicating action $receive_{P}(A)$, and the same as $send_{P}(B)$ and $receive_{P}(B)$, and $send_{P}(C)$ and $receive_{P}(C)$. Alice and Bob generate the private key $K_{a,b}$ by a classical comparison action $cmp(K_{a,b})$. Let Alice and Bob be a system $AB$ and let interactions between Alice and Bob be internal actions. $AB$ receives external input $D_i$ through channel $A$ by communicating action $receive_A(D_i)$ and sends results $D_o$ through channel $B$ by communicating action $send_B(D_o)$.

Then the state transition of Alice can be described by probabilistic quantum process algebra as follows.

\begin{eqnarray}
&&A=\sum_{D_i\in \Delta_i}receive_A(D_i)\cdot A_1\nonumber\\
&&A_1=Set[q]\cdot A_2\nonumber\\
&&A_2=send_Q(q)\cdot A_3\nonumber\\
&&A_3=send_P(A)\cdot A_4\nonumber\\
&&A_4=receive_P(B)\cdot A_5\nonumber\\
&&A_5=receive_P(C)\cdot A_6\nonumber\\
&&A_6=cmp(K_{a,b})\cdot A\nonumber
\end{eqnarray}

where $\Delta_i$ is the collection of the input data.

And the state transition of Bob can be described by probabilistic quantum process algebra as follows.

\begin{eqnarray}
&&B=receive_Q(q)\cdot B_1\nonumber\\
&&B_1=receive_P(A)\cdot B_2\nonumber\\
&&B_2=\boxplus_{\frac{1}{2n},i=0}^{2n-1}M[q]_i\cdot B_3\nonumber\\
&&B_3=send_P(B)\cdot B_4\nonumber\\
&&B_4=send_P(C)\cdot B_5\nonumber\\
&&B_5=cmp(K_{a,b})\cdot B_6\nonumber\\
&&B_6=\sum_{D_o\in\Delta_o}send_B(D_o)\cdot B\nonumber
\end{eqnarray}

where $\Delta_o$ is the collection of the output data.

The send action and receive action of the same data through the same channel can communicate each other, otherwise, a deadlock $\delta$ will be caused. We define the following communication functions.

\begin{eqnarray}
&&\gamma(send_Q(q),receive_Q(q))\triangleq c_Q(q)\nonumber\\
&&\gamma(send_P(A),receive_P(A))\triangleq c_P(A)\nonumber\\
&&\gamma(send_P(B),receive_P(B))\triangleq c_P(B)\nonumber\\
&&\gamma(send_P(C),receive_P(C))\triangleq c_P(C)\nonumber
\end{eqnarray}

Let $A$ and $B$ in parallel, then the system $AB$ can be represented by the following process term.

$$\tau_I(\partial_H(\Theta(A\between B)))$$

where $H=\{send_Q(q),receive_Q(q),send_P(A),receive_P(A),send_P(B),receive_P(B),send_P(C),\\receive_P(C)\}$ and $I=\{Set[q], \boxplus_{\frac{1}{2n},i=0}^{2n-1}M[q]_i, c_Q(q), c_P(A),\\ c_P(B),c_P(C), cmp(K_{a,b})\}$.

Then we get the following conclusion.

\begin{theorem}
The basic COW protocol $\tau_I(\partial_H(\Theta(A\between B)))$ exhibits desired external behaviors.
\end{theorem}

\begin{proof}
We can get $\tau_I(\partial_H(\Theta(A\between B)))=\sum_{D_i\in \Delta_i}\sum_{D_o\in\Delta_o}receive_A(D_i)\leftmerge send_B(D_o)\leftmerge \tau_I(\partial_H(\Theta(A\between B)))$.
So, the basic COW protocol $\tau_I(\partial_H(\Theta(A\between B)))$ exhibits desired external behaviors.
\end{proof}

\subsection{Verification of SSP Protocol}\label{VSSP6}

The famous SSP protocol\cite{SSP} is a quantum key distribution protocol, in which quantum information and classical information are mixed. We take an example of the SSP protocol to illustrate the usage of probabilistic quantum process algebra in verification of quantum protocols.

\begin{figure}
  \centering
  \includegraphics{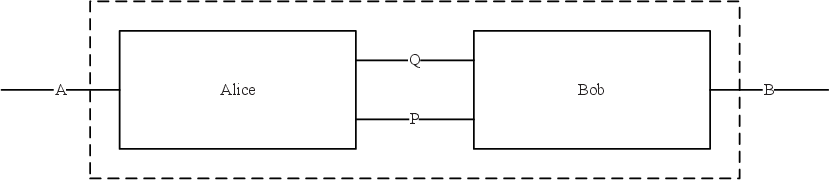}
  \caption{The SSP protocol.}
  \label{SSP}
\end{figure}

The SSP protocol is used to create a private key between two parities, Alice and Bob. SSP is a protocol of quantum key distribution (QKD) which uses six states. Firstly, we introduce the basic SSP protocol briefly, which is illustrated in Figure \ref{SSP}.

\begin{enumerate}
  \item Alice create two string of bits with size $n$ randomly, denoted as $B_a$ and $K_a$.
  \item Alice generates a string of qubits $q$ with size $n$, and the $i$th qubit in $q$ is one of the six states $\pm x$, $\pm y$ and $\pm z$.
  \item Alice sends $q$ to Bob through a quantum channel $Q$ between Alice and Bob.
  \item Bob receives $q$ and randomly generates a string of bits $B_b$ with size $n$.
  \item Bob measures each qubit of $q$ according to a basis by bits of $B_b$, i.e., $x$, $y$ or $z$ basis. And the measurement results would be $K_b$, which is also with size $n$.
  \item Bob sends his measurement bases $B_b$ to Alice through a public channel $P$.
  \item Once receiving $B_b$, Alice sends her bases $B_a$ to Bob through channel $P$, and Bob receives $B_a$.
  \item Alice and Bob determine that at which position the bit strings $B_a$ and $B_b$ are equal, and they discard the mismatched bits of $B_a$ and $B_b$. Then the remaining bits of $K_a$ and $K_b$, denoted as $K_a'$ and $K_b'$ with $K_{a,b}=K_a'=K_b'$.
\end{enumerate}

We re-introduce the basic SSP protocol in an abstract way with more technical details as Figure \ref{SSP} illustrates.

Now, we assume a special measurement operation $Rand[q;B_a]=\sum^{2n-1}_{i=0}Rand[q;B_a]_i$ which create a string of $n$ random bits $B_a$ from the $q$ quantum system, and the same as $Rand[q;K_a]=\sum^{2n-1}_{i=0}Rand[q;K_a]_i$, $Rand[q';B_b]=\sum^{2n-1}_{i=0}Rand[q';B_b]_i$. $M[q;K_b]=\sum^{2n-1}_{i=0}M[q;K_b]_i$ denotes the Bob's measurement operation of $q$. The generation of $n$ qubits $q$ through two unitary operators $Set_{K_a}[q]$ and $H_{B_a}[q]$. Alice sends $q$ to Bob through the quantum channel $Q$ by quantum communicating action $send_{Q}(q)$ and Bob receives $q$ through $Q$ by quantum communicating action $receive_{Q}(q)$. Bob sends $B_b$ to Alice through the public channel $P$ by classical communicating action $send_{P}(B_b)$ and Alice receives $B_b$ through channel $P$ by classical communicating action $receive_{P}(B_b)$, and the same as $send_{P}(B_a)$ and $receive_{P}(B_a)$. Alice and Bob generate the private key $K_{a,b}$ by a classical comparison action $cmp(K_{a,b},K_a,K_b,B_a,B_b)$. Let Alice and Bob be a system $AB$ and let interactions between Alice and Bob be internal actions. $AB$ receives external input $D_i$ through channel $A$ by communicating action $receive_A(D_i)$ and sends results $D_o$ through channel $B$ by communicating action $send_B(D_o)$.

Then the state transition of Alice can be described by probabilistic quantum process algebra as follows.

\begin{eqnarray}
&&A=\sum_{D_i\in \Delta_i}receive_A(D_i)\cdot A_1\nonumber\\
&&A_1=\boxplus_{\frac{1}{2n},i=0}^{2n-1}Rand[q;B_a]_i\cdot A_2\nonumber\\
&&A_2=\boxplus_{\frac{1}{2n},i=0}^{2n-1}Rand[q;K_a]_i\cdot A_3\nonumber\\
&&A_3=Set_{K_a}[q]\cdot A_4\nonumber\\
&&A_4=H_{B_a}[q]\cdot A_5\nonumber\\
&&A_5=send_Q(q)\cdot A_6\nonumber\\
&&A_6=receive_P(B_b)\cdot A_7\nonumber\\
&&A_7=send_P(B_a)\cdot A_8\nonumber\\
&&A_8=cmp(K_{a,b},K_a,K_b,B_a,B_b)\cdot A\nonumber
\end{eqnarray}

where $\Delta_i$ is the collection of the input data.

And the state transition of Bob can be described by probabilistic quantum process algebra as follows.

\begin{eqnarray}
&&B=receive_Q(q)\cdot B_1\nonumber\\
&&B_1=\boxplus_{\frac{1}{2n},i=0}^{2n-1}Rand[q';B_b]_i\cdot B_2\nonumber\\
&&B_2=\boxplus_{\frac{1}{2n},i=0}^{2n-1}M[q;K_b]_i\cdot B_3\nonumber\\
&&B_3=send_P(B_b)\cdot B_4\nonumber\\
&&B_4=receive_P(B_a)\cdot B_5\nonumber\\
&&B_5=cmp(K_{a,b},K_a,K_b,B_a,B_b)\cdot B_6\nonumber\\
&&B_6=\sum_{D_o\in\Delta_o}send_B(D_o)\cdot B\nonumber
\end{eqnarray}

where $\Delta_o$ is the collection of the output data.

The send action and receive action of the same data through the same channel can communicate each other, otherwise, a deadlock $\delta$ will be caused. We define the following communication functions.

\begin{eqnarray}
&&\gamma(send_Q(q),receive_Q(q))\triangleq c_Q(q)\nonumber\\
&&\gamma(send_P(B_b),receive_P(B_b))\triangleq c_P(B_b)\nonumber\\
&&\gamma(send_P(B_a),receive_P(B_a))\triangleq c_P(B_a)\nonumber
\end{eqnarray}

Let $A$ and $B$ in parallel, then the system $AB$ can be represented by the following process term.

$$\tau_I(\partial_H(\Theta(A\between B)))$$

where $H=\{send_Q(q),receive_Q(q),send_P(B_b),receive_P(B_b),send_P(B_a),receive_P(B_a)\}$ and $I=\{\boxplus_{\frac{1}{2n},i=0}^{2n-1}Rand[q;B_a]_i, \boxplus_{\frac{1}{2n},i=0}^{2n-1}Rand[q;K_a]_i, Set_{K_a}[q], \\ H_{B_a}[q], \boxplus_{\frac{1}{2n},i=0}^{2n-1}Rand[q';B_b]_i, \boxplus_{\frac{1}{2n},i=0}^{2n-1}M[q;K_b]_i, c_Q(q), c_P(B_b),\\ c_P(B_a), cmp(K_{a,b},K_a,K_b,B_a,B_b)\}$.

Then we get the following conclusion.

\begin{theorem}
The basic SSP protocol $\tau_I(\partial_H(\Theta(A\between B)))$ exhibits desired external behaviors.
\end{theorem}

\begin{proof}
We can get $\tau_I(\partial_H(\Theta(A\between B)))=\sum_{D_i\in \Delta_i}\sum_{D_o\in\Delta_o}receive_A(D_i)\leftmerge send_B(D_o)\leftmerge \tau_I(\partial_H(\Theta(A\between B)))$.
So, the basic SSP protocol $\tau_I(\partial_H(\Theta(A\between B)))$ exhibits desired external behaviors.
\end{proof}

\subsection{Verification of S09 Protocol}\label{VS096}

The famous S09 protocol\cite{S09} is a quantum key distribution protocol, in which quantum information and classical information are mixed. We take an example of the S09 protocol to illustrate the usage of probabilistic quantum process algebra in verification of quantum protocols.

\begin{figure}
  \centering
  \includegraphics{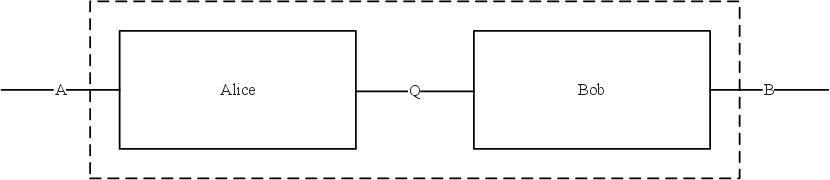}
  \caption{The S09 protocol.}
  \label{S09}
\end{figure}

The S09 protocol is used to create a private key between two parities, Alice and Bob, by use of pure quantum information. Firstly, we introduce the basic S09 protocol briefly, which is illustrated in Figure \ref{S09}.

\begin{enumerate}
  \item Alice create two string of bits with size $n$ randomly, denoted as $B_a$ and $K_a$.
  \item Alice generates a string of qubits $q$ with size $n$, and the $i$th qubit in $q$ is $|x_y\rangle$, where $x$ is the $i$th bit of $B_a$ and $y$ is the $i$th bit of $K_a$.
  \item Alice sends $q$ to Bob through a quantum channel $Q$ between Alice and Bob.
  \item Bob receives $q$ and randomly generates a string of bits $B_b$ with size $n$.
  \item Bob measures each qubit of $q$ according to a basis by bits of $B_b$. After the measurement, the state of $q$ evolves into $q'$.
  \item Bob sends $q'$ to Alice through the quantum channel $Q$.
  \item Alice measures each qubit of $q'$ to generate a string $C$.
  \item Alice sums $C_i\oplus B_{a_i}$ to get the private key $K_{a,b}=B_b$.
\end{enumerate}

We re-introduce the basic S09 protocol in an abstract way with more technical details as Figure \ref{S09} illustrates.

Now, we assume a special measurement operation $Rand[q;B_a]=\sum^{2n-1}_{i=0}Rand[q;B_a]_i$ which create a string of $n$ random bits $B_a$ from the $q$ quantum system, and the same as $Rand[q;K_a]=\sum^{2n-1}_{i=0}Rand[q;K_a]_i$, $Rand[q';B_b]=\sum^{2n-1}_{i=0}Rand[q';B_b]_i$. $M[q;B_b]=\sum^{2n-1}_{i=0}M[q;B_b]_i$ denotes the Bob's measurement operation of $q$, and the same as $M[q';C]=\sum^{2n-1}_{i=0}Rand[q';C]_i$. The generation of $n$ qubits $q$ through two unitary operators $Set_{K_a}[q]$ and $H_{B_a}[q]$. Alice sends $q$ to Bob through the quantum channel $Q$ by quantum communicating action $send_{Q}(q)$ and Bob receives $q$ through $Q$ by quantum communicating action $receive_{Q}(q)$, and the same as $send_{Q}(q')$ and $receive_{Q}(q')$. Alice and Bob generate the private key $K_{a,b}$ by a classical comparison action $cmp(K_{a,b},B_b)$. We omit the sum classical $\oplus$ actions without of loss of generality. Let Alice and Bob be a system $AB$ and let interactions between Alice and Bob be internal actions. $AB$ receives external input $D_i$ through channel $A$ by communicating action $receive_A(D_i)$ and sends results $D_o$ through channel $B$ by communicating action $send_B(D_o)$.

Then the state transition of Alice can be described by probabilistic quantum process algebra as follows.

\begin{eqnarray}
&&A=\sum_{D_i\in \Delta_i}receive_A(D_i)\cdot A_1\nonumber\\
&&A_1=\boxplus_{\frac{1}{2n},i=0}^{2n-1}Rand[q;B_a]_i\cdot A_2\nonumber\\
&&A_2=\boxplus_{\frac{1}{2n},i=0}^{2n-1}Rand[q;K_a]_i\cdot A_3\nonumber\\
&&A_3=Set_{K_a}[q]\cdot A_4\nonumber\\
&&A_4=H_{B_a}[q]\cdot A_5\nonumber\\
&&A_5=send_Q(q)\cdot A_6\nonumber\\
&&A_6=receive_Q(q')\cdot A_{7}\nonumber\\
&&A_7=\boxplus_{\frac{1}{2n},i=0}^{2n-1}M[q';C]_i\cdot A_8\nonumber\\
&&A_{8}=cmp(K_{a,b},B_b)\cdot A\nonumber
\end{eqnarray}

where $\Delta_i$ is the collection of the input data.

And the state transition of Bob can be described by probabilistic quantum process algebra as follows.

\begin{eqnarray}
&&B=receive_Q(q)\cdot B_1\nonumber\\
&&B_1=\boxplus_{\frac{1}{2n},i=0}^{2n-1}Rand[q';B_b]_i\cdot B_2\nonumber\\
&&B_2=\boxplus_{\frac{1}{2n},i=0}^{2n-1}M[q;B_b]_i\cdot B_3\nonumber\\
&&B_3=send_Q(q')\cdot B_4\nonumber\\
&&B_4=cmp(K_{a,b},B_b)\cdot B_{5}\nonumber\\
&&B_{5}=\sum_{D_o\in\Delta_o}send_B(D_o)\cdot B\nonumber
\end{eqnarray}

where $\Delta_o$ is the collection of the output data.

The send action and receive action of the same data through the same channel can communicate each other, otherwise, a deadlock $\delta$ will be caused. We define the following communication functions.

\begin{eqnarray}
&&\gamma(send_Q(q),receive_Q(q))\triangleq c_Q(q)\nonumber\\
&&\gamma(send_Q(q'),receive_Q(q'))\triangleq c_Q(q')\nonumber
\end{eqnarray}

Let $A$ and $B$ in parallel, then the system $AB$ can be represented by the following process term.

$$\tau_I(\partial_H(\Theta(A\between B)))$$

where $H=\{send_Q(q),receive_Q(q),send_Q(q'),receive_Q(q')\}$ and $I=\{\boxplus_{\frac{1}{2n},i=0}^{2n-1}Rand[q;B_a]_i, \\ \boxplus_{\frac{1}{2n},i=0}^{2n-1}Rand[q;K_a]_i, Set_{K_a}[q], H_{B_a}[q], \boxplus_{\frac{1}{2n},i=0}^{2n-1}Rand[q';B_b]_i, \boxplus_{\frac{1}{2n},i=0}^{2n-1}M[q;B_b]_i,  \\ \boxplus_{\frac{1}{2n},i=0}^{2n-1}Rand[q';C]_i, c_Q(q), c_Q(q'), cmp(K_{a,b},B_b)\}$.

Then we get the following conclusion.

\begin{theorem}
The basic S09 protocol $\tau_I(\partial_H(\Theta(A\between B)))$ exhibits desired external behaviors.
\end{theorem}

\begin{proof}
We can get $\tau_I(\partial_H(\Theta(A\between B)))=\sum_{D_i\in \Delta_i}\sum_{D_o\in\Delta_o}receive_A(D_i)\leftmerge send_B(D_o)\leftmerge \tau_I(\partial_H(\Theta(A\between B)))$.
So, the basic S09 protocol $\tau_I(\partial_H(\Theta(A\between B)))$ exhibits desired external behaviors.
\end{proof}

\subsection{Verification of KMB09 Protocol}\label{VKMB096}

The famous KMB09 protocol\cite{KMB09} is a quantum key distribution protocol, in which quantum information and classical information are mixed. We take an example of the KMB09 protocol to illustrate the usage of probabilistic quantum process algebra in verification of quantum protocols.

\begin{figure}
  \centering
  \includegraphics{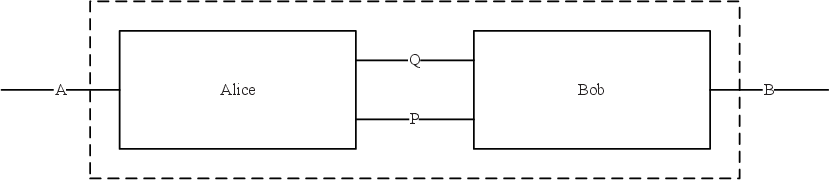}
  \caption{The KMB09 protocol.}
  \label{KMB09}
\end{figure}

The KMB09 protocol is used to create a private key between two parities, Alice and Bob. KMB09 is a protocol of quantum key distribution (QKD) which refines the BB84 protocol against PNS (Photon Number Splitting) attacks. The main innovations are encoding bits in nonorthogonal states and the classical sifting procedure. Firstly, we introduce the basic KMB09 protocol briefly, which is illustrated in Figure \ref{KMB09}.

\begin{enumerate}
  \item Alice create a string of bits with size $n$ randomly, denoted as $K_a$, and randomly assigns each bit value a random index $i=1,2,...,N$ into $B_a$.
  \item Alice generates a string of qubits $q$ with size $n$, accordingly either in $|e_i\rangle$ or $|f_i\rangle$.
  \item Alice sends $q$ to Bob through a quantum channel $Q$ between Alice and Bob.
  \item Alice sends $B_a$ through a public channel $P$.
  \item Bob measures each qubit of $q$ by randomly switching the measurement basis between $e$ and $f$. And he records the unambiguous discriminations into $K_b$, and the unambiguous discrimination information into $B_b$.
  \item Bob sends $B_b$ to Alice through the public channel $P$.
  \item Alice and Bob determine that at which position the bit should be remained. Then the remaining bits of $K_a$ and $K_b$ is the private key $K_{a,b}$.
\end{enumerate}

We re-introduce the basic KMB09 protocol in an abstract way with more technical details as Figure \ref{KMB09} illustrates.

Now, we assume a special measurement operation $Rand[q;K_a]=\sum^{2n-1}_{i=0}Rand[q;K_a]_i$ which create a string of $n$ random bits $K_a$ from the $q$ quantum system. $M[q;K_b]=\sum^{2n-1}_{i=0}M[q;K_b]_i$ denotes the Bob's measurement operation of $q$. The generation of $n$ qubits $q$ through a unitary operator $Set_{K_a}[q]$. Alice sends $q$ to Bob through the quantum channel $Q$ by quantum communicating action $send_{Q}(q)$ and Bob receives $q$ through $Q$ by quantum communicating action $receive_{Q}(q)$. Bob sends $B_b$ to Alice through the public channel $P$ by classical communicating action $send_{P}(B_b)$ and Alice receives $B_b$ through channel $P$ by classical communicating action $receive_{P}(B_b)$, and the same as $send_{P}(B_a)$ and $receive_{P}(B_a)$. Alice and Bob generate the private key $K_{a,b}$ by a classical comparison action $cmp(K_{a,b},K_a,K_b,B_a,B_b)$. Let Alice and Bob be a system $AB$ and let interactions between Alice and Bob be internal actions. $AB$ receives external input $D_i$ through channel $A$ by communicating action $receive_A(D_i)$ and sends results $D_o$ through channel $B$ by communicating action $send_B(D_o)$.

Then the state transition of Alice can be described by probabilistic quantum process algebra as follows.

\begin{eqnarray}
&&A=\sum_{D_i\in \Delta_i}receive_A(D_i)\cdot A_1\nonumber\\
&&A_1=\boxplus_{\frac{1}{2n},i=0}^{2n-1}Rand[q;K_a]_i\cdot A_2\nonumber\\
&&A_2=Set_{K_a}[q]\cdot A_3\nonumber\\
&&A_3=send_Q(q)\cdot A_4\nonumber\\
&&A_4=send_P(B_a)\cdot A_5\nonumber\\
&&A_5=receive_P(B_b)\cdot A_6\nonumber\\
&&A_6=cmp(K_{a,b},K_a,K_b,B_a,B_b)\cdot A\nonumber
\end{eqnarray}

where $\Delta_i$ is the collection of the input data.

And the state transition of Bob can be described by probabilistic quantum process algebra as follows.

\begin{eqnarray}
&&B=receive_Q(q)\cdot B_1\nonumber\\
&&B_1=receive_P(B_a)\cdot B_2\nonumber\\
&&B_2=\boxplus_{\frac{1}{2n},i=0}^{2n-1}M[q;K_b]_i\cdot B_3\nonumber\\
&&B_3=send_P(B_b)\cdot B_4\nonumber\\
&&B_4=cmp(K_{a,b},K_a,K_b,B_a,B_b)\cdot B_5\nonumber\\
&&B_5=\sum_{D_o\in\Delta_o}send_B(D_o)\cdot B\nonumber
\end{eqnarray}

where $\Delta_o$ is the collection of the output data.

The send action and receive action of the same data through the same channel can communicate each other, otherwise, a deadlock $\delta$ will be caused. We define the following communication functions.

\begin{eqnarray}
&&\gamma(send_Q(q),receive_Q(q))\triangleq c_Q(q)\nonumber\\
&&\gamma(send_P(B_b),receive_P(B_b))\triangleq c_P(B_b)\nonumber\\
&&\gamma(send_P(B_a),receive_P(B_a))\triangleq c_P(B_a)\nonumber
\end{eqnarray}

Let $A$ and $B$ in parallel, then the system $AB$ can be represented by the following process term.

$$\tau_I(\partial_H(\Theta(A\between B)))$$

where $H=\{send_Q(q),receive_Q(q),send_P(B_b),receive_P(B_b),send_P(B_a),receive_P(B_a)\}$ and $I=\{\boxplus_{\frac{1}{2n},i=0}^{2n-1}Rand[q;K_a]_i, Set_{K_a}[q], \boxplus_{\frac{1}{2n},i=0}^{2n-1}M[q;K_b]_i, c_Q(q), c_P(B_b),\\ c_P(B_a), cmp(K_{a,b},K_a,K_b,B_a,B_b)\}$.

Then we get the following conclusion.

\begin{theorem}
The basic KMB09 protocol $\tau_I(\partial_H(\Theta(A\between B)))$ exhibits desired external behaviors.
\end{theorem}

\begin{proof}
We can get $\tau_I(\partial_H(\Theta(A\between B)))=\sum_{D_i\in \Delta_i}\sum_{D_o\in\Delta_o}receive_A(D_i)\leftmerge send_B(D_o)\leftmerge \tau_I(\partial_H(\Theta(A\between B)))$.
So, the basic KMB09 protocol $\tau_I(\partial_H(\Theta(A\between B)))$ exhibits desired external behaviors.
\end{proof}

\subsection{Verification of S13 Protocol}\label{VS136}

The famous S13 protocol\cite{S13} is a quantum key distribution protocol, in which quantum information and classical information are mixed. We take an example of the S13 protocol to illustrate the usage of probabilistic quantum process algebra in verification of quantum protocols.

\begin{figure}
  \centering
  \includegraphics{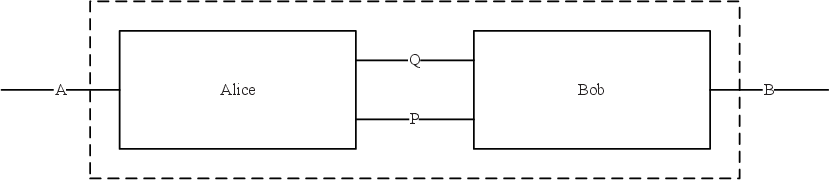}
  \caption{The S13 protocol.}
  \label{S13}
\end{figure}

The S13 protocol is used to create a private key between two parities, Alice and Bob. Firstly, we introduce the basic S13 protocol briefly, which is illustrated in Figure \ref{S13}.

\begin{enumerate}
  \item Alice create two string of bits with size $n$ randomly, denoted as $B_a$ and $K_a$.
  \item Alice generates a string of qubits $q$ with size $n$, and the $i$th qubit in $q$ is $|x_y\rangle$, where $x$ is the $i$th bit of $B_a$ and $y$ is the $i$th bit of $K_a$.
  \item Alice sends $q$ to Bob through a quantum channel $Q$ between Alice and Bob.
  \item Bob receives $q$ and randomly generates a string of bits $B_b$ with size $n$.
  \item Bob measures each qubit of $q$ according to a basis by bits of $B_b$. And the measurement results would be $K_b$, which is also with size $n$.
  \item Alice sends a random binary string $C$ to Bob through the public channel $P$.
  \item Alice sums $B_{a_i}\oplus C_i$ to obtain $T$ and generates other random string of binary values $J$. From the elements occupying a concrete position, $i$, of the preceding strings, Alice get the new states of $q'$, and sends it to Bob through the quantum channel $Q$.
  \item Bob sums $1\oplus B_{b_i}$ to obtain the string of binary basis $N$ and measures $q'$ according to these bases, and generating $D$.
  \item Alice sums $K_{a_i}\oplus J_i$ to obtain the binary string $Y$ and sends it to Bob through the public channel $P$.
  \item Bob encrypts $B_b$ to obtain $U$ and sends to Alice through the public channel $P$.
  \item Alice decrypts $U$ to obtain $B_b$. She sums $B_{a_i}\oplus B_{b_i}$ to obtain $L$ and sends $L$ to Bob through the public channel $P$.
  \item Bob sums $B_{b_i}\oplus L_i$ to get the private key $K_{a,b}$.
\end{enumerate}

We re-introduce the basic S13 protocol in an abstract way with more technical details as Figure \ref{S13} illustrates.

Now, we assume a special measurement operation $Rand[q;B_a]=\sum^{2n-1}_{i=0}Rand[q;B_a]_i$ which create a string of $n$ random bits $B_a$ from the $q$ quantum system, and the same as $Rand[q;K_a]=\sum^{2n-1}_{i=0}Rand[q;K_a]_i$, $Rand[q';B_b]=\sum^{2n-1}_{i=0}Rand[q';B_b]_i$. $M[q;K_b]=\sum^{2n-1}_{i=0}M[q;K_b]_i$ denotes the Bob's measurement operation of $q$, and the same as $M[q';D]=\sum^{2n-1}_{i=0}M[q';D]_i$. The generation of $n$ qubits $q$ through two unitary operators $Set_{K_a}[q]$ and $H_{B_a}[q]$, and the same as $Set_{T}[q']$. Alice sends $q$ to Bob through the quantum channel $Q$ by quantum communicating action $send_{Q}(q)$ and Bob receives $q$ through $Q$ by quantum communicating action $receive_{Q}(q)$, and the same as $send_{Q}(q')$ and $receive_{Q}(q')$. Bob sends $B_b$ to Alice through the public channel $P$ by classical communicating action $send_{P}(B_b)$ and Alice receives $B_b$ through channel $P$ by classical communicating action $receive_{P}(B_b)$, and the same as $send_{P}(B_a)$ and $receive_{P}(B_a)$, $send_{P}(C)$ and $receive_{P}(C)$, $send_{P}(Y)$ and $receive_{P}(Y)$, $send_{P}(U)$ and $receive_{P}(U)$, $send_{P}(L)$ and $receive_{P}(L)$. Alice and Bob generate the private key $K_{a,b}$ by a classical comparison action $cmp(K_{a,b},K_a,K_b,B_a,B_b)$. We omit the sum classical $\oplus$ actions without of loss of generality. Let Alice and Bob be a system $AB$ and let interactions between Alice and Bob be internal actions. $AB$ receives external input $D_i$ through channel $A$ by communicating action $receive_A(D_i)$ and sends results $D_o$ through channel $B$ by communicating action $send_B(D_o)$.

Then the state transition of Alice can be described by probabilistic quantum process algebra as follows.

\begin{eqnarray}
&&A=\sum_{D_i\in \Delta_i}receive_A(D_i)\cdot A_1\nonumber\\
&&A_1=\boxplus_{\frac{1}{2n},i=0}^{2n-1}Rand[q;B_a]_i\cdot A_2\nonumber\\
&&A_2=\boxplus_{\frac{1}{2n},i=0}^{2n-1}Rand[q;K_a]_i\cdot A_3\nonumber\\
&&A_3=Set_{K_a}[q]\cdot A_4\nonumber\\
&&A_4=H_{B_a}[q]\cdot A_5\nonumber\\
&&A_5=send_Q(q)\cdot A_6\nonumber\\
&&A_6=send_P(C)\cdot A_7\nonumber\\
&&A_7=send_Q(q')\cdot A_8\nonumber\\
&&A_8=send_P(Y)\cdot A_9\nonumber\\
&&A_9=receive_P(U)\cdot A_{10}\nonumber\\
&&A_{10}=send_P(L)\cdot A_{11}\nonumber\\
&&A_{11}=cmp(K_{a,b},K_a,K_b,B_a,B_b)\cdot A\nonumber
\end{eqnarray}

where $\Delta_i$ is the collection of the input data.

And the state transition of Bob can be described by probabilistic quantum process algebra as follows.

\begin{eqnarray}
&&B=receive_Q(q)\cdot B_1\nonumber\\
&&B_1=\boxplus_{\frac{1}{2n},i=0}^{2n-1}Rand[q';B_b]_i\cdot B_2\nonumber\\
&&B_2=\boxplus_{\frac{1}{2n},i=0}^{2n-1}M[q;K_b]_i\cdot B_3\nonumber\\
&&B_3=receive_P(C)\cdot B_4\nonumber\\
&&B_4=receive_Q(q')\cdot B_5\nonumber\\
&&B_5=\boxplus_{\frac{1}{2n},i=0}^{2n-1}M[q';D]_i\cdot B_6\nonumber\\
&&B_6=receive_P(Y)\cdot B_7\nonumber\\
&&B_7=send_P(U)\cdot B_8\nonumber\\
&&B_8=receive_P(L)\cdot B_9\nonumber\\
&&B_9=cmp(K_{a,b},K_a,K_b,B_a,B_b)\cdot B_{10}\nonumber\\
&&B_{10}=\sum_{D_o\in\Delta_o}send_B(D_o)\cdot B\nonumber
\end{eqnarray}

where $\Delta_o$ is the collection of the output data.

The send action and receive action of the same data through the same channel can communicate each other, otherwise, a deadlock $\delta$ will be caused. We define the following communication functions.

\begin{eqnarray}
&&\gamma(send_Q(q),receive_Q(q))\triangleq c_Q(q)\nonumber\\
&&\gamma(send_Q(q'),receive_Q(q'))\triangleq c_Q(q')\nonumber\\
&&\gamma(send_P(C),receive_P(C))\triangleq c_P(C)\nonumber\\
&&\gamma(send_P(Y),receive_P(Y))\triangleq c_P(Y)\nonumber\\
&&\gamma(send_P(U),receive_P(U))\triangleq c_P(U)\nonumber\\
&&\gamma(send_P(L),receive_P(L))\triangleq c_P(L)\nonumber
\end{eqnarray}

Let $A$ and $B$ in parallel, then the system $AB$ can be represented by the following process term.

$$\tau_I(\partial_H(\Theta(A\between B)))$$

where $H=\{send_Q(q),receive_Q(q),send_Q(q'),receive_Q(q'),send_P(C),receive_P(C),send_P(Y),\\receive_P(Y),send_P(U),receive_P(U),send_P(L),receive_P(L)\}$

 and $I=\{\boxplus_{\frac{1}{2n},i=0}^{2n-1}Rand[q;B_a]_i, \boxplus_{\frac{1}{2n},i=0}^{2n-1}Rand[q;K_a]_i, Set_{K_a}[q], \\H_{B_a}[q], \boxplus_{\frac{1}{2n},i=0}^{2n-1}Rand[q';B_b]_i, \boxplus_{\frac{1}{2n},i=0}^{2n-1}M[q;K_b]_i, \boxplus_{\frac{1}{2n},i=0}^{2n-1}M[q';D]_i, c_Q(q), c_P(C),\\c_Q(q'), c_P(Y), c_P(U), c_P(L), cmp(K_{a,b},K_a,K_b,B_a,B_b)\}$.

Then we get the following conclusion.

\begin{theorem}
The basic S13 protocol $\tau_I(\partial_H(\Theta(A\between B)))$ exhibits desired external behaviors.
\end{theorem}

\begin{proof}
We can get $\tau_I(\partial_H(\Theta(A\between B)))=\sum_{D_i\in \Delta_i}\sum_{D_o\in\Delta_o}receive_A(D_i)\leftmerge send_B(D_o)\leftmerge \tau_I(\partial_H(\Theta(A\between B)))$.
So, the basic S13 protocol $\tau_I(\partial_H(\Theta(A\between B)))$ exhibits desired external behaviors.
\end{proof}


\end{document}